\documentclass[twoside,b5paper,10pt,english]{book}
\usepackage{graphicx}
\usepackage{epigraph}
\usepackage{dcolumn}

\graphicspath{ {figures/} }
\usepackage{array}
\usepackage{watermark}
\usepackage{fancybox}
\usepackage[dvipsnames,usenames]{color}
\usepackage{babel}
\usepackage[Lenny]{fncychap}
\usepackage[pagestyles]{titlesec}
\usepackage{bookmark}
\usepackage{afterpage}
\usepackage{bm}
\usepackage{physics}
\usepackage{mathtools}
\usepackage[normalem]{ulem}

\usepackage{titlesec}
\titleclass{\part}{top}
\titleformat{\part}[display]
  {\huge\bfseries\centering}{\partname~\thepart}{20pt}{}
\titlespacing*{\part}{0pt}{40pt}{80pt}

\usepackage[b5paper,left=2.5cm,right=1.5cm,top=2.5cm]{geometry}
\usepackage[a4,center,pdflatex]{crop}
\usepackage{hyperref}
\hypersetup{draft}
\usepackage{amsmath,amssymb,amsfonts,bm,cancel,physics,nicefrac}
\usepackage{cite}
\usepackage{pdfpages}
\usepackage[font=footnotesize,labelfont=scriptsize]{caption,subfig}
\newpagestyle{mystyle_empt}{%
\sethead[][][ ]
{}{}{}
}
\newpagestyle{mystyle}{%
\sethead[\thepage][][ Chapter\ \thechapter.  \chaptertitle]
{Section \thesection. \sectiontitle}{}{\thepage}
\setheadrule{1pt}
}
\newpagestyle{mystyle2}{%
\sethead[\thepage][][ \chaptertitle]
{\chaptertitle}{}{\thepage}
\setheadrule{1pt}
}
\newpagestyle{mystyle3}{%
\sethead[\thepage][][ Appendix\ \thechapter.  \chaptertitle]
{Appendix\ \thechapter.  \chaptertitle}{}{\thepage }
\setheadrule{1pt}
}
\newpagestyle{mystyle4}{%
\sethead[\thepage][][ Chapter\ \thechapter.  \chaptertitle]
{Chapter\ \thechapter.  \chaptertitle}{}{\thepage }
\setheadrule{1pt}
}
\renewcommand{\sectionmark}[1]{\gdef\rightmark{#1}}

\usepackage{tocbibind}
\addto{\captionsenglish}{}

\begin{document}
\frontmatter
\begin{titlepage}
\begin{center}

\thiswatermark{\put(5,-650){\includegraphics[width=1.1\textwidth]{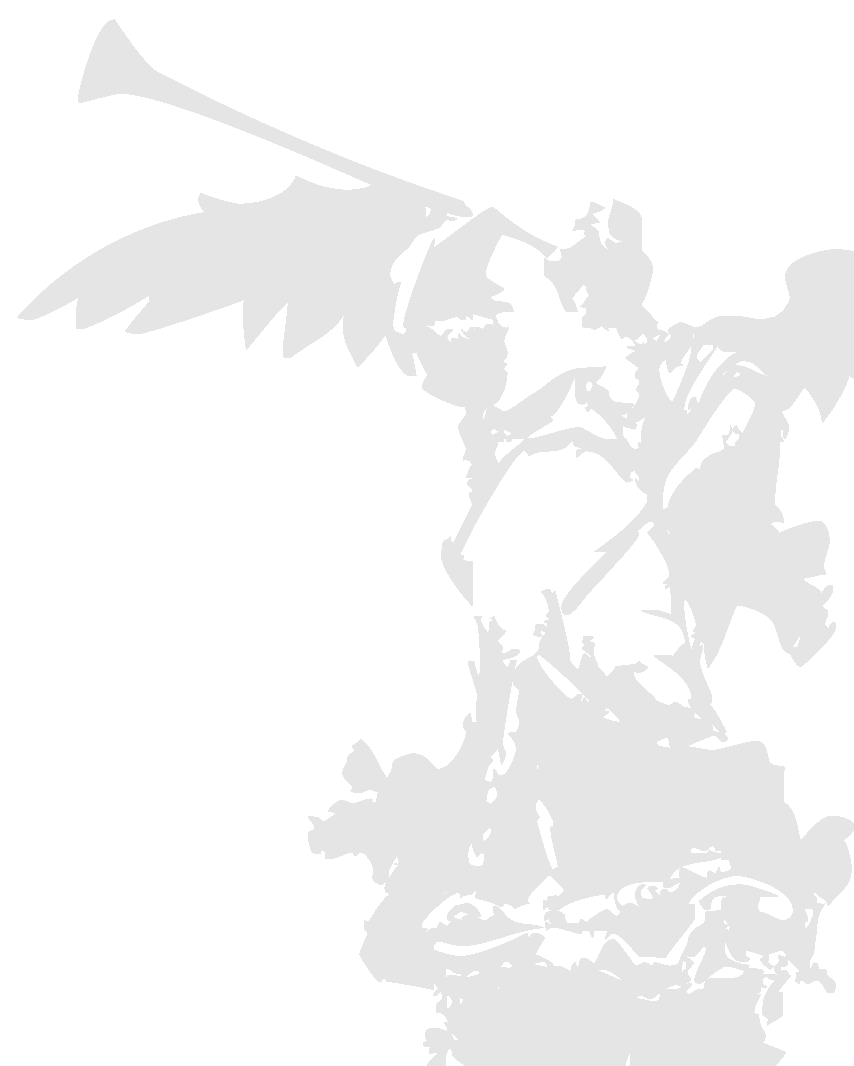}}}

\includegraphics[width=0.45\textwidth]{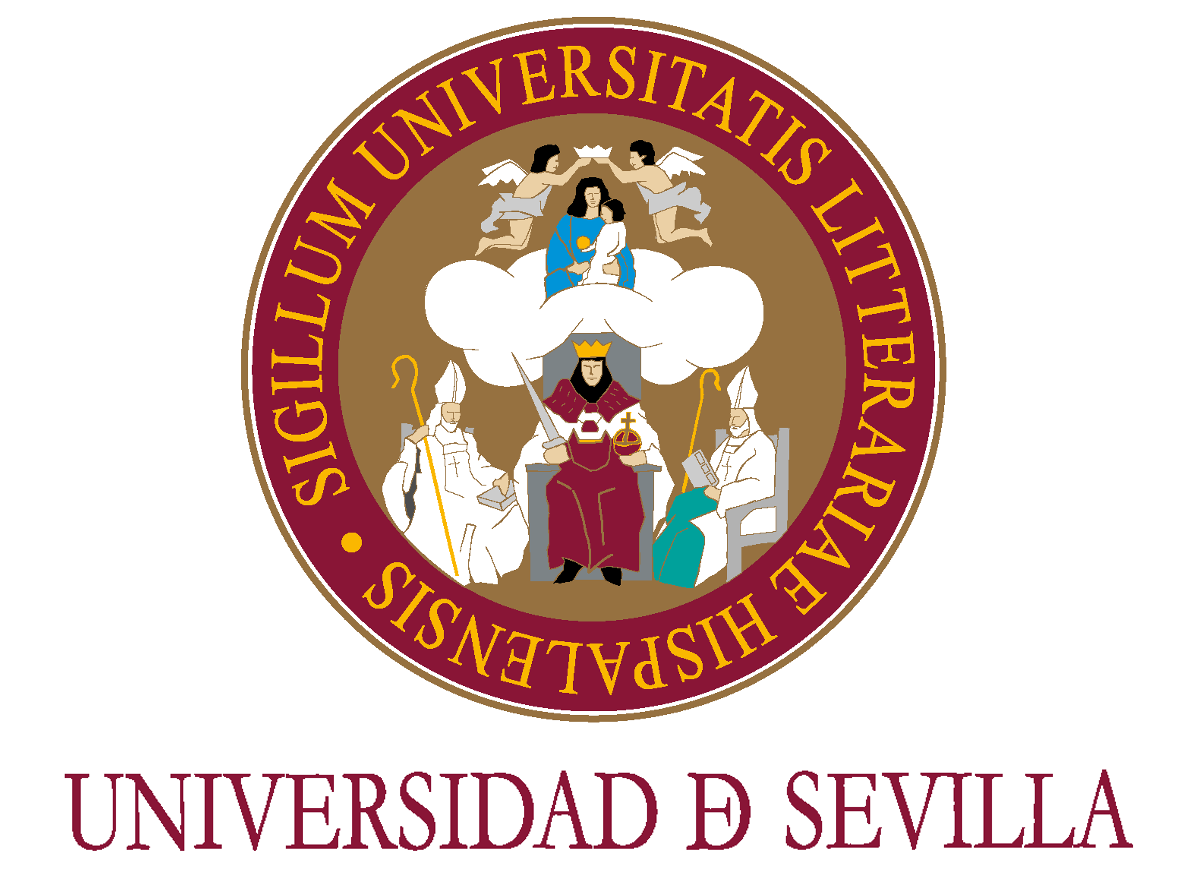}\\[1cm]

 \textsc{\Large Departamento de F\'{\i}sica At\'{o}mica \\Molecular y Nuclear}\\[1.3cm]

 \textsc{\Large Tesis Doctoral}\\[1.3cm]

 \hrulefill \\[0.6cm]
 { \huge \bfseries 
    Control and optimisation of irreversible processes in non-equilibrium systems
 }\\[0.6cm]

 \hrulefill \\[2.0cm]

\begin{minipage}{0.4\textwidth}
\begin{flushleft} \large
\emph{Doctorando:}\\
Antonio Patrón Castro \\
$\quad$
\\ $\quad$
\end{flushleft}
\end{minipage}
\begin{minipage}{0.4\textwidth}
\begin{flushright} \large

\emph{Directores:} \\
Antonio Prados Monta\~{n}o \\
Carlos A. Plata Ramos
\end{flushright}

\end{minipage}
 
\vfill
 
{\large Julio, 2024}

\end{center}

\end{titlepage}

\thispagestyle{empty}




{\clearpage \thispagestyle{empty}}
\chapter*{List of publications}
This thesis includes the research contained in the following works:
\begin{itemize}
    \item  \textbf{A. Patrón}, B. Sánchez-Rey and A. Prados, \textit{Strong nonexponential relaxation and memory effects in a fluid with nonlinear drag}, Physical Review E \textbf{104}, 064127 (2021)
    \item \textbf{A. Patrón}, A. Prados and C. A. Plata, \textit{Thermal brachistochrone for harmonically confined Brownian particles}, European Physics Journal Plus \textbf{137}, 1011 (2022)
    \item \textbf{A. Patrón}, B. Sánchez-Rey, E. Trizac and A. Prados, \textit{Nonequilibrium attractor for nonlinear stochastic dynamics}, Europhysics Letters \textbf{145}, 21001 (2024)
    \item \textbf{A. Patrón}, B. Sánchez-Rey and A. Prados, \textit{Kinetic glass transition in  granular gases and nonlinear molecular fluids}, Physical Review E \textbf{109}, 044137 (2024)
    \item \textbf{A. Patrón}, C. A. Plata, and A. Prados, \textit{Minimum time connection between non-equilibrium steady states: the Brownian gyrator}, Journal of Physics A: Mathematical and Theoretical \textbf{57}, 495004 (2024)
\end{itemize}
Other works that are not included in the thesis are:
\begin{itemize}
    \item \textbf{A. Patrón}, B. Sánchez-Rey, C. A. Plata and A. Prados, \textit{Non-equilibrium memory effects: granular fluids and beyond}, Europhysics Letters \textbf{143}, 61002 (2023)
    \item \textbf{A. Patrón}, A. D. Chepelianskii, A. Prados, and E. Trizac, \textit{On the optimal relaxation rate for the Metropolis Algorithm in one dimension}, Journal of Statistical Mechanics: Theory and Experiment \textbf{2025}, 013214 (2025)
\end{itemize}

{\clearpage \thispagestyle{empty}}
\addtocontents{toc}{\protect\setcounter{tocdepth}{-1}}
\tableofcontents
\addtocontents{toc}{\protect\setcounter{tocdepth}{3}}


{\clearpage \thispagestyle{empty}}
\mainmatter

\chapter{Introduction}
\chaptermark{Introduction}
\label{ch:introduction}


\setlength{\epigraphwidth}{5in} 
\setlength\epigraphrule{0pt}
\epigraph{\small \itshape We want to feel we are in control of our own existence. In some ways we are, in some ways we're not. We are ruled by the forces of chance and coincidence.}{--- Paul Auster, \textit{The New York Trilogy}}

Although the words of the recently deceased American writer may seem discouraging, we must bear in mind that they were meant to apply to the matters of life, understood in terms of human affairs. Concerning nature and its physical reality, since the dawn of man---as brilliantly depicted in Stanley Kubrick's ebony monolith scene from \textit{2001: A Space Odyssey}, humanity has engaged in the endeavour of not only understanding nature, but also bringing its intricate laws under control, for the sake of profiting from them. In fact, it is sometimes the plausible benefits and applications that give birth to major discoveries or even new branches of science. As a major historical example, we put forward Sadi Carnot, who highly contributed to the development of the field of thermodynamics during the 19th century with the aim of taming the power of fire and increasing the efficiency of steam engines~\cite{carnot_reflexions_1872}.

Nevertheless, we may agree with Paul Auster in one thing: there are indeed situations that are ruled by forces of chance. For the physical systems that we consider throughout this thesis, this apparent randomness is not quantum in nature, but stems from the lack of knowledge of the vast number of agents affecting the behaviour of the system of concern. A prototypical example is given by Brownian motion: the apparent random motion of microscopic particles immersed in a liquid~\cite{einstein_investigations_1956}. Albert Einstein theorised that such motion is due to the numerous collisions that the microscopic particles suffer from the liquid molecules, and he related the statistical properties of the displayed random motion---moments of a certain probability distribution---with the physical properties of the background liquid and its molecules. The significance of his theory---which was corroborated experimentally by Jean Baptiste Perrin at the beginning of the 20th century~\cite{perrin_mouvement_1909}---lies in its validation of the atomistic view on reality.
The branch of physics that explains the macroscopic behaviour of matter---in the previous example, the physical properties of the fluid: viscosity, temperature...---based on the microscopic dynamics of its many constituents---the atoms or molecules constituting the background liquid---is known as statistical mechanics, and it sets the main mathematical framework that we employ in our research. It was founded in between the 19th and 20th centuries by the well-known physicists William Thomson~\cite{thomson_dynamical_1851}, James Clerk Maxwell~\cite{maxwell_illustrations_1860}, Ludwig Boltzmann~\cite{boltzmann_further_1966} and Josiah Willard Gibbs~\cite{gibbs_fundamental_1885}, and it provided a microscopic explanation for the physical concepts of work and heat drawn from classical thermodynamics.


Since its very beginnings, many researchers have found applications of statistical mechanics to different branches of physics. In the following, we give a non-exhaustive list of major historical examples. Lev Landau developed from it his phenomenological theory of phase transitions~\cite{landau_theory_1937}, which has played a fundamental role in many subsequent developments of physics. In quantum mechanics, we highlight Satyendra Nath Bose, Albert Einstein, Enrico Fermi and Paul Dirac, who determined the statistical properties of the two major classes of fundamental particles: bosons~\cite{perez_einstein_2010} and fermions~\cite{fermi_quantizzazione_1926,dirac_theory_1926}. And in astrophysics, Subrahmanyan Chandrasekhar determined the maximum limiting mass that a white dwarf star may attain without undergoing gravitational collapse into a neutron star or a black hole~\cite{chandrasekhar_maximum_1931}. In the modern days, the field has extended its applicability to many other fields of knowledge that lie out of the scope of physics, such as biology~\cite{friedrich_approaching_2011,dill_molecular_2010,phillips_physical_2012}, neuroscience~\cite{schneidman_weak_2006,herz_modelling_2006}, and even economy~\cite{bachelier_theorie_1900,bouchaud_theory_2003}, computer science~\cite{anderson_statistical_1986} and social sciences~\cite{castellano_statistical_2009,galam_sociophysics_2012,contucci_statistical_2020}.

Statistical mechanics may be divided into two fields. On the one hand, we have equilibrium statistical mechanics, which provides the connection between the microscopic and macroscopic properties of matter in thermodynamic equilibrium, and it thus deals with systems that have already attained its time-independent equilibrium state. On the other hand, we have non-equilibrium statistical mechanics, which is the main focus of this thesis. The latter deals with (i) transient behaviour, i.e. systems whose statistical properties evolve with time, and (ii) out-of-equilibrium systems, which correspond to those systems for which their corresponding stationary states, if they exist, do not correspond to equilibrium ones. Biological processes occurring inside living bodies belong to this class of non-equilibrium systems. As beautifully pointed out by Erwin Schrödinger~\cite{schrodinger_what_1974}; \textit{life is a thermodynamic process}. Living bodies continuously fight against the second law of thermodynamics by creating negative entropy production, in order to carry out the numerous phenomena---chemical reactions, transport of substances...---that allow the overall system to be regarded as a living being. Following this line of reasoning, he established an analogy between thermodynamic equilibrium and death itself.

In this thesis, we work with different physical systems that fall under the field of statistical mechanics. For each particular system, we assume that we may control its dynamical evolution by varying a set of external parameters, which induce forces acting on the system---e.g. the temperature of a thermal bath, the amplitude of an electric or magnetic field, or the intensity of a laser. It is in the non-equilibrium process of dragging the system out of its stationary state by suitably varying the control parameters where we find the two main lines that our research encompasses. On the one hand, following the classical approach employed in physics problems, we are interested in studying how the system of concern responds upon perturbing it from its stationary---either equilibrium or non-equilibrium---state. Among the possible non-equilibrium phenomena that may occur in the system, the focus of this venue of research is specifically put on the emergence and characterisation of memory effects and glassy behaviour. On the other hand, we may follow a sort-of reverse engineering approach in which, given an initial stationary state and a target state that we wish our system to reach, we want to determine how we have to vary the external, controllable parameters with time in order to drive the system between those states. Furthermore, among all the possible ways that there may be of achieving such a task, our second venue of research is devoted to studying those ways in which the connection time between states is minimised, thus accelerating the relaxation dynamics and beating, when possible, the characteristic relaxation timescales of the system. Both venues of research are relevant for the design and optimal performance of irreversible heat engines, which have attracted great interest in recent years due to their applicability at the nanoscopic scale~\cite{schmiedl_efficiency_2008,blickle_realization_2012,martinez_brownian_2016,martinez_colloidal_2017,plata_building_2020,nakamura_fast_2020,dinis_thermodynamics_2016}. The work carried out in this thesis is thus divided into two parts, corresponding to the two aforementioned lines of research. 

In the following, this introductory chapter is devoted to put forward the fundamentals of the general framework, as well as the main theoretical tools that have been used in our research. In Sec.~\ref{ch1_sec:non-eq}, we provide a brief overview of the main mathematical framework that we employ in our research, that corresponding to stochastic dynamics---whose tools are widely used within the field of non-equilibrium statistical mechanics. Then, in Sec.~\ref{ch1_sec:glass}, we introduce the main aspects of the memory effects and glassy behaviour that we analyse in the first part of this thesis. Finally, Sec.~\ref{ch1_sec:shortcuts} is devoted to some introductory remarks on the field of swift state-to-state transformations, in which our second line of research concerning time-optimisation problems is embedded.

\section{Overview of stochastic dynamics. Fokker-Planck and Langevin descriptions}\label{ch1_sec:non-eq}

The focus of our research concerns physical systems whose degrees of freedom---e.g. the components of either the position or the velocity of a particle---evolve with a certain degree of randomness in time. Let us consider a system whose state is characterised by the variables $\bm{r}^{\sf{T}} \equiv (r_1,r_2,...,r_d)$, to which we refer as the state vector, with $d$ accounting for its dimensionality.\footnote{Throughout this thesis, bold symbols denote column vector quantities, while blackboard bold symbols denote matrices: e.g. $\bm{a}$ is a column vector, while $\bm{b}^{\sf{T}}$ is a row vector. If they both have the same dimension, $c = \bm{b}^{\sf{T}}\bm{a} = \bm{a}^{\sf{T}}\bm{b}$ is a number stemming from the scalar product between vectors, while $\mathbb{D} = \bm{a} \bm{b}^{\sf{T}}$ accounts for a square matrix.} In the previous example of Brownian motion, $\bm{r}$ would stand for the $d$ components of either the position and/or the velocity of the microscopic particle immersed in the background fluid. We say that the time-dependent function $\bm{r} = \bm{r}(t)$ is a stochastic process if it constitutes a collection of random variables\footnote{We skip the most cumbersome technicalities on what random variables and/or stochastic processes are. Concerning the former, just to bear in mind that they correspond to mathematical objects defined on a sample space---the set of all possible values, which may be either discrete or continuous---and by a probability distribution over such space. See references~\cite{van_kampen_stochastic_1992,wilkinson_stochastic_2011,feller_introduction_1968,gardiner_stochastic_2009}, which are the main ones employed throughout this section, for further details.} indexed by time.

Now, given the stochastic process $\bm{r}(t)$, let us say that we have taken $n$ measurements of it at distinct times $\{\bm{r}_j,t_j\}$, with ${j = 1,\ldots,n }$ and $t_1 < t_2 < ... < t_n$. If we were to wonder what is the probability that, at the $(n+1)$-th measurement, the stochastic process attains the state vector $\bm{r}_{n+1}$ at the time $t_{n+1}$, in principle such probability would depend on the entire prior history of the process---i.e. all the previous measurements at times $t_j$, $j=1,\ldots,n$. The latter is quantified by $P(\bm{r}_{n+1},t_{n+1}|\bm{r}_{n},t_{n};\bm{r}_{n-1},t_{n-1};...;\bm{r}_1,t_1)$, which stands as the conditional probability of attaining $(\bm{r}_{n+1},t_{n+1})$ given the prior sequence of measurements. For all the systems considered in this thesis, we further assume that they belong to a subclass of stochastic processes known as Markov processes, which satisfy the property
\begin{equation}\label{ch1_eq:markov-property}
    P(\bm{r}_{n+1},t_{n+1}|\bm{r}_{n},t_{n};\bm{r}_{n-1},t_{n-1};...;\bm{r}_1,t_1) = P(\bm{r}_{n+1},t_{n+1}|\bm{r}_{n},t_{n}),
\end{equation}
also known as the Markov property. That is, for Markov processes, the future behaviour of the system under study does not depend at all on any information regarding the past, but only on its current state. In other words, the future is independent on the past, given the present. This memory-less property of Markov processes resembles the determinism found in both classical and quantum mechanics---in the latter, concerning the time-evolution of the wavefunction, aside from its collapse after a measurement~\cite{penrose_emperors_1989}---in the sense that the future behaviour of the system is uniquely determined by the current, initial state, plus the knowledge of all the external agents acting on it.\footnote{This analogy between classical and stochastic dynamics will be invoked again in later on, when comparing the Langevin equation with Newtonian mechanics.} As in Eq.~\eqref{ch1_eq:markov-property} only the current $(\bm{r}_{n},t_{n})$ and following $(\bm{r}_{n+1},t_{n+1})$ states are involved, then we may set $n=1$ without loss of generality. Now, provided that we know the probability distribution function (PDF) for the current measurement, $P(\bm{r}_1,t_1)$, we may relate it to the joint probability for both the current and following states, $P(\bm{r}_2,t_2,\bm{r}_1,t_1)$ by means of Bayes' theorem~\cite{feller_introduction_1968},
\begin{equation}
    P(\bm{r}_2,t_2,\bm{r}_1,t_1) = P(\bm{r}_{2},t_{2}|\bm{r}_{1},t_{1}) P(\bm{r}_1,t_1).
\end{equation}
And in fact, following from the above, we may obtain the PDF for the following state, $P(\bm{r}_2,t_2)$, by integrating over all the possible current states,
\begin{equation}\label{ch1_eq:transition-prob-1}
    P(\bm{r}_2,t_2) = \int d\bm{r}_1 P(\bm{r}_2,t_2,\bm{r}_1,t_1) = \int d\bm{r}_1 P(\bm{r}_{2},t_{2}|\bm{r}_{1},t_{1}) P(\bm{r}_1,t_1).
\end{equation}
Equation~\eqref{ch1_eq:transition-prob-1} is enticing, as it provides the conditional probability $P(\bm{r}_{2},t_{2}|\bm{r}_{1},t_{1})$ with a new physical meaning: it corresponds to the transition probability between the states $(\bm{r}_1,t_1)$ and $(\bm{r}_2,t_2)$. By further assuming our Markov processes to be stationary, such transition probability becomes invariant upon a time shift, i.e.
\begin{equation}\label{ch1_eq:stationary}
    P(\bm{r}_{2},t_{2} + \tau|\bm{r}_{1},t_{1} + \tau) = P(\bm{r}_{2},t_{2}|\bm{r}_{1},t_{1}).
\end{equation}
Stationary Markov processes are specially appealing in physics, as they allow describing fluctuations in equilibrium systems. Equation~\eqref{ch1_eq:stationary} implies that $P(\bm{r}_{2},t_{2}|\bm{r}_{1},t_{1})$ does not depend on the independent measured times $t_1$ and $t_2$, but on their difference $t_2 - t_1$. Thus, we may introduce the notation
\begin{equation}
    P(\bm{r}_{2},t_{2}|\bm{r}_{1},t_{1}) = \mathcal{T}_{\tau}(\bm{r}_2|\bm{r}_1), \quad \tau \equiv t_2 - t_1,
\end{equation}
such that Eq.~\eqref{ch1_eq:transition-prob-1} becomes
\begin{equation}\label{ch1_eq:transition-prob-2}
    P(\bm{r}_2,t_2) = \int d\bm{r}_1 \mathcal{T}_{\tau}(\bm{r}_2|\bm{r}_1) P(\bm{r}_1,t_1).
\end{equation}
Equation~\eqref{ch1_eq:transition-prob-2} is exact, and it only stems from the assumption of the stationary condition~\eqref{ch1_eq:stationary} for the stochastic process considered. In the following, we work with general states $\bm{r}'$ and $\bm{r}$ instead of the labelled ones $\bm{r}_1$ and $\bm{r}_2$, as it becomes a more intuitive notation. To be consistent, we also identify $t_1$ and $t_2$ with $t$ and $t + \tau$, respectively.

For all the systems that we consider, the PDF $P(\bm{r},t)$ is the main quantity of interest, as it accounts for the physical state and the statistical information of the system. Provided an initial condition $P_i(\bm{r}) \equiv P(\bm{r},0)$ for the dynamics, Eq.~\eqref{ch1_eq:transition-prob-2} determines the overall behaviour of the system at any other time $t$. Nevertheless, working with Eq.~\eqref{ch1_eq:transition-prob-2} becomes rather cumbersome in most applications---as it constitutes an integral equation. Thus, further assumptions on the transition probabilities $\mathcal{T}_{\tau}(\bm{r}|\bm{r}')$ are required in order to gain analytical insight.
For many physical scenarios, the amplitude of the jumps between the---very close---time measurements of a Markov process $\bm{r}(t)$ is usually small. For instance, for the case of Brownian motion, we do not expect the position $\bm{r}(t)$ of the Brownian particle to change dramatically in between subsequent, close in time, measurements. In this regard, we consider the following assumptions on the transition probabilities $\mathcal{T}_{\tau}(\bm{r}|\bm{r}')$:
\begin{enumerate}
    \item The continuum time limit $\tau \rightarrow 0$, i.e. the time between consecutive measurements vanishes. 
    \item The transition probabilities are sharply peaked functions of the distance $|\bm{r}-\bm{r}'|$, such that only transitions between neighbouring states take place.
    \item The PDF $P(\bm{r},t)$ is a smooth function of $\bm{r}$.
\end{enumerate}
On the one hand, the first assumption allows us to Taylor expand both $\mathcal{T}_{\tau}(\bm{r}|\bm{r}')$ and $P(\bm{r},t+\tau)$ from Eq.~\eqref{ch1_eq:transition-prob-2} in powers of $\tau$ and retain up to linear terms. On the other hand, following the second and third assumptions, we may further expand the remaining terms in powers of the small amplitudes $|\bm{r}-\bm{r}'|$---what is also known as the Kramers-Moyal expansion~\cite{van_kampen_stochastic_1992}---and, by retaining up to quadratic terms, we obtain the second-order partial differential equation
\begin{equation}\label{ch1_eq:fokker-planck}
    \frac{\partial P(\bm{r},t)}{\partial t} = \nabla_{\bm{r}}^{\sf{T}}
    \left[ \bm{A}(\bm{r},t)P(\bm{r},t)
    +\mathbb{D}(\bm{r},t)\nabla_{\bm{r}}P(\bm{r},t)\right],
\end{equation}
which is known as the Fokker-Planck equation. In the above, the vector function (column matrix) $\bm{A}(\bm{r},t)$ and the $d \times d$ symmetric matrix function $\mathbb{D}(\bm{r},t)$ are related to the so-called jump moments of the Kramers-Moyal expansion, which correspond to certain integral functions over the transition probabilities. We highlight that the Fokker-Planck equation may also be rewritten as a continuity-like equation of the form
\begin{equation}\label{ch1_eq:continuity-eq}
    \frac{\partial P(\bm{r},t)}{\partial t} + \nabla_{\bm{r}}^{\sf{T}} \bm{J}(\bm{r},t) = 0, \quad \bm{J}(\bm{r},t) \equiv -  \bm{A}(\bm{r},t)P(\bm{r},t)
    - \mathbb{D}(\bm{r},t)\nabla_{\bm{r}}P(\bm{r},t),
\end{equation}
with $\bm{J}(\bm{r},t)$ corresponding to the net probability flux at state $\bm{r}$ at time $t$. The prototypical physical interpretation for the Fokker-Planck equation is intimately related to Brownian motion. With $P(\bm{r},t)$ being the probability distribution function for the position $\bm{r}$ of the Brownian particle at any time $t$, we identify $\bm{A}(\bm{r},t)$ as the deterministic ``force'' term---i.e. the one accounting for the external forces acting on the Brownian particle, other than the collisions between the liquid molecules---and $\mathbb{D}(\bm{r},t)$ as the ``noise" term, referring to the thermal fluctuations that turn the dynamics stochastic. 

It is said that the Fokker-Planck equation characterises the continuous-state Markov process $\bm{r}(t)$ at the ensemble level---i.e. by considering a great number of trajectories of the same random process, the Fokker-Planck equation allows to accurately describe their overall statistical behaviour through the PDF $P(\bm{r},t)$---in fact, such description is deterministic with respect to the time evolution of the PDF. But one may also resort to a mathematically equivalent approach, which consists in characterising the Markov process at the level of individual trajectories. Such approach corresponds to the so-called Langevin equation~\cite{kramers_brownian_1940}, which is given by
\begin{equation}\label{ch1_eq:langevin}
    \dot{\bm{r}} = -\left[\bm{A}(\bm{r},t) - (\alpha -1)\mathbb{B}(\bm{r},t)(\nabla_{\bm{r}}^{\sf{T}}\mathbb{B}(\bm{r},t))^{\sf{T}}\right] + \mathbb{B}(\bm{r},t) \bm{\eta}(t),
\end{equation}
and constitutes a stochastic differential equation.\footnote{In fact, the first of its kind, which marked the beginning of the field of stochastic differential equations~\cite{van_kampen_stochastic_1992,gardiner_stochastic_2009}.}
In the above, $\bm{\eta}(t)$ stands for the unit, Gaussian white-noise in $n_s$ dimensions,\footnote{In most cases we have $n_s = d$, as we have in the forthcoming chapters.} which satisfies the statistical properties
\begin{equation}\label{ch1_eq:stat_prop}
    \left< \eta_j (t) \right> = 0, \quad \left< \eta_j (t) \eta_l (t')\right> = \delta_{jl} \delta (t-t'), \quad j,l=1,2,...n_s,
\end{equation}
where $\left< ... \right>$ constitutes the ensemble average---i.e. average over a great number of realisations of the noise, $\delta_{jl}$ and $\delta (t-t')$ correspond to the Kronecker and Dirac delta functions, respectively, and $\mathbb{B}(\bm{r},t)$ corresponds to a $d \times n_s$ matrix related to $\mathbb{D}(\bm{r},t)$ via
\begin{equation}
    \mathbb{D}(\bm{r},t) = \frac{1}{2}\mathbb{B}(\bm{r},t)\mathbb{B}^{\sf{T}}(\bm{r},t).
\end{equation}
The parameter $\alpha \in [0,1]$ is known as the multiplicative-noise parameter, which takes into account the fact that there are multiple interpretations for the numerical integration of the Langevin equation equivalent to the Fokker-Planck equation from Eq.~\eqref{ch1_eq:fokker-planck}---as it corresponds to a stochastic differential equation, which cannot be integrated using ordinary methods from Riemannian calculus. Let us note that, in the additive noise scenario, for which $\mathbb{B}(\bm{r},t)$ is independent of $\bm{r}$, we have that $\nabla_{\bm{r}}^{\sf{T}}\mathbb{B}(\bm{r},t) = 0$ and the Langevin equation~\eqref{ch1_eq:langevin} becomes independent of $\alpha$. It is only in this case in which all the interpretations of the Langevin equation become equivalent, regardless of the value of $\alpha$. See Appendix~\ref{app:stochastic-integration} for further details on the above discussion, which becomes relevant for the simulations we carry out in further chapters. 

The physical meaning of the Langevin equation~\eqref{ch1_eq:langevin} is quite transparent when interpreting $\bm{r}$ as the $d$ components of the velocity of a free particle immersed in a fluid---what is known as underdamped Brownian motion for a free particle. In that case, Eq.~\eqref{ch1_eq:langevin} resembles Newton's second law of motion, in which its right-hand side accounts for the net force acting on the Brownian particle. Specifically, the term within brackets stands for the deterministic contribution to the force, while $\mathbb{B}(\bm{r},t) \bm{\eta}(t)$ accounts for the forces due to the random collisions with the particles of the background fluid. 

For the majority of the systems considered in this work, our starting point constitutes either a Fokker-Planck or a Langevin equation. On the one hand, the Fokker-Planck equation allows us to extract information concerning the moments of the PDF $P(\bm{r},t)$ in a very straightforward way while, on the other hand, the Langevin description is useful for numerical simulation purposes.

\subsection{Stationary solutions in the long-time limit}\label{ch1_subsubsec:long-time-lim}

In many physical scenarios, the corresponding Fokker-Planck equations present a fundamental feature: in the $t \rightarrow +\infty$ limit, all of their solutions tend to a stationary solution.\footnote{Note that it does not need to be the case generally, and it depends on the explicit forms of the terms $\bm{A}(\bm{r})$ and $\mathbb{D}(\bm{r})$. In fact, free Brownian motion constitutes a remarkable counterexample.} For the systems considered in this thesis, not only they do present stationary states in the long time limit, but such states are unique for each system. Such stationary states are accounted for by the stationary PDF $P_{\text{s}}(\bm{r})$, which is obtained by imposing
\begin{equation}\label{ch1_eq:fp-stationary}
    \frac{\partial P_{\text{s}}(\bm{r})}{\partial t} = 0 \ \Longrightarrow \ \nabla_{\bm{r}}^{\sf{T}} J_{\text{s}}(\bm{r}) = 0, \quad J_{\text{s}}(\bm{r}) \equiv -  \bm{A}_{\text{s}}(\bm{r})P_{\text{s}}(\bm{r})
    - \mathbb{D}_{\text{s}}(\bm{r})\nabla_{\bm{r}}P_{\text{s}}(\bm{r}),
\end{equation}
with $J_{\text{s}}(\bm{r})$ being the net stationary probability flux at the state $\bm{r}$, and
\begin{equation}
    \bm{A}_{\text{s}}(\bm{r}) = \lim_{t\to +\infty} \bm{A}(\bm{r},t), \quad \mathbb{D}_{\text{s}}(\bm{r}) = \lim_{t\to +\infty} \mathbb{D}(\bm{r},t).
\end{equation}
Most of the systems considered in the following chapters are in contact with a unique thermal bath, characterised with an inverse temperature $\beta \equiv (k_BT)^{-1}$. For those thermal systems, we may identify the stationary PDF $P_{\text{s}}(\bm{r})$ with the equilibrium canonical distribution $P_{\text{eq}}(\bm{r}) \propto \text{exp}(-\beta H(\bm{r}))$, with $H(\bm{r})$ being a Hamiltonian-like function accounting for the dynamics of $\bm{r}$---depending on the context, $H(\bm{r})$ might not be a Hamiltonian function but a function playing an analogous role; e.g. if $\bm{r}$ stands for the position of an overdamped $d$-dimensional system, $H(\bm{r})$ could play the role of a confining potential $U(\bm{r})$. Equilibrium systems fulfil the so-called fluctuation-dissipation relation
\begin{equation}\label{ch1_eq:fluctuation-dissipation}
    \bm{A}(\bm{r},t)=\beta(t) \, \mathbb{D}(\bm{r},t)\,\nabla_{\bm{r}} H(\bm{r}),
\end{equation}
which, by inserting it into Eq.~\eqref{ch1_eq:fp-stationary}, implies that the net equilibrium current at any state $\bm{r}$ vanishes, $\bm{J}_{\text{eq}}(\bm{r}) = 0$. Fokker-Planck systems presenting a general stationary distribution $P_{\text{s}}(\bm{r})$ but not satisfying the fluctuation-dissipation relation may present a non-zero value of their corresponding stationary net flux, $\bm{J}_{\text{s}}(\bm{r}) \neq \bm{0}$. Such non-zero fluxes may take place if the system has additional forces acting on it. Examples of the latter include self-propelled---or active---particles, molecular motors, or systems in contact with more than one thermal bath at different temperatures---in chapter~\ref{ch:brownian_gyrator} we explore a specific model concerning this scenario.

Stochastic systems attaining the stationary state in the long term present an additional feature that becomes useful in different applications. Let us introduce the Fokker-Planck operator
\begin{equation}\label{ch1_eq:fp-operator}
    \mathbb{L}_\text{FP}[P] \equiv \nabla_{\bm{r}}^{\sf{T}}
    \left[ \bm{A}_{\text{s}}(\bm{r})P
    +\mathbb{D}_{\text{s}}(\bm{r})\nabla_{\bm{r}}P\right] \ \Longrightarrow \ \frac{\partial P}{\partial t} = \mathbb{L}_\text{FP}[P].
\end{equation}
If the stationary solution exists, $P_{\text{s}}(\bm{r})$ allows to define an inner product for real-valued functions defined in a Hilbert space~\cite{bender_advanced_1999}. If the Fokker-Planck operator is self-adjoint with respect to such inner product---or if it can become a self-adjoint one by means of a similarity transformation, we may write the general solution of the Fokker-Planck equation~\eqref{ch1_eq:fokker-planck} as
\begin{equation}\label{ch1_eq:eigen-expansion}
    P(\bm{r},t) = P_{\text{s}}(\bm{r}) + \sum_{\nu > 0} \mathcal{A}_{\nu} \phi_{\nu}(\bm{r}) e^{-\lambda_{\nu}t},
\end{equation}
with $\lambda_{\nu}$ and $\phi_{\nu}(\bm{r})$ being the eigenvalues and eigenvectors of the operator $\mathbb{L}_{\text{FP}}[P]$\cite{risken_fokker_1989}. The self-adjoint property of $\mathbb{L}_{\text{FP}}$ guarantees that the set of eigenfunctions is complete, and the eigenvalues are all real with $\lambda_{\nu} \geq 0$. In particular, for $\nu = 0$---our choice for the labeling is $\lambda_0 < \lambda_1 < \lambda_2...$, the $\lambda_0 = 0$ eigenvalue corresponds to the stationary PDF $P_{\text{s}}(\bm{r})$, as it follows from imposing $\mathbb{L}_{\text{FP}}[P_{\text{s}}]=0$ in Eq.~\eqref{ch1_eq:fp-operator}.

The eigenfunction expansion~\eqref{ch1_eq:eigen-expansion} results advantageous in studying the relaxation dynamics of the Fokker-Planck equation towards the stationary solution. In fact, by assuming non-degeneracy for the eigenvalues $\lambda_{\nu}$, $\lambda_1$ characterises the relevant relaxation timescale of the dynamics for long enough times: it is the relevant rate at which the system relaxes towards the stationary state. Thus, having control over $\lambda_1$ or its corresponding coefficient $\mathcal{A}_1$ becomes appealing for both time-optimisation problems and the emergence of strong non-equilibrium phenomena, as discussed in Sec.~\ref{ch1_sec:glass}.

\subsection{Gaussian processes}\label{ch1_subsubsec:ornstein-uhlenbeck}

By definition, the Fokker-Planck equation~\eqref{ch1_eq:fokker-planck} constitutes a partial differential equation that is linear in the PDF $P(\bm{r},t)$, which guarantees both the existence and uniqueness of its solutions---provided that the vector $\bm{A}(\bm{r},t)$ and matrix $\mathbb{D}(\bm{r},t)$ are sufficiently smooth functions of their arguments~\cite{bender_advanced_1999}. In the following, we employ the adjective ``linear'' with an alternative meaning. We say that the Fokker-Planck---or Langevin---equation is linear if $\bm{A}(\bm{r},t)$ is a linear function of $\bm{r}$ and $\mathbb{D}(\bm{r},t)$ is homogeneous,
\begin{equation}\label{ch1_eq:coeffs-ou}
    \bm{A}(\bm{r},t) = \mathbb{A}(t) \bm{r}, \quad \mathbb{D}(\bm{r},t) = \mathbb{D}_0(t),
\end{equation}
with $\mathbb{A}(t)$ and $\mathbb{D}_0(t)$ being two $d \times d$ matrices independent of the state variables $\bm{r}$. If $\mathbb{A}(t)$ and $\mathbb{D}_0(t)$ are time-independent, the corresponding Fokker-Planck equation presents a stationary solution $P_{\text{s}}(\bm{r})$ that is Gaussian. If $\mathbb{A}(t)$ and/or $\mathbb{D}_0(t)$ depend on time, the PDF $P(\bm{r},t)$ becomes also Gaussian, if it was so initially. 

These Gaussian processes play a major role for the systems considered in chapters~\ref{ch:thermal_brachistochrones} and \ref{ch:brownian_gyrator}, and they constitute one of the few examples in which the Fokker-Planck equation~\eqref{ch1_eq:fokker-planck} can be solved analytically. In order to do so, let us start by introducing the average value of an arbitrary function $f(\bm{r})$ over the PDF $P(\bm{r},t)$ as
\begin{equation}
    \left< f(\bm{r}) \right> \equiv \int d\bm{r} \ f(\bm{r}) P(\bm{r},t).
\end{equation}
The equation governing its time evolution may be obtained applying the time derivative onto the above, such that we may substitute the Fokker-Planck equation~\eqref{ch1_eq:fokker-planck}, giving\footnote{In order to obtain Eq.~\eqref{ch1_eq:evol-eq-f}, we have to integrate by parts the resulting integrals until obtaining the corresponding averages. The resulting boundary terms vanish by imposing that $P(\bm{r},t)$ and its derivatives must decay fast enough towards zero for large $|\bm{r}|$.}
\begin{equation}\label{ch1_eq:evol-eq-f}
    \frac{d}{dt}\left< f(\bm{r})\right> = - \left<\bm{A}^{\sf{T}}(\bm{r},t) \nabla_{\bm{r}}f(\bm{r})\right> + \left< \nabla_{\bm{r}}^{\sf{T}} \left[\mathbb{D}(\bm{r},t) \nabla_{\bm{r}}f(\bm{r}) \right] \right>.
\end{equation}
The above relation holds regardless of the functional forms of $\bm{A}(\bm{r},t)$ and $\mathbb{D}(\bm{r},t)$. In particular, for the linear case~\eqref{ch1_eq:coeffs-ou}, we may obtain the evolution equations for the first moments,
\begin{equation}\label{ch1_eq:first-moments}
    \frac{d}{dt}\left< \bm{r} \right> = - \mathbb{A}(t) \left< \bm{r} \right>,
\end{equation}
and for the second ones,
\begin{equation}\label{ch1_eq:second-moments}
    \frac{d}{dt}\left<\mathbb{R}_2\right> = -\left[   \mathbb{A}(t)\left<\mathbb{R}_2\right> + \left<\mathbb{R}_2\right>\mathbb{A}^{\sf{T}}(t) \right] + 2\mathbb{D}_0(t),
\end{equation}
with $\left<\mathbb{R}_2\right> \equiv \left< \bm{r} \bm{r}^{\sf{T}}\right>$ being a $d \times d$ symmetric matrix of elements $\left<\mathbb{R}_2\right>_{jl} = \left<r_jr_l \right>$. If $\mathbb{A}(t)$ is symmetric, the term within brackets in the above reduces to the anticommutator $\left\{\mathbb{A}(t),\left<\mathbb{R}_2\right>\right\}$ between the matrices $\mathbb{A}(t)$ and $\left<\mathbb{R}_2\right>$. 

Now, let us assume that the initial state of the dynamics $P(\bm{r},0)$ is Gaussian with null mean $\left<\bm{r}\right>(t=0) = \bm{0}$. Then, the linearity of the evolution equations ensures that $P(\bm{r},t)$ retains the Gaussian form with $\left<\bm{r}\right>(t) = \bm{0}$ at all times,\footnote{This may be intuitively viewed as a direct consequence of the conservation of the Gaussian property when summing random Gaussian variables.} and it can be written as
\begin{equation}\label{ch1_eq:solution-pdf-ou}
    P(\bm{r},t) = \frac{1}{(2\pi)^{d/2} \sqrt{\text{det}\left<\mathbb{R}_2\right> }}\text{exp}\left(-\frac{1}{2}\bm{r}^{\sf{T}}\left<\mathbb{R}_2\right>^{-1}\bm{r}\right),
\end{equation}
with $\left<\mathbb{R}_2\right>^{-1}$ being the inverse of the matrix $\left<\mathbb{R}_2\right>$. That is, for the Gaussian processes we are considering, the probability distribution function $P(\bm{r},t)$---and thus, the system dynamics---is completely characterised by the second moment matrix $\left<\mathbb{R}_2\right>$. This reduces the infinite dimensional problem of determining the PDF $P(\bm{r},t)$ to a finite dimensional one for the second moments---to be concrete, it is reduced to a $d(d+1)/2$ dimensional problem, as $\left<\mathbb{R}_2\right>$ is symmetric.

\subsection{Addition of a Boltzmann collision term}\label{ch1_subsubsec:collision}
There are two models considered in this thesis---specifically in chapters~\ref{ch:memory_effects} and~\ref{ch:glass_transition}---where an additional source of complexity is considered. Let us put forward the situation of having not only one Brownian particle, but an ensemble of Brownian particles, corresponding to hard-spheres of radii $\sigma$ and number density $n_{\text{par}}$, which are immersed in a background fluid that acts as a thermal bath. In this case, apart from the Fokker-Planck contribution to the dynamics---i.e. the thermal fluctuations due to the random collisions between the background fluid and Brownian particles, we must also take into account the possible collisions between different Brownian particles. Therefore, in order to accurately account for the dynamics of the overall system, we need to additionally resort to the tools from kinetic theory~\cite{chapman_mathematical_1990,sharipov_rarefied_2016,brilliantov_kinetic_2004,poschel_granular_2001}.

Kinetic theory was developed in between the 18th and 19th centuries, and it is currently considered as a branch of non-equilibrium statistical mechanics. It explains the thermodynamic properties of gases, by modelling them as an ensemble of particles that interact between them and with the walls of the container through collisions. For the aforementioned two models considered here, the state variables $\bm{r}$ are the velocities $\bm{v}$ of the individual Brownian particles constituting the gas. By further assuming the overall gas to be isotropic and homogeneous, the positions of the Brownian particles are not required for their statistical description. One of the main assumptions employed in kinetic theory constitutes the so-called molecular chaos hypothesis---also referred to as \textit{Stosszahlansatz} in German, as written by Paul and Tatiana Ehrenfest~\cite{ehrenfest_conceptual_2014}, which states that: (i) only binary collisions between hard-sphere Brownian particles take place, and (ii) the velocities of two colliding Brownian particles are statistically independent just before the collision.\footnote{Despite being a heuristic approximation, which becomes useful to simplify collision calculations, the molecular chaos hypothesis is crucial for explaining the irreversibility of macroscopic processes. It is at the core of the derivation of the H-theorem by Ludwig Boltzmann~\cite{boltzmann_further_1966}, which demonstrates that the entropy of an isolated system never decreases, thus providing a statistical-mechanical basis for the second law of thermodynamics.} Thus, we assume that the fluid of Brownian particles is sufficiently dilute such that its dynamics is accurately described at the one-particle level---i.e. neglecting correlations between the velocities of different particles. In this regard, the PDF $P(\bm{r},t)$ shall be taken as the one-particle velocity distribution function (VDF) $f(\bm{v},t)$, which provides the number of particles per unit volume with velocities $\bm{v}$ at time $t$, and it is normalised to the number density $n_{\text{par}}$.

The time-evolution of the one-particle VDF, for systems governed by both stochastic and kinetic dynamics, is given by the Enskog-Fokker-Planck (EFP) equation
\begin{equation}\label{ch1_eq:boltzmann-fokker-planck}
    \frac{\partial f(\bm{v},t)}{\partial t} = \nabla_{\bm{v}}^{\sf{T}}
    \left[ \bm{A}(\bm{v},t)f(\bm{v},t)
    + \mathbb{D}(\bm{v},t)\nabla_{\bm{v}}f(\bm{r},t)\right] + J_{\alpha}[\bm{v}|f,f].
\end{equation}
In the above, the first term on the right-hand side stems from the Fokker-Planck equation \eqref{ch1_eq:fokker-planck}---taking into account the interaction with the background bath and the action of external fields, while the new term $J_{\alpha}[\bm{v}|f,f]$ constitutes the Enskog collision operator, which accounts for the---either elastic or inelastic---binary collisions between hard-sphere Brownian particles, 
\begin{align}\label{ch1_eq:boltzmann}
    J_{\alpha}[\bm{v}_1|f,f] \equiv  \ \sigma^{d-1} g(\sigma) \!\! \int \!\! &d \bm{v}_2 \! \int \!\!
    d \widehat{\bm{\sigma}}\ \Theta (\bm{v}_{12}^{\sf{T}}\ \widehat{\bm{\sigma}})  \bm{v}_{12}^{\sf{T}}\ \widehat{\bm{\sigma}} \;\!\! \nonumber
    \\
    & \times \left[f(\bm{v}_1',t)f(\bm{v}_2',t)\!-\!f(\bm{v}_1,t)f(\bm{v}_2,t)\right].
\end{align}
In the latter, the unit vector $\widehat{\bm{\sigma}}$ accounts for the direction that connects the two centres of the spheres from particle $1$ to particle $2$, $g(\sigma) = \lim_{r\rightarrow \sigma^+} g(r)$ is the contact value of the pair correlation function $g(r)$, $\Theta$ is the Heaviside function, and ($\bm{v}_1$, $\bm{v}_2$) and ($\bm{v}_1'$, $\bm{v}_2'$) are the pre-collisional and post-collisional velocities of two colliding Brownian particles, respectively, which satisfy the collision rule
\begin{equation}\label{ch1_eq:collision-rules}
    \bm{v}_1' = \bm{v}_1 - \frac{1+\alpha}{2}(\bm{v}_{12}^{\sf{T}}\ \widehat{\bm{\sigma}})\widehat{\bm{\sigma}}, \quad \bm{v}_2' = \bm{v}_2 + \frac{1+\alpha}{2}(\bm{v}_{12}^{\sf{T}}\ \widehat{\bm{\sigma}})\widehat{\bm{\sigma}},
\end{equation}
with $\bm{v}_{12} \equiv \bm{v}_1 - \bm{v}_2$ being the relative velocity. The parameter $\alpha \in [0,1]$ is the normal restitution coefficient, which accounts for the degree of inelasticity of the collisions---i.e. how much energy is dissipated in collisions~\cite{brilliantov_kinetic_2004}. Specifically, after each collision, the variation of the kinetic energy $\Delta H$ reads
\begin{equation}\label{ch1_eq:var-kin-en}
    \Delta H \equiv \frac{1}{2}m(\bm{v}_1'^2 + \bm{v}_2'^2 - \bm{v}_1^2 - \bm{v}_2^2) = -\frac{1}{4}m(1-\alpha^2)(\bm{v}_{12}^{\sf{T}}\ \widehat{\bm{\sigma}})^2 \leq 0,
\end{equation}
such that the rate of energy dissipation is proportional to $1-\alpha^2$. The case $\alpha = 1$ is specially appealing, as it constitutes the elastic limit in which binary collisions conserve the kinetic energy. In the elastic case, the Enskog collision operator $J_1[\bm{v}|f,f]$ admits the equilibrium Maxwellian distribution
\begin{equation}\label{ch1_eq:maxwellian}
    f_{\text{eq}}(\bm{v}) =n_{\text{par}}\left(\frac{m}{2\pi k_B T_{\text{s}}}\right)^{d/2} \exp \left(-\frac{mv^2}{2k_B T_{\text{s}}}\right),
\end{equation}
as its eigenfunction with null eigenvalue, with $T_{\text{s}}$ being the stationary value of the bath temperature. Provided that the fluctuation-dissipation relation~\eqref{ch1_eq:fluctuation-dissipation} is satisfied for $f_{\text{eq}}(\bm{v})$---with $H(\bm{v})= mv^2/2$ corresponding to the kinetic energy in this case, the equilibrium Maxwellian VDF constitutes the only stationary solution of the EFP equation.

Let us note that, for $\alpha \neq 1$, energy is continuously dissipated in collisions, and such kinetic systems constitute intrinsically non-equilibrium systems that do not attain the canonical equilibrium distribution in the long term. The latter does not rule out the tendency towards a non-equilibrium stationary state, but it depends on how the dissipative dynamics is balanced out with the energy injection mechanism---for EFP systems, the Fokker-Planck thermal fluctuations. In chapter~\ref{ch:glass_transition} we study a specific system, the uniformly driven granular gas, in which such balance allows the system to reach a stationary state for long enough times.

\section{Glassy behaviour and memory effects}\label{ch1_sec:glass}
Characterising the dynamical behaviour of mesoscopic systems is a current challenge in non-equilibrium statistical mechanics. In this regard, the existence of global attractors of the dynamics constitutes a key element. On the one hand, these may belong to the class of stationary states, which can be either equilibrium or non-equilibrium steady states (NESS), and generally correspond to the long-time limiting behaviour of many physical systems. On the other hand, these attractors may correspond to time-dependent ``hydrodynamic'' states, in the sense that the dynamics of the system is fully accounted by a reduced set of variables, which are referred to as ``hydrodynamic'' or ``macroscopic''. As an example of the latter, isolated granular gases reach the so-called homogeneous cooling state (HCS). In such state, the description of the system is fully accounted for by the granular temperature $T(t)$, which decays---as a consequence of energy dissipation because of inelastic collisions---as a power-law $T(t)\propto t^{-2}$ known as Haff's law~\cite{haff_grain_1983,brey_homogeneous_1996,poschel_granular_2001,garzo_granular_2019,soria_universal_2012,maynar_fluctuating_2009}.

It is experimentally known for a wide variety of fluids that, when quenching them to low enough temperatures, their crystallisation may be avoided due to the ``freezing'' of their dynamical configurational arrangements in a disordered structure\cite{hunter_physics_2012,biroli_perspective_2013,stillinger_glass_2013,lubchenko_theory_2015,bomont_reflections_2017,weeks_introduction_2017}. In such scenario, the system of concern enters into a metastable supercooled regime---corresponding to a non-equilibrium state. Therein, a dramatic slowing down of the dynamics takes place, which sets the ground for the emergence of a wide variety of non-equilibrium phenomena. This regime is referred to commonly as glassy behaviour~\cite{anderson_anomalous_1972,sherrington_solvable_1975,binder_spin_1986}. The presence of glassy behaviour is usually associated with systems with many strongly interacting degrees of freedom, giving rise to a complex energy landscape with multiple minima separated by energy barriers~\cite{stillinger_glass_2013,lubchenko_theory_2015,nagel_experimental_2017}.

The typical phenomenology of glassy behaviour includes non-exponential relaxation functions for the relevant variables~\cite{williams_non-symmetrical_1970,palmer_models_1984,kob_dynamics_1990,brey_stretched_1993,brey_low-temperature_1996,angell_relaxation_2000,brey_slow_2001,richert_physical_2010,paeng_ideal_2015,lahini_nonmonotonic_2017,kringle_structural_2021,nishikawa_relaxation_2021}. For many macroscopic physical systems, the equilibrium state is determined by the values of a few macroscopic variables---e.g. the volume, pressure and temperature of a gas. Exponential relaxation functions for such variables are typically expected in quasi-equilibrium situations---i.e. when the system is slightly perturbed from equilibrium, since the evolution equations of the system can be linearized on the separation to the equilibrium state. Maybe they are exactly linear on the variables themselves, and these alone completely determine the state of the system---for instance, we put forward Newton's cooling law for the temperature, $\dot{T}\propto -T$~\cite{sarafian_alternate_2021}. However, in general non-equilibrium states, the system may not be characterised entirely by these macroscopic variables. In this case, their evolution equations may depend on additional variables, which can make the overall beahaviour of the system depend on its entire prior thermal history, thus giving rise to complex, non-exponential, relaxation responses. 

Among other relevant glassy features in non-equilibrium physics we include memory effects. These relate to how systems retain a record of their prior history, which significantly influences their subsequent dynamical behaviour, and are intimately related to the phenomenon of ageing~\cite{cugliandolo_evidence_1994,keim_memory_2019}. It is said that a system displays ageing when its relaxation dynamics is not invariant under a time shift, when being previously aged for a long enough waiting time. Memory effects thus emerge in physical systems when their time evolution depends on how they were initially prepared---or, equivalently, how they were previously aged. In chapter~\ref{ch:memory_effects}, we specifically study in detail the emergence of two memory effects of interest within a specific physical system: the Mpemba and Kovacs effects, which we thoroughly introduce below.

\subsection{Mpemba effect}\label{ch1_subsec:mpemba}
It was already known in Ancient Greece that, in order to freeze water fast, it is better to heat it up first~\cite{aristotle_meteorology}. It was not until 1963 when high-school student Erasto B. Mpemba brought the attention of the scientific community to this phenomenon---while making ice cream~\cite{mpemba_cool_1969}. This effect, which contradicts Newton's cooling law, was named after the young scientist as the Mpemba effect, and it has attracted the attention of the non-equilibrium statistical mechanics community in the last decade. The current approach for the Mpemba effect consists in the following: given two samples of the same system at different temperatures, for which both of them relax towards a common stationary state at a lower temperature, then the sample that is initially further from the stationary state---i.e. the initially hotter sample---reaches it faster than the initially closer one---i.e. the initially colder sample~\cite{gomez_mpemba_2021,patron_nonequilibrium_2023,santos_mpemba_2024}. We could also have the opposite situation, in which the initially colder sample heats sooner than the initially hotter one towards a common stationary state at a higher temperature, which gives rise to the inverse Mpemba effect. Despite the obvious resemblance to the original Mpemba effect, we regard both memory effects as independent, as in many cases the mechanisms involved in the relaxation towards a stationary temperature from hotter or colder samples might be asymmetric~\cite{gal_precooling_2020,kumar_exponentially_2020,lapolla_faster_2020,ibanez_heating_2023}. Nevertheless, both effects have been observed in a wide variety of systems~\cite{lu_nonequilibrium_2017,lasanta_when_2017,klich_mpemba_2019,torrente_large_2019,baity-jesi_mpemba_2019,biswas_mpemba_2020,mompo_memory_2021,santos_mpemba_2020,biswas_mpemba_2023,takada_mpemba_2021,gomez_gonzalez_mpemba-like_2021,santos_mpemba_2024}.

There are two main approaches that have been employed in theoretical studies in order to characterise the Mpemba effect: the entropic~\cite{lu_nonequilibrium_2017}---or stochastic---and the thermal~\cite{lasanta_when_2017}---or kinetic---approaches. Although they may follow the same mathematical formalism, they differ in the choice of the magnitude accounting for the distance to the stationary state. In the entropic approach, such distance is defined in terms of a functional of the PDF---e.g. the Kullback-Leibler divergence~\cite{kullback_information_1951}, while in the thermal approach it corresponds to the kinetic temperature, which is proportional to the average kinetic energy of the system.

On the one hand, following the entropic approach, one starts by considering that the system of concern may be modelled via a Markovian stochastic process for the relevant variables $\bm{r}$, such that its description is accurately accounted by a PDF $P(\bm{r},t)$. The time evolution of such distribution is governed by either a Fokker-Planck equation, as given by Eq.~\eqref{ch1_eq:fokker-planck}, or an analogous equation.\footnote{For example, there are many studies whose starting point constitutes a master equation for the PDF, which lies out of the traditional scope of the situations considered in this thesis.} Then, distance to the stationary state $P_{\text{s}}(\bm{r})$ is defined in terms of a functional of our PDF, for instance, the Kullback-Leibler divergence
\begin{equation}\label{ch1_eq:kullback-leibler}
    D_{\text{KL}}[P|P_{\text{s}}] \equiv \int d\bm{r} P(\bm{r},t)\ln \left(\frac{P(\bm{r},t)}{P_{\text{s}}(\bm{r})} \right),
\end{equation}
or other distances in probability space, such as the $\mathcal{L}^1$ and $\mathcal{L}^2$ norms. Assuming that the corresponding Fokker-Planck operator~\eqref{ch1_eq:fp-operator} is self-adjoint, we may expand $P(\bm{r},t)$ in eigenmodes, as depicted in Eq.~\eqref{ch1_eq:eigen-expansion}. The entropic Mpemba effect takes place when, under certain conditions, the amplitude of the slowest, non-zero relaxation mode $\lambda_1$ depends on the temperature as a non-monotonic function. Moreover, it is said in this context that the entropic Mpemba effect is strong if, by appropriately choosing the initial conditions and system parameters, the coefficient $\mathcal{A}_1$ corresponding to the eigenvalue $\lambda_1$ vanishes, and thus the next eigenvalue $\lambda_2$ dominates the relaxation dynamics, which thus becomes exponentially faster~\cite{lu_nonequilibrium_2017,klich_mpemba_2019,kumar_exponentially_2020}.

\begin{figure}
    \centering
    \includegraphics[width=3.5in]{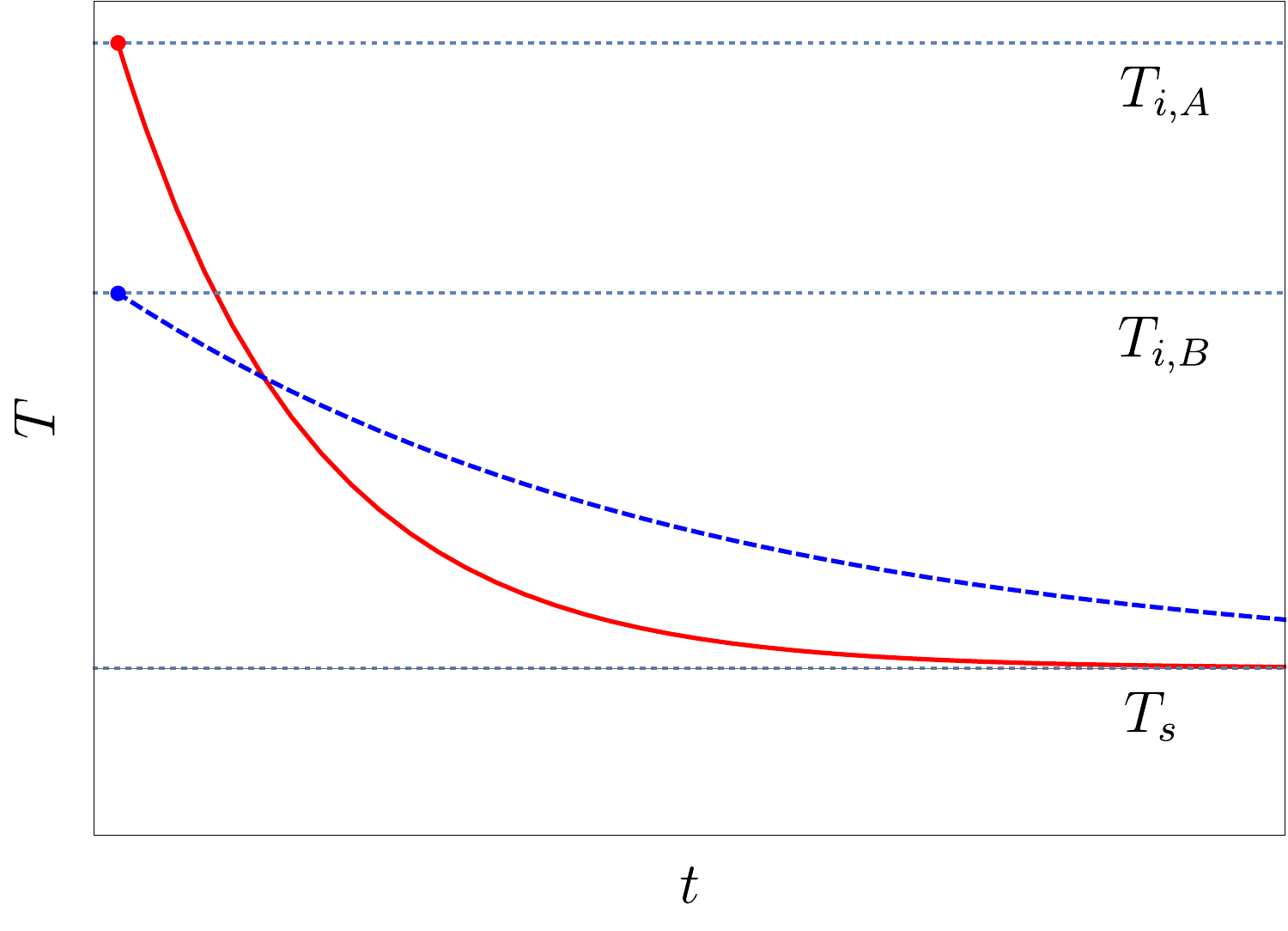}
    \caption{\label{ch1_fig:mpemba-sketch} Qualitative sketch of the Mpemba memory effect, following the thermal approach. The hotter sample A, with initial temperature $T_{i,A}$, is further away from the stationary state at the common bath temperature $T_{\text{s}}$ than the colder sample B, with initial temperature $T_{i,B} < T_{i,A}$. The thermal Mpemba effect emerges when the time evolution of the initially hotter sample (red solid line) overtakes that of the initially colder one (blue dashed).
      }
  \end{figure}

On the other hand, the thermal approach for the Mpemba effect is the one we follow in this thesis---specifically, in chapter~\ref{ch:memory_effects}. Originally introduced for a granular gas~\cite{lasanta_when_2017}, the thermal approach is based on describing the system in terms of the one-particle VDF $f(\bm{v},t)$---which evolves following a kinetic-like equation such as the EFP equation~\eqref{ch1_eq:boltzmann-fokker-planck}, and monitor the relaxation towards the stationary state via the kinetic temperature, which is much closer to an experimentally measurable observable than the Kullback-Leibler distance~\eqref{ch1_eq:kullback-leibler} used in the entropic approach. Figure~\ref{ch1_fig:mpemba-sketch} shows a qualitative sketch of the Mpemba effect following the thermal approach.\footnote{Let us remark that the qualitative sketch would be similar had we employed the entropic approach, upon substituting the temperatures $T$ with the appropriate values of the functional of the PDF.} In order for the thermal Mpemba effect to emerge---i.e. for the red, solid and blue, dashed lines to cross---it is required to have previously prepared at least one of the samples in a non-equilibrium state serving as the initial condition for the dynamics, in which the values of other physical quantities different to the temperature are relevant. The strength of the thermal Mpemba effect---and also for the Kovacs effect, following the same approach later on---is measured by how such initially prepared non-equilibrium state differs from a stationary state at the same temperature. For the case of kinetic fluids, the excess kurtosis $a_2$ accurately accounts for such deviations---see chapter~\ref{ch:memory_effects} for further details. The thermal approach has been employed in a wide range of systems, including granular gases, constituted by either smooth~\cite{lasanta_when_2017} or rough~\cite{torrente_large_2019} inelastic hard spheres, molecular fluids with non-linear drag forces~\cite{santos_mpemba_2020,megias_thermal_2022}, and spin glasses~\cite{baity-jesi_mpemba_2019}---where it is the internal energy the magnitude that displays the Mpemba crossing.

Lastly to mention, regardless of the employed approach, whenever the Mpemba effect takes place, the sample of the system that is initially further from the stationary state takes a shortcut, thus relaxing faster than the initially closer one towards the stationary. Following this, there emerges a natural connection with the general field of \textit{swift state-to-state transformations} (SST)~\cite{guery-odelin_driving_2023}, which is thoroughly introduced in Sec.~\ref{ch1_sec:shortcuts} and lies at the core of the problems studied in chapters~\ref{ch:thermal_brachistochrones} and \ref{ch:brownian_gyrator} from the second part of this thesis. 

\subsection{Kovacs effect}\label{ch1_subsec:kovacs}

The second relevant memory effect that we study here is the so-called Kovacs effect, which was first reported by A. J. Kovacs when studying the relaxation of the volume for polymeric glasses~\cite{kovacs_transition_1963,kovacs_isobaric_1979}. Let us consider a physical quantity $K$ for a system in contact with a thermal bath. Given a fixed value of the bath temperature $T$, the quantity $K$ attains a stationary value denoted by $K_{\text{s}}(T)$, which we assume to be a monotonous function of $T$---the higher the temperature, the larger the stationary value $K_{\text{s}}(T)$. The Kovacs effect stems from the non-monotonous response of the system to the two-jump protocol for the bath temperature that we describe below.

\begin{figure}
    \centering
    \includegraphics[width=3.5in]{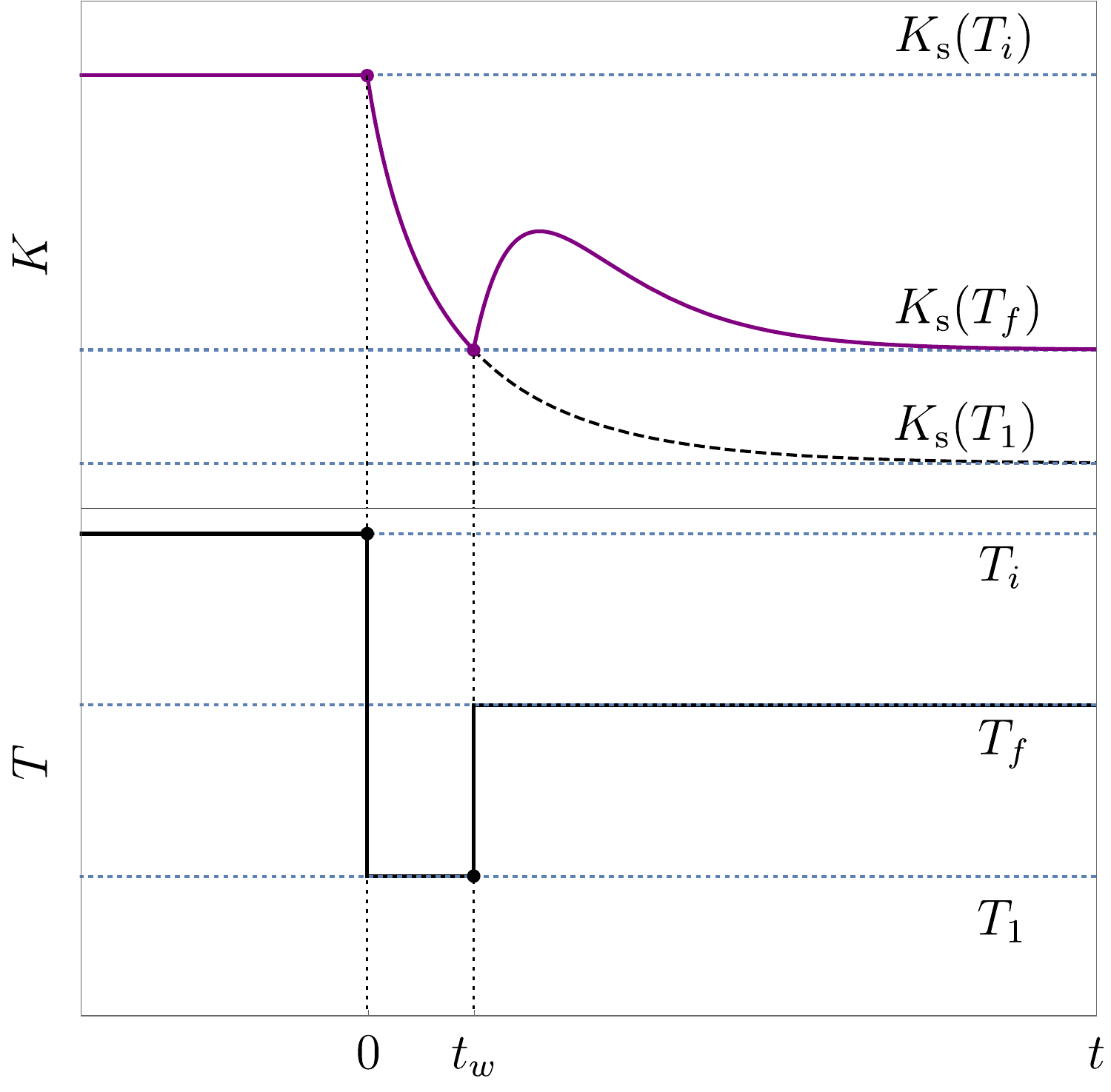}
    \caption{\label{ch1_fig:kovacs-sketch} Qualitative sketch of the Kovacs effect. Specifically, we plot the time evolution of a physical observable $K$ in the top panel, when submitting the system to the two-jump protocol for the bath temperature shown in the bottom panel. The relaxation from $T_i$ to $T_1$ (dashed line) is interrupted after a waiting time $t = t_w$, when the observable $K$ has its stationary value at the final temperature $T_f$, $K(t_w) = K_{\text{s}}(T_f)$. However, $K(t)$ deviates from $K_{\text{s}}(T_f)$ in a non-monotonic way, by reaching a maximum before returning thereto. The latter shows the need of additional variables in order to completely determine the state of the system.}
  \end{figure} 

We start by driving the system to a stationary state at temperature $T_i$ for $t<0$. Then, at $t=0^+$, we suddenly quench the bath temperature to a new value $T_1 < T_i$, and let the system relax within the time window $0<t<t_w$. At $t=t_w$, we measure the instantaneous value of $K$, $K(t_w)$, and abruptly change once again the bath temperature to the value $T_f$ such that its corresponding stationary value is the one that has been measured, that is, $K(t_w) = K_{\text{s}}(T_f)$---thus having $T_1 < T_f < T_i$. The Kovacs effect emerges when, for $t>t_w$, $K$ departs from the corresponding stationary value $K_{\text{s}}(T_f)$ and evolves in a non-monotonic form, before reaching again such value for long enough times. The emergence of the Kovacs effect implies that the pair of variables $(T,K)$ does not suffice to determine the state of the system, i.e. there are more ``hidden" variables at play. In Fig.~\ref{ch1_fig:kovacs-sketch}, we display a sketch of the aforementioned two-jump protocol---also known as the Kovacs protocol---and the associated Kovacs hump for the system.

Apart from polymeric glasses, the Kovacs effect has been repeatedly reported in a wide variety of contexts~\cite{berthier_surfing_2002,buhot_kovacs_2003,bertin_kovacs_2003,arenzon_kovacs_2004,cugliandolo_memory_2004,mossa_crossover_2004,aquino_kovacs_2006,prados_kovacs_2010,bouchbinder_nonequilibrium_2010,diezemann_memory_2011,chang_kovacs_2013,ruiz-garcia_kovacs_2014,prados_kovacs-like_2014,trizac_memory_2014,brey_memory_2014}. For systems displaying glassy behaviour, the Kovacs hump is often attributed to the complex energy landscape, typical for those systems. Nevertheless, the Kovacs effect has also been observed in systems presenting a much simpler energy landscape, such as the granular gas or the molecular fluid we consider in chapters~\ref{ch:memory_effects} and \ref{ch:glass_transition}. Therein, the energy landscape corresponds to the kinetic energy, which only presents one energy minimum.
  
For non-equilibrium systems described by a master equation, there are general results for the shape of the Kovacs hump in the linear response regime~\cite{prados_kovacs_2010}---i.e. when submitting the system to a small perturbation around the stationary state. Assuming that (i) the stationary state corresponds to an equilibrium one, for which the stationary solution corresponds to the equilibrium canonical distribution $P_{\text{eq}}(\bm{r})\propto e^{-\beta H(\bm{r})}$, and (ii) that the detailed balance condition---which is similar to the fluctuation-dissipation relation \eqref{ch1_eq:fluctuation-dissipation} for Fokker-Planck systems---is satisfied with respect to $P_{\text{eq}}(\bm{r})$, then the form of the Kovacs hump for the average energy $E(t) \equiv \left<H \right>(t)$ is related to that for the ``direct'' relaxation function from $T_i$ to $T_f$---i.e. following a one-jump protocol. In fact, from its explicit expression, one deduces that: (i) the Kovacs hump is always positive, i.e. $E(t) \geq E_{\text{eq}}(T_f)$ and (ii) it only presents one maximum. 

  Although there are plenty of studies of the Kovacs effect belonging to the non-linear regime---large values for the bath temperature jumps, the aforementioned behaviour described in linear response theory has been found in many systems~\cite{kovacs_transition_1963,kovacs_isobaric_1979,berthier_surfing_2002,buhot_kovacs_2003,bertin_kovacs_2003,cugliandolo_memory_2004,arenzon_kovacs_2004,mossa_crossover_2004,aquino_kovacs_2006,prados_kovacs_2010,bouchbinder_nonequilibrium_2010,diezemann_memory_2011,chang_kovacs_2013,ruiz-garcia_kovacs_2014,peyrard_memory_2020,mandal_memory_2021}. Hence, the terming ``normal Kovacs effect" has been coined to those scenarios in which the relevant observable $K(t)$ accounting for the Kovacs hump is always positive---within the time window $t>t_w$---and it only presents one maximum.\footnote{This normal Kovacs hump stems from the structure of the direct relaxation function which, in the linear response regime, can be written as a sum of exponentially decreasing modes with positive coefficients.} All the other scenarios differing from the normal Kovacs effect---i.e. negative Kovacs function $K(t)\leq 0$~\cite{prados_kovacs-like_2014,trizac_memory_2014}, multiple extrema~\cite{lasanta_emergence_2019}...---are referred to as ``anomalous'' Kovacs responses.

\subsection{Glass transition and hysteresis cycles}\label{ch1_subsec:glass-transition}

As mentioned earlier, a wide class of fluids enters into a metastable supercooled state---which differs from the crystallised, solid state---when being submitted to low enough temperatures. Such state involves a dramatic slowing down of the dynamics that prevents the system from reaching equilibrium on experimental timescales, since it gets trapped in a local minimum of the energy landscape, with high energy barriers and not enough thermal energy to surpass them. Above the melting point temperature $T_m$ in a solid-liquid transition, density fluctuations of the liquid typically relax on a timescale of the order of picoseconds. However, in the metastable supercooled regime, relaxation times increase exponentially, such that they become up to 14 orders of magnitude longer when the temperature is around $2T_m/3$~\cite{debenedetti_supercooled_2001}. At this stage, the liquid does not flow anymore, and it is then said that the glass transition takes place: configurational rearrangements cease, the liquid structure becomes ``frozen"  and the system  gets trapped in a non-equilibrium state that is solid, yet disordered, called the glassy state~\cite{angell_formation_1995,dyre_colloquium_2006,berthier_theoretical_2011,wolynes_structural_2012,hunter_physics_2012,biroli_perspective_2013,stillinger_glass_2013,lubchenko_theory_2015,bomont_reflections_2017,weeks_introduction_2017,dauchot_glass_2022,novikov_temperature_2022,barrat_computer_2023,berthier_modern_2023}.

Despite the great effort devoted to the research of glassy systems in the past decades, the glass transition continues to be an open problem in the non-equilibrium statistical mechanics community. There is not yet a conclusive answer to the fundamental query of whether the glass transition constitutes a purely dynamical, {kinetic}, phenomenon or it is instead the consequence of an underlying phase transition---as it has been predicted in certain theoretical studies~\cite{berthier_theoretical_2011,wolynes_structural_2012,stillinger_glass_2013,bomont_reflections_2017,dauchot_glass_2022,novikov_temperature_2022}. Many analyses have addressed the rich phenomenology that {accompanies} the glass transition from different and complementary perspectives. Specifically, spin models emphasise the characterisation of potential energy landscapes with numerous energy minima connected by complex dynamics pathways~\cite{nishikawa_relaxation_2022,ros_complex_2019,ediger_glass_2021}, while kinetically constrained models point towards the fact that relaxation events are cooperative because of the presence of geometric frustration~\cite{ritort_glassy_2003,tarjus_frustration-based_2005}. Nevertheless, the overall development of a successful theory to explain all the phenomenological observations in a unified and satisfactory manner is still in progress.

\begin{figure}
    {\centering 
    \includegraphics[width=2.65in]{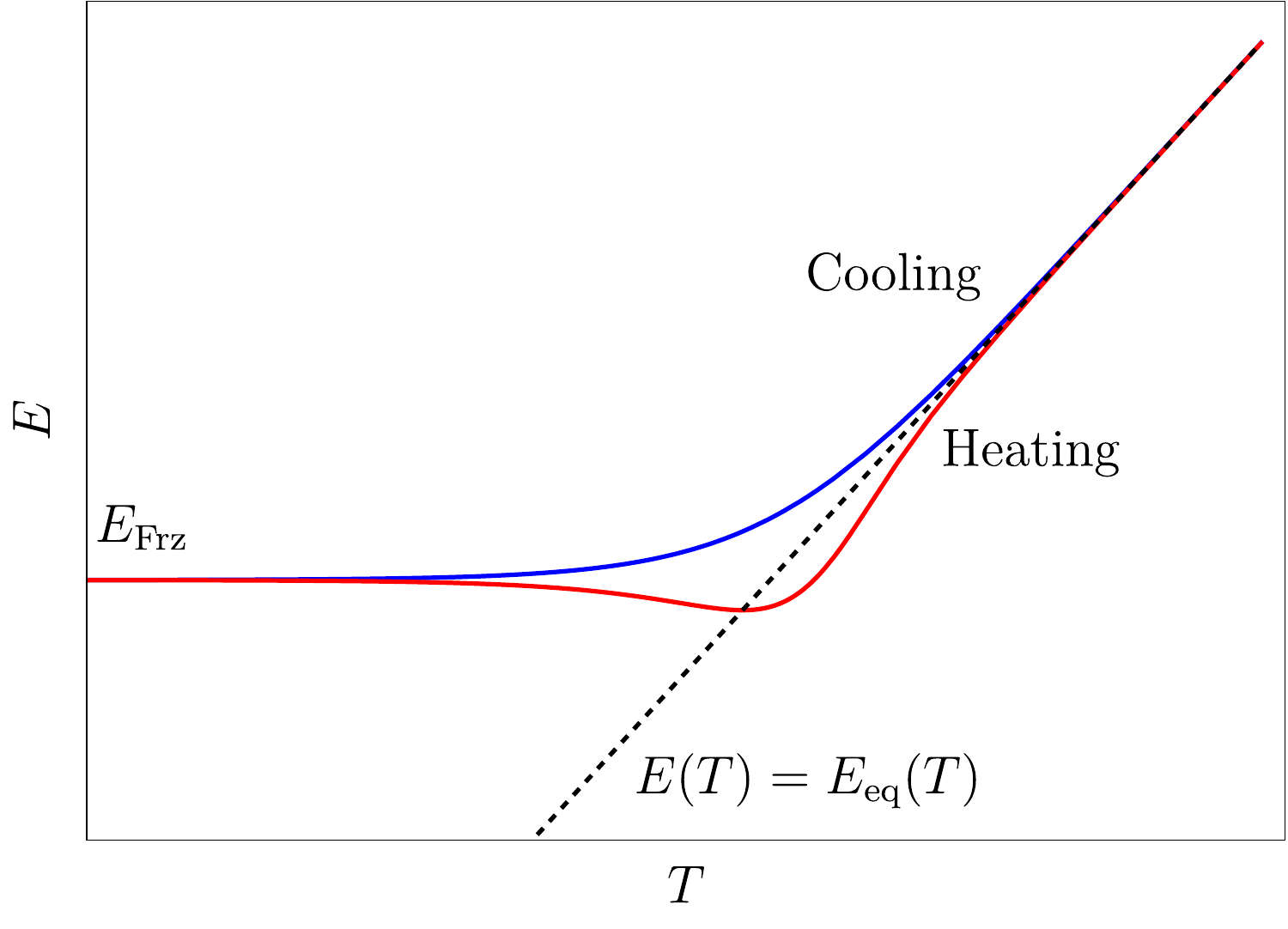}
    \includegraphics[width=2.65in]{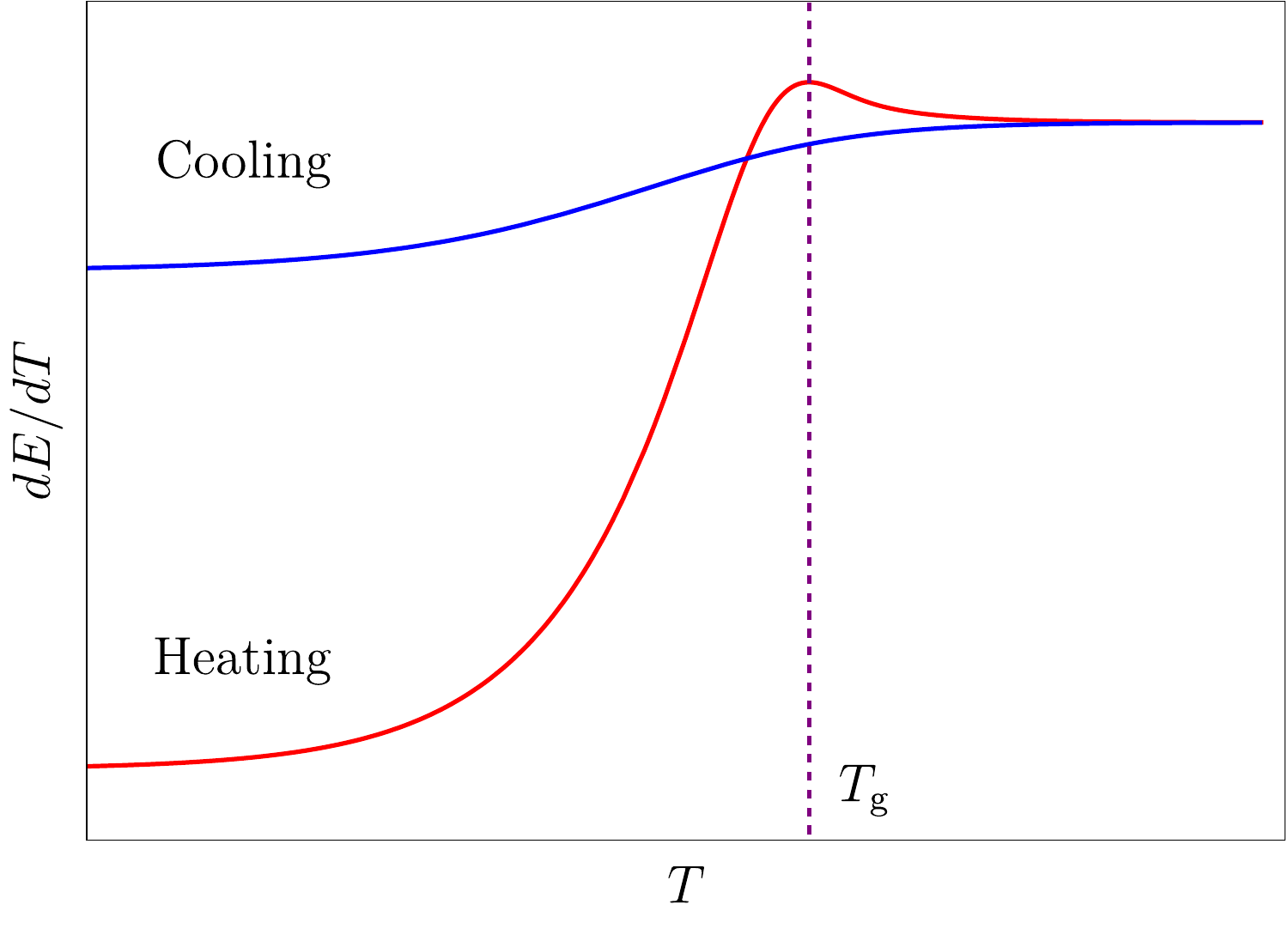}}
    \caption{Sketch of the hysteretic behaviour for a physical system described in terms of the average energy $E(T)$. Left panel corresponds to the hysteresis cycle: a cooling protocol (blue line) followed by a reheating one (red line)---while the right panel shows the behaviour of the apparent heat capacity $dE/dT$ along the cycle with the same colour code. The black, dashed line in the left panel corresponds to the equilibrium curve $E(T) = E_{\text{eq}}(T)$. The purple, dashed line on the right panel accounts for the temperature $T_g$ where $dE/dT$ reaches a maximum along the heating protocol.}
        \label{ch1_fig:hysteresis}
\end{figure}

Glassy systems exhibit hysteretic behaviour when submitted to cooling protocols followed by reheating. In the following discussion and in Fig.~\ref{ch1_fig:hysteresis}, we exemplify the observed behaviour with the average energy $E(t)$, but one must bear in mind that other physical quantities might be the relevant ones. We consider the system to be initially at equilibrium at a sufficiently high temperature $T$, which is then continuously cooled down at a constant rate $r_c$. For high values of $T$, the relaxation timescale $\tau$ is much shorter than the cooling time $r_c^{-1}$, such that the system does not deviate from the instantaneous equilibrium curve, $E(t) = E_{\text{eq}}(T)$. As the temperature decreases, $\tau$ increases and $E(t)$ deviates from the $E_{\text{eq}}(T)$, finally reaching a plateau ``frozen'' value $E_{\text{Frz}}$ when $\tau$ exceeds $r_c^{-1}$---even diverging. This purely kinetic phenomenon is termed the kinetic glass transition---in other contexts, it is also referred to as the laboratory glass transition~\cite{ediger_supercooled_1996,gupta_laboratory_2007,debenedetti_supercooled_2001}. The temperature of the {kinetic} glass transition---actually a range of temperatures---at which the system departs from equilibrium and gets frozen increases with the cooling rate $r_c$. Consequently, the properties of the glass depend on the cooling protocol.

For the reheating process, the system starts from the frozen state at sufficiently low values of the bath temperature $T$, which is now continuously heated up at a constant rate $r_h$. For short times, the temperature is still very low, such the relaxation time $\tau \gg r_h^{-1}$ and the system barely departs from the frozen state. As time progresses, and when $\tau$ becomes of the order of $r_h^{-1}$ and keeps decreasing, the system eventually departs from the frozen state and attains the instantaneous equilibrium curve $E(t) = E_{\text{eq}}(T)$. Typically, $E(t)$ overshoots the equilibrium curve before returning thereto~\cite{brey_normal_1993,brey_dynamical_1994}. This entails that the apparent heat capacity $dE/dT$\footnote{We employ the term \textit{apparent} because the system departs from equilibrium and therefore $dE/dT$ is here a dynamical quantity, depending on the rate of variation of the temperature, and it is not equal to the heat capacity in the thermodynamic sense.} displays a non-monotonic behaviour with a marked peak at a certain temperature $T_g$ which, in some contexts, has been employed to mark the temperature $T_g$ at which the glass transition takes place~\cite{angell_formation_1995,dyre_colloquium_2006,gao_calorimetric_2013,tropin_modern_2016,richet_thermodynamics_2021}. Nevertheless, there is not a unique way of defining the glass transition temperature, as we discuss in chapter~\ref{ch:glass_transition}.

\section{Swift state-to-state transformations}\label{ch1_sec:shortcuts}

The term \textit{swift state-to-state transformations} (SST)~\cite{guery-odelin_driving_2023} refers to the collection of techniques employed in order to drive stochastic thermodynamic systems towards a desired target state. It is customary to pay attention to those drivings that, apart from attaining the task, accelerate the relaxation dynamics of the system in comparison with a reference protocol---typically a direct-step protocol, in which the system relaxes within the ``natural" relaxation timescale of its dynamics.

These techniques were first devised for quantum systems---where the terming \textit{shortcuts to adiabaticity} (STA) was widely used instead~\cite{funo_shortcuts_2020}, in which the original goal consisted in driving the system of concern from an initial energy eigenstate $\left.|n(0)\right>$ of an initial Hamiltonian $\hat{H}(0)$ to the corresponding eigenstate $\left.|n(t_f)\right>$ of a final Hamiltonian $\hat{H}(t_f)$ in a short, finite time $t_f$.\footnote{Although such connection may be achieved by virtue of the adiabatic theorem~\cite{griffiths_introduction_2017}, in which the system remains at the instantaneous eigenstate $\left.|n(t)\right>$ of the Hamiltonian $\hat{H}(t)$ at all times, it only holds for infinitely slow driving processes.} Due to the versatility of the mathematical framework employed in many of these techniques---as most of them apply for systems described in terms of continuity-like equations, it was not long after when these ideas were extended to alternative contexts, such as those within the frameworks of classical mechanics~\cite{patra_shortcuts_2017} and stochastic thermodynamics~\cite{guery-odelin_nonequilibrium_2014}. For the latter, the terming \textit{engineering swift equilibration}~\cite{martinez_engineered_2016} and \textit{shortcuts to isothermality}~\cite{li_shortcuts_2017} are employed in the literature instead for tackling specific problems. The SST techniques constitute a broader class that includes the aforementioned ones. As throughout this thesis we deal with different situations that do not always involve “equilibration” nor ``ısothermality”, we employ the terming SST for the remainder of this and the subsequent chapters.

Our interest is focused on non-equilibrium systems, where the PDF $P(\bm{r},t)$ constitutes the state of the system that we aim at driving from an initial shape $P(\bm{r},0)$ to a final shape $P(\bm{r},t_f)$ in a finite time $t_f$. In some situations though, it is analytically more plausible to work out a closed set of differential equations for the moments or cumulants of $P(\bm{r},t)$---as is the case for the Gaussian processes introduced in Sec.~\ref{ch1_subsubsec:ornstein-uhlenbeck}, without resorting to truncation schemes---instead of directly working with the Fokker-Planck (or master) equation.

\subsection{Brief revision and thesaurus\label{ch1_subsec:thesaurus}}
In the following, we enumerate and describe in a non-exhaustive manner some of the main techniques and concepts employed in the design of shortcuts, specially focusing the discussion on their applications for systems evolving under stochastic dynamics.
\begin{itemize}
    \item \textbf{Inverse engineering}: The typical approach employed in order to tackle problems in physics constitutes that of determining the time evolution of a certain system given a driving protocol for the external agents. In inverse engineering techniques, we follow the opposite approach; our starting point is the specific evolution that we want to impose to the system, and the evolution equations allow us to determine how we must engineer the external drivings in order to achieve it~\cite{guery-odelin_shortcuts_2019}. For instance, for Fokker-Planck systems, we would impose a desired functional form of $P(\bm{r},t)$, and determine how we may engineer either $\bm{A}(\bm{r},t)$, the Hamiltonian $H(\bm{r},t)$ and/or $\mathbb{D}(\bm{r},t)$---as they may be related by means of the fluctuation-dissipation relation~\eqref{ch1_eq:fluctuation-dissipation}---in order for $P(\bm{r},t)$ to be a solution of the corresponding Fokker-Planck equation~\eqref{ch1_eq:fokker-planck}. In fact, if the initial and target states in a connection problem constitute equilibrium---not necessarily Gaussian---states of the same shape---i.e. the Hamiltonians $H_i(\bm{r}) \equiv H(\bm{r},0)$ and $H_f(\bm{r}) \equiv H(\bm{r},t_f)$ belong to the same family of functions, $H_f(\bm{r})=H_i[(\bm{r}-\bm{\mu}_f)/\sigma_f]$ with $\bm{\mu}_f$ and $\sigma_f$ constants---a simple solution may be found by imposing the shape~\cite{plata_taming_2021,guery-odelin_driving_2023}
    \begin{equation}\label{ch1_eq:ansatz}
        P(\bm{r},t) = \frac{1}{\sigma (t) \mathcal{Z}_i}\text{exp}\left[-\beta H_i\left( \frac{\bm{r}-\bm{\mu}(t)}{\sigma(t)} \right) \right], \quad \mathcal{Z}_i = \int d\bm{r}e^{-\beta H_i(\bm{r})},
    \end{equation}
    with $\bm{\mu}(t)$ and $\sigma(t)$ being continuous functions of time, satisfying the initial $\bm{\mu}(0) = \bm{0}$, $\sigma(0) = 1$, final $\bm{\mu}(t_f) = \bm{\mu}_f, \sigma(t_f) = \sigma_f$ and boundary conditions $\dot{\bm{\mu}}(0) =\dot{\bm{\mu}}(t_f)=\bm{0}$, $\dot{\sigma}(0) = \dot{\sigma}(t_f) = 0$. In this case, the time evolution of $H(\bm{r},t)$ is determined by the functions $\bm{\mu}(t)$ and $\sigma(t)$ by introducing Eq.~\eqref{ch1_eq:ansatz} into the corresponding Fokker-Planck equation. 
    \item \textbf{Counterdiabatic (CD) driving}: Adiabatic protocols are those for which, when driving the system sufficiently slowly\footnote{We remark that this is the definition employed within the context of quantum mechanics, and it bears no relation with the thermodynamic definition of null-heat processes.}---by varying for instance the Hamiltonian function, the PDF remains at the instantaneous stationary distribution
    \begin{equation}
        P_{\text{s}}(\bm{r},t) = \frac{1}{\mathcal{Z}(t)}e^{-\beta H(\bm{r},t)}, \quad \mathcal{Z}(t) = \int d\bm{r}e^{-\beta H(\bm{r},t)}.
    \end{equation}
    Following CD methods, we seek to construct an auxiliary term $H_{\text{CD}}(\bm{r},t)$ such that, under the full Hamiltonian function $H_{\text{full}}(\bm{r},t) = H(\bm{r},t) + H_{\text{CD}}(\bm{r},t)$, the system still attains the instantaneous stationary form $P_{\text{s}}(\bm{r},t)$ without the time evolution of $H(\bm{r},t)$ being slow~\cite{guery-odelin_shortcuts_2019}. Let us note that, although the approaches are different in essence, the idea of CD driving may constitute a particular case of inverse engineering methods, specifically when imposing the instantaneous equilibrium form for the probability distribution function $P(\bm{r},t)$.
    \item \textbf{Fast-forward (FF)}: Similar to the CD techniques, the FF approach is based on adding an auxiliary Hamiltonian contribution $H_{\text{FF}}(\bm{r},t)$ to the dynamics. But contrary to the former, following the full Hamiltonian, the system does not remain at the instantaneous stationary distribution at all times, but only at the initial and final ones~\cite{guery-odelin_shortcuts_2019}. The main idea is to consider a certain reference process $P_{r}(\bm{r},t)$---solution of the Fokker-Planck equation, which does not need to be a slow---adiabatic---one, connecting two given states. Then, a time distortion $\Lambda (t)$ is considered for such reference, such that $P(\bm{r},t) = P_{r}(\bm{r},\Lambda (t))$, and we seek for the additional contribution $H_{\text{FF}}(\bm{r},t)$ that we need to supply the system with in order to follow the prescription $P_{r}(\bm{r},\Lambda (t))$~\cite{plata_taming_2021}. Following this approach, the time-frames of the reference process are fixed, but played at a different rate when applying the FF driving protocol; either faster $(\dot{\Lambda} > 1)$, slower $(0<\dot{\Lambda}<1)$ and even backwards in time $(\dot{\Lambda}<0)$. One may thus combine simple reference processes with different time distortions in order to build a connection between arbitrary initial and final distributions. Let us remark that the FF approach converges to the CD driving in the limit in which the duration of the reference process tends to infinity, such that it becomes adiabatic.
    \item \textbf{Optimal control theory (OCT)}: So far, in the preceding approaches we have dealt with the matter of how to connect arbitrary states in a finite time. In principle, there are many ways of designing shortcuts that allow the desired connection---i.e. there are multiple choices of either the CD and FF Hamiltonian contributions achieving the task. It is therefore natural to further ask which driving protocols, among all the possible ones connecting two states, minimise the connection time $t_f$ between states or the energetic cost---e.g. irreversible work, heat, entropy production---required for the task. These questions lie at the core of OCT~\cite{pontryagin_mathematical_1987,liberzon_calculus_2012}, which constitutes the main venue of SST employed in our research, and which we thoroughly describe in Sec.~\ref{ch1_subsec:control-theory}.

    \item \textbf{Speed limits}: The acceleration of the dynamics entailed by the shortcuts to adiabaticity always comes at a price. And this price becomes relatively high when considering optimal protocols: generally speaking, in order to optimise a certain quantity, the more resources we need to invest.
    To put forward an example, the minimum irreversible work $W_{\text{irr}}$ in non-equilibrium processes is typically proportional to the inverse of the connection time $t_f$, thus diverging for sufficiently short values of the latter~\cite{schmiedl_optimal_2007}. These trade-off relations between different quantities involved in driving protocols are encapsulated in the so-called speed limits~\cite{ito_stochastic_2018,shiraishi_speed_2018,nicholson_timeinformation_2020,ito_stochastic_2020}. These date back to Schrödinger's time-energy uncertainty relation in quantum mechanics~\cite{griffiths_introduction_2017}, although they have been extended over the years to other contexts, such as Markov systems~\cite{ito_stochastic_2018,shiraishi_speed_2018}. Moreover, there are also approaches resorting to information geometry concepts that hold for general dynamics~\cite{ito_stochastic_2020,nicholson_timeinformation_2020}, which we briefly explore in Sec.~\ref{ch1_subsec:speed-limits}.
\end{itemize}


\subsection{Optimal control theory}\label{ch1_subsec:control-theory}
As briefly mentioned earlier, SST techniques concern the problem of externally driving the system between given initial and target states. Our interest goes now to those situations in which, assuming that we may exert some degree of control on the system---such that we succeed in connecting arbitrary states, we want to determine which connections are optimal.

Optimisation problems are ubiquitous in physics. Furthermore, optimality constitutes a universal principle in nature, as most of the fundamental physical theories are someway governed by the solutions of optimisation problems. In the following, we briefly illustrate this statement by mentioning some paradigmatic examples. Both classical mechanics and quantum field theory can be formally built upon the minimisation of an action functional~\cite{landau_mechanics_1976,srednicki_quantum_2007}. In general relativity, particles move along the geodesics, which are the paths that minimise the spacetime distance between points~\cite{hobson_general_2006}. In geometrical optics, light rays follow the paths in a medium that take the shortest time---what is known as Fermat's principle, from which Snell's laws of reflection and refraction follow~\cite{born_principles_1999}. Phase transitions in materials can be understood from the minimisation of a free energy functional within Landau's theory~\cite{landau_theory_1937}.

The aforementioned optimisation problems fall under the umbrella of variational calculus: they constitute problems in which we search for the function that optimises---i.e. either minimises or maximises---a certain functional. Optimisation problems of the variational type date back centuries. To provide some historical examples, we have Dido's isoperimetric problem (around 850 B.C), Galileo Galilei's catenary problem (1630s) or Daniel Bernoulli's brachistochrone---curve that takes the shortest time---problem\footnote{Brachistochrone problems lie at the core of the second part of this thesis.} (1696). It was not until 1756 when Leonhard Euler and Joseph-Louis Lagrange set the analytical foundations of what is currently known as calculus of variations~\cite{liberzon_calculus_2012}. They jointly determined that the solutions of a given variational problem must obey a system of second-order differential equations known as the Euler-Lagrange equations, provided suitable initial and boundary conditions. For the case of classical mechanics, the Euler-Lagrange equations reduce to Newton's second law of motion, while for quantum field theory, they reduce to the field equations---Klein-Gordon or Dirac equations. Despite the clear versatility in tackling different optimisation problems, variational calculus suffers from two main drawbacks: 
\begin{enumerate}
    \item First, as the Euler-Lagrange equations constitute second-order differential equations, the solutions must be twice-differentiable functions, thus leaving out of the picture continuous functions with discontinuous derivatives, or purely discontinuous functions. 
    \item And secondly, it does not deal with non-holonomic constraints---i.e. those involving inequalities. For instance, the fact that solutions may be restricted to a certain subset of the overall space.
\end{enumerate}
These drawbacks were finally dealt with around the 1950s, when Richard Bellman in the United States and Lev Pontryagin in the Soviet Union independently developed new mathematical methods that extended the applicability of calculus of variations to a broader class of optimisation problems~\cite{liberzon_calculus_2012}. These methods correspond to Bellman's Dynamic Programming, on the one hand, and Pontryagin's Maximum Principle (PMP), on the other hand. Both approaches provide equivalent methods that set the founding ground of optimal control theory. In analogy with classical mechanics, i.e. by drawing an equivalence between variational calculus and Lagrangian mechanics, the Bellman and Pontryagin methods would correspond to the Hamilton-Jacobi and Hamiltonian formulations, respectively. In the following, we will focus on PMP, which corresponds to the main tool that we employ for tackling the optimisation problems in chapters~\ref{ch:thermal_brachistochrones} and \ref{ch:brownian_gyrator}.

\subsubsection{Pontryagin's Maximum Principle}
We consider a $d$-dimensional system determined by a set of dynamical variables $\bm{z}^{\sf{T}} \equiv (z_1,z_2,...,z_d)$, which obey the dynamic system of equations
\begin{equation}\label{ch1_eq:dynamical-system}
    \dot{\bm{z}} = \bm{f}(\bm{z};\bm{k}),
\end{equation}
with $\bm{f}^{\sf{T}} \equiv (f_1,f_2,...,f_d)$ being the evolution equations and $\bm{k}^{\sf{T}} \equiv (k_1,k_2,...k_M)$ the set of $M$ controllable parameters affecting the dynamics. The values that these parameters may attain can be bounded, such that they belong to a set known as the control set, defined in the $M$-dimensional space and denoted here by $U$. Let us remark that both the dynamical variables and the control parameters are functions of time, $\bm{z} = \bm{z}(t), \bm{k} = \bm{k}(t)$, and in order for Eq.~\eqref{ch1_eq:dynamical-system} to be complete, we need to provide $\bm{z}(t)$ with suitable initial conditions. Now, we are interested in optimising a cost functional of the form
\begin{equation}\label{ch1_eq:functional}
    J[\bm{z};\bm{k}] \equiv \int_0^{t_f}dt \ \mathcal{L}(\bm{z};\bm{k}),
\end{equation}
which is a functional of the control parameters $\bm{k}$, with $\mathcal{L}$ being commonly known as the Lagrangian function---analogously to the original Lagrangian from variational calculus. We may thus formulate our control problem in the following way: given the dynamical system from Eq.~\eqref{ch1_eq:dynamical-system}, we are looking for the optimal protocols $\bm{k}^*(t), \ 0\leq t\leq t_f$ that optimise the functional from Eq.~\eqref{ch1_eq:functional}. The connection time $t_f \in [0,+\infty)$ is free, i.e. we do not fix the time at which the protocol ends. In order to do so, we introduce an additional variable $z_0$ with $z_0(0) = 0$ such that
\begin{equation}
    \dot{z}_0 = \mathcal{L}(\bm{z};\bm{k}).
\end{equation}
Let us note that $z_0(t_f) = J[\bm{z};\bm{k}]$, according to Eq.~\eqref{ch1_eq:functional}. Next, we introduce the momenta $\psi_0$ and $\bm{\psi}^{\sf{T}} \equiv (\psi_1,\psi_2,...,\psi_d)$, which correspond to the conjugate variables to $z_0$ and $\bm{z}$, respectively, and the so-called Pontryagin's Hamiltonian function $\Pi (\bm{z},\psi_0,\bm{\psi},\bm{k})$, which is given by
\begin{equation}\label{ch1_eq:pontryagins-hamiltonian}
    \Pi (\bm{z},\psi_0,\bm{\psi};\bm{k}) \equiv \psi_0 \mathcal{L}(\bm{z};\bm{k}) + \bm{\psi}^{\sf{T}} \bm{f}(\bm{z};\bm{k}). 
\end{equation}
By definition, the above function does not depend on $z_0$. 

PMP states a necessary condition for having an optimal control $\bm{k}^*(t)$ that minimises the functional from Eq.~\eqref{ch1_eq:functional}, within the considered class of admissible controls---i.e. those lying within the control set $U$. In the following, we present PMP for the case of basic fixed-endpoint control problems, for which $\bm{z}(t_f)$ is fixed. Let $\bm{k}^*(t)$ be an admissible control and $\bm{z}^*(t)$ the associated solution of the evolution equations. If $\bm{k}^*(t)$ has to yield a solution of the optimisation problem, there must exist a solution $(\psi_0^*,\bm{\psi}^*(t)^{\sf{T}}) \neq (0,\bm{0}^{\sf{T}})$ for all $t \in [0,t_f]$ such that
\begin{enumerate}
    \item $\bm{z}^*(t)$ and $\bm{\psi}^*(t)$ satisfy Hamilton's canonical equations
    \begin{equation}\label{ch1_eq:hamilton-canonical}
        \dot{\bm{z}}^* = \nabla_{\bm{\psi^*}}\Pi (\bm{z}^*,\psi_0^*,\bm{\psi}^*;\bm{k}^*), \quad \dot{\bm{\psi}^*} = -\nabla_{\bm{z}^*}\Pi (\bm{z}^*,\psi_0^*,\bm{\psi}^*;\bm{k}^*),
    \end{equation}
    with $\psi_0^*\leq 0$ being a constant---which is known as the abnormal multiplier, that we will normalise to $-1$ when different from zero.
    \item For all $t \in [0,t_f]$, it is at the point $\bm{k}^* = \bm{k}^*(t)$ where $\Pi (\bm{z}^*,\psi_0^*,\bm{\psi}^*;\bm{k}^*)$ attains its maximum, i.e.
    \begin{equation}
        \bm{k}^* = \text{arg}\left[ \sup_{\bm{k}} \Pi(\bm{z}^*,\psi_0^*,\bm{\psi}^*;\bm{k})\right].
    \end{equation}
    \item The maximum value of Pontryagin's Hamiltonian, $\Pi^* \equiv \Pi (\bm{z}^*,\psi_0^*,\bm{\psi}^*;\bm{k}^*)$, is $0$ for all $t \in [0,t_f]$.
\end{enumerate}
We highlight that PMP provides necessary conditions for optimality. Henceforth, it only allows to distinguish optimal control candidates, but without asserting whether they are unique, nor even if an actual optimal solution exists. There are also situations in which PMP allows to determine if a certain candidate does not correspond to an optimal solution: if the condition $(\psi_0^*,\bm{\psi}^*(t)^{\sf{T}}) \neq (0,\bm{0}^{\sf{T}})$---also known as the non-triviality condition---is broken for some $t$, then the linearity of Hamilton's canonical equations~\eqref{ch1_eq:hamilton-canonical} ensures that $(\psi_0^*,\bm{\psi}^*(t)^{\sf{T}}) = (0,\bm{0}^{\sf{T}})$. In such a  case, the candidate is not a solution of the optimisation problem.

The second condition of PMP is mathematically equivalent to 
\begin{equation}\label{ch1_eq:euler-lagrange}
    \nabla_{\bm{k^*}}\Pi (\bm{z}^*,\psi_0^*,\bm{\psi}^*;\bm{k}^*) = 0,
\end{equation}
for those optimal protocols $\bm{k}^*(t)$ lying within the boundaries of the control set $U$, or for unconstrained problems. The above may be mapped onto the Euler-Lagrange equations from variational calculus, which entails that the solutions expected from it must correspond to twice-differentiable functions for both the control parameters $\bm{k}^*(t)$ and the dynamical variables $\bm{z}^*(t)$, as the boundaries of the control set are not taken into account. However, there are situations in which the optimal protocol, either partially or completely, belongs to the boundaries of the control set, such that Eq.~\eqref{ch1_eq:euler-lagrange} does not hold anymore. This constitutes the main advantage of PMP over variational calculus, and we comment further on it in the following sections.

The third condition from PMP stems from the fact that we are considering the final time $t_f$ to be free. Had it not be the case, and $\Pi^*$ would only attain just a constant---not necessarily null. A similar condition applies to the case of having a non-fixed value of $\bm{z}(t_f)$, but a set of possible values where $\bm{z}(t_f)$ belongs, known as the target set $S_1$. For the so-called basic variable-endpoint control problems, the above conditions for PMP must be complemented with an additional, fourth one, known as the transversality condition: the vector $\bm{\psi}^*(t)$ must be orthogonal to the tangent space to $S_1$ at $\bm{z}^*(t_f)$, which is mathematically expressed as
\begin{equation}\label{ch1_eq:transversality}
    \bm{\psi}^*(t_f)^{\sf{T}} \bm{u} = 0, \quad \forall \bm{u} \in T_{\bm{z}^*(t_f)}S_1,
\end{equation}
with $T_{\bm{z}^*(t_f)}S_1$ referring to the aforementioned tangent space. However, in this thesis we only consider fixed-endpoint control problems, such that the above condition does not apply, and the values of the conjugate momenta at the endpoints $\bm{\psi}^*(t_f)$ are free.

\subsubsection{Physical insights from variational calculus}\label{ch1_subsubsec:insights-variational}
We have asserted in the preceding section that PMP is a generalisation of variational calculus that applies to a broader scope of scenarios, including discontinuous functions and non-holonomic constraints. However, the connection between the two fields may seem obscure. Therefore, in this section, we shed some light on the latter by arriving at the same mathematical postulates for PMP from the point of view of variational calculus---which is a more intuitive approach for physicists, who are familiar with classical mechanics.

Recall the constrained cost functional $J[\bm{z};\bm{k}]$ from Eq.~\eqref{ch1_eq:functional} that we wish to optimise. The terming ``constrained'' stems from the fact that the functions $\bm{z}(t)$ and $\bm{k}(t)$ are related by means of the dynamical equations \eqref{ch1_eq:dynamical-system}, which might be considered as constraints on the dynamical evolution of the system. In order to account for them, we introduce the unconstrained cost functional
\begin{equation}\label{ch1_eq:unconstrained-cost}
    \tilde{J}[\bm{z};\bm{k}] \equiv \int_0^{t_f}dt \ \tilde{\mathcal{L}}(\bm{z},\dot{\bm{z}},\bm{\psi},t;\bm{k}) = J[\bm{z};\bm{k}] + \int_0^{t_f} dt \ \bm{\psi}^{\sf{T}}\left[ \dot{\bm{z}} - \bm{f}(\bm{z},t;\bm{k})\right],
\end{equation}
with 
\begin{equation}\label{ch1_eq:unconstrained-lagrangian}
    \tilde{\mathcal{L}}(\bm{z},\dot{\bm{z}},\bm{\psi},t;\bm{k}) \equiv \mathcal{L}(\bm{z},t;\bm{k}) + \bm{\psi}^{\sf{T}} \left[ \dot{\bm{z}} - \bm{f}(\bm{z},t;\bm{k})\right]
\end{equation}
being a new ``unconstrained" Lagrangian function, and $\bm{\psi}^{\sf{T}} \equiv (\psi_1,\psi_2,...,\psi_d)$ the so-called Lagrange multipliers, which account for each of the dynamical constraints. We remark that we have considered an additional explicit dependence on time $t$ on all the relevant functions from Eqs.~\eqref{ch1_eq:unconstrained-cost} and \eqref{ch1_eq:unconstrained-lagrangian}, which is done for the sake of generality.

Following the variational approach, optimal solutions are obtained by imposing that the first variation of the functional $\tilde{J}$ vanishes upon varying the state parameters $(\bm{z},\bm{k},\bm{\psi},t)$ as
\begin{align*}
    \bm{z} & \longrightarrow \bm{z}' = \bm{z} + \delta \bm{z}(\boldsymbol{z},\boldsymbol{k},\pmb{\psi},t),
    \\
    \bm{k} & \longrightarrow \bm{k}' = \bm{k} + \delta \bm{k}(\boldsymbol{z},\boldsymbol{k},\pmb{\psi},t),
    \\
    \bm{\psi} & \longrightarrow \bm{\psi}' = \bm{\psi} + \delta \bm{\psi}(\boldsymbol{z},\boldsymbol{k},\pmb{\psi},t),
    \\
    t & \longrightarrow t' = t + \delta t(\boldsymbol{z},\boldsymbol{k},\pmb{\psi},t),
\end{align*}
From now on we will omit the arguments of all functions. We must outline that, due to the variation of the time interval, we must consider the identities
\begin{subequations}
\begin{align}
    &\int_{t_1'}^{t_2'}dt' = \int_{t_1}^{t_2}dt \left( 1 + \frac{d}{dt}\delta t \right), \quad \dot{\bm{z}}' = \frac{d\bm{z}'}{dt'} = \frac{d \bm{z}}{dt } \left( 1 - \frac{d}{dt}\delta t \right) + \frac{d}{dt}\delta \bm{z}.
\end{align}
\end{subequations}
With these, the first variation of $\tilde{J}[\bm{z};\bm{k}]$, $\delta \tilde{J}[\bm{z};\bm{k}] \equiv \tilde{J}'[\bm{z};\bm{k}]-\tilde{J}[\bm{z};\bm{k}]$ is given by
\begin{align}
    \delta \tilde{J}[\bm{z};\bm{k}] & = \int_{t_1}^{t_2}dt \left \lbrace  \left[ \frac{\partial \tilde{\mathcal{L}}}{\partial t} - \frac{d}{dt} \left( \tilde{\mathcal{L}} - \bm{\psi}^{\sf{T}} \dot{\bm{z}} \right) \right] \delta t + (\nabla_{\bm{k}}^{\sf{T}}\tilde{\mathcal{L}}) \delta \bm{k}  + \left[\dot{\bm{z}} - \bm{f} \right]^{\sf{T}} \delta \bm{\psi} \right. \nonumber
    \\
    & \left.  + \left[ \nabla_{\bm{z}}\tilde{\mathcal{L}} - \frac{d}{dt} \nabla_{\dot{\bm{z}}}\tilde{\mathcal{L}} \right]^{\sf{T}} \delta \bm{z} \right \rbrace + \left( \tilde{\mathcal{L}} - \bm{\psi}^{\sf{T}} \dot{\bm{z}} \right) \delta t \biggr\rvert_{t_1}^{t_2} + \bm{\psi}^{\sf{T}} \delta \bm{z} \biggr\rvert_{t_1}^{t_2}.
\end{align}
Let us note that
\begin{equation}
    \tilde{\mathcal{L}} - \bm{\psi}^{\sf{T}} \dot{\bm{z}} = \mathcal{L}  - \bm{\psi}^{\sf{T}} \bm{f} = - \Pi,
\end{equation}
with $\Pi$ being Pontryagin's Hamiltonian function from Eq.~\eqref{ch1_eq:pontryagins-hamiltonian}. Thus, we identify $\psi_0 = -1$ and the Lagrange multipliers $\bm{\psi}$ from our constrained variational problem with the conjugate variables from Pontryagin's procedure. The optimal solution $\bm{k}^*(t)$ is attained for $\delta \tilde{J}[\bm{z}^*;\bm{k}^*]=0$. As all the variations $(\delta \bm{z}, \delta \bm{k}, \delta \bm{\psi}, \delta t)$ are arbitrary and independent, the accompanying terms must vanish, thus giving the following set of equations:
\begin{itemize}
    \item Euler-Lagrange equations for the dynamical variables $\bm{z}$,
    \begin{equation}
        \nabla_{\bm{z}^*}\tilde{\mathcal{L}} - \frac{d}{dt} \nabla_{\dot{\bm{z}}^*}\tilde{\mathcal{L}}=0 \ \Longrightarrow \ \dot{\bm{\psi}}^* = -\nabla_{\bm{z}^*}\Pi,
    \end{equation}
    which we identify as Hamilton's canonical equations for the conjugate variables $\bm{\psi}$.
    \item Euler-Lagrange equations for the conjugate variables $\bm{\psi}$,
    \begin{equation}
       \dot{\bm{z}}^* - \bm{f}^* = 0 \ \Longrightarrow \ \dot{\bm{z}}^* = \nabla_{\bm{\psi}^*}\Pi,
    \end{equation}
    which we identify as Hamilton's canonical equations for the dynamical variables $\bm{z}$.
    \item Euler-Lagrange equations for the control parameters
    \begin{equation}
        \nabla_{\bm{k}^*}\tilde{\mathcal{L}} = 0 \ \Longrightarrow \ \nabla_{\bm{k}^*}\Pi = 0, 
    \end{equation}
    which corresponds to the Hamiltonian maximization condition from Eq.~\eqref{ch1_eq:euler-lagrange}.
    \item The time-derivative of Pontryagin's Hamiltonian,
    \begin{equation}
        \frac{d \Pi}{dt} = -\frac{\partial \tilde{\mathcal{L}}}{\partial t} = \frac{\partial \Pi}{\partial t}.
    \end{equation}
    The above entails that Pontryagin's Hamiltonian becomes a constant of motion if it does not depend explicitly on time, implying that both the Lagrangian $\mathcal{L}$ and the dynamic equations $\bm{f}$ must not depend explicitly on time either, which is the situation we have considered for the premises of PMP. It is true though that Pontryagin's procedure allows us to ``eliminate" the time dependence by introducing an appropriate coordinate to the dynamic system~\cite{liberzon_calculus_2012}. 
    \item Boundary conditions for the Lagrange multipliers: 
    \begin{equation}
        (\bm{\psi}^*)^{\sf{T}} \delta \bm{z}^* \biggr\rvert_{t_1}^{t_2} = 0.
    \end{equation}
    For most of the optimisation problems--such as the ones under consideration in this thesis, the initial conditions are fixed, such that $ \delta \bm{z}^*(t_1) = 0$. For variable-endpoint problems, $ \delta \bm{z}^*(t_2) \neq 0$, and the above expression reduces to the transversality condition from Eq.~\eqref{ch1_eq:transversality}.
    \item Boundary conditions for the Hamiltonian:
    \begin{equation}
        -\Pi^* \ \delta t^* \biggr\rvert_{t_1}^{t_2} = 0.
    \end{equation}
    Given that the initial time is usually fixed, for the free-time optimisation problems considered, the above implies $\Pi^*(t_f) = 0$. This, together with the fact that $\Pi^*$ is a constant of motion for time-independent dynamics, we have that $\Pi^* = 0$ at all times, which corresponds to one of the conditions for PMP.
\end{itemize}
Pontryagin's procedure it is thus equivalent to that from variational calculus, for those optimisation problems for which Eq.~\eqref{ch1_eq:euler-lagrange} holds---i.e. twice-differentiable functions, and protocols belonging within the boundaries of the control set. In the following, last subsection, we explicitly work out a specific class of problems for which variational calculus does not suffice to determine the optimal solution.

\subsubsection{Time-optimisation problems and linear systems}\label{ch1_subsubsec:bang-bang}

Let us put forward the relevant figure of merit that we minimise in the optimisation problems considered in this thesis. This is the connection time $t_f$ between states, that is, we address here the brachistochrone problem from a very generic point of view. In terms of a functional cost~\eqref{ch1_eq:functional}, it is obtained by taking $\mathcal{L}(\bm{z},\bm{k})=1$, such that
\begin{equation}\label{ch1_eq:time-functional}
    J[\bm{z};\bm{k}] = \int_0^{t_f}dt = t_f.
\end{equation}
Now, we say that a dynamical system belongs to the class of ``linear-Pontryagin'' systems if its corresponding dynamical equations are given by
\begin{equation}\label{ch1_eq:linear-pontryagin}
    \dot{\bm{z}} = \bm{f}(\bm{z};\bm{k})= \mathbb{A} \bm{z} + \mathbb{B} \bm{k},
\end{equation}
with $\mathbb{A}$ and $\mathbb{B}$ being two matrices of constant coefficients, and dimensions $d \times d$ and $d \times M$, respectively. The control set $U$ is further assumed to be an $M$-dimensional hyperrectangle of the form
\begin{equation}\label{ch1_eq:control-set}
    U = \{\bm{k}\in \mathbb{R}^M|k_j\in [k_{j,\text{min}},k_{j,\text{max}}], j=1,2,...,M \},
\end{equation}
which entails that the control variables $k_j$ are bounded from above and below by $k_{j,\text{max}}$ and $k_{j,\text{min}}$, respectively. Combining Eqs.~\eqref{ch1_eq:pontryagins-hamiltonian} and \eqref{ch1_eq:linear-pontryagin}, Pontryagin's Hamiltonian function $\Pi (z_0,\bm{z},\psi_0,\bm{\psi},\bm{k})$ reads
\begin{equation}
    \Pi (z_0,\bm{z},\psi_0,\bm{\psi};\bm{k}) = \psi_0 + \bm{\psi}^{\sf{T}} \mathbb{A} \bm{z} + \bm{\psi}^{\sf{T}} \mathbb{B}\bm{k},
\end{equation}
and Hamilton's canonical equations for the conjugate variables give
\begin{equation}
    \dot{\bm{\psi}} = -\nabla_{\bm{z}}\Pi (z_0,\bm{z},\psi_0,\bm{\psi};\bm{k}) = - \mathbb{A}^{\sf{T}} \bm{\psi}.
\end{equation}
In order to determine the optimal protocol $\bm{k}^*(t)$, one may initially assume that it must belong within the boundaries of the control set, such that Eq.~\eqref{ch1_eq:euler-lagrange} holds. Explicitly, the $\bm{k}$-gradient of the Hamiltonian reads
\begin{equation}
    \nabla_{\bm{k^*}}\Pi (z_0^*,\bm{z}^*,\psi_0^*,\bm{\psi}^*;\bm{k}^*) = \mathbb{B}^{\sf{T}} \bm{\psi}^* \equiv  \nabla_{\bm{k^*}}\Pi^*.
\end{equation}
Due to the linearity of Pontryagin's Hamiltonian with respect of $\bm{k}^*$, its $\bm{k}$-gradient is independent of $\bm{k}^*$. In this case, the Hamiltonian is maximised when $\bm{k}^*(t)$ lies at the boundaries of the control set $U$ from Eq.~\eqref{ch1_eq:control-set}---corresponding to the vertices of the $M$-dimensional hyperrectanble---depending on the sign of $\nabla_{\bm{k^*}}\Pi^*$, such that optimal protocols for each of the control parameters $k_j$ take the form
\begin{equation}
    k_j^*(t) = \begin{dcases}
        k_{j,\text{min}} &   \text{if} \ \left(\nabla_{\bm{k^*}}\Pi^* \right)_j < 0, \\
        k_{j,\text{max}} &   \text{if} \ \left(\nabla_{\bm{k^*}}\Pi^* \right)_j > 0, \\
        ? &  \text{if} \ \left(\nabla_{\bm{k^*}}\Pi^* \right)_j = 0,
   \end{dcases}
   \quad j=1,2,...,M.
\end{equation}
There are situations for which at least one of the derivatives of the Hamiltonian $\left(\nabla_{\bm{k^*}}\Pi^* \right)_j$ vanishes for a time window, thus not allowing to determine the value of $k_j^*(t)$. Protocols where these situations occur are known as singular protocols, and they are still compatible with the premises of PMP. In this scenario, as there might be degeneracy---i.e. many protocols for which $\left(\nabla_{\bm{k^*}}\Pi^* \right)_j$ vanishes, it becomes necessary to discern whether a unique optimal protocol exists, or whether optimal solutions are degenerate instead.

Let us discuss non-singular protocols. In that case, the protocol alternatively switches between the vertices of the $M$-dimensional hyperrectangle control set $U$, depending on the signs of each of the derivatives of the Hamiltonian $\left(\nabla_{\bm{k^*}}\Pi^* \right)_j$. The times at which the protocol jumps between vertices are those for which the corresponding derivatives $\left(\nabla_{\bm{k^*}}\Pi^* \right)_j$ vanishes---but only at single points. These protocols are known as bang-bang protocols, and they lie at the core of optimal control problems in many contexts~\cite{chen_fast_2010,ding_smooth_2020,prados_optimizing_2021,ruiz-pino_optimal_2022,liberzon_calculus_2012,ding_smooth_2020,martikyan_comparison_2020}. 

The linear-Pontryagin systems considered in this section are specially appealing, since it can be proven that, under general conditions,\footnote{The necessary conditions are encoded within the so-called Kalman's controllability criterion. See Ref.~\cite{liberzon_calculus_2012} for further details.} optimal protocols are of the bang-bang type---implying that there are no singular protocols---and that the number of jumps in between the vertices of $U$ is, at most, equal to $d-1$~\cite{liberzon_calculus_2012,pontryagin_mathematical_1987}. In general, bang-bang protocols may appear in dynamical systems not belonging to the class of linear-Pontryagin ones, or in control problems where the magnitude to optimise does not correspond to the connection time. In those scenarios, it becomes trickier to assert the minimum number of switchings, or whether there are also optimal solutions of the singular type, of the ``Euler-Lagrange''-type---i.e. those satisfying Eqs.~\eqref{ch1_eq:euler-lagrange}, or even mixed protocols involving bang-bang, singular and Euler-Lagrange parts. The latter is further discussed in chapter~\ref{ch:brownian_gyrator}, as it becomes relevant for the specific optimisation problem we study there.

\subsection{\label{ch1_subsec:speed-limits}Speed limits from information geometry}
Information geometry constitutes a branch from information theory that employs tools from differential geometry in order to study the structure of statistical models and to describe information~\cite{amari_information_2016}. It is an enticing field that has found a great potential to tackle many problems in non-equilibrium statistical mechanics in the last years~\cite{crooks_measuring_2007,sivak_thermodynamic_2012,amari_information_2016,ito_stochastic_2018,ito_stochastic_2020,nicholson_timeinformation_2020}. The connection with the latter field dates back to the early days of statistical mechanics, when Willard Gibbs and Ludwig Boltzmann introduced their conceptions of entropy in terms of the probability distribution for the microstates~\cite{jaynes_gibbs_1965}. More than half a century later, in 1965, Rolf Landauer proposed the so-called Landauer principle, which states the minimum energy required to erase a bit of information for a computational system immersed in a thermal bath~\cite{bennett_landauer_2003}. 

In the modern days, the motivations for studying information geometry and its applications in statistical physics are twofold. On the one hand, the improvement of experimental techniques makes it possible to probe the relationship between different thermodynamic and information observables on a more detailed and microscopic level~\cite{berut_experimental_2012,gavrilov_direct_2017}. On the other hand, there are new theoretical proposals based on understanding information as a physical observable, as important as matter or energy~\cite{szilard_uber_1929,ito_stochastic_2020,berut_experimental_2012,gavrilov_direct_2017}.

In the following, we introduce the main quantities and concepts drawn from information geometry that we employ in our research, specifically in chapter~\ref{ch:thermal_brachistochrones}, in order to provide the optimal protocols that we determine there with further physical insights.

\begin{itemize}
    \item \textbf{Fisher information}: Given a stochastic system described in terms of a PDF $P(\bm{r},t)$, the Fisher information $I(t)$ is defined as
    \begin{equation}\label{ch1_eq:fisher-info}
        I(t) \equiv \int d\bm{r} \  \left( \partial_t \ln P(\bm{r},t) \right)^2 P(\bm{r},t),
    \end{equation}
    and it measures the maximum amount of information that may be obtained by measuring the change of any physical quantity over time. In fact, it determines the intrinsic speed $v_{I}(t) \equiv \sqrt{I(t)}$ at which the system evolves that, by means of the Cramér-Rao bound~\cite{ito_stochastic_2020}, sets the upper speed limit for the relaxation of any observable.
    \item \textbf{Thermodynamic length}: The Fisher information \eqref{ch1_eq:fisher-info} defines a metric in a Riemannian manifold, which thus allows to introduce the notion of distance. Such distance---commonly referred to as the thermodynamic or statistical length---is introduced as
    \begin{equation}
    \label{ch1_eq:length-def}
        \mathcal{L}(t) \equiv \int_0^{t}dt' v_I(t') = \int_0^{t}dt' \ \sqrt{I(t')},
    \end{equation}
    and it measures the length of the path traced by the PDF upon varying the time coordinate.
    \item \textbf{Thermodynamic cost}: Similarly to the thermodynamic length, we may introduce the so-called information cost or divergence
    \begin{equation}
    \label{ch1_eq:cost-def}
        C(t) \equiv  \frac{1}{2}\int_0^{t}dt' v_I(t')^2 = \frac{1}{2}\int_0^{t}dt' \ I(t'),
    \end{equation}
    which provides a measure of the total information that can be extracted of the system over a time interval~\cite{ito_stochastic_2020}. Given its clear analogy with the kinetic energy, it may also be understood as an account for the amount of resources involved in the process.
    \item \textbf{Cauchy-Schwarz speed limit}: As mentioned in Sec.~\ref{ch1_subsec:thesaurus}, speed limits are bounds on the rate of evolution of dynamical systems. They involve trade-off relationships between the evolution speed, or the connection time between initial and final states, and information geometry quantities, which account for the cost required for driving the system 
    or the length of the followed path---among others. 
    
    Taking into account the definitions of  thermodynamic length and cost in Eqs.~\eqref{ch1_eq:length-def} and \eqref{ch1_eq:cost-def}, direct application of the Cauchy-Schwarz inequality entails that the connection time between two states must verify the inequality~\cite{ito_stochastic_2018,ito_stochastic_2020}
    \begin{equation}\label{ch1_eq:speed-limit}
        t_f\geq t_{\text{geo}}\equiv \frac{\mathcal{L}^2}{2\mathcal{C}},
    \end{equation}
    where $t_{\text{geo}}$ is evaluated over the specific path swept by the system in the connection. The quantity $t_{\text{geo}}$ is thus an information geometry lower bound for the connection time. The inequality~\eqref{ch1_eq:speed-limit} is saturated only over the geodesic, that is, $t_f=t_{\text{geo}}$ only for the path connecting the initial and target points that minimises $\mathcal{L}$~\cite{ito_stochastic_2020}.
\end{itemize}

\part{Non-equilibrium attractors and glassy behaviour in \mbox{stochastic} and kinetic systems}

\setlength{\epigraphwidth}{5in} 
\setlength\epigraphrule{0pt}
\epigraph{\small \itshape In all of my universe I have seen no law of nature, unchanging and inexorable. This universe presents only changing relationships which are sometimes seen as laws by short-lived awareness. These fleshy sensoria which we call self are ephemera withering in the blaze of infinity, fleetingly aware of temporary conditions which confine our activities and change as our activities change. If you must label the absolute, use its proper name: Temporary.}{--- Leto Atreides II, God Emperor of Dune}

\label{part:noneq}
\chapter{Memory effects in a molecular fluid with non-linear drag}
\chaptermark{Memory effects in a molecular fluid}
\label{ch:memory_effects}
\newcommand{\st}{\text{s}}

Throughout this part of the thesis, we study the behaviour of different physical systems that are in contact with an externally controllable thermal bath, which governs the time evolution of their dynamics. In this chapter, we start by considering a certain model rooted in the theory of non-linear Brownian motion~\cite{klimontovich_nonlinear_1994}: a molecular fluid with non-linear drag force. This model emerges as a first order correction of the Langevin equation when the masses of both the Brownian particles and the background fluid ones are comparable~\cite{ferrari_particles_2007,ferrari_particles_2014}, and it has been previously employed to accurately describe the behaviour of mixtures of ultracold atoms~\cite{hohmann_individual_2017}.\footnote{The stochastic description applies as long as the balance between low densities and low temperatures allows for neglecting quantum effects.}

Herein, having started initially from equilibrium, we specifically study the relaxation of the aforementioned model towards a new equilibrium state, when suddenly changing the bath temperature from an initial to a final value. We show that, when quenching the bath temperature to low enough values, the system displays glassy behaviour: a slow non-exponential relaxation function and memory effects, which present universal features, in the sense that they do not depend---upon a suitable rescaling of the relevant variables---on the initial and final temperatures, nor on the degree of non-linearity of the drag force, nor on the collision rate. We show that this behaviour is linked to the existence of a \textit{long-lived non-equilibrium state} (LLNES) ruling the dynamics of the system at low temperatures, and we specifically study and characterise the emergence of the Mpemba and Kovacs memory effects.

The structure of this chapter goes as follows. In Sec.~\ref{ch2_sec:model}, we introduce the model of concern and its mathematical framework, providing the main approximation schemes that we work with. Section~\ref{ch2_sec:relaxation} is devoted to the study of the relaxation of the temperature of the system, initially starting with a perturbative approach in the non-linearity in Sec.~\ref{ch2_subsec:perturbation}, and following with an asymptotic analysis for a quench to low temperatures in Sec.~\ref{ch2_subsec:quench}. It is in the latter section that we uncover and characterise the LLNES. Finally, in Sec.~\ref{ch2_sec:memory}, we study the emergence of the Mpemba and Kovacs effects when driving the system towards the LLNES.

\section{\label{ch2_sec:model}Model: molecular fluid with non-linear drag}\sectionmark{Model: molecular fluid}
From a phenomenological standpoint, the model that we consider below constitutes the simplest model of non-linear Brownian motion, in which the drag coefficient has the minimal dependence on the velocity of the Brownian particle~\cite{klimontovich_statistical_1995,lindner_diffusion_2007,goychuk_nonequilibrium_2021}. Let us take the Markov process $\bm{r}$ to stand for the components of the velocity of a $d$-dimensional particle, whose statistical behaviour is accurately described by the one-particle VDF $f(\bm{v},t)$. The time evolution of the VDF is governed by a EFP equation~\eqref{ch1_eq:boltzmann-fokker-planck} with $\alpha = 1$, with the force and noise terms 
having the explicit forms
\begin{equation}\label{ch2_eq:nonlinear-drag}
    \bm{A}(\bm{v},\cancel{t}) = \zeta(v)\bm{v}, \quad \mathbb{D}(\bm{v},\cancel{t}) = \frac{k_BT_s}{m}\zeta(v) \mathbb{I}_d, \quad \zeta(v) = \zeta_0 \left(1 + \gamma \frac{mv^2}{k_BT_s}\right),
\end{equation}
with $\zeta(v)$ being the non-linear drag coefficient and $\mathbb{I}_d$ being the $d \times d$ identity matrix. Given that the Hamiltonian function of the system corresponds to the kinetic energy $H(\bm{v}) = mv^2/2$, the explicit forms of $\bm{A}(\bm{v})$ and $\mathbb{D}(\bm{v})$ guarantee that the fluctuation-dissipation relation \eqref{ch1_eq:fluctuation-dissipation} is satisfied. Hence, the latter ensures that the EFP equation only admits the equilibrium Maxwellian VDF $f_{\text{eq}}(\bm{v})$ from Eq.~\eqref{ch1_eq:maxwellian} as its stationary solution.

On the fundamental level, the aforementioned model arises when considering the behaviour of an ensemble of Brownian particles of mass $m$ and number density $n_{\text{par}}$ immersed in an isotropic and uniform background fluid at equilibrium with temperature $T_{\text{s}}$, the particles of which have masses $m_{\text{bf}}$~\cite{ferrari_particles_2007,ferrari_particles_2014}. In the $m_{\text{bf}}/m \rightarrow 0$ limit, which is known as the Rayleigh limit, the drag coefficient $\zeta_0$ is constant, thus implying that the drag force exerted on the Brownian particle becomes linear in its velocity, $\bm{A}(\bm{v})=\zeta_0\bm{v}$. In real-world scenarios, we have that $m_{\text{bf}}/m \neq 0$, and henceforth, it is worth analysing corrections to the Rayleigh limit. The explicit form of $\zeta(v)$ from Eq.~\eqref{ch2_eq:nonlinear-drag} is obtained by considering first-order corrections, in which we retain up to linear terms in $m_{\text{bf}}/m$ while neglecting higher-order ones, and it is referred to as the quasi-Rayleigh limit. On the one hand, $\zeta_0 \propto T_{\st}^{1/2}$ accounts for the zero-velocity drag coefficient obtained in the Rayleigh limit, and it depends on the background fluid density $n_{\text{bf}}$. On the other hand, the non-linear parameter $\gamma$ is expressed as an integral form including the differential cross-section for the collision between a Brownian and a background fluid particle, whose typical values are bounded by $\gamma \lesssim 0.2$~\cite{ferrari_particles_2007,ferrari_particles_2014,hohmann_individual_2017}. These are the typical values that we employ throughout this thesis. Furthermore, for the case in which both the Brownian and the background fluid particles are three-dimensional hard spheres, it has been shown that $\gamma = m_{\text{bf}}/(10m)$~\cite{ferrari_particles_2007,ferrari_particles_2014,hohmann_individual_2017}. Let us note that for the case of self-diffusion---i.e. $m=m_{\text{bf}}$---we have that $\gamma = 0.1$, which is close to its upper bound.


The model described above may be regarded as a mixture of two fluids, in which, due to the high differences between the values of the masses of their respective particles, their dynamics can be decoupled. In that case, we have (i) a ``slow fluid" of heavy particles, the dynamics of which are the focus of our study and (ii) a ``fast fluid" of lighter particles, which we assume to be at equilibrium---i.e. its relaxation timescale is much smaller than those characterising the dynamical evolution of the slow fluid, thus becoming a thermal bath for it. In fact, this is the physical situation explored in Ref.~\cite{hohmann_individual_2017} for a mixture of ultracold gases. Specifically, they study the behaviour of an ensemble of $^{133}$Cs atoms moving inside a background cloud of $^{87}$Rb atoms. Despite the fact that the experiments are performed in the ultracold regime---involving temperatures in the $\mu$K range, the low densities involved make it possible to neglect quantum effects and to employ the Fokker-Planck equation \eqref{ch1_eq:fokker-planck}---for the one-particle VDF---for describing the $^{133}$Cs ensemble.

The kinetic temperature $T(t)$ is defined as usual in statistical mechanics, proportional to the average kinetic energy 
\begin{equation}\label{ch2_eq:kinetic-temperature}
    T(t) \equiv \frac{m}{dk_B}\left\langle  v^2\right\rangle = \frac{m}{dn_{\text{par}}k_B}\int d\bm{v}\ v^2 f(\bm{v},t).
\end{equation}
By multiplying by $v^2$ and integrating over $\bm{v}$ in the EFP equation \eqref{ch1_eq:boltzmann-fokker-planck}, the evolution equation for the kinetic temperature is obtained as
\begin{equation}\label{ch2_eq:evolution-kin-temp}
    \frac{\dot{T}}{\zeta_0} = 2(T_{\st}-T)\left[1+\gamma (d+2)\frac{T}{T_{\st}}\right]-2\gamma(d+2)\frac{T^2}{T_{\st}}a_2,
\end{equation}
where we have introduced the excess kurtosis
\begin{equation}\label{ch2_eq:excess-kurtosis-1}
    a_2 = \frac{d}{d+2}\frac{\left\langle v^4\right\rangle}{\left\langle v^2\right\rangle^2}-1,
\end{equation}
which measures the deviations from the Maxwellian distribution $f_{\text{eq}}(\bm{v})$ from Eq.~\eqref{ch1_eq:maxwellian}. We highlight that Eq.~\eqref{ch2_eq:evolution-kin-temp} applies for both the Fokker-Planck and EFP equations---and thus, it could have been obtained from Eq.~\eqref{ch1_eq:evol-eq-f} as well, since the kinetic energy becomes a collisional invariant for elastic collisions~\cite{resibois_classical_1977}. Let us further note that, in the Rayleigh limit, i.e. $\gamma = 0$, Eq.~\eqref{ch2_eq:evolution-kin-temp} provides the exponential solution
\begin{equation}\label{ch2_eq:exponential-temperature}
    T(t)=T_{\st} + (T_i-T_\st)e^{-2\zeta_0 t},
\end{equation}
with $T_i \equiv T(0)$ being the initial value of the kinetic temperature, and with relaxation rate $(2\zeta_0)^{-1}$. Remarkably, the EFP equation may be solved analytically in this case, with the resulting VDF being Gaussian at all times if it was initially so, as it was the case for Gaussian processes from Sec~\ref{ch1_subsubsec:ornstein-uhlenbeck}.\footnote{The latter stems from the fact that the logarithm of a Gaussian VDF can be written as a linear combination of collisional invariants, and thus $J_1[\bm{v}|f,f]=0$ at all times~\cite{cercignani_boltzmann_1988,guery-odelin_nonequilibrium_2014}.} The time dependence of the system is thus completely determined by the kinetic temperature $T(t)$. Herein, the system does not deviate from the instantaneous equilibrium distribution for a temperature $T(t)$ at time $t$, and thus, we do not expect the emergence of non-equilibrium phenomena here. However, for $\gamma >0$, the dynamic evolution of the temperature is coupled to that of the excess kurtosis. Therefore, the temperature alone it does not determine the dynamical behaviour of the system, nor its relaxation follows an exponential form. Moreover, the evolution equation for the excess kurtosis is, in turn, coupled to that for the sixth-degree cumulant, thus leading in general to an infinite hierarchy of differential equations for the cumulants of the VDF~\cite{santos_mpemba_2020}.

\subsection{\label{ch2_subsec:sonine}Evolution equations and the Sonine expansion}
Here we derive the evolution equations for the remaining relevant physical variables that characterise the dynamics of our system. For these purposes, we scale the velocities with the thermal velocity $v_{\text{T}}(t)$,
\begin{equation}\label{ch2_eq:scaling-velocities}
    \bm{c}\equiv \frac{\bm{v}}{v_{\text{T}}(t)}, \quad \phi(\bm{c},t)\equiv \frac{(v_{\text{T}}(t))^d}{n_{\text{par}}}f(\bm{v},t), \quad v_{\text{T}}(t) \equiv \sqrt{\frac{2k_BT(t)}{m}},
\end{equation}
where we have introduced the corresponding scaled VDF $\phi(\bm{c},t)$, which is now normalised to unity. The EFP equation for this scaled VDF becomes
\begin{equation}
    \label{ch2_eq:EFP-scaled}
    \partial_t \phi(\bm{c},t) - \nabla_{\bm{c}}^{\sf{T}} \left\{\left[ \frac{\dot{T}}{2T}\bm{c} + \zeta\left(v_{\text{T}}(t)\ c\right) \left(\bm{c}+\frac{T_{\st}}{2T}\nabla_{\bm{c}}\right) \right]\phi(\bm{c},t)\right\} = \nu_s \sqrt{\frac{T}{T_{\st}}} I[\bm{c}|\phi,\phi],
\end{equation}
where
\begin{equation}\label{ch2_eq:collision-frequency}
    \nu_\st \equiv g(\sigma)n\sigma^{d-1}\sqrt{\frac{2k_BT_\st}{m}}
\end{equation}
is the collision frequency at the stationary state and
\begin{equation}\label{ch2_eq:collision-operator-scaled}
    I[\bm{c}_1|\phi,\phi] = \int d \bm{c}_2 \int d \widehat{\bm{\sigma}} \ \Theta (\bm{c}_{12}^{\sf{T}} \ \widehat{\bm{\sigma}}) \ \bm{c}_{12}^{\sf{T}} \ \widehat{\bm{\sigma}} \left[\phi (\bm{c}_1',t)\phi (\bm{c}_2',t)-\phi (\bm{c}_1,t)\phi (\bm{c}_2,t)\right],
\end{equation}
is the dimensionless Enskog operator. 
Taking velocity moments in the scaled EFP equation \eqref{ch2_eq:EFP-scaled} provides an infinite hierarchy of equations for $M_l \equiv \langle c^l \rangle$,
\begin{align}
    \label{ch2_eq:moments-odes}
    \dot{M}_l  = \ l\zeta_0 &\left\{ \left[\gamma (l-2) + \gamma (d+2) \frac{T}{T_s}(1+a_2) - \frac{T_s}{T}\right]M_l \right. \nonumber
    \\
    & \left. - \,\gamma \frac{2T}{T_s}M_{l+2} + \frac{l+d-2}{2}\frac{T_s}{T}M_{l-2} \right\} - \nu_s \sqrt{\frac{T}{T_s}} \mu_l,
\end{align}
where we have introduced the so-called collisional moments
\begin{equation}\label{ch2_eq:collisional-moments}
    \mu_l \equiv - \int d\bm{c}\ c^l I[\bm{c}|\phi,\phi].
\end{equation}
We remark that, by definition, $M_0 = 1$, $M_2 = d/2$ and $M_4 = d(d+2)(1+a_2)/4$. The last relation stems from the definition of the excess kurtosis \eqref{ch2_eq:excess-kurtosis-1}, when introducing the scaled variables from Eq.~\eqref{ch2_eq:scaling-velocities}. In fact, by setting $l=4$ in Eq.~\eqref{ch2_eq:moments-odes}, the evolution equation for the excess kurtosis reads
\begin{align}\label{ch2_eq:evol-eq-kurtosis-1}
    \dot{a}_2 = & \ \zeta_0 \gamma \frac{4T}{T_\st}\left[\frac{2T_\st}{T}(1+a_2)+(d+2)(1+a_2)^2-(d+4)(1+3a_2-a_3)\right] \nonumber
    \\
    & - \zeta_0\frac{4T_\st}{T}a_2 - \frac{4\nu_\st}{d(d+2)}\sqrt{\frac{T}{T_\st}}\mu_4,
\end{align}
where we have introduced the sixth-degree cumulant $a_3$, defined by 
\begin{equation}\label{ch2_eq:sixth-cumulant-1}
    M_6 = \frac{d(d+2)(d+4)}{8}(1+3a_2-a_3).
\end{equation}
In Eq.~\eqref{ch2_eq:evol-eq-kurtosis-1}, we see explicitly that the time evolution of the excess kurtosis is coupled to that of the sixth cumulant.

There are two competing timescales governing the relaxation dynamics: (i) the inverse of the zero-velocity drag coefficient $\zeta_0^{-1}$, which sets the timescale over which the particles feel the action of the background fluid, and (ii) the inverse of the collision frequency at the stationary state $\nu_\st^{-1}$, which sets the characteristic time for binary collisions. In order to quantify their balance, we introduce the dimensionless average time between Brownian-Brownian collisions
\begin{equation}\label{ch2_eq:xi}
    \xi \equiv \zeta_0 \tau_\st, \quad \tau_\st \equiv \sqrt{2}\ \Gamma\left(\frac{d}{2}\right) \pi^{\frac{1-d}{2}}\nu_\st^{-1},
\end{equation}
where $\tau_s$ constitutes the mean free time between collisions at the steady state~\cite{resibois_classical_1977,poschel_granular_2001}. Since both $\zeta_0$ and $\nu_s$ are proportional to $T_s^{1/2}$, the parameter $\xi$ is independent of $T_\st$. This parameter measures the relative relevance of the collisions between Brownian particles and background fluid ones and the binary collisions between Brownian particles. The regime $\xi \gg 1$ ($\xi \ll 1$) implies that the non-linear drag acts over a much longer (shorter) timescale than Brownian-Brownian collisions do. For the case in which both the Brownian and background fluid particles are three-dimensional hard spheres---the latter having number density $n_{\text{bf}}$ and radii $\sigma_{\text{bf}}$, it has been shown that $\xi$ takes the form~\cite{hohmann_individual_2017,santos_mpemba_2020} 
\begin{equation}\label{ch2_eq:xi-form}
    \xi=\frac{2n_{\text{bf}}}{3n_{\text{par}}}\left(1+\frac{\sigma_{\text{bf}}}{\sigma}\right)^{2
  }\frac{\sqrt{5\gamma}}{1+10\gamma}.
\end{equation}
Throughout this thesis, we mainly employ values of $\xi$ of the order of unity, which correspond to the intermediate limit in which both the thermal fluctuations due to the background fluid and Brownian-Brownian binary collisions are equally relevant for the relaxation dynamics of the system.

Equations~\eqref{ch2_eq:evolution-kin-temp}, \eqref{ch2_eq:moments-odes} and \eqref{ch2_eq:evol-eq-kurtosis-1} are exact within the formalism of the EFP equation, but they do not define a closed, finite set of equations. Not only is the dynamic evolution of $M_{l}$ coupled to both $M_{l-2}$ and $M_{l+2}$, but, additionally, the collisional moments $\mu_l$ are non-linear functionals of the scaled VDF. In order to make further analytical progress, we need to resort to approximation schemes to truncate the infinite hierarchy of equations. It is in this regard that, for isotropic states, the scaled VDF can be expanded in a complete set of orthonormal polynomials as
\begin{equation}\label{ch2_eq:sonine-expansion}
    \phi(\bm{c},t) = \frac{e^{-c^2}}{\pi^{d/2}}\left[1 + \sum_{l=2}^{\infty} a_l(t) L_l^{\frac{d-2}{2}}(c^2) \right],
\end{equation}
where $L_l^{(\alpha)}$ are the generalized Laguerre (or Sonine) polynomials~\cite{abramowitz_handbook_1988}. The above expansion is known as the Sonine expansion, and the coefficients with $l=2$ and $l=3$ correspond to the excess kurtosis $a_2$ and the sixth cumulant $a_3$ respectively---also known as the Sonine cumulants or coefficients. Given a positive integer $n$, the $n$-th order Sonine approximation consists in retaining up to the $(n+1)$-th cumulant in the above expansion and neglecting higher order ones, for these are assumed to be small. Moreover, non-linear combinations of the cumulants are also usually neglected, because of their smallness as well. Given that the Sonine expansion is being truncated, we expect these approximation schemes to provide only a qualitative account of the behaviour of the last retained cumulant---i.e. the $(n+1)$-th one---while giving a more accurate, quantitative account for the lower order ones. In the following, we consider two specific Sonine approximation schemes: the first ($n=1$) and the second or extended ($n=2$) Sonine approximations. We also employ the dimensionless variables
\begin{equation}\label{ch2_eq:dimensionless-variables}
    \theta(t) \equiv \frac{T(t)}{T_\st}, \quad t^* \equiv \zeta_0 t,
\end{equation}
and drop the asterisk in the dimensionless time from now on, in order to simplify the notation.
\subsubsection{\label{ch2_subsubsec:first-sonine}First Sonine approximation}
Under the first Sonine approximation, the variables $(\theta,a_2)$ fully characterise the dynamical behaviour of the system, since the sixth cumulant $a_3$ and higher-order ones are neglected. Therefore, we are left with the closed system of equations
\begin{subequations}\label{ch2_eq:first-sonine-edos}
\begin{align}
    \dot{\theta}&= 2(1-\theta)[1+\gamma(d+2)\theta]-2\gamma(d+2)\theta^2a_2,\label{ch2_subeq:temp-first-sonine}\\
    \dot{a}_2&= 8\gamma(1-\theta)-\left[\frac{4}{\theta}-8\gamma+4\gamma(d+8)\theta + \frac{8(d-1)}{d(d+2)}\frac{\sqrt{\theta}}{\xi}\right]a_2, \label{ch2_subeq:kurtosis-first-sonine}
\end{align}
\end{subequations}
which are linear in $a_2$, but non-linear in $\theta$. Note that we have introduced the parameter $\xi$ from Eq.~\eqref{ch2_eq:xi-form}. Equation~\eqref{ch2_subeq:kurtosis-first-sonine} is obtained from Eq.~\eqref{ch2_eq:evol-eq-kurtosis-1} by neglecting the terms proportional to both $a_3$ and $(a_2)^2$, and considering the approximation~\cite{santos_mpemba_2020,megias_thermal_2022}
\begin{equation}\label{ch2_eq:approx-coll-inv-1}
    \mu_4 \approx \frac{\sqrt{2}(d-1) \pi^{\frac{d-1}{2}}}{\Gamma \left( \frac{d}{2} \right) }a_2,
\end{equation}
for the collisional moment $\mu_4$. We expect this approximation scheme to provide a quantitative account for the dimensionless kinetic temperature $\theta$, while being qualitative with respect to the excess kurtosis $a_2$. Nevertheless, due to its simplicity, this scheme allows us to provide further analytical insights on the overall behaviour of the system, as shown in the following sections.

\subsubsection{\label{ch2_subsubsec:second-sonine}Second (extended) Sonine approximation}
This approximation scheme provides a closed set of equations for the variables $(\theta,a_2,a_3)$, while neglecting higher order contributions. Considering Eq.~\eqref{ch2_eq:moments-odes} for $l=4$ and $l=6$, and employing the definition from Eq.~\eqref{ch2_eq:sixth-cumulant-1}, we arrive at the system of equations
\begin{subequations}\label{ch2_eq:second-sonine-eqs}
  \begin{align}
\label{ch2_subeq:temp-second-sonine}
    \dot{\theta}&=2(1-\theta)\left[ 1 +\gamma (d+2)\theta \right] - 2\gamma (d+2)\theta^2 a_2,
\\
\label{ch2_subeq:kurtosis-second-sonine}
    \dot{a}_2 &= 8\gamma (1-\theta)- \left[ \frac{4}{\theta} - 8\gamma + 4\gamma (d+8)\theta + \frac{8(d-1)}{d(d+2)}\frac{\sqrt{\theta}}{\xi} \right]a_2 \nonumber
    \\
    &+ 2\!\left[2\gamma \theta (d+4) + \frac{(d-1)}{d(d+2)}\frac{\sqrt{\theta}}{\xi} \right]\!a_3,
\\
\label{ch2_subeq:sixth-second-sonine}
  \dot{a}_3 &=   12\!\left[-4\gamma+6\gamma\theta+
                 \frac{(d-1)}{d(d+2)(d+4)}
                 \frac{\sqrt{\theta}}{\xi}\right]\!a_{2} \nonumber
                 \\
                 &+6\!\left[4\gamma-\frac{1}{\theta}-\gamma\theta(d+14)-
                   \frac{(d-1)(4d+19)}{2d(d+2)(d+4)}
                   \frac{\sqrt{\theta}}{\xi}
                   \right]\! a_{3},
\end{align}
\end{subequations}
which, again, are linear by construction in the Sonine cumulants but non-linear in $\theta$. In order to obtain both Eqs.~\eqref{ch2_subeq:kurtosis-second-sonine} and \eqref{ch2_subeq:sixth-second-sonine}, we have additionally resorted to the approximations~\cite{santos_mpemba_2020,megias_thermal_2022}
\begin{subequations}\label{ch2_eq:approx-coll-inv}
\begin{align}
    \label{ch2_subeq:approx-coll-inv-1}
    \mu_4 &\approx \frac{\sqrt{2}(d-1) \pi^{\frac{d-1}{2}}}{\Gamma \left( \frac{d}{2} \right) } \left(a_2 - \frac{a_3}{4}\right), 
    \\
    \label{ch2_subeq:approx-coll-inv-2}
    \mu_6 &\approx \frac{3\sqrt{2}(d-1)(9+2d) \pi^{\frac{d-1}{2}}}{4\Gamma \left( \frac{d}{2} \right)}\left(a_2 - \frac{3}{4}a_3 \right),
\end{align}
\end{subequations}
for the collisional moments $\mu_4$ and $\mu_6$, respectively. We expect this approximation scheme to provide a quantitative account of the behaviour of both the kinetic temperature $\theta$ and the excess kurtosis $a_2$, while being only qualitative for $a_3$ in this case.

\subsection{\label{ch2_subsec:robustness}Robustness of the Sonine approximations}
Here, we test the robustness and validity of the two Sonine approximation schemes introduced in the previous section. We compare the numerical integration of the systems of equations \eqref{ch2_eq:first-sonine-edos} and \eqref{ch2_eq:second-sonine-eqs} with Direct Simulation Monte Carlo (DSMC) data, which numerically integrate the EFP equation \eqref{ch1_eq:boltzmann-fokker-planck}~\cite{bird_g_a_molecular_1994,montanero_monte_1996}.\footnote{Throughout this first part of the thesis, we actually employ a numerical algorithm that combines both DSMC and stochastic integration techniques in order to account for both binary collisions and fluctuating white-noise forces respectively. Nevertheless, for the sake of brevity, we refer to this combined method as the DSMC algorithm. See Appendix~\ref{app:simulation-methods} for further details.} Specifically, we consider a $d=2$ dimensional molecular fluid with $\gamma = 0.1$ and $\xi = 1$. The system is initially prepared at an equilibrium state with an initial scaled kinetic temperature $\theta_i \equiv T_i/T_s$, whose values range from $2$ to $100$.

\begin{figure}
{\centering 
\includegraphics[width=2.6in]{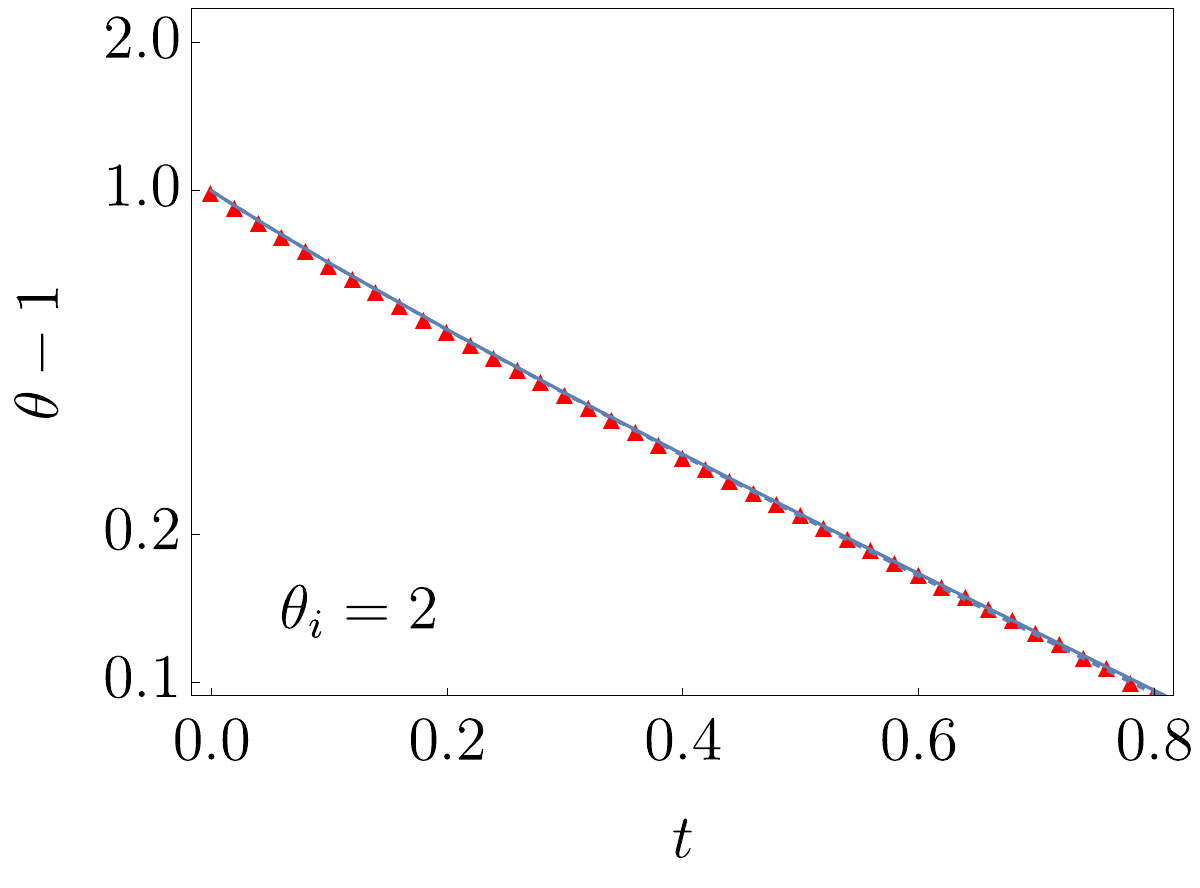}
\includegraphics[width=2.7in]{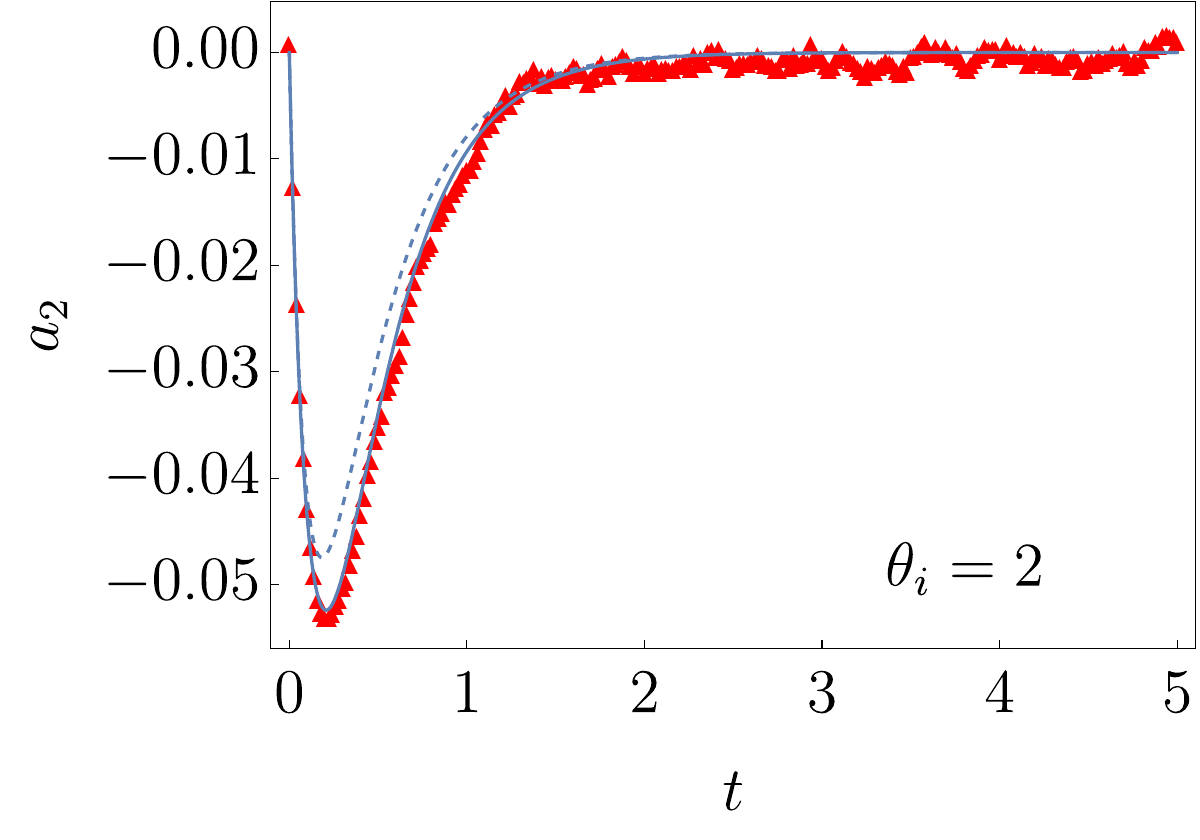}
\\
\includegraphics[width=2.61in]{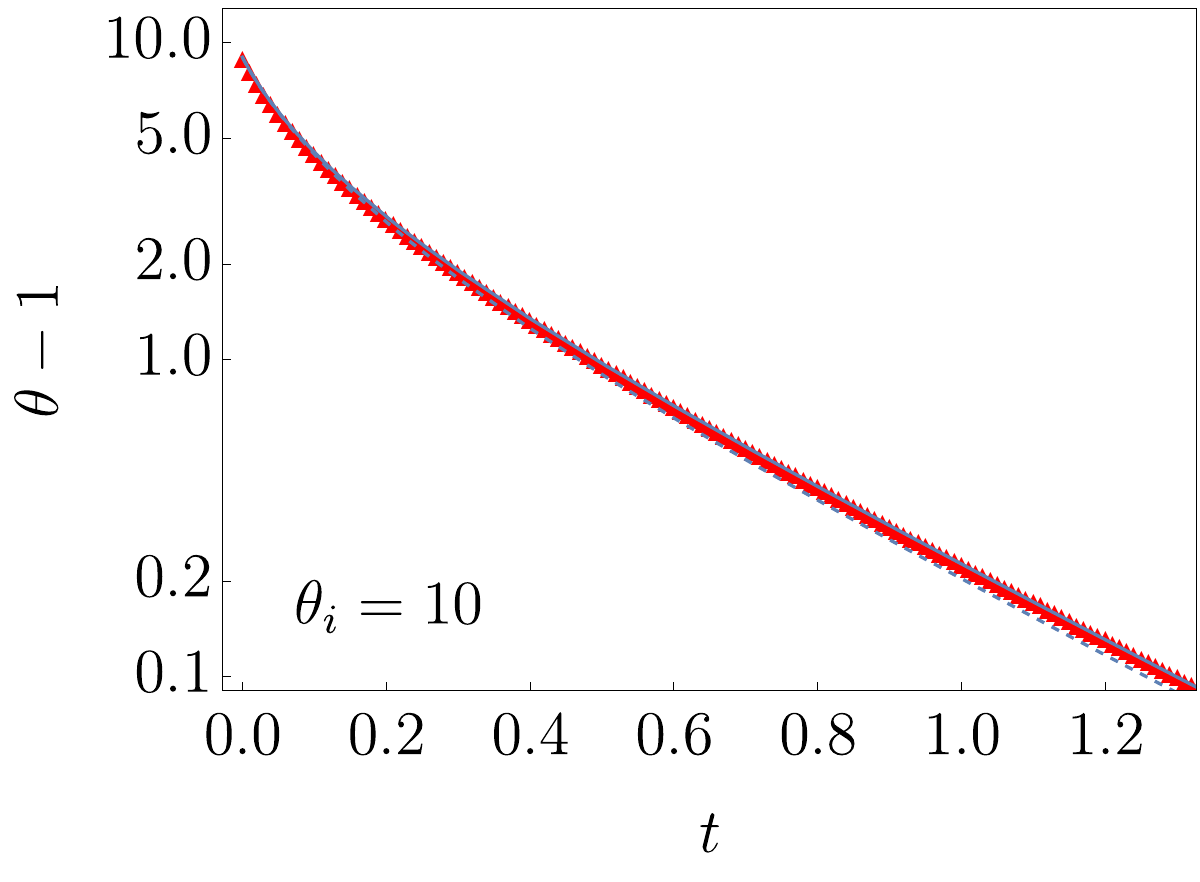}
\includegraphics[width=2.7in]{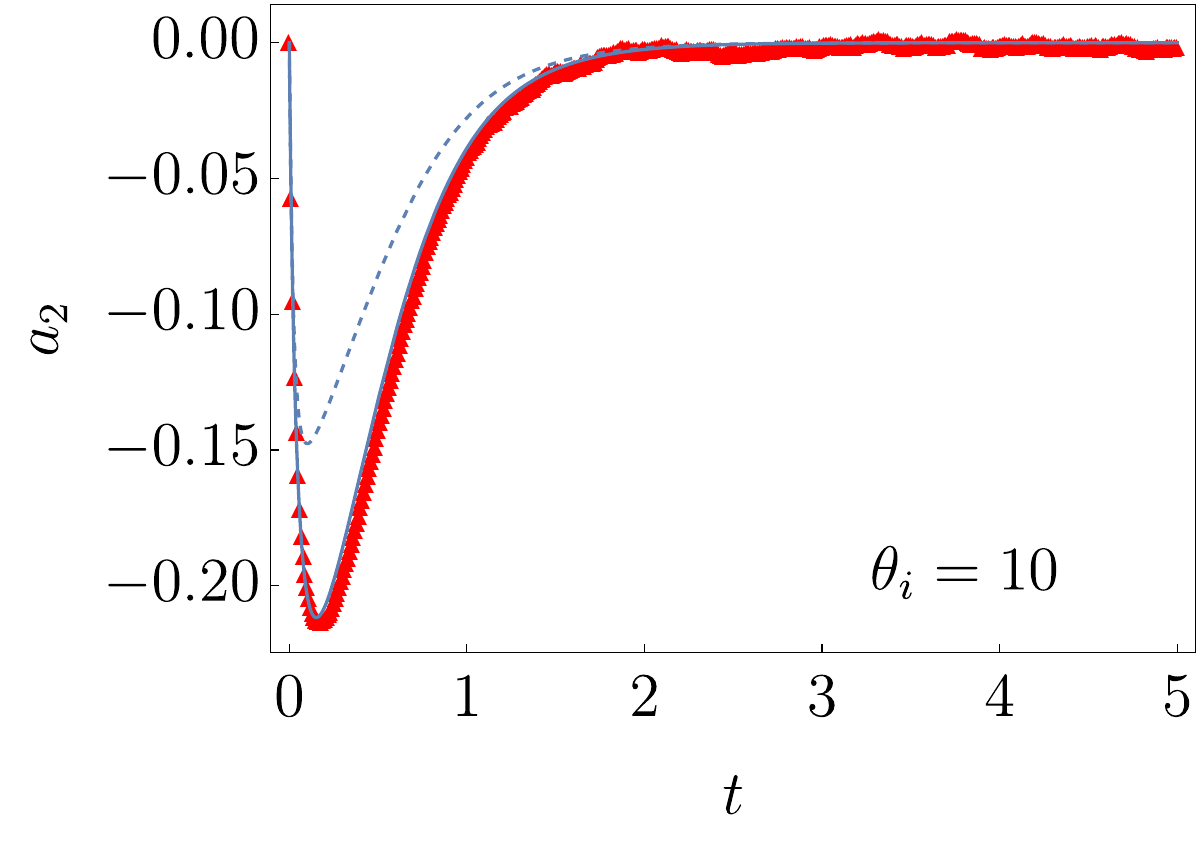}
\\
\includegraphics[width=2.66in]{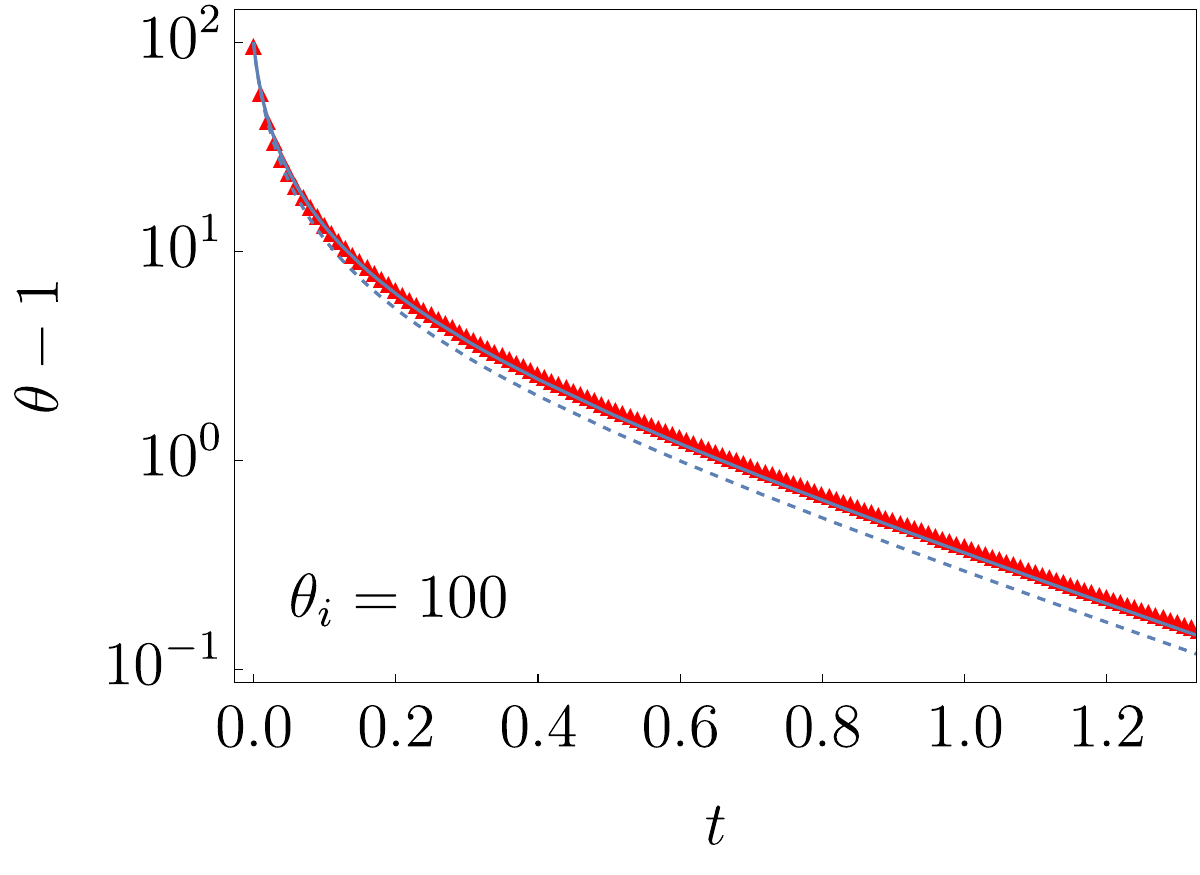}
\includegraphics[width=2.65in]{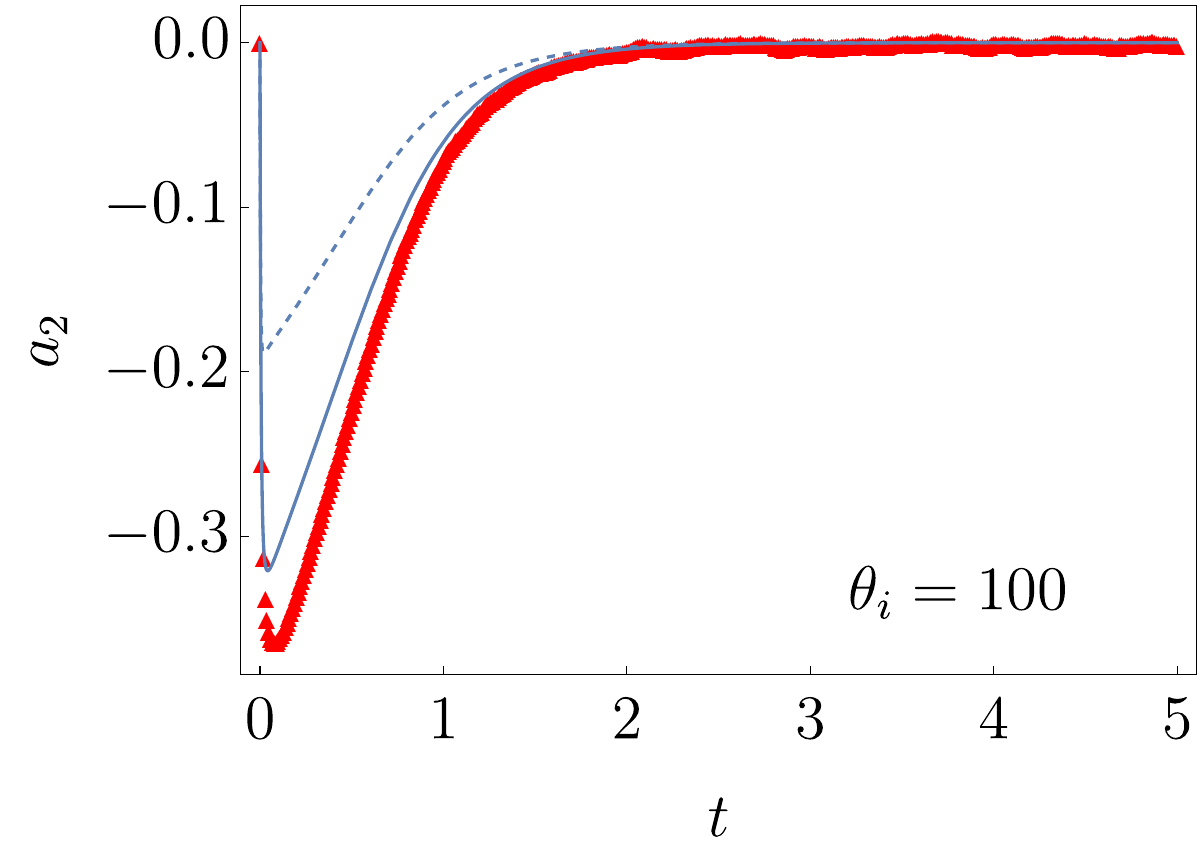}
\\}
\caption{Time evolution of the scaled kinetic temperature (left panels) and the excess kurtosis (right panels). Three different values of the initial scaled temperature $\theta_i$ are considered: $2$, $10$ and $100$. Additional parameters employed are $d=2$, $\gamma = 0.1$ and $\xi = 1$. Symbols (red triangles) correspond to DSMC data, while curves correspond to the numerical integration of Eqs.~\eqref{ch2_eq:first-sonine-edos} for the first Sonine approximation (dashed lines) and Eqs.~\eqref{ch2_eq:second-sonine-eqs} for the extended Sonine approximation (solid lines).}
    \label{ch2_fig:robustness}
\end{figure}

Figure~\ref{ch2_fig:robustness} shows the time evolution of both the kinetic temperature and the excess kurtosis for three specific values of the initial temperature $\theta_i$: $2$, $10$ and $100$. Starting with the kinetic temperature, we may observe that, for moderate values of $\theta_i$, both the first and second Sonine approximations provide a quantitative account of its behaviour. It is only for very deep quenches---e.g. for $\theta_i = 100 \gg 1$---that the first Sonine approximation deviates from the DSMC data, while the second Sonine approximation still gives a good estimate. As we comment on in the following, these discrepancies arise due to the increasing values of the excess kurtosis with $\theta_i$. It is also remarkable to observe that, while the relaxation of the kinetic temperature is basically exponential for $\theta_i = O(1)$, it becomes strongly non-exponential for $\theta_i = 100$. These facts are thoroughly explored in the forthcoming section, constituting the core of the contents of this chapter.

The discrepancies between the first and second Sonine approximations become neater when studying the behaviour of the excess kurtosis. For $\theta_i=2$, both approximations are quite close to the DSMC data, although it may already be observed that the extended approximation is more accurate. As we keep increasing $\theta_i$, the discrepancies between the two schemes increase as well, with the second Sonine approximation always providing the most accurate account of the dynamics. However, for the highest temperature considered, $\theta_i = 100$, the discrepancies from the DSMC start becoming noticeable even for the second Sonine approximation, with a relative error of $10\%$ when estimating the minimum of the excess kurtosis. The latter is further supported by considering the maximum---in absolute value---amplitudes attained by the sixth cumulant $a_3$ for the three different values of $\theta_i$. While for $\theta_i = 2$ it is around $0.02$, thus corroborating the validity of both the first and second Sonine approximations, for $\theta = 10$ it becomes of the order of $0.18$, and for $\theta = 100$, $0.40$, which is even larger than the maximum absolute amplitude attained by the excess kurtosis, thus compromising the validity of both Sonine schemes in this limit.

Overall, the $n$-th order Sonine approximation accurately describes the behaviour up to the previous to last kept cumulant $a_n$ as the initial scaled temperature is increased, only providing a qualitative account of the last kept cumulant $a_{n+1}$. The latter makes it necessary to consider the second Sonine approximation when studying the relaxation of the temperature after a quench. An accurate account of the excess kurtosis is required, since $a_2$ is directly coupled to the dynamical evolution of the kinetic temperature.

\section{\label{ch2_sec:relaxation}Studying the relaxation of the kinetic temperature}
The time origin of our experiment is set at $t=0$. Initially, the molecular fluid is assumed to be at equilibrium with the thermal bath at temperature $T_i$ for $t<0$. Then, at $t=0^+$, the bath temperature is changed to a new value $T_\st$. As the system is no longer in equilibrium, it relaxes towards a new equilibrium state with temperature $T_\st$. The specific goal of this section is carrying out a thorough study on how the kinetic temperature, which is a relevant observable for both theoretical and experimental purposes, relaxes towards $T_\st$. Given the insights gathered in the previous section, there are two main regimes worth exploring: (i) $\theta_i = O(1)$, for which the temperature relaxes exponentially, and where we do not expect large deviations from the equilibrium Maxwellian---under the scaled variables from Eq.~\eqref{ch2_eq:scaling-velocities}---and (ii) $\theta_i \gg 1$, which implies a deep quench to low temperatures, where a non-exponential relaxation window emerges. Throughout this section, we resort to different analytical tools, together with both the first and extended Sonine approximations, in order to explicitly determine the relaxation functions characterising the kinetic temperature.

\subsection{\label{ch2_subsec:perturbation}Perturbative approach in the non-linearity}
As previously mentioned in Sec.~\ref{ch2_sec:model}, the non-linear parameter $\gamma$ is typically small, $\gamma \lesssim 0.2$. The case $\gamma = 0.1$ corresponds to self-diffusion for a gas of three-dimensional hard spheres. Due to its smallness, we may resort to the tools of perturbation theory~\cite{bender_advanced_1999} on the evolution equations to further simplify the analysis. Furthermore, as for $\gamma = 0$ all the Sonine cumulants $a_l$ vanish, for the system does not deviate from the instantaneous Maxwellian distribution on the scaled variables, we expect $a_l$ to be small enough for $\gamma \ll 1$. Therefore, it suffices to carry out the forthcoming analysis under the first Sonine approximation, as the excess kurtosis is expected to remain small for all times.
\subsubsection{\label{ch2_subsubsec:regular-perturbation}Regular perturbation series}
Following a regular perturbative approach in $\gamma$, we write
\begin{subequations}
\label{ch2_eq:regular-perturbative-gamma}
\begin{align}
    \theta(t) &= \theta^{(0)}(t) + \gamma \theta^{(1)}(t) + O(\gamma^2),
    \\
    a_2(t) &= a_2^{(0)}(t) + \gamma a_2^{(1)}(t) + O(\gamma^2),
\end{align}
\end{subequations}
which must be complemented with adequate initial conditions. As we begin our relaxation from an initial equilibrium state with temperature $T_i$, we have $\theta(0) = \theta_i$ and $a_2(0) = 0$. Since these initial conditions must hold regardless of the value of $\gamma$, it follows that 
\begin{equation}\label{ch2_eq:initial-conditions-perturbative}
    \theta^{(0)}(0) = \theta_i, \quad a_2^{(0)}(0) = 0, \quad \theta^{(j)}(0) = a_2^{(j)}(0) = 0, \quad j=1,2,...
\end{equation}
The expansions from Eqs.~\eqref{ch2_eq:regular-perturbative-gamma} are then introduced in the evolution equations \eqref{ch2_eq:first-sonine-edos}, giving rise to the following hierarchy of equations; one set of equations for each order of $\gamma$. To the lowest order $O(\gamma^0) = O(1)$, we have
\begin{subequations}\label{ch2_eq:regular-perturbative-O1}
\begin{align}
\dot{\theta}^{(0)}&= 2(1-\theta^{(0)}),\label{ch2_eq:regular-perturbative-O1-temp}
\\
\dot{a}_2^{(0)}&= -\left[\frac{4}{\theta^{(0)}}+ \frac{8(d-1)}{d(d+2)}\frac{\sqrt{\theta^{(0)}}}{\xi}\right]a_2^{(0)}.\label{ch2_eq:regular-perturbative-O1-a2}
\end{align}
\end{subequations}
The initial conditions \eqref{ch2_eq:initial-conditions-perturbative} entail that 
\begin{equation}
    \theta^{(0)}(t) = 1 + (\theta_i-1)e^{-2t}, \quad a_2^{(0)}(t) = 0,
\end{equation}
which correspond to the solutions obtained for $\gamma = 0$: the purely exponential relaxation from Eq.~\eqref{ch2_eq:exponential-temperature}, which is associated with the system following the instantaneous equilibrium distribution at all times. To the first order, $O(\gamma)$, we have
\begin{subequations}\label{ch2_eq:regular-perturbative-O2}
    \begin{align}
    \dot{\theta}^{(1)}=&-2\theta^{(1)}+2(d+2)\theta^{(0)}(1-\theta^{(0)})-2(d+2)(\theta^{(0)})^2 a_2^{(0)}, \label{ch2_eq:regular-perturbative-O2-temp}
    \\
    \dot{a}_2^{(1)} =& \ 8(1-\theta^{(0)})-\left[\frac{4}{\theta^{(0)}}+\frac{8(d-1)}{d(d+2)}\frac{\sqrt{\theta^{(0)}}}{\xi}\right]a_2^{(1)} \nonumber
    \\
    & -\left[-\frac{4\theta^{(1)}}{(\theta^{(0)})^2}-8+4(d+8)\theta^{(0)}+\frac{4(d-1)\theta^{(1)}}{d(d+2)\xi\sqrt{\theta^{(0)}}} \right]a_2^{(0)}, \label{ch2_eq:regular-perturbative-O2-a2}
    \end{align}
    \end{subequations}
which, given the initial conditions \eqref{ch2_eq:initial-conditions-perturbative}, lead to the solutions
\begin{subequations}\label{ch2_eq:regular-perturbative-O2-sol}
\begin{align}
    \theta^{(1)}(t) &= -2(d+2)(\theta_{i}-1)\left[t+(\theta_{i}-1)\frac{1-e^{-2t}}{2}\right]e^{-2t}, \label{ch2_eq:regular-perturbative-O2-sol-temp}
\\
    a_2^{(1)}(t) &= 8\int_0^tdt'\  \left[1-\theta^{(0)}(t')\right]\exp \left[\hat{I}(t')-\hat{I}(t) \right], \label{ch2_eq:regular-perturbative-O2-sol-a2}
\end{align}
\end{subequations}
where we have defined the integral function $\hat{I}(t)$ as
\begin{align}\label{ch2_eq:integral-function}
    \hat{I}(t) = & \int_0^tdt'\ \left[ \frac{4}{\theta^{(0)}(t')}+\frac{8(d-1)}{d(d+2)}\frac{\sqrt{\theta^{(0)}(t')}}{\xi}\right].
\end{align}
Since Eq.~\eqref{ch2_eq:regular-perturbative-O2-sol-a2} can only be expressed as an integral function, in order to gain more insights, we briefly consider the limit in which Brownian-Brownian collisions are negligible, $\xi \gg 1$, thus recovering the Fokker-Planck description. Specifically, by looking at Eq.~\eqref{ch2_eq:xi-form}, we infer that large values of $\xi$ may be obtained by considering $n_{\text{par}}\ll n_{\text{bf}}$, i.e. that the density of the gas of Brownian particles is much smaller than that for the background fluid. In this case, the integral from Eq.~\eqref{ch2_eq:regular-perturbative-O2-sol-a2} may be explicitly computed, giving
\begin{equation}\label{ch2_eq:a21-explicit}
    a_2^{(1)}(t) = -\frac{4(\theta_{i}-1)}{(e^{2t}+\theta_{i}-1)^2}\left[e^{2t}-(\theta_{i}-1)^2e^{-2t}-4t(1-\theta_{i})+\theta_{i}(\theta_{i}-2)  \right].
\end{equation}
For $\theta_i=O(1)$, i.e. when the initial temperature $T_i$ is of the same order as the new equilibrium temperature $T_\st$, we have that $a_2^{(1)}$ is also of the order of unity and the regular perturbation expansion developed above works well, since $a_2 = \gamma a_2^{(1)} = O(\gamma)$ as assumed at the beginning of this section. On the other hand, when $\theta_i\gg 1$ we see that $a_2^{(1)}(t)$ basically scales with $\theta_i$ and hence the first correction verifies that $\gamma a_2^{(1)}\propto \gamma\theta_i$. Therefore, as the temperatures is increased up to $O(\gamma^{-1})$, $\gamma a_2^{(1)}$ becomes of the order of unity and the regular perturbation expansion breaks down. In fact, $\gamma a_2^{(1)}$ may reach arbitrarily huge values for high enough temperatures, which neatly signals the failure of our perturbative approach in that limit.  This behaviour suggests that the  limit $\theta_i\gg 1$ deserves its own section, in which the combination $\gamma\theta_i$ is expected to play a major role. It seems that $\gamma \theta_i$ rather than just $\gamma$ is the actual perturbative parameter.

Concerning the behaviour of $\theta^{(1)}(t)$, it fails to reproduce the long-time behaviour of the kinetic temperature, as it contains a term proportional to $te^{-2t}$ that induces a non-monotonic behaviour that is not observed in Figure~\ref{ch2_fig:robustness}. This non-monotonicity becomes significant for moderate values of $\gamma$ or $\theta_i$, allowing the perturbative solution to cross the equilibrium curve $\theta = 1$, which constitutes a non-physical scenario. In order to deal with these anomalous contributions, known as secular terms, we develop in the following a multiple scale approach \cite{bender_advanced_1999}.
\subsubsection{\label{ch2_subsubsec:multiple-scales}Multiple-scale analysis}
This perturbative approach is based on assuming the existence of an additional timescale proportional to the perturbation parameter---namely $\tau \equiv \gamma t$---ruling the relaxation dynamics. In this case, both the scaled kinetic temperature, the excess kurtosis, and all the subleading contributions from Eq.~\eqref{ch2_eq:regular-perturbative-gamma} become functions of both the ordinary time $t$ and the new timescale $\tau$. Therefore, we have to redefine the total time derivative of a function $A = A(t,\tau)$ as
\begin{equation}
    \dot{A} = \frac{\partial A}{\partial t} + \frac{d\tau}{dt}\frac{\partial A}{\partial \tau} = \frac{\partial A}{\partial t} + \gamma \frac{\partial A}{\partial \tau},
\end{equation}
following the chain rule. We remark that, with this new redefinition, time derivatives involve a new term of the order of $\gamma$. Combining the above with Eq.~\eqref{ch2_eq:regular-perturbative-gamma}, we obtain a new hierarchy of equations. To the lowest order $O(1)$, we have
\begin{subequations}\label{ch2_eq:multiple-scales-O1}
\begin{align}
\frac{\partial \theta^{(0)}}{\partial t}&= 2(1-\theta^{(0)}),\label{ch2_eq:mulstiple-scales-O1-temp}
\\
\frac{\partial a_2^{(0)}}{\partial t}&= -\left[\frac{4}{\theta^{(0)}}+ \frac{8(d-1)}{d(d+2)}\frac{\sqrt{\theta^{(0)}}}{\xi}\right]a_2^{(0)},\label{ch2_eq:multiple-scales-O1-a2}
\end{align}
\end{subequations}
which are equivalent to Eqs.~\eqref{ch2_eq:regular-perturbative-O1}, but with a partial time derivative, instead of the total one. They provide the solutions
\begin{equation}\label{ch2_eq:multiple-scales-zeroth-order-incomplete}
    \theta^{(0)}(t,\tau)= 1 + [\hat{\theta}(\tau)-1]e^{-2t}, \quad a_2^{(0)}(t,\tau) = \hat{a}_{2}(\tau)\text{exp}\left[-\hat{I}(t,\tau)\right],
\end{equation}
with $\hat{\theta}(\tau)$ and $\hat{a}_{2}(\tau)$ being two functions of $\tau$ still to be determined, and $\hat{I}(t,\tau)$ follows from Eq.~\eqref{ch2_eq:integral-function} with the appropriate $\tau$-dependence. To the first order $O(\gamma)$, we now have
\begin{subequations}
    \label{ch2_eq:multiple-scales-O2}
    \begin{align}
        \frac{\partial \theta^{(0)}}{\partial \tau} + \frac{\partial \theta^{(1)}}{\partial t} = &-2\theta^{(1)} + 2(d+2)\theta^{(0)}(1-\theta^{(0)})-2(d+2)(\theta^{(0)})^2a_2^{(0)}, \label{ch2_eq:multiple-scales-O2-temp}
    \\
        \frac{\partial a_2^{(0)}}{\partial \tau} + \frac{\partial a_2^{(1)}}{\partial t}  =& \ 8(1-\theta^{(0)})-\left[\frac{4}{\theta^{(0)}}+\frac{8(d-1)}{d(d+2)}\frac{\sqrt{\theta^{(0)}}}{\xi}\right]a_2^{(1)} \nonumber
        \\
        & -\left[-\frac{4\theta^{(1)}}{(\theta^{(0)})^2}-8+4(d+8)\theta^{(0)}+\frac{4(d-1)\theta^{(1)}}{d(d+2)\xi\sqrt{\theta^{(0)}}} \right]a_2^{(0)}. \label{ch2_eq:multiple-scales-O2-kurtosis}
    \end{align}
    \end{subequations}
Apart from the partial derivatives, Eqs.~\eqref{ch2_eq:multiple-scales-O2} differ from Eqs.~\eqref{ch2_eq:regular-perturbative-O2} in the additional terms involving partial differentiation with respect to $\tau$ of the previous order. These terms depend on the aforementioned unknown functions $\hat{\theta}(\tau)$ and $\hat{a}_{2}(\tau)$, which we choose such that they cancel the secular terms involving the anomalous relaxation. The secular terms of a certain partial differential equation (PDE) are defined as those which are proportional to the homogeneous solution of such PDE. Focusing on the kinetic temperature, as the homogeneous solution of Eq.~\eqref{ch2_eq:multiple-scales-O2-temp} is given by the exponential $e^{-2t}$---as is for all orders $O(\gamma^n)$---the only secular term is the one involving $(1-\theta^{(0)})$, since it is indeed proportional to $e^{-2t}$ according to Eq.~\eqref{ch2_eq:multiple-scales-zeroth-order-incomplete}. Taking into account that 
\begin{equation}
    \frac{\partial \theta^{(0)}}{\partial \tau} = e^{-2t}\frac{d\hat{\theta}}{d\tau},
\end{equation}
we impose 
\begin{equation}\label{ch2_eq:canceling-secular-terms}
    \frac{d\hat{\theta}}{d\tau} = -2(d+2)(\hat{\theta} - 1),
\end{equation}
such that it cancels out the $(1-\theta^{(0)})$ term from Eq.~\eqref{ch2_eq:multiple-scales-O2-temp}. Given that $\hat{\theta}(0) = \theta_i$, the above yields
\begin{equation}\label{ch2_eq:solution-multiple-scales}
    \hat{\theta}(\tau) = 1+(\theta_i-1)e^{-2(d+2)\tau} \quad \implies \quad \theta^{(0)}(t,\tau) = 1+(\theta_i-1)e^{-2\left[t + (d+2)\tau\right]}.
\end{equation}
Substituting the expression for the original timescale leads to the following zeroth-order expression for the temperature
\begin{equation}
\label{ch2_eq:relaxation-exponential}
    \theta^{(0)}(t) = 1+(\theta_i-1)e^{-2\left[1+\gamma(d+2)\right]t}.
\end{equation}

\begin{figure}
\centering 
\includegraphics[width=3.5in]{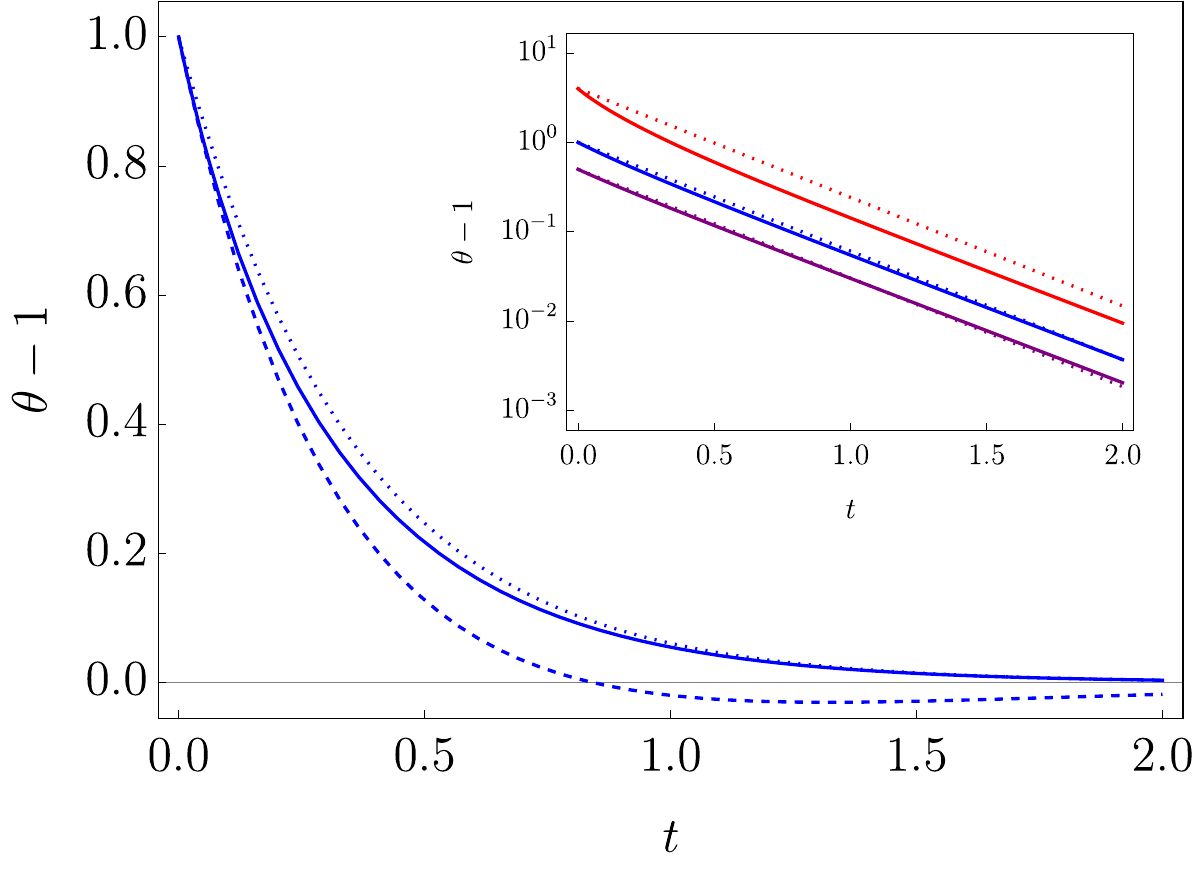}
\caption{Time evolution of the scaled kinetic temperature $\theta$ under the first Sonine approximation for $\theta_i = 2$. Additional parameters employed are $d=2$, $\gamma = 0.1$ and $\xi = 1$. The solid line corresponds to the numerical integration of Eqs.~\eqref{ch2_eq:first-sonine-edos}, while the dashed and dotted ones account for the regular perturbative solution from Eq.~\eqref{ch2_eq:regular-perturbative-O2-sol-temp} and the multiple-scale solution from Eq.~\eqref{ch2_eq:relaxation-exponential}, respectively. Within the inset, we have plotted both solid and dotted lines for different values of $\theta_i$ in logarithmic scale. Specifically, from top to bottom: $\theta_i = 5$ (red), $2$ (blue) and $1.5$ (purple).}
    \label{ch2_fig:perturbative}
\end{figure}
In Figure~\ref{ch2_fig:perturbative}, we compare the numerical integration of the evolution equations provided by the first Sonine approximation with the approximate solutions obtained via our perturbative and multiple scale approaches for the scaled kinetic temperature. On the one hand, the regular perturbative solution fails to reproduce the relaxation behaviour as commented early on, as it involves a non-monotonous hump that overshoots the equilibrium value of the scaled temperature. On the other hand, it is clear that the relaxation is exponential for small $\gamma$ and $\theta_i = O(1)$, thus corroborating our DSMC insights. It is only for $\theta_i=5$ that small discrepancies on the slopes---i.e. the relaxation rates---emerge. Still, the exponential relaxation obtained in Eq.~\eqref{ch2_eq:relaxation-exponential} provides a quantitative account of the dynamics of the kinetic temperature. 

The main conclusion of our perturbative analysis is that the relaxation of the temperature is exponential up to a very good approximation in our system, as long as the initial and the final temperatures are of the same order, i.e. $\theta_i=O(1)$. The latter limits the magnitude of possible memory effects in our system, since exponential relaxation---if exact---prevents the emergence of memory effects, as discussed in Sec.~\ref{ch1_sec:glass}. This is also due to the fact that, as we have demonstrated with our perturbative theory within its limits of applicability, the excess kurtosis is $a_2 = O(\gamma)$. As a consequence, we have to look for other regimes to find ``substantial'' memory effects, as we thoroughly do in the following section.

\subsection{\label{ch2_subsec:quench}Quench to low temperatures}
As discussed in Sec.~\ref{ch1_sec:glass}, glassy behaviour typically emerges when rapidly quenching the fluid of concern to low enough temperatures, such that their dynamical configurational arrangements cease and crystallisation is thus avoided. For the case of our concern, this translates into considering the regime $\theta_i \gg 1$, analogously to the cooling process employed in \cite{prados_kovacs-like_2014} in order to study the Kovacs effect in a uniformly driven granular gas. Given the insights gathered in the preceding section, we expect relatively large values---in absolute value---of the Sonine cumulants in this temperature regime. Hence, we mainly employ in this section the second Sonine approximation scheme to carry out the analysis. 
\subsubsection{\label{ch2_subsubsec:scaling-ev-eqs}{Scaling the evolution equations}}
In order to study the $\theta_i \gg 1$ limit, it is suitable to introduce the relaxation function
\begin{equation}
    Y(t) \equiv \frac{\theta(t)}{\theta_i}.
\end{equation}
By definition, $Y(0)=1$, and $Y(t)$ remains of the order of unity for not too long times. This quantity accounts for the overall relaxation behaviour of the kinetic temperature. Furthermore, had we defined a normalised relaxation function in the standard way,
\begin{equation}\label{ch2_eq:relaxation-function}
    \phi(t) \equiv \frac{T(t)-T_s}{T_i - T_s} = \frac{\theta(t)-1}{\theta_i-1}, \quad \phi (0) = 1, \quad \lim_{t\rightarrow \infty}\phi(t) = 0, 
\end{equation}
we would have that $\phi(t) \simeq Y(t)$ for not too long times, as $\theta(t) \gg 1$ in that regime. In the long-time limit we have that $\lim_{t\rightarrow \infty}Y(t) = \theta_i^{-1} \ll 1$, which clearly differs from $\phi(t)$. Substituting the above scaling into the evolution equations~\eqref{ch2_eq:second-sonine-eqs} gives
\begin{subequations}\label{ch2_eq:evol-eqs-dominant}
\begin{align}
    \label{ch2_subeq:evol-eq-T-dominant}
    \dot{Y} = &-2\gamma \theta_{i} (d+2) Y^2(1+a_2) + O(1),
    \\
    \dot{a}_2  =& -4\gamma \theta_{i} Y \left[ (d+8) (a_2 - a_2^r) - (d+4)(a_3-a_3^r) \right] + O(\gamma) + O(\sqrt{\theta_{i}}/\xi),
    \\
  \dot{a}_3  =& -6\gamma \theta_{i} Y \left[-12(a_2-a_2^r) + (d+14)(a_3-a_3^r)\right]  + O(\gamma) + O(\sqrt{\theta_{i}}/\xi),
\end{align}
\end{subequations}
where
\begin{equation}
  \label{ch2_eq:a2r-a3r}
  a_{2}^{r}\equiv-\frac{2(d+14)}{d^{2}+10d+64}, \qquad a_{3}^{r}\equiv
  -\frac{24}{d^{2}+10d+64}.
\end{equation}
Assuming that $\sqrt{\theta_{i}}/\xi \ll \gamma \theta_i$---such that Brownian-Brownian collisions are subdominant in the relaxation, the dominant terms on the right-hand side of the above system are of the order of $\gamma \theta_i \gg 1$. Let us remark that, under the original variable $\theta$, these dominant contributions stem from the $O(\theta^2)$ terms from Eq.~\eqref{ch2_subeq:temp-second-sonine} and the $O(\theta)$ ones in Eqs.~\eqref{ch2_subeq:kurtosis-second-sonine} and \eqref{ch2_subeq:sixth-second-sonine}. We recall that $\gamma \theta_i$ was the alleged perturbation parameter inferred from Sec.~\ref{ch2_subsec:perturbation}, where it was assumed to be small enough such that the relaxation regime from Eq.~\eqref{ch2_eq:relaxation-exponential} prevailed. It becomes clearer now that the $\gamma \theta_i \gg 1$ regime entails a remarkably different behaviour for both the kinetic temperature and the Sonine cumulants. In fact, the above evolution equations suggest that we introduce a new timescale
\begin{equation}
    s \equiv \gamma \theta_i t,
\end{equation}
which becomes the relevant one for the dynamics, instead of the original timescale $t$. By retaining the dominant terms in $\theta_i$, we arrive at the approximate system
\begin{subequations}\label{ch2_eq:evol-eqs-s-scale}
\begin{align}
  \frac{dY}{ds}&=-2(d+2)Y^{2}(1+a_{2}),
  \label{ch2_subeq:evol-eqs-s-scale-a}\\
  \frac{da_{2}}{ds}&=-4Y \left[(d+8)(a_{2}-a_{2}^{r})-
                     (d+4)(a_{3}-a_{3}^{r}) \right],
  \label{ch2_subeq:evol-eqs-s-scale-b}\\
  \frac{da_{3}}{ds}&=-6Y
  \left[-12(a_{2}-a_{2}^{r})+ (d+14)(a_{3}-a_{3}^{r}) \right],
  \label{ch2_subeq:evol-eqs-s-scale-c}                 
\end{align}
\end{subequations}
Given the above system, we identify the reference values $a_2^r$ and $a_3^r$ from Eq.~\eqref{ch2_eq:a2r-a3r} with the pseudostationary values of the Sonine cumulants in this limit---i.e. those obtained by imposing $da_2/ds=da_3/ds=0$. Specifically, for $d=2$, $a_{2}^{r}\simeq -0.36$ and
$a_{3}^{r}\simeq -0.27$. 

\subsubsection{\label{ch2_subsubsec:llnes}Universal non-exponential relaxation and LLNES}
It is remarkable that the scaled evolution equations \eqref{ch2_eq:evol-eqs-s-scale}, together with the pseudostationary values \eqref{ch2_eq:a2r-a3r} for the Sonine cumulants are independent of both the non-linear parameter $\gamma$, the ratio of temperatures $\theta_i$, and the collision rate $\xi^{-1}$. The latter implies that the relaxation function presents universal features: i.e. when plotting $Y=\theta/\theta_i$ against $s=\gamma \theta_i t$ all the relaxation curves obtained for different values of $\gamma$, $\theta_i$ and $\xi$ should be superimposed onto a unique curve, regardless of the values that such parameters attain, as long as the $\gamma \theta_i \gg 1$ and $\sqrt{\theta_{i}}/\xi \ll \gamma \theta_i$ criterions are met.

\begin{figure}
  \centering
  \includegraphics[width=3.5in]{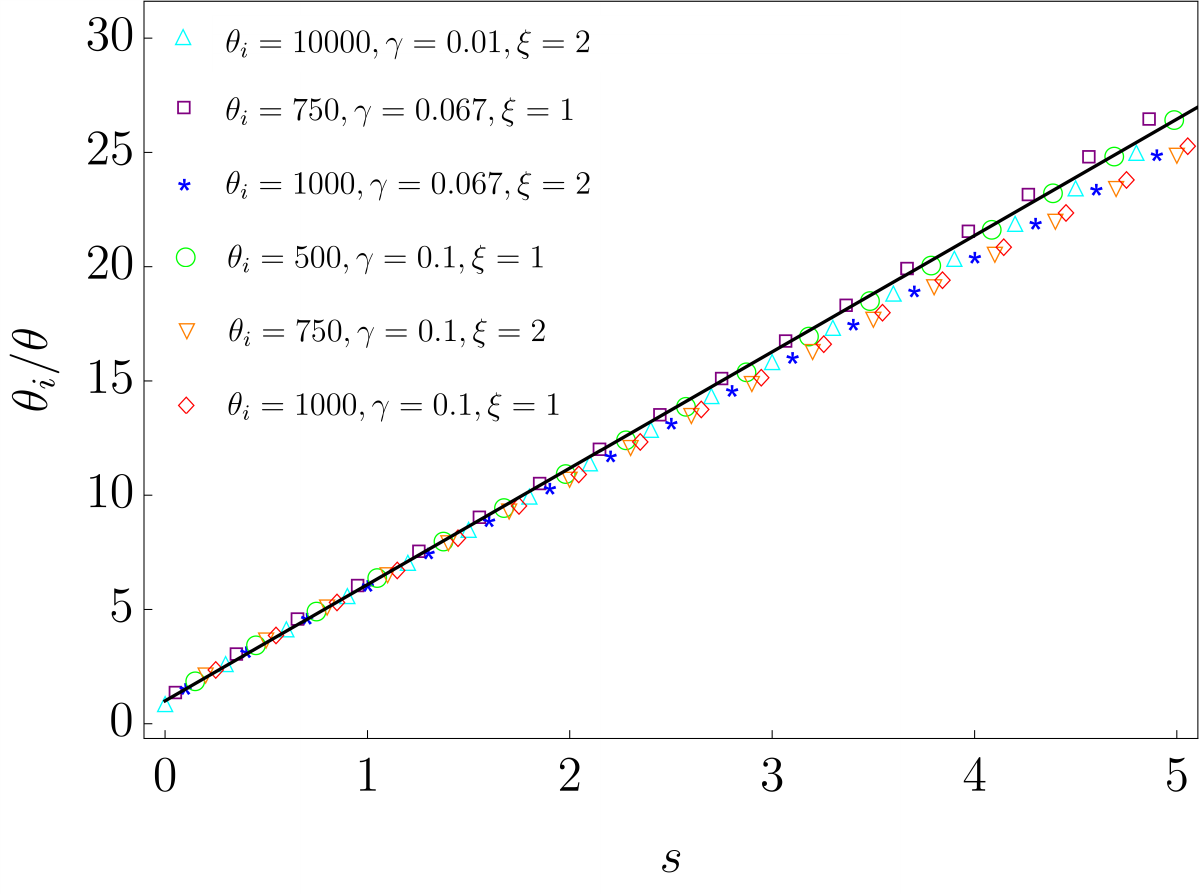}
  \caption{\label{ch2_fig:universal-relaxation} Relaxation function for the kinetic temperature after a
    quench to a low temperature. Specifically, we plot
    $1/Y=\theta_{i}/\theta$ as a function of the scaled timescale
    $s=\gamma\theta_{i}t$. Symbols correspond to DSMC data for different values of the parameters $(\theta_{i},\gamma,\xi)$, as specified in the legend, for $d=2$. Additionally, the black solid curve corresponds to the theoretical prediction from Eq.~\eqref{ch2_eq:algebraic-relaxation}. The linear behaviour of $1/Y$ entails that the kinetic temperature relaxes algebraically as $t^{-1}$.
    }
\end{figure}

In Figure~\ref{ch2_fig:universal-relaxation}, the universality of the relaxation function is validated for different simulation datasets obtained via the DSMC method. Specifically, we plot the inverse relaxation function $1/Y$ versus the timescale $s$ for different values of $\gamma$, $\theta_i$, and $\xi$, such that $50\leq\gamma\theta_i\leq 100$, $0.01\leq\gamma\leq 0.1$, and $1\leq\xi\leq 2$. It is neatly observed that a linear behaviour emerges, regardless of the parameter values employed. The latter implies that $Y(s)$ relaxes algebraically as $s^{-1}$ or, equivalently, as $t^{-1}$. Power-law relaxation functions are common for systems displaying glassy features~\cite{nishikawa_relaxation_2021,chakrabarty_power_2006}. This strongly non-exponential behaviour may be understood in the following way: the Sonine cumulants relax towards their pseudostationary values \eqref{ch2_eq:a2r-a3r} very rapidly, as discussed below. Hence, by substituting $a_2 = a_2^r$ in the evolution equation for $Y$, we get
\begin{equation}\label{ch2_eq:algebraic-relaxation}
    \frac{dY}{ds}=-2(d+2)Y^{2}(1+a_{2}^r) \ \Longrightarrow \ Y(s) = Y_{\text{alg}}(s) \equiv \frac{1}{1+2(d+2)(1+a_2^r)s}.
\end{equation}
This theoretical prediction is also plotted in Figure~\ref{ch2_fig:universal-relaxation}, where the excellent agreement between the theory and the numerics is clearly observed. Our results point towards the existence of a non-equilibrium state, which we refer to as the \textit{long-lived non-equilibrium state} (LLNES), that the system approaches when being submitted to a deep quench of temperatures. Such state is time-dependent, and its time dependency is accounted exclusively by the kinetic temperature, which relaxes algebraically---mainly due to the fact that $\dot{\theta}\propto -\theta^2$ for $\theta \gg 1$, when closely examining Eq.~\eqref{ch2_subeq:temp-second-sonine}---while the Sonine cumulants attain their pseudostationary values, which only depend on the dimensionality of the system. The system remains over the LLNES for most of its relaxation, as Figure~\ref{ch2_fig:universal-relaxation} shows. It is only for very large values of $s$ that $1/Y$ deviates from the linear increase and saturates to its equilibrium value $1/Y(\infty) = \theta_i$.\footnote{Note that this is not observed in Figure~\ref{ch2_fig:universal-relaxation} for the sake of emphasising the linear behaviour.}

\begin{figure}
    {\centering 
    \includegraphics[width=2.65in]{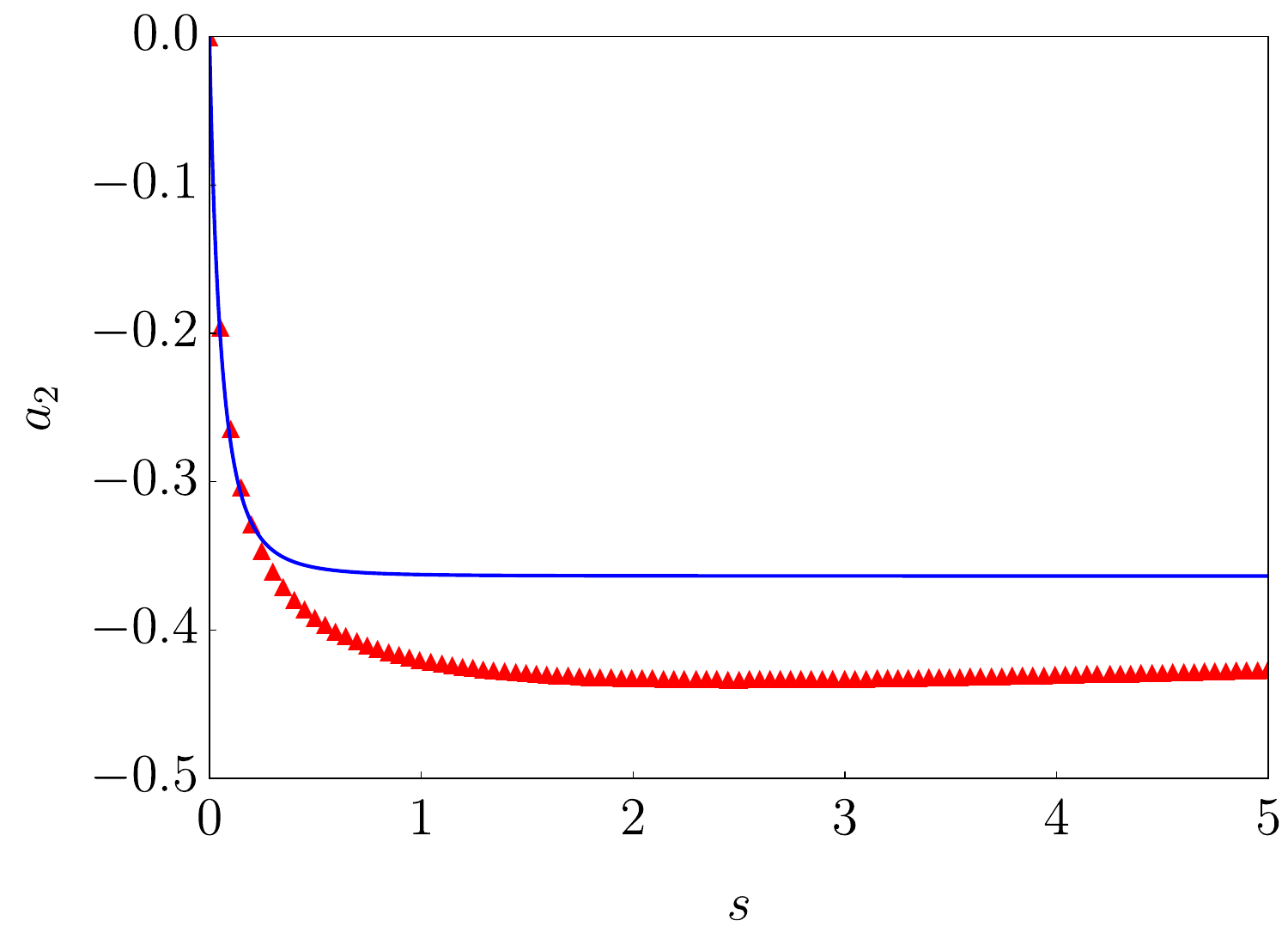}
    \includegraphics[width=2.65in]{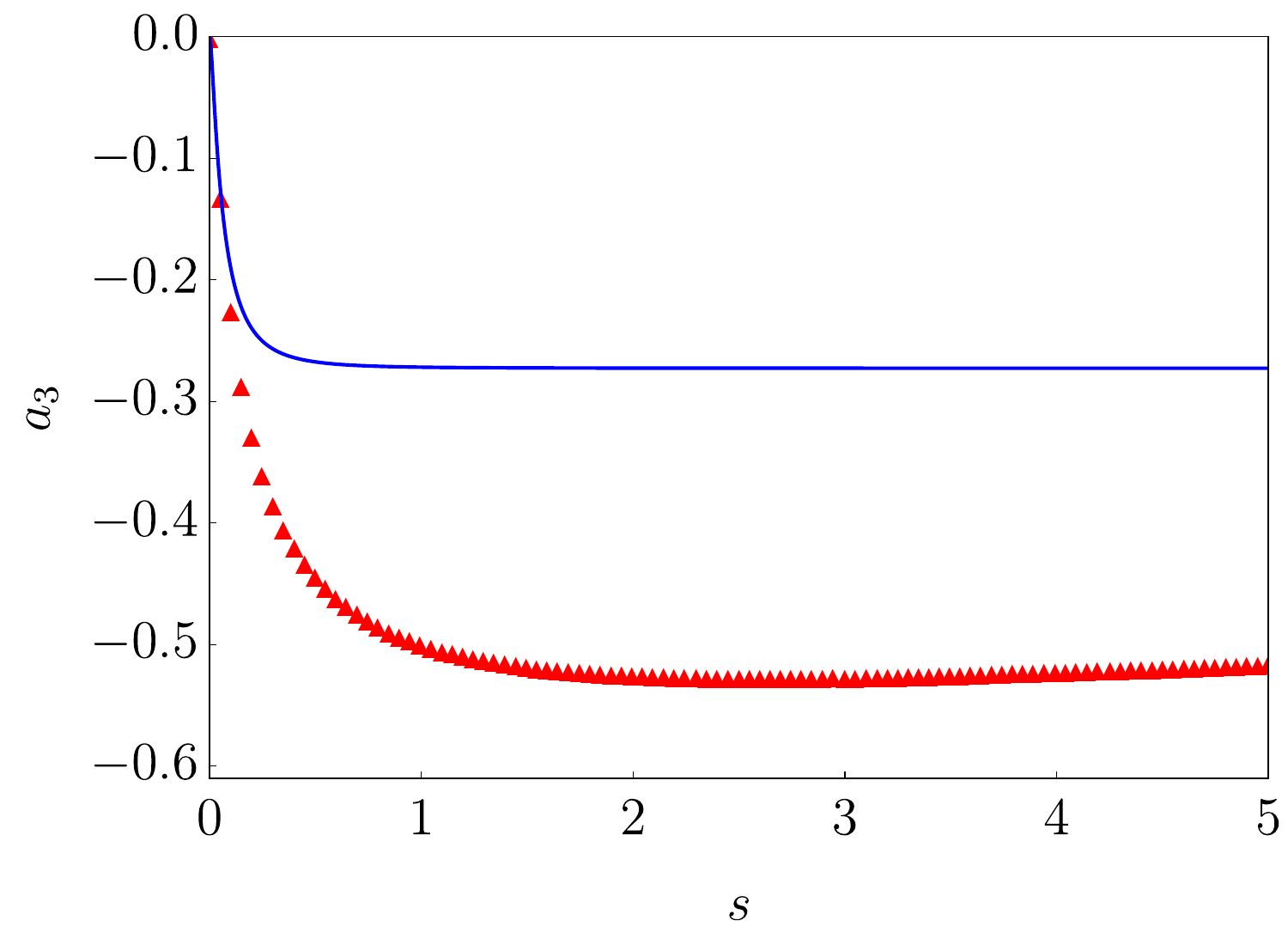}}
    \caption{Relaxation of the excess kurtosis $a_2$ (left) and the sixth cumulant $a_3$ (right) after a quench to a low temperature in the timescale $s=\gamma\theta_{i}t$. Symbols correspond to DSMC data for $\theta_{i} = 1000$, while solid lines correspond to the numerical integration of the scaled evolution equations from Eq.~\eqref{ch2_eq:evol-eqs-s-scale}. Additional employed parameters are $d=2$, $\gamma = 0.1$, and $\xi = 1$.}
        \label{ch2_fig:cumulants-llnes}
\end{figure}

The assumption of the fast relaxation of the Sonine cumulants may be checked by looking into their time evolution over the timescale $s$. The latter is depicted in Figure~\ref{ch2_fig:cumulants-llnes}, which shows the same time window $0\leq s\leq 5$ as in Figure~\ref{ch2_fig:universal-relaxation}. Over there, we plot DSMC simulation data and the numerical integration of the approximate system of equations~\eqref{ch2_eq:evol-eqs-s-scale}. It may be neatly observed that both cumulants rapidly become negative and quite large in absolute value, being approximately constant for $s\geq 1$. However, there appear some discrepancies between the theoretical predictions  and the DSMC data for the plateau values reached by the cumulants. These mainly stem from the Sonine approximation scheme employed, in which the contributions from higher order cumulants $a_n$, with $n\geq 4$, are neglected.\footnote{In chapter~\ref{ch:nonequilibrium_attractor}, we uncover the true form of the VDF at the LLNES, from which the true values of the Sonine cumulants can be determined---and are explicitly shown in Appendix~\ref{app:sonine-cumulants-LLNES}.} Moreover, the numerical plateau reached by the sixth cumulant $a_3$ is larger in absolute value than that for the excess kurtosis $a_2$, implying the failure of the Sonine approximation schemes, where it is implicitly assumed that $|a_n| > |a_{n+1}|$ to truncate the expansion. Nevertheless, these discrepancies have very little impact on the behaviour of the kinetic temperature, as Figure~\ref{ch2_fig:universal-relaxation} shows. 

\section{\label{ch2_sec:memory}Memory effects at the LLNES}
In this last section, we take advantage of the far-from-equilibrium LLNES state uncovered in the preceding section to study the emergence of strong memory effects, specifically the Mpemba and Kovacs effects. We expect the strength of such effects to be intimately dependent on the (absolute) amplitude of the Sonine cumulants: the larger the deviations from the equilibrium Maxwellian VDF, the stronger the possible non-equilibrium phenomena that may emerge.
\subsection{\label{ch2_subsec:extrema}Extrema for the cumulants}
First, it is insightful to study how large the Sonine cumulants---especially the excess kurtosis, as the evolution of the kinetic temperature depends on it---may be in absolute value: i.e. their extreme values, as these provide hints on optimal preparations of our system to maximise the strength of the memory effect of interest. Similar to the preceding section, it is instructive to first work under the first Sonine approximation. Focusing on Eq.~\eqref{ch2_subeq:kurtosis-first-sonine}, at the time that the excess kurtosis reaches one of its extrema, $a_2^{\text{ext}}$, we have that $\dot{a}_2 = 0$, thus implying that $a_2^{\text{ex}}$ satisfies the relation
\begin{equation}\label{ch2_eq:a2-ex}
a_2^{\text{ext}}=\frac{8\gamma(1-\theta)}{\frac{4}{\theta}-8\gamma+4\gamma(d+8)\theta + \frac{8(d-1)}{d(d+2)}\frac{\sqrt{\theta}}{\xi}}.
\end{equation}
For given values of $\gamma$ and $\xi$, $a_2^{\text{ext}}$ becomes a function solely on $\theta$. In fact, the asymptotic behaviour of $a_2^{\text{ext}}(\theta)$, both on the $\theta\rightarrow 0^+$ and $\theta \rightarrow +\infty$ limits becomes independent of $\xi$,
\begin{align}
\label{ch2_eq:extremes-excess-kurtosis}
    a_2^{\text{ext}}&\sim 2\gamma\theta, \quad \theta\to 0^+, \nonumber
    \\
    a_2^{\text{ext}}&\to a_2^{\min}=-\frac{2}{d+8} = a_2^{r'}, \quad \theta\to +\infty,
\end{align}
with $a_2^{r'}$ being the pseudostationary value of the excess kurtosis obtained upon a quench to low enough temperatures, but under the first Sonine approximation.\footnote{We must highlight that, throughout the majority of this thesis, we employ the symbol $\sim$ with the meaning of ``asymptotic to''~\cite{bender_advanced_1999}.} The prime indicates that it is different from the value $a_2^r$ from Eq.~\eqref{ch2_eq:a2r-a3r} under the extended Sonine approximation.

Furthermore, we also have that $a_2^{\text{ext}} = 0$ for both $\theta = 0$ and $\theta = 1$, regardless of $\xi$. Thus, the qualitative behaviour of $a_2^{\text{ext}}(\theta)$ is the following: it vanishes at $\theta = 0$ and $\theta = 1$, thus presenting a maximum for $\theta \in (0,1)$. Such maximum is roughly proportional to $\gamma$ and, in the $\xi \rightarrow \infty$ limit, it can be shown that $a_2^{\text{max}} = \gamma /2$. For $\theta > 1$, the excess kurtosis decreases towards its minimum value, which corresponds to the pseudostationary one attained at the LLNES, which is much larger in absolute value than $a_2^{\text{max}}$. For instance, for $\gamma=0.1$, $d=2$ and $\xi=1$ we have that $a_2^{\text{max}}\simeq 0.04$ whereas $a_2^{r'}=-0.2$.

Under the second Sonine approximation, we impose $\dot{a}_2 = \dot{a}_3 = 0$ in Eqs.~\eqref{ch2_eq:evol-eqs-s-scale} in order to obtain the extrema of $a_2$ and $a_3$, namely $a_2^{\text{ext}}$ and $a_3^{\text{ext}}$. The explicit expressions for these as functions of $\theta$, given fixed values of $\gamma$ and $\xi$ are rather cumbersome and not particularly illuminating, so we avoid writing them here. Still, the qualitative behaviour of $a_2^{\text{ext}}$ and $a_3^{\text{ext}}$ resembles that for the excess kurtosis within the first Sonine approximation:
both $a_2^{\text{ext}}(\theta)$ and $a_3^{\text{ext}}(\theta)$  vanish at $\theta=0$ and $\theta=1$, 
and tend to their minimum (negative) values $a_2^{\min}$ and $a_3^{\min}$ for $\theta\to +\infty$, independently of the values of $\xi$ and $\gamma$, corresponding to the pseudo-stationary values $a_2^r$ and $a_3^r$ from Eq.~\eqref{ch2_eq:a2r-a3r}, respectively. 

\begin{figure}
{\centering 
\includegraphics[width=2.62in]{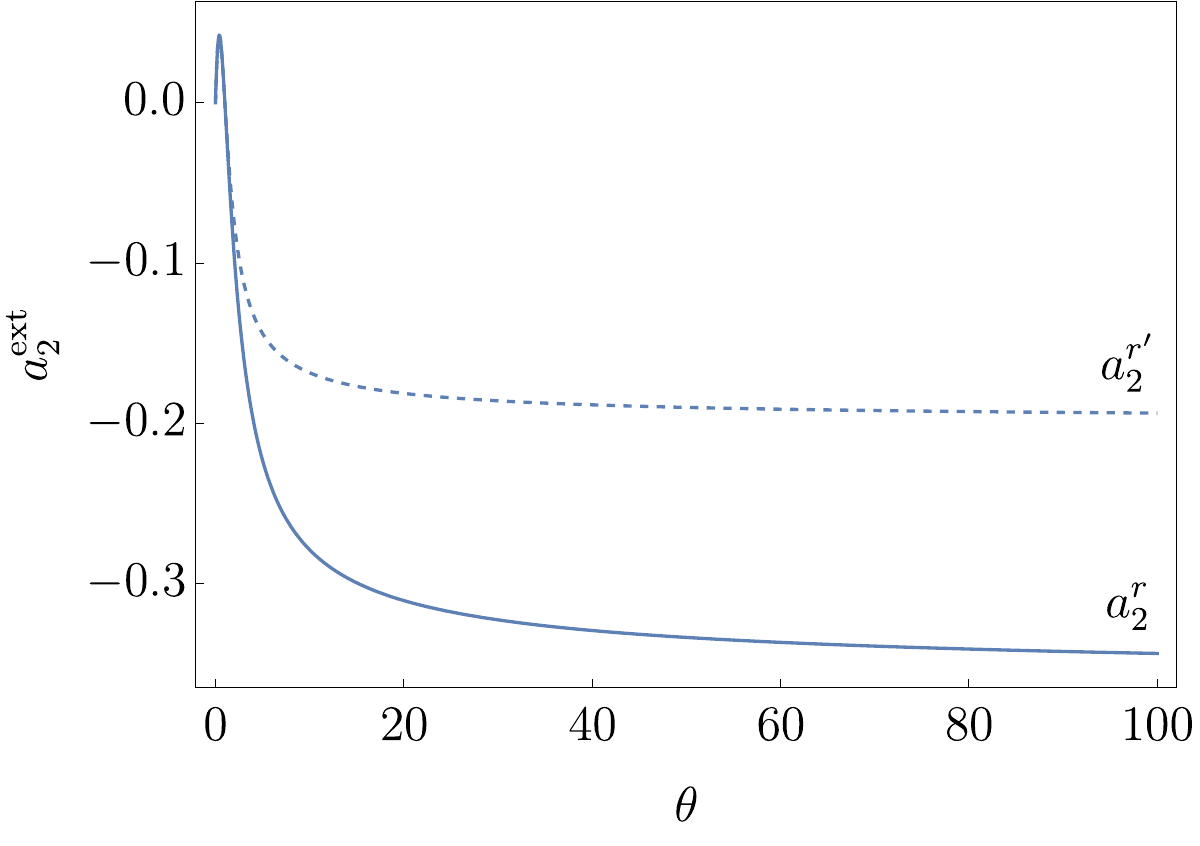}
\includegraphics[width=2.68in]{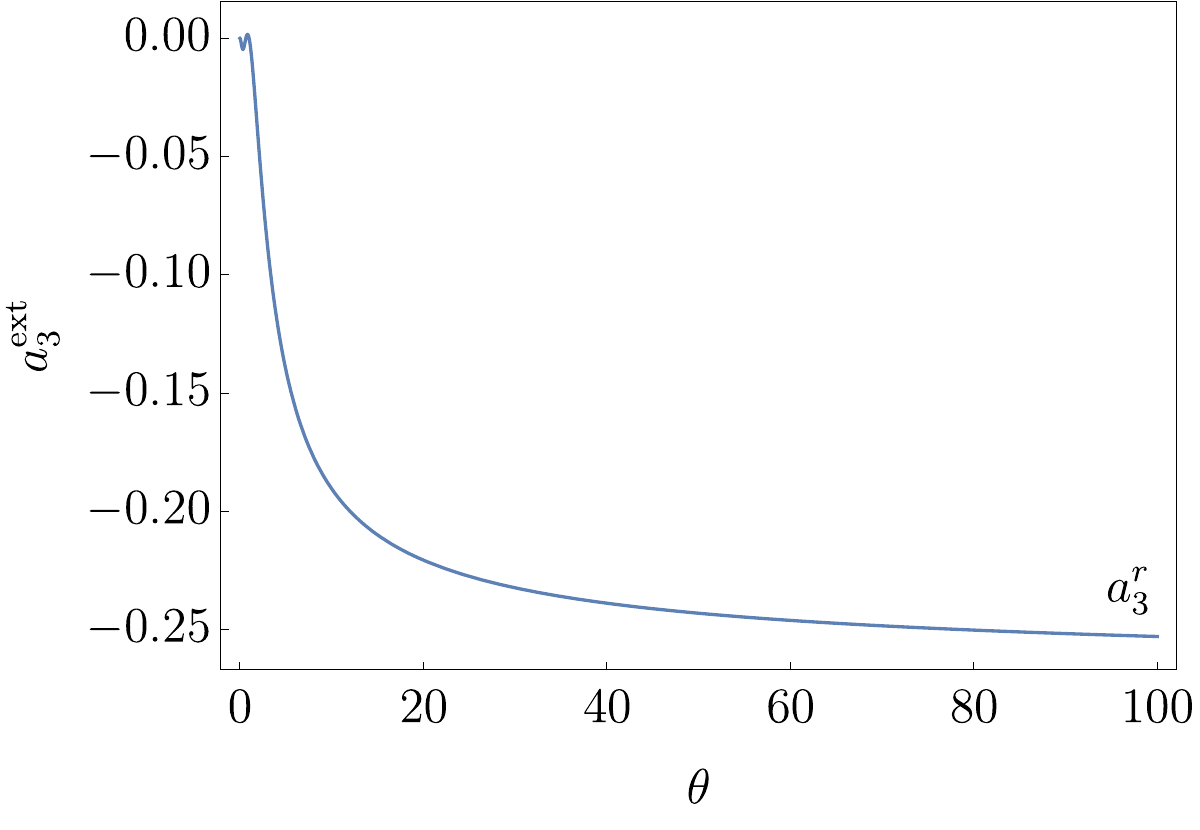}}
\caption{Parametric solutions of the extrema for the cumulants as functions of the dimensionless temperature. Specifically, we plot the extremum for the excess kurtosis $a_{2}$ (left) and the sixth cumulant $a_{3}$ (right), for the employed parameters $d=2$, $\xi = 1$, and $\gamma = 0.1$. Solid curves correspond to the solutions obtained via the extended Sonine approximation, when imposing $\dot{a}_2 = \dot{a}_3 = 0$ in Eq.~\eqref{ch2_eq:evol-eqs-s-scale}, while the dashed curve corresponds to that for the first Sonine approximation, and it is given by Eq.~\eqref{ch2_eq:extremes-excess-kurtosis}. Additionally, we have also marked the corresponding pseudostationary values attained at the LLNES for each approximation scheme.}
    \label{ch2_fig:extremes-cumulants}
\end{figure}

Figure~\ref{ch2_fig:extremes-cumulants} shows the parametric behaviour of the extreme values of both the excess kurtosis $a_2^{\text{ext}}$ and the sixth cumulant $a_3^{\text{ext}}$ as functions of the kinetic temperature $\theta$. As neatly observed, the qualitative behaviour described in the previous paragraph is exactly reproduced in both panels. Although $a_3^{\text{ext}}$ presents more than one root for $\theta < 1$---and henceforth more than one local extrema within the interval $\theta \in (0,1)$---this has little influence on the behaviour of the excess kurtosis, since the extreme values attained for $\theta < 1$ are much smaller than those attained by the excess kurtosis. Also, for both cumulants, their minimum values attained, which are also the largest ones in absolute value, correspond to the pseudostationary values characterising the LLNES.

With all of these, we conclude that the system reaches the largest absolute values for the Sonine cumulants at the LLNES. This entails that the largest deviations from the equilibrium Maxwellian VDF are attained over the LLNES. Therefore, for the forthcoming sections, we choose to prepare the system in the LLNES to observe strong memory effects.

\subsection{\label{ch2_subsec:mpemba}Mpemba effect}
Let us start our analysis with the Mpemba effect. We consider two samples of our molecular fluid: one that is initially hotter (A) at a temperature $\theta_{i,\text{A}}$, and an initially colder one (B) at temperature $\theta_{i,\text{B}} < \theta_{i,\text{A}}$. In the Mpemba effect, the initially hotter sample cools faster towards equilibrium than the colder one. Therefore, the ``cooling rate" of the hotter molecular fluid should be larger. Following the evolution equation for the kinetic temperature \eqref{ch2_subeq:temp-second-sonine}, the Mpemba effect is maximised when the hotter (colder) sample has the largest (smallest) possible value of $a_2$. In such a way, the hotter (colder) sample cools as fast (slow) as possible. A similar direct relation between the cooling rate and the excess kurtosis has been found in other systems described at a kinetic level, both with inelastic and elastic collisions~\cite{lasanta_when_2017,torrente_large_2019,gomez_gonzalez_mpemba-like_2021,santos_mpemba_2020,takada_mpemba_2021}.

In this section, not only do we show that for large enough values of $\Delta a_{2 i}\equiv a_{2 i,\text{A}}-a_{2 i,\text{B}}$ the Mpemba effect emerges, but also (i) how to maximise the amplitude of the effect and (ii) how the system needs to be \textit{aged} previously in order to get the required initial preparation of the samples. As stated above, $a_{2 i,\text{A}}$ ($a_{2 i,\text{B}}$) must take its largest (smallest) possible value to maximise the Mpemba effect. For that, we employ the extreme values of the excess kurtosis derived in Sec.~\ref{ch2_subsec:extrema}.

On the one hand, the minimum value $a_2^{\text{min}}=a_2^r$---maximum in absolute value, since $a_2^r<0$---of the excess kurtosis is obtained for a quench to a very low temperature, i.e. when $\theta_{i}\gg 1$ and the system is cooled to the far-from-equilibrium LLNES state described in the previous section. On the other hand, the maximum value of $a_2$ is obtained following the opposite procedure, i.e. for $\theta_{i}\ll 1$, which corresponds to heating to a much higher temperature. As showed in Sec.~\ref{ch2_subsec:extrema}, $a_2^{\text{max}}$ is proportional to $\gamma$ and much smaller than $|a_2^{\text{min}}|$ in absolute value. Moreover, an even larger absolute value of $a_2^r$ is found in DSMC simulations, as illustrated in Figure~\ref{ch2_fig:cumulants-llnes}.

In order to maximise the Mpemba effect, one may age the two system samples in the following way. The hotter one A is aged by heating it from a much lower temperature, so that $a_2$ takes its maximum value $a_2^{\text{max}}$ and the sample has the largest possible cooling rate. The colder sample B is aged by cooling it from a sufficiently high temperature, so that $a_2$ takes its minimum, pseudostationary, value $a_2^r$, thus reaching the LLNES, and the sample has the smallest possible cooling rate. Still, since $a_2^{\text{max}}$ is quite small compared to $a_2^r$, a practical and very close to optimal procedure is to take the hot sample A at equilibrium, for which $a_{2i,\text{A}}=0$. In this way, the difference $\Delta a_{2i}$ is around 90 per cent of the ideal optimal value $\Delta a_{2i}^{\text{opt}} \equiv a_2^{\text{max}} - a_2^{\text{min}}$. For the remainder of this section, this is the initial preparation that we mainly employ---see Appendix~\ref{app:initial-preparations} for further details on how to initially prepare the system on numerical terms.

\begin{figure}
\centering 
\includegraphics[width=3.5in]{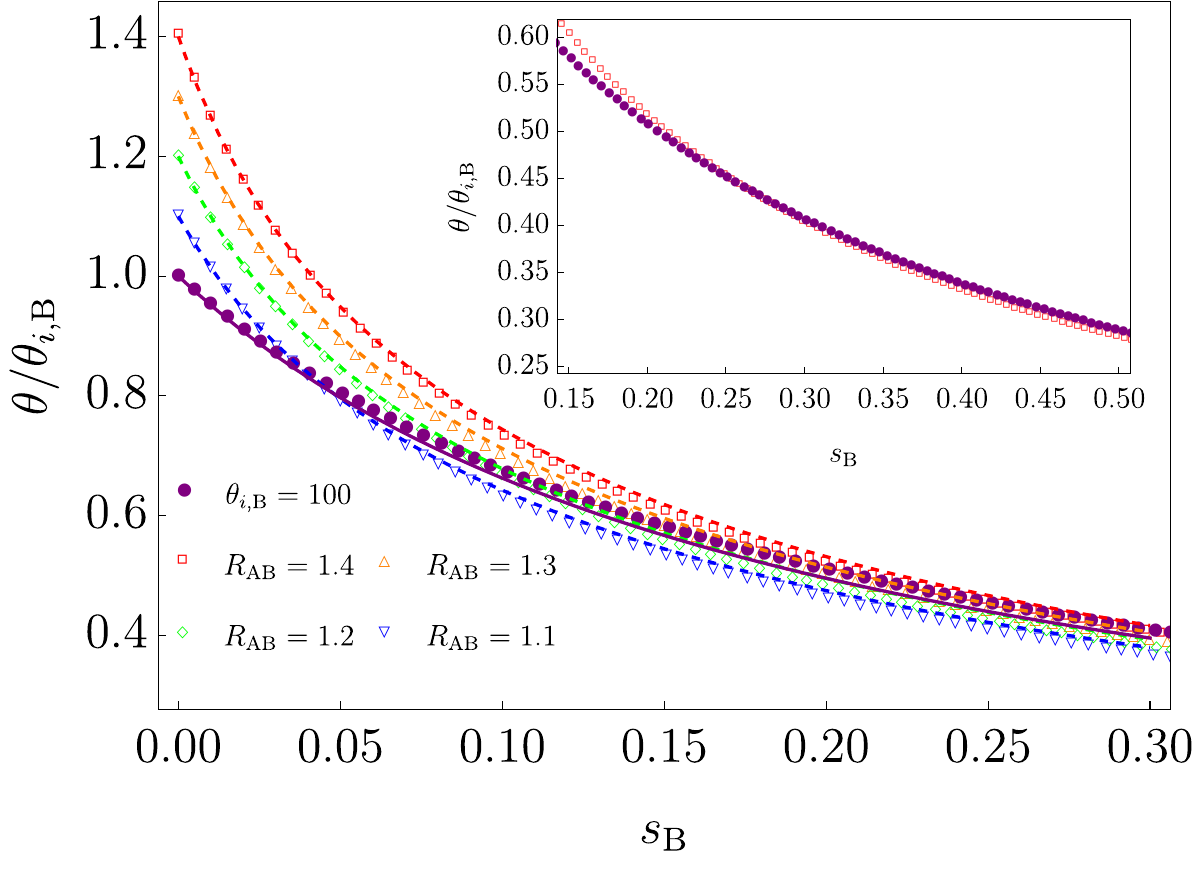}
\caption{Mpemba effect for different initial temperature ratios
    $R_{AB}$. We plot $\theta/\theta_{i,\text{B}}$ as a function of $s_{\text{B}}$ for fixed values of the systems parameters: $d=2$, $\xi = 1$ and $\gamma = 0.1$. Symbols correspond to DSMC simulation data, while lines stem from the numerical integration of Eqs.~\eqref{ch2_eq:evol-eqs-s-scale} with different initial conditions: the hotter samples A (empty DSMC symbols, dashed lines) start from equilibrium with $a_2(0) = 0$, $a_3(0) = 0$, while the colder sample B (filled purple DSMC symbols, solid line) starts from the LLNES with $a_2(0) = a_2^r$, $a_3(0) = a_3^r$. An inset has been plot to illustrate more neatly the Mpemba effect for $R_{AB} = 1.4$ (40\% initial temperature difference).}
    \label{ch2_fig:mpemba-ordinary}
\end{figure}

Both samples of the molecular fluid are put in contact with a common thermal reservoir at a much lower temperature, such that Eqs.~\eqref{ch2_eq:evol-eqs-s-scale} govern their time evolution and, in particular, allow us to describe the \textit{universal} Mpemba effect observed. On the one hand, the initially hotter sample A cools with $a_{2}$ decreasing from zero towards $a_{2}^{r}$, i.e.
\begin{equation}\label{ch2_eq:YA}
  Y_{\text{A}}(s_{\text{A}})=\frac{\theta_{\text{A}}(s_{\text{A}})}{\theta_{i,\text{A}}}=f(s_{\text{A}}),
  \quad s_{\text{A}}=\gamma\theta_{i,\text{A}}t,
\end{equation}
where $f(s)$ is a certain unknown function, independent of  $\theta_{i,\text{A}}$,
the exact form of which is irrelevant for the forthcoming discussion. On the other hand, the
initially colder sample B cools following Eq.~\eqref{ch2_eq:algebraic-relaxation} as it already starts from the LLNES, implying that
\begin{equation}\label{ch2_eq:YB}
  Y_{\text{B}}(s_{\text{B}})=\frac{\theta_{\text{B}}(s_{\text{B}})}{\theta_{i,\text{B}}}=Y_{\text{alg}}(s_{\text{B}}),
  \quad s_{\text{B}}=\gamma\theta_{i,\text{B}}t.
\end{equation}
The Mpemba effect entails that $\theta_{\text{A}}=\theta_{\text{B}}$ for some
crossing time $t_{\times}$---or $s_{\times}$, equivalently. In Figure~\ref{ch2_fig:mpemba-ordinary} we show the emergence of the Mpemba effect. As both the relaxation function $Y$ and the timescale $s$ depend on the initial conditions, for visual comparison we have chosen the scale of the B system, that is, we have plotted $\theta/\theta_{i,B}$ as a function of $s_{\text{B}}=\gamma\theta_{i,\text{B}}t$. By introducing the initial temperature ratio 
\begin{equation}
    R_{\text{AB}}\equiv \frac{\theta_{i,\text{A}}}{\theta_{i,\text{B}}}>1,
\end{equation}
we have that
\begin{equation}
    s_{\text{A}}=R_{\text{AB}}s_{\text{B}}, \quad \frac{\theta_{\text{A}}}{\theta_{i,\text{B}}} = Y_{\text{A}} R_{\text{AB}}. 
\end{equation}
Specifically, we consider one cold sample B, with $\theta_{i,\text{B}}=100$, and four different hot samples A with $R_{\text{AB}}=1.1, 1.2, 1.3$ and $1.4$. Symbols correspond to DSMC simulations of the system and lines to the numerical integration of Eqs.~\eqref{ch2_eq:evol-eqs-s-scale}. The temperature curves cross at a certain time $s_{\text{B},\times}$, corresponding to $t_{\times}$ in the original timescale, $s_{\text{B},\times}=\gamma\theta_{i,\text{B}}t_{\times}$. For $s_{\text{B}}>s_{\text{B},\times}$, the temperature for the initially hotter sample A lies below that for the initially colder one B. The Mpemba effect is even neatly observed for $R_{\text{AB}}=1.2$  (i.e. 20 per cent initial temperature difference). In fact, it is still present up to 40 per cent of initial temperature difference, i.e. $R_{\text{AB}}=1.4$, as illustrated in the inset of the right panel.

\begin{figure}
  \centering
  \includegraphics[width=3.5in]{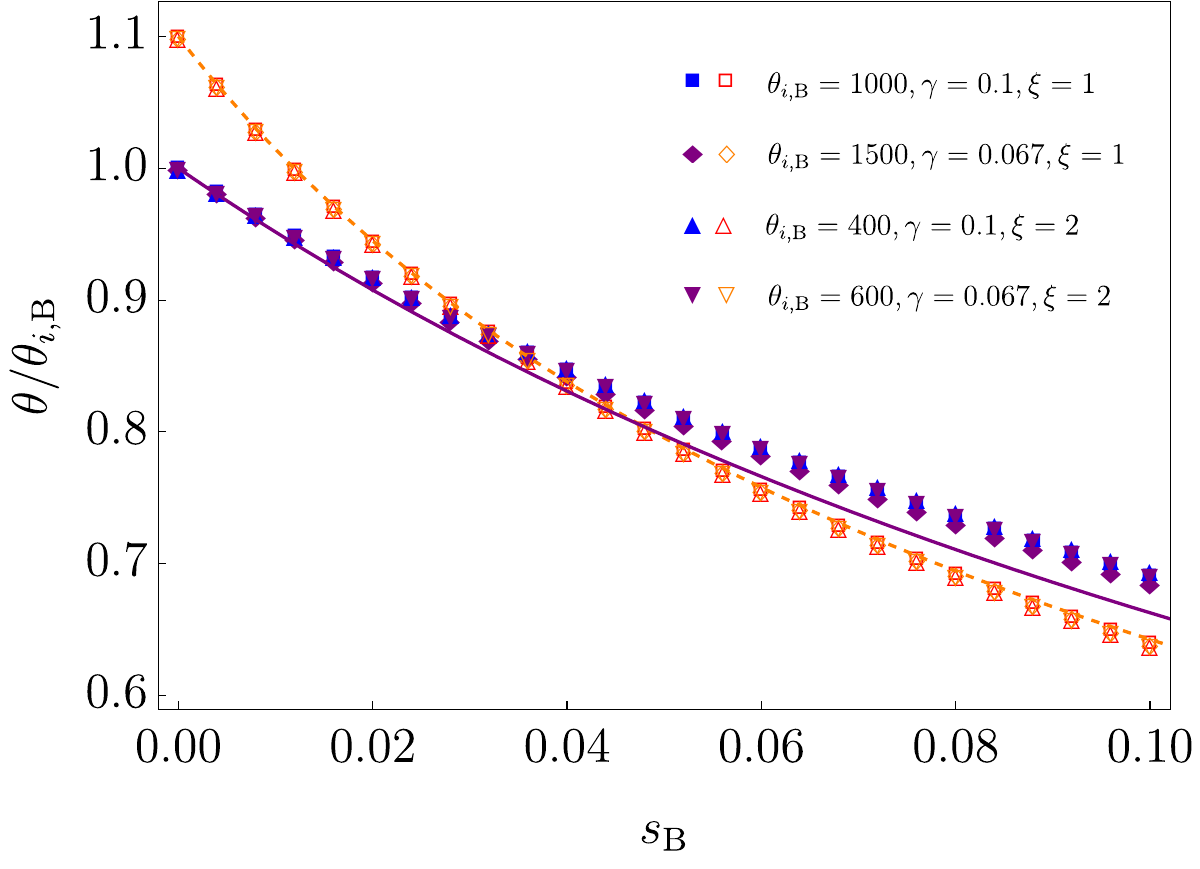}
  \caption{\label{ch2_fig:mpemba-universal} Universal Mpemba effect for different initial preparations of the samples and parameters $(\gamma,\xi)$, for $d=2$. Specifically, we plot $\theta/\theta_{i,\text{B}}$, i.e. the temperature in units of the initial temperature of the colder sample B, as a function of the scaled time $s_{\text{B}}$ for the colder sample, defined in Eq.~\eqref{ch2_eq:algebraic-relaxation}. For a fixed value of the initial temperature ratio $R_{\text{AB}}$, all the curves corresponding to different sets of $(\theta_{i,\text{B}},\gamma,\xi)$ superimpose, both for the hotter (open symbols) and colder samples (filled symbols), A and B respectively. There are eight simulation curves: four corresponding to hot samples A with $R_{\text{AB}}=1.1$ and the corresponding four curves for the cold samples B. Dashed and solid lines are the solutions of Eq.~\eqref{ch2_eq:evol-eqs-s-scale} for ($a_{2}(0)$,$a_{3}(0)$)= ($0,0$) and ($a_2^r$,$a_3^r$), respectively.
    }
\end{figure}

Similar to the non-exponential relaxation \eqref{ch2_eq:algebraic-relaxation} reported at the LLNES, the Mpemba effect also presents universal features. Let us consider a fixed value of the ratio $R_{\text{AB}}$, but different values of the the initial temperatures of the samples $\theta_{i,\text{A}}$ and $\theta_{i,\text{B}}$, the non-linearity parameter $\gamma$,  and the average time between collisions $\xi$, for which the criterions $\gamma \theta_i \gg 1$ and $\sqrt{\theta_i} / \xi \ll \gamma \theta_i$ are met for both samples. 
If we plot $\theta/\theta_{i,\text{B}}$ vs. $s_{\text{B}}$, according to Eqs.~\eqref{ch2_eq:evol-eqs-s-scale}, all the curves corresponding to either the colder or hotter temperatures superimpose towards a unique curve. This is neatly shown in Figure~\ref{ch2_fig:mpemba-universal}, where we have plotted relaxation curves for the fixed value $R_{\text{AB}}=1.1$ and different values of $(\theta_{i,\text{B}},\gamma,\xi)$, as explicitly detailed in the legend. The analytical prediction from Eq.~\eqref{ch2_eq:evol-eqs-s-scale} for the colder sample is slightly under the DSMC data, because of our underestimating the excess kurtosis over the LLNES reported in Fig.~\ref{ch2_fig:cumulants-llnes}. 

In order to quantify the strength of the Mpemba effect, we introduce the parameter $\text{Mp}$~\cite{torrente_large_2019}
\begin{equation}
    \text{Mp} \equiv \max_{t>t_{\times}} \left[\theta_{\text{B}}(t)-\theta_{\text{A}}(t) \right],
\end{equation}
which measures the maximum difference between the relaxation curves once they have crossed each other---recall that both relaxation curves eventually merge at the equilibrium state for long enough times. We have computed the numerical values of $\text{Mp}$ from the DSMC simulation data. Since the strength of the Mpemba effect is proportional to $\theta_{i,\text{B}}$, we have specifically computed $\text{Mp}/\theta_{i,\text{B}}$. For the curves shown in the right panel of Figure~\ref{ch2_fig:mpemba-ordinary}, the values are $\text{Mp}/\theta_{i,\text{B}}=0.059$, $0.045$, $0.034$ and $0.026$ for initial temperature ratios $R_{\text{AB}}=\theta_{i,\text{A}}/\theta_{i,\text{B}}=1.1$, $1.2$, $1.3$ and $1.4$, respectively.  As expected, $\text{Mp}$ decreases with the initial temperature difference $\theta_{i,\text{A}}-\theta_{i,\text{B}}$---or, equivalently, with $R_{\text{AB}}$. Since $\theta_{i,\text{B}}\gg 1$, the values of $\text{Mp}$ for our system are typically larger than unity. In the figure, $\theta_{i,\text{B}}=100$, so $\text{Mp}$ ranges from $2.6$ to $5.9$, values that are higher than those for the large Mpemba-like effect  reported in Ref.~\cite{torrente_large_2019} for a  rough granular gas.

\subsubsection{\label{ch2_subsubsec:inverse-mpemba}Inverse Mpemba effect}
It is also interesting to study the inverse Mpemba effect, in which the initially colder sample B heats sooner than the initially hotter one A, which has also been observed in a wide variety of systems~\cite{lu_nonequilibrium_2017,klich_mpemba_2019,gal_precooling_2020,lasanta_when_2017,torrente_large_2019,santos_mpemba_2020, biswas_mpemba_2020,biswas_mpemba_2023,gomez_gonzalez_mpemba-like_2021,takada_mpemba_2021}. In this case, both samples A and B are put in contact with a thermal reservoir at a larger temperature. If sample A heats slower than sample B, then the inverse Mpemba effect emerges. But heating slower is basically equivalent to cooling faster: in both cases, we want to have $\dot{\theta}_{\text{A}}$ as large as possible. Therefore, from inspection of Eq.~\eqref{ch2_subeq:temp-second-sonine}, we require again to have the initially hotter sample with the maximum possible value of the excess kurtosis $a_2$ and the initially colder one with the minimum possible value. This constitutes exactly the same preparation as for the normal case. 

\begin{figure}
  \centering
  \includegraphics[width=3.5in]{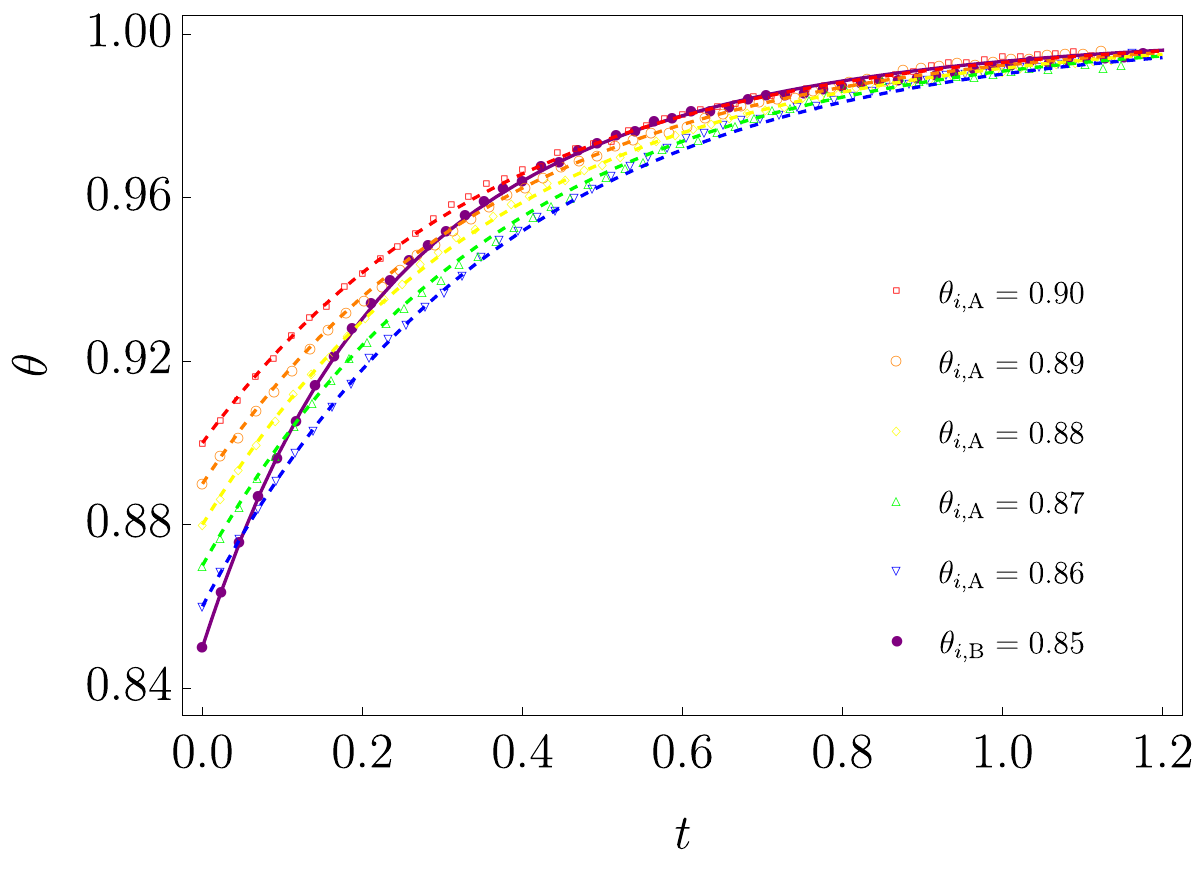}
  \caption{\label{ch2_fig:inverse-mpemba} Inverse Mpemba effect for different initial temperatures for the hotter sample A. Specifically, we consider hotter samples with temperatures $\theta_{i,\text{A}} = 0.86$, $0.87$, $0.88$, $0.89$ and $0.90$, while the colder sample B departs from a temperature of $\theta_{i,\text{B}}= 0.85$. Additional employed parameters are $\xi = 1$ and $d = 2$. Empty (filled) symbols correspond to DSMC data for the hotter (colder) samples, while the dashed (full) lines correspond to the numerical integration of the evolution equations Eqs.~\eqref{ch2_eq:evol-eqs-s-scale} for the hotter (colder) samples.
    }
\end{figure}

Following the above reasoning, we study the inverse Mpemba effect when the initially colder sample B departs from the LLNES while the hotter one A departs from equilibrium. Similarly to what occurred for the ordinary Mpemba effect, the optimal procedure corresponds to follow a heating protocol such that $a_{2i,A}=a_2^{\text{max}}$ but, since $a_{2}^{\text{max}}$ is very small---and of the order of $\gamma$, our initial preparation is nearly optimal and more practical. In Figure~\ref{ch2_fig:inverse-mpemba}, we observe that the initial temperature differences are smaller than those for the normal Mpemba effect. Here, the maximum value of the parameter $R_{\text{AB}}$ is $1.06$, i.e. a 6\% maximum initial temperature difference, whereas in the normal case it was 40\%. Consistently, the strength of the inverse Mpemba effect is smaller than that of the normal one: the values of the $\text{Mp}/\theta_{i,\text{B}}$ parameter range between $0.001$ and $0.013$ in this case.

\subsection{\label{ch2_subsec:kovacs}Kovacs effect}
Finally, let us look into the Kovacs effect. We recall that the Kovacs experiment consists in a two-step protocol: (i) an ageing stage for which the system relaxes towards an equilibrium temperature $T_1$, having started initially at $T_i$, during a time interval $0\leq t\leq t_{w}$, and (ii) a sudden change of the bath temperature at $t=t_{w}$ from $T_1$ to a new value $T_{f}\equiv T(t_{w})$, which coincides with that of the kinetic temperature at that time, as illustrated in Figure~\ref{ch1_fig:kovacs-sketch} from Sec.~\ref{ch1_subsec:kovacs}.

As the dynamical evolution of the kinetic temperature---given by Eq.~\eqref{ch2_subeq:temp-second-sonine}---is determined by both the kinetic temperature itself and the excess kurtosis, the Kovacs hump will emerge if $a_2$ is nonzero at the waiting time $t_{w}$. Also, the larger the absolute value of $a_2$, the greater the magnitude of the expected Kovacs hump. The latter entails that the optimal ageing protocol constitutes a deep quench to a lower temperature $T_1 \ll T_i$, over which the system approaches the LLNES at $t=t_{w}$. Then, for $t>t_{w}$, the evolution equations \eqref{ch2_eq:evol-eqs-s-scale} with  $\theta(t) \equiv T(t)/T_{f}$ characterise the dynamics of the system, given the initial conditions
\begin{equation}\label{ch2_eq:initial-cond-kovacs}
    \theta(t_w) = 1, \quad a_2(t_w) = a_2^r, \quad a_3(t_w) = a_3^r.
\end{equation}

\begin{figure}
  \centering
  \includegraphics[width=3.5in]{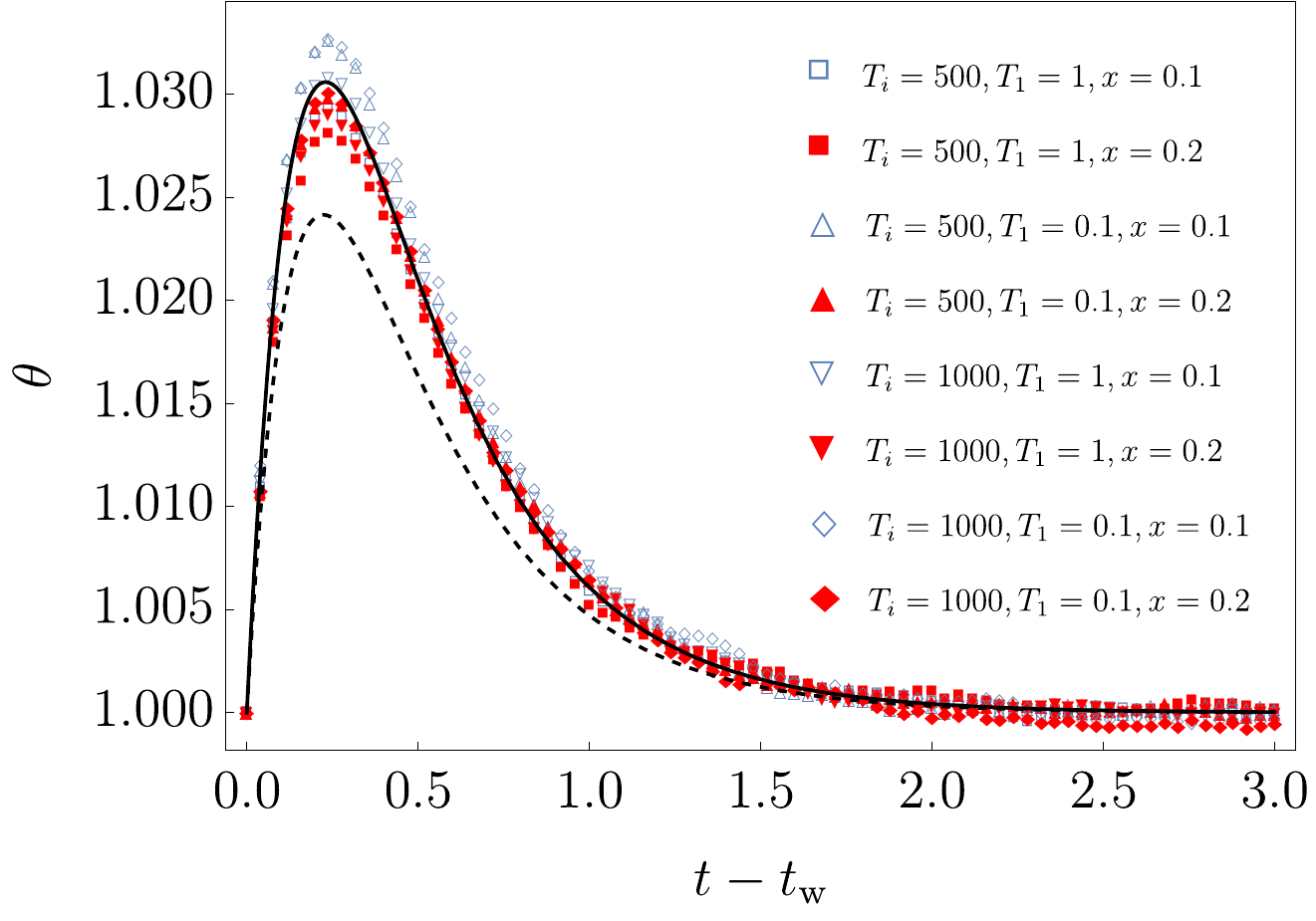}
  \caption{\label{ch2_fig:Kovacs-1} Evolution of the kinetic temperature during the second stage of Kovacs experiment. The parameter values employed are $\gamma = 0.1$, $d=2$,  and $\xi = 1$. Eight simulation curves are shown for different combinations of the initial ($T_{i}$), ageing ($T_1$), and final temperature ($T_f$). 
We write the ageing temperature as $T_{f}=T_{1}+x(T_{i}-T_{1})$, such that the datasets shown correspond to $x=0.2$ (filled symbols) and $0.1$ (open symbols). Curves for smaller values of $x$ are basically superimposed with those for $x=0.1$. The dashed (solid) line corresponds to the numerical integration of  Eqs.~\eqref{ch2_eq:second-sonine-eqs} with the theoretical (simulation) pseudostationary values for the Sonine cumulants at the LLNES.
    }
\end{figure}
The resulting Kovacs hump also has scaling properties, although somehow weaker than those of the temperature relaxation and the Mpemba effect, and hence, we do not refer to them as being universal. Despite the initial conditions and, therefore, the subsequent Kovacs hump not depending on $(T_{i},T_{f},T_{1})$, they do depend on $\gamma$ and $\xi$.
In fact, it is roughly proportional to $\gamma$, as we show below. 
Figure~\ref{ch2_fig:Kovacs-1} presents this scaled Kovacs hump, where we plot $\theta=T/T_{f}$ as a function of $t-t_{w}$, for $t>t_{w}$. Indeed, as mentioned earlier, the triplet $(T_{i},T_{f},T_{1})$ does not affect the Kovacs hump measured in DSMC simulations. Here, for the sake of simplicity, we have taken $T_f$ as unity.\footnote{In the relaxation experiment and the Mpemba memory effect, the unit of temperature is usually taken as the steady temperature $T_s$.} Moreover, our theory quantitatively describes the numerical results: the agreement between the DSMC datasets and our theoretical predictions are very good, especially when the simulation value of $a_2^r$ is employed. Recall that our theory underestimates $|a_2^r|$ by roughly 15 per cent, as shown by Figure~\ref{ch2_fig:cumulants-llnes}.

In order to further study the Kovacs effect on analytical grounds, we develop an additional perturbative approach for the Kovacs function $K(t )\equiv \theta(t)-1$, which accounts for the emergence of the Kovacs hump. See Appendix~\ref{app:perturbative-kovacs} for details. Our approach gives
\begin{align}
\label{ch2_eq:kovacs-hump-first-order}
    K(t) \approx -\gamma a_2^r \frac{2(d+2)}{\lambda_+-\lambda_-}\Bigg[ & \frac{M_{11}+M_{12}+|\lambda_-|}{|\lambda_+|-\hat{\alpha}}\left(e^{-\hat{\alpha} (t-t_w)}-e^{-|\lambda_+|(t-t_w)}\right) \nonumber \\ & - \frac{M_{11}+M_{12}+|\lambda_+|}{|\lambda_-|-\hat{\alpha}}\left(e^{-\hat{\alpha} (t-t_w)}-e^{-|\lambda_-|(t-t_w)}\right)\Bigg],
\end{align}
where $\hat{\alpha}\equiv 2[1+\gamma (d+2)]$, $M_{jl}$ are the elements of the $2\times 2$ matrix $\mathbb{M}$, which are given by
\begin{subequations}\label{ch2_eq:Mij}
\begin{align}
        M_{11} = &-4 \left[1+\gamma(d+6) + \frac{2(d-1)}{d(d+2)\xi} \right], 
        \\
        M_{12} =& \ 2 \frac{a_{3}^r}{a_{2}^r} \left[2\gamma (d+4) + \frac{d-1}{d(d+2)\xi} \right],
        \\
        M_{21} =&\ 12\frac{a_{2}^r}{a_{3}^r}\left[2\gamma + \frac{d-1}{d(d+2)(d+4)\xi} \right],
        \\
        M_{22} =& -6\left[\gamma (d+10) +1 + \frac{(d-1)(4d+19)}{2d(d+2)(d+4)\xi} \right],
\end{align}
\end{subequations}
and its eigenvalues $\lambda_{\pm}$ and corresponding eigenvectors $\bm{u}_{\pm}$ may be expressed as
\begin{equation}\label{ch2_eq:eigenproblem}
    \lambda_{\pm} = \frac{\Tr(\mathbb{M}) \pm \sqrt{[\Tr(\mathbb{M})]^2-4\det(\mathbb{M})}}{2} < 0, \quad \bm{u}_{\pm} = \begin{pmatrix} M_{12} \\ \lambda_{\pm}- M_{11} \end{pmatrix},
\end{equation}
with $\Tr(\mathbb{M})$ and $\det(\mathbb{M})$ being the trace and the determinant of the matrix $\mathbb{M}$, respectively. Since $a_2^r<0$, the predicted Kovacs effect is always normal, as has been shown to be in molecular systems~\cite{prados_kovacs_2010}. We remark that the non-diagonal elements of the matrix $\mathbb{M}$ depend on the Sonine cumulants, in particular on the ratio $a_{3}^r/a_{2}^r$. Had we aged the system differently, $a_2^r$ and $a_3^r$ would have been substituted with $a_2(t_w)$ and $a_3(t_w)$. Nevertheless, since $a_2(t)<0$ when the system is cooled, as observed in Figure~\ref{ch2_fig:cumulants-llnes}, $a_2(t_w)<0$ always and the Kovacs effect remains to be normal.

\begin{figure}
  \centering
  \includegraphics[width=3.5in]{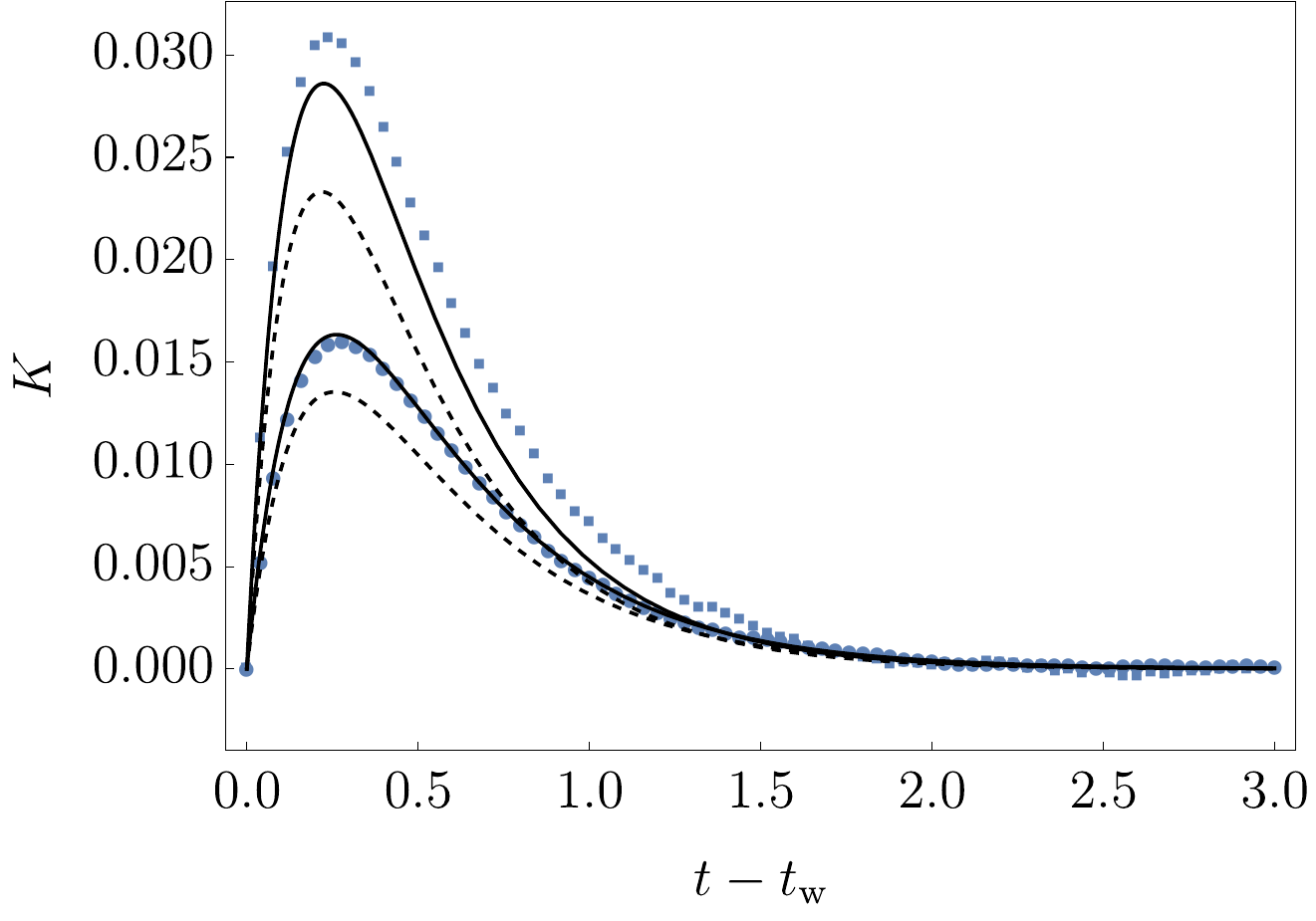}
  \caption{\label{ch2_fig:Kovacs-2} Dependence of the Kovacs hump function on the non-linearity parameter $\gamma$. Two sets of DSMC data are plotted: both correspond to the triplet $(T_i = 1000, T_1 = 0.1, x =0.1)$ for two different values of $\gamma$, specifically $\gamma = 0.1$ (squares) and $0.05$ (circles). Additional parameter values employed are $d=2$  and $\xi = 1$.  The dashed (solid) lines correspond to the first order perturbative expression~\eqref{ch2_eq:kovacs-hump-first-order} with the theoretical (DSMC) values of $a_2^r$ and $a_3^r$.
    }
\end{figure}

In Figure~\ref{ch2_fig:Kovacs-2}, we test the accuracy of our perturbative analysis by comparing our expression \eqref{ch2_eq:kovacs-hump-first-order} for the Kovacs hump function with DSMC simulation data. Specifically, we plot $K(t)$ for two different values of $\gamma$: $\gamma = 0.1$ (filled squares) and $0.05$ (filled circles). Similarly to Figure~\ref{ch2_fig:Kovacs-1}, we write $T_{w}=T_{1}+x(T_{i}-T_{1})$, with the plotted data corresponding to $x=0.1$. We compare the DSMC simulation data with  Eq.~\eqref{ch2_eq:kovacs-hump-first-order}, both employing the theoretical predictions for $a_2^r$  and $a_3^r$ (dashed line) from Eq.~\eqref{ch2_eq:a2r-a3r} and their simulation values (solid line) for each dataset. The mild discrepancies basically stem once again from the difference between the theoretical and DSMC values of the excess kurtosis, as highlighted by the quantitative agreement observed for the solid curves. Our theoretical prediction seems to work better for the $\gamma = 0.05$ case (circles) than for the $\gamma = 0.1$ one (squares), which is consistent with the pertubative approach developed here.

\chapter{Global non-equilibrium attractor for non-linear stochastic dynamics}
\chaptermark{Global non-equilibrium attractor}
\label{ch:nonequilibrium_attractor}
\newcommand{\pdfnondim}{P_i^*}
\newcommand{\LLNES}{\text{L}}

%

As discussed in Sec.~\ref{ch1_subsubsec:long-time-lim}, many physical systems tend in the long time limit, under quite general conditions, to a state in which all trace of initial conditions is lost. Such state is often stationary, either an equilibrium state---such as the Maxwellian VDF \eqref{ch1_eq:maxwellian} or, more generally, the canonical equilibrium---or a non-equilibrium steady state (NESS), but also it could be a time-dependent ``hydrodynamic'' state, in which the system is accurately described in terms of a few macroscopic variables. Such time-dependent states are intimately related to the emergence of non-equilibrium phenomena---in particular glassy behaviour, as briefly mentioned in Sec.~\ref{ch1_sec:glass}.

In chapter~\ref{ch:memory_effects}, we showed for the molecular fluid that there exists a non-equilibrium state, that we referred to as the LLNES, that governs the emergence of glassy behaviour and strong memory effects in such system. In this chapter we demonstrate that the LLNES is not limited to the specific model of the molecular fluid, but it also applies for a wider range of physical systems described at the mesoscopic level by quite general Fokker-Planck or Langevin equations, when quenching the bath temperature to low enough values. Moreover, we prove that the LLNES constitutes a global non-equilibrium attractor of the hydrodynamic type and, similar to the HCS in the context of granular fluids, that it is fully characterised by the variance of the probability distribution function (PDF)---the kinetic temperature, for the case of the VDF for kinetic systems---that decays following a non-exponential relaxation function.

The structure of this chapter goes as follows. In both Secs.~\ref{ch3_sec:emergence} and~\ref{ch3_sec:global_stability}, we work with isotropic systems, as they allow for explicit analytical computations. On the one hand, in Sec.~\ref{ch3_sec:emergence} we analyse the reasons behind the emergence of the LLNES and study its properties. On the other hand, in Sec.~\ref{ch3_sec:global_stability}, we present the proof for the global stability of the LLNES. Lastly, in Sec.~\ref{ch3_sec:robustness} we go beyond the isotropic case and provide numerical evidence for the emergence of the LLNES in more complex potentials, and even for Enskog-Fokker-Planck systems.

\section{\label{ch3_sec:emergence}Emergence and characterisation of the LLNES}
Our starting point constitutes either the Fokker-Planck~\eqref{ch1_eq:fokker-planck} or Langevin~\eqref{ch1_eq:langevin} equation for the Markov process $\bm{r}$, in which the force and noise terms $\bm{A}(\bm{r},\cancel{t})$ and $\mathbb{B}(\bm{r},\cancel{t}) = B(\bm{r}) \mathbb{I}_d$ are general non-linear functions of $\bm{r}$, but satisfying the fluctuation-dissipation relation~\eqref{ch1_eq:fluctuation-dissipation}, thus ensuring that the unique stationary solution corresponds to the equilibrium canonical one $P_{\text{eq}}(\bm{r})\propto \text{exp}(-\beta H(\bm{r}))$. That is, we consider general equilibrium systems in contact with a unique thermal bath. In this section, we thoroughly study the emergence and characterisation of the LLNES for such mesoscopic systems. We start by analysing the physical reasons behind its emergence, followed by characterising the structure of the PDF once the system has reached the LLNES.
\subsection{\label{ch3_subsec:timescale_separation}Timescale separation}
Assuming that the thermal bath is controllable, let us consider the following protocol for the bath temperature $T_{\text{s}}(t)$: initially, the system is prepared at equilibrium at temperature $T_{\text{s}}(t) = T_i$ for $t<0$. Then, we suddenly quench the bath temperature to a very low value, such that $T_{\text{s}}(t)=T_f \ll T_i$ for $t\geq0$. In the subsequent relaxation towards the new equilibrium state at temperature $T_f$, there is a regime in which the noise becomes negligible: since $H(\bm{r})$ is independent of the bath temperature, the fluctuation-dissipation relation \eqref{ch1_eq:fluctuation-dissipation} entails that $B^2(\bm{r})/|\bm{A}(\bm{r})|\propto T_{f}\ll T_{i}$. Therefore, terms containing $B(\bm{r})$ in Eq.~\eqref{ch1_eq:langevin} may be neglected and the Langevin equation reduces to the deterministic, noiseless equation
\begin{equation}\label{ch3_eq:noiseless-evol}
  \dot{\bm{r}}=-\bm{A}(\bm{r}).
\end{equation}
In the following, we determine the conditions under which the initial conditions of the noiseless dynamics are forgotten for long enough times. Specifically, we consider the simple but physically relevant situation of isotropic systems---i.e. with radial symmetry---such that 
\begin{equation}
    \bm{A}(\bm{r})=A(r)\bm{\hat{r}}, \
r=|\bm{r}|, \ \bm{\hat{r}}=\bm{r}/r.
\end{equation}
The deterministic ``force'' $\bm{A}(\bm{r})$ must be confining but otherwise arbitrary in its form. This is indeed the case of a Brownian particle confined in a general isotropic potential $U(r)$, for which the Langevin equation reads
\begin{equation}\label{ch3_eq:LE-overdamped}
  \dot{\bm{r}}=-\gamma^{-1}\,U'(r)\hat{\bm{r}}+\sqrt{2D}\,\bm{\eta}(t),
\end{equation}
where $\gamma$ and $D$ are the friction and diffusion coefficients, and in this case they are
assumed to be position independent. The identifications $H(r)=U(r)$,
$\bm{A}(r)=\gamma^{-1}U'(r)\bm{\hat{r}}$ and $B=\sqrt{2D}$---thus additive noise---within the general
fluctuation-dissipation relation~\eqref{ch1_eq:fluctuation-dissipation} provide the
Einstein relation $\beta\gamma D=1$. This is not the only relevant physical situation we could have considered. We may have also addressed the relaxation of the velocity of an underdamped Brownian particle due to the non-linear drag force stemming from its interaction with the background fluid, as considered in the previous chapter. Therein, the Markov process $\bm{r}$ would account for the velocity of such particle. Let us note that, since $A(r)$ may change its sign, the potential may have several minima---or even a continuous set of minima, such as the ``lemon-squeezer'' potential we introduce in Sec.~\ref{ch3_subsubsec:application-overdamped}.

From the noiseless Langevin equation~\eqref{ch3_eq:noiseless-evol}, the time evolution of the system, having started from $r_i$ is implicitly given by
\begin{equation}\label{ch3_eq:x(t)-implicit}
  t=\int_{r(t)}^{r_i}\frac{dr'}{A(r')}.
\end{equation}
In the following, we assume
\begin{equation}\label{ch3_eq:A(x)-small-large-x}
  \quad \lim_{r\to+\infty}r^{-1}A(r)=+\infty, 
\end{equation}
i.e. $A(r)$ diverges faster than linearly for large r. In the context of overdamped Brownian motion, this condition means that the confining potential $U(r)$ is stronger than harmonic at long distances. This assumption allows to write Eq.~\eqref{ch3_eq:x(t)-implicit} as
\begin{equation}\label{ch3_eq:x(t)-implicit-asymp}
t =
\int_{r(t)}^{+\infty}\frac{dr'}{A(r')}-\int_{r_i}^{+\infty}\frac{dr'}{A(r')}.
\end{equation}
The first and second terms on the right hand side of Eq.~\eqref{ch3_eq:x(t)-implicit-asymp} are the times needed to relax from a very large value of $r$, $r \gg r_i$, to the instantaneous position $r(t)$ and the initial position $r_i$, respectively. Now, recalling that $T_f \ll T_i$, the following timescale separation
\begin{equation}\label{ch3_eq:LLNES-interm-times}
  t_{1}(T_i)\equiv\tau(T_{i}) \ll t_{2}(T_f)\equiv\tau(T_{f})
\end{equation}
applies, where $\tau(T_{\text{s}})$ is the relaxation time to equilibrium at temperature $T_{\text{s}}$. In this way, there appears an intermediate time regime 
\begin{equation}\label{ch3_eq:LLNES-interm-times-2}
    t_{1}(T_i) \ll t \ll t_{2}(T_f),
  \end{equation}
in which the second term on the right hand side of Eq.~\eqref{ch3_eq:x(t)-implicit-asymp} becomes negligible against the first, while the thermal noise is still irrelevant. Over the timescale defined in Eq.~\eqref{ch3_eq:LLNES-interm-times}, we thus have
\begin{equation}\label{ch3_eq:LLNES-def}
  r(t)\sim r_{\text{L}}(t), \quad  \int_{r_{\text{L}}(t)}^{+\infty} \frac{dr}{A(r)}=t.
\end{equation} 
In the above, the state $r_{\text{L}}(t)$ corresponds to a non-equilibrium attractor of the dynamics of the system, implying that all the solutions of the Langevin equation tend to it over the timescale defined in Eq.~\eqref{ch3_eq:LLNES-interm-times-2}, regardless of their initial conditions. We identify $r_{\text{L}}(t)$ with the long-lived non-equilibrium state (LLNES) from the previous chapter---the connection will be further clarified as we progress through this chapter. 

Over this far-from-equilibrium state, independent of  initial conditions, the radial PDF\footnote{In the following, specifically when dealing with isotropic situations, we will refer to either the PDF or the radial PDF interchangeably.} is given by
\begin{equation}\label{eq:P-LLNES-x}
  P_{\text{L}}(r,t)\sim \delta (r-r_{\text{L}}(t)), 
\end{equation}
as we formally show later on in Sec.~\ref{ch3_subsec:fourier}. The function $r_{\text{L}}(t)$ defined by Eq.~\eqref{ch3_eq:LLNES-def} depends on the specific form of the function $A(r)$. However, we may introduce a scaled variable $\bm{c}$
\begin{equation}
    \bm{c}\equiv \frac{\bm{r}}{\expval{r(t)}},
\end{equation}
such that its corresponding PDF is universal and time-independent,
\begin{equation}\label{ch3_eq:P-LLNES-xi}
  P_{\text{L}}(c,\cancel{t})\sim \delta (c-1). 
\end{equation}
We recall that, over the LLNES, $\langle r(t) \rangle=r_{\text{L}}(t)$. Note that the terms containing $B(r)$ in the Langevin equation~\eqref{ch1_eq:langevin} eventually drive the system towards the equilibrium state at temperature $T_{f}$. In other words, the LLNES is ``destroyed'' for long enough times, when $r_{\text{L}}(t)=O(\langle r \rangle_{\text{eq}}(T_{f}))$, i.e.~as $t=O(t_2)$.\footnote{Let us remark that, although $\langle r \rangle_{\text{eq}}(T_{f})$ is finite and positive, for isotropic systems we have that $\langle \bm{r} \rangle = \bm{0}$ always.}

\subsubsection{\label{ch3_subsubsec:application-overdamped}Application to a confined Brownian particle}
Let us now apply the results presented in this section to two different physical situations.
First, we consider the confined Brownian particle from Eq.~\eqref{ch3_eq:LE-overdamped}, specifically for the anharmonic potential
\begin{equation}
\label{ch3_eq:nonlinear-potential}
  U(r)=\frac{1}{2}k r^{2}+\frac{1}{4}\lambda r^{4}, \quad \lambda>0.
\end{equation}
We require that $\lambda>0$ such that the potential is confining: 
\begin{equation}
    A(r)=ar+br^{3}, \quad a\equiv k/\gamma, \; b\equiv \lambda/\gamma.
\end{equation} 
Furthermore, as Eq.~\eqref{ch3_eq:A(x)-small-large-x} holds, we have all the necessary ingredients for the timescale separation. We remark that, for $k>0$ ($a>0$), the force term $A(r)>0$ $\forall r\ne 0$ and $U(r)$ has only one global minimum at the origin, while the case $k<0$ ($a<0$) corresponds to a ``lemon-squeezer'' potential with multiple minima at $r=r_c$---corresponding to two minima for $d=1$, with
\begin{equation}
    r_c\equiv\sqrt{\frac{|a|}{b}}=\sqrt{\frac{|k|}{\lambda}}.
\end{equation}
In either case, the Langevin equation~\eqref{ch3_eq:LE-overdamped} reduces to
\begin{equation}\label{ch3_eq:overdamped-langevin}
  \dot{\bm{r}}=-ar \left(1+\text{sgn}(a)\frac{r^2}{r_c^2}\right)\hat{\bm{r}}+\sqrt{2D}\,\bm{\eta}(t),
\end{equation}
with $\text{sgn}(x)$ being the sign function of $x$. Apart from $r_c$---which sets the length for mechanical equilibrium for $k<0$, there are two additional characteristic lengths in this physical scenario, 
\begin{equation}
    r_{\lambda}\equiv \left(\frac{k_B T}{\lambda} \right)^{1/4}, \quad r_{k}\equiv \left(\frac{k_B T}{|k|} \right)^{1/2},
\end{equation}
which---aside from multiplicative constants---correspondingly give the equilibrium lengths at high and low temperatures. Such lengths stem from the equipartition theorem,
\begin{equation}
    \left< r \ \frac{dU}{dr}\right>_{\text{eq}} = k\left<r^2 \right>_{\text{eq}} + \lambda \left<r^4 \right>_{\text{eq}} = k_B T.
\end{equation}
On the one hand, the quartic term in the above dominates for high initial temperatures, thus providing the length $r_{\lambda}$. On the other hand, for low final temperatures, we may assume that the quadratic term dominates instead---which only holds for $k>0$, thus giving the equilibrium length $r_{k}$. Following this, it is useful for our analysis to introduce the dimensionless temperature 
\begin{equation}
    T^*\equiv \frac{k_B T\lambda}{k^2}=\left(\frac{r_k}{r_\lambda}\right)^4,
\end{equation}
for which the high and low temperature regimes thus correspond to the limits $T^*\gg 1$ and $T^*\ll 1$, respectively. Let us note that $r_c=r_{\lambda}^2/r_k$.\footnote{This is mathematically consistent with the fact that the potential $U(r)$ from Eq.~\eqref{ch3_eq:nonlinear-potential} is characterised by two parameters $k$ and $\lambda$. Thus, only two relevant lengths are required for determining the dynamics.} In order to analyse the emergence of the LLNES in this overdamped system, let us take the particularisation of Eq.~\eqref{ch3_eq:x(t)-implicit-asymp}, which gives
\begin{equation}\label{ch3_eq:explicit-sol-nonlin-pot}
  2at=
  \ln\left(1+\text{sgn}(a)\frac{r_c^2}{r^{2}(t)}\right)-\ln\left(1+\text{sgn}(a)\frac{r_c^2}{r_i^{2}}\right).
\end{equation}
For high enough initial temperatures $T_i^*\gg 1$, 
we may estimate $r_i$ with $r_{\lambda,i}=(k_B T_i/\lambda)^{1/4}$, as the quartic term of the potential dominates. For $r_i \gg r(t)$, there appears an intermediate time window over which initial conditions are forgotten. 
Moreover, if $r(t) \gg r_c$, the resulting behaviour becomes independent of the sign of $a$,
\begin{equation}
    2at \sim \text{sgn}(a)\frac{r_c^2}{r^{2}(t)} - \text{sgn}(a)\frac{r_c^2}{r_i^{2}} \sim \text{sgn}(a)\frac{r_c^2}{r^{2}(t)} = \frac{a}{br^2(t)},
\end{equation}
thus giving
\begin{equation}\label{ch3_eq:LLNES-explicit-short-times-v2}
  r(t)\sim r_{\text{L}}(t)=(2bt)^{-1/2}, \quad  (T_i^*)^{-1/2} \ll 2at\ll 1,
\end{equation}
with $r_{\text{L}}(t)$ only depending on $b=\lambda/\gamma$, i.e.~only on the dominant behaviour of the potential at long distances. 

\begin{figure}
    \centering
    \includegraphics[width=2.65in]{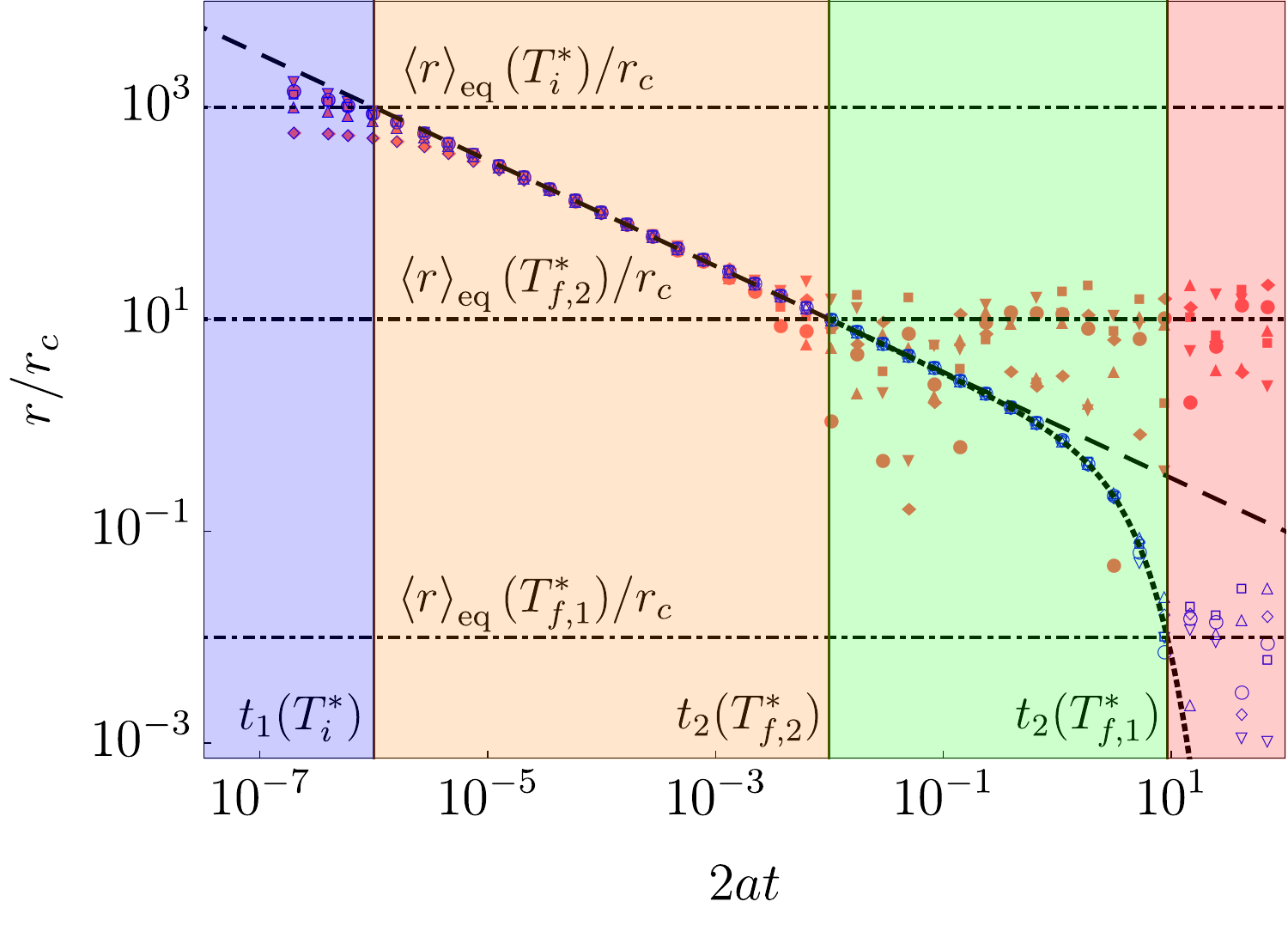}
    \includegraphics[width=2.65in]{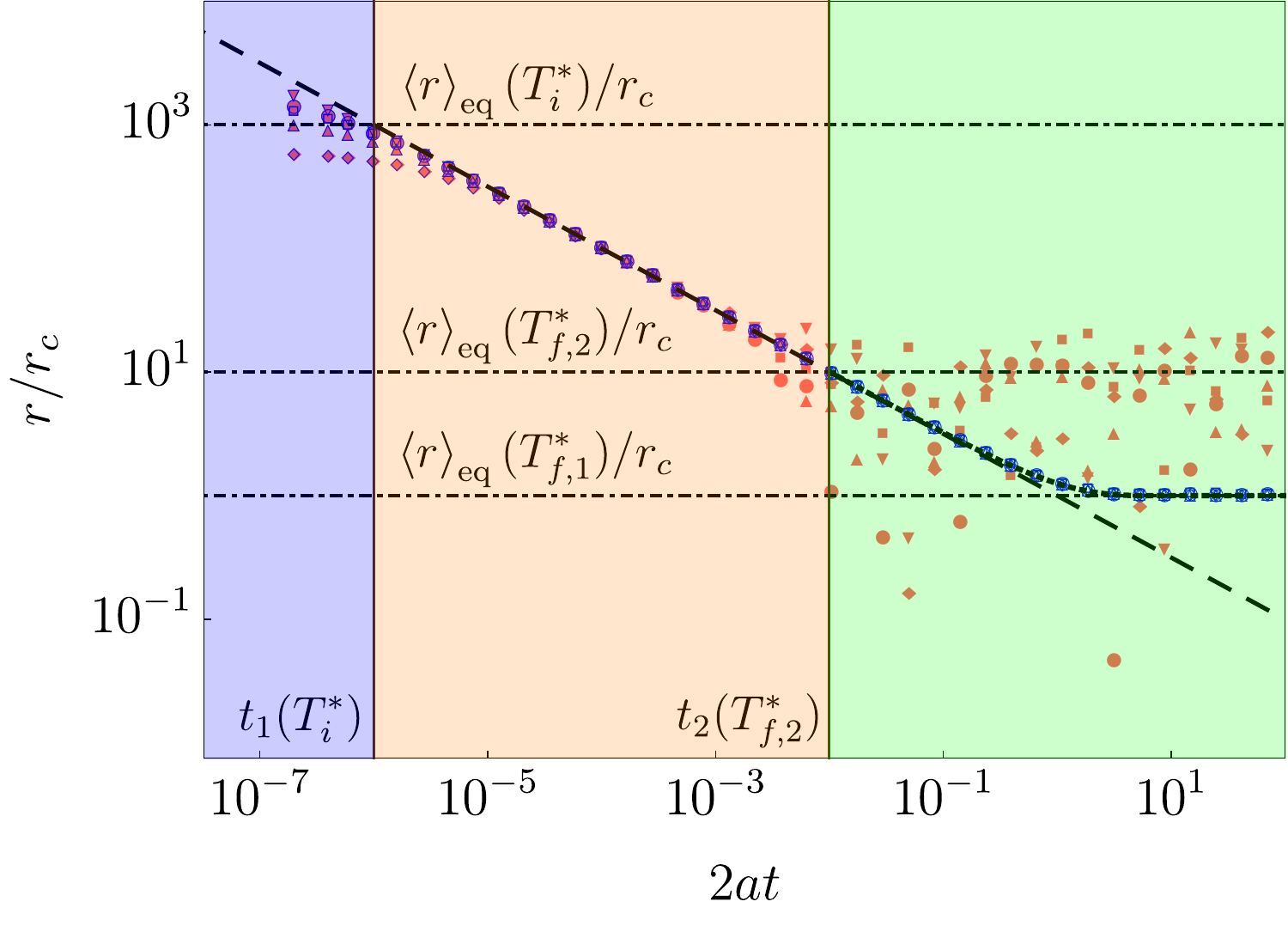}
    \caption{\label{ch3_fig:langevin-traj}Stochastic trajectories for the overdamped Brownian particle in the non-linear potential from Eq.~\eqref{ch3_eq:nonlinear-potential}, for $k>0$ (left) and $k<0$ (right). Specifically, we plot five realisations of the Langevin Equation~\eqref{ch3_eq:overdamped-langevin} for $d=1$, sampled from an initial equilibrium state such that
    $\langle r \rangle_{\text{eq}}(T_i^*)/r_c=10^3$,  to two final states such that $\langle r \rangle_{\text{eq}}(T_{f,1}^*)/r_c=10^{-2}$ (blue empty symbols) and $\langle r \rangle_{\text{eq}}(T_{f,2}^*)/r_c=10$ (red filled symbols).
    Black vertical lines correspond to the characteristic times that delimit the different coloured regions: (1) $t <t_1(T_i^*)$ (blue), where initial conditions still prevail, (2) $t_1(T_i^*) <t<t_2(T_{f,2}^*)$ (orange), where the power law Eq.~\eqref{ch3_eq:LLNES-explicit-short-times-v2} (dashed line) applies for both final temperatures, (3) $t_2(T_{f,2}^*)< t < t_2(T_{f,1}^*)$ (green), where Eq.~\eqref{ch3_eq:LLNES-explicit-specific} (dotted curve) now applies for $T_{f,1}^*$ whereas thermal noise becomes relevant for $T_{f,2}^*$, and (4) $t> t_2(T_{f,1}^*)$ (pink), where noise becomes relevant for $T_{f,1}^*$. 
      }
  \end{figure}

We remark that, in order to derive Eq.~\eqref{ch3_eq:LLNES-explicit-short-times-v2}, we only need to consider the high enough initial temperature: the role of the final temperature is possibly limiting the timescale over which the LLNES is observed. Noise becomes negligible as long as  $r_{\text{L}}(t)$ is much larger than the equilibrium value at the final temperature, $r_{k,f}=(k_B T_f/k)^{1/2}$, which provides the condition $2at \ll (T_f^*)^{-1}$. On the one hand, for $T_f^*=O(1)$ or larger, the LLNES in Eq.~\eqref{ch3_eq:LLNES-explicit-short-times-v2} is restricted to the time window  $(T_i^*)^{-1/2}\ll 2at \ll (T_f^*)^{-1}$. On the other hand, for $T_f^*\ll 1$, the LLNES may be extended to longer times such that $2at=O(1)$, $r(t)$ becomes of the order of $r_c$ and
\begin{equation}\label{ch3_eq:LLNES-explicit-specific}
  r_{\text{L}}(t)=r_c \left[\text{sgn}(a) \left(e^{2at}-1\right)\right]^{-1/2},
\end{equation}
since the second logarithmic terms in Eq.~\eqref{ch3_eq:explicit-sol-nonlin-pot} become negligible. Figure~\ref{ch3_fig:langevin-traj} shows a set of stochastic trajectories for the overdamped Brownian particle for which the behaviours depicted in Eqs.~\eqref{ch3_eq:LLNES-explicit-short-times-v2} and \eqref{ch3_eq:LLNES-explicit-specific} are neatly observed, for both the $k>0$ and $k<0$ cases. Specifically, we consider two sets of 5 trajectories starting from equilibrium at a high temperature $T_i^*$, but relaxing towards different low final temperatures $T_{f,1}^*$ and $T_{f,2}^*$, with $T_{f,1}^* \ll T_{f,2}^* \ll T_i^*$. Initially, both sets rapidly forget their initial conditions and tend towards the universal behaviour from Eq.~\eqref{ch3_eq:LLNES-explicit-short-times-v2}---which constitutes the lowest non-vanishing order of Eq.~\eqref{ch3_eq:LLNES-explicit-specific} for short times. As time progresses, the set of trajectories relaxing towards $T_{f,2}^*$---red filled symbols---eventually departs from the LLNES and reaches equilibrium, while the set relaxing to $T_{f,1}^*$---blue empty symbols---follows the behaviour from Eq.~\eqref{ch3_eq:LLNES-explicit-specific} until reaching equilibrium. Each of the timescales $t_1(T_i^*)$, $t_2(T_{f,1}^*)$ and $t_2(T_{f,2}^*)$ were numerically estimated by matching the equilibrium lengths $r_{\lambda,i}$ and $r_{k,f}$ with the LLNES behaviour from Eqs.~\eqref{ch3_eq:LLNES-explicit-short-times-v2} and \eqref{ch3_eq:LLNES-explicit-specific}.

It is worth noting that, for the $k<0$ case, the system follows the deterministic evolution towards mechanical equilibrium at $r=r_c$. In general, the presence of multiple minima only induces additional characteristic lengths, corresponding to mechanical equilibrium, which have to be taken into account. As long as we initially have sufficiently high temperatures---such that the system does not ``feel'' the presence of those minima, the system ``forgets'' its initial condition and falls on a LLNES regime that is independent of these new lengths. If the final temperature is low enough, the LLNES extends to longer times and the system approaches to its closest mechanical equilibrium, with small thermal fluctuations around it, such that the system seems frozen. We discuss further on this point in Sec.~\ref{ch3_subsec:snowplough}. 

\subsubsection{\label{ch3_subsubsec:application-molecular}Application to a molecular fluid with non-linear drag}
Now we consider another relevant physical system: an isotropic molecular fluid
with non-linear drag force, similar to the one presented in the previous chapter. By ignoring the binary collisions between Brownian particles, the purely Fokker-Planck equation from Eq.~\eqref{ch1_eq:fokker-planck} accurately accounts for the dynamical evolution of the one-particle VDF $f(\bm{v},t)$. The terms $\bm{A}(\bm{v},t)$ and $\mathbb{B}(\bm{v},t)$---or $\mathbb{D}(\bm{v},t)$ equivalently in this case---follow from Eq.~\eqref{ch2_eq:nonlinear-drag} and thus fulfil the detailed balance condition with respect to the equilibrium Maxwellian $f_{\text{eq}}(\bm{v})$. However, here we consider a broader class of velocity-dependent drag coefficients $\zeta(v)$.

In the previous chapter, we considered the non-linear drag coefficient $\zeta(v)$ in Eq.~\eqref{ch2_eq:nonlinear-drag}, which is quadratic in the velocities for large values of $v$. Here, we analyse a more general situation, in which the drag coefficient has a general algebraic behaviour for large velocities, 
\begin{equation}\label{ch3_eq:algebraic-behaviour}
    \zeta^*(v) \equiv \frac{\zeta(v)}{\zeta_0} \sim\gamma \left(\frac{v}{v_{\text{th},f}}\right)^n, \quad v_{\text{th},f}\equiv \left(\frac{2 k_B T_f}{m}\right)^{1/2},
\end{equation} 
with $\gamma$ being the non-linearity parameter and $v_{\text{th},f}$ the thermal velocity at $T_f$---i.e. $v_{\text{T}}(t)$ from Eq.~\eqref{ch2_eq:scaling-velocities} but in the limit $t \rightarrow +\infty$. For $n>1$, there appears a timescale over which the non-linear drag dominates the relaxation and thermal noise becomes negligible. Over this wide time window, initial conditions are forgotten and the LLNES emerges. Specifically, the algebraic behaviour~\eqref{ch3_eq:algebraic-behaviour} dominates the behaviour of the integral from Eq.~\eqref{ch3_eq:LLNES-def}, thus giving
\begin{equation}
  v_{\text{L}}(t)= v_{\text{th},f}(\gamma \zeta_0n t)^{-1/n}, \quad (T_f/T_i)^{n/2}\ll n \gamma\zeta_0 t \ll 1.
\end{equation}
It is worth noting the strong analogy with Eq.~\eqref{ch3_eq:LLNES-explicit-short-times-v2}. With this, the kinetic temperature, which was defined as $T(t)\propto \langle v^2\rangle$, thus shows a slow non-exponential, algebraic, decay as $T(t)\propto t^{-2/n}$. Notice that for $n=2$ we recover the specific non-exponential relaxation function obtained in the previous chapter, which was given by Eq.~\eqref{ch2_eq:algebraic-relaxation}.

Although the current analysis holds exclusively from the Fokker-Planck framework---since it allows to write a noiseless Langevin equation for the velocities $\bm{v}$ of the Brownian particles, under certain circumstances it may be extended to scenarios where binary collisions---either elastic or inelastic---are considered. We comment further on this point in Sec.~\ref{ch3_subsec:molecular_fluid}.

\begin{figure}
    \centering
    \includegraphics[width=3.5in]{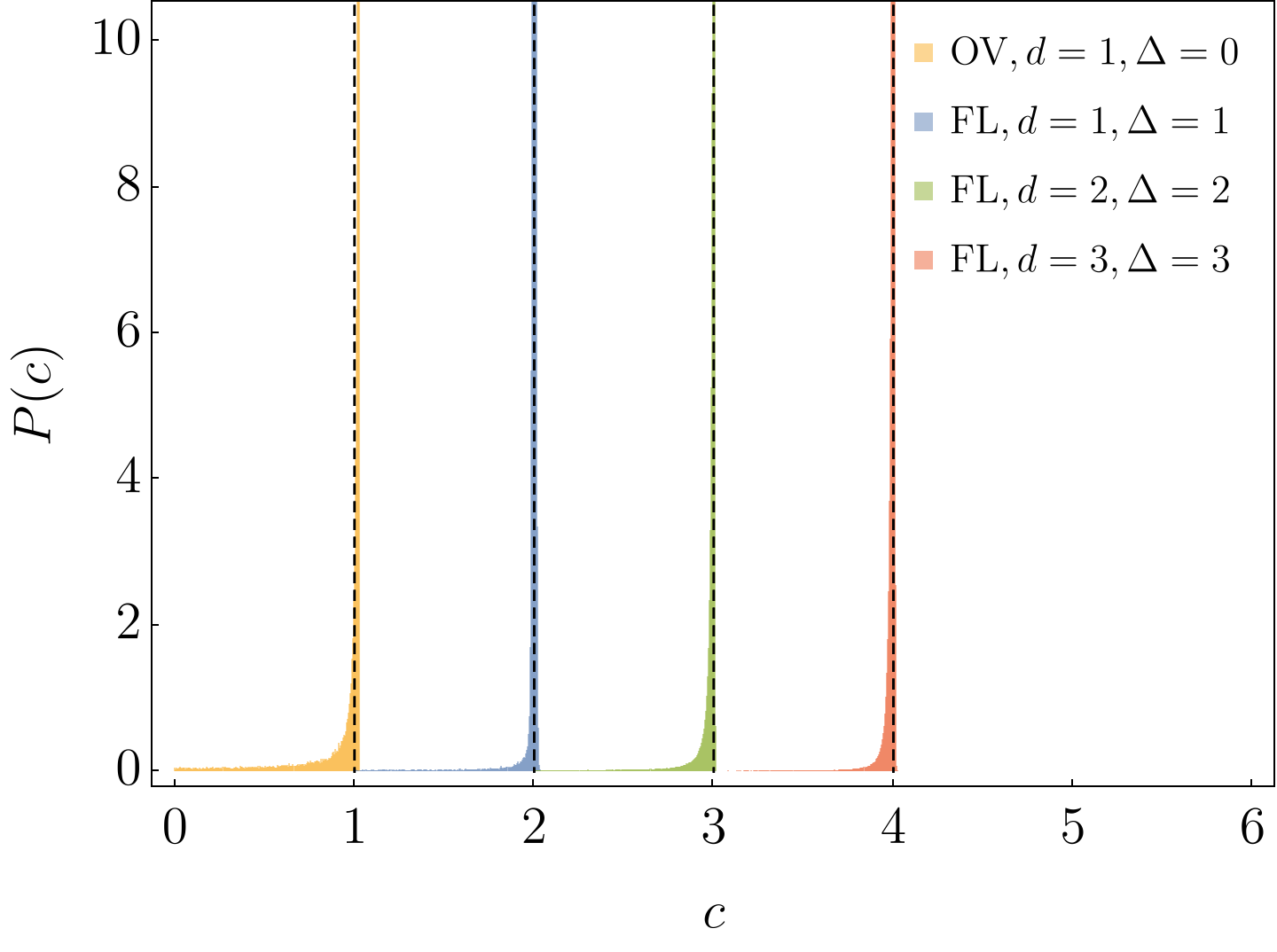}
    \caption{Scaled radial PDF $\tilde{P}(c) \equiv c^{d-1}\Omega_d P(c)$ at the LLNES for different physical
    situations. Specifically, we plot the PDF as a function of the scaled radial variable $c$ for both the overdamped particle in a non-harmonic potential (OV), given by the Langevin equation~\eqref{ch3_eq:overdamped-langevin}, and the molecular fluid with non-linear drag (FL), accounted by the Fokker-Planck equation \eqref{ch1_eq:fokker-planck} with coefficients from Eq.~\eqref{ch2_eq:nonlinear-drag}, for different spatial dimensions. For the former, $\bm{c}=\bm{r}/\expval{r}$, while for the latter, $\bm{c}=\bm{v}/\expval{v}$. In order to appreciate the universal Dirac-delta shape from Eq.~\eqref{ch3_eq:P-LLNES-xi}, each PDF is shifted an offset $\Delta$ to the right, as indicated in the legend. The vertical black dashed lines correspond to the positions of the delta peaks---plus offsets---according to Eq.~\eqref{ch3_eq:P-LLNES-xi}. The plotting range on the vertical axis has been cut to $10$ for visualisation purposes---specifically for appreciating the tails of the distribution.}\label{ch3_fig:deltas}
  \end{figure}

In Figure~\ref{ch3_fig:deltas}, we plot the radial PDF as a function of the scaled variable $\bm{c}$, 
for the two specific examples of physical systems described in this section so far. Data corresponds to the numerical integration of the Langevin equation for $N = 10^5$ stochastic trajectories, where we let each system relax towards the LLNES---see Appendix~\ref{app:initial-preparations} for further details. The delta-peak structure is clearly observed, for one-, two-, and three-dimensional systems. For the non-linear fluid considered, the data shown corresponds to $n=2$. Let us note that in each case, a small tail to the left of the peak is observed for the PDFs. This fact is thoroughly explored in Sec.~\ref{ch3_subsec:scaling}.

\subsection{\label{ch3_subsec:fourier}Fourier transform of the PDF}
In order to provide the argument for the emergence of the Dirac-delta behaviour~\eqref{ch3_eq:P-LLNES-xi} with a more rigorous support, in this section we show that the Fourier transform of the radial PDF $P(r,t)$ actually converges to that for a Dirac-delta distribution. By solving the noiseless Langevin equation~\eqref{ch3_eq:noiseless-evol}, we have that $r(t) = f(r_i,t)$, with $f$ being some unknown function, dependent on the explicit form of $A(r)$. The only property that we impose on $f(r_i,t)$ is that, for long enough times, it loses its dependence on the initial condition $r_i$, thus becoming the function $r_{\text{L}}(t)$ found at the LLNES; i.e. $f(r_i,t) \sim r_{\text{L}}(t)$ for large $t$.

Now, if $P_i(r_i)$ is the initial PDF for the initial conditions $r_i$---as the system was initially equilibrated at temperature $T_i$---then the PDF $P(r,t)$ of $r$ at any time $t$ is
\begin{align}
\label{ch3_eq:pdf-1}
    P(r,t) &= \int_0^{+\infty} dr_i \ \delta \left( r - f(r_i,t) \right) P_i(r_i).
\end{align}
Note that $P(r,t)=0$ for $r>r_{\text{L}}(t)$, because $f(r_i,t)\le r_{\text{L}}(t)$.\footnote{In the following, we assume that $r(t)$ decays towards the closest minimum of the potential in absense of thermal noise, such that $f(r_i,t)$ is monotonous in $r_i$, and $r_{\LLNES}(t)$ may also be obtained as the asymptotic behaviour of $f(r_i,t)$ in the limit $r_i \rightarrow +\infty$.} The argument for the emergence of the Dirac-delta above is equivalent to substitute $f(r_i,t)$ with $r_{\text{L}}(t)$ in Eq.~\eqref{ch3_eq:pdf-1} by considering that $r_{\text{L}}(t)\ll r_i$. The latter is non-rigorous since, for any time $t$, and for a general initial PDF, there is a range of values of $r_i$ close to zero such that $r_i$ is of the order of $r_{\text{L}}(t)$ or smaller.

In order to proceed further, we go to Fourier space by introducing the characteristic function
\begin{equation}
    G(k,t)\equiv \langle e^{ikr}\rangle=\int_0^{+\infty} dr\, e^{ikr} P(r,t)=\int_0^{+\infty} dr_i \,e^{ikf(r_i,t)} P_i(r_i).
\end{equation}
Let us consider now the long-time limit. We split the integration over $r_i$ into two slices: from $0$ to $\varepsilon(t)$, and from $\varepsilon(t)$ to $+\infty$. We choose $\varepsilon(t)$ small, in the sense that $\varepsilon(t)\to 0^+$ in the limit as $t\to +\infty$ but such that $\varepsilon(t)/r_{\text{L}}(t)\gg 1$. Therefore, in the second slice we may safely substitute $f(r_i,t)$ with $r_{\text{L}}(t)$, with vanishing error for long times:
\begin{align}
    G(k,t)&=\int_0^{\varepsilon(t)} dr_i e^{ikf(r_i,t)} P_i(r_i)+\int_{\varepsilon(t)}^{+\infty} dr_i e^{ikf(r_i,t)} P_i(r_i) \nonumber
    \\
    &\sim \int_0^{\varepsilon(t)} dr_i e^{ikf(r_i,t)} P_i(r_i)+e^{ikr_{\text{L}}(t)}\int_{\varepsilon(t)}^{+\infty} dr_i P_i(r_i) \nonumber \\
    & = e^{ikr_{\text{L}}(t)} + \underbrace{\int_0^{\varepsilon(t)} dr_i \left(e^{ikf(r_i,t)}-1\right) P_i(r_i)}_{\Delta G(k,t)}, \quad t\to +\infty.
\end{align}
The first term constitutes the Fourier transform of the Dirac-delta, the second term $\Delta G(k,t)$ can be shown to go to zero:
\begin{align}
    \abs{\Delta G(k,t)}&=\abs{\int_0^{\varepsilon(t)} dr_i \left(e^{ikf(r_i,t)}-1\right) P_i(r_i)} \nonumber
    \\
    &\le \int_0^{\varepsilon(t)} dr_i \abs{e^{ikf(r_i,t)}-1} P_i(r_i) \nonumber
    \\
    &\le 2 \int_0^{\varepsilon(t)} dr_i P_i(r_i) \to 0, \quad t\to +\infty.
\end{align}
In summary, we have rigorously proved that
\begin{equation}
    G(k,t)\sim e^{ik r_{\text{L}}(t)} \ \Longrightarrow \ P(r,t)\sim \delta\left(r-r_{\text{L}}(t)\right), \quad t\to +\infty.
\end{equation}
Following this formal proof, we emphasise that (i) we have not employed the explicit form of $f(r_i,t)$, but only that $f(r_i,t)$ ``forgets'' the initial condition $r_i$ in the long-time limit, and (ii) the shape of the PDF for the initial conditions $P_i(r_i)$ is completely general. We must highlight though that the proof presented here only holds when thermal noise is negligible---i.e. within the timescale separation regime from Eq.~\eqref{ch3_eq:LLNES-interm-times}. In this regard, the $t\to +\infty$ limit shall be understood as $t \gg t_1(T_i)$, and the Dirac-delta shape is attained as long as $t \ll t_2(T_f)$.

\subsection{\label{ch3_subsec:scaling}Scaling behaviour and tails of the distribution}
Following the emergence of tails in the PDFs depicted in Fig.~\ref{ch3_fig:deltas}, here we look into the structure of $P(r,t)$ in more detail. Specifically, we would like to investigate its scaling behaviour when converging towards the Dirac-delta solution. Throughout this section, for the sake of mathematical simplicity, we consider the family of algebraic potentials
\begin{equation}\label{ch3_eq:suppl-U-non-lin}
    U(r) = \frac{1}{n+1}k\, r^{n+1}, \quad n>1, \quad k>0.
\end{equation}
The noiseless Langevin equation for this family of potentials reads
\begin{equation}
\label{ch3_eq:suppl-noiseless-Langevin}
    \dot{r} = -b r^n, \quad b\equiv k/\gamma,
\end{equation}
which, given the initial condition $r_i \equiv r(t=0)$, may be explicitly integrated, giving
\begin{equation}\label{ch3_eq:integration}
    (n-1)bt=r^{1-n}(t)-r_i^{1-n}.
\end{equation}
In the above, the first and second terms on the right hand side are related to the times needed to relax from $r\to+\infty$ to the instantaneous and initial positions $r(t)$ and $r_i$, respectively. Solving Eq.~\eqref{ch3_eq:integration} for $r(t)$ gives
\begin{equation}\label{ch3_eq:suppl-fri-def}
    r(t) = f(r_i,t) = r_i\left[1 + \left(\frac{r_i}{r_{\LLNES}(t)}\right)^{n-1}\right]^{-\frac{1}{n-1}},
\end{equation}
with 
\begin{equation}\label{ch3_eq:suppl-LLNES-algebraic}
    r_{\text{L}}(t) \equiv \left[(n-1) b t\right]^{-\frac{1}{n-1}},
\end{equation}
being the solution at the LLNES, which is attained for long enough times, for which $r(t) \ll r_i$. Let us now introduce the ``length'' scale $\ell_i$ associated with the initial conditions. We use quotation marks here because $r$ may not be a distance. For the usual equilibrium initial conditions, we have that 
\begin{equation}\label{ch3_eq:suppl-elli-def}
    \ell_i\equiv (\beta_i k)^{-\frac{1}{n+1}}=\left(\frac{k_B T_i}{k}\right)^{\frac{1}{n+1}}.
\end{equation} 
Generally, we consider that
\begin{equation}
    P_i(r_i)=\ell_i^{-1} \pdfnondim(r_i^*), \quad r_i^*=\frac{r_i}{\ell_i}, 
\end{equation}
where $\pdfnondim$ is the PDF of the dimensionless variable $r_i^*$, which is kept as completely arbitrary. Accordingly, we have that 
\begin{equation}
    P(r,t)=\ell_i^{-1}P^*(r^*,t), \quad r^*=\frac{r}{\ell_i}.
\end{equation}
Since $r^*$ is very close to $r^*_{\text{L}}(t)$ for long times, we assume that the deviations of $r^*$ from $r^*_{\LLNES}(t)$ are small and scale as some power of $r^*_{\LLNES}(t)$. We thus define a new variable $\xi$ with PDF $\phi(\xi,t)$ as follows:
\begin{equation}
    r^*\equiv r^*_{\LLNES}(t)-\xi \left[r^*_{\LLNES}(t)\right]^{\psi}, \quad 0 \le \xi \le \left[r^*_{\LLNES}(t)\right]^{1-\psi},
    \label{ch3_eq:app-xi-def}
\end{equation}
or equivalently,
\begin{equation}
    \xi = \frac{r^*_{\LLNES}(t)-r^*}{\left[r^*_{\LLNES}(t)\right]^{\psi}} \implies \phi(\xi,t)=\left[r^*_{\LLNES}(t)\right]^{\psi} P^*(r^*,t),
    \label{ch3_eq:app-xi-def2}
\end{equation}
where $\psi>0$ is a parameter to be determined later. Let us note that the range of values of $\xi$ stems from the fact that $0\leq r^* \leq r^*_{\text{L}}(t)$. Now, we use the explicit form of $f(r_i,t)$ in Eq.~\eqref{ch3_eq:suppl-fri-def} and analitically solve Eq.~\eqref{ch3_eq:pdf-1} to write
\begin{align}
\label{ch3_eq:scaled-pdf}
    \phi(\xi,t) &= \frac{r^*_{\LLNES}(t)^{\psi}}{\left[1-h(\xi,t) \right]^{\frac{n}{n-1}}} P_i^*\left( \frac{ r^*_{\LLNES}(t)-\xi r^*_{\LLNES}(t)^{\psi}}{\left[1-h(\xi,t) \right]^{\frac{1}{n-1}}}\right), \qquad 0 \le \xi \le \left[r^*_{\LLNES}(t)\right]^{1-\psi},
\end{align}
with
\begin{equation}
    h(\xi,t) \equiv \left(  1 - \xi \left[r^*_{\LLNES}(t)\right]^{\psi-1}\right)^{n-1}
\end{equation}
being the Jacobian for the transformation of the Dirac-delta from Eq.~\eqref{ch3_eq:pdf-1}. In the long-time limit, $r^*_{\LLNES}(t)\to 0$, and as long as $\xi \left[r^*_{\LLNES}(t)\right]^{\psi-1} \ll 1$, we can introduce the approximation for the function $h(\xi,t)$,
\begin{equation}
    h(\xi,t)=1-\xi (n-1)\left[r^*_{\LLNES}(t)\right]^{\psi-1} + O(\xi \left[r^*_{\LLNES}(t)\right]^{2(\psi-1)}),
\end{equation}
and then
\begin{equation}\label{ch3_eq:scaled-pdf-fin}
    \phi(\xi,t)\sim \frac{\left[r^*_{\LLNES}(t)\right]^{\frac{n-\psi}{n-1}}}{\left[(n-1)\xi\right]^{\frac{n}{n-1}}} P_i^* \left( \frac{\left[r^*_{\LLNES}(t)\right]^{\frac{n-\psi}{n-1}}}{\left[(n-1)\xi\right]^{\frac{1}{n-1}}}\right), \quad \xi \left[r^*_{\LLNES}(t)\right]^{\psi-1}\ll 1.
\end{equation}
This equation suggests that $\psi=n$: with this choice, the scaled PDF becomes a time-independent $\phi(\xi,t) \sim \phi_{\text{st}}(\xi)$, which is given by
\begin{equation}
\label{ch3_eq:scaled-pdf-asymp-2}
\phi_{\text{st}}(\xi) \equiv \left[(n-1)\xi\right]^{-\frac{n}{n-1}} P_i^* \left( \left[(n-1)\xi\right]^{-\frac{1}{n-1}}\right), \quad \xi \left[r^*_{\LLNES}(t)\right]^{n-1}\ll 1.
\end{equation}
Henceforth, the latter implies that the PDF
\begin{equation}
    P^*(r^*,t) = \left[r^*_{\LLNES}(t)\right]^{-n}\phi_{\text{st}}\left(\frac{r^*_{\LLNES}(t)-r^*}{\left[r^*_{\LLNES}(t)\right]^{n}}\right)
\end{equation}
presents scaling properties for long times. 

The time-independent scaled solution $\phi_{\text{st}}(\xi)$ measures the deviations from the Dirac-delta, which can be visualised as the left tails on the left panel in Figure~\ref{ch3_fig:deltas}. Provided that $\phi_{\text{st}}(\xi)$ has finite variance, the found scaling entails that the standard deviation of $r^*$, $\sigma^*_{r^*}(t)$, scales as $\left[r^*_{\LLNES}(t)\right]^n$ and thus $\sigma_{r^*}^*(t)/\left<r^*\right>(t)\propto \left[r^*_{\LLNES}(t)\right]^{n-1}\to 0$ for long times, which is a signature of the emergence of the Dirac-delta distribution. If the first and/or second moments of $\phi_{\text{st}}$ diverge, it can be shown that, depending on the value of $n$ and the choice of $P_i^*$, the standard deviation of $\phi_{\text{st}}$ diverges either logarithmically or algebraically as $r_{\text{L}}^*(t)\rightarrow 0$---see Appendix \ref{app:asymptotic-llnes} for further details. In any case, the standard deviation for $r^*$, $\sigma_{r^*}^*(t)$, decays always in such a way that $\sigma_{r^*}^*(t)/\left<r^*\right>(t) \rightarrow 0$. 

It is interesting to compare the ``width of the Dirac-delta'', as measured by $\sigma^*_{r^*}(t)$, and the width of the final equilibrium distribution, which---in the dimensionless variables we are using here---is given by $\ell_f^*=\ell_f/\ell_i$, where $\ell_f$ is the characteristic width of the final distribution. Note that this is a relevant question: $\ell_f^*$ also tends to zero for a deep quench, since for a final equilibrium canonical distribution we have that $\ell_f= (\beta_f k)^{-\frac{1}{n+1}}$, i.e. $\ell_f^*= (\beta_f/\beta_i)^{-\frac{1}{n+1}}=(T_f/T_i)^{\frac{1}{n+1}}\to 0$. To ensure that the Dirac-delta is in fact much narrower than the final distribution, we have to impose
\begin{equation}\label{ch3_eq:suppl-cond-Dirac-narrow}
    \frac{\sigma^*_{r^*}(t)}{\ell_f^*}\propto {\left[r^*_{\LLNES}(t)\right]^n} \left(\frac{T_f}{T_i}\right)^{-\frac{1}{n+1}}\ll 1 \ \Longrightarrow \ \left[r^*_{\LLNES}(t)\right]^n\ll \left(\frac{T_f}{T_i}\right)^{\frac{1}{n+1}}.
\end{equation}

A remarkable fact is that $\phi_{\text{st}}(\xi)$ depends on the details of the initial distribution, implying that, although the dominant Dirac-delta behaviour is independent on the initial conditions, the intrinsic structure of the PDF always keeps track of the initial state, although its effects are subdominant. Nevertheless, under quite general conditions, the scaling function $\phi_{\text{st}}(\xi)$ also presents universal features. To start with, we have that 
\begin{equation}
\label{ch3_eq:app-small-large-xi}
\lim_{\xi \rightarrow +\infty} \phi_{\text{st}}(\xi) = 0, \quad  \lim_{\xi \rightarrow 0^+} \phi_{\text{st}}(\xi) = 0. 
\end{equation}
As direct consecuences of Eq.~\eqref{ch3_eq:scaled-pdf-fin}, the first condition follows as long as $P^*_i(r_i^*)$ remains finite or $P^*_i(r_i^*)$ diverges slower than ${r_i^*}^{-n}$ for small $r_i^*$; the second condition follows as long as $P^*_i(r_i^*)$ does not have ``fat'' tails, i.e. when  $P^*_i(r_i^*)$ decays faster than algebraic towards zero for large values of its argument. Equation~\eqref{ch3_eq:app-small-large-xi} implies that $\phi_{\text{st}}(\xi)$ presents a non-monotonic behaviour with a maximum at intermediate values of $\xi$. Both the position and the height of such maximum depend on the details of the initial distribution $P^*_i(r_i^*)$. By looking at the definition of $\phi_{\text{st}}(\xi)$ from Eq.~\eqref{ch3_eq:scaled-pdf-asymp-2}, we deduce that its large $\xi$ behaviour involves a power-law tail of the form
\begin{equation}
\label{ch3_eq:tail}
\phi_{\text{st}}(\xi) \sim P^*_i(0) \left[(n-1)\xi\right]^{-\frac{n}{n-1}} \propto {\xi^{-\frac{n}{n-1}}},
\end{equation}
if $P^*_i(0)\ne 0$. This behaviour is universal, in the sense that it only depends on the potential acting on the Brownian particle through $n$, while the initial conditions are completely encoded within the multiplicative constant.

\begin{figure}
    \centering
    \includegraphics[width=3.5in]{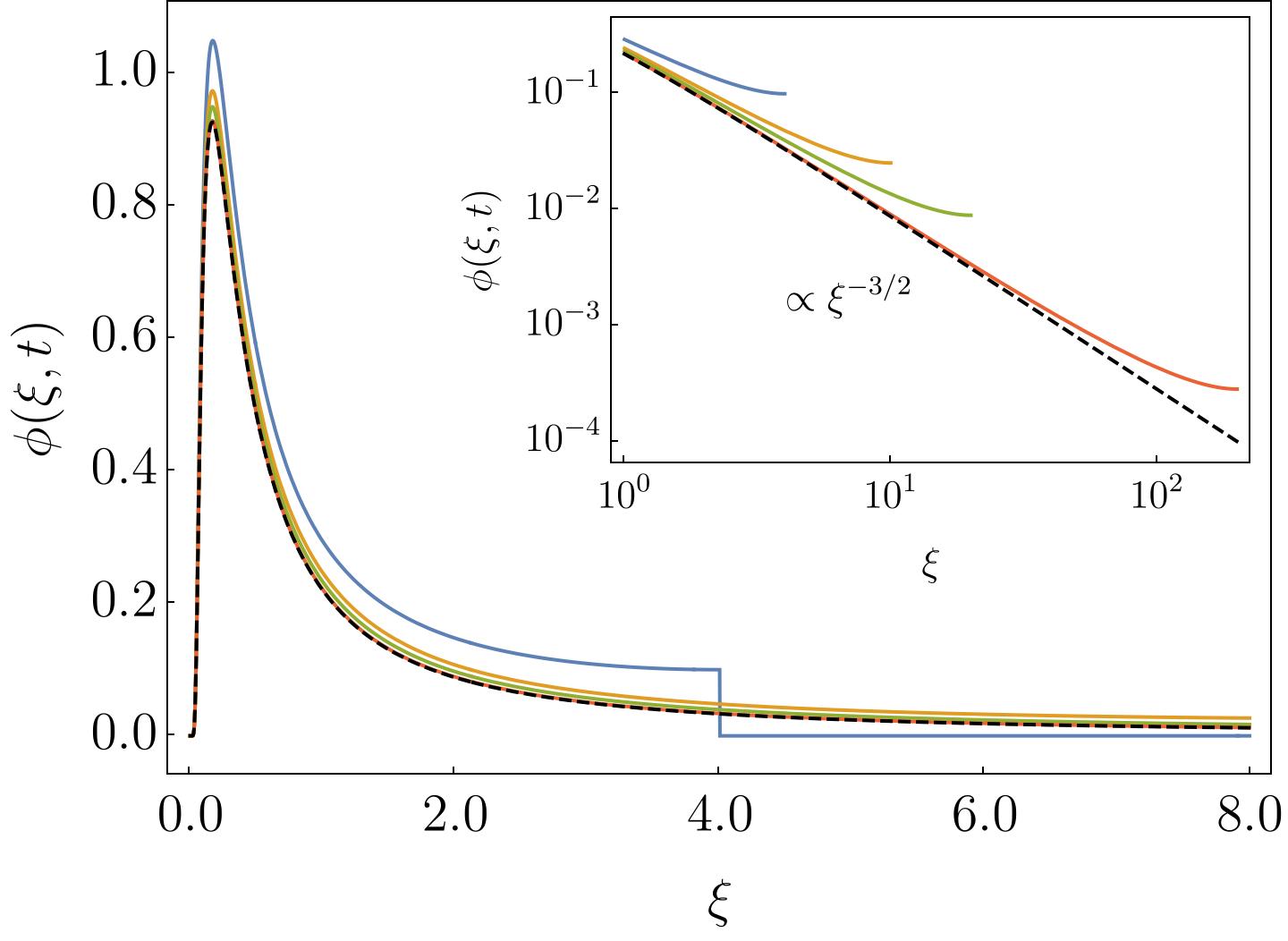}
    \caption{PDF for the scaled variable $\xi$ defined in Eq.~\eqref{ch3_eq:app-xi-def}, with $P_i^*(r_i^*)$ being the radial PDF corresponding normal distribution for $d=1$. Data shown corresponds to the case $n=3$, for which we specifically plot the solid curves obtained from Eq.~\eqref{ch3_eq:scaled-pdf} for different times $t=2$ (blue), $5$ (yellow), $10$ (green) and $100$ (red), together with the  stationary distribution in Eq.~\eqref{ch3_eq:scaled-pdf-asymp-2}, corresponding to the black, dashed curve. The inset is displayed in logarithmic scale, such that the algebraic tails for large $\xi$ from Eq.~\eqref{ch3_eq:tail} are visualised---the dashed straight line with slope $-3/2$. The cuts for each of the PDFs at intermediate times correspond to the upper bound $\xi = [r_{\LLNES}^*(t)]^{1-n}$, following the definition of $\xi$ from Eq.~\eqref{ch3_eq:app-xi-def2}.}\label{ch3_fig:app-scaled-pdf}
  \end{figure}

  Figure~\ref{ch3_fig:app-scaled-pdf} shows the PDF for the scaled variable $\xi$. We observe how $\phi(\xi,t)$ approaches the steady solution $\phi_{\text{st}}(\xi)$ for long enough times. In the inset we employ a logarithmic scale in order to illustrate the tails of the distribution, as given by Eq.~\eqref{ch3_eq:tail}. Note that the PDF departs from the theoretical tail for very large values of $\xi$, with the separation point moving to higher values of $\xi$ as time increases. This is consistent with our theoretical analysis, since the steady distribution in Eq.~\eqref{ch3_eq:scaled-pdf-asymp-2}  is valid only when $\xi\left[r^*_{\LLNES}(t)\right]^{n-1}\ll 1$, i.e. we expect that it no longer describes $P^*(r^*,t)$ for $\xi=O([r^*_{\LLNES}(t)]^{1-n})$. This is further corroborated by the fact that $\xi$ is bounded from above by $[r^*_{\LLNES}(t)]^{1-n}$ by definition \eqref{ch3_eq:app-xi-def2}.

\section{\label{ch3_sec:global_stability}Global stability of the LLNES}
In this section we provide a rigorous proof of the global stability of the LLNES under the timescale separation from Eq.~\eqref{ch3_eq:LLNES-interm-times}. That is, regardless of the initial conditions for the dynamics---encoded by the initial probability distribution $P_i(r_i)$, the LLNES attracts all the solutions of the Fokker-Planck equation in the long-time limit. First, we show how the LLNES emerges as an asymptotic solution of the Fokker-Planck equation, and subsequently, we construct a suitable Lyapunov functional to prove the global stability via an H-theorem.
\subsection{\label{ch3_subsec:asymptotic}Asymptotic solution of the Fokker-Planck equation}
Let us focus on the Fokker-Planck equation from Eq.~\eqref{ch1_eq:fokker-planck}. It is convenient to rescale the Markov process $\bm{r}$ with a characteristic time-dependent ``length'' $\ell(t)$ and accordingly rescale the PDF. For the sake of simplicity, we choose
\begin{equation}
\label{ch3_eq:suppl-scaled}
    \ell(t)=\sqrt{\left< r^2\right>(t)}, \quad \bm{c}\equiv \frac{\bm{r}}{\ell(t)}, \quad \psi(\bm{c},t) = \ell^d(t) P(\bm{r},t),
\end{equation}
with $l(0)$ being proportional to the equilibrium length $\ell_i$ from Eq.~\eqref{ch3_eq:suppl-elli-def}. Now, from the Fokker-Planck equation \eqref{ch1_eq:fokker-planck}, by multiplying by $r^2$ on both sides and integrating over $\bm{r}$, one obtains
\begin{equation}\label{ch3_eq:eq_rara}
    \gamma \frac{d}{dt}\left< r^2 \right>(t)=-2\left[k \left< r^{n+1}\right>(t)-dk_B T_f\right].
\end{equation}
The stationary solution of the above corresponds to  $k \left< r^{n+1}\right>=dk_B T_f$---which follows from the equipartition theorem, and it entails that $\left< r^{n+1}\right>=d\,\ell_f^{n+1}$. The Fokker-Planck equation for the rescaled PDF $\psi$, including the thermal noise, is given by
\begin{equation}
    \partial_t \psi= b\, \ell ^{n-1}\nabla_{\bm{c}}^{\sf{T}} \left\{ \left[c^{n-1}-M_{n+1}\right]\bm{c}\,\psi + \frac{k_BT_f}{k\,\ell^{n+1}}\left[d\bm{c}\psi+\nabla_{\bm{c}}\psi\right] \right\},
    \label{ch3_eq:suppl-FPE-rescaled}
\end{equation}
where $b$ was defined in Eq.~\eqref{ch3_eq:suppl-noiseless-Langevin}, $\hat{M}_n\equiv \left<c^{n}\right>=\left<r^n\right>/\ell^n$, and we have taken into account that, by definition, $\hat{M}_2=1$. Let us note that Eq.~\eqref{ch3_eq:suppl-FPE-rescaled} constitutes an integro-differential equation, since the moments $\hat{M}_n$ are expressed as integrals over $\bm{c}$. We now employ the dimensionless variables $\ell_i$ introduced in the previous section, giving
\begin{equation}\label{ch3_eq:suppl-Y-def}
    \ell(t) = \ell_i \ell^*(t)=\left(\frac{k_B T_i}{k}\right)^{\frac{1}{n+1}}\ell^*(t),
\end{equation}
where $\ell^*(t)$ is of the order of unity for not very long times. Substituting it into Eq.~\eqref{ch3_eq:eq_rara}, we get the dynamical evolution
\begin{equation}\label{ch3_eq:suppl-Y-def-2}
    \ell^*\frac{d\ell^*(t)}{dt}=-b\,\ell_i^{n-1} \left({\ell^*}^{n+1} \hat{M}_{n+1}-d \frac{T_f}{T_i}\right).
\end{equation}
Introducing the above relation into Eq.~\eqref{ch3_eq:suppl-FPE-rescaled} gives
\begin{equation}
\label{ch3_eq:suppl-FPE-rescaled-v2}
    \partial_t \psi = b \,\ell_i^{n-1} {\ell^*}^{n-1}\,
    \nabla_{\bm{c}}^{\sf{T}} \left\{ \left[c^{n-1}-\hat{M}_{n+1}\right]\bm{c}\,\psi + 
    {\ell^*}^{-(n+1)}
    \frac{T_f}{T_i}\left[d\bm{c}\psi+\nabla_{\bm{c}}\psi\right] \right\}.
 \end{equation}
The above suggests the introduction of a new timescale $s$, satisfying
\begin{equation}\label{ch3_eq:FPE-timescale-s}
    ds=b \,\ell_i^{n-1} {\ell^*}^{n-1}
    dt,
\end{equation}
for which the scaled Fokker-Planck equation reads
\begin{equation}
    \label{ch3_eq:suppl-FPE-rescaled-v3}
    \partial_s \psi = \nabla_{\bm{c}}^{\sf{T}} \left\{ \left[c^{n-1}-\hat{M}_{n+1}\right]\bm{c}\,\psi + 
    {\ell^*}^{-(n+1)}
    \frac{T_f}{T_i}\left[d\bm{c}\psi+\nabla_{\bm{c}}\psi\right] \right\}.
\end{equation}

Let us remark that Eq.~\eqref{ch3_eq:suppl-FPE-rescaled-v3} is exact, and thus valid for any temperature ratio $T_f/T_i$. Notwithstanding, since we are interested in a  quench from a high temperature, we have that $T_i \gg T_f$ and the last term on the right hand side of Eq.~\eqref{ch3_eq:suppl-FPE-rescaled-v3} becomes subdominant and thus, it can be neglected---as long as 
${\ell^*}^{-(n+1)}{T_f}/{T_i}\ll 1$. 
Therefore, we have
\begin{equation}
\label{ch3_eq:suppl-FPE-dimensionless-approx}
    \partial_s \psi \sim  \nabla_{\bm{c}}^{\sf{T}}\left\{ \left[c^{n-1}-\hat{M}_{n+1}\right]\bm{c}\,\psi  \right\}, \quad  
    {\ell^*}^{-(n+1)}\frac{T_f}{T_i}\ll 1.
\end{equation}
As the initial condition---equilibrium state at temperature $T_i$---has radial symmetry and the time evolution preserves it, $\psi(\bm{c},t)$ also has radial symmetry at all times. 
Equation~\eqref{ch3_eq:suppl-FPE-dimensionless-approx} admits a stationary solution $\psi_{\text{st}}(\bm{c})$ with radial symmetry, $\psi_{\text{st}}(\bm{c})=\Omega_d^{-1}\tilde{\psi}_{\text{st}}(c)$\footnote{We remark that $\tilde{\psi}_{\text{st}}(c)$ constitutes the radial part of the scaled PDF $\psi_{\text{st}}(\bm{c})$, as the radial and azimutal coordinates are independent---$\Omega_d^{-1}$ accounts for the azimutal PDF, which is uniform for isotropic systems. Such radial part differs from the scaled radial PDF $\psi_{\text{st}}(c)$ by a Jacobian term $c^{d-1}$.}---where $\Omega_d$ constitutes the $d$-dimensional solid angle: 
\begin{equation}\label{ch3_eq:llnes-condition}
     \nabla_{\bm{c}}^{\sf{T}} \left\{ \left[c^{n-1}-\hat{M}_{n+1}\right]\bm{c}\,\tilde{\psi}_{\text{st}}(c)  \right\} = 0 \iff \left[c^{n-1}-\hat{M}_{n+1}\right]\tilde{\psi}_{\text{st}}(c) =0.
\end{equation}
The above equality stems from the fact that there are no normalisable, nor positive definite solutions for $\tilde{\psi}_{\text{st}}$ admitting a non-zero value of $[c^{n-1}-\hat{M}_{n+1}]\tilde{\psi}_{\text{st}}(c)$. Equation~\eqref{ch3_eq:llnes-condition} further entails an infinite hierarchy for the moments $\hat{M}_m$, $m=0,1,2,\ldots$,
\begin{align}
    0 &=\int d\bm{c} \, c^{m}\left[c^{n-1}-\hat{M}_{n+1}\right]\psi_{\text{st}}(\bm{c}) =  \hat{M}_{m+ n-1} - \hat{M}_m \hat{M}_{n+1}, \quad \hat{M}_0=1.
    \label{ch3_eq:suppl-Mm-infinite-hier}
\end{align}
The particular case $m=2$ implies that  $\hat{M}_2=1$, in agreement with the definition of scaled variables from Eq.~\eqref{ch3_eq:suppl-scaled}. The above relation, together with Hölder's inequality~\cite{loeve_probability_1997}
\begin{equation}\label{ch3_eq:holder}
    \hat{M}_r \geq (\hat{M}_s)^{r/s}, \quad r>s>0,
\end{equation}
further entails that
\begin{equation}\label{ch3_eq:moments-LLNES-2}
    \hat{M}_m = 1 \quad \forall m\geq 0,
\end{equation}
and thus,
\begin{equation}\label{ch3_eq:LLNES-solution}
    \tilde{\psi}_{\text{st}}(c) =\tilde{\psi}_{\LLNES}(c)\equiv \delta (c-1), \iff \psi_{\text{st}}(\bm{c})=\psi_{\LLNES}(\bm{c})=\Omega_d^{-1}\tilde{\psi}_{\LLNES}(c),
\end{equation}
where $\psi_{\text{st}}$ corresponds to the scaled PDF at the LLNES.\footnote{Let us note that, since $\tilde{\psi}_{\LLNES}(c)$ involves a Dirac-delta centered at $c=1$, both the radial part of the PDF and the radial PDF become the same.} Once the system has relaxed to the LLNES, we have that $\left< r\right>(t)=\ell(t) \hat{M}_1(t) \to \ell(t)=r_{\LLNES}(t)$, with $r_{\LLNES}(t)$ as given by Eq.~\eqref{ch3_eq:suppl-LLNES-algebraic}. The time evolution of $\ell (t)$ then follows from the limiting behaviour from Eq.~\eqref{ch3_eq:suppl-Y-def-2},
\begin{equation}
    \dot{\ell}^*=-b\,\ell_i^{n-1} {\ell^*}^n \iff \dot{\ell}=-b\,{\ell}^n , \quad {\ell^*}^{-(n+1)}\frac{T_f}{T_i}\ll 1.
\end{equation}
The latter is consistent with the noiseless Langevin equation from Eq.~\eqref{ch3_eq:suppl-noiseless-Langevin}, provided that $t$ is not too long---such that the noise terms may become relevant. The more complete analysis developed here has allowed us to identify the intermediate timescale over which the LLNES is developed, ${\ell^*}^{-(n+1)}{T_f}/{T_i}\ll 1$: the larger the temperature ratio $T_i/T_f$ is, the wider this intermediate timescale---and the more relevant the LLNES---becomes.

\subsection{\label{ch3_subsec:lyapunov}Lyapunov functional and H-theorem}
Within the context of statistical mechanics and thermodynamics, in order to prove the global stability of a relevant stationary state, one introduces an appropriate Lyapunov functional~\cite{lyapunov_general_1992}. A functional of the PDF is said to belong to the class of Lyapunov functionals if it satisfies the following requirements: (i) it is bounded from below by the value attained at the desired stationary state, (ii) it monotonically decreases with time, and (iii) its time derivative vanishes only when the PDF attains the aforementioned stationary state. Finding a suitable Lyapunov functional for the problem of concern is equivalent to proving what is known in the literature as an H-theorem, which owes its name to the Lyapunov functional---also known as H-functional---that Boltzmann introduced to prove that all the solutions of the Boltzmann equation tend in the long-time limit to the equilibrium Maxwellian distribution~\cite{resibois_classical_1977,chapman_mathematical_1990}.

For the LLNES, which constitutes the asymptotic, stationary solution of the scaled Fokker-Planck equation~\eqref{ch3_eq:suppl-FPE-rescaled-v3}, we introduce the functional
\begin{equation}
    H(t) \equiv -\left<\ln c\right> = -\int d\bm{c}\ \psi (\bm{c},t) \ln c,
\end{equation}
which we show below to be a Lyapunov functional. First, following Jensen's inequality, it is readily seen that $H$ is bounded from below by its global minimum, 
\begin{equation}
    H(t)=-\left<\ln(c^2)\right>/2 \geq -\ln \left<c^2\right>/2 =-\ln \hat{M}_2/2=0,
\end{equation}
which is attained at the LLNES:
\begin{equation}
    H_{\LLNES} = -\int d\bm{c}\ \psi_{\LLNES}(\bm{c}) \ln c=-\int_0^{+\infty}dc\, c^{d-1} \,\delta(c-1) \ln c = 0,
\end{equation}
thus fulfilling the first (i)  necessary requirement. Let us now consider the time evolution of $H(t)$. We employ the timescale $s$ defined in Eq.~\eqref{ch3_eq:suppl-FPE-rescaled-v3}, giving
\begin{align}
    \frac{dH}{ds} &= -\int d\bm{c} \ (\partial_s \psi) \ln c \nonumber
    \\
    &=\int d\bm{c}\ \left\{ \left[c^{n-1}-\hat{M}_{n+1}\right]\bm{c}\,\psi + 
    {\ell^*}^{-(n+1)}
    \frac{T_f}{T_i}\left[d\bm{c}\psi+\nabla_{\bm{c}}\psi\right] \right\}^{\sf{T}} \nabla_{\bm{c}} \ln (c)
\end{align}
where we have introduced Eq.~\eqref{ch3_eq:suppl-FPE-rescaled-v3} and integrated by parts, assuming that the boundary terms vanish at infinity. Once again, in the limit $T_i \gg T_f$,  the term proportional to $\nabla_{\bm{c}}\psi$ may be neglected as long as ${\ell^*}^{-(n+1)}(T_f/T_i) \ll 1$, which we comment on below. Therefore, 
\begin{align}
    \frac{dH}{ds} &\sim \int d\bm{c} \ \left[c^{n-1}-\hat{M}_{n+1}\right]\psi \ \bm{c}^{\sf{T}} \nabla_{\bm{c}} \ln c = \int d\bm{c} \ \left[c^{n-1}-\hat{M}_{n+1}\right] \psi \nonumber
    \\
    &= \hat{M}_{n-1} - \hat{M}_{n+1}, \quad {\ell^*}^{-(n+1)}\frac{T_f}{T_i} 
    \ll 1,
\end{align}
where we have used that $\bm{c}^{\sf{T}} \nabla_{\bm{c}}\ln c=1$. Let us note that the difference $\hat{M}_{n-1} - \hat{M}_{n+1}$ vanishes at the LLNES, following Eq.~\eqref{ch3_eq:moments-LLNES-2}. For the very last step of the proof, we resort again to Hölder's inequality from Eq.~\eqref{ch3_eq:holder}:
\begin{equation}
    \hat{M}_{n-1}-\hat{M}_{n+1}\le \hat{M}_{n+1}^{\frac{n-1}{n+1}}-\hat{M}_{n+1}=\hat{M}_{n+1}^{\frac{n-1}{n+1}}\left(1-\hat{M}_{n+1}^{\frac{2}{n+1}}\right)\le \hat{M}_{n+1}^{\frac{n-1}{n+1}}\left(1-\hat{M}_2\right)=0,
\end{equation}
where we have brought to bear that $n>1$ and $\hat{M}_2=1$. With this, we have just proven that $dH/ds \leq 0$, and since $H$ is then a non-increasing function of time bounded from below, $H$ tends to a well-defined limit for $s\gg 1$ at which $dH/ds$ vanishes: therefore, the requirements (ii) and (iii) are fulfilled, and $H$ is a Lyapunov functional over the timescale $s$, implying that all the solutions of the Fokker-Planck equation tend thereto. 

It must be stressed that this H-theorem holds exclusively within the timescale $s$, and ceases to be valid when ${\ell^*}^{-(n+1)}(t)(T_f/T_i)$ becomes of the order of unity. This comes about when $\ell^*$ becomes close to its equilibrium value at $T_f$: when $\ell^*(t)=O(\ell^*_{\text{eq}}(T_f))$, we have that $\ell^*(t)=O((T_f/T_i)^{1/(n+1)})$, recalling the definition of $\ell^*(t)$, as given by Eq.~\eqref{ch3_eq:suppl-Y-def}.  In other words, the LLNES disappears when $\left<r\right>(t)$ becomes of the order of the equilibrium value at the final temperature $T_f$, as stated on physical grounds in Sec.~\ref{ch3_sec:emergence}. 

Therefore, we have rigorously proven that the LLNES ``attracts'' all the solutions of the Fokker-Planck equation over the intermediate time scale for which $s\gg 1$ and thermal noise is negligible, i.e. ${\ell^*}^{-(n+1)}{T_f}/{T_i}\ll 1$. We highlight that this condition is compatible with that for the Dirac-delta being actually narrower than the final equilibrium distribution in Eq.~\eqref{ch3_eq:suppl-cond-Dirac-narrow}, and thus, we can unify both of them as $(T_f/T_i)^{\frac{1}{n+1}} \ll r_{\LLNES}^*(t) \ll (T_f/T_i)^{\frac{1}{n(n+1)}}$.
\section{\label{ch3_sec:robustness}Robustness in more complex scenarios}
Up to now, we have focused our study of the LLNES on isotropic situations, in order to depict its main features in a simple way that allows for analytical computations and rigorous---from a physics perspective---proofs. For the remainder of this chapter, we provide a combination of results and intuitive arguments to prove that the LLNES may emerge in a broader range of scenarios. In order to illustrate and support these insights, we analyse below different relevant physical potentials beyond the isotropic case in which the LLNES emerges: with anisotropy, and interacting degrees of freedom. Additionally, we also show that the LLNES extends to stochastic systems further involving binary collisions---either elastic or inelastic, similar to the molecular fluid we employed in chapter~\ref{ch:memory_effects}.

Giving a rigorous condition for the emergence of the LLNES in such a general situation is a challenging mathematical problem, which lies outside the scope of this thesis. Still, we may gather an intuition about the conditions under which the LLNES emerges by analysing some specific examples, both analytically---when their simplicity makes it possible---and numerically. This is the approach that we follow in the remainder of this section, in which we consider several relevant examples to illustrate some common features appearing in all of them. 
\subsection{\label{ch3_subsec:snowplough}Intuitive idea: ``snowplough'' picture}
Let us recall the $d$-dimensional overdamped system trapped in a general confining potential $U(\bm{r}) = U(x_1,x_2,...,x_d)$. For instance, this is the case of an overdamped Brownian particle confined in an optical trap, for which $\bm{r}$ would account for its cartesian coordinates, or the case of a molecular fluid with non-linear drag at the kinetic level, in which $\bm{r}$ here stands for the velocity instead, and the corresponding ``confining potential" would be the velocity-dependent function whose gradient in velocity space provides the non-linear drag force---as presented at the end of Sec.~\ref{ch3_subsubsec:application-molecular}. In any case, such general potential may have a plethora of minima due to either the form of the external force or as a result of the repulsive interactions between the different degrees of freedom. Such minima define different characteristic lengths---i.e. the positions of those minima with respect to the origin---that, as long as the initial temperature is sufficiently high, become irrelevant for the dynamics of concern---as argued below. 

We stress that the general picture drawn in the following anticipates the common features behind the LLNES in the different examples we investigate afterwards, which hint at---but not prove---the existence of a general framework for it. The emergence and shape of the LLNES seems to be related to the dominant contributions of $U(\bm{r})$ for large $r$, which we refer to as the effective potential $U_{\text{eff}}(\bm{r})$ in the following. The effective potential $U_{\text{eff}}(\bm{r})$ corresponds to the potential that the system feels initially, since the initial condition corresponds to a high temperature and thus the details of the bottom of the potential $U(\bm{r})$ are irrelevant. 
Our analysis of the isotropic case suggests that it is necessary that the force diverges faster than linearly at long distances, i.e. $\lim_{r\to\infty} |\bm{A}(\bm{r})|/r \to \infty$.

\begin{figure}
    \centering 
    \includegraphics[width=5.2in]{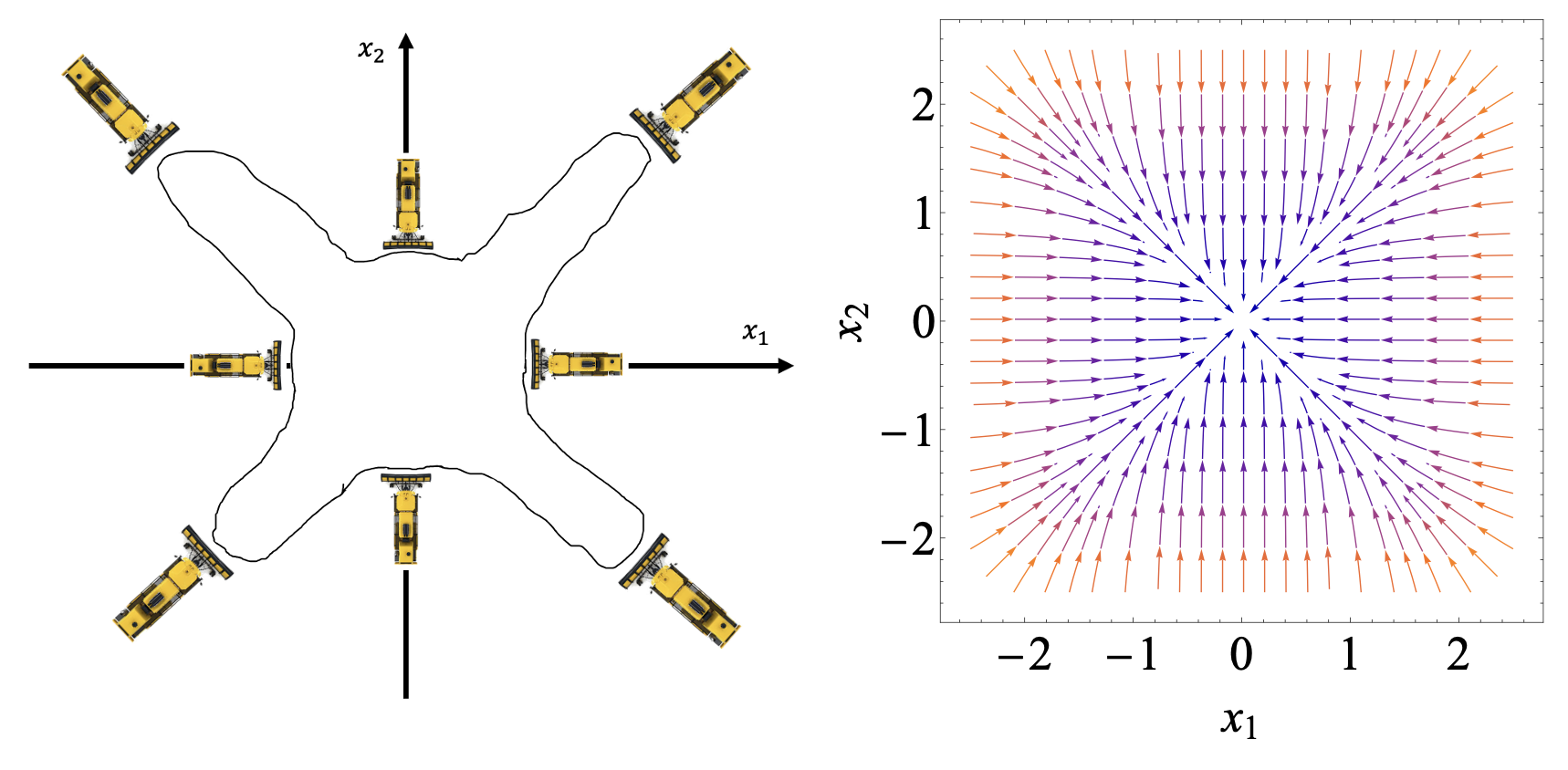}
    \caption{Qualitative sketch of the snowplough picture for a two-dimensional overdamped particle in a confining potential of the form $U(\bm{r}) = x_1^4 + x_2^4$. Left: Snowploughs acting on the snow---i.e probability density---over the $(x,y)$-plane. The fastest ones swap the snow along the $x_1$ and $x_2$ axis, thus accumulating it over the regions that the slowest snowploughs have not reached yet---along the diagonals $x_2 = \pm x_1$. Right: Stream plot of the force $\bm{F} = -\nabla U$.}
        \label{ch3_fig:snowplough}
    \end{figure}

The main intuitive idea stems from the stream plots of the effective force $-\nabla U_{\text{eff}}(\bm{r})$. In the deterministic dynamics, the movement of the particle follows the stream lines, since $\dot{\bm{r}}\propto -\nabla U_{\text{eff}}(\bm{r})$ and thus the probability weight tends to accumulate along the directions in which the force exerted on the system increases the slowest. A graphical image may be given by imagining a ``snowplough'' acting on the directions of the fastest variation of the force, and thus concentrating the probability weight over the slowest ones. For long enough times, the analysis of the specific cases considered below suggests that, regardless of the initial probability---or snow---distribution, all the probability accumulates over the slowest directions, the initial conditions become immaterial, and the Dirac-delta PDF characterising the LLNES emerges for faster-than-linearly diverging forces. Figure~\ref{ch3_fig:snowplough} shows a sketch of the snowplough picture for the specific case of a confining potential of the form $U(\bm{r}) = x_1^4 + x_2^4$, similar to the one we employ as our first example in the next section---specifically in Eq.~\eqref{ch3_eq:suppl-potential-anisotropic-1}.

We would like to highlight that the symmetries of the confining potential have an impact on the geometrical structure of the LLNES. In this context, symmetries refer to the presence of cyclic variables in the potential $U(\bm{r})$, when expressed in a certain coordinate system. In previous sections, we have analysed in depth the radially symmetric (or isotropic) situation, where $U$ only depends on the distance to the origin $r$ and the $d$-dimensional solid-angle variables are thus cyclic. That analysis shows that the LLNES has also radial symmetry in this case and, on a physical basis, one expects that this would also be the case for other symmetries. 


\subsection{\label{ch3_subsec:potentials}Various confining potentials}

\subsubsection{\label{ch3_subsubsec:anisotropic}Anisotropic potentials}
So far, our discussion has been focused on overdamped systems with radial symmetry, in which the relevant dynamics is given by the radial coordinate $r\equiv |\bm{r}| =\sqrt{x_1^2 + x_2^2 ... + x_d^2}$. Here, we  show that such a restriction is not necessary for the emergence of the LLNES, by considering some specific examples.

\begin{figure}
    {\centering 
    \includegraphics[width=2.6in]{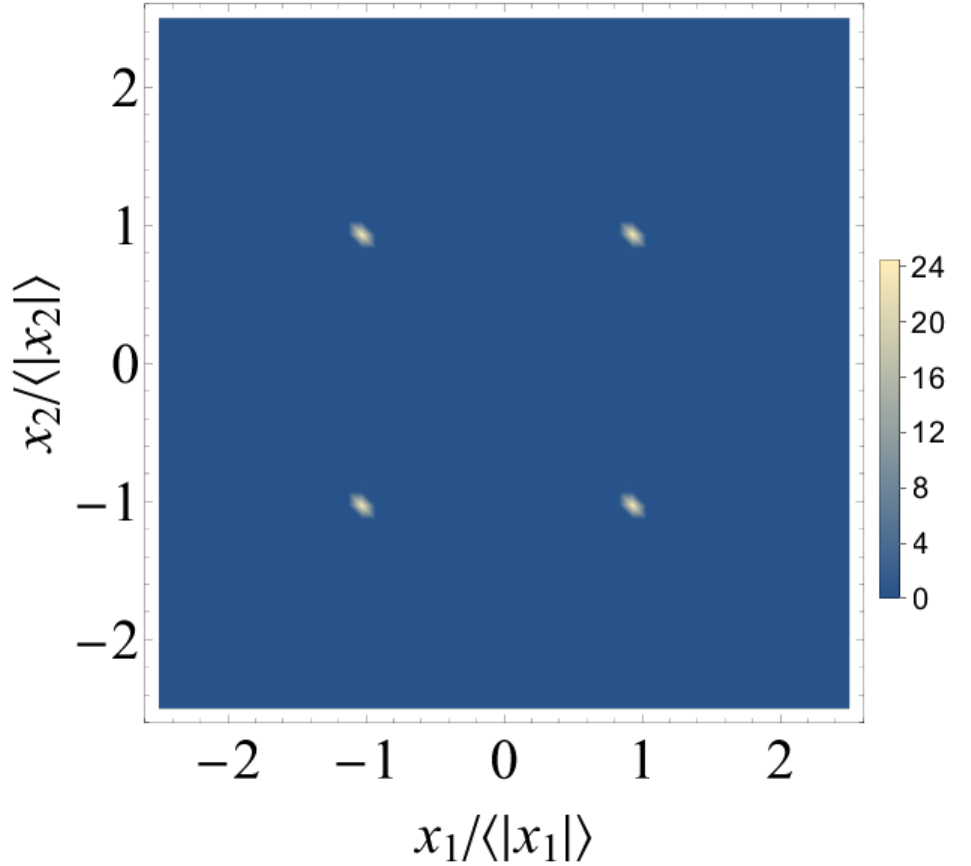}
    \includegraphics[width=2.6in]{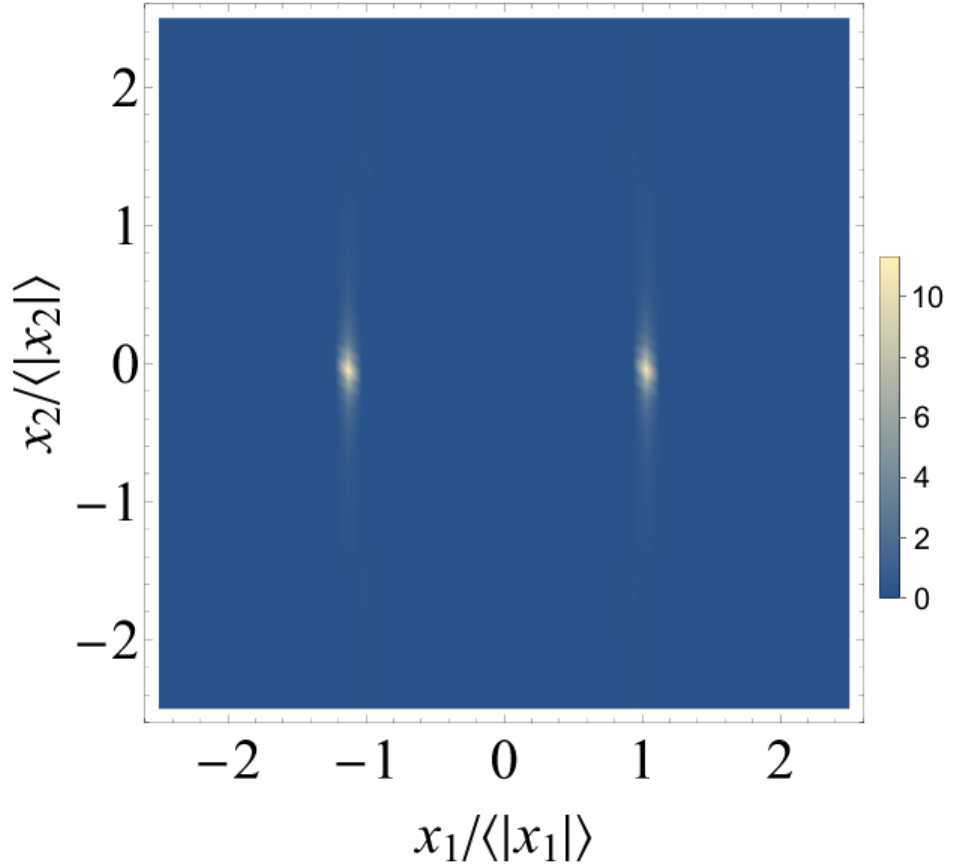}
    \\
    \includegraphics[width=2.6in]{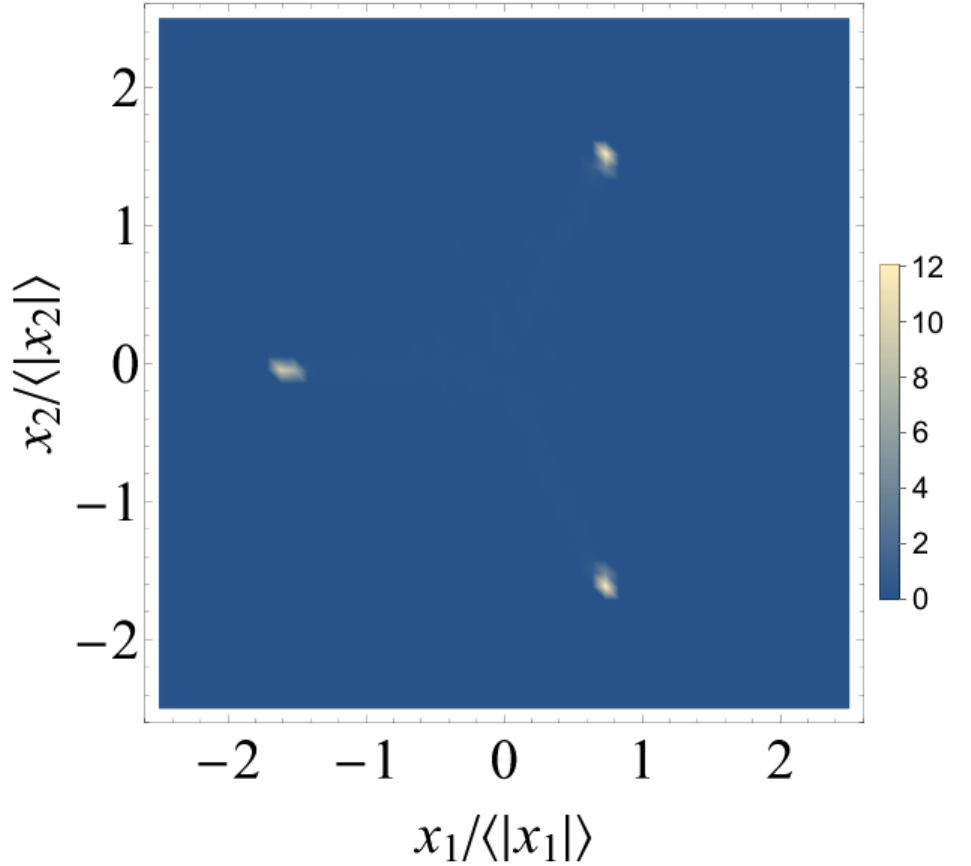}
    \includegraphics[width=2.6in]{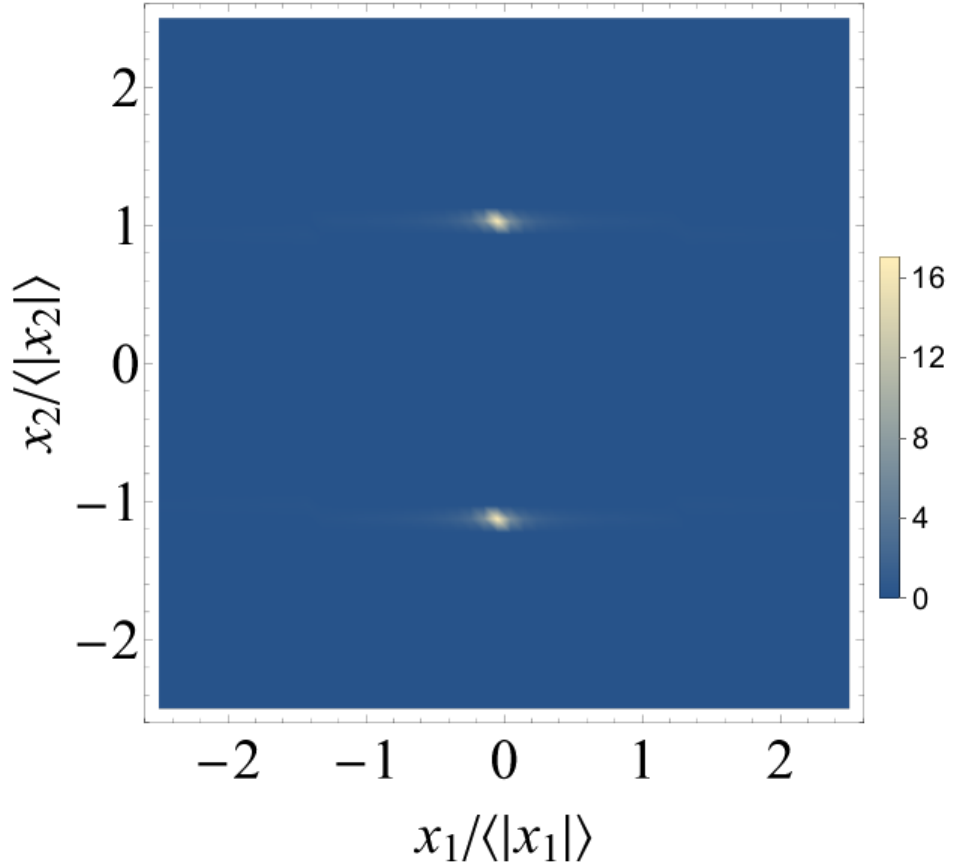}
    \\}
    \caption{Density plots of the PDF for different anisotropic potentials. Specifically, these are presented on the $(x_1/\left<|x_1|\right>,x_2/\left<|x_2|\right>)$ plane. The graphs correspond to the potentials in Eq.~\eqref{ch3_eq:suppl-potential-anisotropic-1} (top left), \eqref{ch3_eq:suppl-potential-anisotropic-2} (top right), \eqref{ch3_eq:suppl-potential-anisotropic-3} (bottom left), and \eqref{ch3_eq:suppl-potential-anisotropic-4} (bottom right). For the latter, the 3D PDF has been plotted onto the $x_3 = 0$ plane. The density plots of the PDF are shown for a long time, once the system has reached the LLNES. To obtain the plotted PDFs, we have numerically solved the corresponding Fokker-Planck equation, including the noise term, by considering an ensemble of trajectories relaxing from very high initial temperature $T_i$ to a low enough final temperature $T_f$ such that $T_i/T_f = 10^6$.}
        \label{ch3_fig:anisotropic-LLNES}
\end{figure}

\begin{figure}
    {\centering 
    \includegraphics[width=2.6in]{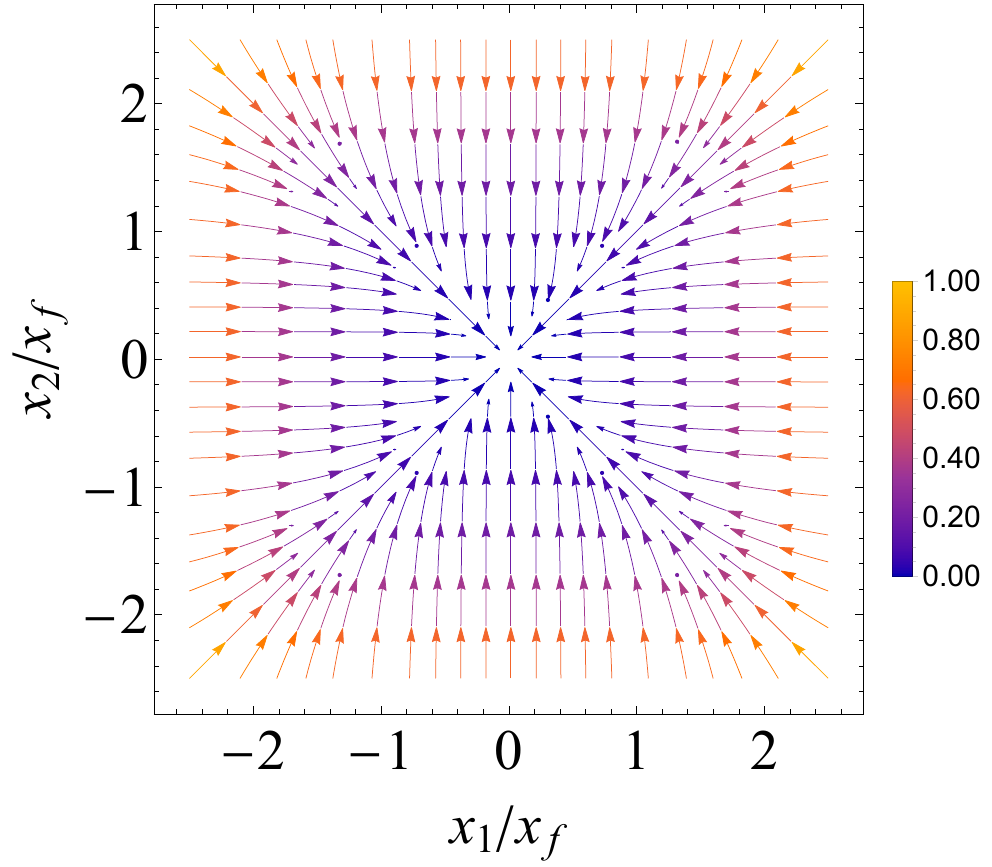}
    \includegraphics[width=2.6in]{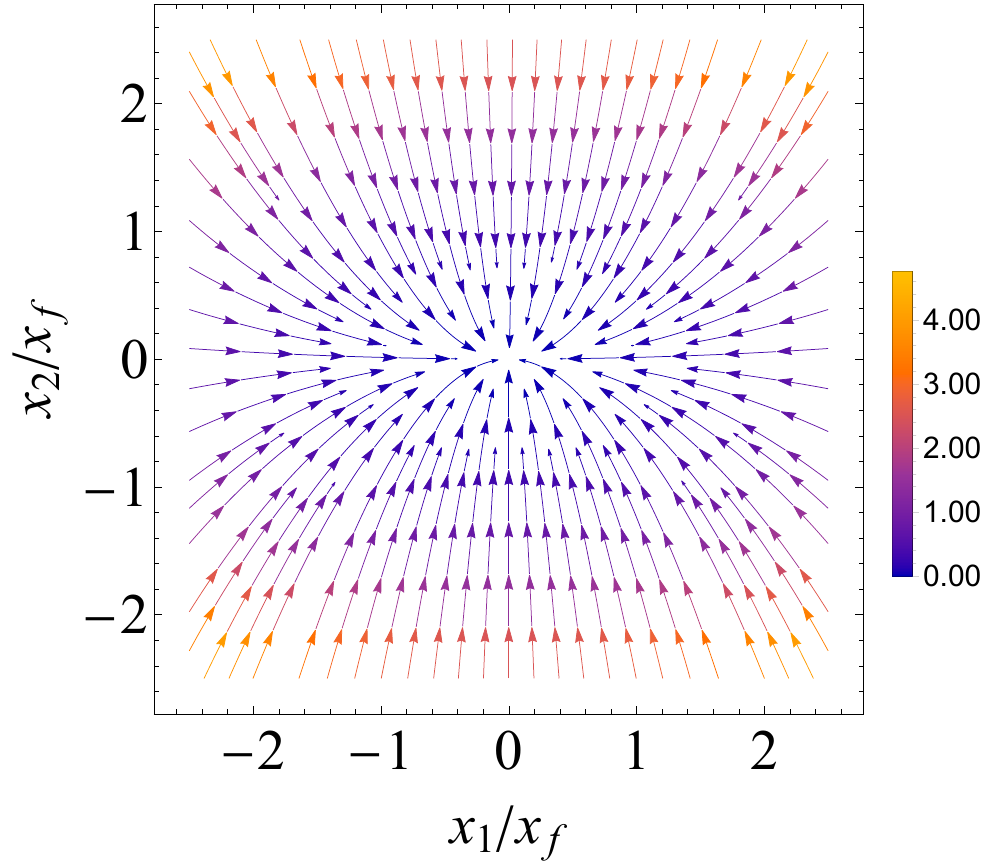}
    \\
    \includegraphics[width=2.6in]{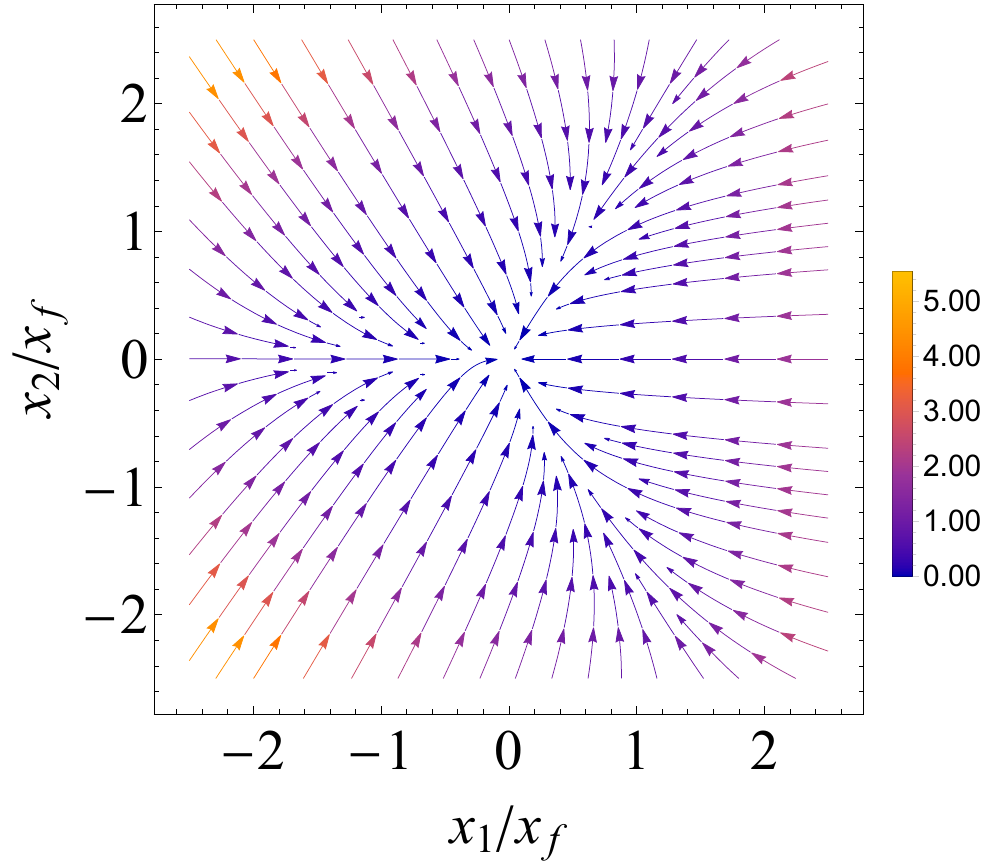}
    \includegraphics[width=2.6in]{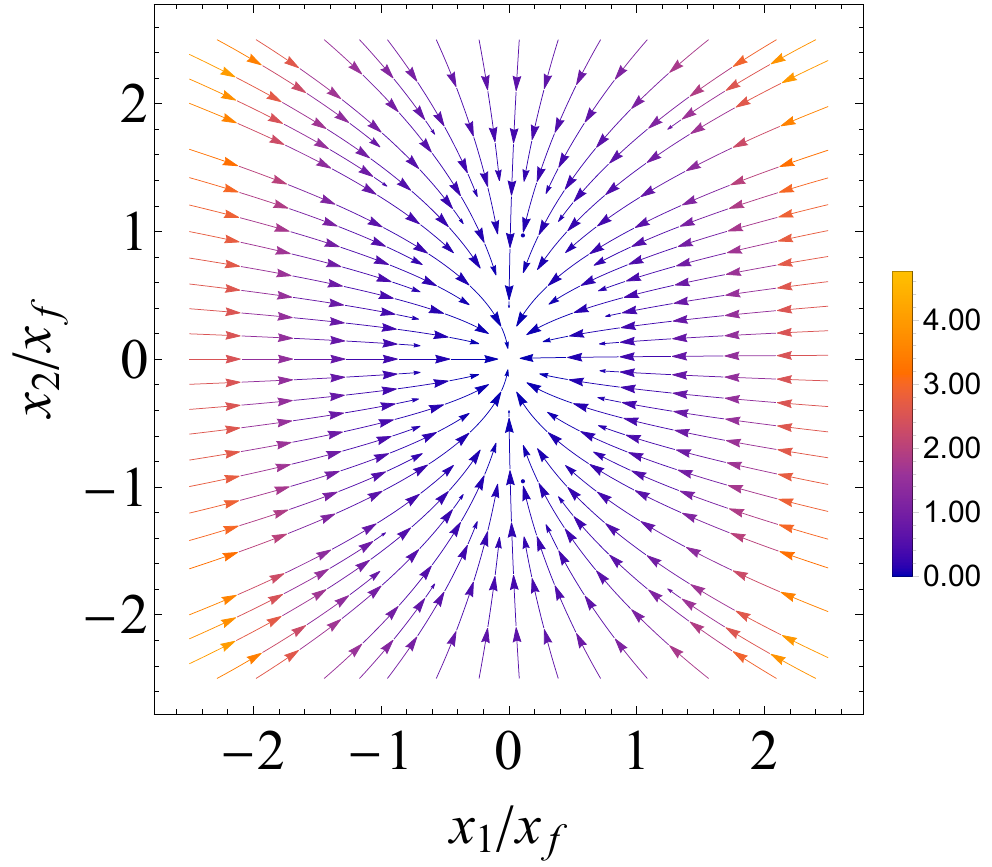}
    \\}
    \caption{Stream plots of the force $\bm{F} = -\nabla U$ for each of the anisotropic potentials considered in figure~\ref{ch3_fig:anisotropic-LLNES}. These stream plots are presented on the $(x_1/x_f,x_2/x_f)$ plane, where $x_f\equiv (k_BT_f/k)^{1/4}$ constitutes a characteristic length for equilibrium at $T_f$---for the upper-left panel, we considered $k_1 = k_2 = k$. The four panels follow the same ordering as those from figure~\ref{ch3_fig:anisotropic-LLNES}. For the bottom right panel, the 3D stream plot is shown onto the $x_3 = 0$ plane.}
        \label{ch3_fig:streamplots}
\end{figure}

As a first example, let us consider a two-dimensional confining potential of the form
\begin{equation}
\label{ch3_eq:suppl-potential-anisotropic-1}
    U(x_1,x_2) = \frac{1}{4}\left(k_1x_1^4 + k_2x_2^4\right), \quad k_1,k_2>0,
\end{equation}
which gives the following set of decoupled noiseless Langevin equations,
\begin{equation}
    \dot{x}_j= -b_j x_j^3, \quad b_j\equiv \frac{k_j}{\gamma}, \quad j=1,2,
\end{equation}
whose solutions are given by
\begin{equation}
    x_j(t)= x_{j,i}\left(1+2x_{j,i}^2b_j t\right)^{-1/2} \sim  x_{j,\LLNES}(t) \equiv  \text{sign}(x_{j,i}) (2b_j t)^{-1/2}.
\end{equation}
In the above, $x_{j,i}$ corresponds to the initial value of $x_j(t)$. The two coordinates, $x_1$ and $x_2$ tend independently to their respective LLNES states, $\delta(x_j- \text{sgn}(x_{j})x_{j,\LLNES})/2$,\footnote{Note that $\delta(x_j- \text{sgn}(x_{j})x_{j,\LLNES})$ is normalised to $2$ for any $x_{j,\LLNES}>0$, since the integration interval over $x_j$ may be split into two intervals $(-\infty,0)$ and $[0,+\infty)$, over which we have two independent Dirac-deltas.} implying that
\begin{equation}
\label{ch3_eq:pdf-llnes-anisotropic-1}
    P_{\LLNES}(x_1,x_2;t) = \frac{1}{4}\delta(x_1- \text{sgn}(x_{1})x_{1,\LLNES}(t))\delta(x_2- \text{sgn}(x_{2})x_{2,\LLNES}(t)).
\end{equation}
The upper left panels of Figs.~\ref{ch3_fig:anisotropic-LLNES} and \ref{ch3_fig:streamplots} show the numerical PDF at the LLNES, which accurately resembles the theoretical prediction \eqref{ch3_eq:pdf-llnes-anisotropic-1}, and the corresponding streamplot of the potential, respectively. Equation~\eqref{ch3_eq:pdf-llnes-anisotropic-1} holds regardless of the values of $b_1$ and $b_2$, i.e. of $k_1$ and $k_2$. However, these constants define two different timescales ($\tau_1 \propto b_1^{-1}$ and $\tau_2 \propto  b_2^{-1}$, respectively) that must be taken into account to discern whether the system has reached the LLNES or not. For instance, in the extreme case where $b_1 \gg b_2$, by the time that $x_1(t)$ reaches its long-time behaviour, $x_2(t)$ still has not forgotten the trace of the initial conditions. Thus, it is the slowest degree of freedom that dictates whether the system has already reached the LLNES or not.\footnote{However, it is the fastest degree of freedom the one that dictates the timescale separation: as it relaxes the fastest, it could happen that it reaches the regime where thermal noise becomes relevant, with the slowest degree of freedom not even having reached the LLNES yet.}

The previous example involves uncoupled degrees of freedom, which explains why the resulting PDF at the LLNES is expressed as the product of the PDF's for each degree of freedom independently, as Eq.~\eqref{ch3_eq:pdf-llnes-anisotropic-1} clearly shows---being thus a direct extension of the situation found for one-dimensional or isotropic systems. Let us now consider a more involved situation, in which the potential couples the different degrees of freedom:
\begin{equation}\label{ch3_eq:suppl-potential-anisotropic-2}
    U(x_1,x_2) = \frac{1}{4}k(x_1^2 + 2x_2^2)^2,
\end{equation}
which provides the following noiseless Langevin equations,
\begin{equation}\label{ch3_eq:eqs-anisotropic-2}
    \dot{x}_1 = - b x_1(x_1^2+2x_2^2), \quad \dot{x}_2 = -2b x_2(x_1^2+2x_2^2), \quad b\equiv \frac{k}{\gamma}.
\end{equation}
The above system can also be analytically integrated.\footnote{It is straightforward to first solve the coupled system from Eq.~\eqref{ch3_eq:eqs-anisotropic-2} for the parametric equation $x_2(x_1)$, $dx_2/dx_1 = 2x_2/x_1$, giving $(x_1/x_{1,i})^2=x_2/x_{2,i}$. Then, by substituting back onto Eq.~\eqref{ch3_eq:eqs-anisotropic-2}, $x_1(t)$ and $x_2(t)$ can be solved independently.} In the long-time regime, one has
\begin{equation}
    x_1(t) \sim \text{sign}(x_{1,i}) \ (2bt)^{-1/2} \equiv \ x_{\LLNES}(t), \quad x_2(t) \sim x_{\LLNES}(t)^2 \left(\frac{x_{2,i}}{x_{1,i}^2}\right).
\end{equation}
Thus, we have that $x_1(t)$ forgets the initial conditions but $x_2(t)$  never does. 
Nevertheless, taking into account that $x_{1,i} \sim x_{2,i} \gg 1$, and that $x_2(t)$ decays much faster than $x_1(t)$, by the time that $x_1(t)$ reaches the LLNES, $x_2(t) \ll x_1(t)$. Therefore, we conclude that
\begin{equation}
\label{ch3_eq:pdf-llnes-anisotropic-2}
    P_{\LLNES}(x_1,x_2;t) = \frac{1}{2}\delta(x_1-\text{sign}(x_1)x_{\LLNES}(t))\delta(x_2).
\end{equation}
Similarly to the previous example, both the above PDF and the corresponding streamplots of the potential are plotted on the upper right panels of Figs.~\ref{ch3_fig:anisotropic-LLNES} and \ref{ch3_fig:streamplots}, respectively.

The previous examples are rather simple, in the sense that their corresponding noiseless Langevin equations can be analytically solved. The general ``snowplough'' picture described in Sec.~\ref{ch3_subsec:snowplough} indeed applies to these examples, as detailed below together with the remainder of the examples considered in this section. Nevertheless, the intuitive snowplough picture can be further tested by considering a more involved potential:
\begin{equation}
\label{ch3_eq:suppl-potential-anisotropic-3}
    U(\rho,\phi) = \frac{1}{4}k\rho^4 \left[2 + \cos(3\phi)\right]^2,
\end{equation}
where we are employing polar coordinates, $x_1 = \rho\cos \phi$ and $x_2 = \rho\sin \phi$. This case is not analytically solvable, but the directions with slowest increase of the force are those satisfying $\cos3\phi =-1$, i.e. $\phi = \pi/3$, $\pi$ and $5\pi/3$. We thus expect the probability to be accumulated on the vertices of an equilateral triangle, thus emerging the following PDF 
\begin{equation}
    P_{\LLNES}(\rho,\phi;t) \propto \delta (\rho - \rho_{\LLNES}(t))\left[\delta \left(\phi - \frac{\pi}{3}\right) + \delta \left(\phi - \pi\right) + \delta \left(\phi - \frac{5\pi}{3}\right)\right],
\end{equation}
at long times, with $\rho_{\LLNES}(t)\equiv (2bt)^{-1/2}$, $b\equiv k/\lambda$, being the algebraic decay solution characterising the LLNES. The PDF and the streamplots for this example are displayed in the bottom left panel of 
Figs.~\ref{ch3_fig:anisotropic-LLNES} and \ref{ch3_fig:streamplots}, respectively.

All the cases considered above correspond to two-dimensional systems. Lastly, we move to higher dimensions by looking into the three-dimensional potential
\begin{equation}
\label{ch3_eq:suppl-potential-anisotropic-4}
    U(x_1,x_2,x_3) = \frac{1}{4}k(2x_1^2 + x_2^2 + 3x_3^2)^2.
\end{equation}
Similarly to the potential in Eq.~\eqref{ch3_eq:suppl-potential-anisotropic-2}, this example can be solved analytically, giving that $x_1(t),x_3(t) \ll x_2(t)$ by the time $x_2(t)$ reaches the LLNES, thus providing the solution
\begin{equation}
    P_{\LLNES}(x_1,x_2,x_3;t) =  \frac{1}{2}\delta(x_1)\delta(x_2-\text{sign}(x_2)x_{\LLNES}(t))\delta(x_3).
\end{equation}
Following the same arguments as those employed for the potential from Eq.~\eqref{ch3_eq:suppl-potential-anisotropic-3}, the direction of the slowest increase of the force corresponds to the $x_2$-axis, i.e. $x_1 = x_3 = 0$. For this last case, the PDF and the corresponding streamplots of the potential are shown at the bottom right panels of Figs.~\ref{ch3_fig:anisotropic-LLNES} and \ref{ch3_fig:streamplots}, respectively.

Overall, Figs.~\ref{ch3_fig:anisotropic-LLNES} and~\ref{ch3_fig:streamplots} show the density plots of the PDF for each of the potentials presented in this section, and the plots showing the stream lines of the corresponding ``force'', respectively. For the former, similarly to Fig.~\ref{ch3_fig:deltas}, data corresponds to the numerical integration of the corresponding Langevin equation for $N = 10^5$ stochastic trajectories, as thoroughly described in Appendix~\ref{app:simulation-methods}. It is clearly observed that the directions over which the Dirac-delta peaks emerge are those for which the force increases the slowest. These stream plots neatly illustrate that each of these directions have its own basin of attraction, thus supporting---on an intuitive basis---the ``snowplough'' picture described before in Sec.~\ref{ch3_subsec:snowplough}. The snowplough sweeps the snow (probability weight) on the directions of fastest variation of the force and accumulates it on the slowest ones.

\subsubsection{\label{ch3_subsubsec:interactions}Interacting degrees of freedom}
In some of the previous examples we have encountered potentials in which there were couplings between the different degrees of freedom. Here, we consider the case of having a general family of pair-wise potentials, in order to show that the LLNES is resilient to the inclusion of interactions. Let us consider the following potential for two interacting degrees of freedom,
\begin{equation}
\label{ch3_eq:interaction-potential}
    U(x_1,x_2) = \frac{1}{4}k(x_1^4 + x_2^4) + \frac{1}{n}c\, |x_1 - x_2|^n,
\end{equation}
that is, two particles, with positions $x_1$ and $x_2$, which are both confined in the same one-dimensional quartic potential while interacting with a potential that depends on the relative distance between them. The $c$ constant determines whether the interaction is attractive ($c>0$) or repulsive ($c<0$), while $n$ quantifies the strength of the interaction. The system is once more submitted to a quench, the initial temperature being much larger than the final one. 

\begin{figure}
    {\centering 
    \includegraphics[width=2.65in]{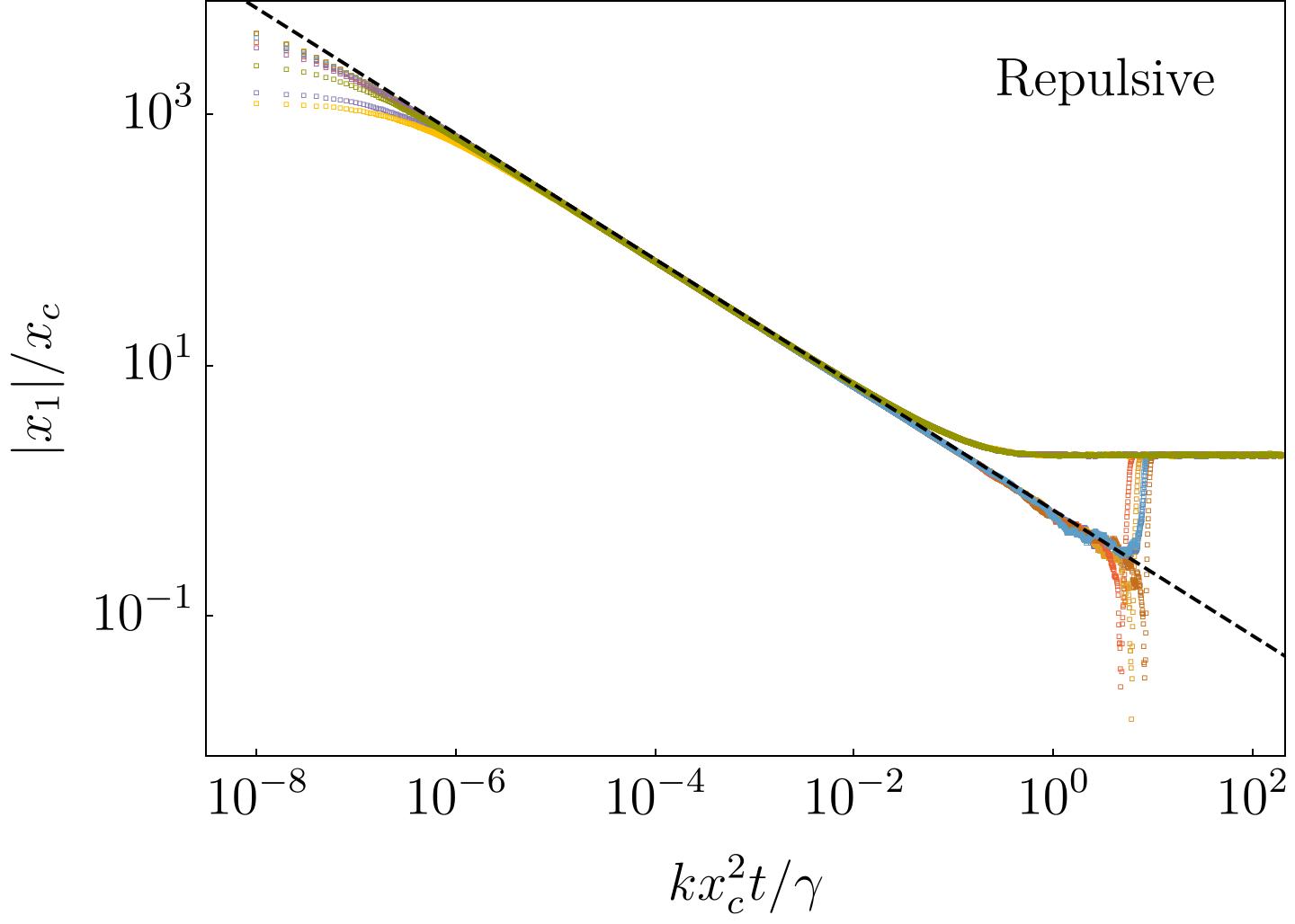}
    \includegraphics[width=2.65in]{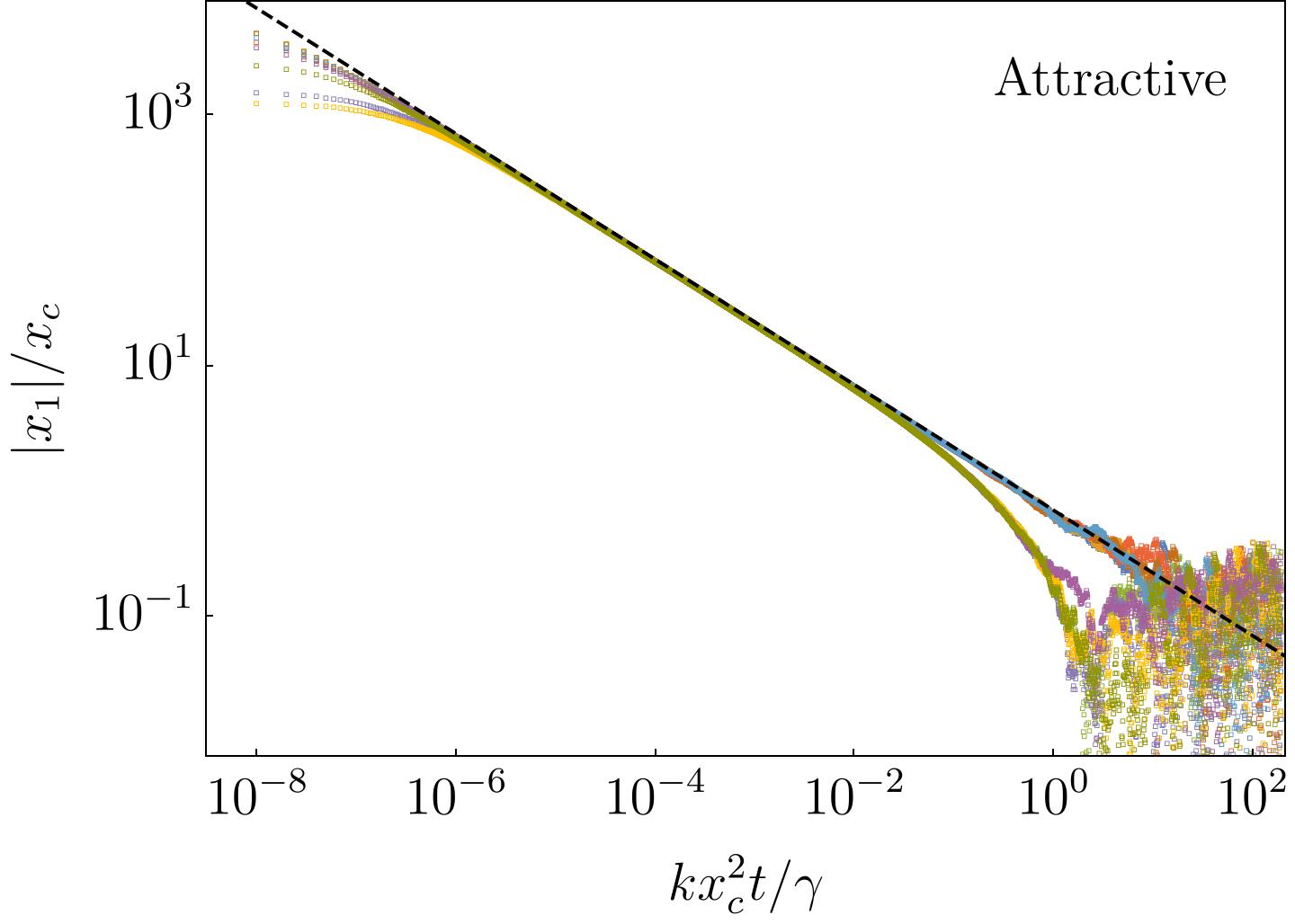}}
    \caption{Emergence of the LLNES for a system of two interacting particles. Specifically, we plot the modulus of the position of one of the particles, for the potential given in Eq.~\eqref{ch3_eq:interaction-potential} with $n=2.5$, as a function of time. The left and right panels correspond to repulsive ($c<0$) and attractive ($c>0$) interactions. In both panels, it is shown the relaxation from an initial equilibrium state with $k_BT_i/kx_c^4 = 10^{12}$ to a final equilibrium state with $k_BT_f/kx_c^4 = 10^{-2}$. The different coloured lines correspond to ten stochastic trajectories, corresponding to numerically integrating the Langevin equation with noise, with initial values sampled from the equilibrium distribution at $T_i$. The black, dashed line corresponds to the solution at the LLNES, given by Eq.~\eqref{ch3_eq:LLNES-explicit-short-times-v2}, which is independent of the value of $c$.}
        \label{ch3_fig:interactions}
    \end{figure}

First, we consider the repulsive case. Here, we restrict to $n < n_{\text{crit}} = 4$: otherwise, the interaction potential becomes so strong that the particles are no longer confined equilibrium ceases to exist for the repulsive case. For sufficiently high temperatures, the system initially does not feel the effect of the interactions and thus relaxes as in the non-interacting case. The repulsive interaction induces an additional characteristic length $x_c \equiv (|c|/k)^{1/(4-n)}$ that becomes significant only when the system approaches the bottom of the potential.

In the attractive case, the resulting potential is confining for all values of $n$. The effect of interactions now entails the emergence of a new characteristic timescale corresponding to \textit{sticking}: particles tend to move together, i.e. relax along the line $x_1=x_2$---note that this direction constitutes the predilect region over which probability accumulates, following the snowplough picture. On the one hand, if $n>n_{\text{crit}}=4$, the particles rapidly stick and afterwards relax together towards the LLNES. On the other hand, if $n<n_{\text{crit}}=4$, interactions are subdominant with respect to the confining potential. Thus, the relaxation is again similar to that of the non-interacting case initially, with the interaction becoming relevant only close to the bottom of the potential.
    
Figure~\ref{ch3_fig:interactions} shows the relaxation for one of the particles, when the system is quenched from a sufficiently high temperature. On the left and right panels, the repulsive and attractive cases are shown, respectively. As commented above, repulsive interactions introduce the new characteristic length $x_c$, related to that of  mechanical equilibrium of the system. In this sense, the role of repulsive interactions is similar to that of multistability, note the analogy with the time evolution displayed in the right panel of Fig.~\ref{ch3_fig:langevin-traj}. For attractive interactions, those trajectories for which the sign of the initial conditions $x_{1,i}$ and $x_{2,i}$ are equal continue together over the LLNES until noise becomes relevant: instead, those trajectories for which the sign of the initial conditions $x_{1,i}$ and $x_{2,i}$ are different deviate from it to switch to the other side of the potential and facilitate the sticking of the two particles. A part of this switching behaviour is still noiseless and resembles the extension to longer times of the LLNES in  Eq.~\eqref{ch3_eq:LLNES-explicit-specific} from Sec.~\ref{ch3_subsubsec:application-overdamped}.

\subsection{\label{ch3_subsec:molecular_fluid}Extension to Enskog-Fokker-Planck systems}
In Sec.~\ref{ch3_subsubsec:application-molecular}, we considered the case of a molecular fluid of Brownian particles in the colisionless limit, such that it suffices to employ the non-linear Fokker-Planck or Langevin frameworks for its description. In this last section, we show how the LLNES may emerge for systems additionally involving binary collisions---either elastic or inelastic, the framework of which differs from the one developed in previous sections. Specifically, we consider physical systems described at the kinetic level via the one-particle VDF $f(\bm{v},t)$, which obeys the Enskog-Fokker-Planck equation (EFP)~\eqref{ch1_eq:boltzmann-fokker-planck}, but with the terms $\bm{A}(\bm{v},t)$ and $\mathbb{D}(\bm{v},t)$ being drawn from the molecular fluid presented in Sec.~\ref{ch3_subsubsec:application-molecular}.

In the following, we resort to the scaling and dimensionless variables that we employed for the description of the molecular fluid with non-linear drag from chapter~\ref{ch:memory_effects}. Specifically, from Sec.~\ref{ch2_subsec:sonine}, we rescale the velocities $\bm{v}$ and the VDF with respect to the thermal velocity $v_{\text{T}}(t)$ following Eq.~\eqref{ch2_eq:scaling-velocities}, and we employ the dimensionless temperature $\theta$ and time as defined in Eq.~\eqref{ch2_eq:dimensionless-variables}. The resulting evolution equation for the scaled VDF $\phi(\bm{c},t)$ reads~\cite{santos_mpemba_2020,patron_strong_2021}
\begin{align}
\label{ch3_eq:suppl-evol-eq-non-lin-fluid-scaled}
    \partial_{t} \phi =&\nabla_{\bm{c}}^{\sf{T}} \left\{\left[ \frac{\dot{\theta}}{2\theta}\bm{c} + \zeta^*(\sqrt{\theta}c)\left(\bm{c}+\frac{1}{2\theta}\nabla_{\bm{c}}\right) \right]\phi\right\} + \xi^{-1} \theta^{1/2} J^*_{\alpha}[\bm{c}| \phi,\phi]. 
\end{align}
Let us remark that the above equation resembles Eq.~\eqref{ch2_eq:EFP-scaled} for the case of a quadratic non-linear drag. The term $J^*_{\alpha}[\phi,\phi]$ stands for the dimensionless collision operator---which reduces to the dimensionless Enskog operator $I[\bm{c}|\phi,\phi]$ from Eq.~\eqref{ch2_eq:collision-operator-scaled} in the elastic limit $\alpha = 1$, and $\xi$ constitutes the dimensionless parameter from Eq.~\eqref{ch2_eq:xi} accounting for the relative significance of collisions with respect to the non-linear drag. For $\theta \gg 1$, Eq.~\eqref{ch3_eq:suppl-evol-eq-non-lin-fluid-scaled} presents an asymptotic, noiseless solution for $n>1/2$, such that the collision term also becomes subdominant. Such asymptotic solution corresponds to the scaled VDF at the LLNES,
\begin{equation}\label{ch3_eq:LLNES-sol-2}
    \phi_{\LLNES}(\bm{c}) = \frac{1}{M_2^{\frac{d-1}{2}}\Omega_d}\delta (c - \sqrt{M_2}),
\end{equation}
with $M_{m} \equiv \langle c^m \rangle$ and $M_2 = d/2$ by definition. Note that the above resembles the LLNES solution from Eq.~\eqref{ch3_eq:LLNES-solution}, with the only difference being on the factor of $d/2$ stemming from the choice of the scaled variable. Now, if collisions are elastic, the evolution equation for the scaled temperature follows
\begin{equation}\label{ch3_eq:suppl-evol-eq-temp-non-lin-fluid}
    \dot{\theta}=-\frac{4}{d}\theta \left<c^2 \zeta^*(c\sqrt{\theta})\right>+\frac{2}{d}\left(\left< c\frac{\partial \zeta^*(c\sqrt{\theta})}{\partial c}\right>+d\left<\zeta^*(c\sqrt{\theta})\right>\right).
\end{equation}
Recall that, similarly to what occurred with Eq.~\eqref{ch2_eq:evolution-kin-temp}, this equation could have been obtained from Eq.~\eqref{ch1_eq:evol-eq-f}, since the kinetic energy constitutes a collisional invariant. In the $\theta \gg 1$ regime, i.e. when the system approaches the LLNES solution~\eqref{ch3_eq:LLNES-sol-2}, the moments $M_m$ become time independent and equal to $M_m^r \equiv (d/2)^{m/2}$, as our resulting VDF attains the Dirac-delta form. It is worth noting that the values Sonine cumulants---as introduced in Eq.~\eqref{ch2_eq:sonine-expansion} within the Sonine expansion---at the LLNES can also be determined
explicitly following Eq.~\eqref{ch3_eq:LLNES-sol-2}. See Appendix~\ref{app:sonine-cumulants-LLNES} for further details on this. Now, the dominant term in the evolution equation of the scaled temperature~\eqref{ch3_eq:suppl-evol-eq-temp-non-lin-fluid} is given by
\begin{equation}\label{ch3_eq:suppl-evol-eq-temp-non-lin-fluid-dom}
    \dot{\theta}\sim -\frac{4\gamma}{d} \, \theta^{\frac{n+2}{2}}  M_{n+2}^r \propto -\theta^{\frac{n+2}{2}},
\end{equation}
which entails a non-exponential relaxation behaviour for $n>0$. Note that inelastic collisions would introduce an additional term, proportional to $\theta^{3/2}$~\cite{poschel_granular_2001}, on the right hand side of Eq.~\eqref{ch3_eq:suppl-evol-eq-temp-non-lin-fluid}. Therefore, this term is negligible for $\theta\gg 1$ as long as $n>1$.

\chapter{Glass transition in stochastic and kinetic systems}
\chaptermark{Glass transition}
\label{ch:glass_transition}
\newcommand{\hcs}{\text{HCS}}
\newcommand{\Frz}{\text{Frz}}
\newcommand{\BL}{\text{(0)}}
\newcommand{\crit}{\text{crit}}

\newcolumntype{P}[1]{>{\centering\arraybackslash}p{#1}}

In previous chapters we have witnessed the emergence of a plethora of non-equilibrium phenomena for physical systems governed by stochastic and kinetic dynamics, when being submitted to a rapid quench to low enough values of the corresponding bath temperature. As discussed in Sec.~\ref{ch1_sec:glass}, such effects---non-exponential relaxation functions, memory effects, tendency towards global attractors---are hallmarks of what is known in the literature as glassy behaviour. Such behaviour entails a dramatic slow down of the relaxation dynamics of the system, in which configurational rearrangements cease, but the system remains at a non-equilibrium state known as the glassy state.

In this chapter we analyse the emergence of a kinetic glass transition---in the sense of the freezing of the kinetic temperature when continuously cooling the system---in two physical models that are radically different from a fundamental point of view: a uniformly driven granular gas~\cite{van_noije_velocity_1998,van_noije_randomly_1999,montanero_computer_2000,garcia_de_soria_energy_2009,garcia_de_soria_universal_2012,prados_kovacs-like_2014,garzo_transport_2013,maynar_fluctuating_2009} and the molecular fluid with non-linear drag considered in the previous chapters. Despite their profound differences---namely, the energy dissipation mechanism of the dynamics, these two systems tend towards non-equilibrium global attractors when being quenched to low enough temperatures: the HCS for the granular gas, and the LLNES for the molecular fluid. When attaining such states, their relaxation dynamics is fully accounted by the kinetic temperature, which relaxes algebraically (although differently) in both systems. Thus, they are enticing candidates for the emergence of a glass transition.

The structure of this chapter goes as follows. In Sec.~\ref{ch4_sec:two_models}, we present the two models of concern and the dynamical equations characterising their behaviour. Section~\ref{ch4_sec:mathematical_analysis} is devoted to providing the main physical intuition behind the emergence of the kinetic glass transition in both systems, and carrying out a detailed mathematical analysis thereof, by resorting to the tools of singular perturbation theory, specifically boundary layer theory. We study the glass transition by employing different cooling programs, where we investigate also the role of the non-equilibrium attractors---i.e. the LLNES and the HCS---in its emergence. Finally, in Sec.~\ref{ch4_subsec:hysteresis_cycles}, we apply our developed framework to analyse hysteresis cycles in both systems, which display universal features.  

\section{\label{ch4_sec:two_models}Two models: granular gas and molecular fluid}
In this first section, we introduce the dynamical equations accounting for the behaviour of the two physical systems considered in this chapter. We start by giving a thorough description of the uniformly driven granular gas, while for the molecular fluid, we rewrite the dynamical equations depicted in Sec.~\ref{ch2_subsec:sonine} in terms of new scaled variables.

\subsection{\label{ch4_subsec:granular_gas}Uniformly driven granular gas}
We start by introducing the uniformly driven granular gas. It consists of an ensemble of $d$-dimensional hard spheres of mass $m$ and diameter $\sigma$. These hard spheres undergo binary inelastic collisions, in which the tangential component of the relative velocity between two particles remains unaltered, while the normal component is reversed and shrunk by a factor $\alpha$. This parameter $\alpha$ is called the normal restitution coefficient, $0\le \alpha \le 1$, which accounts for the energy dissipation. Elastic collisions---in which the kinetic energy is conserved---are recovered for $\alpha=1$~\cite{poschel_granular_2001,garzo_granular_2019}. Note that without any external driving, the kinetic (or granular) temperature $T(t)$ decays towards zero in the long term. For the uniformly heated granular gas considered here, the system is supplemented with an energy-injection mechanism known as the stochastic thermostat~\cite{van_noije_velocity_1998,van_noije_randomly_1999}, which introduces additional white-noise forces acting on each particle independently. In this scenario, the system reaches a steady state in the long term, because the kinetic energy lost via inelastic collisions is balanced on average by the energy input from the stochastic thermostat.

The model considered here belongs to the class of models introduced in Sec.~\ref{ch1_subsubsec:collision}, such that its statistical behaviour is accurately described in terms of a Enskog-Fokker-Planck (EFP) equation~\eqref{ch1_eq:boltzmann-fokker-planck} for the one-particle VDF $f(\bm{v},t)$.\footnote{A more detailed account  of the kinetic theory of granular fluids described here can be found in the literature---e.g., see Refs.~\cite{van_noije_velocity_1998,van_noije_randomly_1999,montanero_computer_2000,garcia_de_soria_universal_2012,prados_kovacs-like_2014,trizac_memory_2014,garzo_granular_2019}.} Specifically, the corresponding Fokker-Planck coefficients $\bm{A}(\bm{v},t)$ and $\mathbb{D}(\bm{v},t)$ are given by
\begin{eqnarray}
    \bm{A}(\bm{v},\cancel{t}) = \bm{0}, \quad \mathbb{D}(\bm{v},t) = \frac{\tilde{\xi}(t)^2}{2} \mathbb{I}_d.
\end{eqnarray}
In the above, the stochastic thermostat comes into play via the noise, Fokker-Planck term $\mathbb{D}(\bm{v},t)$, with $\tilde{\xi}(t)$ corresponding to the intensity of the noise, which we assume to be externally controllable. The fact that $\bm{A}(\bm{v}) = \bm{0}$ is controversial, since in the elastic limit $\alpha \rightarrow 1$, binary collisions are elastic, and the energy is continuously increased without being balanced via the stochastic thermostat~\cite{marconi_about_2013}. Thus, our model applies as long as $\alpha \neq 1$, such that we may assume that the collision term $J_{\alpha}[\bm{v}|f,f]$ from Eq.~\eqref{ch1_eq:boltzmann} dominates the dissipative dynamics over the drift term $\bm{A}(\bm{v})$, and the latter becomes negligible.

Similarly to how we proceeded in chapter~\ref{ch:memory_effects}, upon the rescaling of the velocities with respect to the thermal velocity $v_{\text{T}}(t)$~\eqref{ch2_eq:scaling-velocities}, the scaled VDF $\phi(\bm{c},t)$ may be expanded in Laguerre---or Sonine---polynomials as in Eq.~\eqref{ch2_eq:sonine-expansion}. Hence, the behaviour of the system is completely characterised by the kinetic temperature $T(t)$ and the Sonine cumulants $a_l(t)$, $l=2,3,...$ Throughout this chapter, here for the granular gas---and later for the molecular fluid, we work under the first Sonine approximation. Therein, we recall that it is only needed to monitor the kinetic temperature $T(t)$ and the excess kurtosis $a_2$, also known as the first Sonine cumulant.

For the analysis that we carry out below, it is useful to introduce a characteristic length $\lambda$ and a characteristic frequency $\nu$ as
\begin{align}\label{ch4_subeq:coll-rate}
    \lambda=\frac{d  \pi^{\frac{1-d}{2}}}{2 n \sigma^{d-1} }\Gamma \left( \frac{d}{2}\right), \quad 
    \nu(T)=\left(1-\alpha^2\right)\lambda^{-1} \left(\frac{k_B T}{m}\right)^{1/2},
\end{align}  
where $\lambda$ gives the mean free path, i.e. the average distance travelled by one particle between collisions~\cite{poschel_granular_2001}, while $\nu(T)$ gives the cooling rate of the granular gas, i.e. the rate at which kinetic energy is dissipated in collisions---recall that the factor $1-\alpha^2$ also appeared for the variation of the kinetic energy after each inelastic collision, following Eq.~\eqref{ch1_eq:var-kin-en}.

In the absence of stochastic thermostat, the granular gas reaches the spatially-uniform hydrodynamic state known as the HCS, for which the scaled VDF $\phi(\bm{c},t)$ becomes stationary, although the original VDF $f(\bm{v},t)$ is non-steady. The dynamical evolution of the system is completely characterised by the granular temperature, which decays algebraically in time, $T(t) \propto t^{-2}$, following Haff's law~\cite{haff_grain_1983,brey_homogeneous_1996,poschel_granular_2001,garzo_granular_2019}. Under the first Sonine approximation, the stationary value of the excess kurtosis at the HCS is given by~\cite{santos_second_2009,van_noije_velocity_1998}
\begin{equation} 
    \label{ch4_eq:a2hcs}
    a_2^{\hcs}=\frac{16(1-\alpha)(1-2\alpha^2)}{25+2\alpha^2 (\alpha-1) +24 d+\alpha (8d-57)}.
\end{equation}

When the stochastic thermostat is present, the granular gas reaches a NESS in the long time limit. The kinetic temperature $T_{\st}$ at the NESS is given in terms of the stochastic strength $\tilde{\xi}$ via the relation~\cite{van_noije_velocity_1998}
\begin{equation}
    \frac{k_B T_{\st}}{m} = \left[\frac{\lambda\, \tilde{\xi}^2}{\left(1-\alpha^2\right)\left(1+\frac{3}{16}a_2^{\st}\right)}\right]^{2/3},
    \label{ch4_eq:Ts-xi}
\end{equation}
where $a_2^{\st}$ is the NESS value of the excess kurtosis~\cite{van_noije_velocity_1998,montanero_computer_2000},
\begin{equation} \label{ch4_eq:a2s}
  a_2^\st=\frac{16(1-\alpha)(1-2\alpha^2)}
{73+56d-24d\alpha-105\alpha+30(1-\alpha)\alpha^2}.
\end{equation}
Such value has the same sign as $a_2^{\hcs}$, attaining a null value at $\alpha = 1/\sqrt{2}$. In the following, we work under the dimensionless variables
\begin{equation}
	\label{ch4_eq:adim_var}
	\theta \equiv
	\frac{T}{T_i}  , \quad \theta_{\st} \equiv \frac{T_{\st}}{T_i}, \quad t^*\equiv \nu(T_i)t,
\end{equation}
with $T_i \equiv T (t = 0)$ being the initial temperature. In order to simplify our notation, we drop the asterisk in the dimensionless time from now on. Our definition of the dimensionless time is equivalent to taking the time unit equal to the characteristic relaxation time of the granular gas at the initial temperature $T_i$. Note that, with our choice of units, the initial value of the dimensionless temperature is always $\theta_i\equiv \theta(0)=1$. 

Now, from the EFP equation~\eqref{ch1_eq:boltzmann-fokker-planck}, the evolution equations for the scaled temperature and the excess kurtosis are derived~\cite{van_noije_velocity_1998,montanero_computer_2000,prados_kovacs-like_2014,trizac_memory_2014,brey_rheological_2012},
\begin{subequations}\label{ch4_eq:granularSonine}
\begin{align}
    \dot{\theta}&=  \theta_{\st}^{3/2} \left(1+\frac{3}{16} a_2^{\st}\right) - \theta^{3/2} \left(1+\frac{3}{16} a_2\right),
    \label{ch4_subeq:granularSonine-theta}\\
    \dot{a}_2&= 2\theta^{1/2} \left\{  \left[1-\left(\frac{\theta_{\st}}{\theta}\right)^{3/2} \right] a_2 +B \, (a_2^{\st} -a_2) \right\},
    \label{ch4_subeq:granularSonine-a2}
\end{align}
\end{subequations}
where we have introduced the parameter
\begin{align}\label{ch4_eq:B-param}
B &\equiv \frac{73+8d(7-3\alpha)+15\alpha[2\alpha(1-\alpha)-7]}
{16(1-\alpha)(3+2d+2\alpha^2)}.
\end{align}
Note that $B$ may be written in terms of $a_2^{\hcs}$ and $a_2^{\st}$, specifically $B = a_2^{\hcs}/(a_2^{\hcs}-a_2^{\st})$ as predicted by Eq.~\eqref{ch4_subeq:granularSonine-a2} for $\theta_{\st}=0$~\cite{prados_kovacs-like_2014,trizac_memory_2014}. We remark that, by setting $\theta_{\st}=0$ and $a_2 = a_2^{\text{HCS}}$ in Eq.~\eqref{ch4_subeq:granularSonine-theta}, we recover Haff's law $\theta(t) \propto t^{-2}$.

\subsection{\label{ch4_subsec:molecular_fluid}Molecular fluid with non-linear drag}

In Sec.~\ref{ch2_sec:model}, we extensively described the molecular-fluid with non-linear drag that we also employ in this chapter. Similarly to the previous model, in this case we also work under the first Sonine approximation, for which the dynamical equations characterising the system were given by Eq.~\eqref{ch2_eq:first-sonine-edos}. Now, by introducing the dimensionless variables
\begin{equation}
    \label{ch4_eq:dimensionless-variables-1}
    \theta \equiv
    \frac{T}{T_i}  , \quad \theta_{\st} \equiv \frac{T_{\st}}{T_i}, \quad 
    t^*\equiv \zeta_{0}(T_i) \; t,
\end{equation}
with $T_i \equiv T(t=0)$ being again the initial temperature, Eq.~\eqref{ch2_eq:first-sonine-edos} may be rewritten as
\begin{subequations}\label{ch4_eq:evol-eqs-first}
    \begin{align}
        \dot{\theta}=   \theta_{\st}^{1/2} \, \bigg[ & 2(\theta_{\st}-\theta) + 2\gamma (d+2)\, \theta 
        - 2\, \gamma \, (d+2)(1+a_2)\frac{\theta^2}{\theta_{\st}} \bigg],\label{ch4_subeq:T-evol}\\
        \dot{a}_2=  \theta_{\st}^{1/2} \; \biggl\{&  8\gamma \left(1-\frac{\theta}{\theta_{\st}}\right) 
         -\left[\frac{4\theta_{\st}}{\theta}-8\gamma+4\gamma(d+8)\frac{\theta}{\theta_{\st}} \right]a_2\biggr\}. \label{ch4_subeq:a2-evol} 
    \end{align}
\end{subequations}
In the above, we have also taken into account that $\zeta_0(T_{\st})=\zeta_0(T_i)\theta_{\st}^{1/2}$---which stems from the definition of the non-linear drag in Eq.~\eqref{ch2_eq:nonlinear-drag}. Let us remark that, as we are neglecting Brownian-Brownian collisions, $\zeta_0$ accounts for the relevant characteristic timescale for the dynamics. This fact is key for the qualitative arguments depicted in the following section.


\section{\label{ch4_sec:mathematical_analysis}Kinetic glass transition}
In this section we thoroughly study the emergence of a kinetic glass transition in the models introduced in the preceding section. Considering a uniform cooling program, we start by commenting on the physical arguments that account for the glass transition, namely, the freezing of the relaxation timescale of the dynamics for each system. Then we provide a more rigorous, mathematical account of it by employing the tools of boundary layer theory~\cite{bender_advanced_1999}. Lastly, we extend the applicability of our insights to a wider range of cooling programs beyond the linear one.
\subsection{\label{ch4_subsec:intuition_glass}Physics behind the emergence of the glass transition}
The possible emergence of a glass transition in a given system is deeply connected with the slowing down of its dynamics, typically as the bath temperature is continuously lowered. For the two systems considered, we assume that the bath temperature $T_{\st}$ is controllable---for the case of the granular gas, it is the intensity of the stochastic thermostat $\tilde{\xi}$ that is controllable. Specifically, we consider a time-dependent driving bath temperature $T_{\st}(t)$ that continuously decreases from its initial value to zero.

The systems are initially prepared in their corresponding stationary states at temperature $T_{\st}(0) = T_i$, thus the initial value of the dimensionless temperature is $\theta_i=\theta(t=0) = \theta_{\st}(t=0) = 1$. Therefrom, we apply a linear cooling program with rate $r_c$,
\begin{equation}
\label{ch4_eq:cooling-programme}
 \frac{d\theta_{\st}}{dt} = -r_c, \quad \theta_{\st}(t)=1-r_c\, t.
\end{equation}
The choice of a linear cooling program is done for the sake of concreteness, but a more general family of protocols is considered in Sec.~\ref{ch4_sec:different_cooling_programs}, which further broadens the scope of our analysis. The characteristic timescale for the cooling process in the linear program, $t_0\equiv r_c^{-1}$, corresponds to the time at which $\theta_{\st}$ vanishes, $\theta_{\st}(t=t_0)=0$. The cooling process is assumed to be slow, i.e. the characteristic cooling time $t_0$ is much longer than the characteristic relaxation time of the system at the initial state. Since the latter is of the order of unity in our dimensionless variables, slow cooling entails that $t_0\gg 1$ or $r_c\ll 1$. 

The emergence of a kinetic glass transition in these systems can be intuitively predicted by physical arguments. We start by considering the case of the granular gas, and then we extrapolate our arguments to the molecular fluid, due to their similarities. For time-independent strength of the stochastic thermostat $\tilde{\xi}$, the granular gas relaxes to the steady state characterized by the ``bath" temperature $\theta_{\st}$ and the stationary excess kurtosis $a_2^{\st}$. The characteristic timescale for this relaxation is proportional to $\nu(T_{\st})^{-1}$ from Eq.~\eqref{ch4_subeq:coll-rate}. Specifically, it is given by 
\begin{equation}\label{ch4_eq:tau-def}
    \tau(\theta_{\st})= \frac{1}{K}\theta_{\st}^{-1/2},
\end{equation}
where $K$ is a constant of the order of unity~\cite{sanchez-rey_linear_2021}---below, we provide its detailed calculation---which is approximately equal to $3/2$. In the low bath temperature limit, $\tau$ algebraically diverges as $\theta_{\st}^{-1/2}$  and, despite our slow cooling, the characteristic timescale for relaxation eventually becomes much longer than the cooling time. Therefore, we expect the system to depart from the stationary curve and get frozen---i.e. a kinetic glass transition shows up.

In order to approximately quantify the above qualitative argument, we may introduce a dimensionless effective timescale
\begin{equation}\label{ch4_eq:s-scale-def}
    s=\int_{t}^{t_0}\frac{dt'}{\tau(\theta_{\st}(t'))}=\frac{1}{r_c}\int_{0}^{\theta_{\st}}\frac{d\theta_{\st}'}{\tau(\theta_{\st}')},
\end{equation}
which measures the number of effective relaxation times remaining from the current time $t$ to the final time of the cooling process $t_0$. Specifically, $s$ as a function of $t$ constitutes a monotonic decreasing function, which starts from $s(t=0)$ and decays zero. Now, as long as $s\gg 1$, we expect the system to be able to follow the instantaneous NESS curve $\theta=\theta_{\st}$ as long as we consider sufficiently slow cooling program. Note that our defining of the timescale $s$ also allows for a more precise definition of slow cooling: a cooling program is said to be slow if $s(t=0)\gg 1$, i.e. the system has time to reach the instantaneous NESS curve before becoming frozen, independently of the initial preparation. When $s$ becomes of the order of unity, the system does not have enough time to relax towards the instantaneous NESS curve and thus it freezes. Following the usual terminology of glassy systems~\cite{scherer_relaxation_1986}, we may introduce a fictive kinetic temperature, as the bath temperature at which the NESS kinetic temperature equals the frozen value. 

The above physical picture implies that we can estimate the fictive temperature $\theta_f$ by imposing
\begin{equation}
    s(\theta_{\st}=\theta_f)=1,
\end{equation}
i.e.
\begin{equation}
    \theta^{\Frz}\equiv \lim_{\theta_{\st}\to 0} \theta \simeq \theta_f.
\end{equation}
Bringing to bear Eqs.~\eqref{ch4_eq:cooling-programme} and \eqref{ch4_eq:tau-def},
\begin{equation}\label{ch4_eq:s-scale-v2}
    s=\frac{K}{r_c}\int_{0}^{\theta_{\st}}d\theta_{\st}' \, \sqrt{\theta_{\st}'}=\frac{2K}{3} \frac{\theta_{\st}^{3/2}}{r_c}.
\end{equation}
Then, the fictive temperature and the kinetic temperature at the frozen state are estimated as
\begin{equation}\label{ch4_eq:thetaf-power-law}
    \theta_f =\left(\frac{3r_c}{2K}\right)^{2/3} , \quad \theta_{\Frz}\propto r_c^{2/3}.
\end{equation}

In summary, the slowing down of the dynamics of the granular gas, due to the algebraic divergence of the relaxation time in Eq.~\eqref{ch4_eq:tau-def} entails that the granular gas is expected to depart from the instantaneous NESS curve $\theta=\theta_{\st}$ as the the intensity of the stochastic thermostat is continuously decreased to zero. Hence, a kinetic glass transition is expected to appear in this system when cooled down to low bath temperatures, and our timescale argument suggests that the system would get frozen for bath temperatures $\theta_{\st}\lesssim \theta_f$, where $\theta_f$ follows the power law~\eqref{ch4_eq:thetaf-power-law} with the cooling rate. Moreover, the kinetic temperature at the frozen state is expected to be approximately equal to $\theta_f$, thus following the same power law with the cooling rate. The correctness of this physical picture is supported by the detailed mathematical theory that is developed in the next sections.

Concerning the molecular fluid, the physical reason for the emergence of a kinetic glass transition in this system as well is completely analogous to that for the granular gas. In the molecular fluid with non-linear drag, the characteristic timescale for relaxation is determined by $\zeta_0^{-1}$, which also diverges as $T_{\st}^{-1/2}$ for low bath temperatures. Therefore, we expect the same scalings with the cooling rate as in the granular gas.

\subsubsection{\label{ch4_subsubsec:relaxation_time_granular}Relaxation time for the granular gas}
In Ref.~\cite{sanchez-rey_linear_2021}, the linear relaxation to the NESS of the uniformly heated granular gas was investigated in detail. The relaxation of the system from the NESS corresponding to a value of the driving intensity $\tilde{\xi}+\delta \tilde{\xi}$ to the NESS corresponding to a driving intensity $\tilde{\xi}$ was considered---i.e. in the so-called linear response regime.

Therefore, the granular (or kinetic) temperature evolves from the initial value $\theta_{\st}+\delta\theta_{\st}$ to the final value $\theta_{\st}$. The relaxation function for the granular temperature can be defined as,
\begin{equation}
    \phi_{\theta}(t)\equiv \frac{\theta(t)-\theta_{\st}}{\delta\theta_{\st}},
\end{equation}
which is normalised in the usual way, $\phi_{\theta}(t=0)=1$. Let us remark the resemblance with the relaxation function defined in Eq.~\eqref{ch2_eq:relaxation-function}. Specifically, this relaxation function was shown to have the form
\begin{equation}
    \phi_{\theta}(t)=a_+ e^{-\lambda_+ t}+a_- e^{-\lambda_{-} t},
\end{equation}
where, due to the normalisation, $a_-=1-a_+$. The above was obtained from linearising and solving the corresponding evolution equations given the perturbation $\tilde{\xi} \rightarrow \tilde{\xi} + \delta \tilde{\xi}$. Both $\lambda_+$ and $\lambda_+$ are proportional to $\theta_{\st}^{1/2}$
\begin{equation}
\lambda_{\pm}=K_{\pm}\theta_{\st}^{1/2},
\end{equation}
with
\begin{subequations}
  \begin{align}
    K_+&=\frac{3}{2}+\frac{9}{32}\frac{1+4B}{4B-3}a_2^{\st}+O(a_2^{\st})^2, \\
    K_-&=2B-\frac{9}{4(4B-3)}a_2^{\st}+O(a_2^{\st})^2,
\end{align}  
\end{subequations}
and 
\begin{equation}
    a_+=1+\frac{9}{4(4B-3)}a_2^{\st}+O(a_2^{\st})^2.
\end{equation}

The characteristic relaxation time of the relaxation is thus given by
\begin{equation}
    \tau \equiv \int_0^{\infty} dt\, \phi_{\theta}(t)=\frac{a_+}{\lambda_+}+\frac{a_-}{\lambda_{-}} .
\end{equation}
Since both $\lambda_+$ and $\lambda_-$ are proportional to $\theta_{\st}^{1/2}$, and both $a_+$ and $a_-$ are independent of the bath temperature, we have the scaling in Eq.~\eqref{ch4_eq:tau-def}, with the parameter $K$ being given by
\begin{equation}
     \frac{1}{K}=\frac{a_+}{K_+}+\frac{a_-}{K_-}.
\end{equation}
In practice, it is straightforward to check numerically that $a_{+} \approx 1 \gg a_{-}$ regardless of the value of $\alpha$, such that $K\approx K_{+} \approx 3/2$, which provides the value $\theta_f = r_c^{2/3}$ for the fictive temperature. Thus, we may state that the fictive temperature is roughly independent of the restitution coefficient.

\subsection{\label{ch4_subsec:perturbative}Regular perturbative expansion and the frozen state}
The fact that the overall behaviour of the system radically changes upon uniformly cooling it down suggests that the tools from singular perturbation theory are useful to tackle the problem analytically. Our analysis below shows that indeed different regions emerge---in terms of different behaviours in the $(\theta,\theta_{\st})$ space, which we will label with the terminology of Ref.~\cite{bender_advanced_1999} for boundary layer problems. First, one has the \textit{outer layer}, inside which the kinetic temperature $\theta$ does not deviate much from the bath temperature $\theta_{\st}$, and a regular perturbation expansion is adequate. Second, one has the \textit{inner layer}, for which the regular perturbation expansion breaks down and it is necessary to rescale the variables to obtain an approximate solution. It is in the inner layer that the kinetic temperature $\theta$ rapidly separates from $\theta_{\st}$ and gets frozen at $\theta^{\Frz}$. Finally, one has the \textit{matching region}, over which the solution continuously changes from the inner to the outer solution.

The generic framework described above is applied to the evolution equations for both the granular gas and the molecular fluid with non-linear drag. As we discuss below, many of the results that we obtain are similar for both models, such that, in order to avoid redundancy, some of the main figures that we plot correspond exclusively to one model---mainly the granular gas.

We are interested in studying the behaviour of the kinetic temperature $\theta$ in terms of the bath temperature $\theta_{\st}$. Therefore, following Eq.~\eqref{ch4_eq:cooling-programme}, we rewrite the evolution equations for both $\theta$ and $a_2$ in terms of derivatives with respect to $\theta_{\st}$,
\begin{align}
    \label{ch4_eq:cooling-program2}
       \frac{d\theta}{dt}= -r_c \frac{d\theta}{d\theta_{\st}}, \quad 
     \frac{da_2}{dt} = -r_c \frac{da_2}{d\theta_{\st}},
     \end{align}
with $r_c \ll 1$. In order to find an approximate solution of the evolution equations for both systems, we introduce the regular perturbation series in powers of the cooling rate,
\begin{subequations}
    \label{ch4_eq:pert-gran}
\begin{align}
    \theta(\theta_{\st}) &= \theta^{(0)}(\theta_{\st}) + r_c \; \theta^{(1)}(\theta_{\st}) + O(r_c^2), \\
    a_2(\theta_{\st}) &= a_2^{(0)}(\theta_{\st}) + r_c \; a_2^{(1)}(\theta_{\st}) + O(r_c^2).
\end{align}
\end{subequations}
Equations~\eqref{ch4_eq:pert-gran} are then inserted into Eqs.~\eqref{ch4_eq:granularSonine} and \eqref{ch4_eq:evol-eqs-first}, for the granular gas and the molecular fluid, respectively---in which we subsequently equal the terms with the same power of $r_c$ and solve for $\{\theta^{(k)},a_2^{(k)}\}$, $k=0,1,\ldots$, where we have to take into account the initial conditions
\begin{equation}
    \theta(1)=1, \quad a_2(1)=a_{2,i}.
    \label{ch4_eq:initial_conditions}
\end{equation}
with $a_{2,i}$ being $a_2^{\st}$ for the granular gas, and zero for the molecular fluid. Now, at lowest order $O(r_c^0) = O(1)$, i.e. those independent of $r_c$, for both systems we have that
\begin{equation}\label{ch4_eq:zeroth-order-sols}
    \theta^{(0)}(\theta_{\st}) = \theta_{\st}, \quad a_2^{(0)}(\theta_{\st}) = a_{2,i},
\end{equation}
The above constitutes the solution we would obtain for $r_c = 0$, i.e. when the system remains at the instantaneous stationary state. At the first order $O(r_c)$, i.e. linear terms in $r_c$, we have
\begin{subequations}\label{ch4_eq:first-order-sols-granular}
    \begin{align}
        \label{ch4_eq:first-order-sols-granular-temp}
        \theta^{(1)}(\theta_{\st}) &= \frac{2}{3\theta_{\st}^{1/2}} \left[1+\frac{3}{16} \; a_2^{\st} \left( 1+\frac{1}{B} \right) \right]^{-1},
        \\
        \label{ch4_eq:first-order-sols-granular-a2}
        a_2 ^{(1)}(\theta_{\st}) &= -\frac{a_2^{\st}}{B\theta_{\st}^{3/2}} \left[1+\frac{3}{16} \; a_2^{\st} \left( 1+\frac{1}{B} \right) \right]^{-1},
\end{align}
\end{subequations}
for the granular gas, and
\begin{subequations}\label{ch4_eq:first-order-sols-molecular}
    \begin{align}
        \theta^{(1)}(\theta_{\st}) &= \frac{1}{2\theta_{\st}^{1/2}}\frac{1+\gamma (d+6)}{\left[1 + \gamma (d+4)\right]^2 - 2\gamma^2 (d+4)}, \\ a_2^{(1)}(\theta_{\st}) &= -\frac{1}{\theta_{\st}^{3/2}}\frac{\gamma}{\left[1 + \gamma (d+4)\right]^2 - 2\gamma^2 (d+4)},
\end{align}
\end{subequations}
for the molecular fluid. The expansions from Eqs.~\eqref{ch4_eq:first-order-sols-granular} and \eqref{ch4_eq:first-order-sols-molecular} break down for low bath temperatures $\theta_{\st}\ll 1$, for which the terms proportional to $r_c$ (i) first become of the order of the leading, independent of $r_c$, contributions and (ii) later diverge in the limit as $\theta_{\st}\to 0$. In particular, (i) implies that both of expansions cease to be valid for $\theta_{\st} = O(r_c^{2/3})$, when the $O(r_c)$ terms are comparable to the $O(1)$ ones. In other words, our perturbative approach is thus limited to high enough bath temperatures, $\theta_{\st}\gg r_c^{2/3}$. 

Note that $r_c^{2/3}$ is precisely the dependence on $r_c$ of the fictive temperature $\theta_f$ from Eq.~\eqref{ch4_eq:thetaf-power-law} by qualitative physical arguments. From a physical standpoint, this marks the beginning of the kinetic glass transition: we expect the system to become ``frozen'' as soon as $\theta_{\st}=O(r_c^{2/3})$. In this region of temperatures, the terms in the perturbative expansions for $\theta$ are proportional to $r_c^{2/3}$, and $a_2$ are independent of $r_c$. Therefore, we expect that
\begin{subequations}\label{ch4_eq:scaling-outer-v1}
 \begin{align}
     \theta^{\Frz}&\equiv \lim_{\theta_{\st}\to 0} \theta \propto r_c^{2/3} ;
    \label{ch4_subeq:freezescaling} \\
    \label{ch4_subeq:scaling-2}
    a_2^{\Frz} &\equiv \lim_{\theta_{\st}\to 0} a_2 = O(1).  
 \end{align} 
\end{subequations}
The latter hints at the fact that all the cumulants of the Sonine expansion become independent of the cooling rate, i.e., the frozen state of the system is unique---something that is numerically confirmed later on. The scalings proposed in Eq.~\eqref{ch4_eq:scaling-outer-v1}, which arise from the presented qualitative arguments, are  quantitatively justified by the boundary layer approach carried out in the following section.

Let us now compare our analytical predictions  with simulation results. On the one hand, for the granular gas, these are obtained from Direct Simulation Monte Carlo (DSMC) integration~\cite{bird_g_a_molecular_1994} of the Boltzmann equation~\eqref{ch1_eq:boltzmann-fokker-planck} governing the dynamics. Unless otherwise specified, for all the simulations of the granular gas performed,  we have employed the system parameters $d=3$, a number of particles of $N=10^5$, and two different values of the restitution coefficient: $\alpha=0.9$ and $\alpha=0.3$, in order to test the robustness of our theoretical approach. On the other hand, for the molecular fluid, we have solved the corresponding Langevin equation for an ensemble of $N=10^5$ stochastic trajectories with model parameters $d=3$ and $\gamma = 0.1$. Further details concerning the numerical simulations are relegated to Appendix~\ref{app:simulation-methods}.

\begin{figure}
    {\centering 
    \includegraphics[width=2.65in]{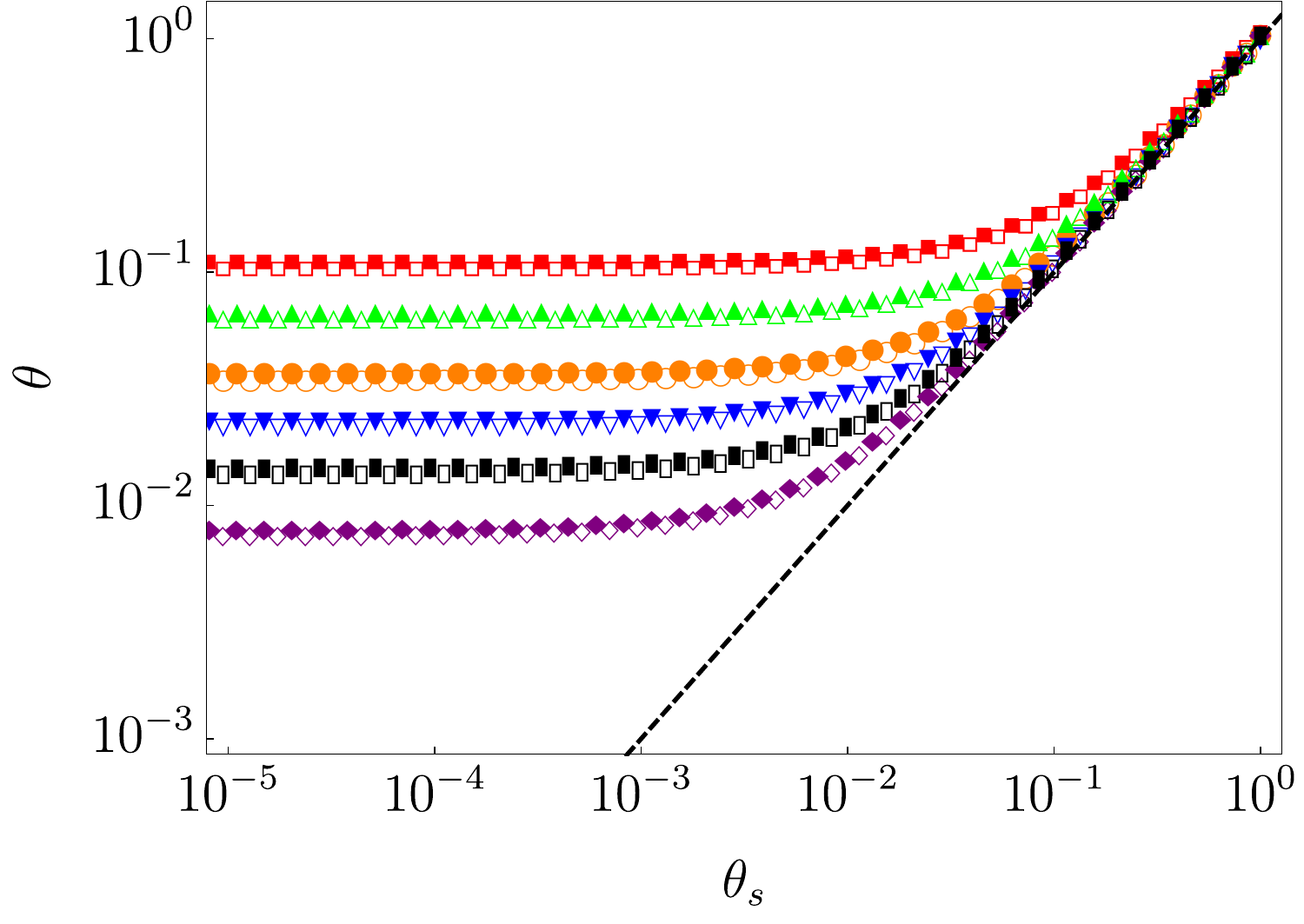}
    \includegraphics[width=2.65in]{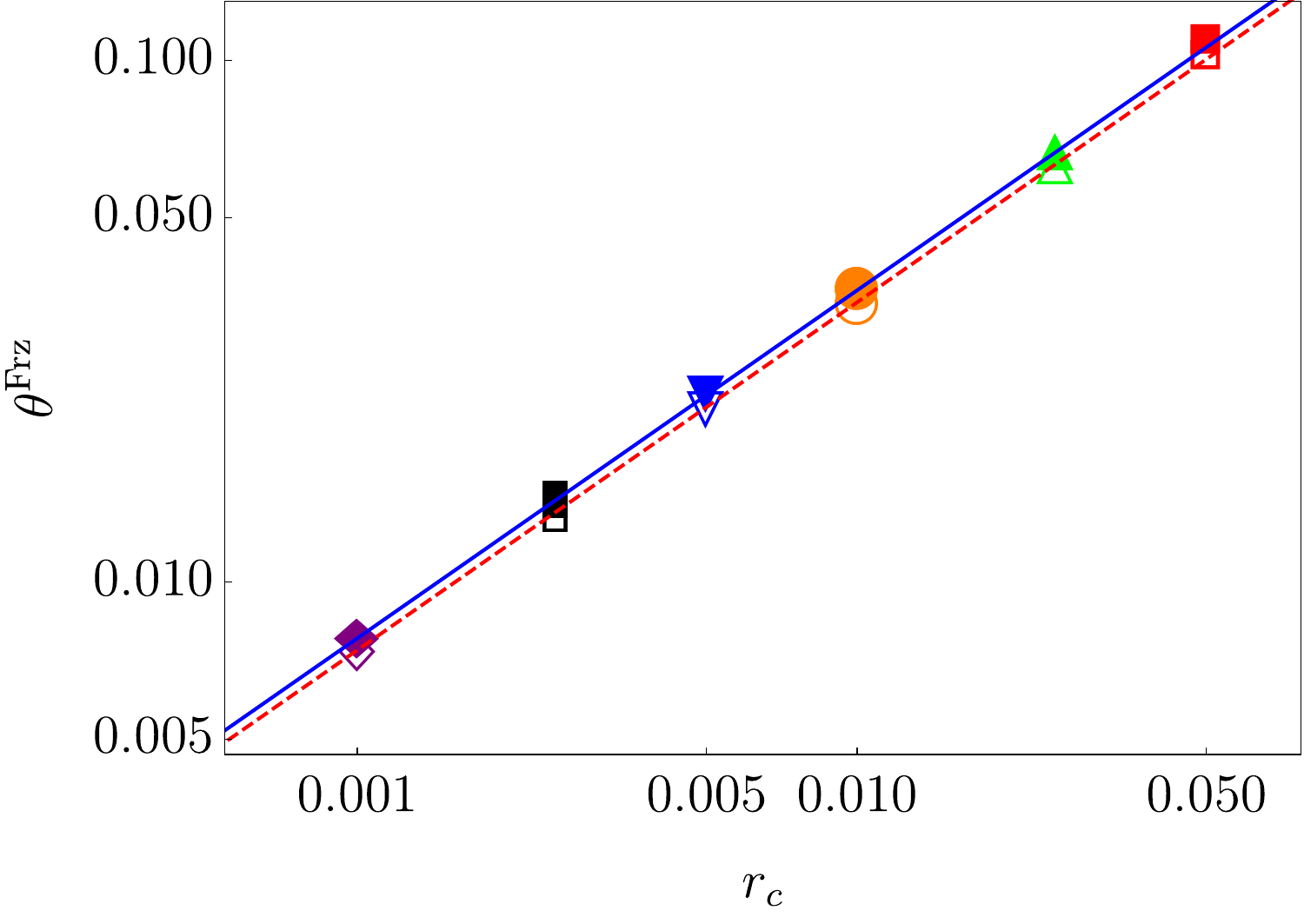}}
    \caption{Left: Time evolution of the dimensionless kinetic (or granular) temperature $\theta$ as a function of the bath temperature $\theta_{\st}$ for the granular gas. Symbols correspond to DSMC  data for the linear cooling protocol \eqref{ch4_eq:cooling-programme} with different cooling rates $r_c$, namely: $r_c$= 0.05 (red squares), $0.025$ (green up triangles), $0.01$ (orange circles), $0.005$ (blue down triangles), $0.0025$ (black rectangles), and $0.001$ (purple diamonds) for two values of the restitution coefficient: $\alpha = 0.9$ (empty symbols)  and  $\alpha = 0.3$ (filled symbols). The dashed line corresponds to the instantaneous NESS curve $\theta = \theta_{\st}$. Right: Limit values of the kinetic temperature at the frozen state {$\theta^{\Frz}$} as a function of {$r_c$.} The plotted points have been extracted from the DSMC data on the left panel. { The lines correspond to the best fits to the function ${\theta^{\Frz}=}a\, r_c^{b}$,} both in excellent agreement with the theoretical prediction \eqref{ch4_subeq:freezescaling}. We have considered a granular gas in the three-dimensional case $d = 3$. The same parameter values are employed in the remainder of the numerical simulations for the granular gas.}
        \label{ch4_fig:frozen-temp-gran}
\end{figure}

Figure~\ref{ch4_fig:frozen-temp-gran} presents numerical results for the linear cooling program~\eqref{ch4_eq:cooling-programme} for the granular gas. On the left panel, we plot the behaviour of the kinetic temperature $\theta$ as a function of the bath temperature $\theta_{\st}$, for different cooling rates. The emergence of a kinetic glass transition is clearly observed in this system. On the right panel, the final granular temperatures $\theta^{\Frz}$ at the frozen state are plotted as a function of the cooling rate $r_c$. They are very well fitted---following the R-squared method on the log-log scale---by the power law  $\theta^{\Frz}=a\, r_c^{b}$ with $a = 0.741$ and $b = 0.666$ for $\alpha = 0.9$ (dashed), and $a = 0.781$ and $b = 0.666$ for $\alpha = 0.3$ (solid). In both cases, the numerical value $b = 0.666$ confirms the scaling {predicted} by our theory in Eq.~\eqref{ch4_subeq:freezescaling}. 

\begin{figure}
    {\centering 
    \includegraphics[width=2.65in]{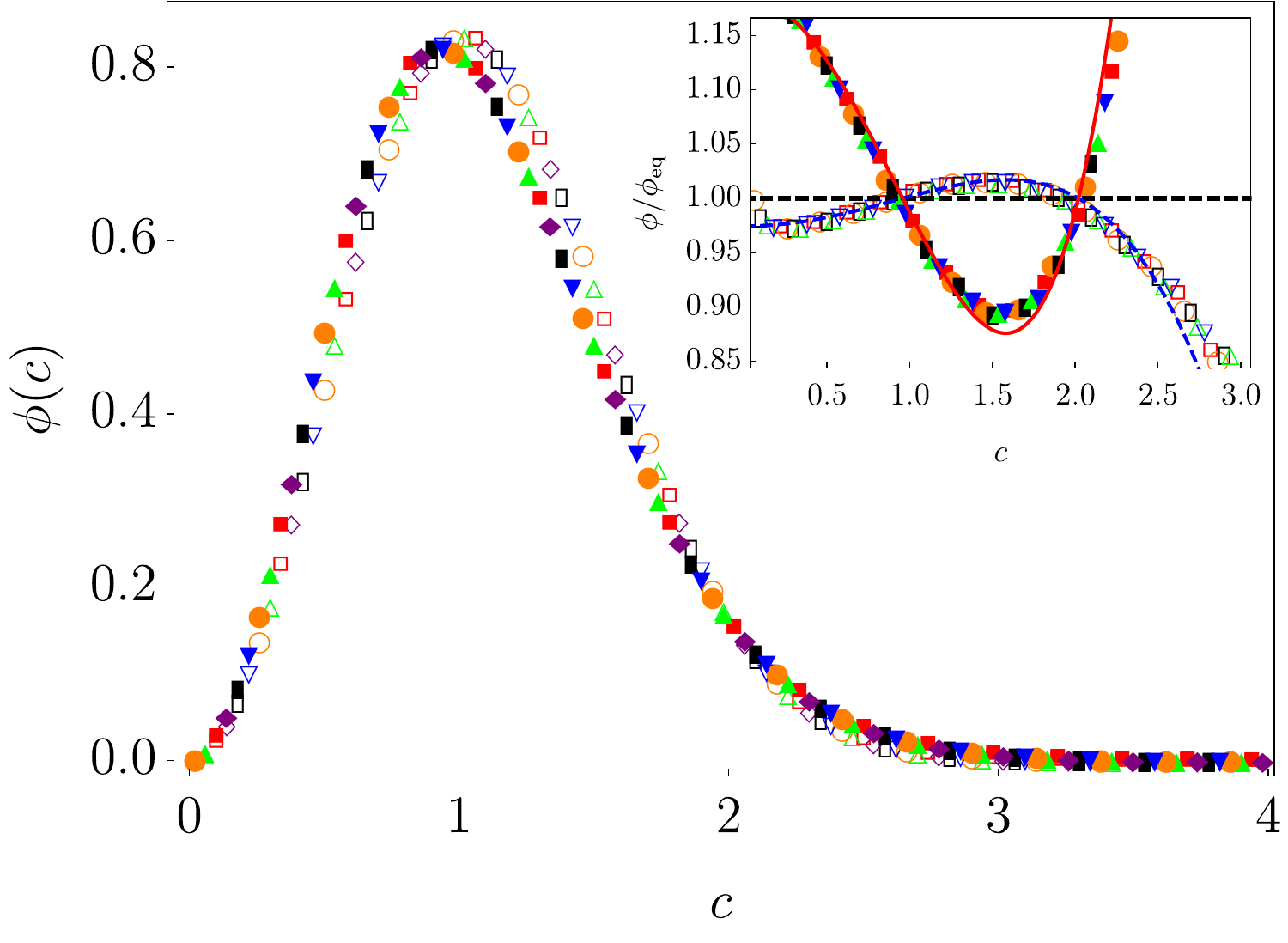}
    \includegraphics[width=2.65in]{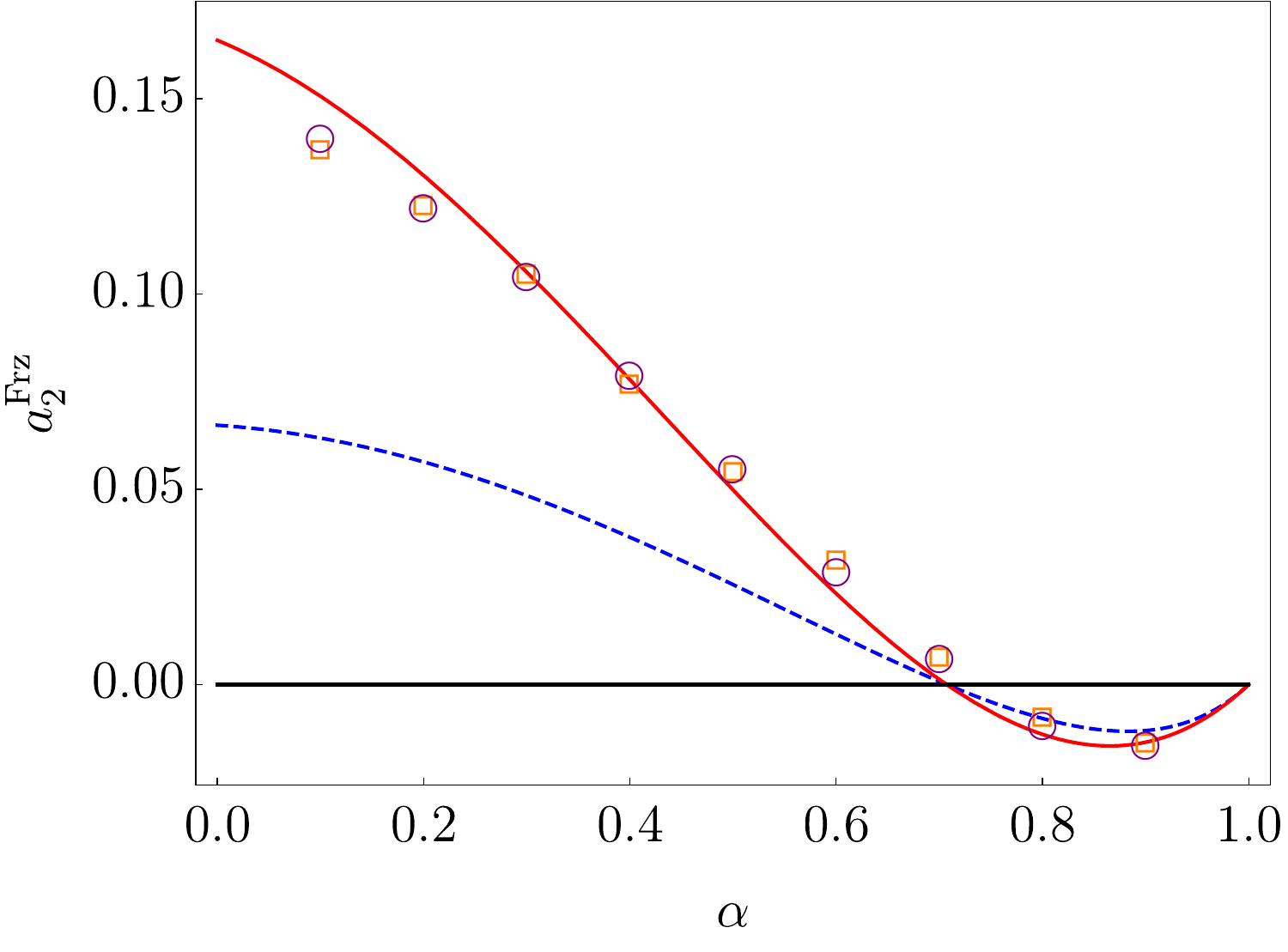}}
    \caption{Universality of the frozen state. Left: Radial VDF at the frozen state for different values of the cooling rate $r_c$. The colour code and symbols are the same as in Fig.~\ref{ch4_fig:frozen-temp-gran}. For each value of $\alpha$, its corresponding VDFs  are superimposed over a unique, universal, curve independent of $r_c$, in agreement with our theoretical prediction. In the inset, we show the radial VDF at the frozen state divided by the equilibrium Maxwellian, with the lines corresponding to the polynomials in Eq.~\eqref{ch2_eq:sonine-expansion} within the first Sonine approximation for  $\alpha=0.9$ (blue dashed) and  $\alpha=0.3$ (red solid), respectively---with $a_2^{\text{Frz}}$ being computed from the VDF. Right: Excess kurtosis at the frozen state $a_2^{\Frz}$ as a function of the restitution coefficient $\alpha$. Here, for the sake of clarity, we show DSMC data corresponding to only two values of the cooling rate, $r_c=0.01$ (squares) and $r_c=0.001$ (circles). The numerical values {of $a_2^{\Frz}$} are compared with both the NESS value $a_2^{\st}$ (blue dashed line) and the HCS value $a_2^{\text{HCS}}$ (red solid line), being very close to the latter.}
        \label{ch4_fig:VDF-graph}
\end{figure}

Note that, although our theoretical argument for the universality of the frozen state above has been carried out within the first Sonine approximation, the numerical results show that this remarkable property holds for the exact (numerical) VDF. In Fig.~\ref{ch4_fig:VDF-graph}, we numerically prove that the frozen state is indeed unique, still for the granular gas. On the left panel, the dimensionless, radial VDFs at the frozen state corresponding to different cooling rates overlap on a universal curve. To neatly visualize the non-Gaussian character of the frozen state, we present (i) the ratio of the radial VDFs over the equilibrium Maxwellian in the inset of the left panel and (ii) the excess kurtosis at the frozen state $a_2^{\Frz}$ as a function of the restitution coefficient $\alpha$ on the right panel. In order to improve the statistics, we have employed a larger system with $N=10^6$ particles for both the inset and the right panel. The inset allows us to illustrate in a neater way the differences between the frozen states corresponding to $\alpha = 0.9$ and $\alpha = 0.3$, since their respective excess kurtosis have opposite signs---Eq.~\eqref{ch2_eq:sonine-expansion} tells us that the plotted ratio is basically $1+a_2^{\text{Frz}} L_2^{1/2}(c^2)$, where $a_2^{\text{Frz}}$ has being computed from the actual VDF. The right panel allows us to check that $a_2^{\Frz}$---and thus the VDF---is indeed independent of $r_c$ for all inelasticities. In addition, this graph shows that $a_{2}^{\Frz}$ is really far from the steady-state kurtosis $a_2^{\st}$ but very close to the HCS values $a_2^{\hcs}$,  which suggests that the HCS has a key role in the frozen state---as further discussed in Sec.~\ref{ch4_sec:different_cooling_programs}.

\begin{figure}
    \centering
    \includegraphics[width=3.5in]{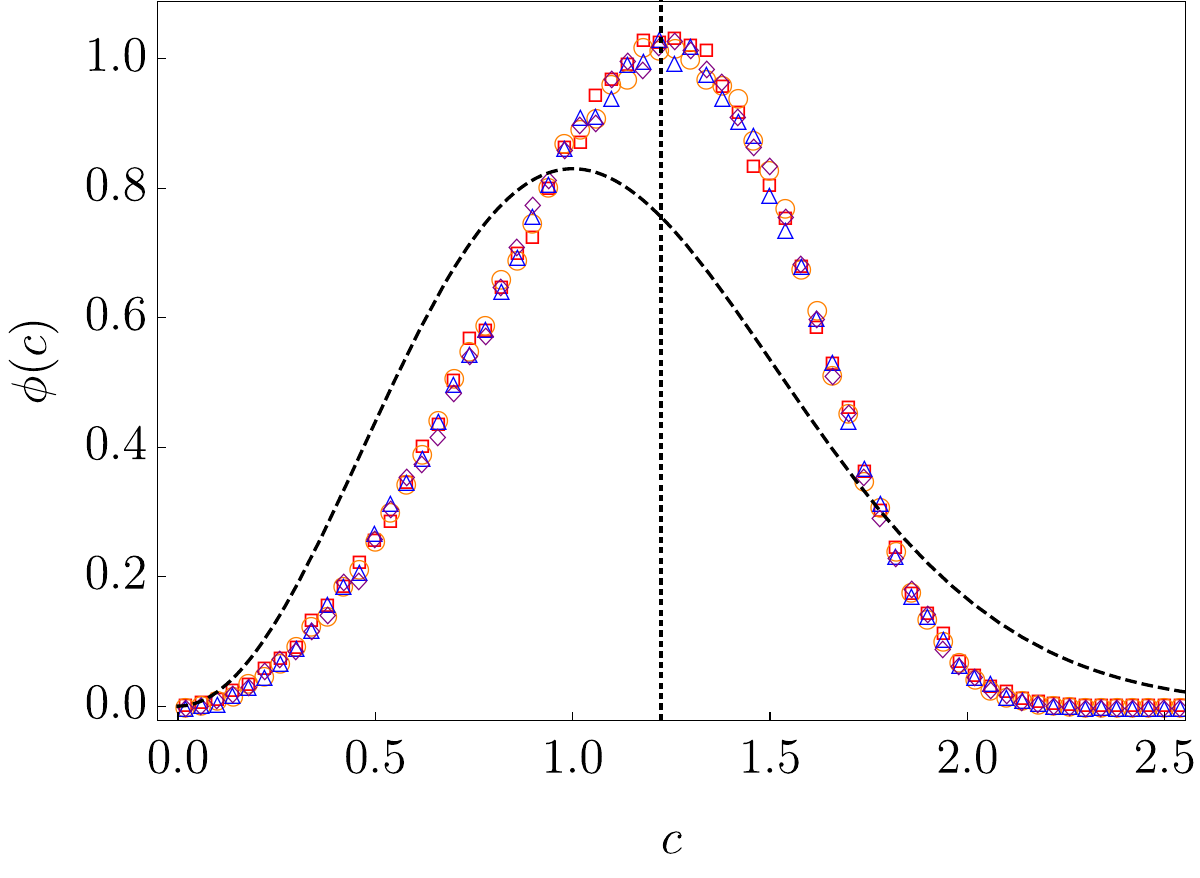}
    \caption{\label{ch4_fig:frozen-temp-molec} Plot of the dimensionless VDF at the frozen state for the non-linear fluid. Symbols correspond to the numerical integration of the Langevin equation {with} $N= 10^5$ stochastic trajectories for different cooling rates: $r_c= 0.005$ (purple diamonds), $0.01$ (blue triangles), $0.05$ (orange circles) and $0.1$ (red squares).  The dashed curve corresponds to the equilibrium Maxwellian, whereas the dotted line stands for the position of the {LLNES} Dirac-delta peak, as given by Eq.~\eqref{ch3_eq:LLNES-sol-2}. Other parameters are $\gamma = 0.1$, $d=3$.
      }
\end{figure}

Lastly, Figure~\ref{ch4_fig:frozen-temp-molec} shows that the frozen state is unique for the molecular gas as well. In particular, we plot the scaled radial VDF for different values of the cooling rate $r_c$, having initially started from the equilibrium state, characterised by the scaled Maxwellian distribution $\phi_{\text{eq}}(c) \propto c^{d-1} e^{-c^2}$ for $d=3$. Additionally, we have plotted the Dirac-delta peak corresponding to the LLNES, as given by Eq.~\eqref{ch3_eq:LLNES-sol-2}, as we expect it to play a role in the emergence of glass transition since $\theta / \theta_{st}\gg 1$ when the kinetic temperature freezes. Contrary to the case for the granular gas, the frozen VDF differs significantly from that of the LLNES. However, it seems that it constitutes an intermediate state, as it clearly deviates from the Maxwellian distribution and it narrows towards the position of the Dirac-delta peak. We further discuss this point in Sec.~\ref{ch4_sec:different_cooling_programs}.

\subsection{\label{ch4_subsec:boundary_layer}Boundary layer approach. Universality}
We are now concerned with the behaviour of both systems for very low bath temperatures, when they are close to their respective frozen states. 
Following the general theory for boundary layer problems~\cite{bender_advanced_1999}, a distinguished limit of the evolution equations in each case is found below by introducing suitably scaled variables for low bath temperatures. In this way, we find an inner expansion, which is afterwards matched with the outer solution derived before in Sec.~\ref{ch4_subsec:perturbative}. Therefrom, an (approximate) solution for all values of the bath temperature, known as a uniform solution~\cite{bender_advanced_1999}, is built.
To start with our boundary layer approach, we define the scaled temperatures
\begin{equation}\label{ch4_eq:scaling-cooling}
	Y \equiv  r_c^{-2/3}\theta, \quad X \equiv r_c^{-2/3}\theta_{\st} ,
\end{equation} 
as suggested by Eq.~\eqref{ch4_eq:scaling-outer-v1}. Interestingly, the evolution equations for both systems become independent of the cooling rate upon the above scaling. On the one hand, for the granular gas, inserting Eqs.~\eqref{ch4_eq:cooling-program2} and \eqref{ch4_eq:scaling-cooling} into the dynamical system \eqref{ch4_eq:first-order-sols-granular} gives
\begin{subequations}\label{ch4_eq:evolscaled-granular}
    \begin{align}
        -\frac{dY}{dX}&=  X^{3/2} \left(1+\frac{3}{16} a_2^{\st}\right) -Y^{3/2} \left(1+\frac{3}{16} a_2\right), \label{ch4_subeq:evolscaled-Y} \\
        -\frac{da_2}{dX}&= 2\, Y^{1/2} \left[ \left(1-\frac{X^{3/2}}{Y^{3/2}}\right) a_2 +B \, (a_2^{\st} -a_2) \right], 
    \end{align}
    \end{subequations}
while on the other hand, the corresponding changes onto the dynamical system~\eqref{ch4_eq:first-order-sols-molecular} for the molecular fluid give
\begin{subequations}
    \label{ch4_eq:ODESinnerc}
    \begin{align}
    -\frac{dY}{dX} =  X^{1/2} \, \bigg\{ 2\, ( X-Y)\left[1+\gamma (d+2)\frac{Y}{X}\right] -2\gamma(d+2)\frac{Y^2}{X} a_2 \bigg\},  \label{ch4_subeq:ODESinnerc-Y}\\
        -\frac{da_2}{dX} = X^{1/2}\bigg\{  8\gamma \left(1-\frac{Y}{X}\right)-\left[\frac{4X}{Y}-8\gamma +4\gamma (d+8)\frac{Y}{X} \right]a_2  \bigg\}.
    \end{align}
\end{subequations}
These equations provide us with the inner solution, which is expected to be valid for $(X,Y,a_2)$ of the order of unity, i.e., close to the frozen state as discussed above. Equations~\eqref{ch4_eq:evolscaled-granular} are complemented with the initial conditions \eqref{ch4_eq:initial_conditions}, which now read
\begin{equation}\label{ch4_eq:boundary}
    Y(r_c^{-2/3}) = r_c^{-2/3}, \quad a_2(r_c^{-2/3}) = a_2^{\st}.
\end{equation}
Note that, using the scaled temperatures $(X,Y)$, all the dependence of the inner solution on the cooling rate $r_c$ takes place through the initial conditions.

\begin{figure}
    \centering
    \includegraphics[width=3.5in]{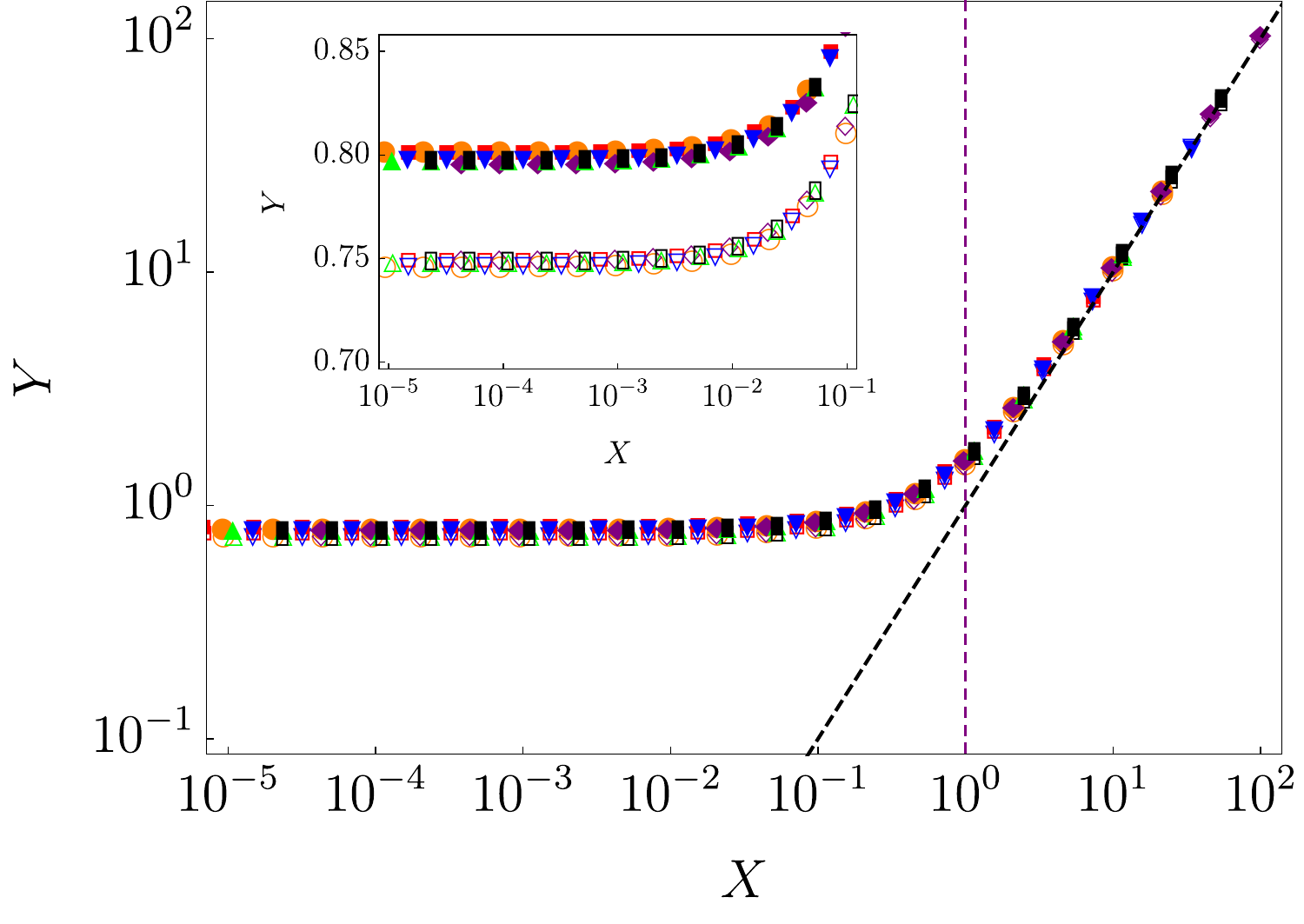}
    \caption{\label{ch4_fig:univ-cooling-gran} Scaled granular temperature $Y$ as a function of the scaled bath temperature $X$. We have employed the linear cooling protocol \eqref{ch4_eq:cooling-programme} with different cooling rates and two values of the restitution coefficient $\alpha$. The colour codes and symbols for the DSMC data are the same as those employed in the left panel of Figure~\ref{ch4_fig:frozen-temp-gran}. The dashed, purple vertical line marks the fictive temperature $X_f = \theta_f / r_c^{2/3} = 1$ from Eq.~\eqref{ch4_eq:thetaf-power-law}. for $K = 3/2$. The inset shows the mild dependence of the frozen temperature $Y^{\text{Frz}}$ on the restitution coefficient, which becomes noticeable only when plotting on linear scale the $Y$ axis.
      }
\end{figure}

Figure~\ref{ch4_fig:univ-cooling-gran} illustrates the same glass transition shown on the left panel of Figure~\ref{ch4_fig:frozen-temp-gran} for the granular gas, but in terms of the scaled variables $X$ and $Y$. For each value of the restitution coefficient $\alpha$, all the curves for different values of the cooling rate $r_c$ collapse onto a unique master curve, independent of $r_c$. Our theoretical prediction for the scaled fictive temperature $X_f$---which stems from Eq.~\eqref{ch4_eq:thetaf-power-law} from previous sections---is also plotted: it is independent of $r_c$ as well, since $\theta_f$ is proportional to $r_c^{2/3}$, as given by Eq.~\eqref{ch4_eq:thetaf-power-law}, and $X_f=r_c^{-2/3} \theta_f$. Our theory thus gives an excellent estimate for the actual fictive temperature of the system. Since the plotted numerical data corresponds to the DSMC integration of the kinetic equation \eqref{ch1_eq:boltzmann-fokker-planck}, not to our perturbation approach, this suggests that the exact solution to the problem presents a universal behaviour in scaled variables. The universal behaviour of the scaled temperature illustrated depends mildly on $\alpha$; the differences between the $\alpha = 0.9$ (open symbols) and $\alpha = 0.3$ (filled symbols) datasets are very small, as shown in the inset for $Y^{\text{Frz}}$. This is due to the smallness of the values of the excess kurtosis in the granular gas, where typically $|a_2|\lesssim 0.15$. Besides, note that the terms containing the excess kurtosis in the evolution equation for the temperature~\eqref{ch4_subeq:granularSonine-theta} are of the form $(1+3a_2/16)$, and thus $3a_2/16\lesssim 0.03$, so the differences between different values of $\alpha$ are expected to be of a few per cents. This mild dependence is even more negligible for the fictive temperature $X_f$, which is approximately 1 up to the third digit---according to the value of the constant $K$ from Eq.~\eqref{ch4_eq:thetaf-power-law}.


\begin{figure}
    \centering
    \includegraphics[width=3.5in]{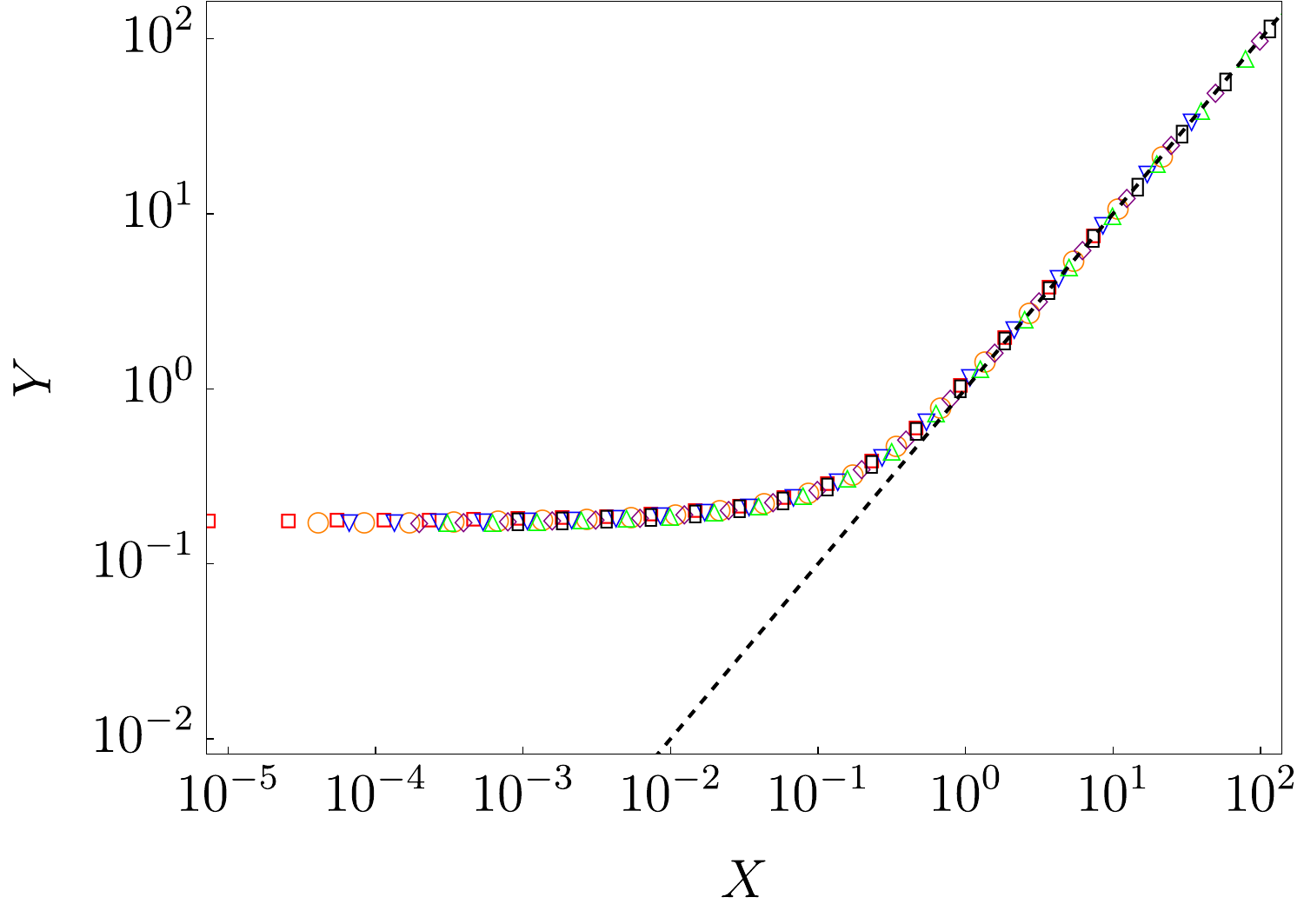}
    \caption{\label{ch4_fig:univ-cooling-mol} Scaled kinetic temperature $Y$ as a function of the scaled bath temperature $X$ for the molecular fluid with non-linear drag, with parameter values $\gamma = 0.1$ and $d=3$. We have employed the linear cooling protocol \eqref{ch4_eq:cooling-programme} with different cooling rates: $r_c= 0.05$ (red squares), $0.025$ (green up triangles), $0.01$ (orange circles), $0.005$ (blue down triangles), $0.0025$ (black rectangles), and $0.001$ (purple diamonds).
      }
\end{figure}

In order to test the robustness of the scaling \eqref{ch4_eq:scaling-cooling} for the molecular fluid, we check our theoretical prediction with numerical data in Fig.~\ref{ch4_fig:univ-cooling-mol}, which is analogous to Fig.~\ref{ch4_fig:univ-cooling-gran} for the granular gas. Datasets were obtained from the numerical simulation of the corresponding Langevin equation. We have also compared the numerical value of $Y^{\Frz}$ and $a_2^{\Frz}$ obtained from simulation data with our theoretical prediction---i.e. the numerical integration of Eq.~\eqref{ch4_eq:ODESinnerc} with the boundary conditions $\lim_{X\to\infty}Y(X)=\infty$,  $\lim_{X\to\infty}a_2(X)=0$ for different values of $r_c$\footnote{In practice, we impose the boundary condition for a sufficiently high enough value of $X$, such that our results do not change upon increasing it.}. For the former approach, we obtained the values $Y^{\Frz} = 0.403$ and $a_2^{\Frz}=-0.146$, while for the latter, we obtained $Y^{\Frz} = 0.397$ and $a_2^{\Frz}=-0.154$. The agreement is excellent for both the kinetic temperature and the excess kurtosis, although the slight mismatch for the $a_2$ was to be expected within the first Sonine approximation, since $a_2^{\Frz}$ is quite large for the non-linear fluid. Moreover, $a_2^{\Frz}$ is not as close to its value at the LLNES, $a_2^r=-0.4$ for $d=3$ as predicted by Eq.~\eqref{ch3_eq:LLNES-sonine}, as it was $a_2^{\Frz}$ to its HCS value in the granular gas. As mentioned earlier, we provide a more detailed discussion on this point in Sec.~\ref{ch4_sec:different_cooling_programs}.

\subsubsection{\label{ch4_subsubsec:uniform-sol}Uniform solution}
In order to understand the universality in the behaviour of our scaled variables, it is useful to build approximate, to the lowest order, expressions over the whole bath temperature range, not only in the boundary layer. For high enough bath temperatures, $\theta_{\st}\gg r_c^{2/3}$, we have the outer expansion in Eq.~\eqref{ch4_eq:pert-gran}. For low enough bath temperatures, $\theta_{\st}=O(r_c^{2/3})$, we have the boundary layer system~\eqref{ch4_eq:evolscaled-granular} for the scaled variables. In boundary layer perturbation theory, the lowest-order perturbative solution---known as the \textit{uniform} solution---is constructed as the sum of the lowest-order outer and inner solutions, minus the common behaviour found in an intermediate matching region~\cite{bender_advanced_1999}. Below we derive such a uniform solution for our cooling protocol.

We denote the outer solution at the lowest order by $(\theta_O,a_{2,O})$. Equation~\eqref{ch4_eq:pert-gran} tells us that
\begin{equation}
\label{ch4_eq:outer-series}
    \theta_O(\theta_{\st})=\theta_{\st}, \quad a_{2,O}(\theta_{\st})=a_{2}^{\st},
\end{equation}
which is the instantaneous NESS curve. Now, let us seek the solution of the inner problem at the lowest order, which we denote by $(Y_I(X),a_{2,I}(X))$. To obtain it, we solve Eq.~\eqref{ch4_eq:evolscaled-granular} with the boundary conditions
\begin{equation}
    \lim_{X\to\infty}Y_I(X) = \infty, \quad \lim_{X\to\infty} a_{2,I}(X) = a_2^{\st},
    \label{ch4_eq:boundary-infty}
\end{equation}
which correspond to the limit as $r_c\to 0$ in Eq.~\eqref{ch4_eq:boundary}. Therefore, $(Y_I(X),a_{2,I}(X))$ does not depend on $r_c$, since neither the evolution equations~\eqref{ch4_eq:evolscaled-granular} nor the boundary conditions~\eqref{ch4_eq:boundary-infty} depend on $r_c$. 

Although it is not possible to write $(Y_I(X),a_{2,I}(X))$ in a simple closed form, an asymptotic analysis for large values of $X$ gives
\begin{equation}
    Y_I(X)\sim X, \qquad a_{2,I}(X)\sim a_2^{\st}, \quad X\gg 1,
    \label{ch4_eq:inner-asymp}
\end{equation}
which is consistent with the tendency of the DSMC data in Fig.~\ref{ch4_fig:univ-cooling-gran} to the instantaneous NESS curve for large $X$. By comparing Eqs.~\eqref{ch4_eq:outer-series} and \eqref{ch4_eq:inner-asymp}, we obtain the common behaviour
\begin{equation}
    \theta_c (\theta_{\st})=\theta_{\st}, \qquad a_{2,c}(\theta_{\st})=a_2^{\st}, \quad \theta_{\st}\ll 1, \; X\gg 1,
    \label{ch4_eq:common-behaviour}
\end{equation}
or, $r_c^{2/3}\ll \theta_{\st}\ll 1$, which is the region at which the outer and inner solution match. The uniform solution is built as~\cite{bender_advanced_1999}
\begin{align}
    \theta^{\BL}(\theta_{\st})&=\theta_O(\theta_{\st})+r_c^{2/3}Y_I(X=r_c^{-2/3}\theta_{\st})-\theta_c(\theta_{\st}),\\
    a_{2}^{\BL}(\theta_{\st})&=a_{2,O}(\theta_{\st})+a_{2,I}(X=r_c^{-2/3}\theta_{\st})-a_{2,c}(\theta_{\st}),
\end{align}
Since the common behaviour~\eqref{ch4_eq:common-behaviour} equals the outer solution~\eqref{ch4_eq:outer-series}, the range of validity of the inner solution  extends to the whole temperature interval. In other words, the uniform solution coincides with the inner solution:
\begin{align}
    \theta^{\BL}(\theta_{\st})&=r_c^{2/3}Y_I(X=r_c^{-2/3}\theta_{\st}), \\
    a_{2}^{\BL}(\theta_{\st})&=a_{2,I}(X=r_c^{-2/3}\theta_{\st}).
\end{align}
The first equation tells us that $Y^{\BL}\equiv r_c^{-2/3}\theta^{\BL}=Y_I$, i.e., we have the universal behaviour in Fig.~\ref{ch4_fig:univ-cooling-gran} over the uniform solution---to the lowest order. From the lowest order solution depicted above, the frozen values of the scaled variables are readily obtained,
\begin{subequations}
\label{ch4_eq:frozen-cond}
    \begin{align}
        Y^{\Frz}&\equiv \lim_{X\to 0}Y^{\BL}(X),
        \\
        a_2^{\Frz}&\equiv \lim_{X\to 0}a_{2}^{\BL}(X).
    \end{align}
\end{subequations}
Our above argument about the independence of $Y^{\BL}(X)$ on the cooling rate is immediately translated to $\theta_{\Frz}=r_c^{2/3}Y^{\Frz}$, which means that $\theta_{\Frz}$ follows the power law behaviour $\theta_{\Frz}\propto r_c^{2/3}$ that we have already checked on the right panel of Figure~\ref{ch4_fig:frozen-temp-gran}. Also, the independence of $a_2^{\Frz}$ on $r_c$ has already been checked on the right panel of Figure~\ref{ch4_fig:VDF-graph}. Moreover, the independence on $r_c$ of the curves $(Y^{\BL},a_2^{\BL})$ as a function of $X$ gives rise to the universal cooling curve in Fig.~\ref{ch4_fig:univ-cooling-gran}. Therefore, our theory explains the observed universal behaviour of the simulation data in scaled variables.

\subsection{\label{ch4_sec:different_cooling_programs}Glass transition for different cooling programs}

Up to this point of the chapter, we have employed linear cooling programs in order to study the emergence of glassy behaviour in both molecular fluids and granular gases. Let us now consider the following family of cooling protocols,
\begin{equation}\label{ch4_eq:general-cooling}
    \frac{d\theta_{\st}}{dt} = -r_c \theta_{\st}^{k}, 
\end{equation}
with $k$ being a real number. The $k=0$ case reduces to the already studied linear cooling {protocol}. Let us note that the characteristic time $t_0$ at which $\theta_s$ vanishes radically changes as a function of $k$, as it remains finite for $k<1$---i.e. $\theta_{\st}$ reaches zero in a finite time, whereas it becomes infinite for $k\geq 1$.

In the following, we still consider that the cooling is slow, in the sense that $r_c\ll 1$. Following the same approach as in Sec.~\ref{ch4_subsec:intuition_glass} for the linear cooling program, we introduce the effective timescale
\begin{equation}
    s = \int_{t}^{t_0} \frac{dt'}{\tau(\theta_{\st}(t'))} = \frac{1}{r_c}\int_0^{\theta_{\st}}d\theta_{\st} \ (\theta_{\st}')^{-k}\tau^{-1}(\theta_{\st}'),
\end{equation}
where we have inserted Eq.~\eqref{ch4_eq:general-cooling} into the second integral from Eq.~\eqref{ch4_eq:s-scale-def}. Assuming again that $s$ is of the order of unity when the system freezes, we arrive at the following estimation of the fictive temperature $\theta_f$;
\begin{equation}\label{ch4_eq:theta-f-scaling-with-k}
    \theta_f = \left[\left(\frac{3}{2}-k\right)\frac{r_c}{K}\right]^{\frac{2}{3-2k}}, \quad \theta_{\text{Frz}} \propto r_c^{\frac{2}{3-2k}},
\end{equation}
where we have taken into account that the characteristic relaxation timescale $\tau$ is still given by Eq.~\eqref{ch4_eq:tau-def}, regardless of the cooling program employed. Thus, the above entails that the system presents a kinetic glass transition as long as $k<k_{\text{crit}}=3/2$. Equation~\eqref{ch4_eq:theta-f-scaling-with-k} generalizes the power law behaviour $r_c^{2/3}$ found in previous sections.

Now, in order to corroborate the scaling found in Eq.~\eqref{ch4_eq:theta-f-scaling-with-k}, we follow a regular perturbation approach similar to those employed in Sec.~\ref{ch4_subsec:perturbative} for both the granular gas and the molecular fluid. The solution to the lowest order corresponds again to the instantaneous stationary solutions $\theta^{(0)} = \theta_{\st}$, $a_2^{(0)}=a_2^{\st}$. The first-order $O(r_c)$ corrections are provided by the equations
\begin{subequations}
    \begin{align}
        -\theta_{\st}^k &= \theta_{\st}^{1/2}\left\{c_1 \theta^{(1)} + c_2\theta_{\st}a_2^{(1)}\right\},
        \\
        0 &=c_3 \frac{\theta^{(1)}}{\theta_{\st}^{1/2}} + c_4\theta_{\st}^{1/2}a_2^{(1)},
    \end{align}
\end{subequations}
with $c_i, \ i=1,..,4$ being constants that depend on the parameters of the specific system of concern. These equations entail the scalings
\begin{equation}
\label{ch4_eq:scaled-varying-k}
    \theta^{(1)} \propto \theta_{\st}^{k-\frac{1}{2}}, \quad a_2^{(1)} \propto \theta_{\st}^{k-\frac{3}{2}}.
\end{equation}
which imply that $\theta^{(1)}\ll\theta^{(0)}=\theta_{\st}$, $a_2^{(1)}\ll a_2^{(0)}=a_2^{\st}$ when $k > k_{\crit}= 3/2$ . Therefore, if we may extrapolate our results to the entire hierarchy of Sonine cumulants---as we did in the preceding sections, for $k>k_{\crit}$ a kinetic glass transition is presented neither by the granular gas nor by the molecular fluid. The cooling is so slow for $k>k_{\crit}$ that both systems remain basically over the instantaneous stationary curve $\{\theta=\theta_{\st}, a_2=a_2^{\st}\}$ for all bath temperatures. Let us remark that the existence of a critical value of $k$ for the emergence of a glass transition has already been reported in simple models~\cite{brey_residual_1991,brey_dynamical_1994}.

Let us now consider the case $k\leq k_{\crit}$. In this case, the regular perturbation approach breaks down for low enough bath temperatures, which marks the onset of the kinetic glass transition. Our regular perturbation approach ceases to be valid when the $O(1)$ terms become comparable with the $O(r_c)$ ones, thus implying
\begin{equation}
    \theta_{\st} = O\left(r_c^{\frac{2}{3-2k}}\right),
    \label{ch4_eq:thetast-func-k}
\end{equation}
which, consistently with our discussion above, only makes sense for $k<k_{\crit}$. Note that for $k=0$ we recover the scaling predicted in Eq.~\eqref{ch4_eq:thetaf-power-law} for the linear cooling program. Equation~\eqref{ch4_eq:thetast-func-k} entails that the kinetic temperature at the frozen state is expected to scale as $\theta^{\Frz}\propto r_c^{\frac{2}{3-2k}}$, consistently with Eq.~\eqref{ch4_eq:theta-f-scaling-with-k}. Interestingly, regardless of the choice of $k$, the frozen state is still universal, in the sense that it is independent of the cooling rate $r_c$, since
\begin{equation}
    a_2 - a_2^{\st}\sim a_2^{(1)} \propto r_c \,\theta_{\st}^{k-\frac{3}{2}} = O(1).
\end{equation}
As in Sec.~\ref{ch4_subsec:boundary_layer}, the above scaling relations suggest the introduction of scaled variables
\begin{equation}
    Y \equiv r_c^{-\frac{2}{3-2k}}\theta , \quad X \equiv r_c^{-\frac{2}{3-2k}}\theta_{\st} .
\end{equation}
In terms of them, the dynamic equations for the cooling protocol become $r_c$-independent. We remark that the evolution equations for the scaled variables $(Y,a_2)$ in both systems are the same as the ones we have written in previous sections---Eqs.~\eqref{ch4_eq:evolscaled-granular} and \eqref{ch4_eq:ODESinnerc}, respectively, with the only change $ d/dX \to X^k d/dX$ on their left-hand sides.

\begin{figure}
    {\centering 
    \includegraphics[width=2.41in]{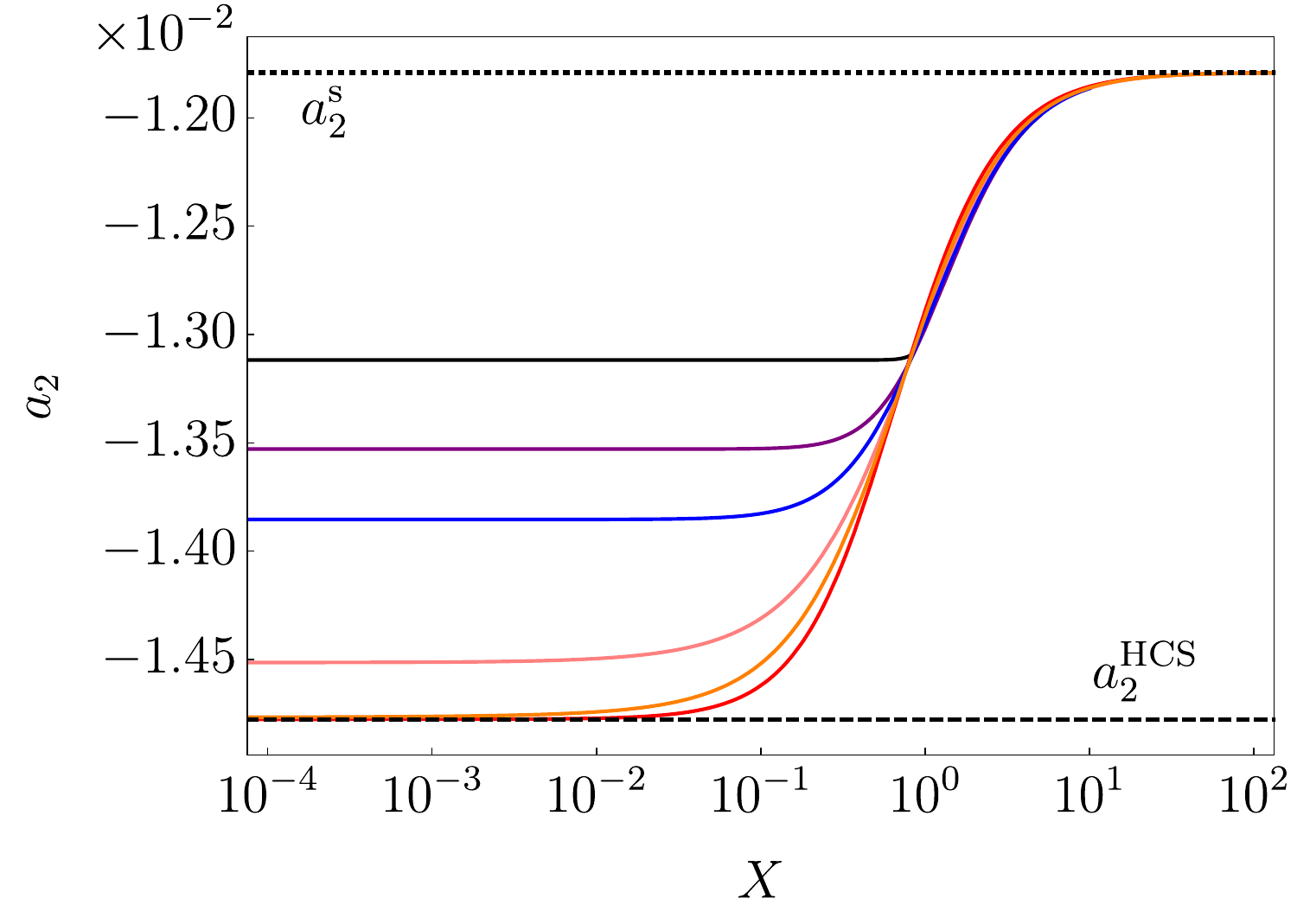}
    \includegraphics[width=2.89in]{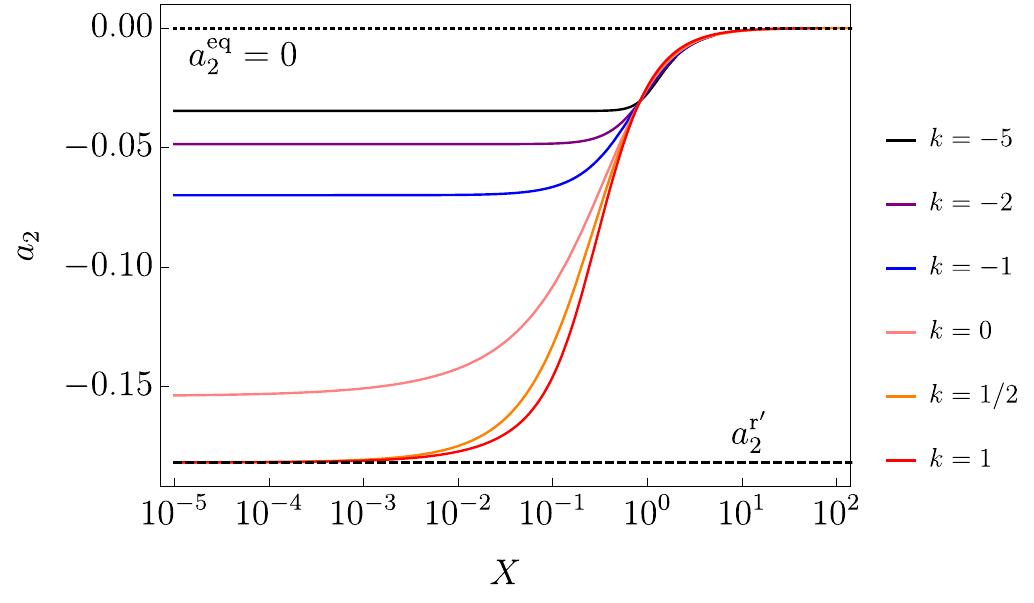}}
    \caption{{Excess kurtosis $a_2$ as a function of the scaled bath temperature $X$ for {the family of cooling programs}~\eqref{ch4_eq:general-cooling} with different values of $k$. Left and right panels correspond to the granular gas and the molecular fluid, respectively. {The solid lines correspond to the numerical integration of the evolution equations in the first Sonine approximation, { for the values of $k$ depicted in the legend---i.e. from top to bottom $k$ increases from $-5$ to $1$.}} In both panels, a dimensionless cooling rate $r_c=0.1$ has been employed---although our results are independent of $r_c$ in the scaled variables. On the left (right) panel, {(i)} the  dotted line represents the value of the excess kurtosis at the NESS $a_2^{\st}$ (equilibrium value of the excess kurtosis $a_2^{\text{eq}}=0$), from which all the curves depart for large $X$, {and (ii)} the  dashed line accounts for the value $a_2^{\text{HCS}}$ at the HCS from Eq.~\eqref{ch4_eq:a2hcs}---and $a_2^{r'}$ at the LLNES from Eq.~\eqref{ch2_eq:extremes-excess-kurtosis}, within the first Sonine approximation.}}
        \label{ch4_fig:varying-k-a2}
\end{figure}

Figure~\ref{ch4_fig:varying-k-a2} shows the numerical integration of the evolution equation for the excess kurtosis $a_2$ within the first Sonine approximation towards its frozen state in a cooling program with rate $r_c$ for different values of $k$ in Eq.~\eqref{ch4_eq:general-cooling}, for both the granular gas and the molecular fluid. {The} excess kurtosis follows a similar trend: on the one hand, for $k\rightarrow -\infty$, the time window over which $\theta_{\st}$ decays towards zero becomes infinitely small, and thus the excess kurtosis does not have time to deviate from its stationary state value and is approximately constant for all $X$.  On the other hand, as the value of $k$ is increased, the time window to relax also increases. The limiting case $k = k_{\crit}$ constitutes the ultimate balance between a sufficiently wide time window to relax, and a fast enough relaxation protocol such that $\theta$ deviates from the $\theta = \theta_{\st}$ behaviour. 

It is worth noting that, for $1/2 \leq k < k_{\text{crit}}$, the excess kurtosis tends to the value over the HCS---for the granular gas---and the LLNES---for the non-linear molecular fluid. The lower bound $k = 1/2$ corresponds to the value above which the deviations from the $\theta = \theta_{\st}$ line become significantly small, but still allowing for the kurtosis to evolve towards the frozen state, as Eq.~\eqref{ch4_eq:scaled-varying-k} tells us. Since we are showing the numerical integration of the evolution equations in the first Sonine approximation---Eqs.~\eqref{ch4_eq:granularSonine} and \eqref{ch4_eq:ODESinnerc} respectively, these limit values of the excess kurtosis correspond to their theoretical estimates in this framework. For the granular gas, this is given by Eq.~\eqref{ch4_eq:a2hcs}, which is quite accurate due to its smallness.\footnote{Recall that it was discussed in Sec.~\ref{ch2_subsec:quench}} For the non-linear fluid, the first Sonine approximation gives ${a_2^{r}}^{\prime}=-2/(d+8)$---as depicted in Eq.~\eqref{ch2_eq:extremes-excess-kurtosis}---which is quite different from its exact value in $a_2^r = -2/(d+2)$, which is proven in Appendix~\ref{app:asymptotic-llnes}---this is reasonable, since the deviations from the Gaussian are much larger in the LLNES than in the HCS.

The above discussion hints at the frozen state corresponding to the HCS and the LLNES for the granular gas and the non-linear molecular fluid, respectively, for $1/2 \leq k < k_{\text{crit}}$, given that the excess kurtosis $a_2$ exactly attains the values attained at each global attractor within the first Sonine approximation. This means that the two model systems, either the granular gas or the molecular fluid, reach the corresponding non-equilibrium state, either the HCS or the LLNES, over a time window of the order of $r_c^{-1}$ when {cooled down} with a program for which $1/2 \leq k < k_{\text{crit}}$. The latter suggests useful applications in optimal control~\cite{aurell_optimal_2011,prados_optimizing_2021,guery-odelin_driving_2023,blaber_optimal_2023}---since cooling programs with $1/2 \leq k < k_{\text{crit}}$ allow to optimally reach either the HCS or the LLNES in a finite time---and also within the study of non-equilibrium effects, as previous work on both systems shows that both the HCS and the LLNES are responsible for the emergence of a plethora of non-equilibrium phenomena, such as the Mpemba and Kovacs effects, as thoroughly discussed in chapter~\ref{ch:memory_effects}.

\section{\label{ch4_subsec:hysteresis_cycles}Hysteresis cycles}
Now let us turn our attention to a reheating protocol from the frozen state with rate $r_h>0$, $d\theta_{\st}/dt=r_h$. First, we consider the paradigmatic case $r_h=r_c=r$. Second, we consider the more general case $r_h \neq r_c$. In both cases, we show that the system does not follow backwards the cooling curve, but crosses the NESS line $\theta=\theta_{\st}$ and afterwards tends thereto from below. This is similar to the hysteresis cycle displayed by glassy systems in temperature cycles---i.e. cooling followed by reheating~\cite{scherer_relaxation_1986,scherer_theories_1990}.

\subsection{\label{ch4_subsubsec:univ_hysteresis}Universal hysteresis cycles}
First, we consider the case $r_c=r_h=r$. In complete analogy with the cooling program, we  define scaled variables as
\begin{equation}\label{ch4_eq:scaling-hysteresis}
	Y \equiv r^{-2/3} \theta , \quad X \equiv r^{-2/3} \theta_{\st} .
\end{equation} 
In terms thereof, the evolution equations in the reheating protocol become independent of $r$. On the one hand, for the granular gas, we have
\begin{subequations}\label{ch4_eq:evolscaled-heating}
    \begin{align}
        \frac{dY}{dX}&=  X^{3/2} \left(1+\frac{3}{16} a_2^{\st}\right) -Y^{3/2} \left(1+\frac{3}{16} a_2\right),
        \label{ch4_subeq:evolscaled-heating-temp}
        \\
        \frac{da_2}{dX}&= 2\, Y^{1/2} \left[  \left(1-\frac{X^{3/2}}{Y^{3/2}}\right) a_2 +B \, (a_2^{\st} -a_2) \right], 
    \end{align}
    \end{subequations}
while for the molecular fluid
\begin{subequations}
    \label{ch4_eq:ODESinnerh}
    \begin{align}
    \frac{dY}{dX} =  X^{1/2} \, \bigg\{  & 2\, ( X-Y)\left[1+\gamma (d+2)\frac{Y}{X}\right]
        -2\gamma(d+2)\frac{Y^2}{X} a_2 \bigg\},  \\
        \frac{da_2}{dX} = X^{1/2}\bigg\{ & 8\gamma \left(1-\frac{Y}{X}\right)  
        -\left[\frac{4X}{Y}-8\gamma +4\gamma (d+8)\frac{Y}{X} \right]a_2  \bigg\}.
    \end{align}
\end{subequations}
Note that the above systems resemble those from Eqs.~\eqref{ch4_eq:evolscaled-granular} and \eqref{ch4_eq:ODESinnerc}, respectively, differing only in a minus sign. Such systems must be complemented with the new initial conditions
\begin{equation}\label{ch4_eq:evolscaled-heating-bc}
    Y(0) = Y^{\Frz}, \quad a_2(0) = a_2^{\Frz},
\end{equation}
which correspond to that of the frozen state from the previously applied cooling program, given by Eq.~\eqref{ch4_eq:frozen-cond} to the lowest order---recall that $r_c=r_h=r$.

A completely similar analysis to that carried out for the cooling program shows that the inner solution for the heating program gives the uniform solution to the lowest order again. The hysteresis cycle is unique in the rescaled axes $Y$ vs.~$X$, since the rate is not present in either the scaled evolution equations or the boundary conditions~\eqref{ch4_eq:evolscaled-heating-bc}.

\begin{figure}
    {\centering 
    \includegraphics[width=2.65in]{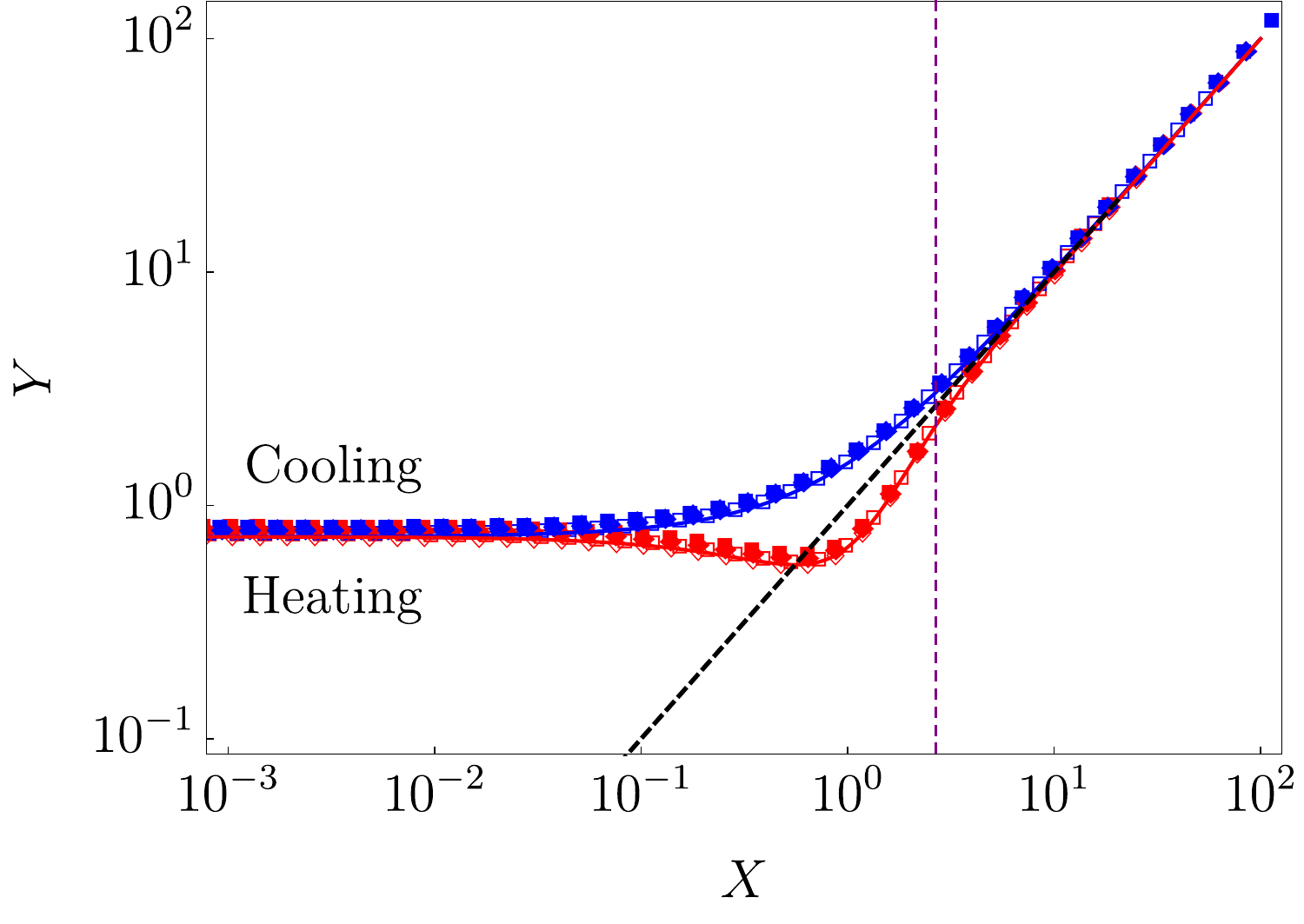}
    \includegraphics[width=2.65in]{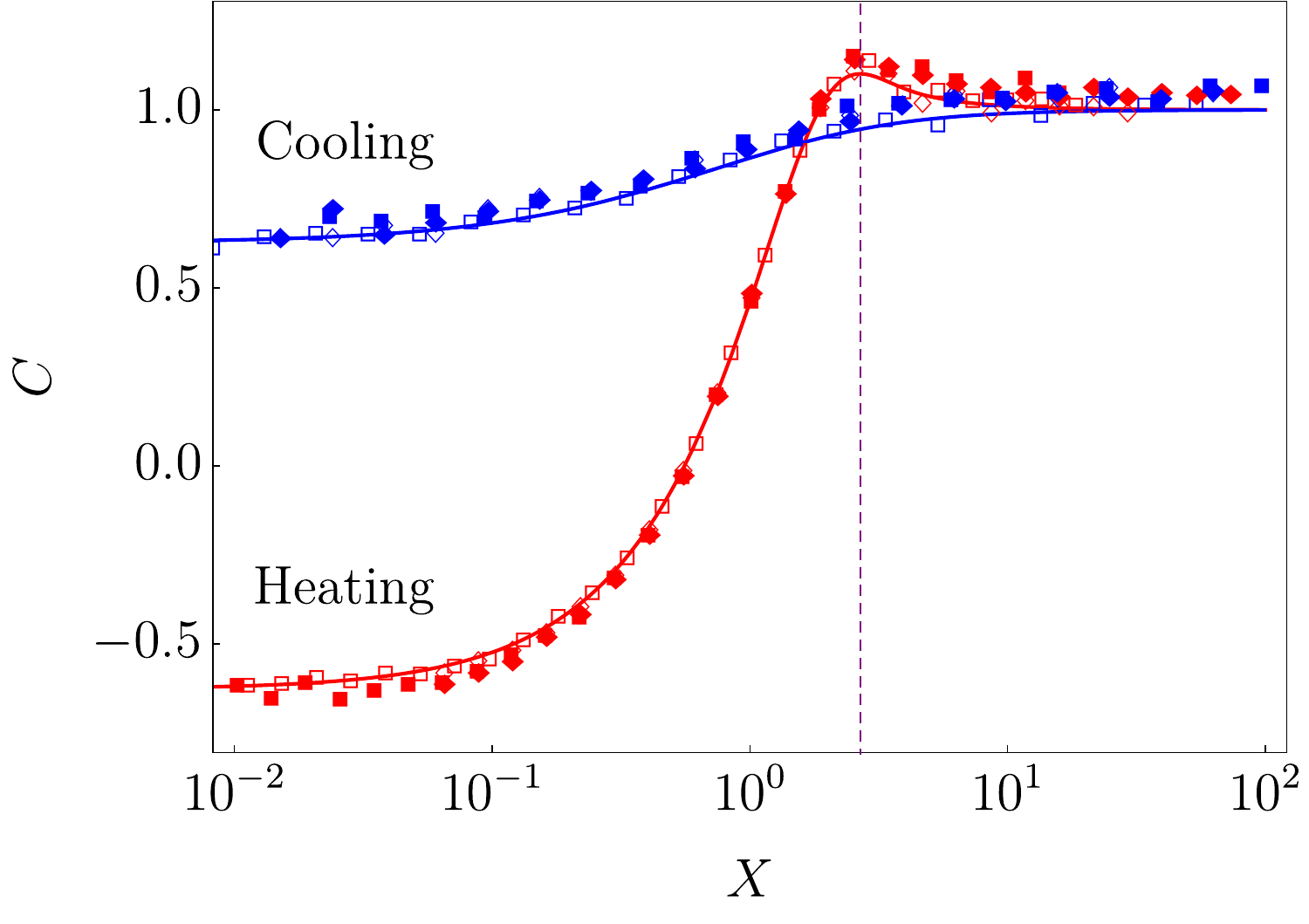}}
    \caption{Hysteresis cycles in the granular gas. The system is first cooled down with rate $r_c$ and later reheated from the frozen state with rate $r_h=r_c=r$. Blue and red symbols and lines correspond to the cooling and heating protocols respectively, as explicitly stated in both panels. Symbols are DSMC data, while the solid curves correspond to the numerical integration of Eqs.~\eqref{ch4_eq:evolscaled-granular} and \eqref{ch4_eq:evolscaled-heating} for cooling and heating, respectively. Specifically, we present results for $r=0.01$ (squares) and $r=0.001$ (diamonds), and for two values of $\alpha$:  $0.9$ (open symbols) and $0.3$ (filled symbols). The dashed  straight line corresponds to the instantaneous NESS curve $Y =X$. The purple vertical line marks the bath temperature $X_g$ at which the heat capacity reaches its maximum in the reheating program---see right panel. Both the solid curves and the purple vertical line were obtained for $\alpha = 0.9$, as they superimpose with the ones corresponding to the $\alpha = 0.3$ case.}
        \label{ch4_fig:hysteresis-gran}
\end{figure}

In Fig.~\ref{ch4_fig:hysteresis-gran}, we numerically check our prediction on the hysteresis cycle being independent of $r$. On the left panel, the hysteresis cycle of the kinetic temperature is shown. DSMC simulation data (symbols) are compared with the boundary layer solution (blue lines) of Eq.~\eqref{ch4_eq:evolscaled-heating}, for different values of the cooling/heating rate $r=r_c=r_h$, {and again for two values of the restitution coefficient, $\alpha = 0.3$ and $0.9$}. It is neatly observed that the boundary layer solution captures very well the numerical results throughout the whole cycle. Remarkably, the heating curve crosses the NESS line $\theta=\theta_{\st}$ (dashed line) and tends thereto from below---this is further analyzed in Sec.~\ref{ch4_subsubsec:normal_heating_curve}. On the right panel, we display the apparent ``heat capacity" $d\theta/d\theta_{\st}=dY/dX$ over the thermal cycle. {In general, the apparent} heat capacity
\begin{equation}
    C \equiv \frac{d\theta}{d\theta_{\st}}=\frac{dY}{dX}
\end{equation}
is non-monotonic in the heating process. In fact, it presents a maximum at a certain value of $\theta_{\st}$ (or $X$), that we refer to as $\theta_{g}$ (or $X_g$),
\begin{equation}
    \frac{dC}{dX}\biggr\rvert_{X=X_g} = \frac{d^2Y}{dX^2}\biggr\rvert_{X=X_g} = 0,
\end{equation}
which has been employed in the literature to define the glass transition temperature~\cite{brey_dynamical_1994,angell_formation_1995,dyre_colloquium_2006,tropin_modern_2016,richet_thermodynamics_2021}, as mentioned in Sec.~\ref{ch1_subsec:glass-transition}. Note that, apart from showing where the maximum of $C$ is attained, $X_g$ marks the temperature upon which $Y(X)$ changes its convexity in the reheating process. In this simple system, the value of the apparent heat capacity  has opposite signs in the limit of very low bath temperatures, as can be observed for small values of $X$ in the right panel of Fig.~\ref{ch4_fig:hysteresis-gran}. This is readily understood from Eqs.~\eqref{ch4_subeq:evolscaled-Y} and \eqref{ch4_subeq:evolscaled-heating-temp}, since
\begin{equation}\label{ch4_eq:C-low-temp-gran}
    \lim_{X\rightarrow 0}C = \pm (Y^{\text{Frz}})^{3/2}\left(1+\frac{3}{16}a_2^{\text{Frz}}\right),
\end{equation}
with the plus and minus signs corresponding to cooling and reheating, respectively. The behaviour of the apparent heat capacity $C$ in Eq.~\eqref{ch4_eq:C-low-temp-gran} has a neat physical meaning. For very low bath temperatures, the term coming from the stochastic driving in the evolution equation for the temperature~\eqref{ch4_subeq:granularSonine-theta} becomes negligible and
\begin{equation}
\frac{d\theta}{dt}\sim -\theta^{3/2}\left(1+\frac{3}{16}a_2\right).
\label{ch4_eq:theta-low-temp}
\end{equation}
That is, the granular gas ``freely cools'', since $\theta$ mononotically decreases with time. Note that, in fact, Eq.~\eqref{ch4_eq:theta-low-temp} is nothing but Haff's law, $\dot\theta\propto -\theta^{3/2}$, and it is equivalent to  Eq.~\eqref{ch4_eq:C-low-temp-gran} for the apparent heat capacity.

\begin{figure}
    {\centering 
    \includegraphics[width=2.65in]{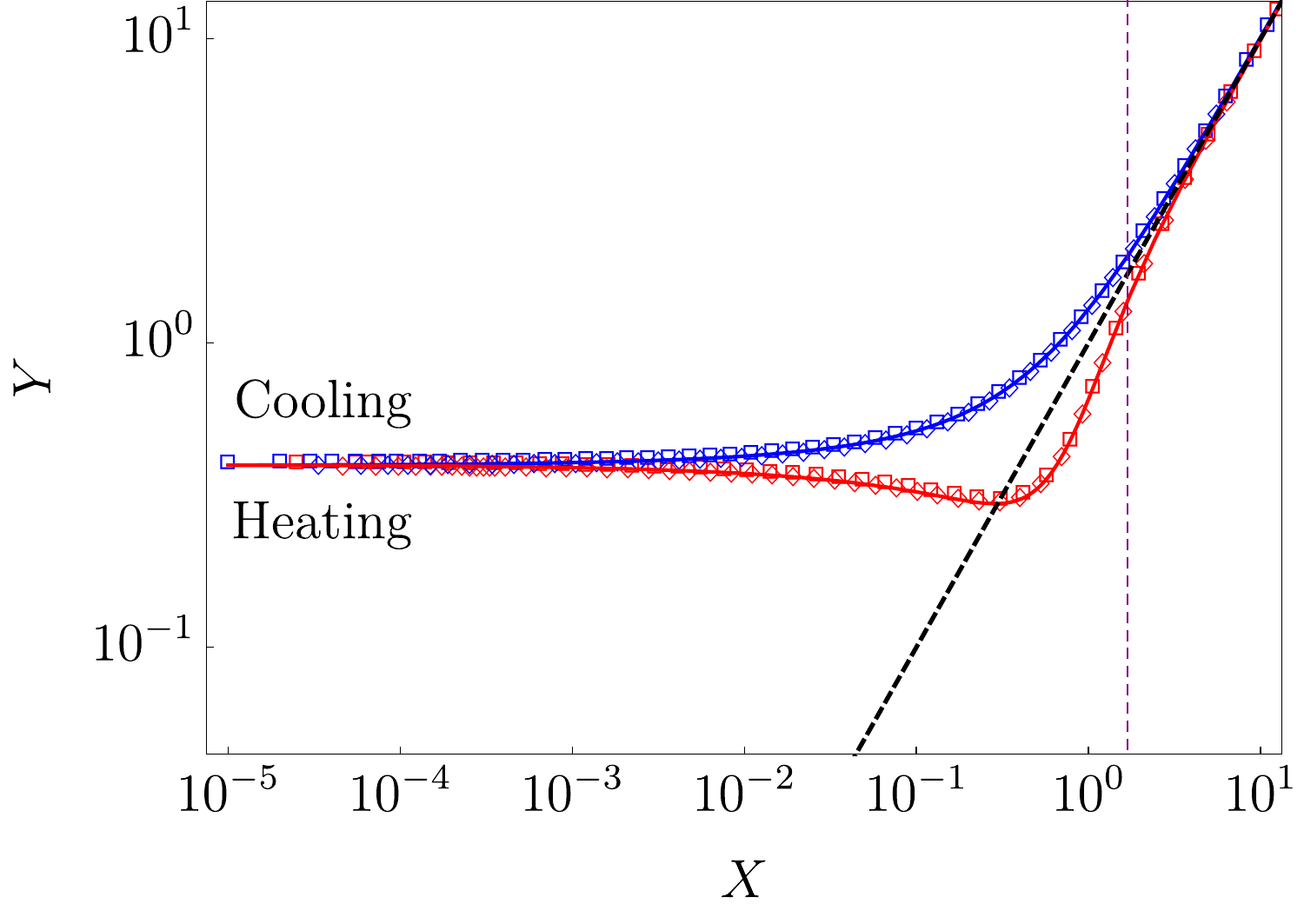}
    \includegraphics[width=2.65in]{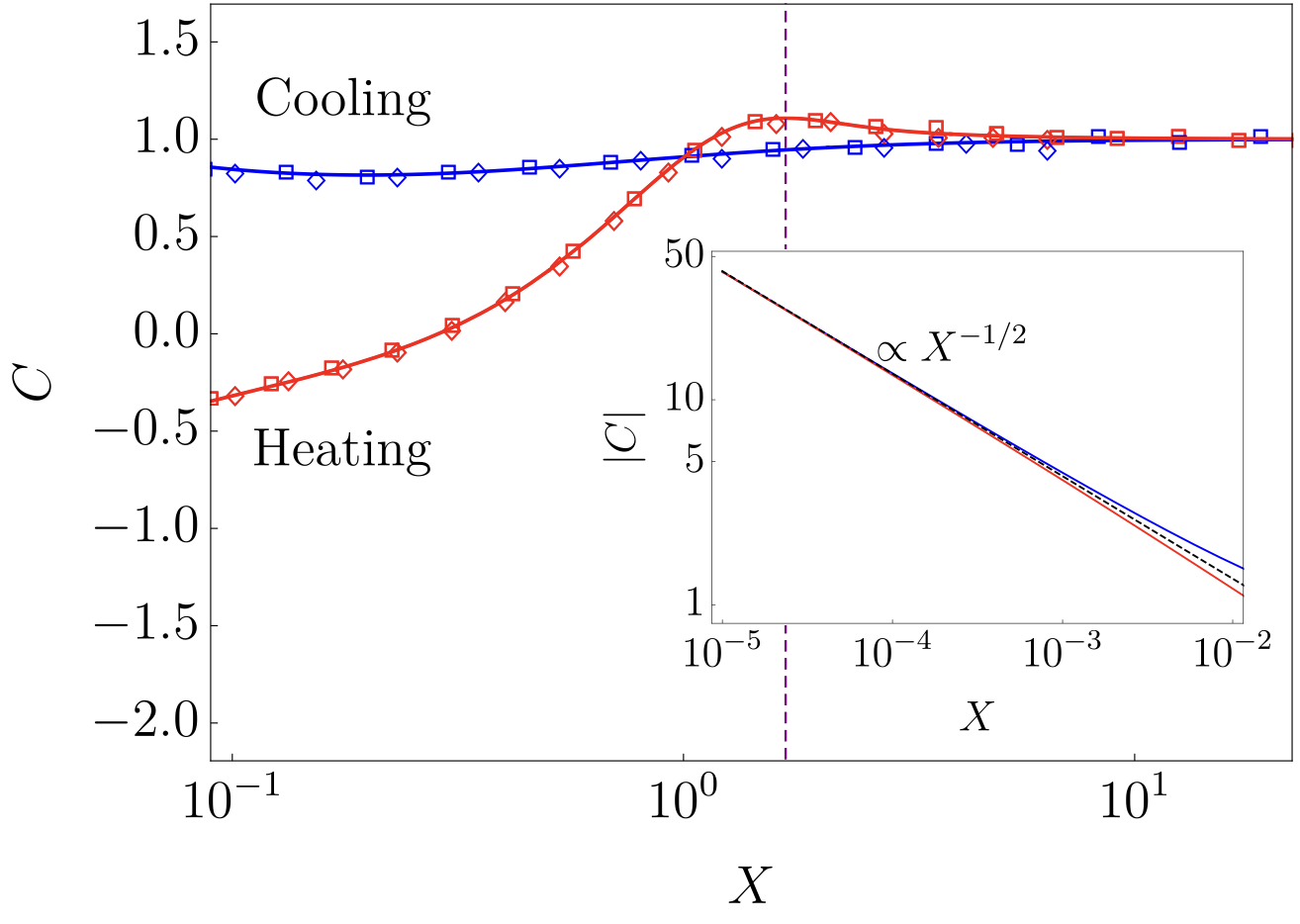}}
    \caption{Hysteresis cycle in the molecular fluid with non-linear drag.  
    Both panels are analogous to the ones from Fig.~\ref{ch4_fig:hysteresis-gran} for the granular gas, with the same values of the cooling and heating rate $r$ in dimensionless variables. Other parameters employed are $\gamma = 0.1$ and $d=3$. The numerical data (symbols) corresponds to the simulation of the corresponding Langevin equation---squares being for $r=0.01$ and diamonds for $r=0.001$, while the theoretical curves (solid lines) correspond to the numerical integration of Eqs.~\eqref{ch4_eq:ODESinnerc} and \eqref{ch4_eq:ODESinnerh}. The vertical lines on both panels mark the bath
    temperature $X_g$ at which the heat capacity reaches its maximum in the reheating program. The inset on the right panel shows the anomalous, diverging behaviour of the absolute value of the apparent heat capacity in logarithmic scale for both cooling and heating, captured by the black dashed line that stands for Eq.~\eqref{ch4_eq:ODESinnerc-approx}}
        \label{ch4_fig:hysteresis-mol}
\end{figure}

Figure~\ref{ch4_fig:hysteresis-mol} is analogous to the plots from Figure~\ref{ch4_fig:hysteresis-gran} for the case of the molecular fluid with non-linear drag. Its left panel shows both the numerical simulations of the Langevin equation and the boundary layer solution for a full hysteresis cycle. Similarly to the granular gas case, our boundary layer solution captures very well the simulation data. On the right panel, the behaviour of the associated apparent heat capacity of the molecular fluid is displayed. In the reheating curve, the typical maximum that may be used to define a glass transition temperature is neatly observed. Interestingly, in the cooling curve, an anomalous behaviour emerges, the apparent heat capacity increases instead of going to a constant. This anomalous behaviour stems from the singular behaviour for small $X$ of the dynamic equation~\eqref{ch4_subeq:ODESinnerc-Y} for $Y$ in the cooling protocol, and it is better discerned in the inset of the right panel in Fig.~\ref{ch4_fig:hysteresis-mol}---which shows a zoom of the very low temperatures region. Specifically, one has that
\begin{align}\label{ch4_eq:ODESinnerc-approx}
C=\frac{dY}{dX} &\sim  2\gamma(d+2)\frac{(Y^{\Frz})^2}{X^{1/2}} (1+a_2^{\Frz}), \quad X\ll 1 ,
\end{align}
which diverges as $X^{-1/2}$. Such diverging behaviour applies as well for the reheating window when considering the $X \ll 1$ limit in Eq.~\eqref{ch4_eq:ODESinnerh} This has to be contrasted with the behaviour for the granular gas: Eq.~\eqref{ch4_eq:C-low-temp-gran} tells us that $C$ goes to a constant for the granular gas---consistently with the results reported in Fig.~\ref{ch4_fig:hysteresis-gran}. 

\subsection{\label{ch4_subsubsec:normal_heating_curve}Normal heating curve}
In order to deepen our understanding of the hysteretic behaviour in reheating, as we cannot gather further analytical insights from the scaled evolution equations, a regular perturbation expansion can be carried out---analogous to the one for the cooling process. By simply substituting $r_c$ with $-r_h$ in Eq.~\eqref{ch4_eq:pert-gran}, we obtain
\begin{subequations}
    \label{ch4_eq:outer-gran}
            \begin{align}
                \theta &= \theta_{\st}- \frac{2\,r_h}{3\theta_{\st}^{1/2}}  \left[1+\frac{3}{16} \; a_2^{\st} \left( 1+\frac{1}{B} \right) \right]^{-1}+O(r_h^2),  \label{ch4_subeq:outer_granular}
                \\
                a_{2} &= a_2^{\st}+\frac{r_h \,a_2^{\st}}{\,B\,\theta_{\st}^{3/2}}  \left[1+\frac{3}{16} \; a_2^{\st} \left( 1+\frac{1}{B} \right) \right]^{-1}+O(r_h^2). 
            \end{align}
\end{subequations}
for the granular gas, and
\begin{subequations}
    \label{ch4_eq:eq:perturbative-mol-heat-o1}
        \begin{align}
            \theta &=\theta_{\st} -  \frac{r_h}{2\theta_{\st}^{1/2}}\, \frac{1+\gamma (d+6)}{\left[1 + \gamma (d+4)\right]^2 - 2\gamma^2
      (d+4)} +O(r_h^2), \label{cu4_subeq:perturbative-mol-heat-o1-a}
            \\
            a_2 &= +\frac{r_h}{\theta_{\st}^{3/2}}\frac{\gamma}{\left[1 + \gamma (d+4)\right]^2 - 2\gamma^2 (d+4)} +O(r_h^2),
        \end{align}
\end{subequations}
for the molecular fluid. These perturbative expressions are expected to be valid for not too low temperatures $\theta_{\st}\gg r_h^{2/3}$, i.e., over the outer layer---using once more the terminology of boundary layer theory. The above equations depend on the heating program $r_h$, but not on the previously applied cooling program with cooling rate $r_c$. In other words, if we start the heating process from different initial frozen temperatures $\theta^{\Frz} = Y^{\Frz} r_c^{2/3}$ corresponding to different values of $r_c$ but reheat with a common rate $r_h$, we expect to approach the behaviours in Eqs.~\eqref{ch4_eq:outer-gran} and ~\eqref{ch4_eq:eq:perturbative-mol-heat-o1} once the system reaches the outer layer.

\begin{figure}
    {\centering 
    \includegraphics[width=2.65in]{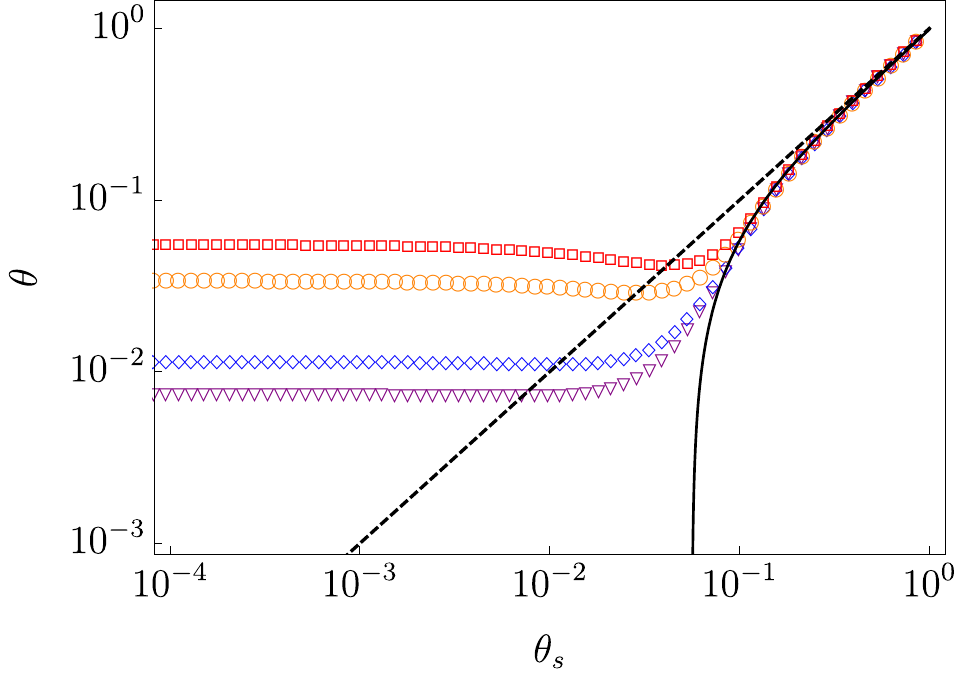}
    \includegraphics[width=2.65in]{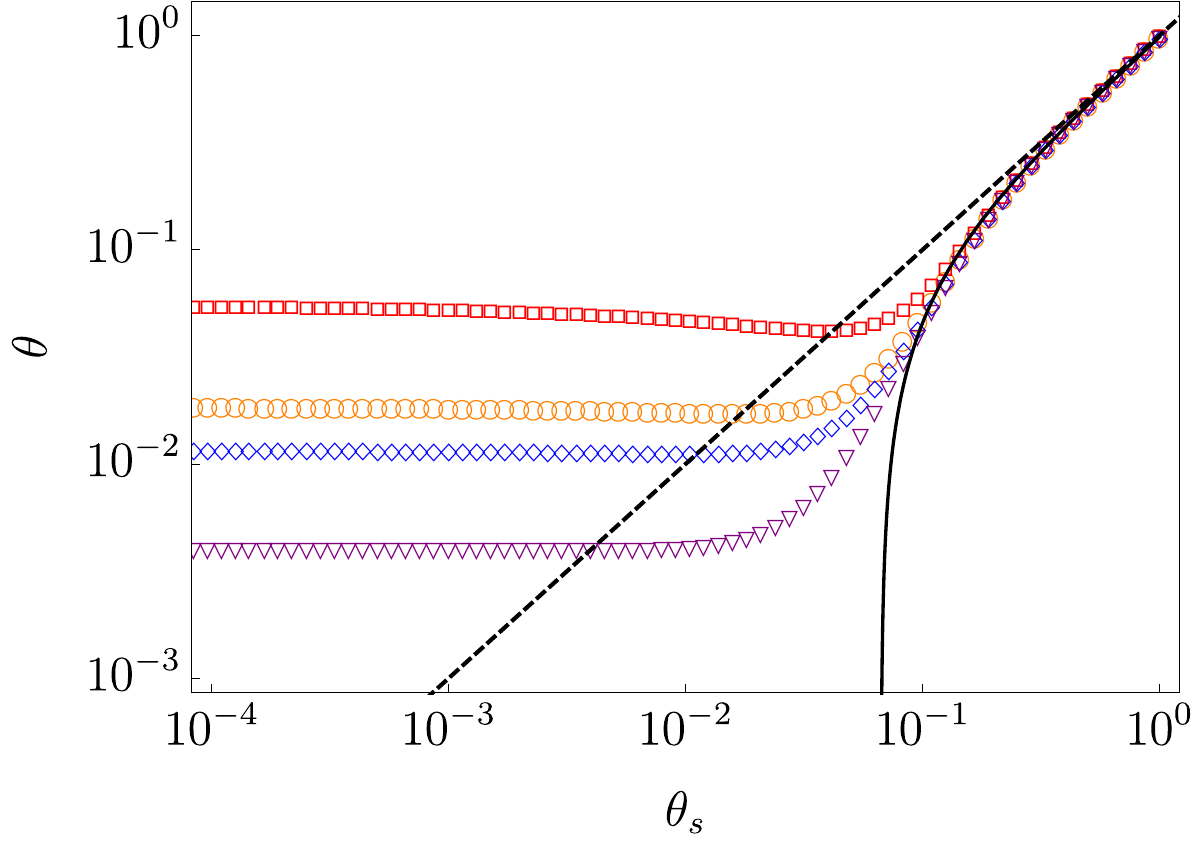}}
    \caption{Hysteresis cycles for reheating with rate $r_h$ from the frozen states corresponding to different cooling rates $r_c$. Specifically, the plots shown are for the granular gas with parameter values $\alpha = 0.9$ and $d = 3$ (left), and a molecular fluid with non-linear drag with parameters $\gamma = 0.1$ and $d = 3$ (right). In both panels, all reheating curves correspond to $r_h = 0.01$, and the different cooling rates employed are: $r_c = 0.05$ (red squares), $0.01$ (orange circles), $0.005$ (blue diamonds) and $0.001$ (purple triangles). Symbols correspond to simulation data---DSMC data for the left panel, and the numerical integration of the corresponding Langevin equation for the right panel---whereas the  solid {lines} correspond to the perturbative expressions for the normal curve in Eq.~\eqref{ch4_eq:outer-gran} (left) and Eq.~\eqref{ch4_eq:eq:perturbative-mol-heat-o1} (right).}
        \label{ch4_fig:univ-heating-curves}
\end{figure}

The behaviour just described above is illustrated in Fig.~\ref{ch4_fig:univ-heating-curves}: despite having different cooling programs, all the reheating curves tend towards a universal curve, independent of $r_c$, for high enough values of the bath temperature. The perturbative expressions from Eqs.~\eqref{ch4_eq:outer-gran} and~\eqref{ch4_eq:eq:perturbative-mol-heat-o1} explain why the kinetic temperature crosses the NESS curve $\theta=\theta_{\st}$ in reheating. The universal curve for the temperature from the outer layer---which corresponds to the perturbative expression~\eqref{ch4_eq:eq:perturbative-mol-heat-o1} for a given $r_h$---is always below the NESS curve---whereas the cooling curves always lie above the NESS curve, as illustrated by Figure~\ref{ch4_fig:frozen-temp-gran}. The latter is not a coincidence, since it stems from the sign of $\theta - \theta_{\st}$ from the perturbative expressions~\eqref{ch4_eq:eq:perturbative-mol-heat-o1} and \eqref{ch4_eq:first-order-sols-granular}, for either the heating or cooling branches respectively.

From a physical standpoint, the overshoot of the instantaneous NESS curve may be understood by taking into account that the kinetic temperature $\theta$ always lags behind the bath temperature $\theta_{\st}$ during the entire time evolution of the hysteresis cycle. In the cooling protocol, this implies that the deviation $\theta- \theta_{\st}$ increases as $\theta_{\st}$ decreases, i.e. as the characteristic relaxation time of $\theta$ increases. In the reheating protocol, $\theta$ initially decreases, specifically as long as $\theta_{\st} \ll \theta$. It is not until $\theta_{\st} \simeq \theta$ that $\theta$ starts to increase but, at the time it does, as its characteristic relaxation time is still large compared to the reheating time, we will have $\theta_{\st}>  \theta$ until it reaches the instantaneous NESS. The latter argument also explains why the crossing points between the NESS curve and the simulation data from Figure~\ref{ch4_fig:univ-heating-curves} are very close to the minimum---i.e. the point at which $d\theta/d\theta_{\st}=0$---of each dataset, since $d\theta / d\theta_{\st} \approx 0$ for $\theta \sim \theta_{\st}$ according to the evolution equation for the temperature~\eqref{ch4_subeq:granularSonine-theta}.

Let us have a more detailed look at the dependence of the reheating curves on the rate $r_c$ of the previous cooling program. In Fig.~\ref{ch4_fig:univ-heating-curves}, it is observed that the reheating curves develop a neat dimple as the cooling rate $r_c$ is increased. This can be understood by going back to the evolution equations for the inner region---Eq.~\eqref{ch4_eq:evolscaled-heating} for the granular gas and Eq.~\eqref{ch4_eq:ODESinnerh} for the molecular fluid---which continue to be valid for $r_c\ne r_h$, but with the new boundary conditions 
\begin{equation}\label{ch4_eq:bc-reheating-rc-ne-rh}
    Y(0) = (r_c/r_h)^{2/3} Y^{\Frz}, \quad a_2(0) = a_2^{\Frz}.
\end{equation}
Note that the boundary conditions in Eq.~\eqref{ch4_eq:evolscaled-heating-bc} are the particularization of Eq.~\eqref{ch4_eq:bc-reheating-rc-ne-rh} to the case $r_h=r_c$. As $r_c/r_h$ increases, we have that $Y(0)=(r_c/r_h)^{2/3} Y^{\Frz}$ also increases and then the initial decrease becomes more noticeable, giving rise to the neat minimum (the dimple) shown by the uppermost curve in Fig.~\ref{ch4_fig:univ-heating-curves}. For small values of $r_c/r_h$, this initial decrease is barely noticeable, and the tendency towards the normal curve is almost horizontal, as shown by the lowermost curve in Fig.~\ref{ch4_fig:univ-heating-curves}. Still, it must be noted that $\theta$ always presents a minimum as a function of $\theta_{\st}$, because ${d\theta}/{d\theta_{\st}}$ is negative for very low temperatures, whereas it is positive, since ${d\theta}/{d\theta_{\st}}\to 1$, for high temperatures.

The approach to a unique curve, independent of the previous cooling program, of the systems upon reheating is similar to the behaviour found in models described by master equations. Therein, it has been analytically proved that there exists a universal \textit{normal} curve that is the global attractor of the dynamics for heating processes~\cite{brey_normal_1993,brey_dynamical_1994,brey_dynamical_1994-1,prados_hysteresis_2000,prados_glasslike_2001}. The expressions in Eqs.~\eqref{ch4_eq:outer-gran} and~\eqref{ch4_eq:eq:perturbative-mol-heat-o1} may be thus regarded as the regular perturbation expansions of a similar normal curve in our systems. 

\begin{figure}
    {\centering 
    \includegraphics[width=2.65in]{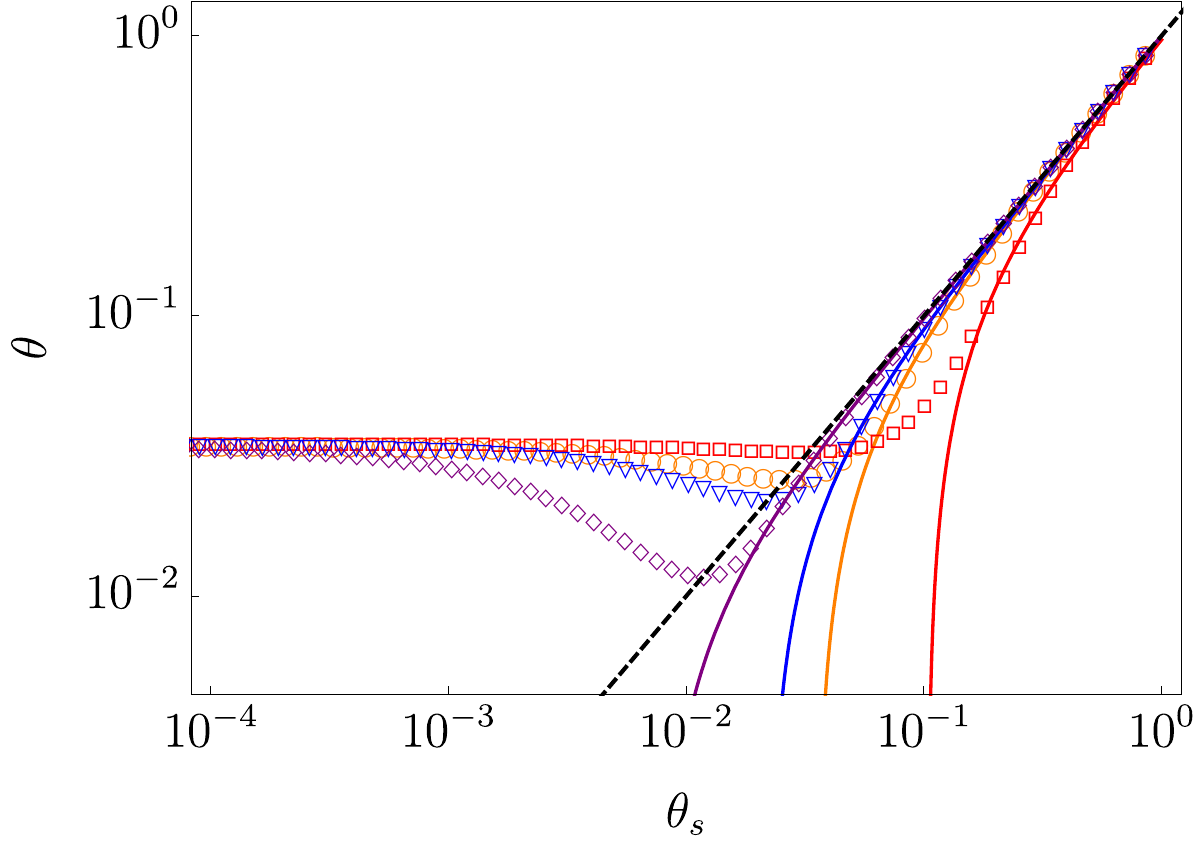}
    \includegraphics[width=2.65in]{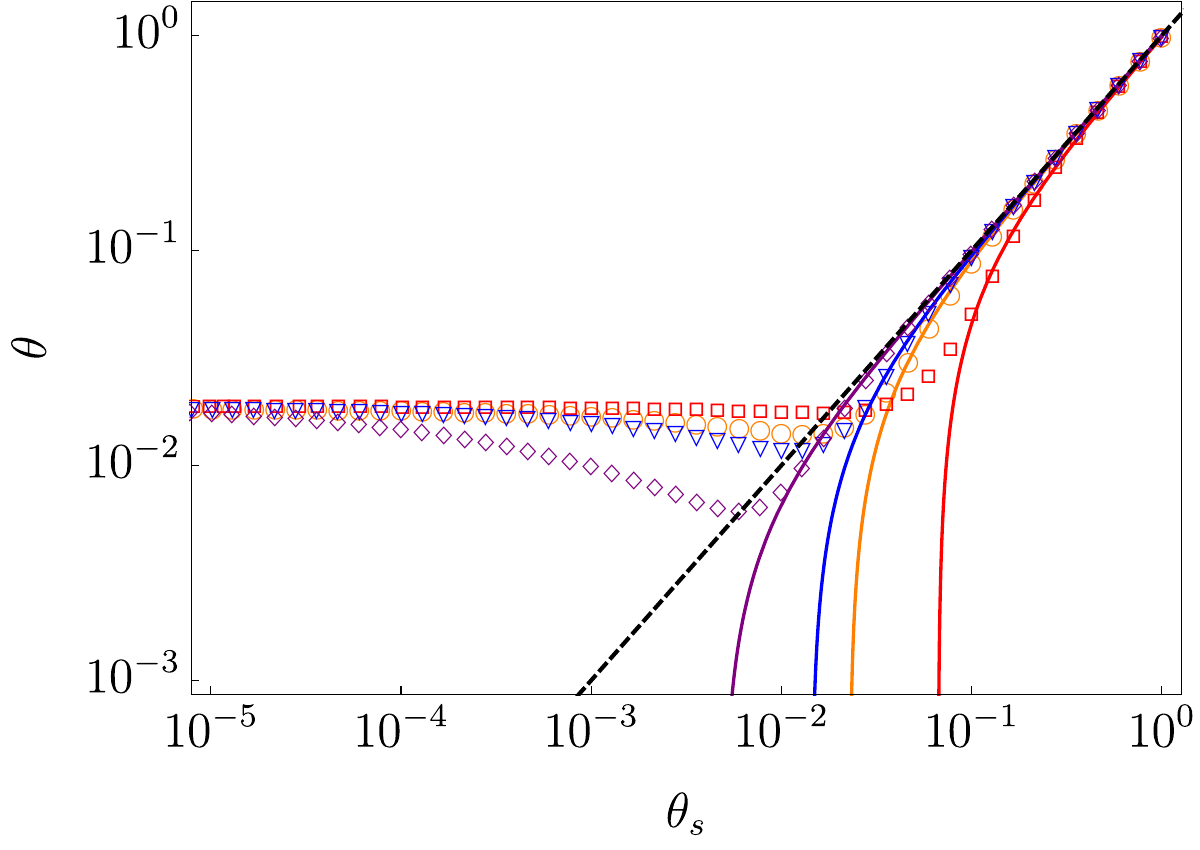}}
    \caption{{Hysteresis cycles for reheating with different rates $r_h$ from the common frozen state corresponding to a given value of $r_c$, for a granular gas with parameter values $\alpha = 0.9$ and $d = 3$ (left) and a molecular fluid with non-linear drag, with parameters $\gamma = 0.1$ and $d = 3$ (right). Specifically, the plotted simulation data---DSMC data for the left panel, and the numerical integration of the corresponding Langevin equation for the right one---correspond to $r_c=0.01$ {with different} reheating rates: $r_h = 0.05$ (red squares), $0.01$ (orange circles), $0.005$ (blue diamonds) and $0.001$ (purple triangles). The solid curves correspond to the expressions from Eq.~\eqref{ch4_eq:outer-gran} (left) and Eq.~\eqref{ch4_eq:eq:perturbative-mol-heat-o1} (right) {for each value of $r_h$; from left to right, $r_h$ increases from $0.001$ to $0.05$.}}}
        \label{ch4_fig:further-univ-heating-curves}
\end{figure}

The tendency towards the normal curve is further illustrated in Fig.~\ref{ch4_fig:further-univ-heating-curves}, in which we consider reheating with different values of $r_h$ from a common frozen state, corresponding to one value of $r_c$. For all the values of $r_h$, the hysteresis cycles cross the instantaneous NESS curve $\theta=\theta_{\st}$, since the corresponding normal curves always lie below it. Moreover, as $r_h$ increases, we go further away from the NESS, since the normal heating curve is more distant from the corresponding NESS curve. Note that the heating curve develops a neat dimple as $r_h$ decreases---consistently with our discussion above, which told us that the minimum gets more marked as $r_c/r_h$ increases.

Our theory predicts that $X_g$ only depends on the ratio $r_c/r_h$, since the evolution in scaled variables in reheating is governed by Eqs.~\eqref{ch4_eq:evolscaled-heating} for the granular gas and Eqs.~\eqref{ch4_eq:ODESinnerh} for the molecular fluid, with the boundary conditions in Eq.~\eqref{ch4_eq:bc-reheating-rc-ne-rh}. Only the latter introduces a dependence on the rates; namely on their ratio $r_c/r_h$. Regardless of the value of such ratio, $X_g$ always remains of the order of unity, which entails that the glass transition temperature $\theta_g=r_h^{2/3} X_g$ is basically proportional to $r_h^{2/3}$ in both systems. We also highlight that, for both systems, $X_g$ is of the same order as the fictive temperature $X_f$ from Eq.~\eqref{ch4_eq:thetaf-power-law}, which is consistent, as both temperatures shall give a qualitative account of the glass transition.

\part{Shortcuts between stationary states for overdamped \mbox{harmonic} dynamics}
\label{part:shortcuts}

\setlength{\epigraphwidth}{5in} 
\setlength\epigraphrule{0pt}
\epigraph{\small \itshape Todo dura siempre más de lo que debería.}{--- Julio Cortázar, \textit{Rayuela}}
\chapter{Thermal brachistochrones for overdamped Gaussian processes}
\chaptermark{Thermal brachistocrone}
\label{ch:thermal_brachistochrones}
\newcommand\mycom[2]{\genfrac{}{}{0pt}{}{#1}{#2}}

The term ``brachistochrone'' stems from Ancient Greek \textit{brákhistos khrónos}---shortest time. The original brachistochrone problem, which refers to the curve of the fastest descent under gravity, is already mentioned in Sec~\ref{ch1_subsec:control-theory}. For the problems considered throughout this second part of the thesis, we use the term brachistochrone as the optimal protocol for the control parameters---recall the notation $\bm{k}$ from Eq.~\eqref{ch1_eq:dynamical-system}---of the system of concern that provides the fastest connection between two given stationary states. The problem of determining the brachistochrone belongs to the field of optimal control theory (OCT) from Sec.~\ref{ch1_subsec:control-theory}, in which we are not only interested in beating the natural relaxation timescale of the system, but also ascertaining by how much we can do it.

In this chapter, we consider a paradigmatic system in non-equilibrium statistical mechanics: the overdamped Brownian dynamics for a system of harmonic oscillators, which are all coupled to a common thermal bath, the temperature $T$ of which we control externally. We employ Pontryagin's Maximum Principle (PMP) to determine the thermal brachistochrone connecting arbitrary equilibrium states. Due to the linearity of the dynamic equations for this system---they belong to the class of linear-Pontryagin systems, optimal protocols are exclusively of the bang-bang type, as highlighted in Sec.~\ref{ch1_subsubsec:bang-bang}, i.e. the bath temperature has to switch alternatively between its minimum $T_{\text{min}}$ and maximum $T_{\text{max}}$ values allowed. We study the minimum connection time over such brachistochrone. Remarkably, it depends on the dimensionality of the physical system, increasing with its dimension $d$. Such unavoidable price to pay when increasing $d$ persists even for almost fully degenerate oscillators---in which the harmonic trap is almost isotropic, and in the limit of having infinite capacity for heating the system, i.e. $T_{\text{max}}\rightarrow +\infty$. We understand the consequences of this phenomenon by (i) considering the unphysical scenario of having negative temperatures for $T_{\max} \to +\infty$ and (ii) on a physical basis from the point of view of information geometry, by comparing with the geometric speed limit from Eq.~\eqref{ch1_eq:speed-limit}.

The structure of the remaining sections of this chapter goes as follows. Section~\ref{ch5_sec:model} is devoted to introducing the model, its mathematical framework, and how to apply PMP for the problem of concern. Then, in Sec.~\ref{ch5_sec:optimal-protocols}, we specifically study how to construct optimal thermal protocols, with special emphasis on the $T_{\text{max}}\rightarrow +\infty$ limit, which allows us to further simplify the calculations. Finally, in Sec.~\ref{ch5_sec:degenerate-case} we study the unavoidable price of increasing the dimensionality of the system in the case of having almost fully degenerate oscillators, and the rich phenomenology that emerges therein.

\section{\label{ch5_sec:model}Model: overdamped harmonic oscillators}
The model that we consider here corresponds to the dynamics of an overdamped Brownian particle in $d$ dimensions in contact with a thermal bath at temperature $T(t)$, submitted to harmonic confinement in each direction---the case $d>3$ can be thought of as several, non-interacting, confined Brownian particles.\footnote{Let us remark that our assumption of uncoupled degrees of freedom implies no loss of generality since, for general harmonic potentials, it is always possible to change variables
to normal modes where our description applies.} Specifically, such dynamics is accounted via a Gaussian process in $d$ dimensions---such as the ones introduced in Sec.~\ref{ch1_subsubsec:ornstein-uhlenbeck}, with the matrices $\mathbb{A}(t)$ and $\mathbb{D}_0(t)$ from Eq.~\eqref{ch1_eq:coeffs-ou} being given by
\begin{equation}\label{ch5_eq:matrices-overdamped}
    \mathbb{A}(\cancel{t}) = \frac{1}{\gamma}\left(\begin{array}{cccc} 
        k_1 & 0 & \cdots & 0 \\
        0 & k_2 & \cdots & 0 \\
        \vdots & \vdots & \ddots & \vdots \\
        0 & 0 & \cdots & k_d 
        \end{array}\right), \quad \mathbb{D}_0(t) = \frac{k_BT(t)}{\gamma} \mathbb{I}_d.
\end{equation}
In this case, $\bm{r}$ stands for the position of the Brownian particle with $r_j$ being its projection onto the $j$-th dimension, $\gamma$ constitutes the friction coefficient 
as similarly defined in Eq.~\eqref{ch3_eq:LE-overdamped} from chapter~\ref{ch:nonequilibrium_attractor}, $k_B$ is the Boltzmann constant, and $k_j$ stands for the stiffness of the harmonic trap along the corresponding direction.\footnote{For the optimisation problem considered here, we keep the stiffnesses of the trap static while varying the bath temperature at will.} Such harmonic trap is accounted via the Hamiltonian
\begin{equation}
    H(\bm{r}) = \frac{1}{2}\sum_{j=1}^{d}k_j r_j^2, \quad k_1 \leq k_2 \leq ... \leq k_d,
\end{equation}
where the ordering of the stiffnesses $k_j$ is chosen for the sake of clarity in the forthcoming discussion, without the loss of generality of our results. The temperature of the bath is externally controlled at will in such a way that it is possible to devise any temperature program. Note that the form of the matrices $\mathbb{A}$ and $\mathbb{D}_0(t)$ from Eq.~\eqref{ch5_eq:matrices-overdamped} and the Hamiltonian $H(\bm{r})$ ensure that the fluctuation-dissipation relation \eqref{ch1_eq:fluctuation-dissipation} is satisfied. Hence, for a fixed temperature, the canonical equilibrium distribution $P_{\text{eq}}(\bm{r}) \propto \text{exp}(-\beta H(\bm{r}))$ corresponds to the only stationary PDF of the Fokker-Planck equation. Moreover, not only is such PDF Gaussian, but factorises into the product of $d$ independent Gaussian distributions, one for each degree of freedom---which is a consequence of the diagonal form of the matrices $\mathbb{A}$ and $\mathbb{D}_0(t)$.

Now, as mentioned in Sec.~\ref{ch1_subsubsec:ornstein-uhlenbeck}, the linearity of the corresponding Fokker-Planck equation guarantees that Gaussian states remain Gaussian for the whole evolution of the system. Therefore, provided a Gaussian initial condition centered at the origin, the latter ensuring that $\left<\bm{r}\right>(t)=\bm{0}$ at all times, it suffices to study the variances---which are also the diagonal second moments of the matrix $\left<\mathbb{R}_2\right>$ from Eq.~\eqref{ch1_eq:second-moments}---of the distribution in each dimension, $z_j (t) \equiv \left<r_j^2\right>(t), \ j=1,...,d$, to fully characterise the dynamics---note that $\left<r_jr_l\right>(t)=0$ for $j\neq l$ at all times in this case. In fact, the specific form of the PDF from Eq.~\eqref{ch1_eq:solution-pdf-ou} simplifies to
\begin{equation}\label{ch5_eq:Gaussian-shape-2}
    P(\bm{r},t)=\prod_{j=1}^d \frac{e^{-\frac{r_j^2}{2z_j(t)}}}{\sqrt{2\pi z_j(t)}}, \quad \forall j, \, \forall t.
\end{equation}
Now, due to the uncoupling of the harmonic confinements---which stems once again from the forms of $\mathbb{A}$ and $\mathbb{D}_0(t)$, the evolution equations for the $z_j(t)$ are also uncoupled,\footnote{We must remark that, although the variances $z_j$ are independent of each other, they are coupled to the same, unique bath temperature $T$. Thus, by varying $T$ we are perturbing all of them at the same time.} and we have that
\begin{equation}
    \label{ch5_eq:Gaussian-shape}
    \gamma \frac{d}{dt} z_j (t) = -2 k_j z_j(t) + 2 k_B T(t). 
\end{equation} 
Note that the above equation constitutes the particularisation of Eq.~\eqref{ch1_eq:second-moments}, by subtitution of Eq.~\eqref{ch5_eq:matrices-overdamped}. Now, in order to simplify the forthcoming discussion, we introduce the dimensionless variables
\begin{equation}
    t^* = \frac{k_1}{\gamma}t, \quad z_j^* = \frac{k_1}{k_B T_i}z_j, \quad T^* = \frac{T}{T_i}, \quad k_j^* = \frac{k_j}{k_1},
\end{equation}
with $T_i$ being the initial temperature, and we drop the asterisks from now on. That is, time is rescaled with the characteristic timescale of the slowest oscillator, the dynamic variables with the equilibrium variance for $k_1$, the temperature with its initial value, and the stiffnesses with the lowest value $k_1$. Then, we have the evolution equations 
\begin{equation}
\label{ch5_eq:ev_z}
\frac{d}{dt} z_j (t) = -2 k_j z_j(t) + 2 T(t), \quad j=1,...,d,
\end{equation}      
with $1=k_1\leq k_2 \leq \cdots \leq k_d$. The system of equations \eqref{ch5_eq:ev_z} constitutes the fundamental law that governs the dynamics of the confined Brownian particle---or the non-interacting Brownian particles in a common heat bath. 

In this chapter, we are interested in devising a protocol for the bath temperature allowing us to connect two arbitrary equilibrium states at different temperatures, $T(0)=T_i=1$ and $T(t_f)=T_f$, in the shortest amount of time. Therefore, we are dealing with a time-optimisation problem, in which our system belongs to the class of linear-Pontryagin systems depicted in Sec.~\ref{ch1_subsubsec:bang-bang}---the evolution equations for the variances~\eqref{ch5_eq:ev_z} constitute a particular case of Eq.~\eqref{ch1_eq:linear-pontryagin}. Thus, PMP guarantees not only the existence of optimal protocols, but also that these belong to the class of bang-bang protocols---i.e. singular and Euler-Lagrange protocols are excluded. That is, the temperature has to alternatively switch between its minimum and maximum values allowed. Hence, it is useful to solve the dynamic equations~\eqref{ch5_eq:ev_z} for a constant value $T$ of the temperature. Its general solution, starting from the initial condition $z_{j,i}$ at $t_i$, follows an exponential relaxation of the form
\begin{equation}
    \label{ch5_eq:ev-op}
    \mathcal{E}_{k_j,T}^{(\Delta t)} \left( z_{j,i}\right) \equiv z_j(t)= \frac{T}{k_j} + \left( z_{j,i} - \frac{T}{k_j} \right) e^{-2k_j \Delta t},
\end{equation} 
where $\Delta t \equiv t-t_i$ and $\mathcal{E}_{k_j,T}^{(\Delta t)}$ stands for the time-evolution operator for a time interval $\Delta t$ of a degree of freedom harmonically confined through a stiffness $k_j$ under constant temperature $T$.

\section{\label{ch5_sec:optimal-protocols}Optimal thermal protocols}
Our goal is to obtain the protocol that connects two equilibrium states in the shortest time. Specifically, in our dimensionless variables, the initial temperature is $T_i=1$, and thus, final values $T_f$ higher or lower than unity represent heating or cooling processes, respectively. The solution to this optimisation problem depends on the constraints considered for the external control, i.e. on the constraints on the temperature $T(t)$. A physical bound arises from below since the absolute temperature cannot be lower than zero. However, experimental limitations may produce the emergence of tighter bounds, in such a way that the temperature cannot exceed certain extreme values, i.e. $0 \leq T_{\min} \leq T \leq T_{\max}$. 

From a theoretical perspective, the shortest finite-time protocol we are looking for  is qualitatively different from the direct step process where at $t=0^+$ the control is switched to the target value for the temperature, $T_f$, which is followed by an exponential relaxation of the variances with natural timescales given by $(2k_j)^{-1}$. While the direct STEP process take is an asymptotic process that takes an infinite time, our optimal protocol performs the connection in a finite time. However, from an applied perspective, the optimal protocol is specially appealing when the optimal connection time beats the aforementioned characteristic timescales.  

Section~\ref{ch5_subsec:finite-Tmax} is devoted to developing the general framework of thermal bang-bang protocols for finite, arbitrary values of both $T_{\max}$ and $T_{\min}$, whereas in Sec.~\ref{ch5_subsec:infinite-capacity-limit} we relax one of these conditions. Therein, we assume an infinite heating power, i.e. the limit as $T_{\max}\to\infty$. In any case, although we put forward the general situation for finite $T_{\min}$, for the sake of simplicity we take $T_{\min}=0$ in the presentation of the results.

\subsection{\label{ch5_subsec:finite-Tmax}Finite heating power}
For the linear-Pontryagin system considered in this chapter, the number of jumps between bounds for bang-bang protocols is $d-1$, entailing $d$ stages in the interval $0<t<t_f$, and thus given by the dimensionality of the problem, as argued in Ref.~\cite{pontryagin_mathematical_1987,prados_optimizing_2021}.\footnote{These $d-1$ jumps between bounds have to be supplemented with two additional jumps: one at $t=0^+$ from $T_i=1$ to either $T_{\max}$ or $T_{\min}$, and the final jump at $t = t_f^-$ to $T_f$.} 
In this chapter, we explore in depth the physical implications of this general result in the relevant context of harmonically trapped Brownian particles. In the following, we resort to our knowing the optimal control being of bang-bang type to obtain both the protocol itself and the optimal connection time.

\begin{figure} 
    \begin{center}
    \includegraphics[width=\textwidth]{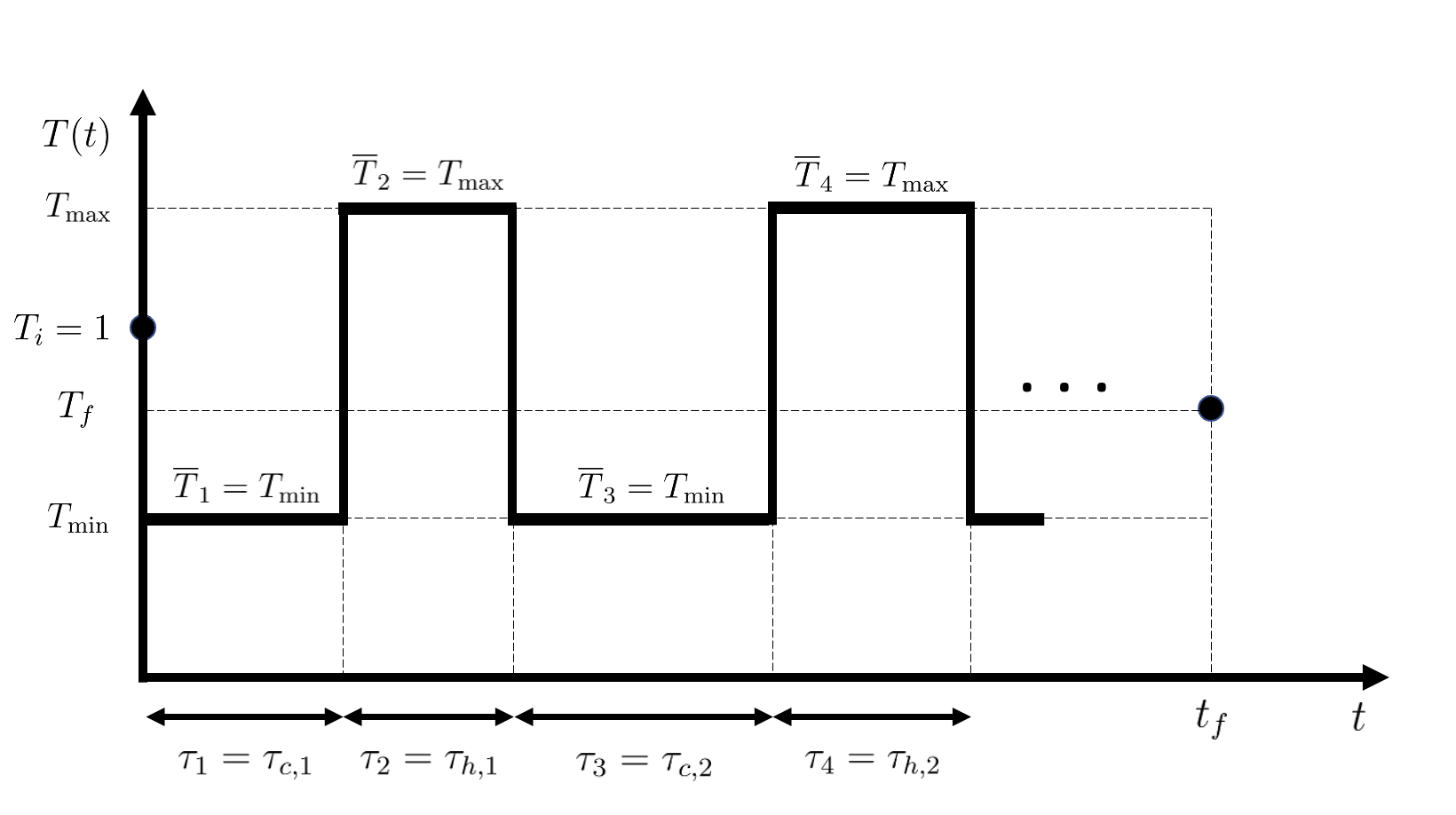}
    \includegraphics[width=\textwidth]{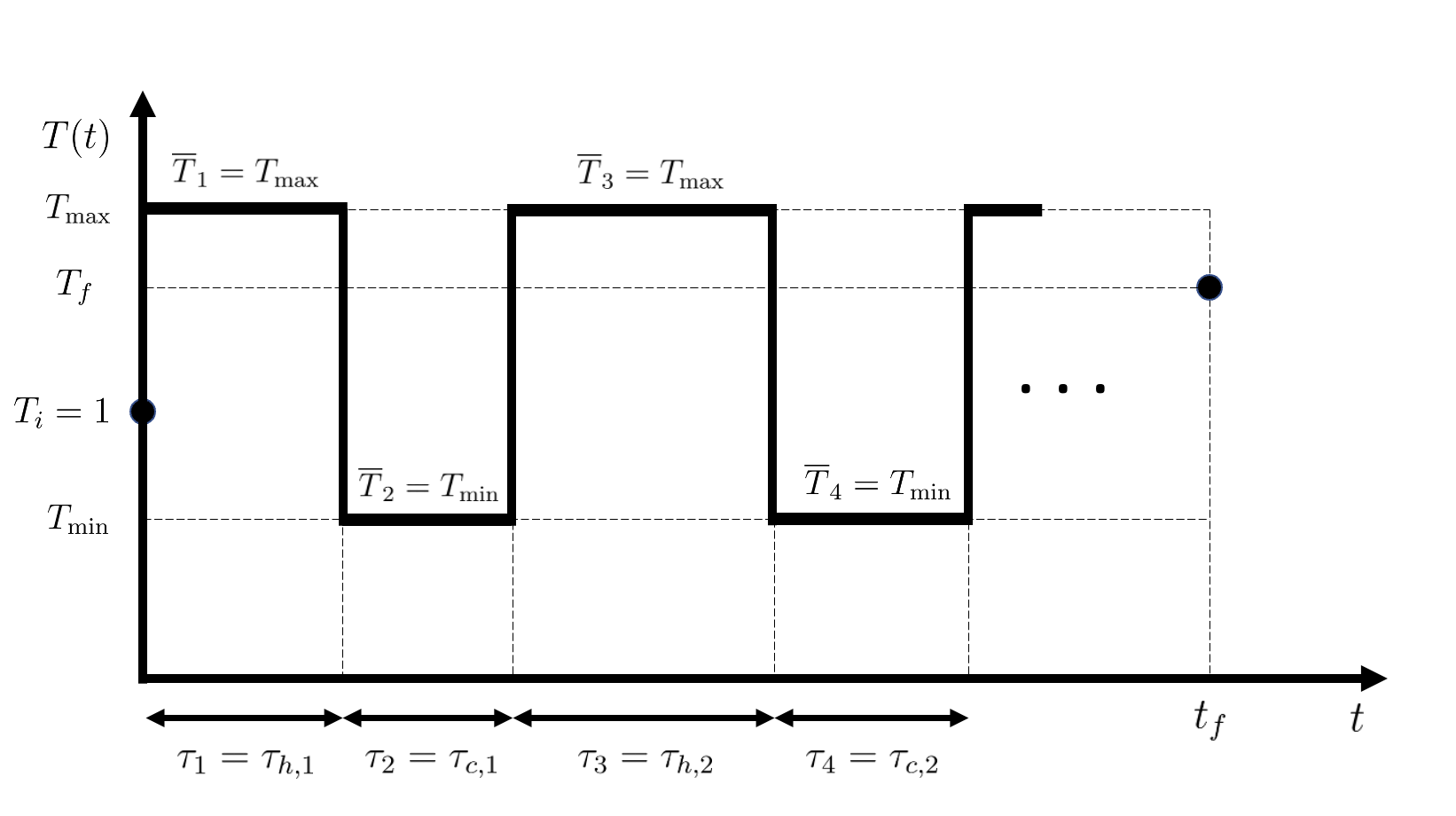}
    \caption{\label{ch5_fig:sketch-Tmax-fin} Sketch of the optimal bang-bang control able to connect two equilibrium states of a $d$-dimensional harmonic oscillator. The optimal control comprises several bangs, i.e. time windows alternating maximum and minimum values of the bath temperature. The character of the first bang---the value of its bath temperature $\overline{T}_1$---is determined by the target temperature $T_f$ being either lower (cooling process, top panel) or higher (heating process, bottom panel) than the initial one $T_i=1$. The different parameters involved in the protocol, namely temperatures and time duration of the bangs, along with the notation used in Eq.~\eqref{ch5_eq:phi-fin} are displayed. The shortest connection time $t_f= \sum_{j=1}^d \tau_j$ stems from the solution of Eq.~\eqref{ch5_eq:comp}.}
    \end{center}
\end{figure}

As discussed above, the shortest connection implies $d$ time windows, with $d-1$ consecutive jumps between the limiting values of the temperature. The value of the temperature along the first time window, either $T_{\max}$ or $T_{\min}$, determines the type of process performed: heating or cooling, respectively. For instance, let us consider a heating process, $T_f>1$: the optimal protocol involves $d$ time windows, alternating heating---at $T_{\max}$---and cooling---at $T_{\min}$---stages, starting with heating at $T_{\max}$. Note that, since the system is assumed to be at equilibrium both at the initial and final times, the boundary conditions $z_{j}(0)=1/k_j$ and $z_{j}(t_f)=T_f/k_j$ for the evolution hold. Let $\tau_{c,j}$ and $\tau_{h,j}$ be the durations of the $j$-th cooling and heating stages, respectively. The optimal bang-bang process is obtained by solving the system of equations
\begin{equation}
\label{ch5_eq:comp-h}
\underbrace{\left( \cdots \circ \mathcal{E}_{k_j,T_{\max}}^{\tau_{h,2}} \circ \mathcal{E}_{k_j,T_{\min}}^{\tau_{c,1}} \circ \mathcal{E}_{k_j,T_{\max}}^{\tau_{h,1}} \right)}_{{\textnormal{composition of }} d {\textnormal{ operators }}}
\left(\frac{1}{k_j}\right)=\frac{T_f}{k_j}, \quad j=1,\ldots,d,
\end{equation} 
for heating processes, $T_f>1$, or
\begin{equation}
\label{ch5_eq:comp-c}
\underbrace{\left( \cdots \circ \mathcal{E}_{k_j,T_{\min}}^{\tau_{c,2}} \circ \mathcal{E}_{k_j,T_{\max}}^{\tau_{h,1}} \circ \mathcal{E}_{k_j,T_{\min}}^{\tau_{c,1}} \right)}_{{\textnormal{composition of }} d {\textnormal{ operators }}}
\left(\frac{1}{k_j}\right)=\frac{T_f}{k_j}, \quad j=1,\ldots,d,
\end{equation} 
for cooling processes, $T_f<1$. Multiplying~\eqref{ch5_eq:comp-h} and \eqref{ch5_eq:comp-c} by $k_j$, we get
\begin{equation}
\label{ch5_eq:comp}
\varphi(k_j,T_{\min},T_{\max},\boldsymbol{\tau})=T_f, \qquad j=1,\ldots,d,
\end{equation}
where the above function can be easily built using the definition of the evolution operator \eqref{ch5_eq:ev-op} and $\boldsymbol{\tau}$ is a vector of dimension $d$ comprising all durations of the elemental stages in the bang-bang protocol. Equation~\eqref{ch5_eq:comp} constitutes a system of $d$ equations with $d$ unknown variables---corresponding to the time durations from $\boldsymbol{\tau}$. Note that, denoting by $\overline{T}_j$---alternatively $T_{\max}$ or $T_{\min}$---and $\tau_j$ the temperature and the duration of the $j$-th elementary stage of the bang-bang process, respectively, the function $\varphi$ can be generally expressed as 
\begin{equation}
\label{ch5_eq:phi-fin}
\varphi(k,T_{\min},T_{\max},\boldsymbol{\tau}) =  \sum_{n=1}^{d+1} \left( \overline{T}_{n-1} - \overline{T}_n\right) \exp \left( -2 k \sum_{m=n}^{d}\tau_m \right),
\end{equation}
for arbitrary $d$. We have defined $\overline{T}_0\equiv 1$ and $\overline{T}_{d+1}\equiv 0$ to give a compact formulation. For an illustration of the optimal control protocol, and the notation employed for our general formulation in Eq.~\eqref{ch5_eq:phi-fin}, see Fig.~\ref{ch5_fig:sketch-Tmax-fin}.

In the following, we explicitly construct the optimal connections for $d=1$, $d=2$ and $d=3$. Not only is this done to be concrete, but also because it is an experimentally relevant situation for a colloidal particle in a harmonic trap. In the following, we keep a finite $T_{\max}$ as a parameter but, as previously mentioned, we choose $T_{\min}=0$ for the sake of simplicity. In each case, we mainly compute the function $\varphi(k,0,T_{\max},\boldsymbol{\tau})$ for both cooling, $T_f<1$, and heating, $T_f>1$, and briefly discuss the results that stem from the found solution when possible. A more thorough discussion on the behaviour of the minimum connection time in each case is carried out for the almost fully degenerate case in Sec.~\ref{ch5_sec:degenerate-case}---which allows for further analytical insights.

\begin{itemize}
    \item $\bm{d = 1}$: Considering first a cooling process, $T_f<1$, it is simple to get
    \begin{equation}
    \varphi(k,0, T_{\max},\boldsymbol{\tau}) =e^{-2 k t_f}
    \end{equation}
    for $d=1$, where $\boldsymbol{\tau}$ is just the duration of the cooling process, $t_f=\tau_{c,1}$. Directly applying Eq.~\eqref{ch5_eq:comp} and solving for $t_f$, we obtain the shortest cooling time  
    \begin{equation}
    \label{ch5_eq:tmin_1d}
    t_f=-\frac{1}{2} \ln T_f
    \end{equation}
    associated to the final temperature $T_f$---we recall that $k_1=1$. Since there is no heating stage for the fastest cooling protocol in the one-dimensional case, Eq.~\eqref{ch5_eq:tmin_1d} is independent of $T_{\max}$.\footnote{Therefore, it coincides with the result for the limit $T_{\max}\to\infty$ in Ref.~\cite{prados_optimizing_2021}.} The protocol leading to the shortest time to cool down the particle is reasonable from a physical point of view: put the system in contact with a thermal bath at zero temperature and wait until the target state is reached. Consistently, the connection time diverges when $T_f \to 0$, and monotonically decreases up to zero for $T_f \to 1 $; a qualitative behaviour that is expected for all $d$.
    
    For heating processes, $T_f>1$, we obtain
    \begin{equation}
    \varphi(k,0, T_{\max},\boldsymbol{\tau})=  \left(1- T_{\max} \right) e^{- 2 k t_f} + T_{\max},
    \end{equation}
    for $d=1$, where the only heating stage of duration $t_f$ represents the whole process, $t_f=\tau_{h,1}$. Solving Eq.~\eqref{ch5_eq:comp} for $t_f$ in this case yields
    \begin{equation}\label{ch5_eq:tf-d1-finiteTm}
    t_f= \frac{1}{2} \ln \left(\frac{T_{\max}-1}{T_{\max}-T_f} \right).
    \end{equation}
    Once more, the qualitative behaviour can be intuitively justified. The shortest time is identically zero just for $T_f=1$. It increases with $T_f$ until it diverges  for $T_f \to T_{\max}$.  Furthermore, in the limit $T_{\max} \to \infty$ the optimal time vanishes for all $T_f>1$. This describes an instantaneous heating process, consequence of the infinite capacity to heat up. This limit is further investigated in Sec.~\ref{ch5_subsec:infinite-capacity-limit}. It is worth remarking that Eq.~\eqref{ch5_eq:tf-d1-finiteTm} holds regardless of the value of $T_{\min}$, since the $d=1$ heating protocol only involves one stage with $T(t) = T_{\max}$.

    \item $\bm{d = 2}$: For $d=2$, two stages or elementary processes are needed. Therefore, the maximum (minimum) value of the temperature plays a role even for cooling (heating). As shown later on, the fact of tuning two variances to their corresponding target states implies an increment of the cost of the optimal process, in terms of the minimum time to achieve the connection---even in the $k_2 \rightarrow k_1$ limit, which we refer to as the almost degenerate case and lies at the core of Sec.~\ref{ch5_sec:degenerate-case}. Hence, the optimal connection time in a system with $d=2$, for arbitrary $k_2$, will be larger or equal than the corresponding one for $d=1$. On the one hand, for finite $k_2$, equality only holds for the trivial value $T_f=1$ which represents no change in the system, and it is of course instantaneous. On the other hand, the convergence to the results corresponding to $d=1$ can be also recovered when the limit $k_2 \to \infty$ is considered. Therein, the trap along the second spatial direction is completely rigid and thus always in equilibrium with vanishing variance. Such a limit is further analysed in Sec.~\ref{ch5_subsec:infinite-capacity-limit} for the case of infinite heating power, $T_{\max} \to \infty$. 

    In the following, for the sake of clarity, we provide the function $\varphi$ concerned in both situations, cooling and heating. In the cooling case, $T_f<1$, we get
    \begin{equation}
        \varphi(k,0, T_{\max},\boldsymbol{\tau})=  e^{- 2 k (\tau_1+\tau_2)} -T_{\max}e^{- 2 k\tau_2}+T_{\max},
    \end{equation}
    whereas for heating, $T_f>1$, we get
    \begin{equation}
    \label{ch5_eq:two-dim-heating}
        \varphi(k,0, T_{\max},\boldsymbol{\tau})= \left(1-T_{\max} \right) e^{- 2 k (\tau_1+\tau_2)} +T_{\max}e^{- 2 k\tau_2}.
    \end{equation}
    The pair $(\tau_1,\tau_2)$ represents $(\tau_{c,1},\tau_{h,1})$ for cooling and  $(\tau_{h,1},\tau_{c,1})$ for heating. In each case, the final connection time is determined as $t_f = \tau_1 + \tau_2$.
    \item $\bm{d=3}$: For a three-dimensional system, three elementary stages or bangs are required. The discussion is similar to the one carried out for the two-dimensional case, but with an additional jump. Herein, we study the effect of adding a new dimension, which involves an extra cost in terms of the optimal connection time. The limit $k_3 \to \infty$ recovers the results for the two-dimensional case. In general, analytic solutions of $\bm{\tau}$ for the system of non-linear equations from Eq.~\eqref{ch5_eq:comp} are generally not possible for $d>1$, but they can always be solved numerically. In Sec.~\ref{ch5_sec:degenerate-case}, we work out a explicit scenario in which some analytical insights can be additionally gathered due to its simplicity.
    
    In the following, the functions $\varphi$ are provided. For cooling, $T_f<1$, one gets
    \begin{equation}
        \varphi(k,0, T_{\max},\boldsymbol{\tau})=  e^{- 2 k (\tau_1+\tau_2+\tau_3)} -T_{\max}e^{- 2 k(\tau_2+\tau_3)}+T_{\max}e^{-2k \tau_3},
    \end{equation}
    while for heating, it is
    \begin{eqnarray}
        \varphi(k,0, T_{\max},\boldsymbol{\tau})&=& \left(1-T_{\max} \right) e^{- 2 k (\tau_1+\tau_2+\tau_3)} +T_{\max}e^{- 2 k(\tau_2+\tau_3)} \nonumber \\
        && -T_{\max}e^{-2k\tau_3}+T_{\max}.
    \end{eqnarray}
    The triplet $(\tau_1,\tau_2,\tau_3)$ stands for $(\tau_{c,1},\tau_{h,1},\tau_{c,2})$ in the cooling case, whereas in the heating case, it becomes $(\tau_{h,1},\tau_{c,1},\tau_{h,2})$. In either case, the final connection time is given by $t_f = \tau_1 + \tau_2 + \tau_3$.
\end{itemize}

\subsection{\label{ch5_subsec:infinite-capacity-limit}Limit of infinite heating power}
In this section, we study the same time optimisation problem for the $d$-dimensional harmonically trapped Brownian particle, but assuming that there is no upper bound for the temperature of the bath, i.e. $T_{\max}\to \infty$. A priori, one could expect this limit to be singular. Nonetheless, the infinite heating power entails a vanishing time for the heating bangs, $\tau_{h,j} \to 0$. In turn, $T_{\max} \tau_{h,j}$ must tend to a certain finite constant, which becomes a new unknown that, in what follows, plays the role that $\tau_{h,j}$ had earlier.

The limit $T_{\max}\to\infty$ could be directly applied to Eq.~\eqref{ch5_eq:comp} but, for the sake of clarity, we choose to introduce it in the evolution operator from the very beginning. To this end, the finite constants
\begin{equation}
\alpha_{h,j} \equiv 2 T_{\max} \tau_{h,j}
\end{equation} 
are defined. In this limit, the evolution operator in Eq.~\eqref{ch5_eq:ev-op} is 
\begin{equation}
\widetilde{\mathcal{E}}_{\alpha_h} \left(z_{j,i}\right)= \lim_{\mycom{T_{\max}\to \infty}{2T_{\max}\tau_{h} \to \alpha_h}} \mathcal{E}_{k_j,T_{\max}}^{\tau_{h}} \left(z_{j,i}\right)= z_{j,i}+ \alpha_h,
\end{equation}
which no longer depends on the elastic constants $k_j$---as explicitly stated in our notation for $\widetilde{\mathcal{E}}$.
The equivalent relations to those in Eqs.~\eqref{ch5_eq:comp-h} and \eqref{ch5_eq:comp-c} are thus
\begin{equation}
\label{ch5_eq:comp-h-inf}
\underbrace{\left( \cdots \circ \widetilde{\mathcal{E}}_{\alpha_{h,2}} \circ \mathcal{E}_{k_j,T_{\min}}^{\tau_{c,1}} \circ \widetilde{\mathcal{E}}_{\alpha_{h,1}} \right)}_{{\textnormal{composition of }} d {\textnormal{ operators }}}
\left(\frac{1}{k_j}\right)=\frac{T_f}{k_j}, \quad j=1,\ldots,d
\end{equation} 
for heating processes, $T_f>1$, and
\begin{equation}
\label{ch5_eq:comp-c-inf}
\underbrace{\left( \cdots \circ \mathcal{E}_{k_j,T_{\min}}^{\tau_{c,2}} \circ \widetilde{\mathcal{E}}_{\alpha_{h,1}} \circ \mathcal{E}_{k_j,T_{\min}}^{\tau_{c,1}} \right)}_{{\textnormal{composition of }} d {\textnormal{ operators }}}
\left(\frac{1}{k_j}\right)=\frac{T_f}{k_j}, \quad j=1,\ldots,d
\end{equation} 
for cooling processes, $T_f<1$. Similarly to Eq.~\eqref{ch5_eq:comp}, multiplying by $k_j$, it is possible to give a compact form for Eqs.~\eqref{ch5_eq:comp-h-inf} and \eqref{ch5_eq:comp-c-inf},
\begin{equation}
\label{ch5_eq:comp-inf}
\widetilde{\varphi}(k_j,T_{\min},\boldsymbol{\tau}_c,\boldsymbol{\alpha}_h)=T_f, \quad j=1,\ldots,d,
\end{equation}
where the vectors $\boldsymbol{\tau}_c$ and $\boldsymbol{\alpha}_h$ contain  the durations of the cooling stages and the intensity of the heating ones, respectively. The sum of the dimension of both vectors is equal to $d$ while the absolute value of their difference is zero (unity) for even (odd) $d$.

\begin{figure} 
\begin{center}
\flushright
\includegraphics[width=0.875\textwidth]{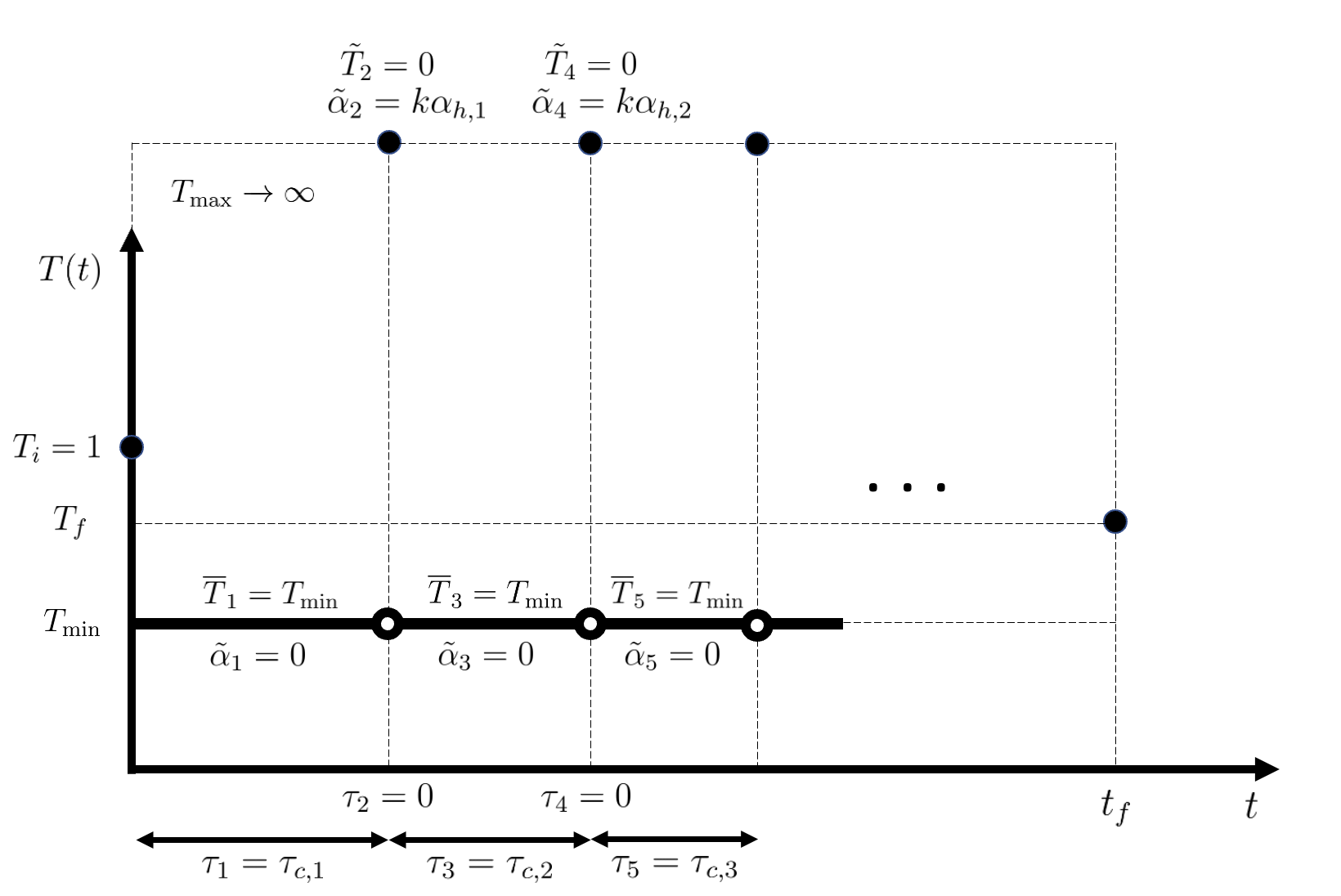}
\includegraphics[width=0.875\textwidth]{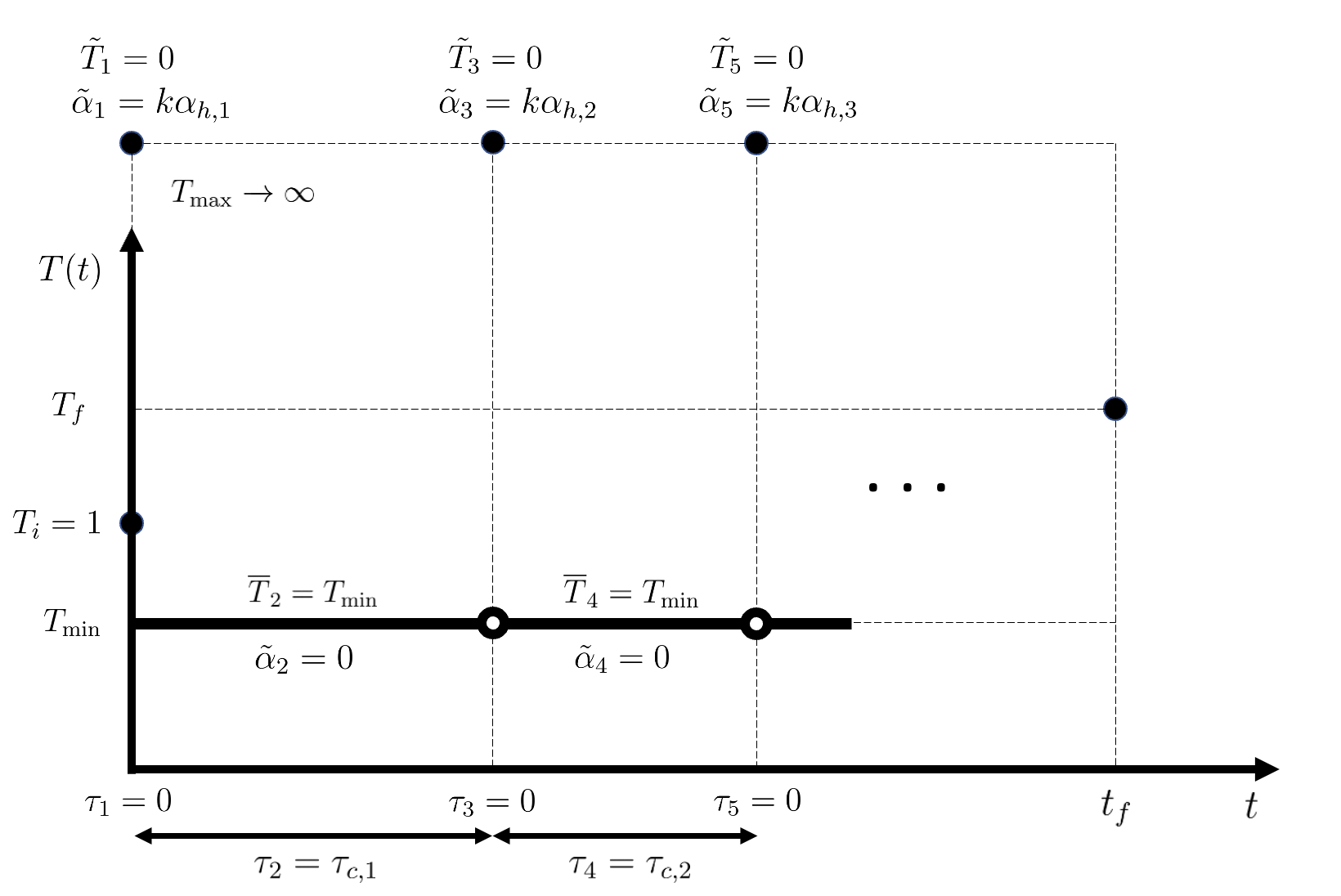}
\caption{\label{ch5_fig:sketch-Tmax-inf} Sketch of the optimal bang-bang control connecting the initial and final states, for infinite heating power $T_{\max}\to\infty$. The alternating elementary bangs are either cooling stages of finite duration or instantaneous heating stages. The top (bottom) panel corresponds to cooling (heating), i.e. $T_f <1$ ($T_f>1$).
The different parameters involved in the protocol, namely temperatures, heating jumps and time spans of the elementary bangs, along with the notation used in Eq.~\eqref{ch5_eq:phi-inf}, are displayed. The shortest connection time $t_f= \sum_{j=1}^d \tau_j$ stems from the solution of Eq.~\eqref{ch5_eq:comp-inf}. Note that $t_f=\tau_{c,1}+\tau_{c,2}+\cdots$, i.e. only the cooling bangs contribute to the minimum connection time.}
\end{center}
\end{figure}

It is possible to give a general expression for $\widetilde{\varphi}$, specifically
\begin{equation}
\label{ch5_eq:phi-inf}
\widetilde{\varphi}(k,T_{\min},\boldsymbol{\tau}_c,\boldsymbol{\alpha}_h) =  \sum_{n=1}^{d+1} \left( \widetilde{T}_{n-1} + \widetilde{\alpha}_n - \widetilde{T}_n\right) \exp \left( -2 k \sum_{m=n}^{d}\tau_m \right),
\end{equation}
where:
\begin{enumerate}
\item $\widetilde{T}_j$ takes values either zero or $T_{\min}$ depending on whether the $j$-th stage corresponds to either heating or cooling, respectively.
\item $\widetilde{T}_0=1$ and $\widetilde{T}_{d+1}=\widetilde{\alpha}_{d+1}=0$.
\item $\widetilde{\alpha}_j$ takes values either sequentially from $\boldsymbol{\alpha}_{h}$ multiplied by $k$ or zero depending on whether the $j$-th stage is of heating or of cooling, respectively.
\item $\tau_j$ takes values either zero or sequentially from $\boldsymbol{\tau}_{c}$ depending on whether the $j$-th stage is of heating or cooling, respectively.
\end{enumerate}  
Thus, the problem of searching the optimal protocol minimising the connection time is reduced to solving the $d$ unknowns in the system of $d$ equations given by Eq.~\eqref{ch5_eq:comp-inf}, with $\widetilde{\varphi}$ given by Eq.~\eqref{ch5_eq:phi-inf}---analogous to the previous case, where we needed to solve Eq.~\eqref{ch5_eq:comp} with $\varphi$ defined in Eq.~\eqref{ch5_eq:phi-fin}. A sketch for the optimal control protocol, and the notation employed in our general formulation in Eq.~\eqref{ch5_eq:phi-inf} is presented in Fig.~\ref{ch5_fig:sketch-Tmax-inf}. In the following, the specific forms of the function $\widetilde{\varphi}$ and their consequences are worked out for $1\leq d\leq 3$.

\begin{itemize}
    \item $\bm{d=1}$: For a cooling process, there is no difference with the case of finite $T_{\max}$. Specifically, it is straightforward to obtain 
    \begin{equation}
    \widetilde{\varphi}(k,0,\boldsymbol{\tau}_c,\boldsymbol{\alpha}_h) = e^{-2k t_f},
    \end{equation}
     where $\tau_1=\tau_{c,1}=t_f$ is the minimum connection time---the duration of the single cooling bang in the protocol. Hence, the minimum time in Eq.~\eqref{ch5_eq:tmin_1d} is re-obtained.
    
    For a heating process, the limit $T_{\max}\to \infty$ indeed induces a change, since
    \begin{equation}
    \widetilde{\varphi}(k,0,\boldsymbol{\tau}_c,\boldsymbol{\alpha}_h) = 1+k\alpha.
    \end{equation}
    In this occasion, Eq.~\eqref{ch5_eq:comp-inf} just gives the correct value of $\alpha_1=\alpha_{h,1}$ to reach the target state.  There is only one heating stage, which is instantaneous $t_f=\tau_{h,1}=\alpha/2T_{\max}=0$. Reasonably, the minimum time to perform a heating process vanishes since one has infinite resources to heat up the system. As we are about to prove, this is no longer the case for $d>1$.  

    \item $\bm{d=2}$: Here, the optimal protocol comprises two bangs or elementary stages, as in the case of finite $T_{\max}$. The main difference stems from  the heating bangs being instantaneous. The role of the duration of the heating $\tau_{h,1}$ is now played by its intensity $\alpha_{h,1}=2T_{\max}\tau_{h,1}$. Using the general formula in Eq.~\eqref{ch5_eq:phi-inf}, one gets
    \begin{equation}\label{ch5_eq:infT-cooling-d2}
    \widetilde{\varphi}(k,0,\boldsymbol{\tau}_c,\boldsymbol{\alpha}_h) = e^{-2kt_f}+k\alpha
    \end{equation}
    for a cooling process, $T_f<1$, while 
    \begin{equation}\label{ch5_eq:infT-heating-d2}
    \widetilde{\varphi}(k,0,\boldsymbol{\tau}_c,\boldsymbol{\alpha}_h) =(1+k \alpha) e^{-2 k t_f}
    \end{equation}
    for a heating process, $T_f>1$. The pairs  $(\widetilde{\alpha}_1,\widetilde{\alpha}_2)$ and $(\tau_1,\tau_2)$  stand for $(0,k \alpha_{h,1} = k \alpha$ and $(\tau_{c,1}=t_f,0)$ in the cooling process and for $(k \alpha_{h,1}=k \alpha,0)$  $(0,\tau_{c,1}=t_f)$ in the heating process, respectively. 

    Solving Eqs.~\eqref{ch5_eq:comp-inf} for a finite value of $k_2$ is possible up to reach the inverse relation between the optimal time and the final temperature, i.e. $T_f$ may be expressed as an analytic function of $t_f$. In particular, for the heating protocol it gives
    \begin{equation}\label{ch5_eq:Tf_analytic-d2-h}
        T_f = \frac{k_2-1}{k_2e^{2t_f}-e^{2k_2t_f}},
    \end{equation}
    whereas for the cooling protocol,
    \begin{equation}\label{ch5_eq:Tf_analytic-d2-c}
        T_f = \frac{k_2e^{-2t_f}-e^{-2k_2t_f}}{k_2-1}.
    \end{equation}
    These formulas can be found in Ref.~\cite{prados_optimizing_2021}. In Fig.~\ref{ch5_fig:tf_vs_Tf_difk}, we illustrate the dependence of $t_f$ on the second elastic constant $k_2$. Therein, it is clearly observed that, at fixed target temperature $T_f$, the minimum connection time monotonically decreases with $k_2$, vanishing in the $k_2\to +\infty$, where we recover the $d=1$ case.

    \begin{figure}
        \centering
        \includegraphics[width=3.5in]{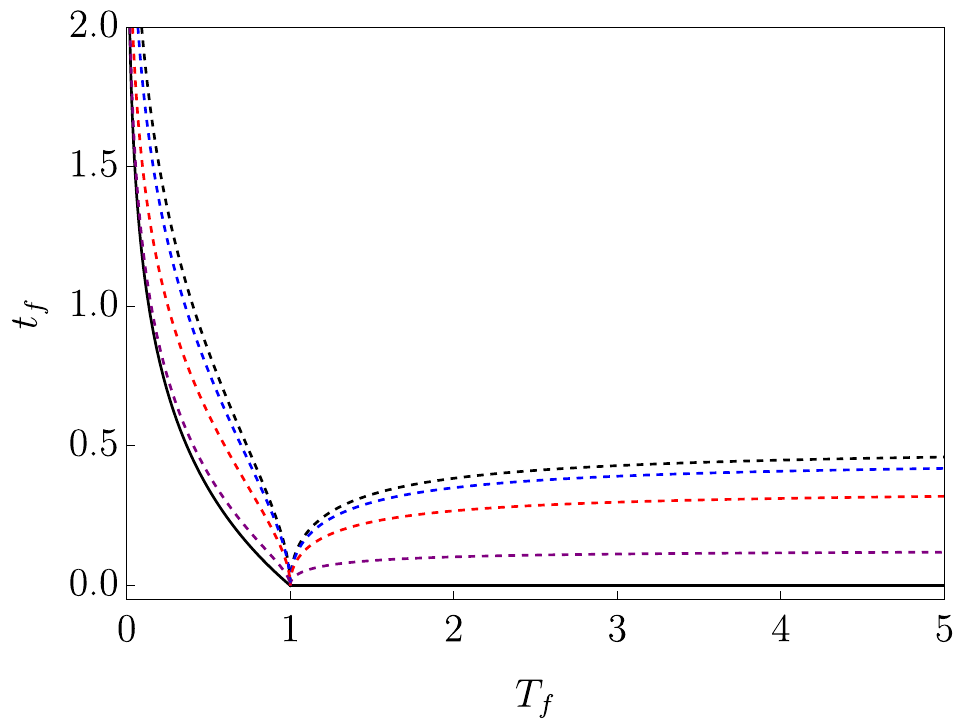}
        \caption{\label{ch5_fig:tf_vs_Tf_difk} Shortest connection time as a function of the final temperature, in the $T_{\max}\rightarrow +\infty$ limit, for different elastic constants  $k_2$ and dimensions $d$. Dashed curves correspond to the $d=2$ case, while the solid ones account for $d=1$. Different colors stand for the different values of $k_2$ considered, namely $k_2=10$ (purple), $k_2=5$ (red), $k_2=1.1$ (blue) and $k_2 \to 1$ (black).}
    \end{figure}

    The limit $k_2 \to k_1 = 1$ is also particularly enticing as, even in this case, there appears a time cost associated with increasing the dimension of the system---a fact that is thoroughly explored in Sec.~\ref{ch5_sec:degenerate-case}. Such feature is especially remarkable in the limit as $T_{\max} \to  \infty$: the minimum time needed to achieve the connection for the two-dimensional case is finite, even when having an infinite heating power---at variance with the one-dimensional case, for which $t_f=0$. In this respect, it is convenient to recall that the fastest cooling rate is reached when the temperature of the thermal bath takes its minimum value, which is  physically bounded by zero. This forbids instantaneous cooling processes, i.e. the cooling power is always limited---which explains the observed asymmetry between cooling and heating processes in this limiting scenario. In Sec.~\ref{ch5_subsec:unphysical-scenario}, this bound is disregarded by assuming an unphysical scenario, where $T_{\min} \to -\infty$. Therein, we show that it is the finite cooling power where finite minimum connection times stem from.

    \item $\bm{d=3}$: Now, three bangs or elementary stages are involved in the optimal protocol. There are three unknowns, two $\alpha_h$'s and one $\tau_c$ for heating processes, and vice versa for cooling processes. Using the general formula in Eq.~\eqref{ch5_eq:phi-inf}, we get
    \begin{equation}\label{ch5_eq:infT-cooling-d3}
    \widetilde{\varphi}(k,0,\boldsymbol{\tau}_c,\boldsymbol{\alpha}_h) = e^{-2k (\tau_{c,1}+\tau_{c,2})} + k\alpha e^{-2k \tau_{c,2}},
    \end{equation}
    for a cooling process, whereas
    \begin{equation}\label{ch5_eq:infT-heating-d3}
    \widetilde{\varphi}(k,0,\boldsymbol{\tau}_c,\boldsymbol{\alpha}_h)= (1 +k \alpha_{h,1})e^{-2k t_f} + k\alpha_{h,2}.
    \end{equation}
    for a heating process. The triplets  $(\widetilde{\alpha}_1,\widetilde{\alpha}_2,\widetilde{\alpha}_3)$ and $(\tau_1,\tau_2,\tau_3)$  stand, respectively, for $(0,k \alpha_{h,1} = k \alpha,0)$ and $(\tau_{1,c},0,\tau_{2,c})$ in the cooling process and for $(k \alpha_{h,1},0,k \alpha_{h,2})$ and  $(0,\tau_{c,1}=t_f,0)$ in the heating one. Similarly to the previous cases, an implicit expression for $t_f$ in terms of $T_f$ can be obtained, for arbitrary $k_2$ and $k_3$ for both heating and cooling---which, once more, can also be found in Ref.~\cite{prados_optimizing_2021}. Specifically, for the heating protocol we have
    \begin{equation}\label{ch5_eq:Tf_analytic-d3-h}
        T_f = \frac{e^{2t_f}(k_3-k_2)-e^{2k_2t_f}k_2(k_3-1)+e^{2k_3t_f}k_3(k_2-1)}{e^{2(1+k_3)t_f}k_2(k_3-1)-e^{2(1+k_2)t_f}k_3(k_2-1)-e^{2(k_2+k_3)t_f}(k_3-k_2)},
    \end{equation}
    whereas for the cooling protocol,
    \begin{equation}\label{ch5_eq:Tf_analytic-d3-c}
        \left(T_f-e^{-2t_f}\right)^{k_3-k_2}  \left(T_f-e^{-2k_2t_f}\right)^{1-k_3}  \left(T_f-e^{-2k_3t_f}\right)^{k_2-1} = k_3^{k_2-1}k_2^{1-k_3}.
    \end{equation}
    Note that the latter corresponds to an implicit equation for both $t_f$ and $T_f$.
\end{itemize}

\section{\label{ch5_sec:degenerate-case}Degenerate case: the unavoidable price for increasing the dimension}~\sectionmark{Degenerate case}
Up to now, we have considered the case in which all the elastic constants $k_j$ are different, where the mathematical problem is completely closed: Eqs.~\eqref{ch5_eq:comp} and \eqref{ch5_eq:comp-inf} are systems of $d$ equations for $d$ unknowns each---i.e. the components of the vector $\boldsymbol{\tau}$ of time spans in Eq.~\eqref{ch5_eq:comp}, and the components of the cooling time spans $\bm{\tau}_c$ and intensity of heatings $\bm{\alpha}_h$ in Eq.~\eqref{ch5_eq:comp-inf}. The optimal connection time is given by the sum of all components of the solution for $\boldsymbol{\tau}$ or $\boldsymbol{\tau}_c$, respectively. 

Now, let us consider the case when some $k_j$, say $n$ of them, are exactly equal---which we refer to as the degenerate case. This is relevant from a physical point of view, since it arises naturally when the harmonic confinement possesses some symmetry, for example cylindrical or spherical in the three-dimensional case. In such a situation, the variances corresponding to dimensions with the same elastic constant fulfil the same mathematical relationship. In such a situation, the problem can be solved by considering that the dimension has been reduced to $d^*=d+1-n<d$, i.e. $d^*$ is the number of different values of the elastic constant $k_j$.\footnote{For the sake of simplicity in the discussion, we consider the case in which only one value of $k_j$ becomes degenerate. The analysis carried out here may be easily generalised to having multiple degenerate values.} The optimal protocol  would thus involve a smaller number $d^*$ of elementary stages or bangs. 

Now, let us address the case in which the $n$ $k_j$'s are arbitrarily close, but not exactly equal. Therein, we need the $d$ bangs to achieve the optimal connection and the situation is quite subtle, as shown below. For the sake of concreteness, we study the almost fully degenerate case $n=d$, defined as the limit where all confinements are almost identical: $k_j \to k_1=1$, $\forall j>1$. This describes a harmonic trap with almost spherical symmetry in $d$ dimensions. This is an experimentally relevant situation, since the elastic constants along orthogonal directions would not be perfectly equal in a real experiment~\cite{rohrbach_stiffness_2005,madadi_polarization-induced_2012,ruffner_universal_2014,yevick_photokinetic_2017,moradi_efficient_2019}. Our use of the word ``almost'' stresses the fact that the $k_j$'s are not exactly equal and dimension reduction is not feasible.

For the almost fully degenerate case, the systems of equations from Eqs.~\eqref{ch5_eq:comp} and \eqref{ch5_eq:comp-inf} are no longer closed if we set all the $k_j$'s equal to unity: there are $d$ unknowns but only one equation. Nevertheless, it is possible to resort to a perturbative approach that allows to find the missing $d-1$ equations required to close the problem.\footnote{If only some  $k_j$'s were almost equal, there would be $d$ unknowns and $d^*<d$ equations. In that case, a similar perturbative approach would provide us with the missing equations.} By setting $k_j=1+\epsilon_j$, with $\epsilon_j\ll 1$ for $j>1$, one gets
\begin{subequations}
\label{ch5_eq:comp-deg}
\begin{eqnarray}
\label{ch5_eq:comp-deg1}
\varphi(1,T_{\min}, T_{\max},\boldsymbol{\tau}) &=& T_f,
 \\
\label{ch5_eq:comp-deg2}\frac{\partial^n}{\partial k^n} \varphi(k,T_{\min}, T_{\max},\boldsymbol{\tau})\biggr\rvert_{k=1} &=& 0, \quad n=1,\ldots,d-1,
\end{eqnarray}
\end{subequations}
for the finite $T_{\max}$ scenario, and
\begin{subequations}
\label{ch5_eq:comp-deg-inf}
\begin{eqnarray}
\label{ch5_eq:comp-deg1-inf}
\widetilde{\varphi}(1,T_{\min},\boldsymbol{\tau}_c,\boldsymbol{\alpha}_h) &=& T_f, \\
\label{ch5_eq:comp-deg2-inf}\frac{\partial^n}{\partial k^n} \widetilde{\varphi}(k,T_{\min},\boldsymbol{\tau}_c,\boldsymbol{\alpha}_h)\biggr\rvert_{k=1} &=& 0, \quad n=1,\ldots,d-1,
\end{eqnarray}
\end{subequations}
in the $T_{\max}\to \infty$ limit.

The solution of \eqref{ch5_eq:comp-deg} for the vector of time spans $\boldsymbol{\tau}$---or the solution of \eqref{ch5_eq:comp-deg-inf} for both the cooling time spans $\boldsymbol{\tau}_c$ and the intensity of heatings $\bm{\alpha}_h$---provides us with the optimal bang-bang protocol in the almost fully degenerate case. Below, we show that this problem does not converge to the solution of the fully degenerate case, i.e. to the one-dimensional solution. This is a remarkable property of the system under study: increasing the dimension comes at an unavoidable price, in which the shortest connection time presents a jump when going from $d$ to $d+1$. This happens even when the elastic constants are almost equal in all directions, and the confinement is arbitrarily close to be spherically symmetric.

In order to illustrate the latter statement, we explicitly revisit the $d>1$ cases within the $T_{\max}\to +\infty$, which were investigated in Sec.~\ref{ch5_subsec:infinite-capacity-limit}.

\begin{itemize}
    \item $\bm{d=2}$: In this case, one needs to solve Eq.~\eqref{ch5_eq:comp-deg-inf}, with $\widetilde{\varphi}$ being given by Eqs.~\eqref{ch5_eq:infT-cooling-d2} and \eqref{ch5_eq:infT-heating-d2}, for either the cooling or heating processes respectively. Following this, we may obtain implicit analytical expressions for the final connection time $t_f$ as functions of the target temperature $T_f$. Explicitly, one has
    \begin{equation}\label{ch5_eq:tfinv_Tmaxinf_2d_c}
    T_f= e^{-2t_f} (1+2t_f),
    \end{equation}
    fog cooling, and 
    \begin{equation}
    \label{ch5_eq:tfinv_Tmaxinf_2d_h}
    T_f=\frac{e^{-2t_f}}{1-2t_f},
    \end{equation}
    for heating. The above expressions can be easily obtained from applying the $k_2 \to 1$ limit on Eqs.~\eqref{ch5_eq:Tf_analytic-d2-h} and \eqref{ch5_eq:Tf_analytic-d2-c} respectively. Note that Eq.~\eqref{ch5_eq:tfinv_Tmaxinf_2d_h} predicts the emergence of a horizontal asymptote for $t_f=1/2$ due to the vanishing of the denominator of its right-hand side, which was anticipated in Fig.~\ref{ch5_fig:tf_vs_Tf_difk} in the limit $k_2 \to 1$.

    \item $\bm{d=3}$: Again, an implicit expression for $t_f$ in terms of $T_f$ can be obtained in the fully degenerate case, by solving Eq.~\eqref{ch5_eq:comp-deg-inf} with $\widetilde{\varphi}$ given by Eqs.~\eqref{ch5_eq:infT-cooling-d3} and \eqref{ch5_eq:infT-heating-d3}, for either the cooling or heating processes respectively, which yields 
    \begin{equation}
        \label{ch5_eq:tfinv_Tmaxinf_3d_c}
        T_f= e^{-2t_f} \left[ 1+2t_f^2 + 2t_f \sqrt{ (1+t_f^2)} \right],
    \end{equation}
    for cooling, and 
    \begin{equation}
        \label{ch5_eq:tfinv_Tmaxinf_3d_h}
        T_f=\frac{e^{-2t_f}(1+t_f)}{1-t_f},
        \end{equation}
    for heating. Similar to the $d=2$ case, the above can also be easily obtained from applying the $k_3 \to k_2 \to 1$ limit on Eqs.~\eqref{ch5_eq:Tf_analytic-d3-h} and \eqref{ch5_eq:Tf_analytic-d3-c} respectively. Additionally, Eq.~\eqref{ch5_eq:tfinv_Tmaxinf_3d_h} predicts a horizontal asymptote at $t_f = 1$.
\end{itemize}

The approach introduced above can be carried out to address an almost fully degenerate case for arbitrary dimension $d$. This is not unphysical: for example, it corresponds to $d$ one-dimensional colloidal particles, each one trapped in its own one-dimensional harmonic potential---with all the stiffnesses of the harmonic wells being almost equal. 

For heating processes, there always emerges an asymptotic value $t_f^{\text{asy}} \equiv \lim_{T_f \to \infty} t_f$ for the minimum connection time.  This value is an upper bound for the minimum connection time for a finite value of the target temperature $T_f>1$ of a heated system, $t_f(T_f)\leq t_f^{\text{asy}}$. The bound $t_f^{\text{asy}}$ monotonically increases with the dimension, presenting finite jumps when going from $d$ to $d+1$---as already discussed for the particular cases $d=1\to 2$ and $d=2\to 3$. In Table~\ref{ch5_tab:asym}, the values for the optimal time for these asymptotes are displayed. These values have been analytically computed solving Eq.~\eqref{ch5_eq:comp-deg-inf}, except for $d=6$, which has been numerically obtained. Note that, for  $1\leq d\leq 3$, $t_f^{\text{asy}}$ follows the  simple formula $t_f^{\text{asy}}=(d-1)/2$: this simple expression is broken for $d=4$.

\begin{table}
    \caption{\label{ch5_tab:asym} Values for the asymptotic value of the minimum heating time $t_f^{\text{asy}}$ in the almost fully degenerate case for a  $d$-dimensional oscillator. Finding the reported values involve solving a polynomial equation of degree $d-1$: therefore, for $d>5$, the solution has been obtained numerically. }
        
        \begin{center}
        \begin{tabular}{cc}
        \hline
        \multicolumn{1}{l}{Dimension $d$} & \multicolumn{1}{c}{
        $t_f^{\text{asy}}$} \\ \hline
        1                     & 0                                            \\
        2                     & $\frac{1}{2}$                                \\
        3                     & 1                                            \\
        4                     & $\frac{2+\sqrt{2}}{2}$                       \\
        5                     & $\frac{3+\sqrt{3}}{2}$                       \\
        6                     & 3.14497                                      \\ \hline
        \end{tabular}
        \end{center}
        \end{table}

\begin{figure}
    \centering
    \includegraphics[width=3.5in]{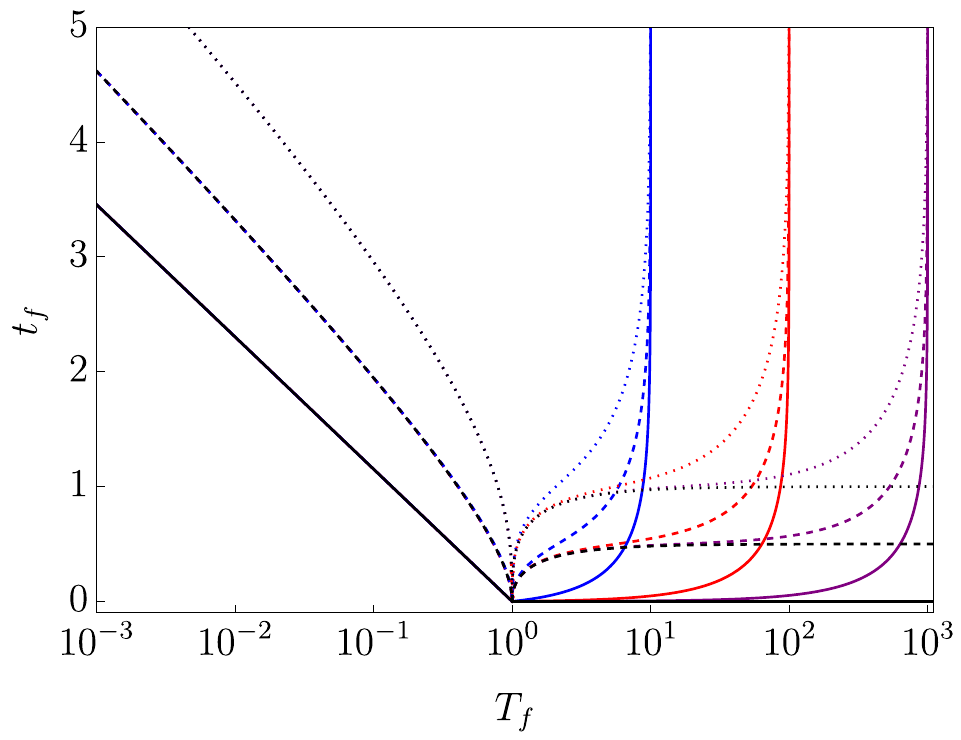}
    \caption{\label{ch5_fig:tf_vs_Tf_difTmax}Shortest connection time as a function of the final temperature for different heating powers $T_{\max}$ and dimensions $d$. Different colours stand for the different values of $T_{\max}$ considered, namely $T_{\max}=10$ (blue), $T_{\max}=100$ (red), $T_{\max}=1000$ (purple), and $T_{\max} \to \infty$ (black). Different dimensions are displayed with different patterns: $d=1$ (solid), $d=2$ (dashed) and $d=3$ (dotted). At the scale of the plot, the results in the cooling region ($T_f<1$) are hardly distinguishable from the limit behaviour  $T_{\max} \to \infty$ given by the black curve. For heating processes, the universal behaviour corresponding to $T_{\max} \to  \infty$ is approached in the regime $T_f/T_{\max}\ll 1$.}
\end{figure}

Figure~\ref{ch5_fig:tf_vs_Tf_difTmax} shows the behaviour of the minimum connection time in the almost fully degenerate case as a function of the final temperature $T_f$, for both heating and cooling processes, different values of $T_{\max}$---keeping $T_{\min}=0$ for consistency, and for $1\leq d \leq 3$. On the one hand, for heating processes, and in the $T_f \ll T_{\max}$ regime, all the curves tend to the universal behaviours for the $T_{\max}\to +\infty$ limit---Eqs.~\eqref{ch5_eq:tfinv_Tmaxinf_2d_h} and \eqref{ch5_eq:tfinv_Tmaxinf_3d_h} for $d=2$ and $d=3$, respectively. It is only when approaching $T_{\max}$ that the connection times diverge. Nevertheless, there is still a finite gap when increasing the dimension, which becomes specially appealing in the $1 \ll T_f \ll T_{\max}$ regime, where all the universal behaviours tend to their corresponding asymptotic values $t_f^{\text{asy}}$ from Table~\ref{ch5_tab:asym}.

\begin{figure}
    \centering
    \includegraphics[width=3.5in]{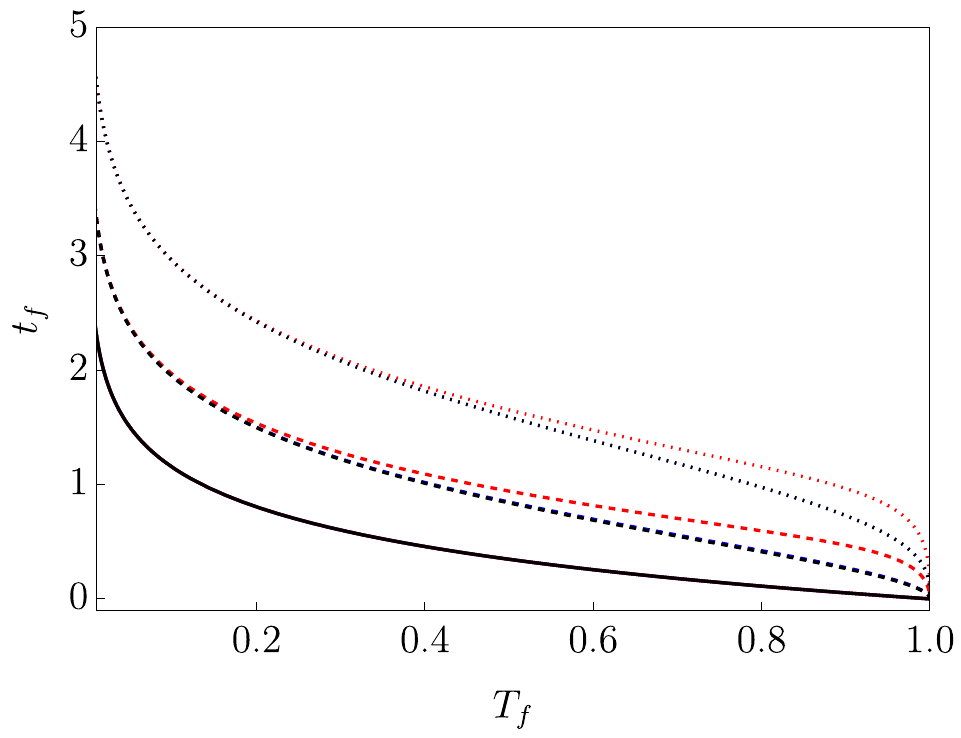}
    \caption{\label{ch5_fig:tf_vs_Tf_difTmax2} Zoom of the cooling branch of the  $t_f$ versus $T_f$ curve. Different heating powers $T_{\max}$ and dimensions $d$ are considered with the following code: $T_{\max}=1.1$ (red), $T_{\max}=10$ (blue), $T_{\max} \to \infty$ (black), and the same pattern code as in Fig.~\ref{ch5_fig:tf_vs_Tf_difTmax} for $d$. Blue and black curves are always indistinguishable. The linear scale in $T_f$ and the particular case $T_{\max}=1.1$ shown here make it possible to discern the effect of $T_{\max}$ on $t_f$ over the cooling branch, which was imperceptible in Fig.~\ref{ch5_fig:tf_vs_Tf_difTmax}.}
\end{figure}

On the other hand, for cooling processes, it seems that the dependence on $T_{\max}$ is negligible,\footnote{For $d=1$, that statement is exact.} since all the curves tend to their corresponding universal behaviours for $T_{\max} \to \infty$---Eqs.~\eqref{ch5_eq:tfinv_Tmaxinf_2d_c} and \eqref{ch5_eq:tfinv_Tmaxinf_3d_c} for $d=2$ and $d=3$, respectively. In Fig.~\ref{ch5_fig:tf_vs_Tf_difTmax2}, we show how such dependence becomes relevant as $T_{\max}$ approaches unity. In any case, the discontinuity in the connection times as we increase the dimension still persist. Contrary to the heating case, the universal behaviours for the cooling processes in the $T_{\max} \to \infty$ limit take into account the asymptotic, divergent behaviour at $T_{\min}=0$, as Fig.~\ref{ch5_fig:tf_vs_Tf_difTmax2} shows for sufficiently small values of $T_f$.

\subsection{\label{ch5_subsec:unphysical-scenario}Unphysical scenario: negative temperatures}
Let us consider the two-dimensional harmonic oscillator, for general values of the boundary temperatures $T_{\max}$ and $T_{\min}$ and for the heating case, $T_f>1$, in the almost fully degenerate case $k_2 \rightarrow k_1=1$. Direct resolution of the evolution equations \eqref{ch5_eq:comp-deg1} over the whole time window leads to the equations
\begin{eqnarray}
    T_f &=& (1-T_{\max})e^{-2(\tau_1+\tau_2)} + \left(T_{\max}-T_{\min}\right)e^{-2\tau_2} +T_{\min},
    \\
    0 &=& (\tau_1+\tau_2)\left(1-T_{\max}\right)e^{-2(\tau_1+\tau_2)} + \tau_2\left(T_{\max}-T_{\min}\right)e^{-2\tau_2}.
\end{eqnarray}
Now, for this unphysical scenario, we assume that $T_c \equiv T_{\max} = - T_{\min}$. This assumption implies that we can heat up the system as much as we can cool it down. We would recover a sort of ``temperature symmetry" that it is not actually present in the physical case. Within this assumption, the evolution equations reduce to
\begin{eqnarray}
\label{ch5_eq:reduced-evol-eqs-Tc-1}
    T_f &=& (1-T_{c})e^{-2(\tau_1+\tau_2)} + 2T_ce^{-2\tau_2} - T_{c},
    \\
    \label{ch5_eq:reduced-evol-eqs-Tc-2}
    0 &=& (\tau_1+\tau_2)(1-T_c)e^{-2(\tau_1+\tau_2)} + 2\tau_2T_ce^{-2\tau_2}.
\end{eqnarray}
\begin{figure}
\centering
\includegraphics[width=3.5in]{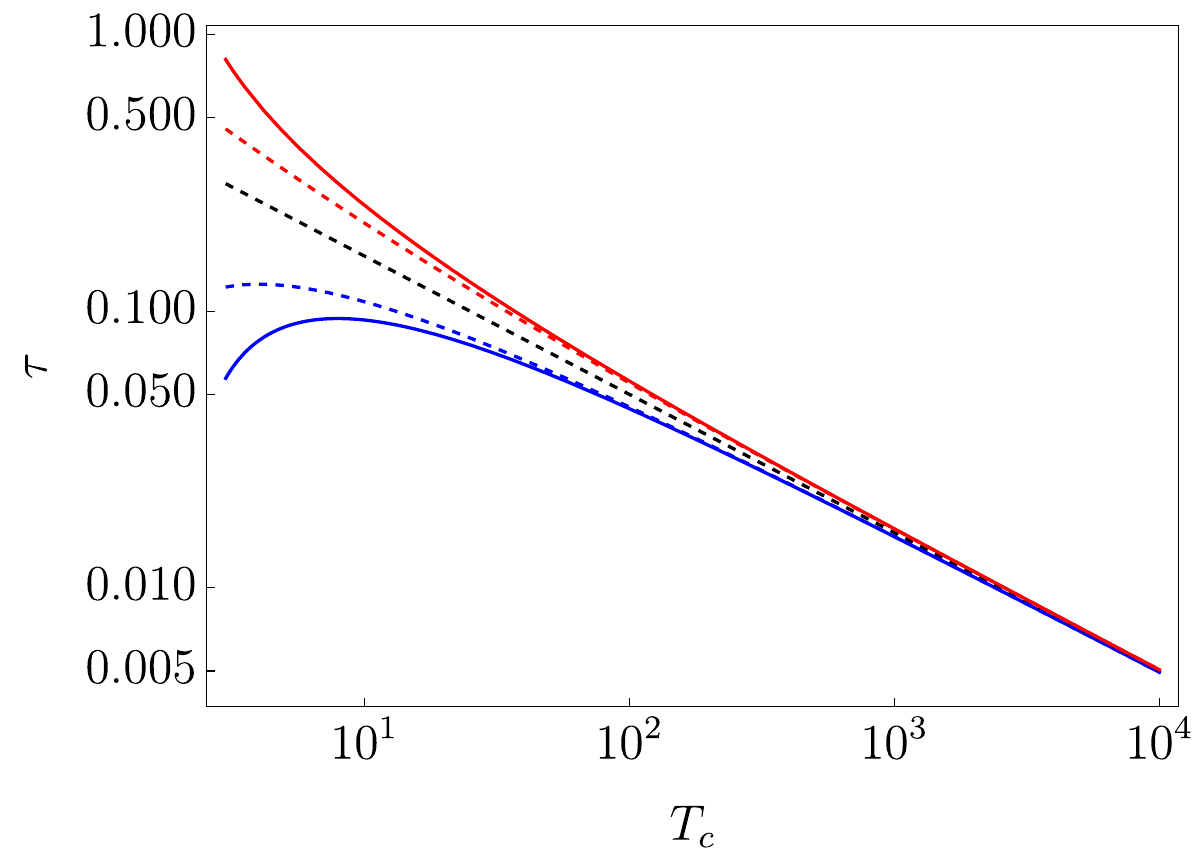}
\caption{Boundary temperature dependence of $\tau_1$ (red) and $\tau_2$ (blue) for the heating protocol in a two-dimensional unphysical harmonic oscillator, for which $T_{\max}=-T_{\min}=T_c$. We consider the almost fully degenerate case $k_2 \rightarrow k_1 = 1$, for $T_f =2$. Solid lines correspond to the numerical solution of Eqs.~\eqref{ch5_eq:reduced-evol-eqs-Tc-1} and \eqref{ch5_eq:reduced-evol-eqs-Tc-2}, while the black dashed line corresponds to the $O(T_c^{-1/2})$ contribution for both time intervals, and the red and blue dashed lines, respectively, to $\tau_1$ and $\tau_2$ up to $O(T_c^{-1})$. All dashed curves are analytical and stem from Eq.~\eqref{ch5_eq:times-unphysical}.} \label{ch5_fig:times-Tc}
\end{figure}
We are interested in the asymptotic behaviour of both $\tau_1$ and $\tau_2$ when $T_c \rightarrow +\infty$. Such infinite power to both heat and cool the system entails vanishing times for both the heating and the cooling windows. However, as Figure~\ref{ch5_fig:times-Tc} shows, the asymptotic behaviour of both $\tau_1$ and $\tau_2$ scales as $T_c^{-1/2}$ in this case. Thus, it is appealing to introduce the constants
\begin{eqnarray}
\label{ch5_eq:alpha-constants}
    \alpha_{h} = 2\sqrt{T_c}\tau_1, \quad \alpha_{c} = 2\sqrt{T_c}\tau_2.
\end{eqnarray}
For $T_c \rightarrow +\infty$, both $\tau_1$ and $\tau_2$ may be expanded in powers of $T_c^{-1/2}$ as
\begin{eqnarray}
    \tau_1 = \frac{\tau_1^{(0)}}{\sqrt{T_c}} + \frac{\tau_1^{(1)}}{T_c} + \frac{\tau_1^{(2)}}{T_c^{3/2}}  \cdots  \ \Rightarrow \ \alpha_h =  \underbrace{2 \tau_1^{(0)}}_{\equiv \alpha_h^{(0)}} + \underbrace{\frac{2\tau_1^{(1)}}{\sqrt{T_c}}}_{\equiv \alpha_h^{(1)}/\sqrt{T_c}} + \underbrace{\frac{2\tau_1^{(2)}}{T_c}}_{\equiv \alpha_h^{(2)}/T_c} \cdots,
    \\
    \tau_2 = \frac{\tau_2^{(0)}}{\sqrt{T_c}} + \frac{\tau_2^{(1)}}{T_c}  + \frac{\tau_2^{(2)}}{T_c^{3/2}}  \cdots \ \Rightarrow \ \alpha_c =  \underbrace{2 \tau_2^{(0)}}_{\equiv \alpha_c^{(0)}} +  \underbrace{\frac{2\tau_2^{(1)}}{\sqrt{T_c}}}_{\equiv \alpha_c^{(1)}/\sqrt{T_c}} + \underbrace{\frac{2\tau_2^{(2)}}{T_c}}_{\equiv \alpha_c^{(2)}/T_c} \cdots.
\end{eqnarray}
With these definitions, $T_c \rightarrow +\infty$ in the reduced evolution equations leads to
\begin{eqnarray}
    T_f &=& 1 + \sqrt{T_c}(\alpha_h^{(0)} - \alpha_c^{(0)}) +(\alpha_h^{(1)} - \alpha_c^{(1)})
    \nonumber
    \\
    &&-\frac{1}{2}\left[(\alpha_h^{(0)})^2 + 2\alpha_h^{(0)} \alpha_c^{(0)} - (\alpha_c^{(0)})^2\right] + O(T_c^{-1/2}), 
    \\
    0 &=& \sqrt{T_c}(\alpha_h^{(0)} - \alpha_c^{(0)}) +(\alpha_h^{(1)} - \alpha_c^{(1)})
    \nonumber
    \\
    &&-\left[(\alpha_h^{(0)})^2 + 2\alpha_h^{(0)} \alpha_c^{(0)} - (\alpha_c^{(0)})^2\right] + O(T_c^{-1/2}), 
\end{eqnarray}
On the one hand, the above system of equations is only consistent if $\alpha_h^{(0)}$ and $\alpha_c^{(0)}$ are equal, such that the $O(\sqrt{T_c})$ terms vanish. On the other hand, we are left with an undetermined system of two equations for the variables $(\alpha_h^{(0)},\alpha_h^{(1)},\alpha_c^{(1)})$,
\begin{eqnarray}
    \label{ch5_eq:alpha-0}
    T_f &=& 1 + (\alpha_h^{(0)})^2,
    \\
    \label{ch5_eq:alpha-0-2}
    0 &=& \alpha_h^{(1)} - \alpha_c^{(1)} - 2(\alpha_h^{(0)})^2.
\end{eqnarray}
In order to close the system, we need to take into account the subdominant contributions in $T_c$. Those of $O(T_c^{-1/2})$ read
\begin{eqnarray}
    0 &=& \alpha_h^{(0),3} - 2\alpha_h^{(0)}(1+\alpha_h^{(1)}) + (\alpha_h^{(2)}-\alpha_c^{(2)}),
    \\
    0 &=& 3\alpha_h^{(0),3} - 2\alpha_h^{(0)}(1+2\alpha_h^{(1)}) + (\alpha_h^{(2)}-\alpha_c^{(2)}),
\end{eqnarray}
from which we obtain 
\begin{equation}
\label{ch5_eq:alpha-1}
    \alpha_h^{(1)} = (\alpha_h^{(0)})^2, \ \alpha_c^{(1)} = -(\alpha_h^{(0)})^2.
\end{equation}
In general, the $O(T_c^{(1-n)/2})$ allows us to solve up to the variables $\alpha_h^{(n)}-\alpha_c^{(n)}$ and $\alpha_h^{(n-1)}$. Finally, up to $O(T_c^{-1})$, the time intervals are given by
\begin{equation}
\label{ch5_eq:times-unphysical}
    \tau_1 = \frac{1}{2}\sqrt{\frac{T_f-1}{T_c}} + \frac{T_f-1}{T_c}, \quad \tau_2 = \frac{1}{2}\sqrt{\frac{T_f-1}{T_c}} - \frac{T_f-1}{T_c},
\end{equation}
which clearly vanish in the $T_c \rightarrow +\infty$ limit. In Figs.~\ref{ch5_fig:times-Tc} and \ref{ch5_fig:alpha-Tc}, the excellent agreement between our asymptotic analysis (dashed lines) and the numerical solution (solid lines) is evident.
\begin{figure}
\centering
\includegraphics[width=3.5in]{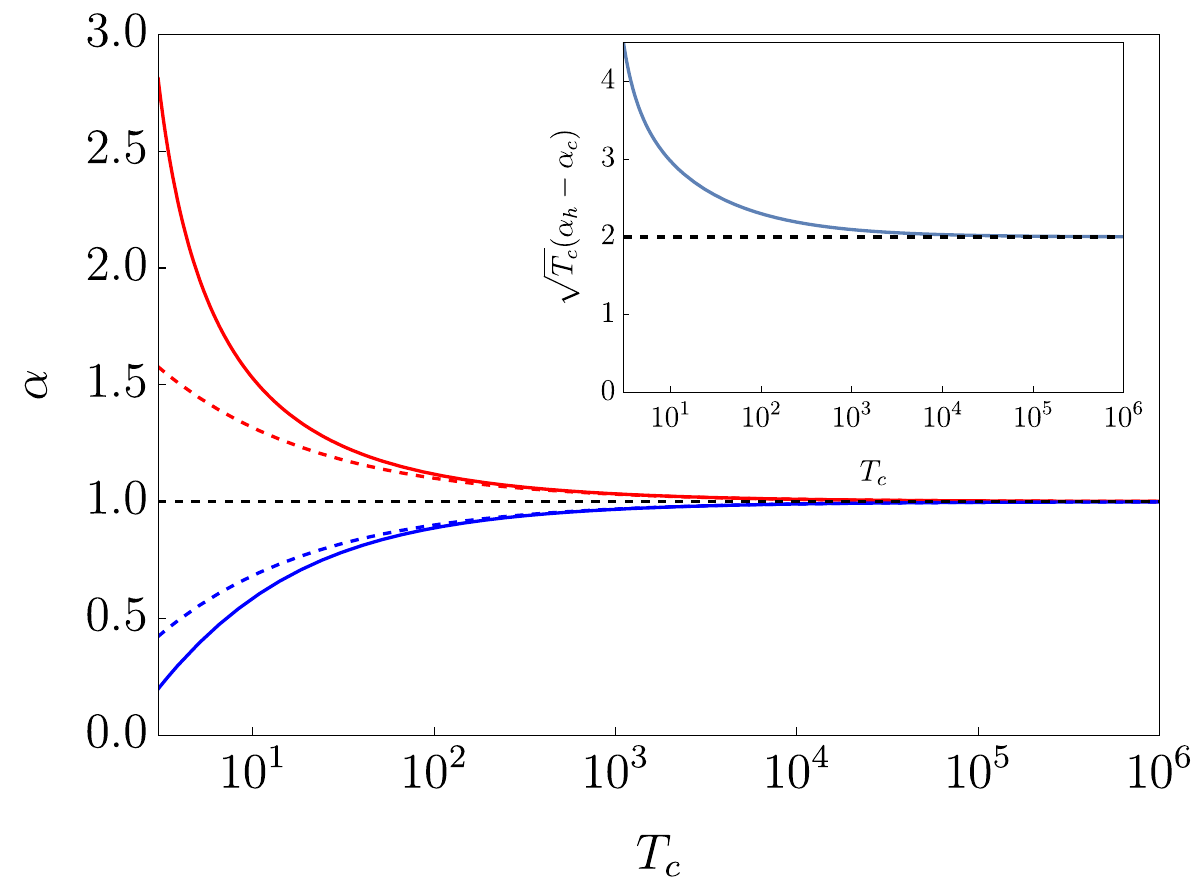}
\caption{Boundary temperature dependence of the constants $\alpha_h$ (red) and $\alpha_c$ (blue) for the heating protocol in a two-dimensional unphysical harmonic oscillator.  As in Fig.~\ref{ch5_fig:times-Tc}, $k_2 \rightarrow k_1 = 1$ and $T_f =2$. Solid lines correspond to the numerical solution of the evolution equations Eqs.~\eqref{ch5_eq:reduced-evol-eqs-Tc-1} and \eqref{ch5_eq:reduced-evol-eqs-Tc-2}, plus the definitions from Eq.~\eqref{ch5_eq:alpha-constants}, while the black dashed line corresponds to the $O(1)$ contribution for both constants, and the red and blue dashed lines, respectively, to $\alpha_h$ and $\alpha_c$ up to $O(T_c^{-1/2})$. All dashed curves are  analytical predictions stemming from Eqs.~\eqref{ch5_eq:alpha-0} and \eqref{ch5_eq:alpha-1}.}\label{ch5_fig:alpha-Tc}
\end{figure}

\subsection{\label{ch5_subsec:information-geometry}Insights from information geometry}

We have shown the emergence of an unavoidable price when adding dimensions to the almost degenerate oscillator under consideration, in terms of the time of the shortest connection. In this final section, we study this expense from the point of view of information geometry, which was briefly introduced in Sec.~\ref{ch1_subsec:speed-limits} from chapter~\ref{ch:introduction}. Specifically, the Fisher information~\eqref{ch1_eq:fisher-info}, the thermodynamic length $\mathcal{L}$~\eqref{ch1_eq:length-def} and its divergence $\mathcal{C}$~\eqref{ch1_eq:cost-def}, are analysed for the optimal bang-bang protocols considered in this chapter.     
\subsubsection{Fisher information}\label{ch5_subsubsec:fisher-info}
Let us start by considering the Fisher information. In its calculation, the Gaussian nature of the distribution plays a remarkably simplifying role. In particular, introducing Eq.~\eqref{ch5_eq:Gaussian-shape-2} into Eq.~\eqref{ch1_eq:fisher-info} leads to
\begin{equation}
\label{ch5_eq:Fisher-gaus}
    I(t) = \frac{1}{2}\sum_{j=1}^d \left[\frac{\dot{z}_j(t)}{z_j(t)}\right]^2,
\end{equation}
where we have made use of the relation
\begin{equation}
    \left<r_j^2r_l^2\right>(t) = \int d\boldsymbol{r} \  r^2_j r^2_l P(\boldsymbol{r},t) = (1+2\delta_{j,l}) z_j(t)z_l(t).
\end{equation}
In the fully degenerate case, all terms in the sum in Eq.~\eqref{ch5_eq:Fisher-gaus} are identical and thus the Fisher information simplifies to
\begin{equation}
\label{ch5_eq:Fisher-deg}
    I(t) = \frac{d}{2} \left[\frac{\dot{z}(t)}{z(t)}\right]^2.
\end{equation}
This expression is also valid for the almost fully degenerate case, for which $k_j=1+ \epsilon_j$, with $\epsilon_j\ll 1$, since the difference with Eq.~\eqref{ch5_eq:Fisher-deg} vanishes in the limit $\epsilon_j\to 0$, $\forall j$. A subtlety should be remarked, though: when evaluating the Fisher information over the optimal protocol, the path swept by the system in probability space---codified in the time evolution of $z(t)$---is different for the fully degenerate and the almost fully degenerate cases. For the former, the optimal path is identical to that for the one-dimensional case, which comprises only one bang. For the latter, the optimal path comprises $d$ bangs, with the upper and lower bounds of the temperature alternating over it.

\subsubsection{Thermodynamic length}\label{ch5_subsubsec:thermo-length}
Now, let us compute the thermodynamic length. We restrict ourselves to the almost fully degenerate case---consistently, our starting point is Eq.~\eqref{ch5_eq:Fisher-deg}. First, let us note that  
\begin{equation}\label{ch5_eq:length-inequal}
    \mathcal{L}(t) =  \sqrt{\frac{d}{2}}\int_0^t ds \ \biggr\rvert\frac{\dot{z}(s)}{z(s)}\biggr\rvert \geq \sqrt{\frac{d}{2}}\biggr\rvert \ln \frac{z(t)}{z(0)}\biggr\rvert \;\Longrightarrow\; \mathcal{L}(t_f) \geq  \sqrt{\frac{d}{2}}\rvert \ln T_f \rvert,
\end{equation}
which bounds $\mathcal{L}(t_f)/\sqrt{d}$ by its optimal value for $d=1$. Now, we derive $\mathcal{L}(t_f)$ for the optimal bang-bang protocol in arbitrary $d$ dimensions. For the sake of simplicity, and consistently with our approach in Sec.~\ref{ch5_sec:optimal-protocols}, we set $T_{\min}=0$. Taking into account the exponential relaxation in the cooling bangs, described by the evolution operator $\mathcal{E}_{1,0}^{\tau_c}$, as given by Eq.~\eqref{ch5_eq:ev-op}, one gets
\begin{equation}
\label{ch5_eq:length-Tmax}
 \mathcal{L}(t_f) = \sqrt{\frac{d}{2}} \left[ \ln T_f + 4 \tau_{c}\right],
\end{equation}
where $\tau_c$ is the total time employed in cooling stages. The explicit calculations leading to the above are rather cumbersome and not particularly illuminating, such that we relegate them to Appendix~\ref{app:thd-length}. Equation~\eqref{ch5_eq:length-Tmax} holds for arbitrary values of $T_{\max}$, being $T_{\min}=0$ and almost fully degeneration the only hypotheses necessary for deriving it. Notably, the limit $T_{\max} \to \infty$ further simplifies Eq.~\eqref{ch5_eq:length-Tmax}, since the heating bangs are instantaneous and thus $\tau_c=t_f$:
\begin{equation}
\label{ch5_eq:length-inf}
 \mathcal{L}(t_f) = \sqrt{\frac{d}{2}} \left[ \ln T_f + 4 t_f \right].
\end{equation}
Substituting Eq.~\eqref{ch5_eq:length-inf} into the inequality in Eq.~\eqref{ch5_eq:length-inequal} yields the following bounds for the connection time:
\begin{equation}
\label{ch5_eq:bounds-1d}
    t_f\geq - \frac{1}{2}\ln T_f , \; T_f<1; \qquad  t_f \geq 0 , \; T_f>1.
\end{equation}
These bounds are precisely the values over the optimal connection for $d=1$.

Apart from the expected multiplicative factor $\sqrt{d}$ in Eq.~\eqref{ch5_eq:length-inf}, the thermodynamic length has an additional increment when going from $d$ to $d+1$ stemming from its dependence with $t_f$. This feature can be understood as a fingerprint of the aforementioned unavoidable cost of increasing the spatial dimension of the system, even in the almost fully degenerate case.  

\subsubsection{Thermodynamic cost}\label{ch5_subsubsec:thd-cost}
When computing the thermodynamic cost, the linearity in the integrand allows us to integrate each dimension separately, making it unnecessary the assumption of fully degenerate systems to derive an analytical expression for $\mathcal C$. By introducing the general formula~\eqref{ch5_eq:Fisher-gaus} into Eq.~\eqref{ch1_eq:cost-def}, we obtain 
\begin{equation}
\mathcal{C} (t_f) =  \sum_{j=1}^d \left[ - k_j \ln T_f + \int_0^{t_f} ds \,\dot{z}_j(s)\frac{T(s)}{z_j^2(s)} \right].
\end{equation}
We have changed to $z_j$ as variable of integration, employed Eq.~\eqref{ch5_eq:ev_z}, and taken into account the initial and final values for the variances, $z_j(0)=1/k_j$ and $z_j(0)=T_f/k_j$. The temperature $T(s)$ alternatively takes the extreme values $T_{\min}$ (cooling bangs) and $T_{\max}$ (heating bangs). For the case of our concern, the cost simplifies to
\begin{equation}
\label{ch5_eq:cost}
\mathcal{C} (t_f) = - \sum_{j=1}^d \left[  k_j \ln T_f + T_{\max} \sum_l \Delta_{h,l} \left(\frac{1}{z_j}\right) \right],
\end{equation}
where $\Delta_{h,l} (1/z_j) $ refers to the change of the inverse of the $j$-th variance over the $l$-th heating bang---only the heating bang contributes to the sum when $T_{\min}=0$.  On the one hand, the first term, $-\sum_{j=1}^d k_j \ln T_f$ increases linearly with $d$ in the almost fully degenerate case. On the other hand, the second term is expected to increase faster than linearly with $d$, 
due to the double sum over $j$ and $l$.

\subsubsection{Speed limit}\label{ch5_subsubsec:speed-limits}
Given the explicit expressions of the thermodynamic length and its divergence, we may now compute the geometrical bound $t_{\text{geo}}$ from the Cauchy-Schwarz inequality~\eqref{ch1_eq:speed-limit}. In general, we have that $t_f \geq t_{\text{geo}}$---$t_f$ being the final connection time over the brachistochrone, with the equality holding exclusively over the geodesic. We remark that only both the heating and cooling procedures with $T_{\max} \rightarrow \infty$ and $T_{\min} = 0$ in the one dimensional case correspond to such geodesics, being the geodesic and the brachistochrone different in general.

\begin{figure}
    \centering
    \includegraphics[width=3.5in]{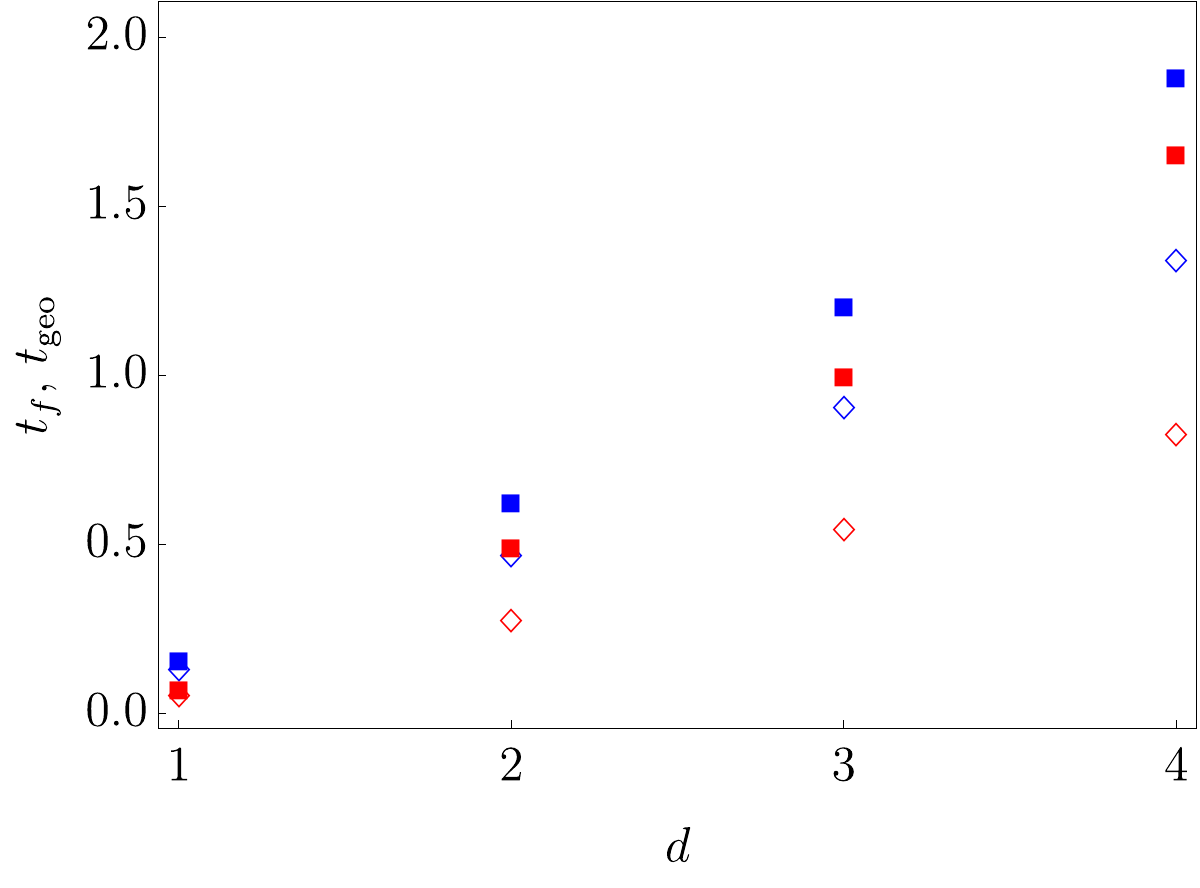}
    \caption{Dimension dependence of both the optimal connection time $t_f$ (filled squares) and the geometric lower bound  $t_{\text{geo}}=\mathcal{L}^2/(2C)$ in Eq.~\eqref{ch1_eq:speed-limit} (empty diamonds) for a general, almost fully degenerate, harmonic oscillator. Two sets of data are shown for different values of $T_{\max}$: $T_{\max} = 5$ (blue) and $T_{\max} = 10$ (red). Additional parameters employed are $T_f = 2$ and $T_{\min}=0$.} \label{ch5_fig:speed-limit-graph}
\end{figure}

As discussed before, both $\mathcal{L}/\sqrt{d}$ and $\mathcal{C}/d$ are expected to be increasing functions of $d$, due to the contribution thereto of the terms that involve details of the optimal connection---specifically,  the terms involving $t_f$ in Eq.~\eqref{ch5_eq:length-inf} and $\sum_j\Delta_h (1/z_j) $ in Eq.~\eqref{ch5_eq:cost}, respectively. Thus, the quantity $t_{\text{geo}} = (\mathcal{L}/\sqrt{d})^2/(2\mathcal{C}/d)$,  is expected to have a non-trivial dependence on the dimension in the almost fully degenerate case we are considering. Therefore, it is worth investigating the dependence on $d$ of the geometric bound $t_{\text{geo}}$. Figure~\ref{ch5_fig:speed-limit-graph} shows that $t_{\text{geo}}$ increases with $d$, which hints once more at the unavoidable cost of increasing the spatial dimension of the system. We must also point out that $t_{\text{geo}}$ decreases with $T_{\max}$, vanishing in the limit as $T_{\max} \rightarrow \infty$: while the optimal connection times $t_f$ tend to the finite values derived in Sec.~\ref{ch5_sec:optimal-protocols}. For $T_{\max} \rightarrow \infty$, $\mathcal L$ remains finite whereas $\mathcal C$ diverges, which implies that $t_{\text{geo}} \rightarrow 0$ for all $d$. Interestingly, this limit behaviour is similar to that found for a uniformly heated granular system in Ref.~\cite{prados_optimizing_2021}, and also for the system we consider in the next chapter, with a different version of the speed limit.

\chapter{Minimum time-connections for a Brownian gyrator}
\chaptermark{Brownian gyrator}
\label{ch:brownian_gyrator}

\newcommand{\rel}{\text{R}}
\newcommand{\irr}{\text{irr}}
In this chapter, we study a paradigmatic example of an out-of-equilibrium system: a Brownian gyrator~\cite{filliger_brownian_2007,dotsenko_two-temperature_2013,cerasoli_asymmetry_2018,baldassarri_engineered_2020}. It corresponds to an overdamped Brownian particle in two dimensions, confined in an elliptic potential, and whose degrees of freedom $\bm{r}^{\sf{T}} = (x,y)$ are submitted to the action of two white-noise forces stemming from two thermal baths at different temperatures $T_x$ and $T_y$, respectively. The two different temperatures make the system reach a non-equilibrium steady-state (NESS) in the long-time limit, and the term ``gyrator'' refers precisely to this system presenting a non-vanishing average torque proportional to the temperature difference $|T_x-T_y|$ at such NESS. 

Specifically, we work with a Brownian gyrator in which the elliptical potential confining the particle is characterised by two controllable parameters $(k,u)$, which respectively account for the diagonal terms $k$---i.e. we have identical stiffnesses in the $(x,y)$ directions---and for the anti-diagonal terms $u$ coupling the two degrees of freedom. We impose that the potential is confining  at all times, which restricts the values of the parameters to a certain region of the $(k,u)$ plane---the control set.
The problem we address here is the following: we consider two NESSs of the Brownian gyrator, corresponding to different values of the parameters characterising the potential, $(k_i,u_i)$ and $(k_f,u_f)$ for the initial and final states, respectively. Then, we employ Pontryagin's Maximum Principle (PMP) to look for the time-dependent protocols $(k(t),u(t))$ that make the connection time between these two states minimum. Throughout the whole chapter, the temperatures $(T_x,T_y)$ of the thermal baths are kept constant. 

The work presented in this chapter shows that there appear two different scenarios for the admissible controls over the brachistochrone: (i) regular bang-bang protocols, in which the controls $(k(t),u(t))$ switch between specific extremal values on the boundary of the control set, and (ii) infinitely degenerate \textit{singular} protocols, which belong to the boundaries of the control set but are more complex. We also analyse the behaviour of the resulting minimum connection time as a function of the chosen initial and target NESS. A very rich phenomenology emerges, including discontinuities in the connection time for a small perturbation of the target state, sets of target states that can be instantaneously reached---in the limit $k_{\max}\to\infty$, and sets of states that cannot be reached in a finite time.  

The structure of this chapter goes as follows. In Sec.~\ref{ch6_sec:model}, we introduce the dynamics of the Brownian gyrator model in detail, including the derivation of a thermodynamic speed limit involving the connection time and the irreversible work spent in the connection. In Sec~\ref{ch6_sec:control-problem}, we rigorously prove that optimal protocols are either of the regular bang-bang or singular type---thus excluding Euler-Lagrange protocols, and we characterise the family of infinitely degenerate singular protocols. Section~\ref{ch6_sec:constructing-bang-bang} is devoted to building the optimal-time bang-bang protocols stemming from PMP. The minimum connection time as a function of the initial and target states is finally analysed in Sec.~\ref{ch6_sec:minimum-time}.

\section{\label{ch6_sec:model}Model: Brownian gyrator}
The Brownian gyrator can be considered as an overdamped Brownian particle submitted to a two-dimensional confining potential $U = U(\bm{r})$ that is additionally in contact with two thermal baths of temperatures $T_x$ and $T_y$, which affect separately each degree of freedom. Specifically, its dynamics is accounted via a two-dimensional Gaussian process such as those discussed in Sec.~\ref{ch1_subsubsec:ornstein-uhlenbeck}, with the matrices $\mathbb{A}(t)$ and $\mathbb{D}_0(t)$ from Eq.~\eqref{ch1_eq:coeffs-ou} being given by
\begin{equation}\label{ch6_eq:matrices-BG}
    \mathbb{A}(t) = \frac{1}{\gamma}\left(\begin{array}{cc} 
        k(t) & u(t) \\
        u(t) & k(t)
        \end{array}\right), \quad \mathbb{D}_0(\cancel{t}) = \frac{k_B}{\gamma} \left(\begin{array}{cc} 
            T_x & 0 \\
            0 & T_y
            \end{array}\right).
\end{equation}
In this case, $\bm{r}^{\sf{T}}=(x,y)$ stands for the position of the Brownian particle, and similarly to the previous chapter, $\gamma$ and $k_B$ correspond to the friction coefficient and the Boltzmann constant, respectively. The confining potential is given by
\begin{equation}
\label{ch6_eq:potential}
    U(\bm{r},t) = \frac{1}{2}\gamma \bm{r}^{\sf{T}} \mathbb{A}(t) \bm{r} = \frac{1}{2}k(t)(x^2+y^2) + u(t)xy,
\end{equation}
which must satisfy the confining conditions
\begin{equation}
    k(t)\geq 0, \qquad \gamma \det(\mathbb{A}(t)) = k(t)^2-u(t)^2\geq 0,
\end{equation}
which ensure the existence of a NESS---for $T_x\ne T_y$; if $T_x=T_y=T$, then the NESS reduces to the canonical equilibrium distribution $P_{\text{eq}}(\bm{r})$ at temperature $T$. 

The parameters $(k,u)$ characterise the harmonic trap: the parameter $k$ corresponds to the stiffness of the potential, which for the sake of simplicity we assume that is identical in both directions, while $u$ accounts for the coupling between the two degrees of freedom. We assume that both parameters can be externally controlled, so they play the role of control functions in our problem. On experimental grounds, the Brownian gyrator model describes the behaviour of a colloidal particle trapped in an elliptical trap, and simultaneously in contact with two thermal baths at different temperatures in the two orthogonal directions. Such thermal baths have been realised in the laboratory with different experimental setups~\cite{ciliberto_heat_2013,chiang_electrical_2017,argun_experimental_2017,cerasoli_spectral_2022}.

Now let us introduce the normal modes $\bm{q}^{\sf{T}}=(q_1,q_2)$ via the linear transformation
\begin{equation}
    \bm{q} = \mathbb{M}\bm{r}, \quad \mathbb{M} \equiv \frac{1}{\sqrt{2}}\left(\begin{array}{cc}
    1 & 1 \\
    1 & -1
    \end{array}\right).
\end{equation}
Note that $\mathbb{M}$ is a symmetric and orthogonal matrix, its orthogonality entails that the Jacobian of the transformation to normal modes equals unity, and thus $P(\bm{q},t) = P(\bm{r},t)$. The latter implies that the Markov process $\bm{q}$ is also Gaussian, and characterised by the matrices
\begin{equation}
    \tilde{\mathbb{A}}(t) = \frac{1}{\gamma}\left(\begin{array}{cc}
        k(t) + u(t) & 0 \\
        0 & k(t) - u(t) 
        \end{array} \right), \quad \tilde{\mathbb{D}}_0 = \frac{k_B}{2\gamma} \left(\begin{array}{cc}
            T_x + T_y & T_x - T_y \\
            T_x - T_y & T_x + T_y
            \end{array}\right),
\end{equation}
which are related to the matrices $\mathbb{A}(t)$ and $\mathbb{D}_0$ by means of the bilinear transformations
\begin{equation}
    \tilde{\mathbb{A}}(t) = \mathbb{M} \mathbb{A}(t) \mathbb{M}, \quad \tilde{\mathbb{D}}_0 = \mathbb{M} \mathbb{D}_0 \mathbb{M},
\end{equation}
respectively. The potential $U$ may also be written in terms of the normal modes as
\begin{eqnarray}
    U(\bm{q},t) = \frac{1}{2}\gamma\bm{q}^{\text{T}} \tilde{\mathbb{A}}(t)\bm{q} = \frac{1}{2}(k(t)+u(t))q_1^2 + \frac{1}{2}(k(t)-u(t))q_2^2.
\end{eqnarray}
Although $U(\bm{q},t)$ may be expressed as the sum of two independent potentials for each normal mode, the corresponding Fokker-Planck equation for $\bm{q}$ presents a cross-derivative term---stemming from the non-diagonal elements of $\tilde{\mathbb{D}}_0$, which is related to the fact that the noises associated with the normal modes are now coupled. This entails that the long-time state for $T_x\neq T_y$ is a NESS, with a non-zero torque proportional to the temperature difference $|T_x-T_y|$~\cite{filliger_brownian_2007,dotsenko_two-temperature_2013}. In fact, the stationary net current $\bm{J}_{\text{s}}(\bm{r})$ is non-zero and also proportional to both $|T_x-T_y|$ and $u$, such that it only vanishes when either $T_x = T_y = T$---i.e. at equilibrium---or $u=0$.\footnote{The case $T_x = T_y = T$ with $u=0$ corresponds to the model presented in the previous chapter for $d=2$, with $k_1 = k_2 = k$ and a fixed value of the bath temperature.} In the latter, each degree of freedom tends independently to equilibrium with its own temperature.

As already discussed in Sec.~\ref{ch1_subsubsec:ornstein-uhlenbeck}, the dynamical behaviour of the system is fully accounted for via the second moments of the PDF. The particularisation of Eq.~\eqref{ch1_eq:second-moments} to the case of our concern provides the evolution equations
\begin{subequations}\label{ch6_eq:BG-evol-eqs-correl-explicit}
    \begin{align}
  \gamma\frac{d\expval{q_1^2}}{dt}  & = -2(k+u)\expval{q_1^2}+k_B (T_x+T_y),\\
  \gamma\frac{d\expval{q_2^2}}{dt}  & = -2(k-u)\expval{q_2^2}+k_B (T_x+T_y),\\
  \gamma\frac{d\expval{q_1 q_2}}{dt} & = -2 k \expval{q_1 q_2}+k_B (T_x-T_y),
    \end{align}
\end{subequations}
where we have taken into account that Eq.~\eqref{ch1_eq:second-moments} also applies for the normal modes $\bm{q}$, with the corresponding matrices $\tilde{\mathbb{A}}(t)$ and $\tilde{\mathbb{D}}_0$.
In the above, it is where we can appreciate the main advantage of employing the normal modes: had we chosen to work with our original variables $\bm{r}^{\sf{T}} = (x,y)$, the dynamical equations for their respective relevant moments would be coupled. However, the above equations for the correlations $\expval{q_j q_k}$ are not completely uncoupled, it has to be taken into account that the controls $(k,u)$ are the same for all of them. For example, if we looked for an inverse engineering solution,  i.e. for the controls $(k(t),u(t))$ stemming from a given time evolution for the correlations, the three functions $\{\expval{q_1^2}(t),\expval{q_2^2}(t),\expval{q_1 q_2}(t)\}$  would not be independent.

For time-independent values of $(k,u)$, one readily obtains the values of the moments at the NESS from Eq.~\eqref{ch6_eq:BG-evol-eqs-correl-explicit}: 
\begin{equation}\label{ch6_eq:stationary-moments}
    \expval{q_1^2}_{\st} = \frac{k_B(T_x + T_y)}{2(k+u)}, \quad \expval{ q_2^2}_{\st} = \frac{k_B(T_x+T_y)}{2(k-u)}, \quad \expval{q_1q_2}_{\st} = \frac{k_B(T_x - T_y)}{2k}.
\end{equation}

\subsection{\label{ch6_subsec:speed-limit}Thermodynamic speed limit}

Let us consider the energetic balance, in average, for our system. In terms of the normal modes just introduced above, the average energy reads
\begin{equation}
    \expval{U}=\frac{1}{2}(k+u)\expval{q_1^2}+\frac{1}{2}(k-u)\expval{q_2^2}.
\end{equation}
The average energy changes in time because both the controls $(k,u)$  and the variances $\expval{q_j^2}$ vary in time. The first contribution corresponds to the average work done on the system, whereas the second corresponds to the average heat exchanged with the thermal baths~\cite{sekimoto_stochastic_2010}:
\begin{equation}
    \frac{d}{dt}\expval{U}=\underbrace{\frac{1}{2}(\dot{k}+\dot{u})\expval{q_1^2}+\frac{1}{2}(\dot{k}-\dot{u})\expval{q_2^2}}_{\expval{\dot{W}}}+\underbrace{\frac{1}{2}(k+u)\frac{d\expval{q_1^2}}{dt}+\frac{1}{2}(k-u)\frac{d\expval{q_2^2}}{dt}}_{\expval{\dot{Q}}}.
\end{equation}
We focus now on the work done on the system in a process where $(k,u)$ are controlled during the time interval $(0,t_f)$:
\begin{align}\label{ch6_eq:evol-av-W}
    \expval{W}&=\int_{0}^{t_f} dt\, \left[ \frac{1}{2}(\dot{k}+\dot{u})\expval{q_1^2}+\frac{1}{2}(\dot{k}-\dot{u})\expval{q_2^2}\right] \nonumber
    \\
    &=\Delta\expval{U}-\int_{0}^{t_f} dt\, \left[ \frac{1}{2}(k+u)\frac{d\expval{q_1^2}}{dt}+\frac{1}{2}(k-u)\frac{d\expval{q_2^2}}{dt}, \right]
\end{align}
where $\Delta\expval{U}\equiv \expval{U}(t=t_f)-\expval{U}(t=0)$. Bringing to bear Eq.~\eqref{ch6_eq:BG-evol-eqs-correl-explicit}, we have
\begin{align}
    k+u &= \frac{k_B(T_x+T_y)}{2\expval{q_1^2}}-\frac{\gamma}{2}\frac{1}{\expval{q_1^2}}\frac{d\expval{q_1^2}}{dt}, &
    k-u &= \frac{k_B(T_x+T_y)}{2\expval{q_2^2}}-\frac{\gamma}{2}\frac{1}{\expval{q_2^2}}\frac{d\expval{q_2^2}}{dt},
\end{align}
which, when inserted into Eq.~\eqref{ch6_eq:evol-av-W}, gives
\begin{align}\label{ch6_eq:evol-av-W-v2}
    \expval{W}&=\Delta \left[ \expval{U}-\frac{k_B(T_x+T_y)}{4}\ln \left(\expval{q_1^2}\expval{q_2^2}\right) \right] \nonumber
    \\
    &+\frac{\gamma}{4}\int_0^{t_f} dt\, \left[\frac{1}{\expval{q_1^2}}\left(\frac{d\expval{q_1^2}}{dt}\right)^2+\frac{1}{\expval{q_2^2}}\left(\frac{d\expval{q_2^2}}{dt}\right)^2\right].   
\end{align}
Equation~\eqref{ch6_eq:evol-av-W-v2} suggests the definitions
\begin{align}
    \mathcal{F}&\equiv \expval{U}-\frac{k_B(T_x+T_y)}{2}\ln \left(\sigma_1\sigma_2\right), \\
    \expval{W_{\irr}}& \equiv \gamma\int_0^{t_f} dt\, \left[\left(\frac{d\sigma_1}{dt}\right)^2+\left(\frac{d\sigma_2}{dt}\right)^2\right], \label{ch6_eq:irr-work}
\end{align}
where
\begin{equation}
    \sigma_j\equiv \sqrt{\expval{q_j^2}}.
\end{equation}
The first contribution to $\expval{W}$ in Eq.~\eqref{ch6_eq:evol-av-W-v2} corresponds to the change of the function of state $\mathcal{F}$, which only depends on the initial and final values of the variances $\expval{q_j^2}$. For $T_x=T_y$, it equals Hemlhotz's free energy for a system of two oscillators; therefore, $\mathcal{F}$ can be considered as its generalisation to a non-equilibrium situation for the Brownian gyrator. The second contribution $\expval{W_{\irr}}$  depends on the protocol employed to connect the initial and final states. In other words, it is a functional of the protocol, which is non-negative and only vanishes when $d{\sigma_j}/dt=0$ for all times---i.e. for an infinitely slow protocol in which the variances have their instantaneous equilibrium values at all times. Therefore, we physically interpret $\expval{W_{\irr}}$ as the \textit{irreversible} contribution to the average work.

Interestingly, the above extension of the irreversible work to the connection between NESSs entails the emergence of a speed limit for it. By using the Cauchy-Schwarz inequality~\cite{loeve_probability_1997,bender_advanced_1999}, we have that
\begin{equation}
    \left| \int_0^{t_f} dt \, \frac{d\sigma_j}{dt} \right|^2 =\left| \sigma_{j,f}-\sigma_{j,i}\right|^2 \leq t_f \int_0^{t_f} dt \, \left(\frac{d\sigma_j}{dt}\right)^2.
\end{equation}
Therefore, we conclude that
\begin{equation}
    \gamma^{-1}t_f \expval{W_{\irr}}\geq \ \left| \sigma_{1,f}-\sigma_{1,i}\right|^2 + \left| \sigma_{2,f}-\sigma_{2,i}\right|^2 .
\end{equation}
For the connection between NESSs in which we are interested in, taking into account Eq.~\eqref{ch6_eq:stationary-moments}, the above inequality becomes
\begin{align}\label{ch6_eq:speed-limit}
    \frac{2\ t_f \expval{W_{\irr}}}{\gamma k_B(T_x+T_y)}\geq   \left(\frac{1}{\sqrt{k_f+u_f}}-\frac{1}{\sqrt{k_i+u_i}}\right)^2+\left(\frac{1}{\sqrt{k_f-u_f}}-\frac{1}{\sqrt{k_i-u_i}}\right)^2.
\end{align}
Equation~\eqref{ch6_eq:speed-limit} is a speed limit inequality for the desired connection between two NESSs of our system. Since the right-hand side only vanishes for $(k_f=k_i,u_f=u_i)$, this speed limit hints at the existence of a minimum time for the connection. This is the problem we address in the following section.

\section{\label{ch6_sec:control-problem}The control problem}

Let us pose the following control problem. Given an initial NESS of the Brownian gyrator, characterised by the control parameters $(k_i,u_i)$, and a final NESS, characterised by the parameters $(k_f,u_f)$, we aim at finding the optimal protocol $(k^*(t),u^*(t))$ that provides the fastest connection between them. In other words, we are interested in obtaining the protocol that connects two arbitrary NESSs in the shortest time $t_f$. 

For our analytical calculations, it is handy to introduce dimensionless variables as follows:
\begin{equation}\label{ch6_eq:dimensionless-variables}
    z_{1,2} \equiv \frac{k_i\expval{q_{1,2}^2}}{k_B(T_x + T_y)}, \quad z_3 \equiv \frac{k_i\expval{q_1 q_2}}{k_B(T_x - T_y)}, \quad k^* \equiv \frac{k}{k_i}, \quad u^* \equiv \frac{u}{k_i}, \quad t^* \equiv \frac{k_i}{\gamma}t.
\end{equation}
We are assuming that $T_x\ne T_y$, which is the relevant situation for our purposes: if $T_x=T_y=T$,  the initial and final states would be equilibrium states at the common temperature $T$. From now on, asterisks are dropped not to clutter our formulas.  The $z$ variables in Eq.~\eqref{ch6_eq:dimensionless-variables} correspond to the dimensionless relevant moments for the normal modes. In dimensionless variables, the evolution equations corresponding to Eq.~\eqref{ch6_eq:BG-evol-eqs-correl-explicit} are
\begin{equation}
\label{ch6_eq:dynamic-equations}
    \frac{dz_j}{dt} = f_j(\bm{z},\bm{\omega}) \equiv -2\omega_j z_j + 1, 
\end{equation}
where we have defined
\begin{equation}\label{ch6_eq:omega-defs}
    \omega_1 = k + u, \quad \omega_2 = k - u, \quad \omega_3 = k.
\end{equation}
We have introduced the compact notation $\bm{z}$ for the moments of the normal modes, such that $\bm{z}^{\sf{T}}\equiv (z_1,z_2,z_3)$. Equation \eqref{ch6_eq:dynamic-equations} is analogous to the dynamical equation for the variance of a one-dimensional Brownian particle confined in a harmonic trap, with each component $\omega_j$ of $\bm{\omega}$---with $\bm{\omega}^{\sf{T}}\equiv (\omega_1,\omega_2,\omega_3)$---playing the role of an effective stiffness of the trap for each relevant moment $z_j$; in fact, equilibrium states fulfil $z_{j,\st}=(2\omega_j)^{-1}$. We must highlight here an interesting feature: due to the linearity of the dynamical equations, regardless of the chosen protocol $(k(t),u(t))$ for the control variables, the time evolution of the system does not depend at all on the values of the bath temperatures $T_x$ and $T_y$, when working with dimensionless variables.

\begin{figure}
  \centering\includegraphics[width=3.5in]{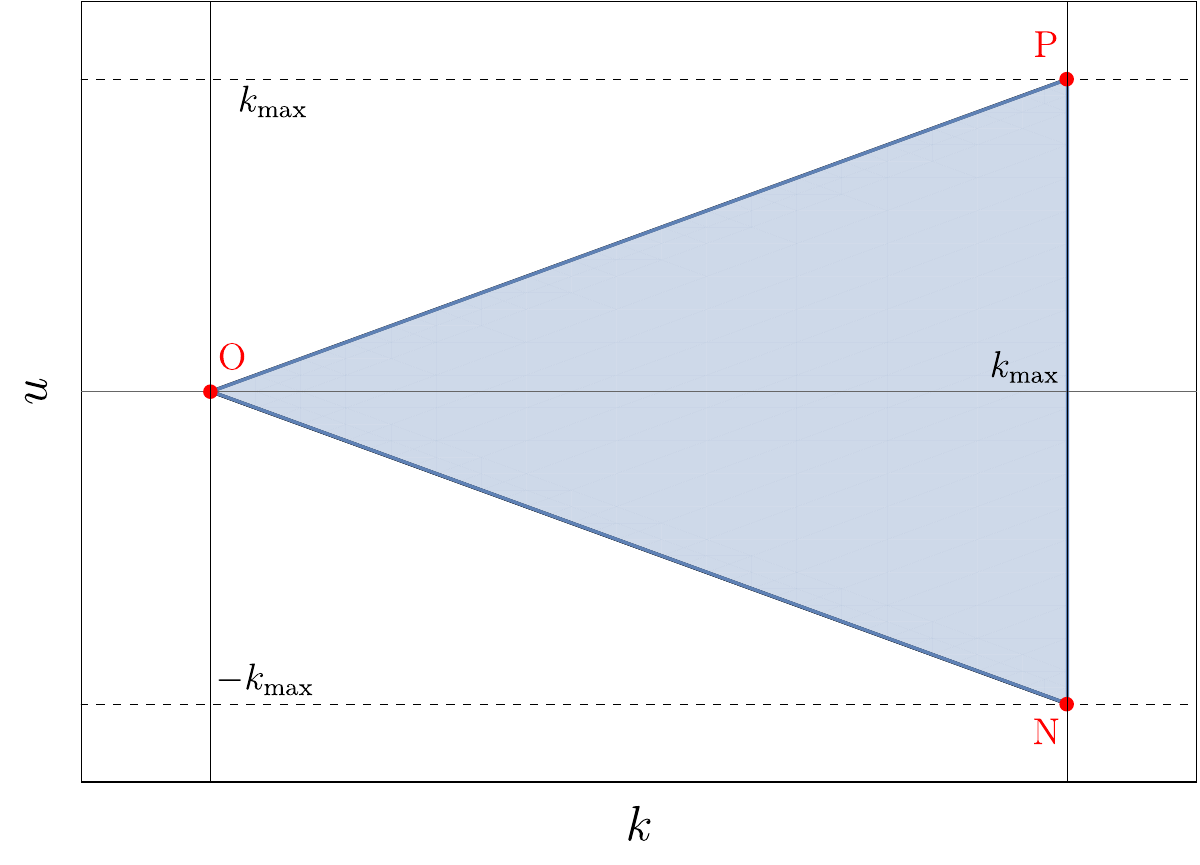}
  \caption{Sketch of the control set in the $(k,u)$ plane. The control set is defined by the inequalities in Eq.~\eqref{ch6_eq:control-set}: the blue, triangle-shaped, area constitutes the region where admissible protocols $(k(t),u(t))$ lie. In red, the vertices of the triangle-shaped set have been depicted, which correspond to the points O $= (0,0)$, P $=(k_{\max},k_{\max})$ and N $=(k_{\max},-k_{\max})$.
  }
  \label{ch6_fig:control-set}
\end{figure}

In terms of the dynamical variables $z_j$, we want to drive our dynamical system \eqref{ch6_eq:dynamic-equations} with initial condition $z_j(0) = (2\omega_{j,i})^{-1}$ towards a final state with $z_j(t_f) = (2\omega_{j,f})^{-1}$ with $\omega_{j,i} \equiv \omega_j(0)$ and  $\omega_{j,f} \equiv \omega_j(t_f)$, in the fastest way. The connection is made by controlling the time dependence of $(k,u)$, keeping the potential confining for all times. Therefore, we have the non-holonomic constraints
\begin{equation}
\label{ch6_eq:control-set}
    -k \leq u \leq k, \quad 0 \leq k \leq k_{\max}.
\end{equation}
The inequalities $-k\leq u\leq k$---equivalent to $k^2-u^2\geq 0$---and $k\geq 0$ ensure that the potential is confining. The additional inequality $k\leq k_{\max}$ accounts for the range of possible values of the strength of the confinement, stemming from experimental limitations on the amplitude of an externally induced electric or magnetic field, or the intensity of optical tweezers. Equation~\eqref{ch6_eq:control-set} defines the control set, i.e. the region in parameter space where all admissible controls $(k(t),u(t))$ must lie. In our case, we have the triangular control set depicted in Fig.~\ref{ch6_fig:control-set}. 

On general grounds, we expect that the fastest SST protocols achieve the desired connection in a finite time. This entails a huge improvement with respect to the direct STEP process, which consists in switching the control parameters from $(1,u_i)$ to $(k_f,u_f)$ instantaneously at $t=0^+$ and letting them constant afterwards. The relaxation to the final NESS corresponding to $(k_f,u_f)$ is exponential in the relevant moments, Eq.~\eqref{ch6_eq:dynamic-equations} tells us that the moment $z_j$ has a characteristic relaxation time $\tau_j\equiv (2\omega_{j,f})^{-1}$.
{Therefore, we can define the characteristic relaxation time $t_{\rel}$ for the STEP process of the Brownian gyrator as
\begin{equation}\label{ch6_eq:characteristic-time-trel}
    t_{\rel} = t_{\rel}(k_f, u_f) = \max\left[\frac{1}{2(k_f-u_f)},\frac{1}{2(k_f+u_f)}\right] = \frac{1}{2(k_f-|u_f|)},
\end{equation}
which gives the relaxation timescale for the slowest mode. From an experimental point of view, it is attractive to implement protocols that beat the natural timescale for relaxation, which one may estimate as $3 t_{\rel}$;\footnote{An exponential relaxation of the form $\exp(-t/t_{\text{R}})$ is completed up to $95 \%$ after $t = 3t_{\text{R}}$.} from a theoretical point of view, the system only reaches the target NESS for infinite time in the STEP process.}

\subsection{\label{ch6_subsec:pontryagin}Application of Pontryagin's Maximum Principle}

As discussed in Sec.~\ref{ch1_subsec:control-theory}, PMP provides necessary conditions for the optimal protocols $(k^*(t),u^*(t))$ that minimise a certain functional of the system variables and the controls. For the case of our concern, such functional corresponds to the connection time $t_f$, as given by Eq.~\eqref{ch1_eq:time-functional}. This minimisation has to be done while keeping $(k(t),u(t))$ in the control set, i.e. $(k,u)$ must fulfil the non-holonomic constraints~\eqref{ch6_eq:control-set} for all times. Now, we define an additional variable $z_0$, such that  $z_0(0) = 0$ and 
\begin{equation}
    \dot{z}_0 = f_0(\bm{z},\bm{\omega}) = 1 \implies  z_0(t_f) = \int_0^{t_f}dt \ f_0(\bm{z},\bm{\omega}) = t_f.
\end{equation}
Next, we introduce the conjugate momenta $(\psi_0,\bm{\psi}^{\sf{T}}) \equiv (\psi_0,\psi_1,\psi_2,\psi_3)$ and Pontryagin's Hamiltonian as
\begin{align}
    \Pi(\bm{z},\psi_0,\bm{\psi},k,u) =& \psi_0 f_0(\bm{z},\bm{\omega}) + \bm{\psi}^{\sf{T}}\bm{f}(\bm{z},\bm{\omega})=\psi_0+\sum_{j=1}^3 \psi_j f_j(\bm{z},\bm{\omega})
    \label{ch6_eq:Pontryagin-Hamiltonian-def}
\end{align}
where $\bm{f}^{\sf{T}}(\bm{z},\bm{\omega}) \equiv (f_1(\bm{z},\bm{\omega}),f_2(\bm{z},\bm{\omega}),f_3(\bm{z},\bm{\omega}))$ is the right-hand side of the dynamic equations~\eqref{ch6_eq:dynamic-equations}. Note that the Hamiltonian does not depend on $z_0$ by construction. Hamilton's canonical equations for both the dynamic and conjugate variables give
\begin{align}\label{ch6_eq:canonical}
    \dot{z}_j = \frac{\partial \Pi}{\partial \psi_j} = f_j(\bm{z},\bm{\omega})=-2\omega_j z_j+1, \quad \dot{\psi}_j = - \frac{\partial \Pi}{\partial z_j} = 2\omega_j \psi_j, \quad j=1,2,3.
\end{align}
Following the discussion from Sec.~\ref{ch1_subsec:control-theory}, PMP entails that we can classify the time-optimal protocols into three distinct categories:
\begin{enumerate}
    \item \textbf{Euler-Lagrange protocols}: They  correspond to situations for which the maximum of $\Pi$ is attained inside the control set, so it is necessary that
    \begin{equation}\label{ch6_eq:E-L-eq}
        \frac{\partial \Pi}{\partial k} = 0, \qquad \frac{\partial \Pi}{\partial u} = 0,
    \end{equation}
    which is equivalent to Eq.~\eqref{ch1_eq:euler-lagrange}. The solutions of the above correspond to twice-differentiable functions that might also be determined via the variational calculus approach discussed in Sec.~\ref{ch1_subsubsec:insights-variational} in the open interval $(0,t_f)$. Still, within PMP framework, sudden changes of the control parameters at both the initial and final times for the protocol, i.e. at $t = 0^+$ and $t=t_f^-$, respectively, are allowed.
    \item \textbf{Bang-bang protocols}: They correspond to situations in which the maximum of $\Pi$ is attained over the boundaries of the control set, so that Eq.~\eqref{ch6_eq:E-L-eq} no longer holds. Specifically, we refer to a bang-bang protocol when the maximum is reached at the vertices of the control set, as explained below.  
        
    For the case of our concern, the boundaries of the control set correspond to the sides of the triangle depicted in Fig.~\ref{ch6_fig:control-set}. The specific point at which $\Pi$ reaches its maximum value depends on the sign of the \textit{switching functions}
    \begin{subequations}\label{ch6_eq:switching-functions}
    \begin{align}
        \phi_{\text{OP}}(\bm{z},\bm{\psi})& \equiv \frac{\partial \Pi}{\partial k} + \frac{\partial \Pi}{\partial u}=-2\left(2\psi_1 z_1+\psi_3 z_3\right), 
        \label{ch6_eq:switching-functions-OP}   \\
        \phi_{\text{ON}}(\bm{z},\bm{\psi}) &\equiv \frac{\partial \Pi}{\partial k} - \frac{\partial \Pi}{\partial u}=-2\left(2\psi_2 z_2+\psi_3 z_3\right), 
        \label{ch6_eq:switching-functions-ON}   \\
        \phi_{\text{NP}}(\bm{z},\bm{\psi}) &\equiv  \frac{\partial \Pi}{\partial u}=2\left(\psi_2 z_2-\psi_1 z_1\right),
        \label{ch6_eq:switching-functions-NP}
    \end{align}   
    \end{subequations}
    which give the component of $\nabla_{\bm{k}}\Pi$ along the lines OP, ON, and NP, respectively, where O $\equiv (0,0)$, P $\equiv (k_{\max},k_{\max})$ and N $\equiv (k_{\max},-k_{\max})$ are the vertices of the triangle already introduced in Fig.~\ref{ch6_fig:control-set}. The sign of $\phi_{\text{AB}}$ gives the direction in which Pontryagin's Hamiltonian $\Pi$ increases, $\phi_{AB}>0$ means that $\Pi$ increases from A to B. Note that (i) the switching functions only depend explicitly on $\bm{z}$ and $\bm{\psi}$, and not on $(k,u)$, because the control functions enter linearly in Pontryagin's Hamiltonian, and (ii) the possible signs of the switching functions are restricted by the relation $\phi_{\text{OP}}-\phi_{\text{ON}}=2\phi_{\text{NP}}$.
        
    At any given time, the sign of the switching functions will determine which of the three vertices of the control set gives the maximum value of $\Pi$. Specifically, for $\phi_{\text{OP}}<0$ and $\phi_{\text{ON}}<0$, the maximum is attained at  O, for $\phi_{\text{OP}}>0$ and $\phi_{\text{NP}}>0$,  the maximum is attained at  P, and for $\phi_{\text{ON}}>0$ and $\phi_{\text{NP}}<0$, the maximum is attained at  N. Let us note that we only need to determine the signs of two switching functions in order to ascertain which point the bang-bang protocol chooses. For visualisation purposes, this is schematically represented in Fig.~\ref{ch6_fig:switching-funcs-regular}. The times at which these switching functions change their sign determine the switching times between the different vertices during a bang-bang protocol.  
    \begin{figure}
    \centering\includegraphics[width=4.5in]{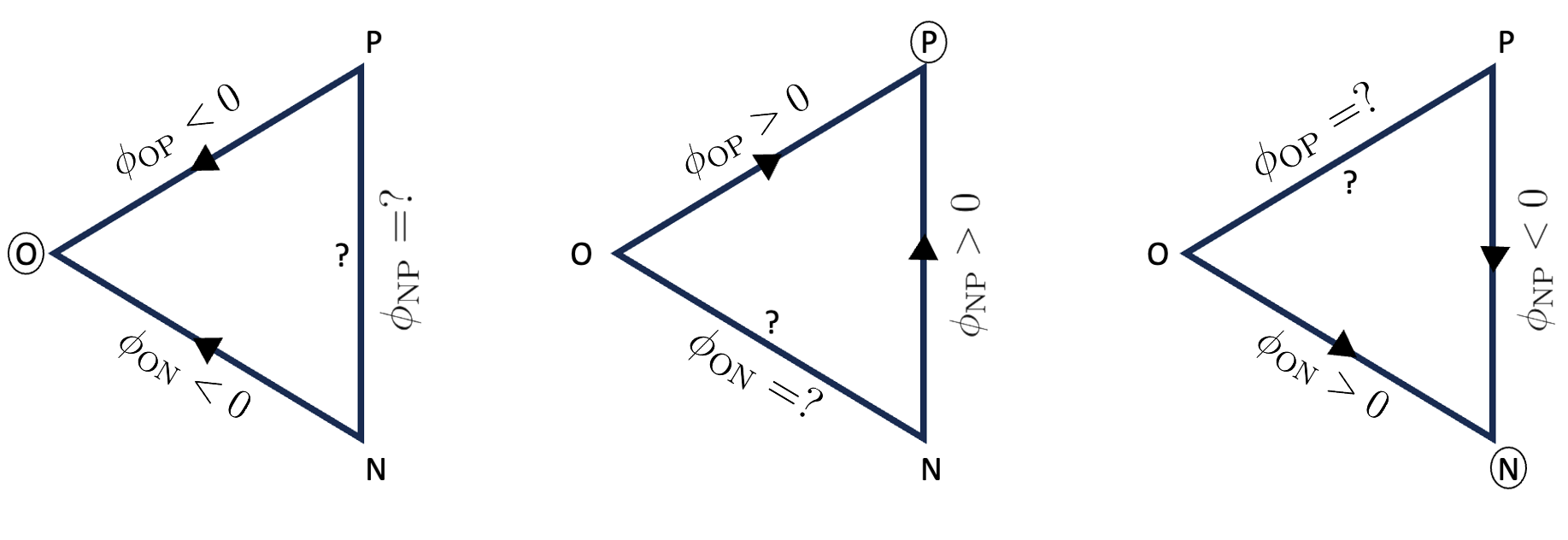}
    \caption{Schematic representation of the different situations for a bang-bang protocol. The arrows mark the direction of the gradient of Pontryagin's Hamiltonian along each of the edges of the triangular set from Fig.~\ref{ch6_fig:control-set}---i.e. the signs of the switching functions from Eq.~\eqref{ch6_eq:switching-functions}. From left to right, the maximum of Pontryagin's Hamiltonian over the boundary is attained at the vertex O, P, and N.
    }
    \label{ch6_fig:switching-funcs-regular}
    \end{figure}  
    \item \textbf{Singular protocols}: The above picture for bang-bang protocols breaks down if (at least) one of the switching functions in Eq.~\eqref{ch6_eq:switching-functions} vanishes during a finite time interval $[t_1,t_2]$ with $0\leq t_1 < t_2 \leq t_f$. We will refer to these protocols as \textit{singular} protocols: any point along the singular branch---i.e.~the branch of the control set over which the corresponding switching function identically vanishes in a finite interval---is a candidate for the optimal protocol, since Pontryagin's Hamiltonian is constant over the singular branch. Similarly to the bang-bang case, Fig.~\ref{ch6_fig:switching-funcs-singular} depicts a schematic representation of the three possible singular situations. 
    \begin{figure}
    \centering\includegraphics[width=4.5in]{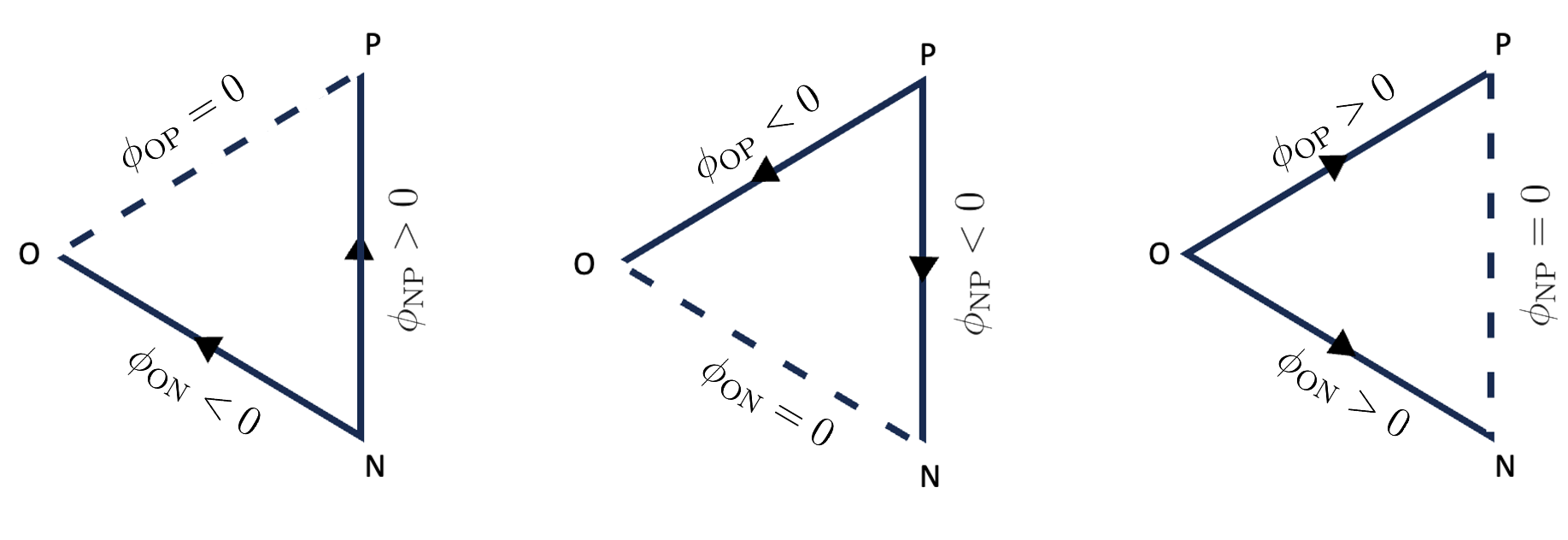}
    \caption{Schematic representation for the different situations for a singular protocol. The dashed lines mark the edge over which Pontryagin's Hamiltonian is constant, due to the vanishing of the corresponding switching function, whereas the arrows have the same meaning as in Fig.~\ref{ch6_fig:switching-funcs-regular}, they mark the gradient of Pontryagin's Hamiltonian  along each of the edges of control set. The relation $\phi_{\text{OP}}-\phi_{\text{ON}}=2\phi_{\text{NP}}$ restricts the possible arrow configurations.
    }
    \label{ch6_fig:switching-funcs-singular}
    \end{figure}
\end{enumerate}
We must emphasise that, although our dynamical system is linear in the control variables, and thus so it is Pontryagin's Hamiltonian function, our system does not belong to the class of linear-Pontryagin systems discussed in Sec.~\eqref{ch1_subsubsec:bang-bang}. In the Brownian gyrator, the dynamical equations \eqref{ch6_eq:dynamic-equations} do not have the structure from Eq.~\eqref{ch1_eq:linear-pontryagin}, due to the products $k z_j$ and $u z_j$---the equations of which are sometimes called bilinear~\cite{liberzon_calculus_2012}. Therefore, protocols belonging to any of the three categories above, Euler-Lagrange, bang-bang, and singular, are candidates for the brachistochrone. Also, optimal-time protocols could also combine different categories, concatenating time intervals corresponding to different classes of protocols. For example, we could have an optimal protocol starting as an Euler-Lagrange solution, then becoming bang-bang at some time $t_1 \in [0,t_f]$, and then  becoming singular at some other time $t_2 \in [t_1,t_f]$. In principle, PMP does not exclude these combinations for a general problem, so we need to thoroughly study the behaviour of the dynamical system and their conjugate variables in order to check which category combinations are possible in the Brownian gyrator's brachistochrone.

\subsection{\label{ch6_subsec:insight-1d}Insights from uncoupled oscillators}
Let us consider the $u(t)=0$ case for the Brownian gyrator, which corresponds to connecting arbitrary stationary states for two uncorrelated overdamped oscillators coupled to two different thermal baths. Such states are thus characterised uniquely by the final value of the control parameter $k(t)$, i.e. $k_f$. As $u_i = u_f = 0$, all the dynamical variables of our original system degenerate towards a unique one $z_j(t) = z(t)$, $j=1,2,3$, whose dynamical evolution is given by
\begin{equation}
\label{ch6_eq:equilibrium-eq}
    \dot{z} = -2k(t)z + 1,
\end{equation}
with $k(t) \in [0,k_{\max}]$ once again. Due to the simplicity of this situation---since we are dealing only with one dynamical variable---we derive the minimum connection times by other means. Equation~\eqref{ch6_eq:equilibrium-eq} can be integrated analytically, thus providing an integral form for the final connection time
\begin{equation}
\label{ch6_eq:connection-time-integral}
    t_f = \int_{z_i}^{z_f} \frac{dz}{1-2 k(z)z},
\end{equation}
with $z_i \equiv z(t=0) = 1/2$---recall, once more, that $k_i = 1$ in our scaled units---and $z_f = z(t=t_f) = 1/(2k_f)$. Now, $t_f$ does not diverge as long as the denominator $1-2k(z)z \neq 0 \ \forall z \in [z_i,z_f]$. This implies that the evolution of $z$ must be monotonic for the optimal time connection: given an initial $z_i$, if $z_f > z_i$ ($z_f < z_i$), then we need $\dot{z} > 0$ ($\dot{z} < 0$). Thus, we have two possibilities:
\begin{itemize}
    \item $2k(z)z > 1$: In this case, the integrand becomes negative during the entire integration interval. The connection time will only be positive if $z_f < z_i$, implying that $k_f > 1$. The integral from Eq.~\eqref{ch6_eq:connection-time-integral} is thus bounded from below by
    \begin{equation}
        t_f = \int_{z_i}^{z_f} \frac{dz}{1-2 k(z)z} \geq \int_{z_f}^{z_i} \frac{dz}{2 k_{\max}z-1} = \frac{1}{2k_{\max}} \ln \left(k_f \frac{k_{\max}-1}{k_{\max}-k_f} \right),
    \end{equation}
    which corresponds to the final connection time for a one-bang protocol with $k(t) = k_{\max}$.
    \item $2k(z)z < 1$: Here, the integrand is positive, thus implying that $z_f > z_i$ and $k_f < 1$. In this case, the integral from Eq.~\eqref{ch6_eq:connection-time-integral} is bounded from below by
    \begin{equation}\label{ch6_eq:1d-bang-O}
        t_f = \int_{z_i}^{z_f} \frac{dz}{1-2 k(z)z} \geq \int_{z_i}^{z_f}dz = z_f - z_i = \frac{1}{2}\left(\frac{1}{k_f} - 1 \right),
    \end{equation}
    which corresponds to the final connection time for a one-bang protocol with $k(t) = 0$.
\end{itemize}

Given the above, we have proved that optimal protocols achieving the fastest connection between stationary states for uncorrelated oscillators are also of the bang-bang type, specifically corresponding to one-bang protocols of the type $k(t) = 0$, for $k_f<1$, and $k(t) = k_{\max}$, for $k_f>1$, for all $t \in (0,t_f)$. Although the reasoning presented here cannot be straightforwardly extended to the $u(t) \neq 0$ scenario, it serves as a simple example of a non-linear-Pontryagin system in which the optimal protocol happens to be of bang-bang type.

\subsection{\label{ch6_subsec:existence-euler-lagrange}On the existence of Euler-Lagrange solutions}
In this section we explicitly show that Euler-Lagrange solutions---i.e those for which the optimal control $(k^*(t),u^*(t))$ lies in the interior of the control set---do not exist for the time-optimisation problem of concern. We recall that our dynamical system does not belong to the class of linear-Pontryagin systems, since the evolution equations do not have the form~\eqref{ch1_eq:linear-pontryagin}. Therefore, the question of whether we may rule out or not Euler-Lagrange solutions is non-trivial.

\subsubsection{Pure Euler-Lagrange solutions}
We start by considering ``pure'' Euler-Lagrange solutions, i.e. those for which the control variables lie in the interior of the control set for all $t \in (0,t_f)$. Therefore, Eq.~\eqref{ch6_eq:E-L-eq} must be satisfied at all times: specifically, we have that
    \begin{align}\label{ch6_eq:pontry-der}
    \frac{\partial \Pi}{\partial k}  &= -2(\psi_1z_1 + \psi_2z_2 + \psi_3z_3) = 0,
    &
    \frac{\partial \Pi}{\partial u} & = -2(\psi_1z_1 - \psi_2z_2) = 0.
    \end{align}
Since these two equalities hold $\forall t \in (0,t_f)$, their time derivatives must also vanish. On the one hand, the first time derivative gives
\begin{subequations}
    \begin{align}
        &\frac{d}{dt}\left(\psi_1z_1 + \psi_2z_2 + \psi_3z_3 \right) = \psi_1 + \psi_2 + \psi_3 = 0,\label{ch6_subec:derk}
        \\
        &\frac{d}{dt}\left(\psi_1z_1 - \psi_2z_2\right) = \psi_1 - \psi_2 = 0.\label{ch6_subec:deru}
    \end{align}
\end{subequations}
Let us start with Eq.~\eqref{ch6_subec:deru}. Together with Eq.~\eqref{ch6_eq:pontry-der} and the fact that $\Pi=0$ over the optimal solution, they imply that $\psi_1=\psi_2 \ne 0$ and thus $z_1 = z_2$---we cannot have $\psi_1=\psi_2=0$, because then Eq.~\eqref{ch6_subec:derk} implies that $\psi_3=0$, and $\Pi=0$ in turn implies that $\psi_0=0$.\footnote{Recall that one of the conditions in PMP is that $(\psi_0,\bm{\psi}^{\sf{T}})\ne (0,0,0,0)$, the so-called non-triviality condition~\cite{liberzon_calculus_2012}, as discussed in Sec.~\ref{ch1_subsec:control-theory}.} Moreover, their initial and final values must coincide as well,
\begin{subequations}
\begin{align}
    z_1(0) &= \frac{1}{2(1+u_i)} = \frac{1}{2(1-u_i)} = z_2(0) & \Longleftrightarrow \quad u_i &= 0,
    \\
    z_1(t_f) &= \frac{1}{2(k_f+u_f)} = \frac{1}{2(k_f-u_f)} = z_2(t_f) & \Longleftrightarrow \quad  u_f &= 0.
\end{align}
\end{subequations}
Thus, regardless of the form of the optimal control $(k,u)$, Euler-Lagrange protocols only allow to connect uncorrelated states. Now, Eq.~\eqref{ch6_subec:derk} provides further insights. Recalling again that $\Pi(t) = 0$ over the optimal solution, we have
\begin{align}\label{ch6_eq:second-condition-pmp}
    0 = \Pi &= \psi_0 + \psi_1\left[-2(k+u)z_1 + 1\right] + \psi_2\left[-2(k-u)z_2 + 1\right] + \psi_3\left[-2kz_3 + 1\right] \nonumber
    \\
    &= \psi_0 + \cancelto{0}{(\psi_1 + \psi_2 + \psi_3)} -2k\cancelto{0}{\left(\psi_1z_1 + \psi_2z_2 + \psi_3z_3 \right)} -2u\cancelto{0}{\left(\psi_1z_1 - \psi_2z_2\right)} .
\end{align}
Therefore, $\psi_0 = 0$ for Euler-Lagrange protocols. Although this does not contradict PMP, it will become useful later in this section to disregard ``mixed'' protocols. 

The second time derivatives of Eq.~\eqref{ch6_eq:pontry-der} must vanish too:
\begin{subequations}
    \begin{align}
        &\frac{d}{dt}\left(\psi_1 + \psi_2 + \psi_3 \right) = 2k\cancelto{0}{(\psi_1 + \psi_2 + \psi_3)} + 2u\cancelto{0}{(\psi_1-\psi_2)} = 0, \label{ch6_subeq:derrk}
        \\
        &\frac{d}{dt}\left(\psi_1 - \psi_2\right) = 2u(\psi_1 + \psi_2) = 0. \label{ch6_subeq:derru}
    \end{align}
\end{subequations}
Equation~\eqref{ch6_subeq:derrk} identically vanishes due to the relations \eqref{ch6_subec:derk} and \eqref{ch6_subec:deru} and, thus, does not provide any further information. On the other hand, Eq.~\eqref{ch6_subeq:derru} leaves us with two different cases:
\begin{enumerate}
    \item $\psi_1 + \psi_2=0$. This again leads to  $\psi_0=\psi_1=\psi_2=\psi_3=0$, which makes us ignore this possibility.
    \item $u=0$, for all $t\in(0,t_f)$. Recalling the dynamical system from Eq.~\eqref{ch6_eq:dynamic-equations}, together with the fact that $u_i = u_f = 0$, we have that all the dynamical variables $z_j$ degenerate onto the same one, and the structure of our optimisation problem radically changes, as we only need to study the behaviour of one dynamical variable now. In such scenario, as shown in Sec.~\ref{ch6_subsec:one-bang}, optimal-time protocols belong to the subclass of one-bang protocols, for which the control parameter $k$ takes either the value $0$ or $k_{\max}$---depending on the choice of the target state---during the entire time window.
\end{enumerate}


\subsubsection{Mixed Euler-Lagrange solutions}

Once we have disregarded ``pure'' Euler-Lagrange solutions, we consider here ``mixed'' solutions involving at least one time interval of Euler-Lagrange type. In the following, we investigate whether such Euler-Lagrange intervals can be mixed with regular bang-bang ones---we do not consider mixing with singular intervals, the impossibility of which is shown in Sec.~\ref{ch6_subsec:characterising-singular} when characterising optimal singular protocols. 

Let us consider a bang-bang window starting from O at some time $t_0$. Then, the switching functions must verify $\phi_{\text{OP}}(t_0)<0$ and $\phi_{\text{ON}}(t_0)<0$---as shown in Fig.~\ref{ch6_fig:switching-funcs-regular}. Making use of Eq.~\eqref{ch6_eq:switching-functions}, we have the constraints
\begin{equation}\label{ch6_eq:inequalities-1}
    2\psi_1(t_0)z_1(t_0) + \psi_3(t_0)z_3(t_0) >0, \quad 2\psi_2(t_0)z_2(t_0) + \psi_3(t_0)z_3(t_0) >0.
\end{equation}
Taking into account that both $k$ and $u$ vanish at the vertex O, Pontryagin's Hamiltonian function at $t=t_0$ is $\Pi = \psi_0 + \psi_1(t_0) + \psi_2(t_0) + \psi_3(t_0) = 0$, which entails that
\begin{align}\label{ch6_eq:constraint-2}
    \psi_0 &= -\psi_1(t_0) - \psi_2(t_0) - \psi_3(t_0) 
    < \frac{z_3(t_0)}{2z_1(t_0)}\psi_3(t_0) + \frac{z_3(t_0)}{2z_2(t_0)}\psi_3(t_0) - \psi_3(t_0) \nonumber 
    \\
    &= \psi_3(t_0) \left(\frac{z_3(t_0)}{2z_1(t_0)} + \frac{z_3(t_0)}{2z_2(t_0)} - 1 \right),
\end{align}
where we have substituted the inequalities from Eq.~\eqref{ch6_eq:inequalities-1}. Now we have two possibilities: (i) either $t_0=0$, implying that we start the optimal protocol at O, or (ii) $t_0 \ne 0$, corresponding to the scenario for which we had a previous Euler-Lagrange window from which we switched towards O. In the former, we substitute the initial state into \eqref{ch6_eq:constraint-2}, giving
\begin{equation}
    \psi_0 < \psi_3(t_0)\left(\frac{1+u_i}{2} + \frac{1-u_i}{2} - 1 \right) = 0 \ \Rightarrow \ \psi_0 < 0.
\end{equation}
In the latter, we take into account that $z_1(t_0)=z_2(t_0)=z_3(t_0)$ in the initial Euler-Lagrange window, so now Eq.~\eqref{ch6_eq:constraint-2} implies
\begin{equation}
    \psi_0 < \psi_3(t_0)\left(\frac{1}{2} + \frac{1}{2} - 1 \right) = 0 \ \Rightarrow \ \psi_0 < 0.
\end{equation}
In both cases,  $\psi_0 < 0$:  this is inconsistent with the fact that, during an Euler-Lagrange window, $\psi_0$ must be zero. Since $\psi_0$ is constant and cannot change during the entire protocol, this rules out the possibility of having mixed Euler-Lagrange protocols with O time windows. 

We proceed in a similar manner for the possible switching between P/N and Euler-Lagrange windows. Let us now consider a P window, starting again at some time $t_0$. Then, the switching functions must verify $\phi_{\text{OP}}(t_0)>0$ and $\phi_{\text{NP}}(t_0)>0$, which, using again Eq.~\eqref{ch6_eq:switching-functions}, imply the constraints
\begin{equation}\label{ch6_eq:inequalities-2}
    2\psi_1(t_0)z_1(t_0) + \psi_3(t_0)z_3(t_0) <0, \quad \psi_1(t_0)z_1(t_0) - \psi_2(t_0)z_2(t_0) < 0.
\end{equation}
Now, Pontryagin's Hamiltonian at time $t_0$ reads
\begin{equation}
    0 = \Pi = \psi_0 + \psi_1(t_0) + \psi_2(t_0) + \psi_3(t_0) - 2k_{\max}(2\psi_1(t_0)z_1(t_0) + \psi_3(t_0)z_3(t_0)),
\end{equation}
which entails
\begin{align}
    \psi_0 &= -\psi_1(t_0) - \psi_2(t_0) - \psi_3(t_0) + 2k_{\max}(2\psi_1(t_0)z_1(t_0) + \psi_3(t_0)z_3(t_0))\nonumber
    \\ 
    &< -\psi_1(t_0)\left( 1+\frac{z_1(t_0)}{z_2(t_0)}\right) - \psi_3(t_0)+ 2k_{\max}(2\psi_1(t_0)z_1(t_0) + \psi_3(t_0)z_3(t_0)),
\end{align}
where we have substituted the inequalities from Eq.~\eqref{ch6_eq:inequalities-2}. On the one hand, had we initially started at vertex P, i.e. $t_0 = 0$, then we would have
\begin{equation}
    \psi_0 < 2(k_{\max}-1)\left(2\psi_1(t_0)z_1(t_0) + \psi_3(t_0)z_3(t_0) \right) < 0 \ \Rightarrow \ \psi_0 < 0,
\end{equation}
{since $k_i=1<k_{\max}$.} On the other hand, had we previously applied an Euler-Lagrange window, then
\begin{equation}
    \psi_0 < {(2k_{\max}z_1(t_0)-1)}(2\psi_1(t_0) + \psi_3(t_0)) < 0 \ \Rightarrow \ \psi_0 < 0,
\end{equation}
{since $(2k_{\max}z_1(t_0)-1)>0$.} Once again, as in both cases we have that $\psi_0 < 0$, then we may disregard mixed protocols involving Euler-Lagrange and P windows. Let us remark that, due to the symmetry between the P and N points, the above proof extends to N time windows as well, and thus, we may conclude that there cannot be mixed protocols involving Euler-Lagrange and bang-bang windows.

\subsection{\label{ch6_subsec:characterising-singular}Characterisation of singular protocols}

As mentioned in Sec.~\ref{ch6_subsec:pontryagin}, singular protocols are those for which at least one of the switching functions $\phi_{\text{OP}}$, $\phi_{\text{ON}}$ or $\phi_{\text{NP}}$ identically vanishes during a time window within {$(t_1,t_2)\in(0,t_f)$  and, in addition, the maximum of Pontryagin's Hamiltonian is attained over the corresponding edge of the triangular control set. In this section we show that singular protocols, if they exist, extend to the whole time interval $(0,t_f)$, i.e. switchings between regular and singular protocols is not possible.}

Let us start by considering the edge $\overline{\text{OP}}$ and its corresponding switching function $\phi_{\text{OP}}$, which is given by Eq.~\eqref{ch6_eq:switching-functions-OP}.  Since this function is identically zero in $(t_1,t_2)$, its time derivatives must also cancel:
\begin{align}
    \dot{\phi}_{\text{OP}} &= -2(2\psi_1 + \psi_3) = 0,
    &
    \ddot{\phi}_{\text{OP}} &= -4k(4\psi_1 + \psi_3) = 0.
\end{align}
{Since $k(t)$ does not identically vanish over the edge $\overline{\text{NP}}$}---otherwise we would recover the scenario depicted in Sec.~\ref{ch6_subsec:insight-1d}, the only available solution is $\psi_1(t) = \psi_3(t) = 0$, $\forall t\in(t_1,t_2)$. However, recalling Hamilton's canonical equations \eqref{ch6_eq:canonical}, we have that { $\dot{\psi}_1=2(k+u)\psi_1$ and $\dot{\psi}_3=2k\psi_3$.} Therefore, if $\psi_{1,3}$ vanishes at some specific time, $\psi_{1,3}(t)$ must be zero $\forall t$. Hence, $\psi_1(t) = \psi_3(t) = 0$, $\forall t\in(0,t_f)$, and  $\phi_{\text{OP}}(t)$ also vanishes over the whole time interval: the singular protocol over the edge $\overline{\text{OP}}$ extends to all times in $(0,t_f)$. Now, invoking that $\Pi = 0$ from the second condition of PMP, we have that
\begin{align}
    0 = \Pi &= \psi_0 + \cancelto{0}{\psi_1}(-4kz_1+1) + \psi_2 + \cancelto{0}{\psi_3}(-2kz_3+1) \nonumber 
    \\
    &\Rightarrow \ \psi_2 = -\psi_0 > 0, \; \forall t\in (0,t_f).
\end{align}
That is, $\psi_2(t)$ should be constant for singular protocols, and moreover, it cannot cancel. { This is consistent with the evolution equations: over the $\overline{\text{OP}}$ edge, the evolution equation for $\psi_2$ reduces to $\dot{\psi}_2=0$.}

It is straightforward to extend the above analysis to the $\overline{\text{ON}}$ edge and the $\phi_{\text{ON}}$ switching function, due to the symmetry between the P and N points. In such a case, the same results apply, upon exchanging the behaviour found above for the $\psi_1$ and $\psi_2$ momenta. { Wrapping things up, we have shown that singular protocols over the $\overline{\text{OP}}$ and $\overline{\text{ON}}$ edges may appear, but they have to extend to the whole time interval $(0,t_f)$.}

Let us now consider the last edge $\overline{\text{NP}}$ and the corresponding switching function, given by Eq.~\eqref{ch6_eq:switching-functions-NP}. Again, since $\phi_{\text{NP}}$ identically vanishes in $(t_1,t_2)$, taking time derivatives we get 
\begin{align}\label{ch6_subeq:PN-branch}
    \dot{\phi}_{\text{NP}} &= -2(\psi_1 - \psi_2) = 0, 
    &
    \ddot{\phi}_{\text{NP}} &= -4u(\psi_1 + \psi_2) = 0.
\end{align}
Similarly, since $u(t)$ does not identically vanish over the edge $\overline{\text{NP}}$, we have that $\psi_1(t)=\psi_2(t)=0$ for all $t\in(t_1,t_2)$. Again, from the evolution equations for the momenta, we conclude that $\psi_1(t)=\psi_2(t)=0$ for all $t\in(0,t_f)$, the singular time window extends to the whole interval. Invoking again that Pontryagin's Hamiltonian vanishes over the time optimal protocol,
\begin{align}
        0 = \Pi &= \psi_0 + \cancelto{0}{\psi_1}(-2(k_{\max}+u)z_1 + 1) \nonumber
        \\
        &+ \cancelto{0}{\psi_2}(-2(k_{\max}-u)z_1 + 1) + \psi_3(-2k_{\max}z_3+1),
\end{align}
we have
\begin{align}
 \psi_3 = \frac{\psi_0}{2k_{\max}z_3-1}<0,
\end{align}
{where we have taken into account that $z_3(t) \geq (2k_{\max})^{-1}$, which stems from the evolution equation for $z_3$, Eq.~\eqref{ch6_eq:dynamic-equations}.}
    
Following the preceding discussion, we conclude that the singular protocols corresponding to the $\overline{\text{OP}}$, $\overline{\text{ON}}$, and $\overline{\text{NP}}$ edges share similar features: (i) they correspond to ``pure'' protocols that last for the whole interval $(0,t_f)$, (ii) they involve the vanishing of two conjugate momenta during the entire protocol, and (iii) they are characterised by the action of an unique control variable---either $k(t)$, for the $\overline{\text{OP}}$ and $\overline{\text{ON}}$ edges, or $u(t)$, for the  $\overline{\text{NP}}$ edge. 

An additional interesting feature is that, for each singular protocol, the connection time is always fixed---regardless of the choice of $k(t)$ or $u(t)$ over the considered edge. The physical reason for this behaviour is that the time evolution of one of the variances sets the connection time, independently of the others. More specifically, for the $\overline{\text{OP}}$ edge, we have that $\dot{z}_2 = 1$, thus giving
\begin{equation}\label{ch6_eq:edge-op-time}
t_f^{({\text{OP}})} = z_2(t_f) - z_2(t_0) = \frac{1}{2}\left(\frac{1}{k_f-u_f} -\frac{1}{1-u_i}\right). 
\end{equation}
Similarly, for the $\overline{\text{OP}}$ branch, we have $\dot{z_1} = 1$, implying that
\begin{equation}\label{ch6_eq:edge-on-time}
t_f^{({\text{ON}})} = z_1(t_f) - z_1(t_0) = \frac{1}{2}\left(\frac{1}{k_f+u_f} -\frac{1}{1+u_i}\right). 
\end{equation}
And finally, for the $\overline{\text{NP}}$, we have $\dot{z}_3 = -2k_{\max}z_3+1$, which gives
\begin{equation}
t_f^{({\text{NP}})} = \frac{1}{2k_{\max}} \ln \left(k_f \frac{k_{\max}-1}{k_{\max}-k_f} \right).
\end{equation}
Each of the singular branches have still two dynamical variables whose behaviour is determined via one control parameter $k(t)$ or $u(t)$. In principle, since there are no further restrictions on the values of $k$ or $u$, there is an infinite number of possible optimal protocols achieving the desired connection in the given connection time $t_f$. One could take such degeneracy as an opportunity for optimising an additional figure of merit, such as the irreversible work or the information cost.

\section{\label{ch6_sec:constructing-bang-bang}Constructing optimal bang-bang protocols}

We devote this section to building optimal bang-bang protocols with different numbers of bangs---i.e. different number of switchings between the vertices O, P, N of the control set. Here, we use the general notation V to refer to one of these vertices. { We will show that (i) one-bang protocols only allow for connecting the initial point $\text{I}\equiv (1,u_i)$ with some isolated points $(k_f,u_f)$ within the control set, (ii) two-bang protocols allow for connecting I with some specific curves in the $(k_f,u_f)$ plane, (iii) three-bang protocols allow for connecting I with all the points of $(k_f,u_f)$ within the control set. 
}

As the controls $(k,u)$ take constant values at any of the vertices, the evolution equations~\eqref{ch6_eq:dynamic-equations} can be analytically solved. Let us assume that the switching to a certain vertex V occurs at a certain time $t_0$, for which $z_j(t_0)=z_{j,0}$, and denote by $\omega_{j,\! V}$ the value of the $j$-th effective stiffness at the considered vertex; the time evolution of $z_j$ is given by 
\begin{equation}\label{ch6_eq:operator-k-finite}
    \mathcal{E}^{(\tau)}_{\omega_{j,\text{V}}}(z_{j,0}) \equiv z_j(t) = \frac{1}{2\omega_{j,\! V}} + \left(z_{j,0}-\frac{1}{2\omega_{j,\! V}}\right)e^{-2\tau\omega_{j,\! V}},
\end{equation}
with $\tau \equiv t - t_0\ge 0$. Following a similar approach as in the previous chapter, we have defined the operator $\mathcal{E}^{(\tau)}_{\omega_{j,\text{V}}}(z_{j,0})$, which generates the time evolution of the considered moment $z_j$ during a time interval $\tau$, with constant driving $\omega_{j,\! V}$. This operator has a well-defined  limit for $\omega_{j,\! V} \to 0$,
\begin{equation}\label{ch6_eq:operator-k-0}
    \mathcal{E}_0^{(\tau)}(z_{j,0}) = \lim_{\omega_{j,\! V} \to 0}\mathcal{E}^{(\tau)}_{\omega_{j,\text{V}}}(z_{j,0}) = z_{j,0} + \tau,
\end{equation}
$z_j$ linearly increases with time. This will be useful for our analysis, since at each of the vertices (O, P, N) at least one $\omega_{j,\! V}$ vanishes and then the corresponding moment linearly increases. { In fact, at the origin O we have $\omega_{j,\text{O}}=0$ and thus $\mathcal{E}^{(\tau)}_{\omega_{j,\text{O}}}=\mathcal{E}_0^{(\tau)}$, $\forall j$.}

A bang-bang protocol $V_1 V_2 V_3$\ldots, where $V_1 V_2 V_3$\ldots is any permutation of the vertices of the control set, such that $V_{j+1}\ne V_j$, is obtained by sequentially applying the time evolution operators corresponding to each of the vertices $V_j$ during a certain time window $\tau_j$. The total duration of the bang-bang protocol is thus given by $t_f=\sum_j \tau_j$, and the final value of the variables $z_j$ is 
\begin{equation}\label{ch6_eq:general-bang-bang}
    z_j(t_f) = \underbrace{\left( \cdots \circ \mathcal{E}_{\omega_{j,V_3}}^{\tau_3} \circ \mathcal{E}_{\omega_{j,V_2}}^{\tau_2} \circ \mathcal{E}_{\omega_{j,V_1}}^{\tau_1} \right)}_{{\textnormal{composition of }} M {\textnormal{ operators }}}
\left(z_{j,i}\right), \quad j=1,2,3.
\end{equation}
The goal now is to look for a consistent choice of the sequence $V_1 V_2 V_3$\ldots with the right values of the corresponding time intervals $\tau_1$, $\tau_2$, $\tau_3$, \ldots to satisfy the boundary conditions for the moments, i.e. to drive the system from the initial NESS with $z_j(0)= (2\omega_{j,i})^{-1}$ to the target NESS with $z_j(t_f)= (2\omega_{j,f})^{-1}$.

In order to simplify our analysis, we consider the limit $k_{\max} \rightarrow +\infty$, i.e. the limit of having infinite capacity for compression---we recall that the constraint $k\leq k_{\max}$ was of practical nature, whereas the constraints $k\geq 0$ and $k^2-u^2\geq 0$ ensure that the potential is confining for all times. The infinite capacity for compression entails vanishing times $\tau \rightarrow 0$ for the windows corresponding to the points P and N, as they involve exponential decays for the dynamic variables with characteristic timescales of the order of $1/k_{\max}$. Therefore, $k_{\max} \tau$ must remain finite, and this makes it possible to introduce a parameter $\xi$ that substitutes $\tau$ for the determination of the optimal control. Specifically, we make 
\begin{equation}\label{ch6_eq:quenching-factor}
    k_{\max} \tau = -\frac{1}{2}\ln \xi \ \Rightarrow \ \xi = e^{-2k_{\max} \tau},
\end{equation}
with $\xi \in [0,1]$ being the quenching factor corresponding to the relaxing dynamic variable. {
Let us consider the vertex P, thereat we have $\omega_{1,\text{P}}=2 k_{\max}$, $\omega_{2,\text{P}}=0$, $\omega_{3,\text{P}}=k_{\max}$. Therefore, for $k_{\max}\to\infty$, while keeping $\xi$ constant, one has the following instantaneous evolutions, from $t_0$ to $t_0^+=t_0+\tau$:
\begin{equation}
    z_1(t_0^+)= z_{1,0}\,\xi^2 , \quad z_2(t_0^+)= z_{2,0}, \quad z_3(t_0^+)=z_{3,0}\,\xi.
\end{equation}
If we define the operator
\begin{equation}
    \tilde{\mathcal{E}}_{\xi}(z_{j,0}) \equiv z_{j,0}\xi,
\end{equation}
the above instantaneous evolutions can be rewritten as
\begin{equation}\label{ch6_eq:operator-kmax-inf}
    z_1(t_0^+)= \tilde{\mathcal{E}}_{\xi^2}(z_{1,0}) , \quad z_2(t_0^+)= \tilde{\mathcal{E}}_{1}(z_{2,0}), \quad z_3(t_0^+)=\tilde{\mathcal{E}}_{\xi}(z_{3,0}).
\end{equation}
For the vertex N, the situation is completely analogous, but the evolutions for $z_1$ and $z_2$ are exchanged because $\omega_{1,\text{N}}=0=\omega_{2,\text{P}}$ and $\omega_{2,\text{N}}=2k_{\max}=\omega_{1,\text{P}}$.

Bang-bang controls in the limit $k_{\max}\to\infty$ are constructed analogously to Eq.~\eqref{ch6_eq:general-bang-bang}, but with the operators $\mathcal{E}_{\omega_{j,\text{V}}}^{\tau}$, with V being either P or N, being substituted with the corresponding operator from Eq.~\eqref{ch6_eq:operator-kmax-inf}. Our discussion above entails that we thus have a concatenation of operators $\mathcal{E}_0^{(\tau)}$, corresponding to the vertex O, and operators $\tilde{\mathcal{E}}_{\xi^2}$, $\tilde{\mathcal{E}}_{\xi}$, and $\tilde{\mathcal{E}}_{1}$, corresponding to the vertices P or N, in the time evolution of the moments.
}

\subsection{\label{ch6_subsec:one-bang}One-bang protocols}

This is the most basic case, for which the optimal protocol remains at one of the vertices, O, P or N, during the entire time interval $(0,t_f)$. In the following, we analyse the evolution corresponding to each of the three vertices separately. 

We start our analysis with the vertex O, for which $\omega_{j,\text{O}}=0$ for all $j$. Therefore, we have 
\begin{equation}\label{ch6_eq:bang-bang-O}
    z_j(t_f) = \frac{1}{2\omega_{j,f}} =\mathcal{E}_0^{(t_f)}\left(\frac{1}{2\omega_{j,i}}\right) = \frac{1}{2\omega_{j,i}} + t_f, \quad j=1,2,3.
\end{equation}
Note that the above corresponds to an algebraic system of three equations with just one variable; the connection time. Thus, it will only be consistent for specific choices of both the initial and final conditions. Recalling the definition of the $\omega_j$'s, Eq.~\eqref{ch6_eq:omega-defs}, the above system presents the following solutions: (i) $t_f = 0$ for $(k_f,u_f) = (1,u_i)$---i.e. the initial and final states are the same, ad thus the system does not evolve, (ii) $t_f \rightarrow + \infty$ for $(k_f,u_f) = (0,0)$---i.e. the target stationary state becomes the point O, which is reached in an infinite amount of time, as the optimal protocol becomes a direct step protocol, and (iii) $t_f = (k_f^{-1} - 1)/2$ for $u_f = u_i = 0$ and $k_f < 1$. The latter case corresponds to connecting uncorrelated states of the Brownian gyrator, for which the dynamic variables degenerate into a unique one.

Now, me move onto the analysis of a one-bang protocol at the vertex P. In this case, we have the instantaneous evolution found in Eq.~\eqref{ch6_eq:operator-kmax-inf},
\begin{subequations}\label{ch6_eq:bang-bang-P}
    \begin{align}
        z_1(0^+) &= \frac{1}{2(k_f+u_f)}=\tilde{\mathcal{E}}_{\xi^2}\left(\frac{1}{2(1+u_i)}\right) = \frac{\xi^2}{2(1+u_i)}, \label{ch6_subeq:bang-bang-P-z1}
        \\
        z_2(0^+) &= \frac{1}{2(k_f-u_f)} = \tilde{\mathcal{E}}_{1}\left(\frac{1}{2(1-u_i)}\right) = \frac{1}{2(1-u_i)},
        \\
        z_3(0^+) &= \frac{1}{2k_f} = \tilde{\mathcal{E}}_{\xi}\left(\frac{1}{2} \right) = \frac{\xi}{2}.
    \end{align}
\end{subequations}
Note that $t_f=0$ for this protocol, in the limit $k_{\max}\to\infty$ we are considering. In the above, we have taken into account the explicit expressions for the initial and final values of the moments $\omega_j$ of Eq.~\eqref{ch6_eq:omega-defs}. Equation~\eqref{ch6_eq:bang-bang-P} presents the following solutions: (i) $\xi = 1$ for $(k_f, u_f) = (1, u_i)$---i.e., again, the initial and final states are the same, and (ii) $\xi = (1+u_i)/(1-u_i)$ for $u_i<0$, which corresponds to a final point P$^*$ with coordinates
\begin{equation}\label{ch6_eq:P*}
    k_{\text{P}^*}=\frac{1-u_i}{1+u_i}, \qquad u_{\text{P}^*}=-u_i\, k_{\text{P}^*}, \quad u_i<0.
\end{equation}
The point $\text{P}^*$ is the only non-trivial NESS, i.e. different from the initial state, reachable with a one-bang protocol at vertex P. Note that $|u_i|<1$, since otherwise we would obtain an unphysical solution.

Lastly, the one-bang N protocol is completely analogous to the P one; we have the same algebraic equations as in Eqs.~\eqref{ch6_eq:bang-bang-P}, but upon the variable changes $(z_1,z_2) \to (z_2,z_1$) and $u_{i,f} \to -u_{i,f}$. Thus, the corresponding solutions in this case are (i) the trivial solution $\xi = 1$, i.e. again $(k_f, u_f) = (1, u_i)$, and (ii) $\xi = (1-u_i)/(1+u_i)$ for $u_i > 0$,  which corresponds to a final point $\text{N}^*$ such that
\begin{equation}\label{ch6_eq:N*}
    k_{\text{N}^*}=\frac{1+u_i}{1-u_i}, \qquad u_{\text{N}^*}=-u_i\, k_{\text{N}^*}, \quad u_i>0.
\end{equation}
The symmetry between the P and N protocols will be further exploited when studying higher-order bang-bang protocols in the forthcoming sections. Note that both $\text{P}^*$ and $\text{N}^*$ tend to the initial point in the limit $u_i\to 0$, i.e. for an uncorrelated initial state.

\subsection{\label{ch6_subsec:two-bang}Two-bang protocols}

Optimal two-bang protocols involve a switching time ${t_1\in (0,t_f)}$ that splits the whole time interval into two different time windows, in which the control parameters correspond to one of the vertices, i.e. O, P, or N. Permutations of two points avoiding consecutive repetition give rise to six possible protocols: OP, ON, PO, NO, PN and NP. These protocols involve the action of two operators of the type \eqref{ch6_eq:operator-k-0} or \eqref{ch6_eq:operator-kmax-inf} on each of the
dynamic variables. 

\begin{figure} 
\centering \includegraphics[width=4.5in]{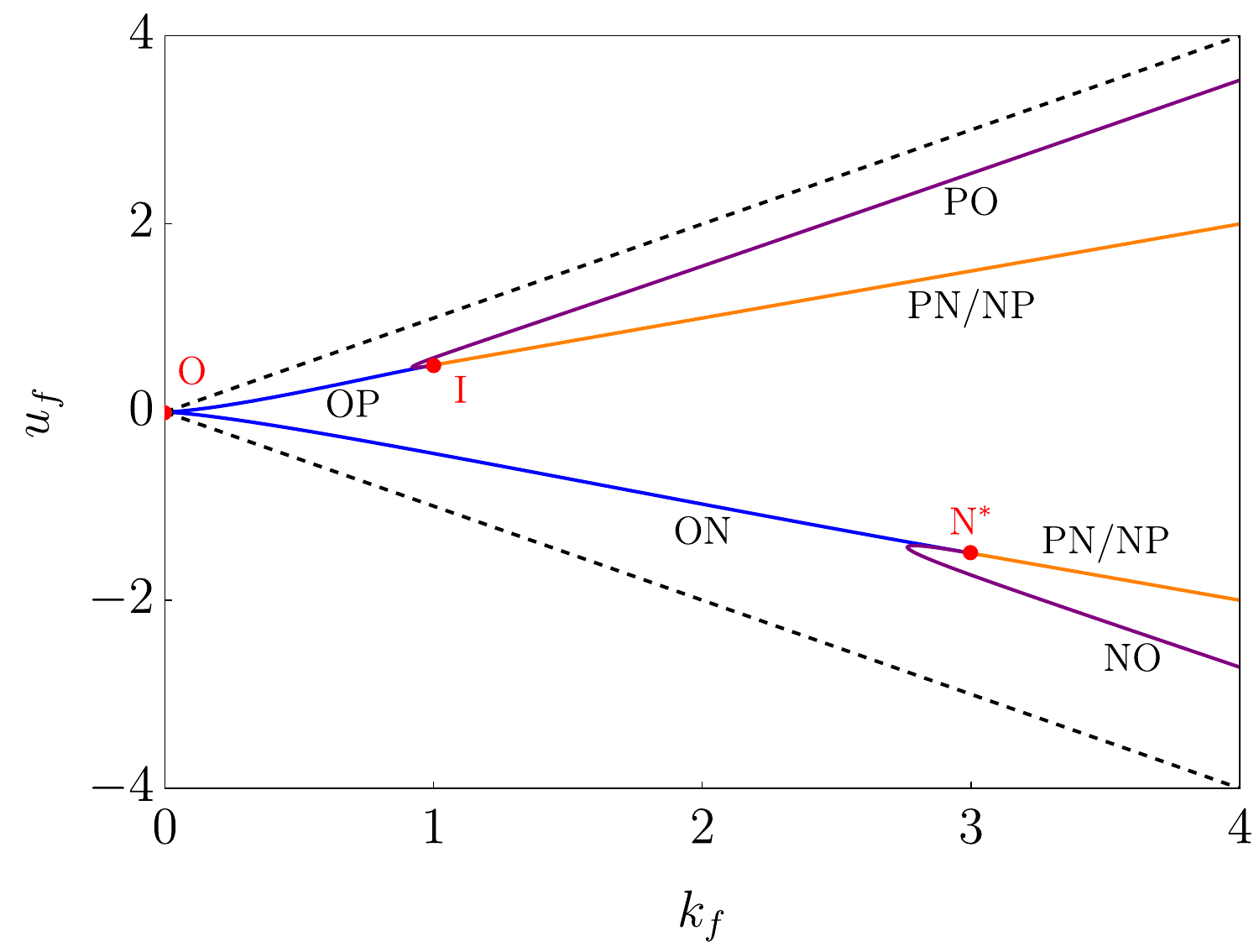}
\\
\centering \includegraphics[width=4.5in]{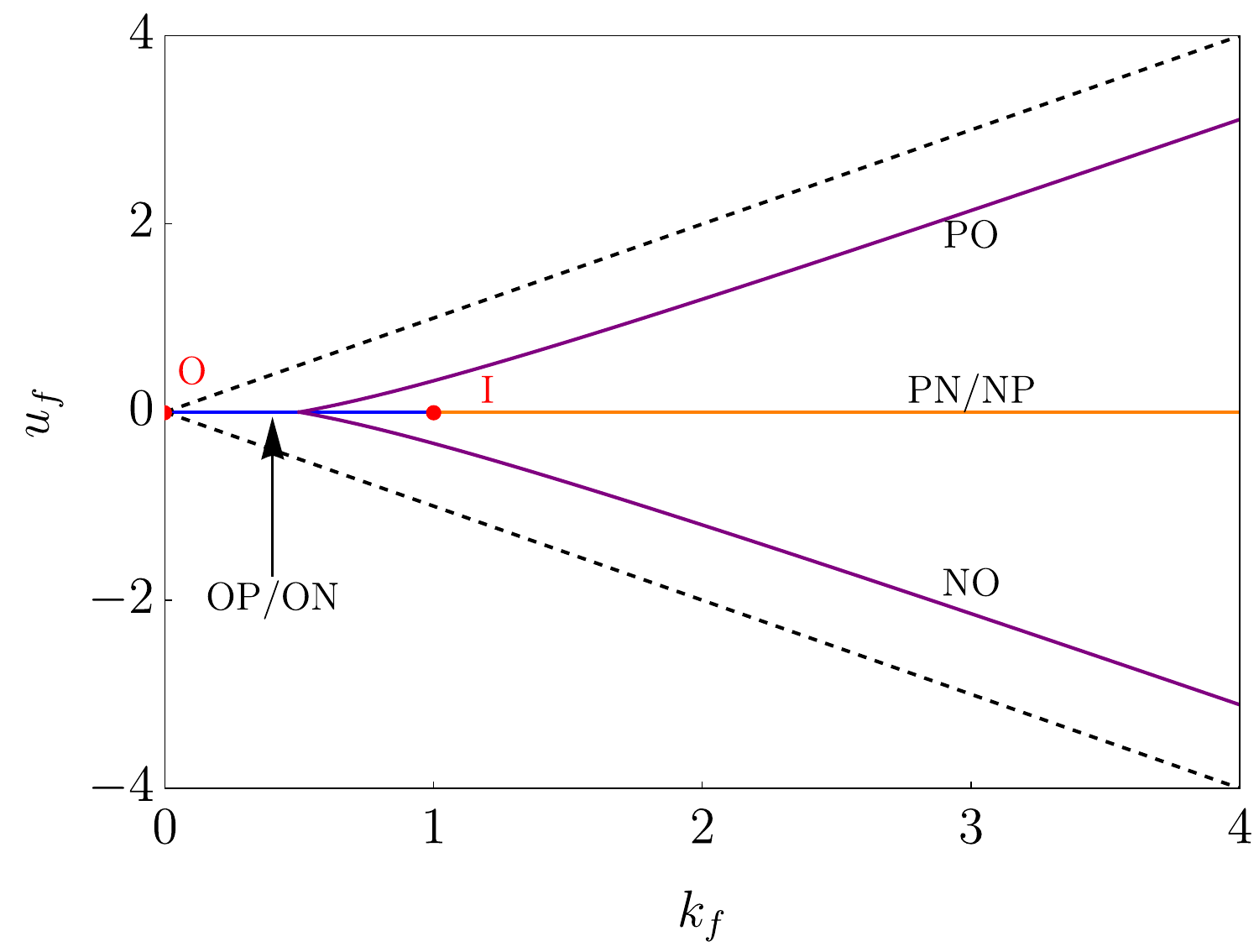}
\caption{Reachable curves in the $(k_f,u_f)$ plane with two-bang protocols. The curves have been obtained in the infinite capacity of compression, $k_{\max}\to\infty$. Specifically, the top panel corresponds to $u_i=0.5$ and the bottom panel to $u_i=0$---recall that $k_i=1$ in our dimensionless variables. Dashed lines represent the boundaries of the control set, given by the constraints $k_f>0$, $k_f^2-u_f^2 > 0$. The red points mark the reachable points with one-bang protocols. On the top panel, we have the initial state $I=(1,u_i)$, and the point $\text{N}^*$ defined in the text. On the bottom panel, I and $\text{N}^*$ merge into a unique point, { and the plot is symmetric with respect to the $u_f=0$ axis upon exchanging N$\leftrightarrow$P. We have also added the origin O as a red point in both panels, although it is attained in an infinite amount of time.}
}
    \label{ch6_fig:two-bang-protocols}
\end{figure}

The two-bang protocols involve the action of two successive evolution operators, which in turn introduce two unknowns---$(t_f,\xi)$ for the protocols OP, ON, PO and NO, and $(\xi_1,\xi_2)$ for the protocols PN and NP. Since the boundary conditions for the target state give us three algebraic equations, one for each $z_{j,f}$, there will be solutions for the unknowns only for specific choices of the initial and target NESSs. However, for a given fixed initial condition, there will be curves of the form $f(k_f,u_f)=0$ consistent with each of the algebraic equations, instead of the isolated points found in the one-bang case. The systems of algebraic equations characterising each of the two-bang protocols are given by:

\begin{itemize}
    \item \textbf{OP}: Variables $(t_f, \xi)$.
    \begin{subequations}
        \begin{align}
            \frac{1}{2(k_f + u_f)} &= \left(\tilde{\mathcal{E}}_{\xi^2}\circ \mathcal{E}_0^{(t_f)}\right)\left(\frac{1}{2(1+u_i)}\right)= \left(\frac{1}{2(1+u_i)} + t_f\right)\xi^2,
            \\
            \frac{1}{2(k_f - u_f)} &= \left(\tilde{\mathcal{E}}_1\circ \mathcal{E}_0^{(t_f)}\right)\left(\frac{1}{2(1-u_i)}\right) = \frac{1}{2(1-u_i)} + t_f,
            \\
            \frac{1}{2k_f} &= \left(\tilde{\mathcal{E}}_{\xi}\circ \mathcal{E}_0^{(t_f)}\right)\left(\frac{1}{2}\right)= \left(\frac{1}{2} + t_f\right)\xi.
        \end{align}
    \end{subequations}
    \item \textbf{ON}: Variables $(t_f, \xi)$.
    \begin{subequations}
        \begin{align}
            \frac{1}{2(k_f + u_f)} &= \left(\tilde{\mathcal{E}}_1\circ \mathcal{E}_0^{(t_f)}\right)\left(\frac{1}{2(1+u_i)}\right) = \frac{1}{2(1+u_i)} + t_f,
            \\
            \frac{1}{2(k_f - u_f)} &= \left(\tilde{\mathcal{E}}_{\xi^2}\circ \mathcal{E}_0^{(t_f)}\right)\left(\frac{1}{2(1-u_i)}\right)= \left(\frac{1}{2(1-u_i)} + t_f\right)\xi^2,
            \\
            \frac{1}{2k_f} &= \left(\tilde{\mathcal{E}}_{\xi}\circ \mathcal{E}_0^{(t_f)}\right)\left(\frac{1}{2}\right)= \left(\frac{1}{2} + t_f\right)\xi.
        \end{align}
    \end{subequations}
    \item \textbf{PO}: Variables $(t_f, \xi)$.
    \begin{subequations}
        \begin{align}
            \frac{1}{2(k_f+u_f)} &= \left(\mathcal{E}_0^{(t_f)}\circ \tilde{\mathcal{E}}_{\xi^2}\right)\left(\frac{1}{2(1+u_i)}\right)= \frac{\xi^2}{2(1+u_i)} + t_f,
            \\
            \frac{1}{2(k_f-u_f)} &= \left(\mathcal{E}_0^{(t_f)}\circ \tilde{\mathcal{E}}_1\right)\left(\frac{1}{2(1-u_i)}\right)= \frac{1}{2(1-u_i)} + t_f,
            \\
            \frac{1}{2k_f} &=\left(\mathcal{E}_0^{(t_f)}\circ \tilde{\mathcal{E}}_{\xi}\right)\left(\frac{1}{2}\right)= \frac{\xi}{2} + t_f.
        \end{align}
    \end{subequations}
    \item \textbf{NO}: Variables $(t_f, \xi)$.
    \begin{subequations}
        \begin{align}
            \frac{1}{2(k_f+u_f)} &= \left(\mathcal{E}_0^{(t_f)}\circ \tilde{\mathcal{E}}_1\right)\left(\frac{1}{2(1+u_i)}\right)= \frac{1}{2(1+u_i)} + t_f,
            \\
            \frac{1}{2(k_f-u_f)} &= \left(\mathcal{E}_0^{(t_f)}\circ \tilde{\mathcal{E}}_{\xi^2}\right)\left(\frac{1}{2(1-u_i)}\right)= \frac{\xi^2}{2(1-u_i)} + t_f,
            \\
            \frac{1}{2k_f} &=\left(\mathcal{E}_0^{(t_f)}\circ \tilde{\mathcal{E}}_{\xi}\right)\left(\frac{1}{2}\right)= \frac{\xi}{2} + t_f.
        \end{align}
    \end{subequations}
    \item \textbf{PN and NP}: Variables $(\xi_1, \xi_2)$, and thus a null connection time.
    \begin{subequations}\label{ch6_subeq:PN-protocol}
        \begin{align}
            \frac{1}{2(k_f+u_f)} &= \left(\tilde{\mathcal{E}}_1\circ \tilde{\mathcal{E}}_{\xi_1^2}\right)\left(\frac{1}{2(1+u_i)}\right) = \frac{\xi_1^2}{2(1+u_i)}, 
            \\
            \frac{1}{2(k_f-u_f)} &= \left( \tilde{\mathcal{E}}_{\xi_2^2}\circ\tilde{\mathcal{E}}_1\right)\left(\frac{1}{2(1-u_i)}\right) = \frac{\xi_2^2}{2(1-u_i)},
            \\
            \frac{1}{2k_f} &= \left(\tilde{\mathcal{E}}_{\xi_2}\circ \tilde{\mathcal{E}}_{\xi_1}\right)\left(\frac{1}{2}\right) = \frac{\xi_1\xi_2}{2}.
        \end{align}
    \end{subequations}
\end{itemize}
Figure~\ref{ch6_fig:two-bang-protocols} show the curves in the $(k_f,u_f)$-plane that are reachable with two-bang protocols, for two different choices of the initial condition $u_i$. First, it is worth highlighting that we have fixed the initial condition $u_i$ to be non-negative, without loss of generality. Due to the symmetry between the P and N vertices, the figure for $u_i<0$ would be obtained by reflecting the figure for $u_i>0$ through the $u_f=0$ axis, with the exchange N$\leftrightarrow$P in all protocols. This symmetry is particularly evident in the $u_i = 0$ case, whose figure is symmetric with respect to the $u_f = 0$ axis. We have also chosen the color code of each curve for the same reason: for example, the blue curves, corresponding to the protocols OP and ON, are exchanged upon the flipping of the figure, and they merge for the case $u_i = 0$.\footnote{In fact, they become a one-bang protocol at vertex O. As $u_i = u_f = 0$, this scenario is equivalent to the case of uncorrelated oscillators from Sec.~\ref{ch6_subsec:insight-1d}, for which the connection time is given by Eq.~\eqref{ch6_eq:1d-bang-O}.} The same applies for the purple curves, corresponding to the protocols PO/NO, and the orange ones, accounting for the PN/NP protocols. Moreover, we have also marked with red dots the points reachable with a one-bang protocol, described in Sec.~\ref{ch6_subsec:one-bang}. Note that $\text{P}^*$ does not appear, since we are considering $u_i > 0$. On the top panel, we have the initial state $\text{I} = (1, u_i)$ and the point $\text{N}^*$, while on the bottom panel the initial state and the point $\text{N}^*$ have merged into a unique point. We have also added the origin O, although such point is only attained in an infinite amount of time. We note that these points constitute limiting values for all the two-bang protocols: OP starts from the origin O and ends at the initial state I, ON starts at O and ends at the point $\text{N}^*$, both PO and PN/NP start at I, and both NO and PN/NP start at the point $\text{N}^*$. These are not coincidences: the one-bang solutions must belong to a subgroup of the two-bang ones, they are only attained when (at least) one of the operators corresponding to one of the time windows of the considered two-bang protocol becomes the identity operator. For instance, let us consider the two-bang protocol ON. When the operator corresponding to the N time window becomes the identity, the ON protocol reduces to the O protocol that makes it possible to reach the origin; when the operator corresponding to the O time window becomes the identity, the ON protocol reduces to the N protocol that makes it possible to reach the point N$^*$.

\subsection{\label{ch6_subsec:three-bang}Three-bang protocols}

In the last case, we look into is three-bang optimal protocols, which split the time interval $(0,t_f)$ into three time windows with two switching times. In each time window, the controls $(k,u)$ attain the values corresponding to one of the three vertices: O, P, and N. Permutations of these points---eliminating those having two identical consecutive vertices---give rise to twelve three-bang protocols: OPO, POP, ONO, NON, OPN, ONP, PON, NOP, PNO, NPO, PNP, NPN. The systems of algebraic equations characterising each of the three-bang protocols are given by:
\begin{itemize}
    \item \textbf{OPO}: Variables $(\tau_1, \tau_2, \xi)$, with $\tau_1 + \tau_2 = t_f$.
    \begin{subequations}
        \begin{align}
            \frac{1}{2(k_f + u_f)} &= \left(\mathcal{E}_0^{(\tau_2)} \circ \tilde{\mathcal{E}}_{\xi^2}\circ \mathcal{E}_0^{(\tau_1)}\right)\left(\frac{1}{2(1+u_i)}\right)= \left(\frac{1}{2(1+u_i)} + \tau_1\right)\xi^2 +\tau_2,
            \\
            \frac{1}{2(k_f - u_f)} &= \left(\mathcal{E}_0^{(\tau_2)} \circ\tilde{\mathcal{E}}_1\circ \mathcal{E}_0^{(\tau_1)}\right)\left(\frac{1}{2(1-u_i)}\right) = \frac{1}{2(1-u_i)} + \tau_1 + \tau_2,
            \\
            \frac{1}{2k_f} &= \left(\mathcal{E}_0^{(\tau_2)} \circ\tilde{\mathcal{E}}_{\xi}\circ \mathcal{E}_0^{(\tau_1)}\right)\left(\frac{1}{2}\right)= \left(\frac{1}{2} + \tau_1\right)\xi + \tau_2.
        \end{align}
    \end{subequations}
    \item \textbf{ONO}: Variables $(\tau_1, \tau_2, \xi)$, with $\tau_1 + \tau_2 = t_f$.
    \begin{subequations}
        \begin{align}
            \frac{1}{2(k_f + u_f)} &= \left(\mathcal{E}_0^{(\tau_2)} \circ\tilde{\mathcal{E}}_1\circ \mathcal{E}_0^{(\tau_1)}\right)\left(\frac{1}{2(1+u_i)}\right) = \frac{1}{2(1+u_i)} + \tau_1 + \tau_2,
            \\
            \frac{1}{2(k_f - u_f)} &= \left(\mathcal{E}_0^{(\tau_2)} \circ \tilde{\mathcal{E}}_{\xi^2}\circ \mathcal{E}_0^{(\tau_1)}\right)\left(\frac{1}{2(1-u_i)}\right)= \left(\frac{1}{2(1-u_i)} + \tau_1\right)\xi^2 +\tau_2,
            \\
            \frac{1}{2k_f} &= \left(\mathcal{E}_0^{(\tau_2)} \circ\tilde{\mathcal{E}}_{\xi}\circ \mathcal{E}_0^{(\tau_1)}\right)\left(\frac{1}{2}\right)= \left(\frac{1}{2} + \tau_1\right)\xi + \tau_2.
        \end{align}
    \end{subequations}
    \item \textbf{POP} (Singular): Variables $(t_f, \xi_1, \xi_2)$.
    \begin{subequations}
        \begin{align}
            \frac{1}{2(k_f+u_f)} &= \left(\tilde{\mathcal{E}}_{\xi_2^2} \circ \mathcal{E}_0^{(t_f)}\circ \tilde{\mathcal{E}}_{\xi_1^2}\right)\left(\frac{1}{2(1+u_i)}\right)= \left(\frac{\xi_1^2}{2(1+u_i)} + t_f\right) \xi_2^2,
            \\
            \frac{1}{2(k_f-u_f)} &= \left(\tilde{\mathcal{E}}_1 \circ \mathcal{E}_0^{(t_f)}\circ \tilde{\mathcal{E}}_1\right)\left(\frac{1}{2(1-u_i)}\right)= \frac{1}{2(1-u_i)} + t_f,
            \\
            \frac{1}{2k_f} &=\left(\tilde{\mathcal{E}}_{\xi_2}\circ \mathcal{E}_0^{(t_f)}\circ \tilde{\mathcal{E}}_{\xi_1}\right)\left(\frac{1}{2}\right)= \left(\frac{\xi_1}{2} + t_f\right) \xi_2.
        \end{align}
    \end{subequations}
    \item \textbf{NON} (Singular): Variables $(t_f, \xi_1, \xi_2)$.
    \begin{subequations}
        \begin{align}
            \frac{1}{2(k_f+u_f)} &= \left(\tilde{\mathcal{E}}_1 \circ \mathcal{E}_0^{(t_f)}\circ \tilde{\mathcal{E}}_1\right)\left(\frac{1}{2(1+u_i)}\right)= \frac{1}{2(1+u_i)} + t_f,
            \\
            \frac{1}{2(k_f-u_f)} &= \left(\tilde{\mathcal{E}}_{\xi_2^2} \circ \mathcal{E}_0^{(t_f)}\circ \tilde{\mathcal{E}}_{\xi_1^2}\right)\left(\frac{1}{2(1-u_i)}\right)= \left(\frac{\xi_1^2}{2(1-u_i)} + t_f\right) \xi_2^2,
            \\
            \frac{1}{2k_f} &=\left(\tilde{\mathcal{E}}_{\xi_2}\circ \mathcal{E}_0^{(t_f)}\circ \tilde{\mathcal{E}}_{\xi_1}\right)\left(\frac{1}{2}\right)= \left(\frac{\xi_1}{2} + t_f\right) \xi_2.
        \end{align}
    \end{subequations}
    \item \textbf{OPN and ONP}: Variables $(t_f, \xi_1, \xi_2)$.
    \begin{subequations}
        \begin{align}
            \frac{1}{2(k_f + u_f)} &= \left(\tilde{\mathcal{E}}_{1}\circ \tilde{\mathcal{E}}_{\xi_1^2}\circ \mathcal{E}_0^{(t_f)}\right)\left(\frac{1}{2(1+u_i)}\right)= \left(\frac{1}{2(1+u_i)} + t_f\right)\xi_1^2,
            \\
            \frac{1}{2(k_f - u_f)} &= \left(\tilde{\mathcal{E}}_{\xi_2^2}\circ \tilde{\mathcal{E}}_1\circ \mathcal{E}_0^{(t_f)}\right)\left(\frac{1}{2(1-u_i)}\right) = \left(\frac{1}{2(1-u_i)} + t_f\right)\xi_2^2,
            \\
            \frac{1}{2k_f} &= \left(\tilde{\mathcal{E}}_{\xi_2}\circ \tilde{\mathcal{E}}_{\xi_1}\circ \mathcal{E}_0^{(t_f)}\right)\left(\frac{1}{2}\right)= \left(\frac{1}{2} + t_f\right)\xi_1\xi_2.
        \end{align}
    \end{subequations}
    \item \textbf{PON}: Variables $(t_f, \xi_1, \xi_2)$.
    \begin{subequations}
        \begin{align}
            \frac{1}{2(k_f+u_f)} &= \left(\tilde{\mathcal{E}}_{1}\circ \mathcal{E}_0^{(t_f)}\circ \tilde{\mathcal{E}}_{\xi_1^2}\right)\left(\frac{1}{2(1+u_i)}\right)= \frac{\xi_1^2}{2(1+u_i)} + t_f,
            \\
            \frac{1}{2(k_f-u_f)} &= \left(\tilde{\mathcal{E}}_{\xi_2^2}\circ\mathcal{E}_0^{(t_f)}\circ \tilde{\mathcal{E}}_1\right)\left(\frac{1}{2(1-u_i)}\right)= \left(\frac{1}{2(1-u_i)} + t_f\right)\xi_2^2,
            \\
            \frac{1}{2k_f} &=\left(\tilde{\mathcal{E}}_{\xi_2}\circ \mathcal{E}_0^{(t_f)}\circ \tilde{\mathcal{E}}_{\xi_1}\right)\left(\frac{1}{2}\right)= \left(\frac{\xi_1}{2} + t_f\right)\xi_2.
        \end{align}
    \end{subequations}
    \item \textbf{NOP}: Variables $(t_f, \xi_1, \xi_2)$.
    \begin{subequations}
        \begin{align}
            \frac{1}{2(k_f+u_f)} &= \left(\tilde{\mathcal{E}}_{\xi_2^2}\circ\mathcal{E}_0^{(t_f)}\circ \tilde{\mathcal{E}}_1\right)\left(\frac{1}{2(1+u_i)}\right)= \left(\frac{1}{2(1+u_i)} + t_f\right)\xi_2^2,
            \\
            \frac{1}{2(k_f-u_f)} &= \left(\tilde{\mathcal{E}}_{1}\circ \mathcal{E}_0^{(t_f)}\circ \tilde{\mathcal{E}}_{\xi_1^2}\right)\left(\frac{1}{2(1-u_i)}\right)= \frac{\xi_1^2}{2(1-u_i)} + t_f,
            \\
            \frac{1}{2k_f} &=\left(\tilde{\mathcal{E}}_{\xi_2}\circ \mathcal{E}_0^{(t_f)}\circ \tilde{\mathcal{E}}_{\xi_1}\right)\left(\frac{1}{2}\right)= \left(\frac{\xi_1}{2} + t_f\right)\xi_2.
        \end{align}
    \end{subequations}
    \item \textbf{PNO and NPO}: Variables $(t_f, \xi_1, \xi_2)$.
    \begin{subequations}
        \begin{align}
            \frac{1}{2(k_f+u_f)} &= \left(\mathcal{E}_0^{(t_f)}\circ\tilde{\mathcal{E}}_1\circ \tilde{\mathcal{E}}_{\xi_1^2}\right)\left(\frac{1}{2(1+u_i)}\right) = \frac{\xi_1^2}{2(1+u_i)} + t_f, 
            \\
            \frac{1}{2(k_f-u_f)} &= \left(\mathcal{E}_0^{(t_f)}\circ \tilde{\mathcal{E}}_{\xi_2^2}\circ\tilde{\mathcal{E}}_1\right)\left(\frac{1}{2(1-u_i)}\right) = \frac{\xi_2^2}{2(1-u_i)} + t_f,
            \\
            \frac{1}{2k_f} &= \left(\mathcal{E}_0^{(t_f)}\circ\tilde{\mathcal{E}}_{\xi_2}\circ \tilde{\mathcal{E}}_{\xi_1}\right)\left(\frac{1}{2}\right) = \frac{\xi_1\xi_2}{2} + t_f.
        \end{align}
    \end{subequations}
    \item \textbf{PNP and NPN}: Variables $(\xi_1, \xi_2, \xi_3)$, and thus a null connection time.
    \begin{subequations}
        \begin{align}
            \frac{1}{2(k_f+u_f)} &= \left(\tilde{\mathcal{E}}_{\xi_3^2} \circ\tilde{\mathcal{E}}_1\circ \tilde{\mathcal{E}}_{\xi_1^2}\right)\left(\frac{1}{2(1+u_i)}\right) = \frac{\xi_1^2\xi_3^2}{2(1+u_i)}, 
            \\
            \frac{1}{2(k_f-u_f)} &= \left(\tilde{\mathcal{E}}_1\circ \tilde{\mathcal{E}}_{\xi_2^2}\circ\tilde{\mathcal{E}}_1\right)\left(\frac{1}{2(1-u_i)}\right) = \frac{\xi_2^2}{2(1-u_i)},
            \\
            \frac{1}{2k_f} &= \left(\tilde{\mathcal{E}}_{\xi_3} \circ\tilde{\mathcal{E}}_{\xi_2}\circ \tilde{\mathcal{E}}_{\xi_1}\right)\left(\frac{1}{2}\right) = \frac{\xi_1\xi_2\xi_3}{2},
        \end{align}
    \end{subequations}
\end{itemize}
As clearly observed in the above list, the symmetry between P and N entails that many of the aforementioned three-bang protocols have the same algebraic systems; specifically the pairs (OPN, ONP), (PNO, NPO), and (PNP, NPN). Thus, they allow to reach the same final states with the same connection time, so in practice we only need to deal with nine different three-bang protocols. Although the majority of these protocols constitute regular bang-bang protocols---i.e. their switching times are completely determined via the switching functions from Eq.~\eqref{ch6_eq:switching-functions}, the protocols POP and NON constitute pure singular bang-bang protocols, as proven in Appendix~\ref{app:singular-bang-bang}. 

As was the case for one- and two-bang protocols, the evolution equations provide three algebraic equations from the three boundary conditions at the final time for the moments $z_{j,f}$ in the target NESS. Since the three-bang protocols introduce three unknown parameters, this algebraic system provides a solution for them and thus univocally characterises the three-bang protocols. However, the fact that the ``quenching'' parameters for the P and N vertices must verify $0\le \xi \le 1$ and that the lengths $\tau$ of the time windows for the O vertex must verify $\tau \ge 0$ impose restrictions that make each V$_1$V$_2$V$_3$ protocol reach a certain region of the $(k_f,u_f)$ plane---given an initial condition $u_i$. 

\begin{figure}
\centering\includegraphics[width=4.5in]{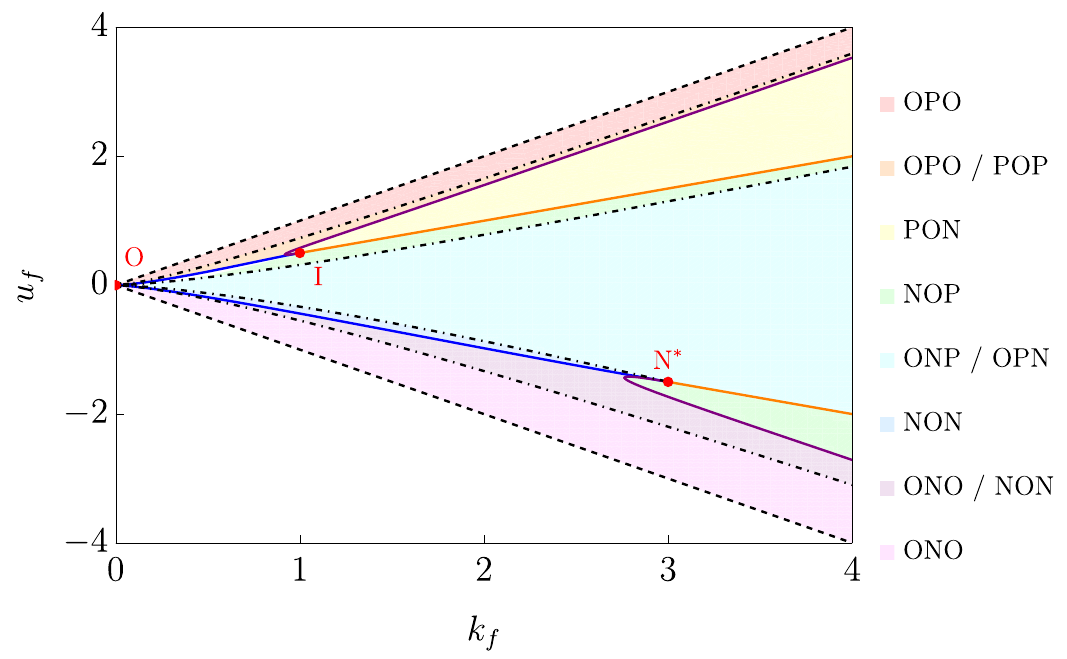}
  \caption{{
  Reachable regions in the $(k_f, u_f)$-plane by means of  optimal three-bang protocols, from an initial NESS with $u_i = 0.5$. The regions are labelled in the legend, ordered from top to bottom. The red points and the solid curves are the same ones as in the left panel of Fig.~\ref{ch6_fig:two-bang-protocols}. Specifically, the solid curves correspond to the lines reachable by two-bang protocols. The additional dotted curves also mark the boundary of certain optimal regions reachable by three-bang protocols, over these curves the three-bang protocols cease to exist, but they do not reduce to two-bang protocols.}
  }
  \label{ch6_fig:three-bangs-uipos}
\end{figure}

\begin{figure}
\centering\includegraphics[width=4.5in]{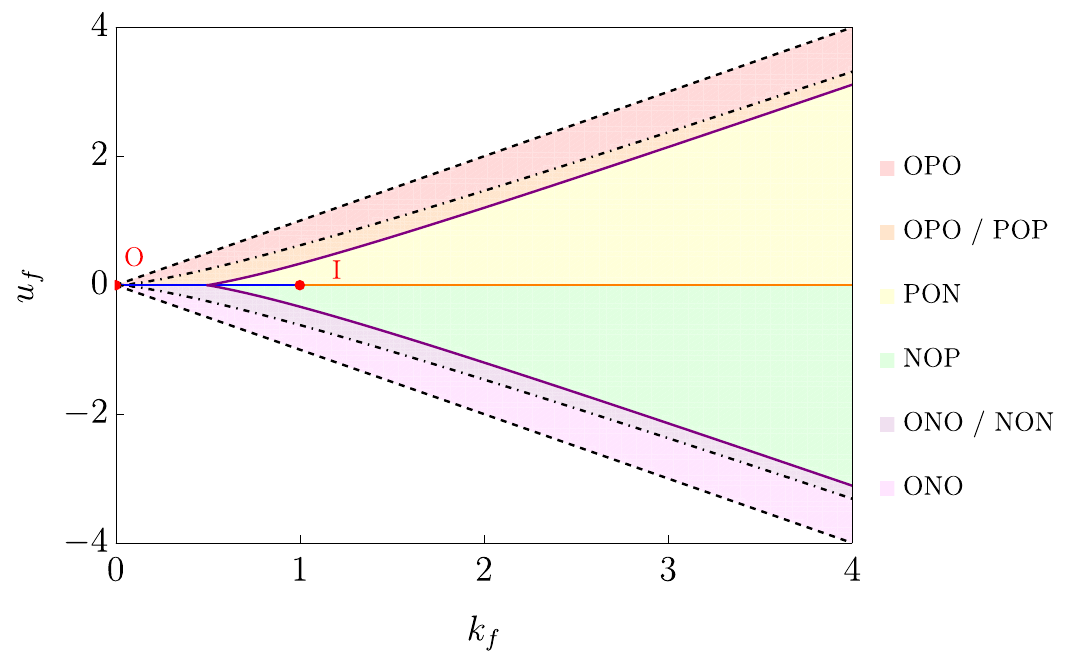}
  \caption{Same as in Fig.~\ref{ch6_fig:three-bangs-uipos}, but for a different initial state, now corresponding to $u_i=0$. { Again, note the symmetry with respect to the $u_f=0$ axis, upon exchanging N$\leftrightarrow$P.}
  }
  \label{ch6_fig:three-bangs-ui0}
\end{figure}

Figures~\ref{ch6_fig:three-bangs-uipos} and \ref{ch6_fig:three-bangs-ui0} corroborate the insights above for the same values of the initial condition considered in Fig.~\ref{ch6_fig:two-bang-protocols}: $u_i = 0.5$ for Fig.~\ref{ch6_fig:three-bangs-uipos}, and $u_i = 0$ for Fig.~\ref{ch6_fig:three-bangs-ui0}. In both figures, we plot the $(k_f, u_f)$ regions attained via three-bang protocols where the connection time is minimum. {This is an important clarification, since there are several regions corresponding to different three-bang protocols that intersect, as it is the case 
for the NOP and ONP/OPN protocols in Fig.~\ref{ch6_fig:three-bangs-uipos}, where the NOP region (green) also belongs to the ONP/OPN one (cyan), but NOP becomes the optimal protocol in that common region because its connection time is shorter.} That is the main reason why the PNO/NPO protocols neither appear in Fig.~\ref{ch6_fig:three-bangs-uipos} nor in Fig.~\ref{ch6_fig:three-bangs-ui0}: they are always suboptimal, i.e. give longer connection times, with respect to other intersecting bang-bang protocols.

The curves corresponding to two-bang protocols from Fig.~\ref{ch6_fig:two-bang-protocols} play for three-bang protocols a role similar to that played by the points corresponding to one-bang protocols for two-bang ones: the curves from  Fig.~\ref{ch6_fig:two-bang-protocols} delimit the boundaries of certain regions reachable by the different three-bang protocols. This is again logical: two-bang protocols are a limiting case for three-bang protocols.\footnote{In terms of the algebraic systems of equations displayed above for the three-bang protocols, which are characterised by three variables, it is equivalent to the case in which at least one of such variables attains its lower or upper bound---if such variable corresponds to the length of a time window $\tau$, its lower bound would be $\tau = 0$, while if it corresponds to a quenching parameter $\xi$, its lower and upper bounds correspond to $\xi = 0$ and $\xi = 1$, respectively.} However, there are additional dotted curves that also delimit regions reached by three-bang protocols and do not correspond to two-bang protocols. At these dotted curves, the algebraic systems for the three-bang protocols cease to have physical solutions because one (or more) of the unknown parameters becomes complex.

It is worth emphasising that three-bang protocols already make it possible to reach any point in the $(k_f, u_f)$-plane---thus any target NESS of the Brownian gyrator---within the boundaries of the control set from Eq.~\eqref{ch6_eq:control-set}. In principle, we could keep constructing higher-order-bang protocols, but on a physical basis one expects the three-bang protocols to give the minimum connection time, since they comprise the minimum number of time windows. In this sense, it must be stressed that we have not been able to find shorter connections with a higher number of bangs.\footnote{Still, from a strict mathematical viewpoint, we cannot assure that higher-order-bang protocols are suboptimal against our three-bang solutions.}

We devote the forthcoming section to the detailed study of the connection time for the optimal three-bang protocols presented in this section. Interestingly, these ``simple'' protocols already beat the characteristic relaxation timescale of the dynamics for certain sets of initial and final states.

\section{\label{ch6_sec:minimum-time}Behaviour of the minimum connection time}

In the previous section we have shown that, given a fixed initial state $\text{I}=(1,u_i)$ in our dimensionless variables, we are able to reach any final NESS of the Brownian gyrator, characterised by one point in the $(k_f,u_f)$-plane within the control set, by means of a three-bang protocol. Now, we turn our attention into the original matter of concern, which is the minimum connection time between two NESSs. Therefore, we study here the behaviour of the connection time over the aforementioned three-bang protocols, candidates to be the brachistochrone, as a function of the target state.

\begin{figure} 
\centering \includegraphics[width=4.5in]{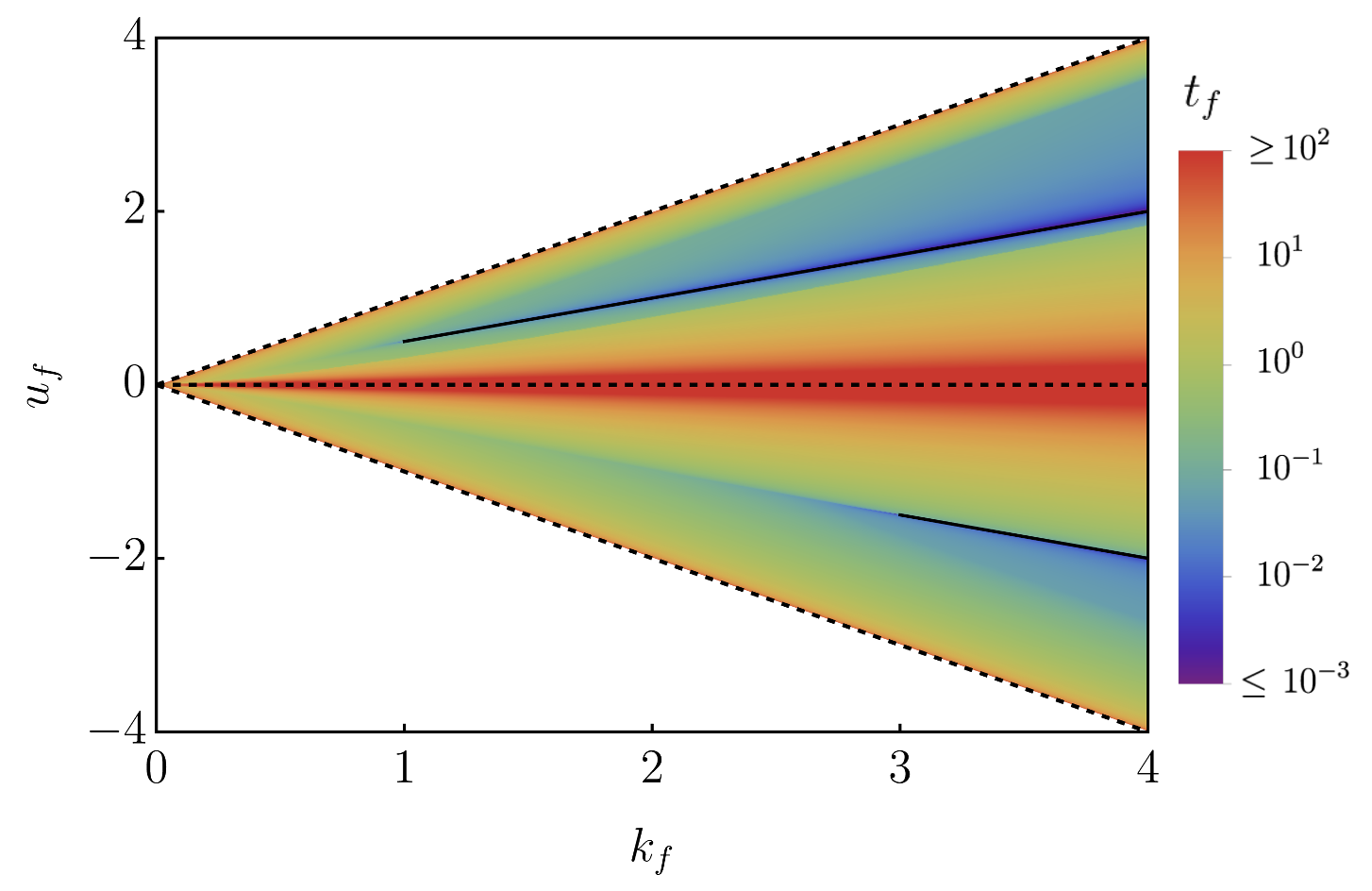}
\\
\centering \includegraphics[width=4.5in]{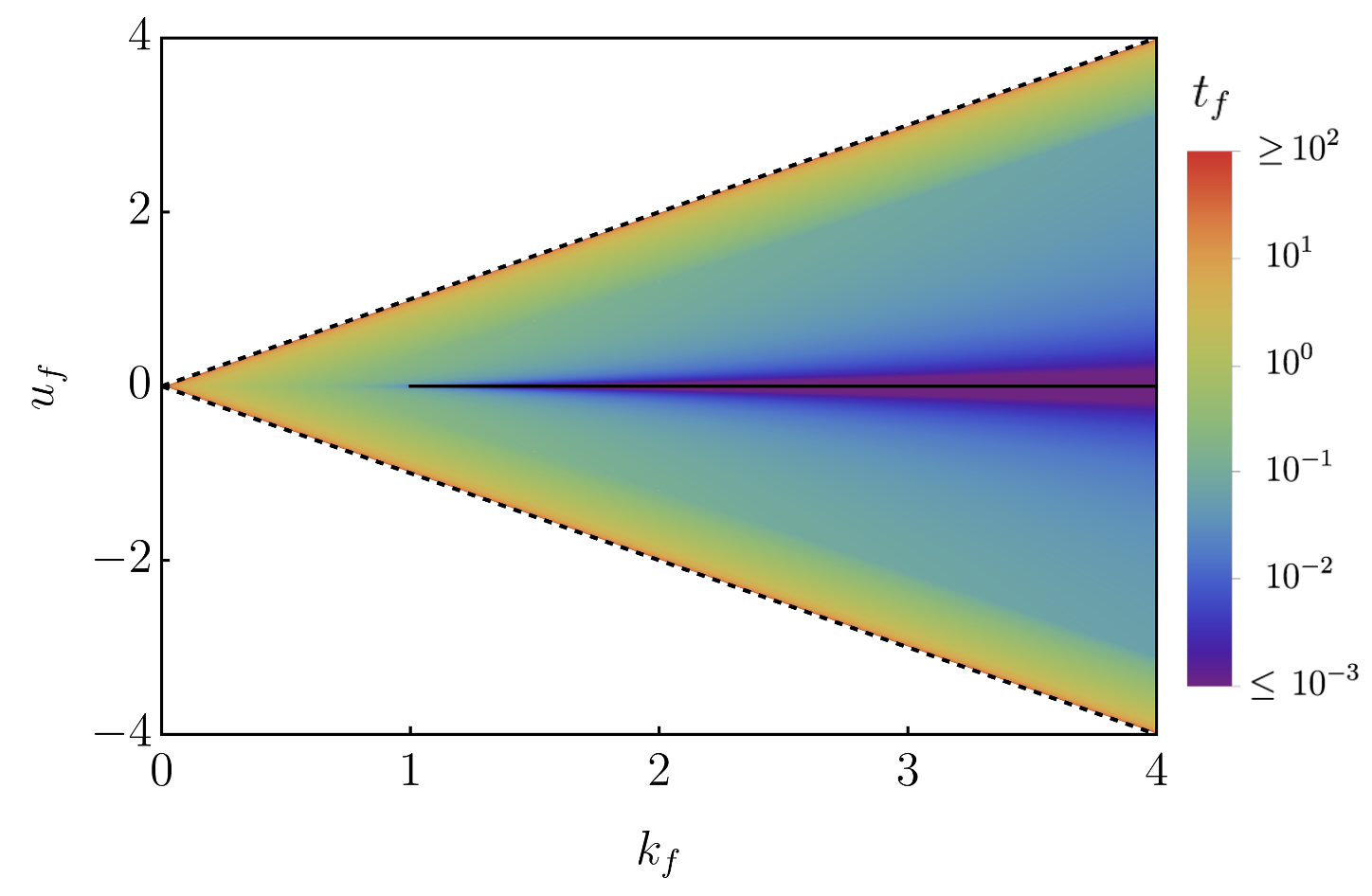}
\caption{Density maps of the minimum connection time $t_f$ on the phase-plane $(k_f,u_f)$. Note the logarithmic scale for $t_f$. Similarly to Fig.~\ref{ch6_fig:two-bang-protocols}, we have considered different initial states in the top and bottom panels: $u_i = 0.5$ (top) and $u_i = 0$ (bottom); recall that $k_i=1$ in our dimensionless units. For the top panel, the corresponding optimal three-bang protocols providing the plotted minimum connection time $t_f$ are shown in Fig.~\ref{ch6_fig:three-bangs-uipos}; for the bottom panel, in Fig.~\ref{ch6_fig:three-bangs-ui0}. Dashed lines represent the sets of points at which the connection time diverges, whereas the black solid ones account for the points with null connection time in the $k_{\max}\to +\infty$ limit. Note that the timescale has been chosen to be equal in both panels.
}
    \label{ch6_fig:heat-maps-nonull-ui}
\end{figure}

Figure~\ref{ch6_fig:heat-maps-nonull-ui} shows density maps for the connection time $t_f$ over each of the optimal three-bang protocols built in Sec.~\ref{ch6_subsec:three-bang}. The behaviour of the minimum connection time $t_f$ radically changes from the $u_i \neq 0$ case to the $u_i = 0$ case. The most remarkable change refers to the final states lying on the line $u_f = 0$. On the one hand, for $u_i = 0$, points lying on the $u_f = 0$ line are reached in (i) a finite time for $k_f < 1$ and (ii) in a null connection time for $k_f > 1$.\footnote{If $k_{\max}$ were finite, the corresponding minimum connection time would be of the order of $1/k_{\max}$, similarly to what occurred in the previous chapter for $d=1$: for large, but finite $T_{\max}$, the connection time is of the order of $1/T_{\max}$.} {The case $u_i=u_f=0$ is equivalent to connecting equilibrium states of two uncoupled oscillators with different bath temperatures $T_x\ne T_y$, for which the minimum connection time has been explicitly derived in Sec.~\ref{ch6_subsec:insight-1d}.} On the other hand, for $u_i \neq 0$, states lying in the line $u_f = 0$ can only be reached in an infinite time. It is straightforward to show from the systems of equations displayed in Sec.~\ref{ch6_subsec:three-bang} that 
\begin{equation}\label{ch6_eq:diverging-1}
    t_f \propto u_f^{-2}, \qquad |u_f| \ll 1, \quad u_i \neq 0,
\end{equation}
which correspond to the relevant OPN/ONP three-bang protocols for $|u_f| \ll 1$.

The above has striking implications. From a physical point of view, if we initially start from a state in which the two coordinates of the Brownian gyrator are uncoupled, $u_i = 0$, which is equivalent to having two independent oscillators at equilibrium with their respective baths at temperatures $T_x$ and $T_y$, any NESS of the Brownian gyrator, with $u_f\ne 0$ can be reached in a finite time. However, had we started from an initial NESS with $u_i\ne 0$, it would take an infinite amount of time to reach any uncoupled final state with $u_i=0$. Hence, the candidates OPN/ONP to optimal-time SST protocol  would present no advantage with respect to the direct STEP protocol in this case. Recalling Eq.~\eqref{ch6_eq:characteristic-time-trel}, the relaxation time for the STEP process for $u_f=0$ results to be $t_{\rel}(k_f,0) = (2k_f)^{-1}$, which is finite. 

Apart from the $u_f = 0$ line, the connection times also diverge along the $u_f = \pm k_f$ lines, which bound the control set of admissible controls, both for $u_i=0$ and $u_i\ne 0$. Similarly to the previous analysis carried out for the $u_f = 0$ line, asymptotic expressions for the minimum connection time along the lines $u_f = \pm k_f$ can be derived from the algebraic systems that we explicitly write in Sec.~\ref{ch6_subsec:three-bang}. Specifically, along the $u_f = k_f$ line, the OPO protocol gives
\begin{equation}\label{ch6_eq:divergence-2}
    t_f(k_f,u_f) = \frac{1}{2(k_f - u_f)} - \frac{1}{2(1-u_i)} \sim \frac{1}{2(k_f - u_f)}, \qquad k_f-u_f \ll 1-u_i
\end{equation}
while for the $u_f = -k_f$ line, the ONO protocol gives
\begin{equation}\label{ch6_eq:divergence-3}
    t_f(k_f,u_f) = \frac{1}{2(k_f + u_f)} - \frac{1}{2(1+u_i)} \sim \frac{1}{2(k_f + u_f)}, \quad k_f+u_f \ll 1+u_i.
\end{equation}
Let us note that the minimum times in Eqs.~\eqref{ch6_eq:divergence-2} and \eqref{ch6_eq:divergence-3} have the same divergence of the characteristic relaxation time for the STEP process in Eq.~\eqref{ch6_eq:characteristic-time-trel}, which entails this property to be a consequence of the physical bounds of our model: the fact that $k_f^2 - u_f^2 > 0$ for the system to attain a NESS.\footnote{These diverging behaviours for the connection time resemble those found in the previous chapter, in the limits $T_f \to T_{\min}$ and $T_f \to T_{\max}$.} Moreover, the connection times for the OPO and ONO protocols are the same as those for the singular protocols over the $\overline{\text{OP}}$ and $\overline{\text{ON}}$ lines---Eqs.~\eqref{ch6_eq:edge-op-time} and \eqref{ch6_eq:edge-on-time}, respectively.

\begin{figure}[h!]
{\centering \includegraphics[width=2.6in]{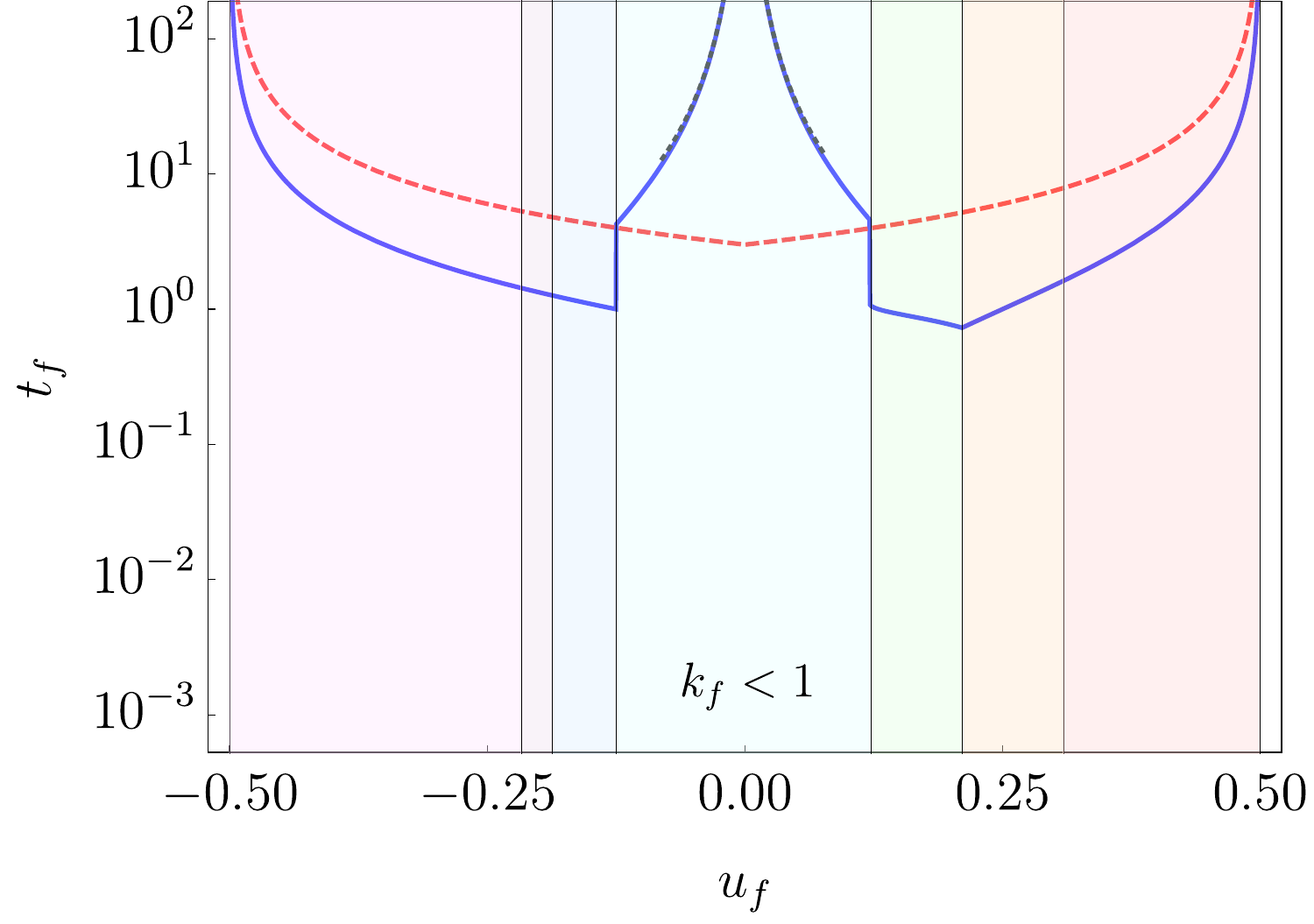} \includegraphics[width=2.6in]{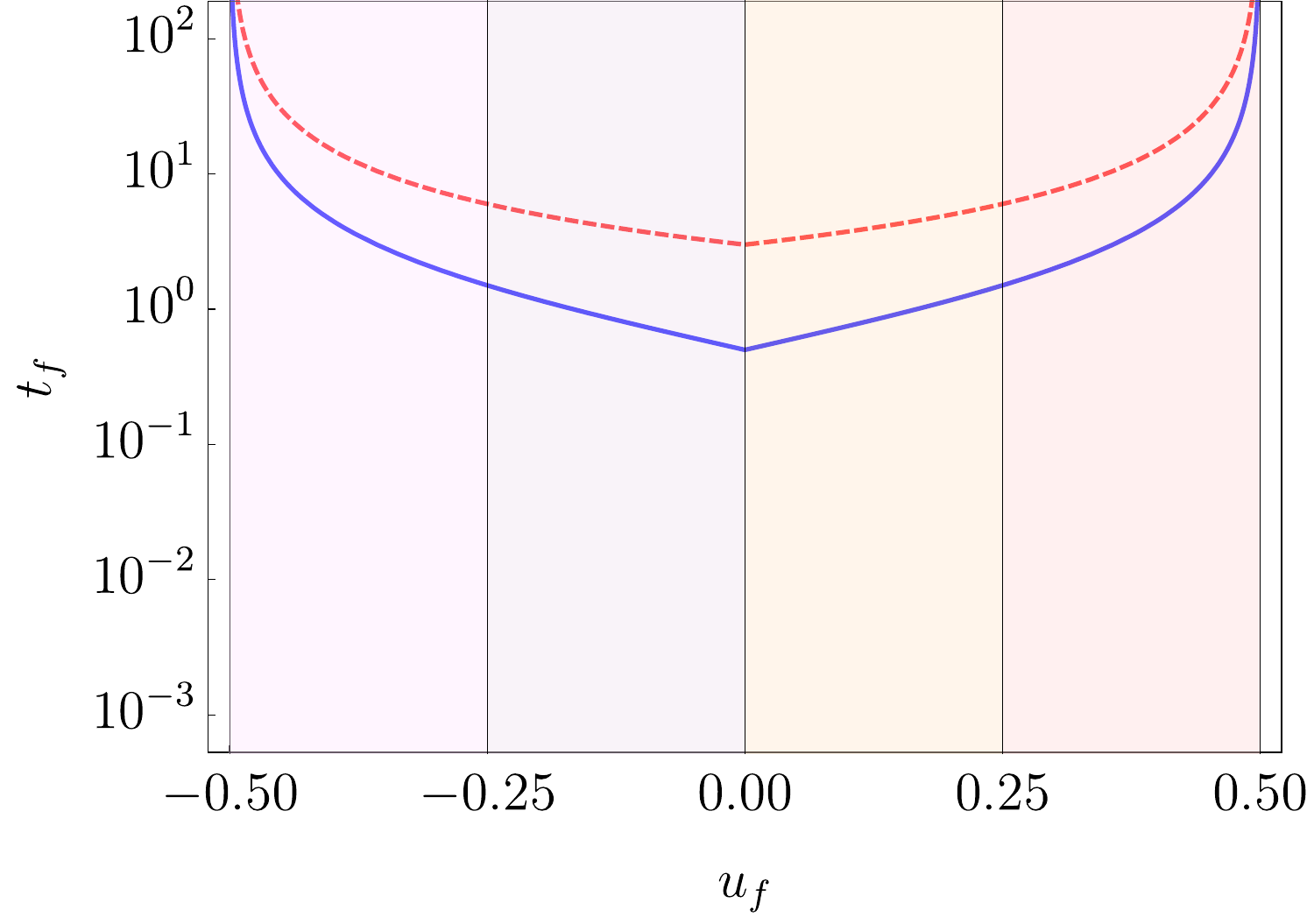}
\\
\includegraphics[width=2.6in]{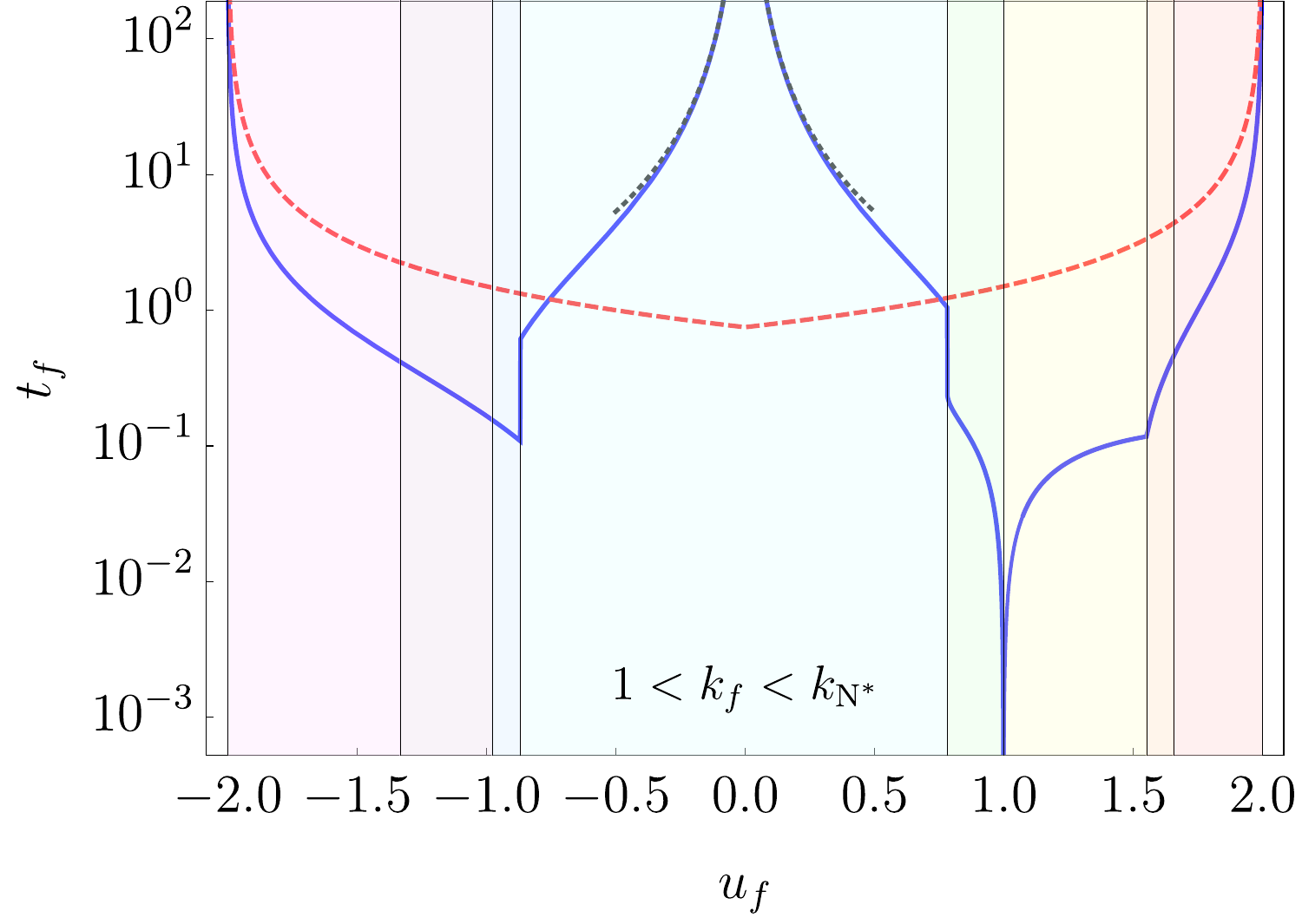}
\includegraphics[width=2.6in]{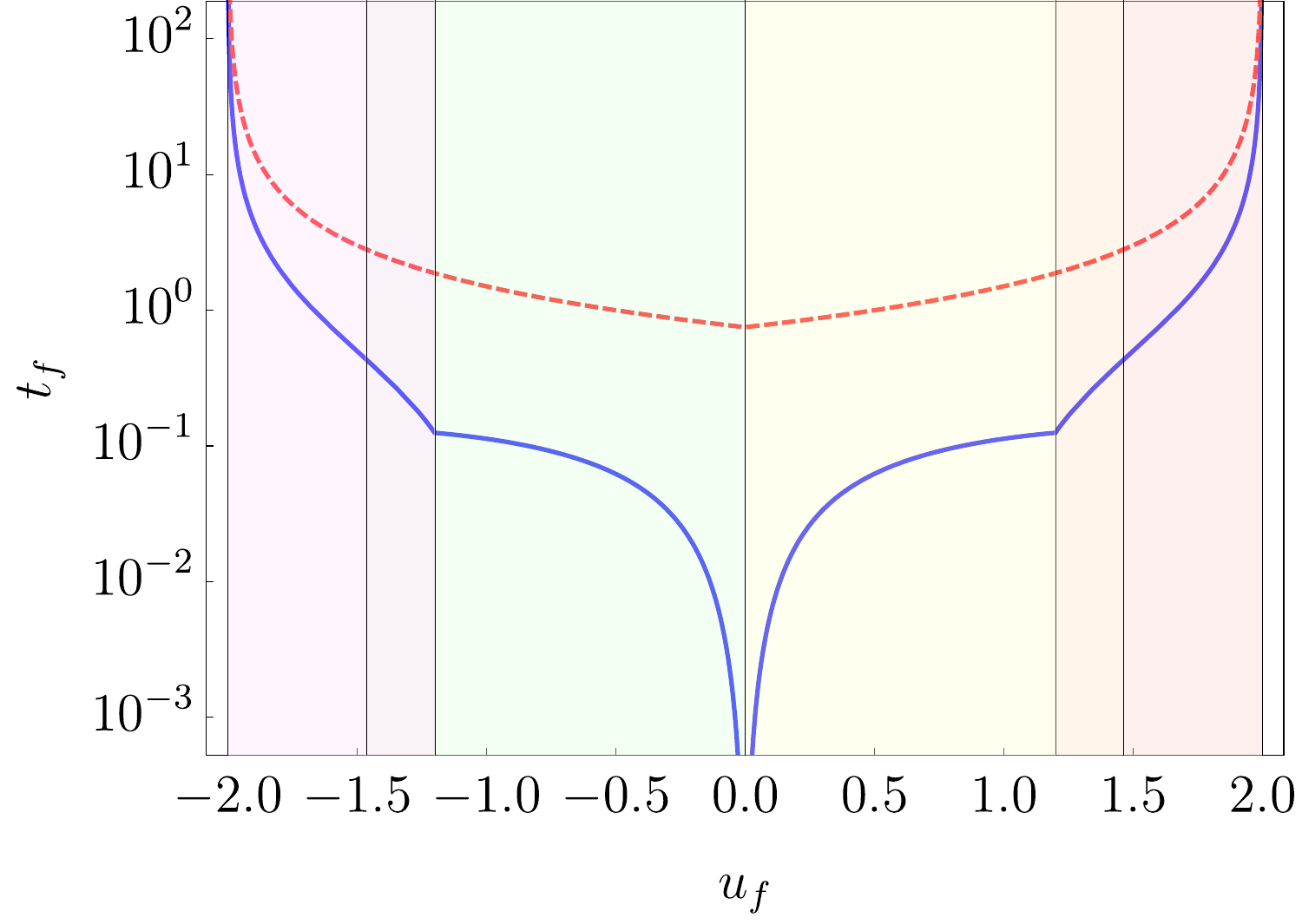}
\\
\includegraphics[width=2.6in]{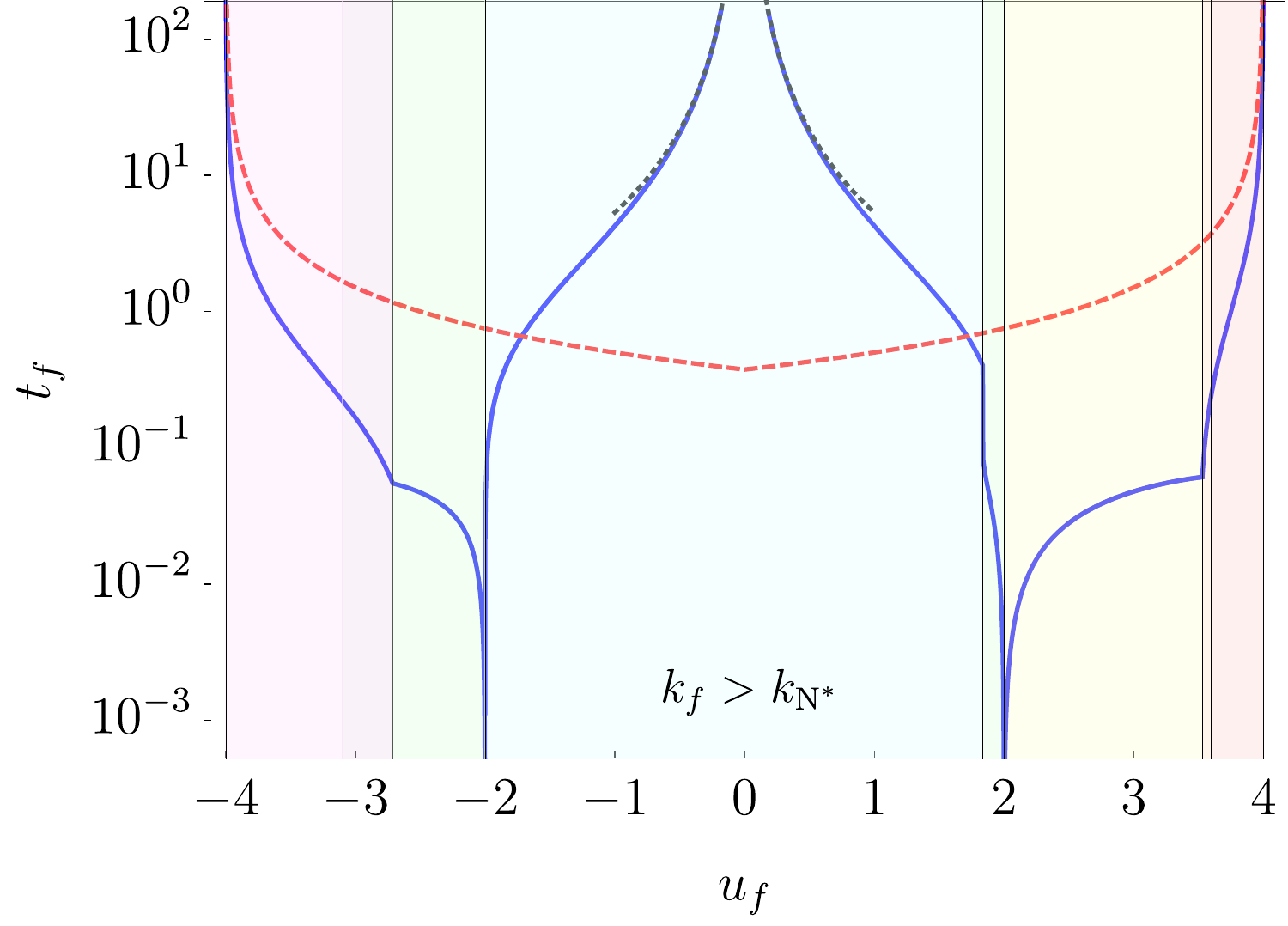} \includegraphics[width=2.6in]{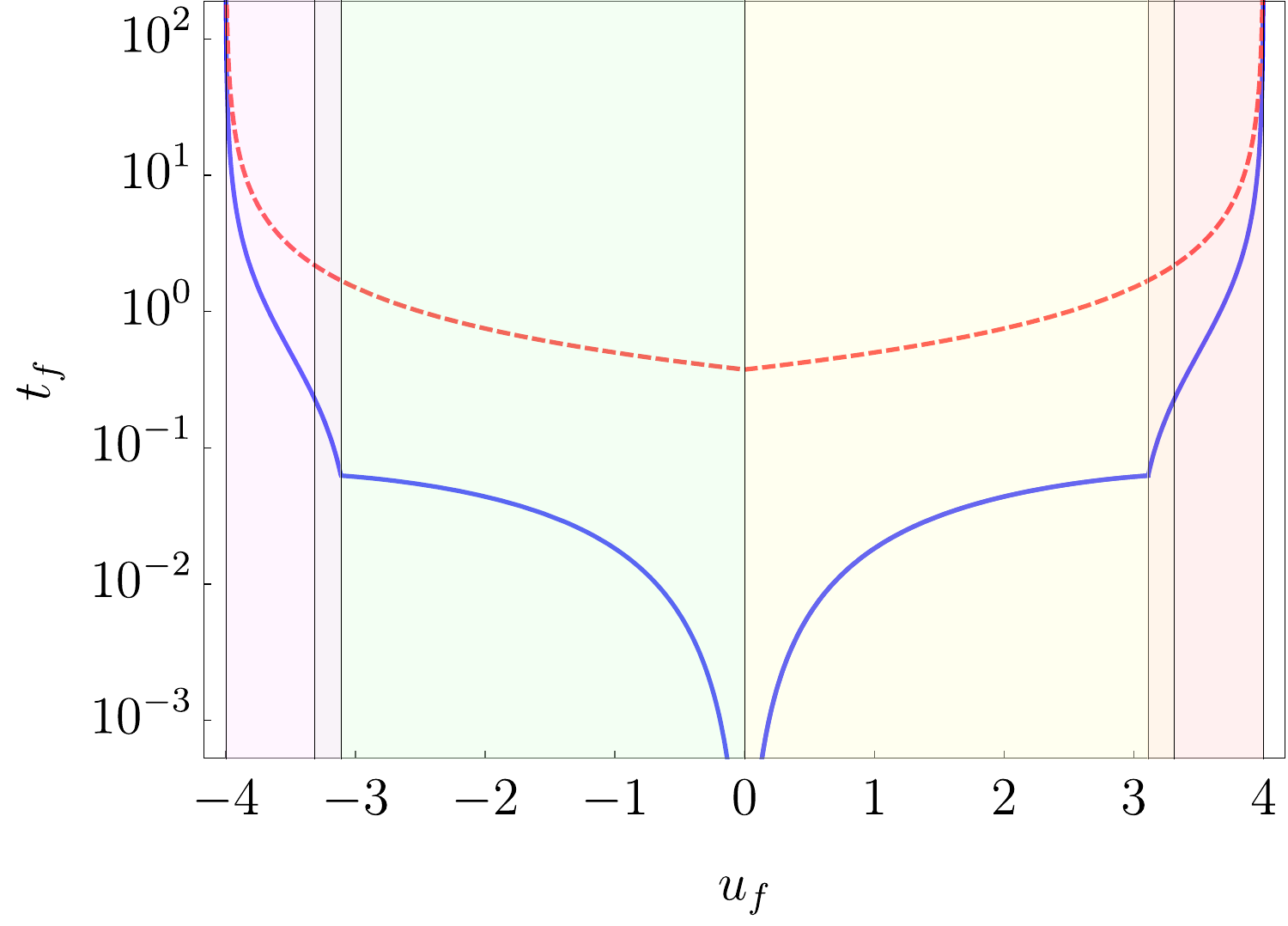}}
\caption{Minimum connection time $t_f$ as a function of $u_f$, for different fixed values of $k_f$. In other words, we are representing vertical slices of the two panels of Fig.~\ref{ch6_fig:heat-maps-nonull-ui}---$u_i=0.5$ with $k_{\text{N}^*}=3$ (left) and $u_i=0$ (right)---at several values of $k_f$. From top to bottom, we show the results for $k_f=0.5$, $2$, and $4$. The blue solid line corresponds to $t_f$, whereas the red dashed curve corresponds to three times the characteristic relaxation time, $3 t_{\rel}$, for the STEP protocol defined in Eq.~\eqref{ch6_eq:characteristic-time-trel}. The {dotted} curves on the left panels correspond to the asymptotic behaviour from Eq.~\eqref{ch6_eq:diverging-1} for $|u_f|\ll 1$. The coloured regions on the background follow the colour code from Figs.~\ref{ch6_fig:three-bangs-uipos} (left) and \ref{ch6_fig:three-bangs-ui0} (right). Note that the timescale has been chosen to be  equal in all the panels.
}
    \label{ch6_fig:time-slices}
\end{figure}
In order to improve our understanding on the global landscape of minimum connection times, several sections of Figs.~\ref{ch6_fig:heat-maps-nonull-ui} for  different fixed values of $k_f$ are plotted in Fig.~\ref{ch6_fig:time-slices}. Again, this is done for the two choices of initial NESSs we have repeatedly employed: $u_i = 0.5$ and $u_i = 0$. Recall that, due to the symmetry between the P and N vertices, the results corresponding to the $u_i = 0.5$ would be equivalent to those for $u_i = -0.5$, upon reflection through the $u_f = 0$ axis. In Fig.~\ref{ch6_fig:time-slices}, the solid lines correspond to the minimum connection times already shown in Figs.~\ref{ch6_fig:heat-maps-nonull-ui}, for the considered value of $k_f$. The background colour follows the same colour code as in Figs.~\ref{ch6_fig:three-bangs-uipos} and \ref{ch6_fig:three-bangs-ui0}, thus showing which three-bang protocol the minimal time connection belongs to. The red dashed lines correspond to {three times the characteristic relaxation time $t_{\rel}$} for the STEP process, thus allowing us to quantify the efficiency of the minimum-time protocols in beating the STEP one.

As a general trend, the minimum connection time decreases with $k_f$. This is reasonable from a physical point of view: high values of $k_f$ can only be reached by employing bang-bang protocols in which the P and N time windows dominate the dynamics over the O one. Since the former involve vanishing---i.e. of the order of $1/k_{\max}$--- time windows, the resulting minimum connection time is small. 

The choice for the three values of $k_f$ displayed in Fig.~\ref{ch6_fig:time-slices} has not been arbitrary. For $u_i \neq 0$ (left panels), three qualitatively different behaviors are observed for the connection time. For $k_f < k_i=1$ (top leftp), minimum connection times are non-zero and finite---with the exception of the asymptotic divergences when approaching $u_f = 0$ and $u_f = \pm k_f$. For $1<k_f<k_{\text{N}^*}$ (middle left), where N$^*$ is the only point reached by a one-bang N protocol, the upper PN/NP line in the left panel of Fig.~\ref{ch6_fig:two-bang-protocols} comes into play with a zero connection time. Lastly, for $k_f>k_{\text{N}^*}$ (bottom left), the lower branch of the PN/NP line comes into play, which also involves a vanishing $t_f$, providing a second instantaneous brachistochrone. Recall that the PN/NP line of Fig.~\ref{ch6_fig:two-bang-protocols} describes the set of final NESS attainable with a PN/NP two-bang protocol,

We may understand the difference between the phenomenology above for the cases $k_f<k_i=1$ and $k_f>k_i=1$ on a physical basis. Let us start by considering $k_f < 1$: this requires a decompression to attain a NESS with a target value of the stiffness that is smaller than the initial one. Then, regardless of the specific target value of $k_f$, the corresponding bang-bang optimal protocol must include a time window in the O vertex, and therefore the connection time is non-vanishing. Next, let us move onto the case $1\leq k_f < k_{\text{N}^*}$: there appear states with $u_f>0$ reachable by instantaneous---in the limit $k_{\max}\to\infty$---two-bang PN/NP protocols, with the P window being the dominating one{, because we are considering $u_i>0$ and $u_{\text{P}}=k_{\max}>0$. These are the states over the positive branch, with $u_f>0$, of the PN/NP line. Finally, for $k>k_{\text{N}^*}$, there appear states with $u_f<0$ reachable by instantaneous PN/NP protocols, with the N window being dominating now. This is reasonable: N$^*$ is the only point that can be reached from the initial one with a one-bang N protocol, so states close to it over the negative branch, with $u_f<0$, of the PN/NP line are expected to be dominated by the N time window.}

The central OPN/ONP region, which includes $u_f=0$, in the left panels of Fig.~\ref{ch6_fig:time-slices} deserves further attention, as it presents interesting features. First, the minimum connection time $t_f$ diverges for $u_f\to 0$, which implies that, from a practical perspective, optimal bang-bang protocols become inefficient with respect to the direct STEP protocol, because the $3 t_{\rel}$ curve lies below $t_f$ over most the  region OPN/ONP. Second,  the minimum connection time $t_f$ presents discontinuities at the boundaries of the region OPN/NOP: on the left, for $u_f<0$, when the region NON emerges; on the right, for $u_f>0$, when the region NOP emerges. {Such discontinuities stem from the emergence of the NON and NOP regions, since the OPN/ONP protocols continue to exist in such regions but give longer connection times than the NON and NOP protocols. Therefore, the latter, which appear abruptly, become the optimal ones.}

The right panels of Fig.~\ref{ch6_fig:time-slices}, for which $u_i = 0$, are simpler to understand. There are fewer three-bang regions stemming from the scheme in Fig.~\ref{ch6_fig:three-bangs-ui0}, since the existing OPN/ONP regions for $u_i\ne 0$ disappear in the case $u_i=0$: as we have already commented, this vanishing stems from the degeneration of the PN/NP lines into one unique line $u_f=0$, $k_f\ge 1$, as seen in Fig.~\ref{ch6_fig:two-bang-protocols}. Therefore, there are only two cases of interest: (i) $k_f < 1$, for which the minimum connection time is always non-zero, due to the presence of time window corresponding to the vertex O in the optimal protocols, and (ii) $k_f > 1$, for which the minimum connection time vanishes along the $u_f = 0$ line---i.e. the PN/NP line, with a vanishing time in the limit $k_{\max}\to\infty$. In this case, the minimum connection time from optimal control systematically beats the characteristic relaxation time of the STEP protocol.\footnote{The combination of PN/NP protocols in this case must be equivalent to the one-bang protocol with $(k,u) = (k_{\max},0)$, as deduced in Sec.~\ref{ch6_subsec:insight-1d}.}

For the bang-bang protocols analysed in this chapter, it can be shown that the average irreversible work $\expval{W_{\irr}}$ linearly increases with $k_{\max}$ in the limit $k_{\max}\to\infty$ that we have considered to simplify our discussion. This entails that the bound given by Eq.~\eqref{ch6_eq:speed-limit} becomes very loose, since it vanishes as $k_{\max}^{-1}$.  We provide a detailed derivation of the latter in Appendix~\ref{ch6_subsec_energy-balance}. This behaviour is similar to the one found in other systems for thermal control, when the upper bound of bath temperature goes to infinity and the corresponding bang become instantaneous~\cite{prados_optimizing_2021,patron_thermal_2022}.

\bookmarksetup{startatroot}

\chapter{Conclusions}
\chaptermark{Conclusions}
\label{ch:conclusions}
\sectionmark{Conclusions}
This thesis is devoted to the study of physical systems embedded within the field of non-equilibrium statistical mechanics. Specifically, the state of the systems of interest constitutes a stochastic process that can be externally driven by a set of controllable parameters. On the one hand, for systems in contact with a thermal bath, we have studied the emergence of strong memory effects and glassy behaviour upon varying the bath temperature, and how these are related to the existence of non-equilibrium attractors governing the dynamics. On the other hand, for overdamped harmonic systems, we have studied the problem of minimising the connection time between arbitrary stationary---equilibrium or non-equilibrium---states, by suitably varying either the bath temperature or the stiffnesses of the potential. 

In this final chapter we enumerate the main conclusions of the work presented in this thesis, as well as some perspectives for future research. In the following, such conclusions are divided into two parts, corresponding to the two lines of research that this thesis encompasses. 


\subsubsection*{Part I: Non-equilibrium attractors and glassy behaviour}

\begin{enumerate}
    \item We have put forward a physical model of a molecular fluid with non-linear drag, which consists of an ensemble of $d$-dimensional Brownian particles immersed in a background fluid acting as a thermal bath for the former. The dynamics of the system is accurately accounted by the Enskog-Fokker-Planck (EFP) equation for the one-particle velocity distribution function (VDF).
    \item We have studied how the kinetic temperature of the molecular fluid relaxes towards equilibrium when suddenly changing the value of the bath temperature. Two different regimes emerge: (i) a quasi-equilibrium regime for small variations of the bath temperature, in which the kinetic temperature relaxes exponentially, and (ii) a strong non-equilibrium regime, in which the temperature relaxes algebraically when quenching the bath temperature to low enough values.
    \item The non-exponential relaxation regime is intimately related the existence of a \textit{long-lived non-equilibrium state} (LLNES) that the system tends to approach, in which the Sonine cumulants become large---in absolute value. Such state is universal, in the sense that it is independent of the non-linearity of the system, the ratio of initial and final temperatures, and the collision rate. We have employed the LLNES to study the emergence and characterisation of strong Mpemba and Kovacs effects, which display universal and scaling features, respectively.
    \item We have shown that the LLNES is not specific for the molecular fluid, but emerges for a general class of overdamped, non-linear stochastic systems. Provided that (i) the force term in the Fokker-Planck/Langevin equations diverges faster than linearly and (ii) a sufficiently deep quench to low values of the bath temperature is done, there emerges a timescale separation in which the initial conditions of the dynamics are forgotten but thermal noise is still negligible, such that the probability distribution function (PDF) attains a Dirac-delta form, which characterises the LLNES.
    \item We have carried out a thorough analysis of the LLNES for isotropic, algebraic systems. Specifically, (i) we have shown that the PDF at the LLNES presents scaling properties and fat tails and (ii) we have introduced a Lyapunov functional in order to show that the LLNES constitutes a global non-equilibrium attractor of the dynamics over an intermediate timescale, by means of an H-theorem.
    \item We have provided numerical evidence beyond the isotropic and algebraic cases, together with intuitive arguments, to show that the LLNES emerges in more complex scenarios. These include potentials with multiple minima, anisotropies, interacting degrees of freedom, and even the addition of a Boltzmann collision term---either elastic or inelastic---to the Fokker-Planck dynamics.
    \item In order to gain further insights on the glassy behaviour inherent to the LLNES, we have put forward another physical model to compare the molecular fluid with: a uniformly heated granular gas, rooted in the kinetic theory of dissipative dynamics. When turning off the stochastic thermostat, it is known that the granular gas approaches the so-called \textit{homogeneous cooling state} (HCS), characterised by stationary values of the Sonine cumulants and a non-exponential relaxation function for the temperature, given by Haff's law---in strong resemblance with the LLNES for the molecular fluid with non-linear drag.
    \item By considering a linear cooling program for the bath temperature, and resorting to the tools of boundary layer theory, we have shown that both the molecular fluid and the granular gas display a kinetic glass transition with similar properties: (i) a unique frozen state, which is independent of the cooling rate for each system, and (ii) the same scaling behaviour of the kinetic temperature with the cooling rate.
    \item Furthermore, by considering a more general family of cooling programs for the bath temperature, we have analytically shown that the frozen state may correspond to the LLNES for the molecular fluid and the HCS for the granular gas, if the cooling program fulfils: (i) that is fast enough for the system to deviate from the instantaneous NESS curve, and (ii) that it lasts for enough time for both systems to reach their respective non-equilibrium attractors.
    \item  By considering a linear reheating program for the bath temperature from the frozen state, we have shown that both systems display hysteretic behaviour, which preserves the same scaling properties as the previously applied cooling programs. Moreover, the heating curves for the kinetic temperature in both systems tend to a common normal heating curve---regardless of the initial value of the frozen temperature.
    \item Finally, we briefly comment on some possibilities for future research:
    \begin{enumerate}
        \item As stressed throughout this thesis, we have considered the emergence of the LLNES for situations in which the state vector $\bm{r}$ stood as either the position of a confined overdamped Brownian particle or the velocity of a Brownian particle immersed in an isotropic background fluid. In a generic situation, the components of $\bm{r}$ may include both positions and velocities, and thus, the framework developed in chapter~\ref{ch:nonequilibrium_attractor} may apply also to underdamped systems, although it is not trivial to ascertain the necessary conditions that both the non-linear drag and the confining potential must satisfy in order to guarantee the required timescale separation for the emergence of the LLNES.
        
        \item It is worth investigating the possible deeper connections between the LLNES and other hydrodynamic states found in different out-of-equilibrium systems. For instance, quasi-elastic one-dimensional granular systems have been shown to display Dirac-delta PDFs~\cite{benedetto_kinetic_1997,barrat_velocity_2002,baldassarri_influence_2002,garcia_de_soria_universal_2012} resembling that of the LLNES. On another note, the LLNES reported in this thesis displays some similarities with the quasi-stationary states observed in some systems with long-range interactions, such as the HMF model~\cite{latora_non-gaussian_2001,rapisarda_nonextensive_2005}.
        \item Testing the emergence of the LLNES in real experiments is an interesting prospect for future work. In particular, it seems worth exploring the relevance of the LLNES to control the time evolution of mesoscopic systems, such as biomolecules or memory devices.  In this regard, it must be stressed that the two specific examples considered in this thesis---the confined overdamped Brownian particle and the molecular fluid---describe actual physical systems. Current techniques make it possible to control the shape of the potential confining a colloidal particle immersed in a fluid~\cite{ciliberto_experiments_2017,martinez_colloidal_2017}, and the Langevin equation for the velocity with non-linear drag has been successfully employed to describe mixtures of ultracold atoms~\cite{hohmann_individual_2017}, as mentioned in chapter~\ref{ch:memory_effects}. 
        \item The universality of the frozen state, in the sense of its independence of the cooling rate in scaled variables, is an appealing feature of the {kinetic} glass transition found in this work---both for the smooth granular gas and the molecular fluid with (quadratic) non-linear drag coefficient. The possible extension of this property to other systems, for example rough granular fluids~\cite{brilliantov_translations_2007,kremer_transport_2014,torrente_large_2019,garzo_impact_2018}, molecular fluids with more complex non-linearities~\cite{klimontovich_nonlinear_1994,lindner_diffusion_2007,casado-pascual_directed_2018,goychuk_nonequilibrium_2021,patron_strong_2021}, or binary mixtures~\cite{serero_hydrodynamics_2006,khalil_homogeneous_2014,gomez_gonzalez_mpemba-like_2021,gomez_gonzalez_time-dependent_2021}, is an interesting prospect for future work.
    \end{enumerate}
\end{enumerate}

\subsubsection*{Part II: Shortcuts between stationary states}

\begin{enumerate}
    \item We have put forward a paradigmatic model system that is significant for actual experiments---since it describes an optically trapped colloidal particle. This is a $d$-dimensional harmonic oscillator, the stiffnesses of which are kept constant but the temperature of the bath in which it is immersed can be externally controlled.  Specifically, the system is initially at equilibrium at temperature $T_i$ and we want to drive it to a final equilibrium state with temperature $T_f$. It is important to stress that this is a relevant physical situation, which corresponds to isochoric (zero mechanical work) protocols. 
    \item As the above system belongs to the class of linear-Pontryagin systems depicted in Sec.~\ref{ch1_subsubsec:bang-bang}, we have been able to rigorously prove that the thermal brachistochrone is a protocol of the bang-bang type, which comprises $d$ alternating heating and cooling time windows with the maximum ($T_{\max}$) and minimum ($T_{\min}$) values available for the bath temperature. 
    \item A relevant result of our approach is the finite increment of the minimum connection time when moving from dimension $d$ to $d+1$, even when all the elastic constants are (almost) equal---what we have called the almost fully degenerate case. For a particle confined in a perfectly isotropic $d$-dimensional harmonic well---what we call the fully degenerate case, the minimum connection time equals that of the one-dimensional case. 
    To shed further light to this regard, we have resorted to information geometry concepts, such as the thermodynamic length and cost. Not only do thermodynamic length and cost share the remarkable feature of additional expenses when considering systems with increasing dimensionality, but also the geometric time bound---$t_{\text{geo}}$ from Eq.~\eqref{ch1_eq:speed-limit}---increases with dimension.
    \item Then we have considered another paradigmatic model in non-equilibrium statistical mechanics: the Brownian gyrator. It consists of a Brownian particle trapped in a two-dimensional elliptic potential, characterised by two controllable parameters $(k,u)$, and additionally in contact with two thermal baths at different fixed temperatures in the two orthogonal directions. The parameter $k$ corresponds to the stiffness of the potential in the two directions, which we assume to be equal, whereas $u$ couples the two equations of motion. The control set is defined assuming the potential is confining. Under such constraints, the Brownian gyrator has a NESS for each pair of constant values of $(k,u)$.
    \item We have argued that the protocols providing the minimum connection time between two NESSs are of bang-bang type, i.e. the controls lie on the boundaries of the control set. In other words, the optimal time protocols do not have time windows with Euler-Lagrange solutions, for which $(k,u)$ would belong in the interior of the control set. 
    \item We have focused on three-bang regular protocols, i.e. protocols with three time windows, with each time window corresponding to the controls being given by one of the vertices of the control set. These three-bang protocols (i) can be univocally determined as a function of the target state $(k_f,u_f)$, for an arbitrary given initial state $(k_i,u_i)$, and (ii) connect all possible initial states with all possible final states. Simpler one-bang and two-bang protocols only make the connection to specific target points and lines, respectively, of the control set. The connection time landscape radically changes from the $u_i \neq 0$ to the $u_i = 0$ cases, with the former being much more involved than the latter.
    \item A rich phenomenology emerges when studying the behaviour of the minimum connection time as a function of the target state $(k_f,u_f)$. The latter includes: (i) connections that imply an infinite amount of time along the boundary lines $k_f = \pm u_f$ and the axis $u_f = 0$---the latter for $u_i\neq 0$, (ii) instantaneous connections in the $k_{\max}\to\infty$ limit, which result from the combination of P and N vertices, and (iii) finite discontinuities in the minimum connection time for $u_i\neq 0$ as a consequence of one of two coexisting protocols ceasing to exist.
    \item Finally, we conclude this part with some perspectives for future work:
    \begin{enumerate}
        \item We have worked with systems in the overdamped regime, so it would be interesting to extend the ideas developed here to the underdamped regime. The Gaussian behaviour persists in the underdamped regime, which should allow for obtaining a simple dynamical system involving not only the variances of the position coordinates but also the rest of second moments, namely variances of velocity coordinates and cross terms. Still, the possible existence of oscillatory modes---depending on the relative values of the natural frequencies and the damping constant---makes the problem an interesting challenge from the point of view of control theory.
        \item The current state of the art in optical trapping allows for the experimental implementation of the optimal protocols developed here for a Brownian gyrator, following the techniques employed in its recent experimental realisations~\cite{ciliberto_heat_2013,argun_experimental_2017,chiang_electrical_2017,cerasoli_spectral_2022}. The fact that bang-bang protocols involve instantaneous switchings of the control variables could be circumvented by engineering either the potential or the bath temperature to vary over a timescale much shorter than that corresponding to the timescale of the dynamics of the corresponding Brownian object. Hence, experimental realisation of a controlled Brownian gyrator deserves further investigation.
        \item The results presented in this part are remarkably relevant for devising optimal mesoscopic heat engines. Since the emergence of stochastic thermodynamics~\cite{sekimoto_stochastic_2010,peliti_stochastic_2021}, the goal of building functional Brownian heat engines has been a persistent aspiration that has been addressed from both theoretical and experimental perspectives~\cite{blickle_realization_2012,martinez_colloidal_2017,plata_building_2020,nakamura_fast_2020,zhang_optimization_2020,tu_abstract_2021}. Nevertheless, there is no a universal optimal mesoscopic design yet. Many classical thermodynamic cycles, such as Stirling's \cite{blickle_realization_2012,muratore-ginanneschi_efficient_2015,krishnamurthy_micrometre-sized_2016} and Otto's \cite{deng_boosting_2013,abah_shortcut--adiabaticity_2019}, comprise isochoric branches. Thus, since isochoric processes resemble processes with constant stiffness in our system, the results derived in this thesis with extra significance for future applications.
    \end{enumerate}
\end{enumerate}

\appendix
\chapter{Simulation methods}
\label{app:simulation-methods}
In this appendix we provide details concerning the simulation methods employed throughout this thesis---specifically, in chapters~\ref{ch:memory_effects}-\ref{ch:glass_transition}---in order to check the validity of our theoretical predictions. Although the majority of the simulations carried out for either the molecular fluid or the granular gas involve binary collisions, both elastic and inelastic, and fluctuating, thermal noise---which involves white-noise and drag forces---we may decouple both numerical contributions to study them independently. In this regard, we first study in Sec.~\ref{subapp:dsmc} how to numerically solve Boltzmann-like equations via the DSMC method, and then we devote Sec.~\ref{app:stochastic-integration} to brief introductory remarks on the stochastic integration of Langevin dynamics with multiplicative thermal noise. Finally, we briefly describe in Sec.~\ref{app:initial-preparations} the initial preparations we have employed in order to study the emergence of non-equilibrium effects in chapters~\ref{ch:memory_effects}-\ref{ch:glass_transition}.

\section{Direct Simulation Monte Carlo (DSMC)}\label{subapp:dsmc}
DSMC techniques constitute stochastic Monte Carlo methods employed for the simulation of sufficiently dilute gases at the molecular level---thus allowing to numerically solve the Boltzmann equation---for finite Knudsen number, i.e. for physical scenarios in which the mean free path of a molecule is of the same order or greater than the dominant characteristic length of the system. 

Initially proposed by G. A. Bird~\cite{bird_g_a_molecular_1994}, DSMC methods are state-of-the-art numerical simulation methods for the majority of applications within the fields of kinetic theory and rarefied gas dynamics~\cite{sharipov_rarefied_2016}. Their accuracy has been corroborated by comparisons with known exact solutions of the Boltzmann equation~\cite{cercignani_boltzmann_1988,nambu_direct_1980}, and with detailed formal proofs~\cite{nambu_direct_1980,wagner_convergence_1992}. However, the DSMC methods differ from the Boltzmann equation when determining the transport coefficients of dilute gases with different pair interactions~\cite{montanero_monte_1996}. Overall, these methods consist in creating a Markov chain for the state variables characterising the system---namely, the positions $\bm{r}$ and velocities $\bm{v}$ of the $N$ particles of the gas. In this regard, we may identify the probability of having two arbitrary particles colliding with the transition probabilities of the chain.

In the original version of the DSMC method, similar to the Molecular Dynamics techniques~\cite{frenkel_understanding_2023}, we consider an ensemble of $N$ hard-sphere particles of radii $\sigma$ and number density $n_{\text{par}}$ that are placed in a closed volume, which is split into cells (or boxes) of size $L$. Such size needs to be much smaller than both the mean free path $\lambda$ and the so-called hydrodynamic length $l_{\text{h}}$, over which the changes in the hydrodynamic fields become noticeable. Nevertheless, as we work under isotropy and homogeneity assumptions throughout this thesis---for which the description of the system is accounted via the one-particle VDF, spatial coordinates do not play a role in our case. The velocities of the particles are updated after a timestep $\Delta t$, which must be sufficiently smaller than both the mean free time $\tau$ and the hydrodynamic time $t_{\text{h}}$, which are given by
\begin{equation}
    \tau \equiv \lambda \left(\frac{2k_BT_{\text{s}}}{m}\right)^{-1/2}, \quad t_{\text{h}} \equiv l_{\text{h}} \left(\frac{2k_BT_{\text{s}}}{m}\right)^{-1/2},
\end{equation}
respectively. In the above, $T_{\text{s}}$ corresponds to the stationary value of the kinetic temperature, and $m$ the mass of the particles. This process is carried out in two steps: (i) first, velocities freely evolve following ballistic motion\footnote{If there are any other external forces acting on the particles, such as white-noise or drag forces---see Sec.~\ref{app:stochastic-integration} for further details, these act over a timescale that may be larger or smaller than $\Delta t$, i.e. either acting every certain number of collisions or in between collisions.} and (ii) after a timestep $\Delta t$ two particles are randomly chosen to collide, and their post-collisional velocities are assigned. We have carried out this two-step process for all the molecular fluid and the granular gas simulations, by employing the following algorithm:
\begin{enumerate}
    \item Choose a random unit vector $\hat{\bm{\sigma}}$.
    \item Choose at random two candidate particles $j$ and $l$ to collide, with $j,l = 1,2,... N$.
    \item We accept the collision with a probability $\omega_{jl}$ that is equal to the collision rate times the timestep,
        \begin{equation}\label{app_eq:transition-prob}
            \omega_{jl} = \frac{1}{2}\Omega (d)N n_{\text{par}} \sigma^{d-1} \Theta (\bm{v}_{jl}^{\text{T}}\ \hat{\bm{\sigma}}) |\bm{v}_{jl}^{\text{T}}\ \hat{\bm{\sigma}}| \Delta t,
        \end{equation}
    with $\Omega (d)$ being the $d$-dimensional solid angle and $\bm{v}_{jl}\equiv \bm{v}_j - \bm{v}_l$ the relative velocity between the chosen particles.
    \item If the collision is accepted, then assign the corresponding post-collisional velocities to each particle,
        \begin{equation}
            \bm{v}_j' = \bm{v}_j - \frac{1+\alpha}{2}(\bm{v}_{jl}^{\text{T}}\ \hat{\bm{\sigma}}) \hat{\bm{\sigma}}, \quad \bm{v}_l' = \bm{v}_l + \frac{1+\alpha}{2}(\bm{v}_{jl}^{\text{T}}\ \hat{\bm{\sigma}}) \hat{\bm{\sigma}},
        \end{equation}
    with $\alpha \in [0,1]$ being the normal restitution coefficient, in order to account for the possible inelasticity of the binary collisions.
\end{enumerate}
Let us remark from Eq.~\eqref{app_eq:transition-prob} that, as $\omega_{jl}$ stands as a probability, the timestep $\Delta t$ must be at least inverse proportional to $N$ in order to be properly normalised. Now, considering either $\tau$---or $t_{\text{h}}$ or any other relevant timescale, if additional external forces are considered---as the characteristic timescale of the dynamics, we have that, after a number $N_{\text{coll}} \equiv \left \lfloor{\tau/\Delta t}\right \rfloor$ of collision trials, we may compute the physical quantities of interest---the kinetic temperature, the Sonine cumulants, etc, at a given time $\tau$ as averages over the ensemble of particles.

\section{Stochastic integration}\label{app:stochastic-integration}
Let us consider a Markov process $\bm{r}^{\sf{T}}\equiv (r_1,r_2,...,r_d)$, whose time-evolution is accounted via the Langevin equation~\eqref{ch1_eq:langevin}. For all the stochastic systems considered in this thesis, the noise term matrix $\mathbb{B}(\bm{r})$ is diagonal, such that we may write the corresponding Langevin equation for the components of $\bm{r}$ as~\cite{mannella_fast_1989}
\begin{equation}
    \label{start}
        \dot{r}_j = a_j(\bm{r}) + b_j(\bm{r}) \eta_j(t), \quad j = 1,...,d, 
\end{equation}
with $a_j(\bm{r})$ corresponding to the components of the term within brackets in Eq.~\eqref{ch1_eq:langevin}---i.e. the $\alpha$-dependent one, and $b_j(\bm{r})$ being the diagonal components of $\mathbb{B}(\bm{r})$ respectively. Herein, $\eta_j(t)$ stands for the Gaussian white-noise unit, which satisfies the statistical properties from Eq.~\eqref{ch1_eq:stat_prop}. We are interested in the numerical integration of Eq.~\eqref{start}, which becomes essential in order to corroborate most of the theoretical findings presented throughout this thesis. In order to do so, we may start by integrating it over an integration timestep $h$, starting from a current state at time $t_0$, which gives
\begin{equation}
    \label{start-2}
        \int_{t_0}^{t_0+h} dt \ \dot{r}_j = r_j(t_0+h) - r_j(t_0) = \int_{t_0}^{t_0+h} dt \ a_j(\bm{r}(t)) + \int_{t_0}^{t_0+h} dt \ b_j(\bm{r}(t)) \eta_j(t).
\end{equation}
In order to deal with the terms on the right-hand side, we introduce the Taylor expansions of the functions $a_j(\bm{r}(t))$ and $b_j(\bm{r}(t))$ around the current state $\bm{r}(t_0)$,
\begin{subequations}
    \label{taylor}
    \begin{align}
        a_j(\bm{r}(t)) &= a_j(\bm{r}(t_0)) + \sum_{l=1}^N \left[r_l(t)-r_l(t_0)\right]\left.\frac{\partial a_j}{\partial r_l}\right|_{t=t_0} \nonumber
        \\
        &+ \frac{1}{2}\sum_{l=1}^N\sum_{k=1}^N \left[r_l(t)-r_l(t_0)\right]\left[r_k(t)-r_k(t_0)\right]\left.\frac{\partial^2 a_j}{\partial r_l \partial r_k}\right|_{t=t_0} + ...
        \\
        b_j(\bm{r}(t)) &= b_j(\bm{r}(t_0)) + \sum_{l=1}^N \left[r_l(t)-r_l(t_0)\right]\left.\frac{\partial b_j}{\partial r_l}\right|_{t=t_0} \nonumber
        \\
        &+ \frac{1}{2}\sum_{l=1}^N\sum_{k=1}^N \left[r_l(t)-r_l(t_0)\right]\left[r_k(t)-r_k(t_0)\right]\left.\frac{\partial^2 b_j}{\partial r_l \partial r_k}\right|_{t=t_0} + ...
    \end{align}
\end{subequations}
such that we may substitute them back into Eq.~\eqref{start-2}. The latter involves an endless recursive process, since the $\left[r_j(t)-r_j(t_0)\right]$ terms from the above expansions may be substituted again by Eq.~\eqref{start-2}, and so on. We truncate this process up to the desired accuracy in terms of powers of the timestep $h$. In order to simplify the forthcoming formulae, we will resort to the notation
\begin{equation}
    a_j \equiv a_j(\bm{r}(t_0)), \quad a_{j,l} \equiv \left.\frac{\partial a_j}{\partial r_l}\right|_{t=t_0}, \quad a_{j,lk}=\left.\frac{\partial^2 a_j}{\partial r_l \partial r_k}\right|_{t=t_0},
\end{equation}
which similarly applies for $b_j(\bm{r}(t))$. Thus, up to order $O(h)$ in the timestep, Eq.~\eqref{start-2} reads
\begin{align}\label{app_eq:01-stoc}
    r_j(t_0+h) - r_j(t_0) = &\  a_j h + b_j \int_{t_0}^{t_0+h} dt \ \eta_j(t) \nonumber
    \\
    &+ \sum_{l}b_{j,l} b_l \int_{t_0}^{t_0+h} dt \int_{t_0}^t dt' \ \eta_j(t) \eta_l(t') + O(h^{3/2}).
\end{align}
In the above, if we ignore the integral term in the second line---or if we consider a constant, additive, noise term $b(\bm{r}) = b_0$---we recover the so-called Euler-Maruyama method for the approximate numerical integration of stochastic differential equations~\cite{faniran_numerical_2015}. The integrals from Eq.~\eqref{app_eq:01-stoc} are ill-defined for the stochastic white-noise forces $\eta_j(t)$. Although having a clear physical interpretation, it is more mathematically rigorous to work with the so-called Wiener process $W_j(t)$, which relates to $\eta_j(t)$ by means of $dW_j(t) = \eta_j(t)dt$. By taking $W_j(t_0) = 0$, it can be shown that $W_j(t)$ constitutes a Gaussian random process, with the statistical properties
\begin{equation}
    \left<W_j(t)\right> = 0, \quad \left< W_j(t)W_l(t') \right> = \text{min}(t,t') \delta_{jl}.
\end{equation}
In principle, in order to tackle the integrals from Eq.~\eqref{start-2}, we would initially proceed \textit{a la Riemann}: we would first discretise the integration interval $0 \equiv t_0 \leq t_1 \leq t_2 ... \leq t_{n-1} \leq t_n \equiv h$ and then introduce an intermediate point $\tau_k \in [t_{k-1},t_k]$ for each time sub-interval. In this way, the integral of an arbitrary function $F(t)$ times a white-noise force $\eta_j(t)$ reads
\begin{equation}
    \int_{t_0}^{t_0+h} dt F(t) \eta_j(t) = \int_{t_0}^{t_0+h} dW_j(t) \ F(t) = \lim_{n \rightarrow \infty} \sum_{k=1}^n F(\tau_k) \left[W_j(t_k) - W_j(t_{k-1})\right].
\end{equation}
If $W_j(t)$ was a continuous function of time, then any choice of the intermediate time $\tau_k$ would yield the same value for the integral, as small variations of $\tau_k$ along the $[t_{k-1},t_k]$ vanish in the $n\rightarrow \infty$ limit. However, $W_j(t)$ is continuous but nowhere differentiable, such that the choice of $\tau_k$ will affect the final result. By writing $\tau_k = \alpha t_{k-1} + (1-\alpha)t_k$ with $\alpha \in [0,1]$, there are three main choices of the "multiplicative-noise" parameter $\alpha$ that constitute the usual conventions employed in stochastic calculus: $\alpha = 0$, $1/2$ and $1$, corresponding to the Itô, Stratonovich and Kilmontovich conventions, respectively. However, one must bear in mind that, for each physical situation of concern, the correct convention for the stocahstic integration of the Langevin equation is dictated by physics, not mathematics~\cite{mannella_ito_2012,van_kampen_stochastic_1992}.

On the one hand, for all the simulations carried out in chapters~\ref{ch:memory_effects}-\ref{ch:glass_transition} we have employed the Itô convention, mainly because it is more appealing for numerical integration schemes---i.e. it is more intuitive to update functions at a new time step ($F(t_{k})$) by employing the values they had at the beginning of the previous time step ($F(t_{k-1}), \alpha = 0$). On the other hand, the Stratonovich convention becomes more enticing for pure mathematical applications, for it can be shown that one may recover the usual tools---chain rule, fundamental theorems...---from ordinary integral calculus. Following the Itô convention, Eq.~\eqref{start} up to $O(h^{3/2})$ may be written as~\cite{mannella_fast_1989}
\begin{align}\label{app_eq:stoch-int-032}
    r_j(t_0+h) &- r_j(t_0) =\ \sqrt{h} Y_{1} b_j + a_j h \nonumber
    \\
    & + \sum_{l}\left[ \frac{h}{2} b_{j,l} b_l (Y_{1}^2-1)+ b_{j,l}a_l h^{3/2} Y_{1} + (a_{j,l}b_l - b_{j,l}a_l) Z(h)\right] \nonumber
    \\
    &  +  \sum_{l,k}\left[ \frac{1}{2}b_{j,lk}b_lb_k \left(\frac{h^{3/2}}{3}Y_1^3-Z(h) \right) + \frac{h^{3/2}}{2}b_{j,l}b_{l,k}b_k Y_1 \left(\frac{Y_1^2}{3}-1\right) \right]\nonumber
    \\
    &  + O(h^{2}),
\end{align}
where
\begin{equation}
    Z(h) = \frac{h^{3/2}}{2}\left(Y_1 + \frac{Y_2}{\sqrt{3}}\right),
\end{equation}
and with $Y_1$ and $Y_2$ being two uncorrelated random variables that are distributed following a standard normal distribution. The choice of retaining up to $O(h^{3/2})$ terms stems from the fact that the functions $a_j(\bm{r})$, $b_j(\bm{r})$ and their derivatives may attain large values for the purposes of our work. Specifically, the noise-term for the molecular fluid with non-linear drag diverges as $T_{\text{s}}^{-1}$ for small enough temperatures, thus requiring higher accuracy when integrating its corresponding Langevin equation.

\section{Numerical preparations}\label{app:initial-preparations}
In chapters~\ref{ch:memory_effects}-\ref{ch:glass_transition}, we employ DSMC and stochastic integration simulations in order to numerically reproduce the strong memory effects predicted for the molecular fluid with non-linear drag, the PDF for the LLNES in various Fokker-Planck systems, and the glass transition displayed for both the granular gas and the molecular fluid, respectively. 

For most of the simulations carried out, the system of concern has to be initially prepared at a stationary state with bath temperature $\theta_i$. On the one hand, for the molecular fluid, as stationary states correspond to equilibrium ones, it suffices to generate $d N$ uncorrelated Gaussian random variables with null mean and variance equal to $\theta_{i}$---corresponding to the velocity components for the ensemble of $N$ $d$-dimensional hard-sphere particles---in order to sample the Maxwellian distribution. On the other hand, for the granular gas, we also start from a Maxwellian distribution with temperature $\theta_i$, and we let it relax---following the DSMC and stochastic integration algorithms---towards the actual stationary state. To assert whether we have reached such stationary state, it suffices to keep track of both the kietic temperature and the excess kurtosis, until they reach their plateau values.

To observe the strong Mpemba and Kovacs effects, it is required to drive the molecular fluid as close as possible to the LLNES. Starting from an equilibrium state at $\theta_i \gg 1$ and by keeping track of the excess kurtosis, we let the system relax for a time $t^*$ until the excess kurtosis is high enough in absolute value---in fact, it reaches values higher than the pseudostationary value $a_2^r$ predicted under the extended Sonine approximation~\eqref{ch2_eq:a2r-a3r}, and we measure the current value of the kinetic temperature $\theta_i(t^*)$.  Now, on the one hand, for the Mpemba effect, $\theta_i(t^*)$ constitutes the initial temperature for the cold sample B. By rescaling the temperatures involved with respect of $\theta_i(t^*)$ we ensure that the cold sample departs from the LLNES with $\theta_{i,\text{B}}=1$ for $t>t^*$, while for the hot sample A we only need it to depart from equilibrium with temperature $\theta_{i,\text{A}}=R_{\text{AB}}$ towards the same final stationary state. On the other hand, for the emergence of the Kovacs effect, it suffices to rescale the temperature to its current value $\theta_i(t^*)$, and set the stationary temperature to unity as well.

In chapter~\ref{ch:nonequilibrium_attractor}, we showed different snapshots of the PDF for different physical systems reaching the LLNES. In order to obtain them, we initially prepared the system of concern at equilibrium at a sufficiently high temperature, and then let it relax towards a sufficiently low final temperature. Similarly to how we proceeded in order to study strong memory effects in chapter~\ref{ch:memory_effects}, we kept track of the excess kurtosis, and take the snapshot of the PDF once it satisfies $|a_2(t^*) - a_2^r|\leq \epsilon$, but with $a_2^r$ being given by the true value attained at the LLNES---Eq.~\eqref{ch3_eq:LLNES-sonine} from Appendix ~\ref{app:sonine-cumulants-LLNES}. 

The glass transition and the hysteresis cycles from chapter~\ref{ch:glass_transition} involve time-dependent linear protocols for the bath temperature $\theta_{\text{s}} = \theta_{\text{s}}(t) = 1 \pm rt$, with $t_0 = r^{-1}$ setting the relevant characteristic timescale for the dynamics. Numerically speaking, the bath temperature comes into play when performing the stochastic integration calculations from Eq.~\eqref{app_eq:stoch-int-032}, as the functions $a_j(\bm{r})$ and $b_j(\bm{r})$ from Eq.~\eqref{start} explicitly depend on it. Thus, at every timestep $h$ we just update the value $\theta_s \to \theta_s \pm r t$. For the cooling procedure, this iterative process is repeated until reaching $t=t_0$---in fact, for the molecular fluid, the process must stop before reaching such limiting value, since the functions $a_j(\bm{r})$ and $b_j(\bm{r})$ diverge for $\theta_{\text{s}}(t_0) = 0$. For the heating process, we proceed in a similar manner, where we take our initial state to be the final state from the previously applied cooling procedure.

\chapter{Perturbative approach to the Kovacs effect}
\label{app:perturbative-kovacs}\sectionmark{Perturbative approach to the Kovacs effect}

In order to provide further analytical insight into the emergence of the Kovacs effect, we resort once again to the tools of perturbation theory. A perturbation theory in the Sonine cumulants is not expected to bear good results, since the pseudostationary values $a_2^r$ and $a_3^r$ are not so small in absolute value as compared with unity. However, taking into account that $\gamma \leq 0.1$, we may develop a perturbation theory in the coupled parameter $\gamma a_0$, with $a_0$ being a parameter of the same order as $a_2^r$ and $a_3^r$; i.e. $a_2^r/a_0$ and $a_3^r/a_0$
 are of the order of unity. By defining
 \begin{equation}\label{ch2_eq:cumulants-rescaled}
     A_2(t) \equiv \frac{a_2(t)}{a_2^r}, \quad A_3(t) \equiv \frac{a_3(t)}{a_3^r},
 \end{equation}
 which are $O(1)$, we may write the perturbative expansions
\begin{subequations}
\label{ch2_eq:Kovacs-perturbative-a3}
\begin{align}
    \theta(t) &= \theta^{(0)}(t) + \gamma a_{0}\,\theta^{(1)}(t) + O((\gamma a_{0})^2),
    \\
    A_2(t) &= A_2^{(0)}(t) + \gamma a_{0} A_2^{(1)}(t) + O((\gamma a_{0})^2),
    \\
    A_3(t) &= A_3^{(0)}(t) + \gamma a_{0} A_3^{(1)}(t) + O((\gamma a_{0})^2).
\end{align}
\end{subequations}

Given that the initial conditions \eqref{ch2_eq:initial-cond-kovacs} must hold regardless of the value of the perturbation parameter $\gamma a_0$, it follows that
\begin{subequations}\label{ch2_eq:initial-conditions-perturbative-kovacs}
\begin{align}
    \theta^{(0)}(t_{w}) &=  A_2^{(0)}(t_{w}) = A_3^{(0)}(t_{w}) = 1, \\
    \theta^{(j)}(t_{w}) &= A_2^{(j)}(t_{w}) = A_3^{(j)}(t_{w}) = 0, \quad j=1,2,\ldots
\end{align}
\end{subequations}
The perturbative expansions \eqref{ch2_eq:Kovacs-perturbative-a3} lead to the following hierarchy of equations: to the lowest order $O(1)$, we have
\begin{subequations}\label{ch2_eq:Kovacs-perturbative-order-0}
    \begin{align}
        \dot{\theta}^{(0)}= &\ 2(1-\theta^{(0)})\left[1+\gamma (d+2)\theta^{(0)}\right],
        \\
        \dot{A}_2^{(0)}= &\ \frac{8\gamma}{a_{2}^r}(1-\theta^{(0)}) -\left[\frac{4}{\theta^{(0)}}-8\gamma+4\gamma(d+8)\theta^{(0)}+ \frac{8(d-1)}{d(d+2)}\frac{\sqrt{\theta^{(0)}}}{\zeta_{0}^*}\right]A_2^{(0)} \nonumber
        \\
        & + 2 \left[ 2\gamma \theta^{(0)} (d+4) + \frac{(d-1)}{d(d+2)}\frac{\sqrt{\theta^{(0)}}}{\xi} \right]\frac{a_{3}^r}{a_{2}^r}A_3^{(0)},
        \\
        \dot{A}_3^{(0)}= &\ 12\left[-4\gamma + 6\gamma \theta^{(0)} + \frac{(d-1)\sqrt{\theta^{(0)}}}{d(d+2)(d+4)\xi}\right]\frac{a_2^r}{a_3^r}A_2^{(0)} \nonumber
        \\
        & + 6 \left[4\gamma - \frac{1}{\theta^{(0)}} - \gamma \theta^{(0)}(d+14) - \frac{(d-1)(4d+19)\sqrt{\theta^{(0)}}}{2d(d+2)(d+4)\xi}\right]A_3^{(0)},
    \end{align}
\end{subequations}
while to the first order $O(\gamma a_0)$, we write for the kinetic temperature
\begin{align}   
      \dot{\theta}^{(1)}=-2\theta^{(1)}\left[1+\gamma (d+2)\theta^{(0)}\right]+2\gamma (d+2)\theta^{(1)}[1-\theta^{(0)}] -2(d+2) \frac{a_2^r}{a_0}(\theta^{(0)})^2 A_2^{(0)}.
      \label{ch2_eq:theta^1-a3}
\end{align}
We do not provide the evolution equations for both $A_2^{(1)}(t)$ and $A_3^{(1)}(t)$, since these do not explicitly appear on the evolution equation of $\dot{\theta}^{(1)}$, which is the quantity of interest here. Let us focus on the $O(1)$ contributions. Given the initial conditions \eqref{ch2_eq:initial-conditions-perturbative-kovacs}, we have that $\theta^{(0)}(t) = 1$. This is consistent with the fact that, for either $\gamma = 0$ ---i.e. linear drag at the Rayleigh limit--- or $a_0 = 0$ ---i.e. the system is at equilibrium--- there is no Kovacs effect, and the temperature remains at its equilibrium value for $t>t_{w}$. The lowest order $A_2^{(0)}(t)$ and
$A_3^{(0)}(t)$ are obtained by solving
\begin{equation}\label{ch2_eq:a2-a3-linear}
    \frac{d}{dt}\bm{A}^{(0)} = \mathbb{M} \bm{A}^{(0)},
\end{equation}
where the vector $\bm{A}^{(0)}$ and the matrix $\mathbb{M}$ are defined as
\begin{equation}\label{ch2_eq:a2-a3-vector-matrix}
        \bm{A}^{(0)} \equiv \begin{pmatrix}
        A_2^{(0)} \\
        A_3^{(0)}
        \end{pmatrix}, \quad \mathbb{M} \equiv \begin{pmatrix}
        M_{11} & M_{12} \\
        M_{21} & M_{22}
        \end{pmatrix}.
\end{equation}
In the above, $M_{ij}$ are the elements of the $2 \times 2$ matrix $\mathbb{M}$ from Eq.~\eqref{ch2_eq:Mij}, and $\lambda_{\pm}$ its corresponding eigenvalues, introduced in Eq.~\eqref{ch2_eq:eigenproblem}. Following this, the solution of Eq.~\eqref{ch2_eq:a2-a3-linear} is given by
\begin{equation}
\label{ch2_eq:vector-cumulant}
    \bm{A}^{(0)}(t) = c_+ \bm{u_+} e^{\lambda_+ (t-t_w)} + c_- \bm{u_-} e^{\lambda_- (t-t_w)},
\end{equation}
with the constants $c_+$, $c_-$ being determined by imposing the initial conditions \eqref{ch2_eq:initial-conditions-perturbative-kovacs},
\begin{equation}\label{ch2_eq:constants}
    c_+ = \frac{M_{11}+M_{12}-\lambda_-}{(\lambda_+-\lambda_-)M_{12}}, \quad c_- = \frac{\lambda_+-M_{11}-M_{12}}{(\lambda_+-\lambda_-)M_{12}}.
\end{equation}
Once the lowest order is completed, we may determine $\theta^{(1)}(t)$ from Eq.~\eqref{ch2_eq:theta^1-a3}. Taking into account that $\theta^{(0)}(t) = 1$, Eq.~\eqref{ch2_eq:theta^1-a3} simplifies to 
\begin{equation}\label{ch2_eq:theta1-kovacs-ode}
    \dot{\theta}^{(1)}=-2\theta^{(1)}\left[1+\gamma (d+2)\right]-2(d+2) \frac{a_2^r}{a_0} A_2^{(0)}.
\end{equation}
As $\theta^{(1)}(t_{w})=0$, we may integrate the above to obtain
\begin{equation}\label{ch2_eq:kov-pert-01}
    \theta^{(1)}(t) = -2(d+2)\frac{a_2^r}{a_0} e^{-\hat{\alpha} t}\int_{t_w}^t e^{\hat{\alpha} t'} A_2^{(0)}(t') dt',
\end{equation}
where $\hat{\alpha}\equiv 2[1+\gamma (d+2)]$ and
\begin{align} 
        A_2^{(0)}(t) = \frac{1}{\lambda_+-\lambda_-}\Big[(M_{11}+M_{12}-\lambda_-)e^{\lambda_+(t-t_w)} +(\lambda_+-M_{11}-M_{12})e^{\lambda_-(t-t_w)}\Big].
        \label{ch2_eq:exponentials-a20}
\end{align}
Introducing the latter onto Eq.~\eqref{ch2_eq:kov-pert-01} provides us with the Kovacs function $K(t) = \theta(t) -1 \approx \gamma a_0 \theta^{(1)}$ up to order $O(\gamma a_0)$, corresponding to Eq.~\eqref{ch2_eq:kovacs-hump-first-order} in the main text.

\chapter{\label{app:asymptotic-llnes}Asymptotic behaviours at the LLNES}
The emergence of the Dirac-delta behaviour at the LLNES was justified for those situations for which the moments of the stationary scaled distribution $\phi_{\text{st}}(\xi)$ remain finite, such that $\sigma_{r^*}^*(t)/\left<r^*\right>(t)\propto \left[r^*_{\LLNES}(t)\right]^{n-1}\to 0$ for long times. In order to generalise the latter to those situations in which the standard deviation of $\phi(\xi,t)$, $\sigma_{\xi}(t)$, diverges, we shall analyse the moments of $\phi(\xi,t)$,
\begin{equation}
    \left< \xi^m \right>(t) \equiv \int_0^{[r^*_{\LLNES}]^{1-n}}d\xi \ \xi^m \phi(\xi,t).
\end{equation}
The scaled PDF $\phi(\xi,t)$ is well-defined in the $r^*_{\LLNES} \rightarrow 0$ limit---or equivalently, in the limit $t\rightarrow +\infty$, since it attains the stationary PDF $\phi_{\text{st}}(\xi)$, and
\begin{align}\label{app_eq:moments-llnes}
    \left< \xi^m \right>(t) &\sim \int_0^{[r^*_{\LLNES}]^{1-n}}d\xi \ \xi^m \phi_{\text{st}}(\xi) \nonumber
    \\
    &= \int_0^{[r^*_{\LLNES}]^{1-n}}d\xi \ \frac{\xi^m}{[(n-1)\xi]^{\frac{n}{n-1}}}P_i^*\left([(n-1)\xi]^{-\frac{1}{n-1}}\right), \quad r^*_{\LLNES} \ll 1.
\end{align}
Plausible divergences of the above integral stem from the fact that $\phi_{\text{st}}(\xi)$ presents fat tails. Therefore, we are interested in studying the dominant contribution stemming from the fat tail, which is around $\xi = [r^*_{\LLNES}]^{1-n}$. Upon the change of variables $\omega = [(n-1)\xi]^{-1/(n-1)}$ the integral from Eq.~\eqref{app_eq:moments-llnes} becomes
\begin{equation}
    \left< \xi^m \right>(t) = \frac{1}{(n-1)^m} \int_{\omega^*_{\LLNES}}^{+\infty}d\omega \ \left(\frac{1}{\omega}\right)^{m(n-1)} P^*_i(\omega), \quad \omega^*_{\LLNES} \equiv \frac{r^*_{\LLNES}}{(n-1)^{\frac{1}{n-1}}}. 
\end{equation}
In the following, we further assume that the initial PDF $P^*_i(r^*_i)$ can be factorised as
\begin{equation}
    P^*_i(r^*_i) = [r_i^*]^z \hat{P}^*_i(r^*_i),
\end{equation}
with $\hat{P}^*_i(0)\neq 0$ and $z\geq 0$, such that we take into account possible Jacobian terms. Thus, we are left with the final expression for the moments,
\begin{equation}\label{app_eq:moments-llnes-2}
    \left< \xi^m \right>(t) = \frac{1}{(n-1)^m} \int_{\omega^*_{\LLNES}}^{+\infty}d\omega \ \left(\frac{1}{\omega}\right)^{m(n-1)-z} \hat{P}^*_i(\omega),
\end{equation}
whose behaviour we study now depending on the values of $m$, $n$ and $z$. 
\begin{itemize}
    \item For $m(n-1)-z<1$, the integral from Eq.~\eqref{app_eq:moments-llnes-2} is well-defined in the $r^*_{\LLNES}$ limit,
    \begin{equation}
        \left< \xi^m \right>(t) = \frac{1}{(n-1)^m} \int_{0}^{+\infty}d\omega \ \left(\frac{1}{\omega}\right)^{m(n-1)-z} \hat{P}^*_i(\omega) \equiv \left< \xi^m \right>_{\text{st}},
    \end{equation}
    for which we recover the case considered in the main text.
    \item For $m(n-1)-z=1$, Eq.~\eqref{app_eq:moments-llnes-2} diverges. Integration by parts shows that the moments diverge logarithmically as
    \begin{equation}
        \left< \xi^m \right>(t) \sim -\frac{\hat{P}_i(0)}{(n-1)^m} \log (r^*_{\LLNES}), \quad r^*_{\LLNES} \ll 1.
    \end{equation}
    \item Finally, for $m(n-1)-z>1$, the integral Eq.~\eqref{app_eq:moments-llnes-2} diverges as well, but algebraically as
    \begin{equation}
        \left< \xi^m \right>(t) \sim \frac{1}{(n-1)^m}\frac{\hat{P}_i(0)}{m(n-1)-z-1} \left(\frac{1}{r^*_{\LLNES}}\right)^{m(n-1)-z-1}, \quad r^*_{\LLNES} \ll 1.
    \end{equation}
\end{itemize}
In principle, as we are exclusively interested in the behaviour of the standard deviation $\psi_{\xi}$, we only need to consider the $m = 1$ and $m = 2$ cases for the moments. In fact, if diverging, it is straightforward to show that the overall behaviour is dominated by $\left< \xi^2\right>$, since it always diverges faster than $\left< \xi\right>^2$. Hence, for the original standard deviation
\begin{equation}
    \sigma^*_{r^*}(t) = [r^*_{\LLNES}]^n \sigma_{\xi}(t),
\end{equation}
we identify three possible behaviours:
\begin{itemize}
    \item A \textbf{regular regime} for $2(n-1)-z<1$ for which $\sigma_{\xi}(t)$ is finite in the $r^*_{\LLNES}\rightarrow 0$ limit, such that
    \begin{equation}
        \sigma^*_{r^*}(t) = [r^*_{\LLNES}]^n \sigma_{\xi}(+\infty) \propto [r^*_{\LLNES}]^n,
    \end{equation}
    which, again, corresponds to the case considered in the main text.
    \item A \textbf{logarithmic regime} for $2(n-1)-z=1$ for which $\sigma^*_{r^*}(t)$ attains the form
    \begin{equation}
        \sigma^*_{r^*}(t) \propto [r^*_{\LLNES}]^n \sqrt{-\log(r^*_{\LLNES})}.
    \end{equation}
    \item And an \textbf{algebraic regime} for $2(n-1)-z>1$, characterised by the behaviour
    \begin{equation}
        \sigma^*_{r^*}(t) \propto [r^*_{\LLNES}]^{\frac{3+z}{2}},
    \end{equation}
    which, interestingly, is universal in the sense that it is independent of the non-linearity exponent $n$.
\end{itemize}
Note that, in each regime, we have that $\sigma_{r^*}^*(t)/\left<r^*\right>(t)\to 0$, which verifies our assumptions of the tendency towards the Dirac-delta shape characterising the LLNES.

\chapter{\label{app:sonine-cumulants-LLNES}Sonine cumulants at the LLNES}
We recall that, in order to characterise the deviations from the Maxwellian equilibrium distribution $f_{\text{eq}}(\bm{v})$~\eqref{ch1_eq:maxwellian} in chapter~\ref{ch:memory_effects}, we expanded the scaled VDF $\phi(\bm{c},t)$ as introduced in Eq.~\eqref{ch2_eq:scaling-velocities} in Laguerre polynomials via the Sonine expansion~\eqref{ch2_eq:sonine-expansion}, with $a_l(t)$ corresponding to the Sonine cumulants. In chapter~\ref{ch:memory_effects}, we obtained explicit expressions for the excess kurtosis $a_2$ and the sixth cumulant $a_3$ at the LLNES within the first and second Sonine approximation schemes. As highlighted in that chapter, such values differed from the actual numerical values obtained via DSMC, mainly because of the assumption of the smallnes of the cumulants that it is implicit in any Sonine scheme.

Now, taking into account the explicit expression for the VDF at the LLNES~\eqref{ch3_eq:LLNES-sol-2}, we may obtain analytically the actual pseudostationary values $a_l^r$ of such cumulants. The calculation is straightforward, as the Laguerre polynomials $L_{l}^{\alpha}(x)$ are orthogonal over the interval $[0,+\infty)$ with respect to the scalar product with weight $x^{\alpha}e^{-x}$,
\begin{equation}
    \int_0^{+\infty}dx \ x^{\alpha}e^{-x} L_{n}^{\alpha}(x) L_{m}^{\alpha}(x) = \frac{\Gamma (n+\alpha +1)}{n!} \delta_{n,m},
\end{equation}
such that, by integrating Eq.~\eqref{ch2_eq:sonine-expansion} with $a_l(t) = a_l^r$ and introducing Eq.~\eqref{ch3_eq:LLNES-sol-2}, we arrive at
\begin{equation}
    a_l^r = \frac{l!\ \Gamma(d/2)}{\Gamma(l+d/2)}L_{l}^{\frac{d-2}{2}}(M_2).
\end{equation}
In order to visualise the tendency to the LLNES and its being a far-from-equilibrium state, we resort once again to the previously employed excess kurtosis $a_2$ and the sixth cumulant $a_3$, whose reference values ---which are obtained by substituting $l = 2$ and $3$ in the above--- are
\begin{equation}
    \label{ch3_eq:LLNES-sonine}
        a_{2}^r \equiv -\frac{2}{d+2}, \quad a_{3}^r \equiv -\frac{16}{(d+2)(d+4)}.
\end{equation}

\begin{figure}
    {\centering 
    \includegraphics[width=2.65in]{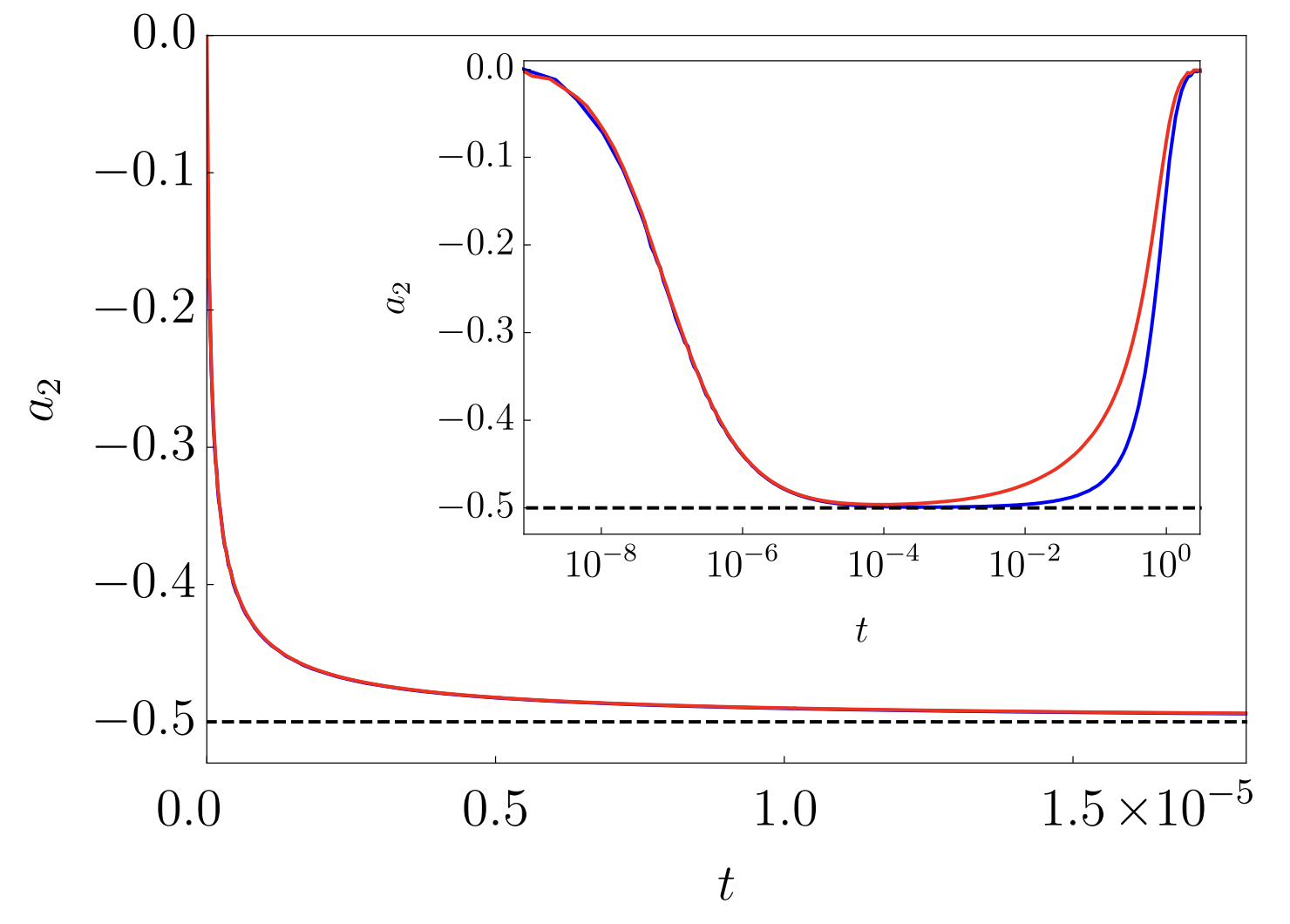}
    \includegraphics[width=2.65in]{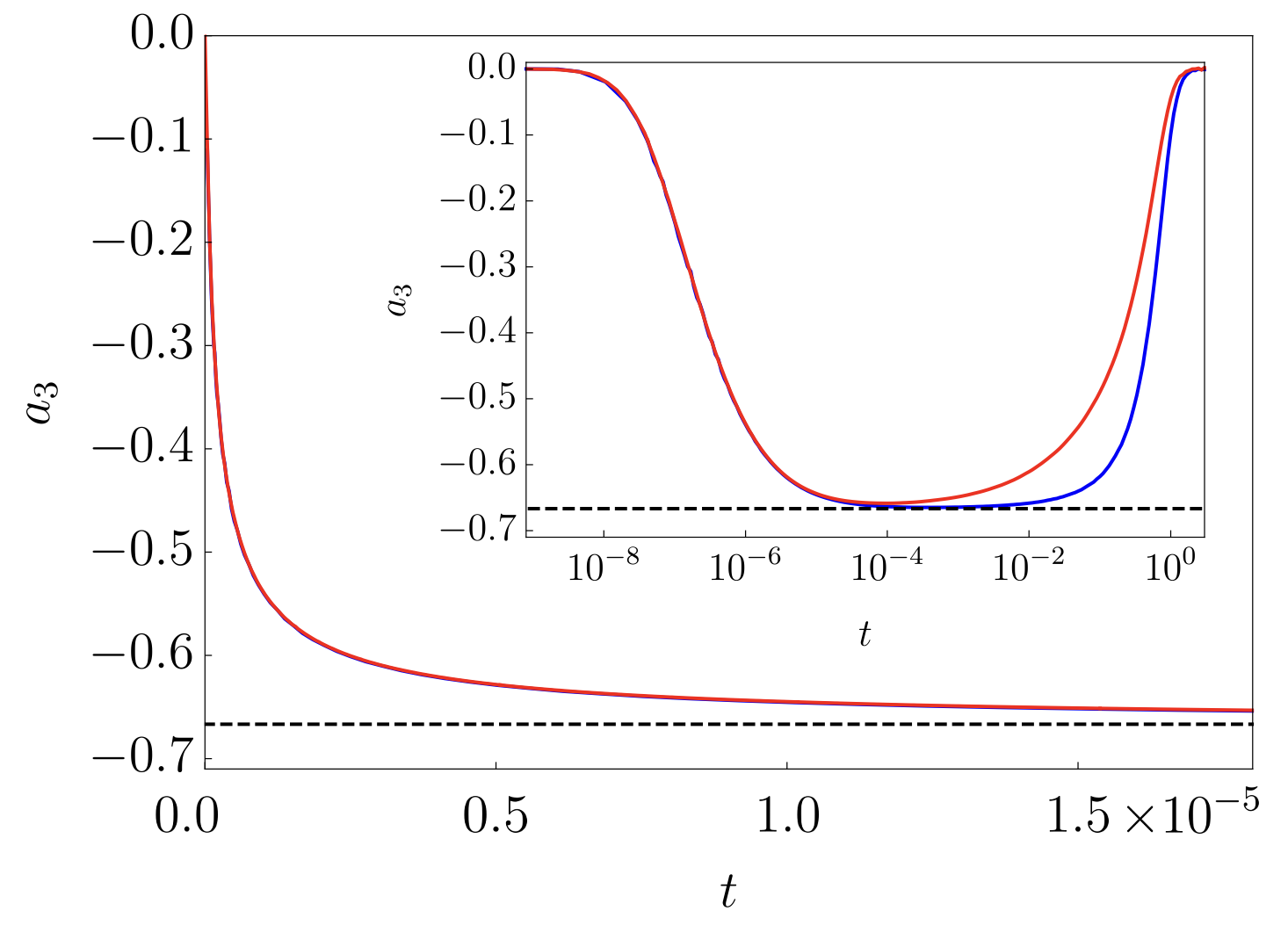}}
    \caption{Time evolution of the  excess kurtosis $a_2$ (left) and the sixth cumulant $a_3$ (right), obtained via DSMC. The main panels show their relaxation towards the LLNES, 
    for $n=2$, $\gamma = 0.1$, $d=2$, $N=10^6$ particles and two values of $\xi$,  $\xi=1$ (red) and  $\xi=10$ (blue). The system relaxes from an initial equilibrium distribution with temperature $T_i$ to a final one $T_f$, such that $\theta_i = T_i/T_f = 10^7$. Black, dashed lines correspond to the asymptotic, pseudostationary values attained at the LLNES, which are given by Eq.~\eqref{ch3_eq:LLNES-sonine}. The relaxation towards the LLNES is independent of the value of $\xi$, i.e. of the relevance of collisions. Both insets show the time evolution in logarithmic scale, in which it is observed that the system eventually returns to equilibrium---at which both $a_2$ and $a_3$ vanish--for longer times. The LLNES lasts longer for $\xi=10$ than for $\xi=1$ because in the former case collisions, which accelerate thermalisation, come into play for longer times.}
        \label{ch3_fig:sonine}
    \end{figure}

Let us note that the above expressions differ from Eqs.~\eqref{ch2_eq:extremes-excess-kurtosis} and \eqref{ch2_eq:a2r-a3r}, corresponding to the pseudostationary values of $a_2$ and $a_3$ within the first and second Sonine approximations, respectively. In fact, the values from Eq.\eqref{ch3_eq:LLNES-sonine} are larger in absolute magnitude regardless of the value of $d$, due to the assumption of the smallness of the cumulants implicit in any Sonine approximation scheme.

Figure~\ref{ch3_fig:sonine} presents the time evolution of both 
Sonine cumulants, which have been obtained via DSMC simulations for an ensemble of $N = 10^6$ particles in each case. We considered a deep quench of bath temperature from an initial to a final equilibrium values $T_i$ and $T_f$, such that $T_i / T_f = 10^7$. It may be observed that, regardless of collisions, which try to thermalise the system towards equilibrium, both cumulants tend to their corresponding LLNES values within an intermediate timescale. It is only for longer times that collisions and noise come into play and make the system go to equilibrium at the final temperature $T_f$, at which both cumulants vanish---as shown in the insets.

\chapter{\label{app:thd-length}Thermodynamic length for almost fully degenerate oscillators}\chaptermark{Thermodynamic length}
The thermodynamic length from Eq.~\eqref{ch5_eq:length-inequal} for the general bang-bang protocol may be split in two terms,
\begin{equation}
\label{app_eq:decomposed-length}
    \mathcal{L}(t_f) = \sqrt{\frac{d}{2}} \left[\sum_{\text{heatings}}\int_{\Delta t_H}dt \ \frac{\dot{z}(t)}{z(t)} - \sum_{\text{coolings}}\int_{\Delta t_C}dt \ \frac{\dot{z}(t)}{z(t)} \right],
\end{equation}
corresponding to the lengths associated to the heating and cooling stages. Given a time $t_J$ where an intermediate heating window starts, we can relate it to the time $t_J'$ when the preceding cooling window started by means of the relation
\begin{equation}
\label{app_eq:useful-relation}
    z(t_J) = z(t_J') \ e^{-2(t_J-t_J')},
\end{equation}
which is a direct consequence of the general form of the evolution equations in the $T_{\min} = 0$ case. Let us start by considering the cooling windows---i.e. the second term on the right-hand side of Eq.~\eqref{app_eq:decomposed-length}. Applying Eq.~\eqref{app_eq:useful-relation}, we have that
\begin{align}
    -\sum_{\text{coolings}}\int_{\Delta t_C}dt \ \frac{\dot{z}(t)}{z(t)} &= \sum_{\text{coolings}} \ln \left( \frac{z(t_C^i)}{z(t_C^f)}\right) \nonumber
    \\
    &= \sum_{\text{coolings}} \ln \left( \frac{z(t_C^f)e^{2\Delta t_C}}{z(t_C^f)}\right) = 2\sum_{\text{coolings}} \Delta t_C = 2 \tau_c,
\end{align}
where $z(t_C^i)$ and $z(t_C^f)$ correspond to the values of the dynamic variable $z$ at the endpoints of the cooling window of length $\Delta t_C$. Now, for the remaining term from Eq.~\eqref{app_eq:decomposed-length}---i.e. the one corresponding to the heating windows, it becomes more intuitive to expand it as
\begin{equation}
    \sum_{\text{heatings}}\int_{\Delta t_H}dt \ \frac{\dot{z}(t)}{z(t)} = \ln \left( \frac{z(t_1)}{z(0)}\right) + \ln \left( \frac{z(t_3)}{z(t_2)}\right) + \ln \left( \frac{z(t_5)}{z(t_4)}\right) + ...,
\end{equation}
where $t_1$, $t_2$... are the corresponding switching times.
Once again, applying Eq.~\eqref{app_eq:useful-relation} leads to
\begin{align}
    \sum_{\text{heatings}}\int_{\Delta t_H}dt \ \frac{\dot{z}(t)}{z(t)} &= \ln \left( \frac{z(t_1)}{z(0)}\right) + \ln \left( \frac{z(t_3)}{z(t_1)e^{-2(t_2-t_1)}}\right) + \ln \left( \frac{z(t_5)}{z(t_3)e^{-2(t_4-t_3)}}\right) + ... \nonumber
    \\
    & = -\ln z(0) + 2(t_2-t_1) + 2(t_4-t_3) + ... + \ln z(t_f) \nonumber
    \\
    & = \ln \left( \frac{z(t_f)}{z(0)}\right) + 2 \sum_{\text{coolings}}\Delta t_C = \ln T_f + 2 \tau_c,
\end{align}
and thus, the statistical length is given by
\begin{equation}
    \mathcal{L}(t) = \sqrt{\frac{d}{2}} \left(\ln T_f + 2t_c + 2t_c\right) = \sqrt{\frac{d}{2}}(\ln T_f + 4\tau_c),
\end{equation}
corresponding to Eq.~\eqref{ch5_eq:length-Tmax}. In the above, we have omitted the absolute value in $\ln T_f$ due to the fact that $T_f > 1$ for the heating protocol considered. The same spirit of the proof employed for the heating protocol also applies for a cooling on, in which we would only need to consider different values of $T(t)$ for the first and last time windows.

\chapter{\label{app:singular-bang-bang}Singular bang-bang protocols for the Brownian gyrator}
\chaptermark{Singular bang-bang protocols}

In this appendix, we rigorously prove that the NON and POP, three-bang protocols presented in Sec.~\ref{ch6_sec:constructing-bang-bang}---and any other higher-order bang-bang protocol involving the NON and/or POP structures---are singular rather than regular bang-bang protocols. Let us start by considering the three-bang protocol POP---due to the symmetry between the P and N points, the results obtained here are also valid for the protocol NON. We do not consider the limit $k_{\max}\rightarrow +\infty$ in this section, since it becomes a singular limit when studying the derivatives of Pontryagin's Hamiltonian function. It suffices to proceed with the forthcoming analysis for $k_{\max}$ finite, and then extend our results to the $k_{\max}\rightarrow +\infty$ regime, which is the one we employ throughout the majority of this work. 

Let us recall that the switching function $\phi_{\text{OP}}(t)$ is given by Eq.~\eqref{ch6_eq:switching-functions-OP}, it accounts for the derivative of Pontryagin's Hamiltonian function along the $\overline{\text{OP}}$ edge of the triangular control set. Since the POP protocol involves two switchings---one switch from P to O at some time $t_1 \in (0,t_f)$ followed by another one from O to P at $t_2 \in (t_1,t_f)$, $\phi_{\text{OP}}(t)$ must have at least two zeros on the entire interval $(0,t_f)$. The protocols starts from vertex P, so we have 
\begin{equation}\label{ch6_eq:bound-initial}
    \phi_{\text{OP}}(0) = -2\left[2\psi_1(0) z_1(0) + \psi_3(0) z_3(0)\right] = -2\left(\frac{\psi_{1,0}}{1+u_i} + \frac{\psi_{3,0}}{2} \right) > 0.
\end{equation}
Now, by integrating Hamilton's canonical equations \eqref{ch6_eq:canonical} at vertex P, we  write our switching function explicitly as
\begin{equation}
    \phi_{\text{OP}}(t) = \phi_{\text{OP}}(0) - \frac{\psi_{1,0}}{k_{\max}}\left(e^{4k_{\max}t}-1\right) -\frac{\psi_{3,0}}{k_{\max}}\left(e^{2k_{\max}t}-1\right), \quad 0\leq t \leq t_1.
\end{equation}
This function constitutes a second degree polynomial for the variable $\text{exp}(2k_{\max}t)$, and thus, it presents two roots at most---i.e. up to two candidates for the switching time $t_1$, for which $\phi_{\text{OP}}(t_1)=0$. The fact that $t_1$ must be positive sets bounds on the sign of the ratio $\psi_{1,0}/\psi_{3,0}$, which must be consistent with that in Eq.~\eqref{ch6_eq:bound-initial}. Once we switch to the point O, by integrating once again the canonical equations at vertex O, we get
\begin{equation}
    \phi_{\text{OP}}(t) = \cancelto{0}{\phi_{\text{OP}}(t_1)} -2\left(2\psi_{1,0}e^{4k_{\max}t_1} + \psi_{3,0}e^{2k_{\max}t_1}\right)(t-t_1), \quad t \geq t_1.
\end{equation}
However, this corresponds to a linear function starting from zero, which does not have any roots apart from the trivial one at $t = t_1$, and thus, the switching back to the last time window at point P is forbidden. {Therefore, a regular three-bang POP protocol is not possible. Still, a three-bang PON protocol would be indeed feasible, since the switching to vertex N is ruled by the switching function $\phi_{\text{ON}}(t)$, for which we do not have any restriction.}

Another possibility to overcome the found issue could be to consider $\phi_{\text{OP}}(t) = 0$ for $t\geq t_1$, which would set the value of the ratio $\psi_{1,0}/\psi_{3,0}$ { from the condition $2\psi_{1,0}e^{4k_{\max}t_1} + \psi_{3,0}e^{2k_{\max}t_1}=0$}. In this case, the time window at point O would become a singular time interval. However, by virtue of our analysis of singular protocols in Sec.~\ref{ch6_subsec:characterising-singular}, if $\phi_{\text{OP}}(t) = 0$ for some time window, then it must be zero fon the whole interval $(0,t_f)$. Henceforth, the protocols POP and NON---and any other higher order protocol involving the POP and/or NON combinations---belong in the class of singular protocols depicted in Fig.~\ref{ch6_fig:switching-funcs-singular}.

\chapter{\label{ch6_subsec_energy-balance}Energetic balance from the Brownian gyrator's speed limit}\chaptermark{Energetic balance}

In this last appendix, we evaluate the contribution to the irreversible work $\expval{W_{\irr}}$ defined in Eq.~\eqref{ch6_eq:irr-work} from the bangs at the different vertices O, P, and N partaking in the bang-bang protocols studied in chapter~\ref{ch:brownian_gyrator}. Throughout the appendix, we consider that the bang starts at a certain time $t_1$ and ends at a certain time $t_2\ge t_1$, so that the duration of the bang is $\tau=t_2-t_1$. It is useful to write $\expval{W_{\irr}}$ as
\begin{equation}
    \expval{W_{\irr}}=\expval{W_{\irr}^{(1)}}+\expval{W_{\irr}^{(2)}}, \quad \expval{W_{\irr}^{(j)}}=\gamma \int_0^{t_f}dt\, \left(\frac{d\sigma_j}{dt}\right)^2. 
\end{equation}

Consistently with the non-dimensionalisation introduced in Eq.~\eqref{ch6_eq:dimensionless-variables}, we introduce dimensionless standard deviations $\sigma_j^*$ as:
\begin{equation}
    \sigma_j=\sqrt{\frac{k_B(T_x+T_y)}{k_i}}\sigma_j^*,
\end{equation}
which implies that $\sigma_j^*=\sqrt{z_j}$. Therefore, we obtain
\begin{equation}
   \expval{W_{\irr}^{(j)}}=\cancel{\gamma} \frac{\cancel{k_i}}{\cancel{\gamma}} \frac{k_B(T_x+T_y)}{\cancel{k_i}}\int_0^{t_f^*} dt^*\, \left(\frac{d\sigma_j^*}{dt^*}\right)^2 ,
\end{equation}
which suggest the following definition for the non-dimensionless work:
\begin{equation}
    \expval{W_{\irr}^{(j)}}=k_B(T_x+T_y)\expval{W_{\irr}^{(j)}}^* \implies \expval{W_{\irr}^{(j)}}^*=\int_0^{t_f^*} dt^*\, \left(\frac{d\sigma_j^*}{dt^*}\right)^2 .
\end{equation}
As in the remainder of the chapter, we drop the asterisks in the following to simplify our notation.

\subsection*{Vertex O}

At the vertex O, we have that $k=u=0$, so the evolution equations for the variances $z_j=\sigma_j^2$ are $\dot{z}_j=1$ and thus
\begin{equation}
    z_j(t)=z_j(t_1)+(t-t_1) \implies \sigma_j(t)=\sqrt{\sigma_j^2(t_1)+(t-t_1)}, \quad t_1\le t\le t_2.
\end{equation}
Therefore, we have that
\begin{equation}
    \expval{W_{\irr}^{(j)}}= \int_{t_1}^{t_2}dt\, \left[\sigma_j^2(t_1)+(t-t_1)\right]^{-1}=\ln \left[\frac{\sigma_j^2(t_2)}{\sigma_j^2(t_1)}\right],
\end{equation}
and, finally
\begin{equation}
    \expval{W_{\irr}}=2\ln \left[\frac{\sigma_1(t_2)\sigma_2(t_2)}{\sigma_1(t_1)\sigma_2(t_1)}\right].
\end{equation}

\subsection*{Vertex P}

At the vertex P, $k=u=k_{\max}$, so the evolution equation for the variances are
\begin{equation}
    \frac{dz_1}{dt}=-4k_{\max}z_1+1, \quad \frac{dz_2}{dt}=1.
\end{equation}
The equation for $z_2$ is the same as for the vertex O, so we do not need to derive the contribution $\expval{W_{\irr}^{(2)}}$. We thus focus on the evolution of $z_1$---or $\sigma_1$---and its corresponding contribution $\expval{W_{\irr}^{(1)}}$:
\begin{equation}
    \sigma_1(t)=\sqrt{\frac{1}{4k_{\max}}+\left(\sigma_1^2(t_1)-\frac{1}{4k_{\max}}\right)e^{-4k_{\max}t}},
\end{equation}
from which we derive
\begin{equation}
    \expval{W_{\irr}^{(1)}}=k_{\max}\left\{\sigma_1^2(t_1)-\sigma_1^2(t_2)+\frac{1}{4k_{\max}}\ln\left[\frac{\sigma_1^2(t_2)}{\sigma_1^2(t_1)}\right]\right\}.
\end{equation}
In the limit $k_{\max}\to\infty$ we have considered, the bang at vertex P becomes instantaneous, $\tau=t_2-t_1\to 0$, but $\xi=e^{-2k_{\max}t/\gamma}<1$ remains finite---as discussed in Sec.~\ref{ch6_sec:constructing-bang-bang}. Then, we get
\begin{equation}
    \expval{W_{\irr}^{(1)}}\sim k_{\max}\,\sigma_1^2(t_1)\left(1-\xi^2\right),
\end{equation}
and this contribution to the irreversible work diverges linearly with $k_{\max}$. An analogous divergent behaviour of the cost of the connection has been found for the thermodynamic geometry cost in other systems~\cite{prados_optimizing_2021,patron_thermal_2022}---although, therein, the parameter that was controlled was the bath temperature instead of the confining potential.

\subsection*{Vertex N}
\sectionmark{Energetic balance}
At the vertex N, $k=-u=k_{\max}$, so the evolution equation for the variances are
\begin{equation}
    \frac{dz_1}{dt}=1, \quad \frac{dz_2}{dt}=-4k_{\max}z_2+1.
\end{equation}
As repeatedly commented along the chapter, there is a symmetry between the P and N vertices. For the calculation of the contribution to the irreversible work we are considering here, we see that we have only to exchange the labels $1\leftrightarrow 2$. Therefore, for the vertex N, it is $\expval{W_{\irr}^{(1)}}$ that coincides with our calculation for vertex O, whereas
\begin{equation}
    \expval{W_{\irr}^{(2)}}=k_{\max}\left\{\sigma_2^2(t_1)-\sigma_2^2(t_2)+\frac{1}{4k_{\max}}\ln\left[\frac{\sigma_2^2(t_2)}{\sigma_2^2(t_1)}\right]\right\},
\end{equation}
which, in the limit $k_{\max}\to\infty$, reduces to
\begin{equation}
    \expval{W_{\irr}^{(2)}}\sim k_{\max}\,\sigma_2^2(t_1)\left(1-\xi^2\right).
\end{equation}
As a consequence of the above results, the irreversible work for any bang-bang protocol involving at least one P or N vertex---i.e. any bang-bang protocol different from the one-bang at vertex O---diverges linearly with $k_{\max}$.

\newpage
\sectionmark{Bibliography}

\bibliography{Bib-tesis} 

\end{document}